\documentclass[10pt, b5paper, twoside, english]{memoir}

\usepackage{mathptmx}
\usepackage{anyfontsize}
\usepackage{t1enc}
\usepackage{graphicx}
\usepackage{epsfig}
\usepackage{amsmath}
\usepackage{amssymb}
\usepackage{amsthm}
\usepackage{slashed}
\usepackage{feynmp}
\usepackage{booktabs}
\usepackage{longtable}
\usepackage[figuresright]{rotating}
\usepackage{microtype} 
\usepackage{array}
\usepackage{tabularx}
\usepackage{multirow}
\usepackage{colortbl}
\usepackage{rotating}
\usepackage{inputenc}
\usepackage{enumitem}
\usepackage{bm}
\usepackage{cite}
\usepackage{url}
\usepackage[b5paper,top=2.5cm,bottom=2.5cm,left=2.5cm,right=2.0cm]{geometry}
\makeindex

\usepackage{color}

\definecolor{greenyellow}   {cmyk}{0.15, 0   , 0.69, 0   }
\definecolor{yellow}        {cmyk}{0   , 0   , 1   , 0   }
\definecolor{goldenrod}     {cmyk}{0   , 0.10, 0.84, 0   }
\definecolor{dandelion}     {cmyk}{0   , 0.29, 0.84, 0   }
\definecolor{apricot}       {cmyk}{0   , 0.32, 0.52, 0   }
\definecolor{peach}         {cmyk}{0   , 0.50, 0.70, 0   }
\definecolor{melon}         {cmyk}{0   , 0.46, 0.50, 0   }
\definecolor{yelloworange}  {cmyk}{0   , 0.42, 1   , 0   }
\definecolor{orange}        {cmyk}{0   , 0.61, 0.87, 0   }
\definecolor{burntorange}   {cmyk}{0   , 0.51, 1   , 0   }
\definecolor{bittersweet}   {cmyk}{0   , 0.75, 1   , 0.24}
\definecolor{redorange}     {cmyk}{0   , 0.77, 0.87, 0   }
\definecolor{mahogany}      {cmyk}{0   , 0.85, 0.87, 0.35}
\definecolor{maroon}        {cmyk}{0   , 0.87, 0.68, 0.32}
\definecolor{brickred}      {cmyk}{0   , 0.89, 0.94, 0.28}
\definecolor{red}           {cmyk}{0   , 1   , 1   , 0   }
\definecolor{orangered}     {cmyk}{0   , 1   , 0.50, 0   }
\definecolor{rubinered}     {cmyk}{0   , 1   , 0.13, 0   }
\definecolor{wildstrawberry}{cmyk}{0   , 0.96, 0.39, 0   }
\definecolor{salmon}        {cmyk}{0   , 0.53, 0.38, 0   }
\definecolor{carnationpink} {cmyk}{0   , 0.63, 0   , 0   }
\definecolor{magenta}       {cmyk}{0   , 1   , 0   , 0   }
\definecolor{violetred}     {cmyk}{0   , 0.81, 0   , 0   }
\definecolor{rhodamine}     {cmyk}{0   , 0.82, 0   , 0   }
\definecolor{mulberry}      {cmyk}{0.34, 0.90, 0   , 0.02}
\definecolor{redviolet}     {cmyk}{0.07, 0.90, 0   , 0.34}
\definecolor{fuchsia}       {cmyk}{0.47, 0.91, 0   , 0.08}
\definecolor{lavender}      {cmyk}{0   , 0.48, 0   , 0   }
\definecolor{thistle}       {cmyk}{0.12, 0.59, 0   , 0   }
\definecolor{orchid}        {cmyk}{0.32, 0.64, 0   , 0   }
\definecolor{darkorchid}    {cmyk}{0.40, 0.80, 0.20, 0   }
\definecolor{purple}        {cmyk}{0.45, 0.86, 0   , 0   }
\definecolor{plum}          {cmyk}{0.50, 1   , 0   , 0   }
\definecolor{violet}        {cmyk}{0.79, 0.88, 0   , 0   }
\definecolor{royalpurple}   {cmyk}{0.75, 0.90, 0   , 0   }
\definecolor{blueviolet}    {cmyk}{0.86, 0.91, 0   , 0.04}
\definecolor{periwinkle}    {cmyk}{0.57, 0.55, 0   , 0   }
\definecolor{cadetblue}     {cmyk}{0.62, 0.57, 0.23, 0   }
\definecolor{cornflowerblue}{cmyk}{0.65, 0.13, 0   , 0   }
\definecolor{midnightblue}  {cmyk}{0.98, 0.13, 0   , 0.43}
\definecolor{navyblue}      {cmyk}{0.94, 0.54, 0   , 0   }
\definecolor{royalblue}     {cmyk}{1   , 0.50, 0   , 0   }
\definecolor{blue}          {cmyk}{1   , 1   , 0   , 0   }
\definecolor{cerulean}      {cmyk}{0.94, 0.11, 0   , 0   }
\definecolor{cyan}          {cmyk}{1   , 0   , 0   , 0   }
\definecolor{processblue}   {cmyk}{0.96, 0   , 0   , 0   }
\definecolor{skyblue}       {cmyk}{0.62, 0   , 0.12, 0   }
\definecolor{turquoise}     {cmyk}{0.85, 0   , 0.20, 0   }
\definecolor{tealblue}      {cmyk}{0.86, 0   , 0.34, 0.02}
\definecolor{aquamarine}    {cmyk}{0.82, 0   , 0.30, 0   }
\definecolor{bluegreen}     {cmyk}{0.85, 0   , 0.33, 0   }
\definecolor{emerald}       {cmyk}{1   , 0   , 0.50, 0   }
\definecolor{junglegreen}   {cmyk}{0.99, 0   , 0.52, 0   }
\definecolor{seagreen}      {cmyk}{0.69, 0   , 0.50, 0   }
\definecolor{green}         {cmyk}{1   , 0   , 1   , 0   }
\definecolor{forestgreen}   {cmyk}{0.91, 0   , 0.88, 0.12}
\definecolor{pinegreen}     {cmyk}{0.92, 0   , 0.59, 0.25}
\definecolor{limegreen}     {cmyk}{0.50, 0   , 1   , 0   }
\definecolor{yellowgreen}   {cmyk}{0.44, 0   , 0.74, 0   }
\definecolor{springgreen}   {cmyk}{0.26, 0   , 0.76, 0   }
\definecolor{olivegreen}    {cmyk}{0.64, 0   , 0.95, 0.40}
\definecolor{rawsienna}     {cmyk}{0   , 0.72, 1   , 0.45}
\definecolor{sepia}         {cmyk}{0   , 0.83, 1   , 0.70}
\definecolor{brown}         {cmyk}{0   , 0.81, 1   , 0.60}
\definecolor{tan}           {cmyk}{0.14, 0.42, 0.56, 0   }
\definecolor{gray}          {cmyk}{0   , 0   , 0   , 0.50}
\definecolor{black}         {cmyk}{0   , 0   , 0   , 1   }
\definecolor{white}         {cmyk}{0   , 0   , 0   , 0   } 

\usepackage{hyperref}
\usepackage{memhfixc}

\newcommand{\mycaption}[2][\@empty]{
\captionnamefont{\small\bfseries\sffamily}
\captiontitlefont{\small}
\ifx \@empty#1 \caption{#2}\else \caption[#2]{#2}\fi}

\usepackage{calc}
\newif\ifNoChapNumber
\makeatletter
\makechapterstyle{VZ34}
{

}
\chapterstyle{VZ34}

\makeevenhead{headings}{\thepage}{}{\small\slshape\leftmark}
\makeoddhead{headings}{\small\slshape\rightmark}{}{\thepage}

\settocdepth{subsection}
\setsecnumdepth{subsection}
\maxsecnumdepth{subsection}
\settocdepth{subsection}
\maxtocdepth{subsection}

\newcolumntype{C}{>{$}c<{$}}

\pdfoutput=1

\begin{document}
\frontmatter
\pagestyle{empty}

\begin{center} 
\begin{figure}[ht!]
\centering
\includegraphics[height=2.8cm]{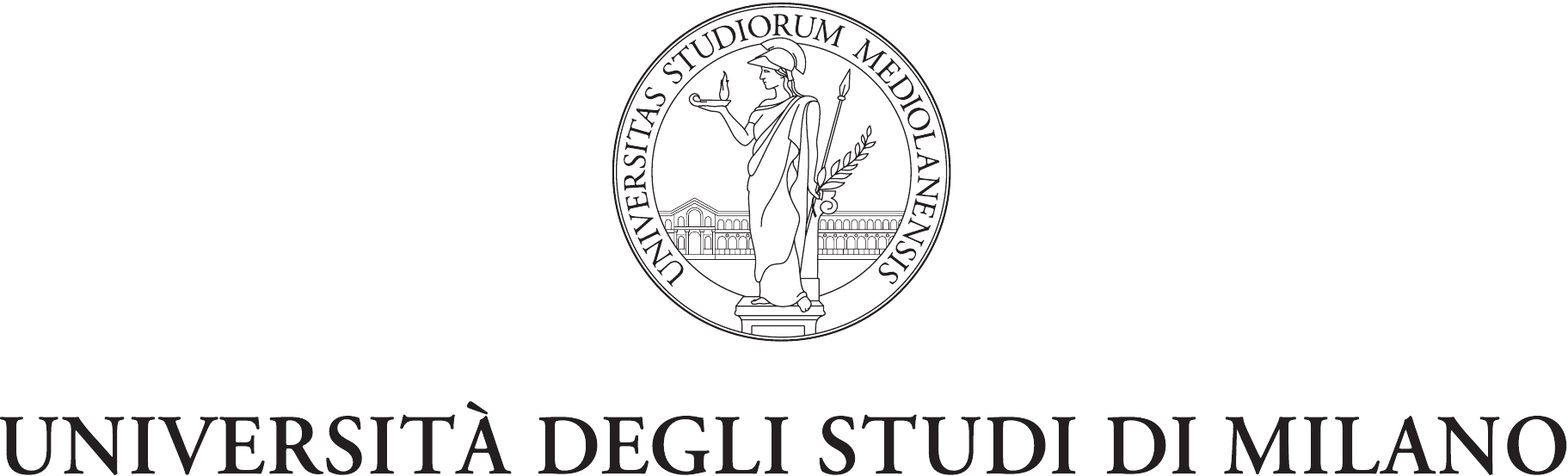}
\end{figure}
Scuola di Dottorato in Fisica, Astrofisica e Fisica Applicata

\medskip
Dipartimento di Fisica  

\medskip
\medskip
\medskip
Corso di Dottorato in Fisica, Astrofisica e Fisica Applicata

\medskip
Ciclo XXVI
\bigskip\vfill
\bigskip
  \bigskip
  {\vfil
    \begin{Huge}
    \begin{bfseries}
      {\renewcommand{\\}{\cr}
     \halign{\hbox to\textwidth{\strut\hfil#\hfil}\cr
     Unbiased spin-dependent\\ Parton Distribution Functions\cr}}
     \end{bfseries}
    \end{Huge}}
  \bigskip
  \bigskip
  \bigskip
\vfill
Settore Scientifico Disciplinare FIS/02

  \bigskip
  \bigskip
  \bigskip
  \bigskip
  \bigskip
  \bigskip
  \bigskip
  \bigskip
  \bigskip 
  \bigskip 
\begin{tabular*}{\textwidth}{@{}l@{\extracolsep{\fill}}l@{}}
    Supervisore: 
     \vtop{\halign{\strut\hfil#\hfil\cr
       Professor Stefano \uppercase{FORTE}\cr}}&
\vspace{0.5cm}\\
    Co-Supervisore: 
     \vtop{\halign{\strut\hfil#\hfil\cr
       Dottor Juan \uppercase{ROJO}\cr}}&
\vspace{0.5cm}\\
    Coordinatore: 
     \vtop{\halign{\strut\hfil#\hfil\cr
       Professor Marco \uppercase{BERSANELLI}\cr}}& \vspace{0.8cm} \\
&Tesi di Dottorato di:\vspace{0.2cm}\\
     &%
     \vtop{\halign{\strut\hfil#\hfil\cr
       Emanuele~R.~\uppercase{NOCERA}\cr}} 
\end{tabular*}
\end{center}
\begin{center}
\vspace{2cm}
    Anno Accademico 2013\\
\bigskip
\bigskip
\end{center}
\clearpage

\pagestyle{empty}
\noindent{\textbf{Commission of the final examination:}}\vspace{0.4cm}

\noindent{External Referee:}

\noindent{Prof. Richard~D.~\uppercase{BALL}}\vspace{0.2cm}

\noindent{External Members:}

\noindent{Prof. Mauro~\uppercase{ANSELMINO}}

\noindent{Prof. Giovanni~\uppercase{RIDOLFI}}\vspace{0.6cm}

\noindent{\textbf{Final examination:}}\vspace{0.4cm}

\noindent{February 28, 2014}

\noindent{Universit\`{a} degli Studi di Milano, Dipartimento di Fisica, Milano, Italy}\vspace{0.6cm}

\noindent{\textbf{MIUR subjects:}} \vspace{0.2cm}

\noindent{FIS/02 - Fisica Teorica, Modelli e Metodi Matematici}\vspace{0.6cm}

\noindent{{\textbf{PACS:}}} \vspace{0.2cm}

\noindent{02.70.Uu, 07.05.Mh, 12.38.-t, 13.60.Hb, 13.88.+e} \vspace{0.6cm}

\noindent{\textbf{Keywords:}} \vspace{0.2cm}

\noindent{Spin, Parton Distribution Functions (PDF), Neural Networks, 
NNPDF, High-energy Physics}
\vspace{2cm}
\begin{figure}[h]
\epsfig{width=0.35\textwidth,figure=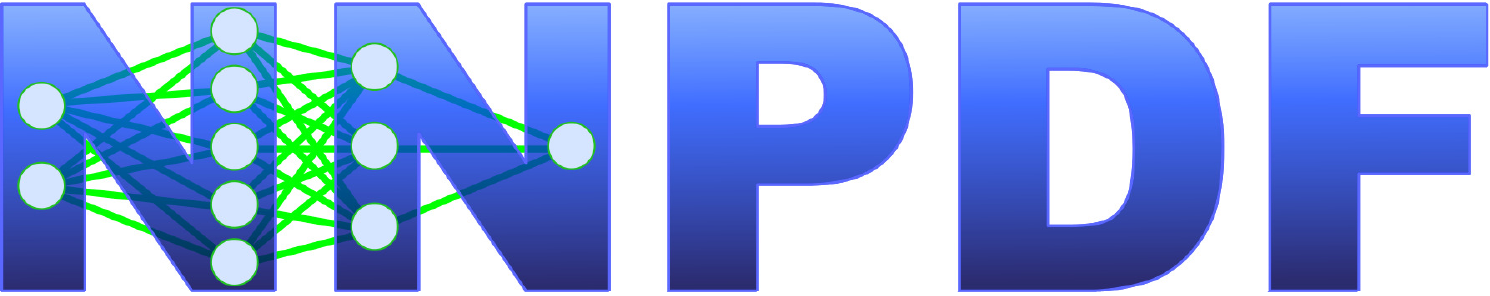}
\end{figure}
\vspace{0.5cm}

\vfill
%

\noindent{\textbf{Internal illustrations:}} \vspace{0.2cm}

\noindent{Emanuele~R.~Nocera, made with \uppercase{ROOT} v5.34/09 
and \texttt{feynMF} v1.5}

\noindent{Fig.~\ref{fig:vectors}: courtesy of Mauro Anselmino}

\noindent{Fig.~\ref{fig:hessemethod}: taken from Ref.~\cite{Campbell:2006wx}}

\noindent{Fig.~\ref{fig:scheme}: courtesy of Stefano Carrazza}
\vspace{0.2cm}

\noindent{\textbf{Template design:}}\vspace{0.2cm}

\noindent{Anna Lisa Varri, Ph.D.}

\noindent{Typeset by Emanuele~R.~Nocera using \TeX Live 2013 on Ubuntu 12.04 LTS}
\clearpage

\vspace{5cm}
{\hfill\textit{To my family}}
\cleardoublepage

\rmfamily
\normalfont

\chapter{Abstract}
\thispagestyle{plain}

We present the first unbiased determination of spin-dependent, or
polarized, Parton Distribution Functions (PDFs) of the proton. A
statistically sound representation of the corresponding uncertainties is
achieved by means of the NNPDF methodology: this was formerly
developed for unpolarized distributions and is now generalized to the
polarized here for the first time. The features of the procedure, based
on robust statistical tools (Monte Carlo sampling for error propagation,
neural networks for PDF parameterization, genetic algorithm for their
minimization, and possibly reweighting for including new data samples
without refitting), are illustrated in detail.
Different sets of polarized PDFs are obtained at next-to-leading order
accuracy in perturbative quantum chromodynamics, based on both 
fixed-target inclusive deeply-inelastic scattering data and the most recent
polarized collider data. A quantitative appraisal on the potential role of
future measurements at an Electron-Ion Collider is also presented. We
study the stability of our results upon the variation of several theoretical
and methodological assumptions and we present a detailed investigation
of the first moments of our polarized PDFs, compared to other recent
analyses.
We find that the uncertainty on the gluon distribution from available data
was substantially underestimated in previous determinations; in
particular, we emphasize that a large contribution to the gluon may arise
from the unmeasured small-x region, against the common belief that this
is actually rather small. We demonstrate that an Electron-Ion Collider
would provide evidence for a possible large gluon contribution to the
nucleon spin, though with a sizable residual uncertainty.

\cleardoublepage

\chapter{List of Publications}
{\textbf{Refereed publications}}
  \begin{itemize}
   \item R.~D.~Ball et al., 
   \textit{Polarized Parton Distributions at an Electron-Ion Collider}, 
   Phys.~Lett. \textbf{B728} (2014) 524
   \newline
   \href{http://arxiv.org/pdf/1310.0461.pdf}{[arXiv:1310.0461]}
   \href{http://www.sciencedirect.com/science/article/pii/S0370269313009982}{DOI: 10.1016/j.physletb.2013.12.023}
   \item R.~D.~Ball et al.,
   \textit{Unbiased determination of polarized parton distributions and their uncertainites}, 
   Nucl.~Phys.~\textbf{B874} (2013) 36
   \newline
   \href{http://arxiv.org/pdf/1303.7236.pdf}{[arXiv:1303.7236]}
   \href{http://www.sciencedirect.com/science/article/pii/S0550321313002691}{DOI: 10.1016/j.nuclphysb.2013.05.007}
   \item M.~Anselmino, M.~Boglione, U.~D'Alesio, S.~Melis, F.~Murgia, E.~R.~Nocera and A.~Prokudin, 
   \textit{General Helicity Formalism for Polarized Semi-Inclusive Deep Inelastic Scattering}, 
   Phys.~Rev.~\textbf{D83} (2011) 114019 
   \newline
   \href{http://arxiv.org/pdf/1101.1011.pdf}{[arXiv:1101.1011]}
   \href{http://prd.aps.org/abstract/PRD/v83/i11/e114019}{DOI: 10.1103/PhysRevD.83.114019}
  \end{itemize}
{\textbf{Publications in preparation}}
  \begin{itemize}
   \item R.~D.~Ball et al.,
   \textit{A first unbiased global extraction of polarized parton distributions}
   \end{itemize}
{\textbf{Publications in conference proceedings}}
  \begin{itemize}
   \item E.~R.~Nocera, \textit{Constraints on polarized parton distributions from open charm and W production data}, 
\textbf{PoS DIS2013} (2013) 211
\newline
\href{http://arxiv.org/pdf/1307.0146.pdf}{[arXiv:1307.0146]}
\item E.~R.~Nocera, \textit{Inclusion of $W^\pm$ single-spin asymmetry data in a polarized PDF determination via Bayesian reweighting}, 
Nuovo~Cim.~\textbf{C36} (2013) 143-147
\newline
\href{http://arxiv.org/pdf/1302.6409.pdf}{[arXiv:1302.6409]}
\href{http://www.sif.it/riviste/ncc/econtents/2013/036/05/article/23}{DOI: 10.1393/ncc/i2013-11592-4}
\item N.~P.~Hartland and E.~R.~Nocera, \textit{A Mathematica interface to NNPDFs}, 
Nucl. Phys. Proc. Suppl.~\textbf{224} (2013) 54-57 
\newline
\href{http://arxiv.org/pdf/1209.2585.pdf}{[arXiv:1209.2585]}
\href{http://www.sciencedirect.com/science/article/pii/S092056321200504X}{DOI: 10.1016/j.nuclphysbps.2012.11.013}
\item E.~R.~Nocera, S.~Forte, G.~Ridolfi and J.~Rojo, 
\textit{Unbiased Polarized Parton Distributions and their Uncertainties}, 
Proceedings of 20th International Workshop on Deep-Inelastic Scattering and Related Subjects (DIS 2012), p.937-942
\newline
\href{http://arxiv.org/pdf/1206.0201.pdf}{[arXiv:1206.0201]}
\href{https://indico.cern.ch/contributionDisplay.py?confId=153252&contribId=273}{DOI:10.3204/DESY-PROC-2012-02/273}
  \end{itemize}
\cleardoublepage

\chapter{Acknowledgements}
\pagestyle{headings}

I would like to thank all those people who have supported me in the course
of my work as a graduate student towards my Ph.D., 
particularly during the completion of
this Thesis. I apologize in advance for missing some of them. 
\vspace{0.1cm}

\noindent First of all, I am grateful to my 
supervisor, prof.~Stefano Forte, who set an example to me of
a scientist and a teacher.
In particular, I acknowledge his wide expertise in physics, 
his patient willingness, his inexhaustible enthusiasm and his 
illuminating ideas, from which I
had the opportunity to benefit (and learn) almost every day 
in the last three years.
\vspace{0.1cm}

\noindent Second, I am in debt to dr.~Juan Rojo, my co-supervisor, 
for having introduced me to the NNPDF methodology, for continuous
assistance with issues about code writing and for encouragement
in pursuing my research with dedication. I owe my gratitude to him also
for the opportunity he will give me to spend a couple of months in Oxford
soon. To both Stefano and Juan I give my thanks for their training 
not only in undertaking scientific research, but also in presenting results: 
in particular, I very much appreciated (and benefitted from) their advices
on talks and proceedings prepared for conferences in which I took part 
as a speaker.
\vspace{0.1cm}

\noindent Besides, I would like to thank prof.~Richard D. Ball, who accepted to 
be the referee of this Thesis, and had carefully reviewed the manuscript:
it has much improved thanks to his corrections and suggestions.
I thank as well the other two external members
of the final examination committee,
prof.~Mauro Anselmino and prof.~Giovanni Ridolfi.
\vspace{0.1cm}

\noindent Additional thanks are due to: the Ph.D. school board, for 
extra financial support provided for housing costs during my stay
in Milan; the \textit{Laboratorio di Calcolo \& Multimedia} (LCM) staff,
for the computational resources provided
to undertake the numerical analyses presented in this Thesis; 
dr.~Francesco Caravaglios, who has allowed for 
sharing with me not only his office, but also his personal ideas on physics, 
mathematics, metaphysics, biology, chemistry: though looking weird
at first, they finally reveal to me his genius.
\vspace{0.1cm}

\noindent I thank all those people I met during the last three years who 
also taught me a lot with their experience and knowledge. 
I would like to mention the members of the 
Department of Physics at the University of Milan, particularly those of 
the Theoretical Division who participated in the weekly journal club:
Daniele Bettinelli, Giuseppe Bozzi, Giancarlo Ferrera, Alessandro Vicini.
I also thank Mario Raciti for giving me the opportunity to work as a teaching
assistant to his course in General Physics for bachelor students in
\textit{Comunicazione Digitale}.
\vspace{0.1cm}

\noindent I aknowledge several physicists I met 
at conferences and workshops, who raised me in the spin physics community:
in particular Alessandro Bacchetta, Elena Boglione, Aurore 
Courtoy, Isabella Garzia, Francesca Giordano, Delia Hasch, 
Stefano Melis, Barbara Pasquini, Alexei Prokudin, Marco Radici, 
Ignazio Scimemi. 
\vspace{0.1cm}

\noindent I very much appreciated the friendly and cheerful environment in the
Milan Ph.D. school, thanks to all my colleagues fellow students. 
I just mention Alberto, Alice, Elena, Elisa, Rosa, Sofia.
Special thanks are deserved by Stefano Carrazza, who 
has rapidly become for me an example of efficiency and hard work
and a friend. 
I really enjoyed our discussions not only about physics and NNPDF code,
but also about motorbikes and trekking. I will never forget our 
journey from Milan to Marseille to attend DIS2013 nor our
excursion to Montenvers, \textit{Grand Balcon Nord} and 
\textit{Plan de l'Aiguille}. 
\vspace{0.1cm}

\noindent During my stay in Milan I was housed in \textit{Centro Giovanile Pavoniano}:
I would like to thank the directorate, in particular Fr.~Giorgio, 
for his kind hospitality, as well as all other guests who contributed to a 
friendly and enjoyable atmosphere.
\vspace{0.1cm}

\noindent Finally, I thank my family, in particular my parents, for 
their support: even though they stopped 
understanding what I am doing long ago, they have given me
the chance to look at the world with curiosity and love. 
\vspace{0.5cm}

\begin{flushright}
Emanuele~R.~Nocera
\end{flushright}
\cleardoublepage


\pagestyle{headings}
\tableofcontents*

\mainmatter
\chapter{Introduction}
\label{sec:intro}

The investigation of the internal structure of nucleons is an 
old and intriguing problem which dates back to almost fifty years ago.
For the past few decades, physicists have been able to describe with 
increasing details the fundamental particles that constitute protons 
and neutrons, which actually make up all nuclei and hence most of the
visible matter in the Universe.
This understanding is encapsulated in the Standard Model, 
supplemented with pertubartive Quantum Chromodynamics (QCD), 
the field theory which currently describes the strong interaction
between the nucleon's fundamental constituents, quarks and gluons.
It is a remarkable property of QCD, known as \textit{confinement}, 
that these are not seen in isolation, 
but only bound to singlet states of the their respective
strong \textit{color} charge.

Protons and neutrons are spin one-half bound states. 
Spin is one of the most 
fundamental concepts in physics, deeply rooted in Poincar\'{e} invariance
and hence in the structure of space-time itself.
The elementary constituents of the nucleon carry spin, 
quarks are spin one-half particles and gluons are spin-one particles.
It is worth recalling that the discovery of the fact that the proton 
has structure - and hence really the birth of strong interaction physics - 
was due to spin, through the measurement of a very unexpected
\textit{anomalous} magnetic moment of the proton by O. Stern and 
collaborators in 1933~\cite{Stern:1933aa}. 
After decades of ever more detailed studies of nucleon structure, 
the understanding of the observed spin of the nucleon in terms of 
their constituents is a major challenge, far from
being succesfully achieved.

\section{Historical overview on the nucleon structure}
\label{sec:histooverview}

The current picture of the nucleon structure is the result of more than
half a century of theoretical and experimental efforts from physicists
around the world. Even though a detailed historical overview is beyond the
scope of this introduction, we find it useful to summarize the main steps
in the building of our knowledge on the proton structure, with
some emphasis on its spin.

Quarks were originally introduced in 1963 by Gell-Mann, Ne'eman and Zweig,
simply based on symmetry considerations~\cite{Gell-Mann:1962aa,Zweig:1962aa,
GellMann:1964nj,Dothan:1965aa}, 
in an attempt to bring order into the large array of strongly-interacting 
particles observed in experiment.
In a few words, they recognised that the known hadrons could be associated 
to some representations of the special unitary $SU(3)$ group.
This led to the concept of quarks as the building blocks of hadrons. 
Mesons were expected to be quark-antiquark bound states,
while baryons were interpreted as bound states of three quarks. 
In Nature there are no indications of the existence of other 
multiquark states: in order to explain this fundamental 
evidence and 
to satisfy the Pauli exclusion principle for baryons, such as 
the $\Delta^{++}$ or the $\Omega^-$ which are made up of three quarks 
of the same flavor, the spin-one-half quarks had to carry a new quantum
number~\cite{Greenberg:1964pe}, later termed \textit{colour}. 
The modern version of this \textit{constituent} quark model 
still successfully describes 
most of the qualitative features of the baryon spectroscopy.

A modern realization of Rutherford's experiment has shown
us that quarks are real. This experiment is the deeply-inelastic scattering 
(DIS) of electrons (and, later, other leptons, including
positrons, muons and neutrinos) off the nucleon, a program that was 
started in the late 1960's at SLAC~\cite{Bloom:1969kc} (for a review see also
Ref.~\cite{Friedman:1972sy}). 
A high-energy lepton interacts with the nucleon, via exchange of a highly 
virtual gauge boson. For a virtuality
of $Q^2 > 1$ GeV$^2$, distances shorter than $0.2$ fm are probed in the proton. 
The early DIS results led to an interpretation 
as elastic scattering of the
lepton off pointlike, spin-one-half, constituents of the 
nucleon~\cite{Bjorken:1968dy,Feynman:1969ej,Bjorken:1969ja,Callan:1969uq},
called \textit{partons}.
At first, this was understood in the so-called \textit{parton model}:
in this model, the nucleon is observed in the so-called 
\textit{infinite momentum frame}, a Lorentz frame in which it is
moving with large four-momentum: partons are assumed to move 
collinearly to the parent hadron, hence their transverse momenta 
and masses can be neglected.  
Lepton-nucleon scattering is then described in the \textit{impulse 
approximation}, \textit{i.e.} partons are treated
as free particles and all partons' self-interactions are neglected.
In the impulse approximation, lepton-nucleon scattering is simply
the incoherent sum of lepton interactions with the individual partons
in the nucleon, which are carrying a fraction $x$ of its four-momentum. 
These interactions can be computed in perturbation theory, and have to
be weighted with the probability that the nucleon contains a parton with
the proper value of $x$. This probability, denoted as $f_{q/p}(x)$,
encodes the momentum density of any parton species $q$, with longitudinal
fraction $x$, in a nucleon $p$, and is called Parton Distribution Function
(PDF). This cannot be computed using 
perturbative theory, since it depends on the non-perturbative process that 
determines the structure of the nucleon; 
hence, it has to be determined from the experiment. 

Partons carrying fractional electric charge 
were subsequently identified with the quarks. The
existence of gluons was proved indirectly from a missing ($\sim 50\%$) 
contribution~\cite{deGroot:1979xx,deGroot:1978hr} to the proton
momentum not accounted for by the quarks. 
Later on, direct evidence for gluons was found in
three-jet production in electron-positron 
annihilation~\cite{Brandelik:1979bd,Berger:1979cj,Bartel:1979ut}. 
From the observed angular distributions
of the jets it became clear that gluons have spin 
one~\cite{Brandelik:1980vs,Ellis:1978wp}.

The successful parton interpretation of DIS assumed that partons 
are almost free (\textit{i.e.}, non-interacting) 
on the short time scales set by the high virtuality of the exchanged photon. 
This implied that the underlying theory of the strong interactions must 
actually be relatively weak on short time or, equivalently, distance 
scales~\cite{Callan:1973pu}. In a groundbreaking development, Gross,
Wilczek and Politzer showed in 1973 that the non-abelian theory  
of quarks and gluons, QCD,
which had just been developed a few months 
earlier~\cite{Fritzsch:1973pi,Gross:1973ju,Weinberg:1973un}, 
possessed this remarkable feature of
\textit{asymptotic freedom}~\cite{Gross:1973id,Politzer:1973fx}, 
a discovery for which they were awarded the 2004 Nobel Prize for
Physics. The interactions of partons at short distances, 
while weak in QCD, were then predicted to
lead to visible effects in the experimentally measured DIS structure 
functions known as 
\textit{scaling violations}~\cite{Georgi:1951sr,Georgi:1976ve}. 
These essentially describe the response of the partonic structure of the proton
to the resolving power of the virtual photon, set by its virtuality $Q^2$. 
The greatest triumph of QCD is arguably the prediction of scaling violations, 
which have been observed experimentally and
verified with great precision. Deeply-inelastic scattering thus paved the 
way for QCD.

Over the following two decades or so, studies of nucleon structure 
became ever more detailed and precise. This was partly due to increased 
luminosities and energies of lepton machines, eventually culminating in the 
HERA electron-proton collider~\cite{DESY-HERA-81-10}. 
Also, hadron colliders entered the scene. 
It was realized, again thanks to asymptotic freedom
and factorization, which follows from it, 
that the partonic structure of the nucleon seen in
DIS is universal, in the sense that a variety 
of sufficiently inclusive hadron collider processes,
characterized by a large scale, admit a factorized 
description~\cite{Drell:1970wh,Berman:1970bf,Berman:1971xz,Jaffe:1972vy,Feynman:1978dt,Ellis:1978sf,Ellis:1978ty,Amati:1978wx,Amati:1978by,Collins:1983ju,Collins:1985ue,Collins:1992xw}. 
This offered the possibility of learning about other aspects of nucleon 
structure (and hence, QCD), for instance about its gluon content which 
is not primarily accessed in DIS. 
Being known with more precision, nucleon structure also 
allowed for new physics studies at hadron colliders,
the outstanding example perhaps 
being the discovery of the $W^\pm$ and $Z^0$ bosons at
CERN's $Sp\bar{p}S$ 
collider~\cite{Arnison:1983rp,Arnison:1983mk,Banner:1983jy}. 
The Tevatron and the Large Hadron Collider (LHC) 
are the most recent continuations of 
this line of research, which has culminated in the discovery of the 
Higgs boson~\cite{Aad:2012tfa,Chatrchyan:2012ufa}, 
announced by ATLAS and CMS collaborations on the $4^{\mathrm{th}}$ July 2012.

Concerning spin physics, a milestone in the study of the nucleon was 
the advent of polarized electron beams
in the early seventies~\cite{prescott:nn}. 
This later on allowed for DIS measurements with polarized
lepton beams and nucleon targets~\cite{Alguard:1976bm}, 
and offered the possibility of studying whether quarks and antiquarks 
show on average preferred spin directions inside a polarized nucleon. 
The program of polarized DIS has been continuing ever since and 
it is now a successful branch of particle physics. 
Its most important result is the finding that
quark and antiquark spins provide an anomalously small
- only about $20\%-30\%$ - amount
of the proton spin~\cite{Ashman:1987hv,Ashman:1989ig}, firstly observed 
by the EMC experiment in the late 1980's.
This finding, which opened a \textit{spin crisis}
in the understanding of the nucleon structure~\cite{Leader:1988vd},
has raised the interest of physicists in clarifying the 
potential role played by new candidates to the nucleon's spin,
like gluons' polarizations and partons' orbital angular momenta.
In parallel, there also was a very important line of research on
polarization phenomena in hadron-hadron reactions in fixed-target kinematics. 
In particular, unexpectedly large single-transverse spin asymmetries were 
seen~\cite{Adams:1991rw,Krueger:1998hz,Bunce:1976yb,Heller:1977mv,Gourlay:1986mf}.

In the last decade, the advent of 
the Relativistic Heavy Ion Collider (RHIC), the first machine to collide
polarized proton beams, started to probe the proton spin in new profound 
ways~\cite{Bunce:2000uv}, complementary, but independent, to polarized DIS. 
In particular, more knowledge on the polarization of gluons in 
the proton and details of the flavor structure of the polarized quarks and 
antiquarks has been recently achieved, as we will discuss in detail in 
this Thesis. However, despite a flurry of experimental and theoretical activity,
a complete and satisfactory understanding of the so-called 
\textit{spin puzzle} is still lacking.

\section{Compelling questions in spin physics}
\label{sec:spinquestions}

The information on the proton spin structure is 
encoded in spin-dependent, or polarized,
Parton Distribution Functions (PDFs) of quarks, antiquarks and gluons
\begin{equation}
\Delta f_{q/p}(x,Q^2)=f_{q/p}^\uparrow(x,Q^2) - f_{q/p}^\downarrow(x,Q^2)
\,\mbox{,}
\label{eq:PDFpolintro}
\end{equation}
which are the momentum densities of partons $q$ with helicity along 
($\uparrow$) or opposite ($\downarrow$) the polarization direction 
of the parent nucleon $p$. 
The $Q^2$ dependence of the parton distributions,
known as  $Q^2$ evolution~\cite{Altarelli:1977zs},
is quantitatively 
predictable in perturbative QCD, thanks to asymptotic freedom.
Physically, it may be thought of the consequence of the fact that
partons are observed with higher resolution when they are probed at 
higher scales;
hence it is more likely that a struck
quark has radiated one or more gluons so that it is effectively resolved 
into several partons, each with lower momentum fraction. 
Similarly, a struck quark may have originated from a gluon
splitting into a quark-antiquark pair.

Polarized inclusive, neutral-current, DIS allows one to only access
the flavor combinations 
$\Delta q^+\equiv \Delta q + \Delta\bar{q}$, 
and the gluon polarization, though the latter is 
mostly determined indirectly by scaling violations.
Of particular interest is the singlet quark antiquark combination
$\Delta\Sigma=\sum_{q=u,d,s}\Delta q^+$, since its integral,
known as the \textit{singlet axial charge}, yields the average of all
quark and antiquark contributions to the proton spin:
\begin{equation}
\langle S_q\rangle\sim\frac{1}{2}\int_0^1dx\Delta\Sigma(x,Q^2)
\,\mbox{.}
\label{eq:singmomentum}
\end{equation}

The anomalously small value observed experimentally for this quantity,
from almost three decades of DIS measurement after the EMC result,
strenghten the common belief that only about a quarter of the proton
spin is carried by quarks and antiquarks. 
The EMC result was followed by an intense scrutiny of the basis of the 
corresponding theoretical framework, which led to the 
realization~\cite{Altarelli:1988nr,Altarelli:1990jp} that
the perturbative behavior of polarized PDFs deviates from parton model
expectations, according to which gluons decouple at large energy scale.
The almost vanishing value measured by EMC for the singlet axial
charge can be explained as a cancellation 
between a reasonably large quark spin contribution, 
\textit{e.g.} $\Delta\Sigma\simeq 0.6 - 0.7$, as expected intuitively, 
and an anomalous gluon contribution, altering Eq.~(\ref{eq:singmomentum}).
A large value of the gluon contribution to the proton spin is required 
to achieve such a cancellation, and QCD predicts that
this contribution grows with the energy scale.
Despite some experimental evidence has suggested that the 
gluon polarization in the nucleon may be rather small, we 
emphasize that it is instead still largely uncertain, as 
we will carefully demonstrate in this Thesis.
Other candidates for carrying the nucleon spin can come from
quark and gluon orbital angular 
momenta~\cite{Jaffe:1989jz,Ji:1996ek,Bakker:2004ib} 
(for a recent discussion on the spin 
decomposition see also Ref.~\cite{Leader:2013jra}).

In any case, the results from polarized inclusive DIS clearly 
call for further investigations:
we summarize in the following some of the outstanding
questions to be aswered in spin physics.

\begin{list}{}{\leftmargin=0pt}

\item {\textbf{With which accuracy do we know spin-dependent parton distributions?}}
The assessment of the singlet and gluon contributions to the proton spin
requires in turn a determination
of polarized parton distributions from available experimental data.
In the last decade, several such determinations have been performed
at next-to-leading order (NLO) in QCD~\cite{Gehrmann:1995ag,Altarelli:1996nm,
Altarelli:1998nb,Gluck:2000dy,Hirai:2003pm,deFlorian:2000bm,Leader:2001kh,
Bluemlein:2002be,Hirai:2008aj,deFlorian:2008mr,deFlorian:2009vb,Blumlein:2010rn,
Leader:2010rb,Jimenez-Delgado:2013boa,Khorramian:2010qa,Arbabifar:2013tma},
which is the current state-of-the-art accuracy for polarized fits,
mostly based on DIS.
Some of them also include a significant amount of data other than DIS, 
namely from semi-inclusive DIS (SIDIS) with identified hadrons in final 
states~\cite{Leader:2010rb,deFlorian:2009vb,Arbabifar:2013tma} 
or from polarized proton-proton collisions~\cite{deFlorian:2009vb}.

However, we notice that they are all based on the \textit{standard}
Hessian methodology for PDF fitting and uncertainty estimation.
This approach is known~\cite{Forte:2010dt} 
to potentially lead to an underestimation 
of PDF uncertainties, due to the limitations in the 
linear propagation of errors and, more importantly, to PDF parametrization
in terms of fixed functional forms.
These issues are especially delicate when the experimental information 
is scarce, like in the case of polarized data.
In particular, in this Thesis we will clearly demonstrate that
a more flexible PDF parametrization is better suited
to analyse polarized experimental data without prejudice.
This will lead to larger, but more faithful, estimates of
PDF uncertainties than those obtained in the other available analyses.
At least, one should conclude that
our knowledge of parton's contribution to the nucleon spin is
much more uncertain than commonly believed, unless one is willing to make
some \textit{a priori} assumptions on their behavior 
in the unmeasured kinematic regions. The two following questions
address in much detail some issues related to quarks and gluons separately.

\item {\textbf{How do gluons contribute to the proton spin?}}
The interest in an accurate determination of the gluon polarization
$\Delta g(x, Q^2)$ is of particular interest for both 
phenomenological and theoretical reasons.

On the phenomenological side, inclusive DIS allows for an indirect
determination of the gluon distribution, through scaling violations.
Since experimental data have a rather limited $Q^2$ lever arm,
it follows that $\Delta g(x, Q^2)$ is only weakly constrained. 
Processes other than inclusive DIS, which
receive leading contributions from gluon initiated 
subprocesses, are better suited to provide direct information on the
gluon distribution. In particular, these include open-charm
production in fixed-target experiments and jet or semi-inclusive production
in proton-proton collisions. However, the kinematic coverage of these 
data is limited: hence the integral of the gluon distribution
can receive large contributions
from the unmeasured region, in particular from the small-$x$ region.

On the theoretical side, it is a remarkable feature of QCD that 
the gluon contribution to the nucleon spin
may well be significant even at large momentum scales. 
The reason is that the integral of 
$\Delta g(x, Q^2)$ evolves as $1/\alpha_s(Q^2)$~\cite{Altarelli:1988nr}, 
that is, it rises
logarithmically with $Q$. This peculiar evolution pattern is a very deep 
prediction of QCD, related to the so-called axial anomaly. 
It has inspired ideas that a reason for the smallness of the quark spin
contribution should be sought in a \textit{shielding} of the quark 
spins due to a particular perturbative
part of the DIS process $\gamma^* g \to q\bar{q}$~\cite{Altarelli:1988nr}.
The associated contributions arise only at order $\alpha_s(Q^2)$;
however, the peculiar evolution of the first moment of the polarized gluon 
distribution would compensate this suppression. To be of any
practical relevance, a large positive gluon 
spin contribution, $\langle \Delta g\rangle > 1$,
would be required
even at low \textit{hadronic} scales of a GeV or so. 
A very large polarization of the confining fields
inside a nucleon, even though suggested by some nucleon 
models~\cite{Jaffe:1995an,Mankiewicz:1996zr,Barone:1998dx,Lee:2000mv}, 
would be a very puzzling phenomenon and would once again challenge our 
picture of the nucleon. 

\item {\textbf{What are the patterns of up, down, and strange quark 
and antiquark polarizations?}}
Inclusive DIS provides information only on the total flavor combinations 
$\Delta q^+\equiv\Delta q + \Delta\bar{q}$, $q=u,d,s$. 
Nevertheless, in order to understand
the proton helicity structure in detail, one needs to learn about 
the various quark and antiquark densities, 
$\Delta u$, $\Delta\bar{u}$, $\Delta d$, $\Delta\bar{d}$ and
$\Delta s$, $\Delta\bar{s}$ separately.
This also provides an important additional test
of the smallness of the quark spin contribution, and could 
reveal genuine flavor asymmetry $\Delta\bar{u}-\Delta\bar{d}$
in the proton sea, claimed by some models of nucleon 
structure~\cite{Dressler:1999zg,Cao:2003zm}. 
These predictions are often related to fundamental concepts such as the
Pauli principle: since the proton has two valence-$u$ quarks which primarily 
spin along with the proton spin direction, $u\bar{u}$ 
pairs in the sea will tend to have the $u$ quark polarized opposite to
the proton. Hence, if such pairs are in a spin singlet, one expects 
$\Delta\bar{u}> 0$ and, by the same
reasoning, $\Delta\bar{d}< 0$. 
Such questions become all the more exciting due to the fact that rather
large unpolarized asymmetries $\bar{u}-\bar{d}\neq 0$ have been observed in 
DIS and Drell-Yan 
measurements~\cite{Amaudruz:1991at,Baldit:1994jk,Hawker:1998ty}. 
Further fundamental questions concern the strange quark polarization. The
polarized DIS measurements point to a sizable negative polarization of 
strange quarks, in line with other observations of significant strange 
quark effects in nucleon structure. 

\item {\textbf{What orbital angular momenta do partons carry?}}
Quark and gluon orbital angular momenta are the other candidates for the 
carriers of the proton spin. Consequently, theoretical work focused also 
on these in the years after the \textit{spin crisis} was announced. 
A conceptual breakthrough was made in the mid 1990s when it was 
realized~\cite{Ji:1996ek} that
a particular class of \textit{off-forward} nucleon matrix elements, 
in which the nucleon has different momentum in the initial and final states, 
measure total parton angular momentum. Simply stated,
orbital angular momentum is $\overrightarrow{r}\times \overrightarrow{p}$, 
where the operator $r$ can be viewed in Quantum Mechanics as
a derivative with respect to momentum transfer.
Thus, in analogy with the measurement of the 
Pauli form factor, it takes a finite momentum transfer on the nucleon 
to access matrix elements with operators containing a factor $r$. 
It was also shown how these \textit{off-forward} distributions, 
referred to as generalized parton distribution functions (GPDs), 
may be experimentally determined from certain exclusive
processes in lepton-nucleon scattering, the prime example being 
Deeply-Virtual Compton Scattering (DVCS) 
$\gamma^* p\to\gamma p$~\cite{Ji:1996ek}. 
A major emphasis in current and future experimental activities
in lepton scattering is on the DVCS and related reactions. 

\item {\textbf{What is the role of transverse spin in QCD?}}
So far, we have only considered the helicity structure of the nucleon, 
that is, the partonic structure we find when we probe the nucleon when its
spin is aligned with its momentum. 
Experimental probes with transversely polarized nucleons could also be 
studied, both at fixed-target and collider facilities, 
and it has been known for a long time now that very interesting spin 
effects are associated with this in QCD. 
Partly, this is known from theoretical studies which revealed that besides
the helicity distributions discussed above, 
for transverse polarization 
there is a new set of parton densities, called 
transversity~\cite{Ralston:1979ys,Jaffe:1991kp}. 
They are defined analogously to Eq.~(\ref{eq:PDFpolintro}), 
but now for transversely polarized partons polarized along or opposite 
to the transversely polarized proton.
Furthermore, if we allow quarks to have an intrinsic Fermi motion in the 
nucleon, they can be interpreted in light of more fundamental 
objects, the so-called Transverse Momentum Dependent parton distribution
functions (TMDs)~\cite{Collins:1981uk}, 
in which the dependence on the intrinsic transverse 
momentum $k_\perp$ is made explicit.
We refer to~\cite{Varenna:2011}
for a comprehensive review on TMDs and the transverse spin structure of the 
proton. Here, we only mention that
the present knowledge of TMDs is comparable to that of PDFs in the early 
1970's and very little is known about transversity.
An intensive experimental campaign is ongoing to take
data in polarized SIDIS and to provide a better determination of these
distributions~\cite{Airapetian:2004zf,Alekseev:2007vi,Alekseev:2009ac,Alekseev:2010ub}.  

\end{list}

\section{Outline of the thesis}
\label{sec:outline}

This Thesis addresses the three first questions in the above list,
presenting a determination of spin-dependent parton distributions
for the proton. In particular, two sets are 
obtained, the first based on inclusive DIS data only, the second
also including the most recent data from polarized proton-proton collisions.
In comparison to other recent analyses,
our study is performed within the NNPDF methodology,
which makes use of robust statistical tools, including Monte Carlo 
sampling for error propagation and parametrization of PDFs in terms of neural 
networks. The methodology has been succesfully applied to the 
unpolarized case~\cite{DelDebbio:2004qj,DelDebbio:2007ee,
Ball:2008by,Ball:2009qv,Ball:2009mk,
Ball:2010de,Ball:2010gb,Ball:2011mu,Ball:2011uy,
Ball:2011eq, Ball:2011gg,Ball:2012cx,Ball:2013hta}
with the goal of providing a faithful representation of
the PDF underlying probability distribution.
This is particularly relevant with polarized data, which are 
rather scarce and, in general, affected by larger uncertainties than those
of their unpolarized counterparts. Unlike all other \textit{standard} fits,
our parton sets do not suffer
from the theoretical bias introduced either by a fixed functional form for
PDF parametrization or by quadratic approximation in the Hessian
propagation of errors.
For this reason, we consider our unbiased determination to be crucial 
for investigating to which extent the common belief
that about a quarter of the nucleon spin is carried by quarks 
and antiquarks, while the gluon contribution is even much smaller,
actually holds.

Our parton determinations are publicly released together with 
computational tools
to use them, including \texttt{FORTRAN}, \texttt{C++} and 
\textit{Mathematica} interfaces. Hence they could be used for any 
phenomenological study of hard scattering processes involving
polarized hadrons in initial states. We should notice that, in addition to 
the investigation of the nucleon spin structure, such studies 
have recently included probes
of different beyond-standard-model (BSM) scenarios~\cite{Bozzi:2004qq}
and possibly the determination of the Higgs boson spin in the diphoton decay channel,
by means of the linear polarization of gluons in an unpolarized 
proton~\cite{Boer:2013fca}. 

In this Thesis, we will not address the 
study of either TMDs or GPDs, but we notice that the methods ilustrated here
apply to the determination of any non-perturbative object from 
experimental data: hence they could be used to 
determine such new distributions in the future,
in so far as experimental data will reach more and more abundance and accuracy.
Also, we do not describe either the apparatus which 
had to be developed to carry out spin physics experiments
or the related technical challenges which had to be faced.
A complete survey on these aspects can be found in 
Refs.~\cite{Varenna:2011,Leader:2001ab,Aidala:2012mv}.

The outline of this Thesis is as follows.

\begin{list}{}{\leftmargin=0pt}

\item {\textbf{Chapter~\ref{sec:chap1}. Polarized Deeply-Inelastic Scattering.}}
We review the theoretical formalism for the description of DIS
with both polarized lepton beams and nuclear targets. In particular 
we derive the expressions for the differential cross-section of the process 
in terms of polarized structure functions. We will restrict our discussion to
the contribution arising from the exchange of a virtual photon
between the lepton and the nucleon.
This is indeed sufficient to describe 
currently available esperimental data, whose energy does not exceed a 
few hundreds of GeV and none of which come from neutrino beams:
hence we do not include either the (suppressed) contribution to 
neutral-current DIS mediated by a $Z^0$ 
boson or charged-current DIS mediated by a $W^\pm$ boson. 
Then, we present the parton model expectations for DIS spin 
asymmetries introducing helicity-dependent, or polarized, 
PDFs and we discuss how they are modified
in the framework of perturbative QCD.
We complete our theoretical overview on polarized DIS with a sketch
of spin sum rules, a summary of the relevant phenomenological relations
between structure functions and measured observables, and the formalism adopted
to take into account kinematic target mass corrections.

\item {\textbf{Chapter~\ref{sec:chap3}. Phenomenology of polarized Parton Distributions.}}
We review how a set of PDFs is usually determined
from a global fit to experimental data. First, we sketch
the general strategy for PDF determination and its main theoretical 
and methodological issues, focusing on those which are peculiar to the 
polarized case. Second, we summarize how some of these problems are
addressed within the NNPDF methodology,  
with the goal of providing a statistically sound determination of PDFs and their 
uncertainties. Finally, we provide an overview on available 
polarized PDF sets.

\item {\textbf{Chapter~\ref{sec:chap2}. Unbiased polarized PDFs from inclusive DIS.}}
We present the first determination of polarized PDFs based on the NNPDF
methodology, \texttt{NNPDFpol1.0}. This analysis includes all available 
data from inclusive, neutral-current, polarized DIS and aims at an unbiased 
extraction of total quark-antiquark and gluon distributions at NLO accuracy. 
We discuss how the statistical distribution of experimental data
is sampled with Monte Carlo generation
of pseudodata. We provide the details of the QCD analysis and discuss the PDF
parametrization in terms of neural networks; we describe
the minimization strategy and the peculiarities in the polarized 
case. We present the \texttt{NNPDFpol1.0} parton
set, illustrating its statistical features and its stability upon the
variation of several theoretical and methodological assumptions. 
We also compare our results
to other recent polarized PDF sets. Finally, we discuss phenomenological
implications for the spin content of the proton and the test of the Bjorken 
sum rule. The analysis presented in this Chapter has been published by
the NNPDF collaboration as a refereed paper~\cite{Ball:2013lla}.

\item {\textbf{Chapter~\ref{sec:chap4}. Polarized PDFs at an Electron-Ion Collider.}}
We investigate the potential impact of inclusive DIS data from a 
future Electron-Ion Collider (EIC) on the determination of polarized PDFs. 
After briefly motivating our study, we illustrate which EIC pseudodata sets 
we use in our analysis and how the fitting procedure needs to be optimized. 
Resulting PDFs are compared to \texttt{NNPDFpol1.0} throughout. 
Finally, we reassess their first moments and we give an estimate of
the charm contribution to the $g_1$ structure function of the proton at an EIC. 
The analysis presented in this Chapter has been published by
the NNPDF Collaboration as a refereed paper~\cite{Ball:2013tyh}.

\item {\textbf{Chapter~\ref{sec:chap5}. Global determination
of unbiased polarized PDFs.}}
We extend the analysis presented in Chap.~\ref{sec:chap2} 
in order to include in our parton 
set, on top of inclusive DIS data, also recent measurements of open-charm
production in fixed-target DIS, and of jet and $W$ production in 
polarized proton-proton collisions. Hence, we present the first global
determination of polarized PDFs based on the NNPDF methodology:
\texttt{NNPDFpol1.1}. After motivating our analysis, we review 
the theoretical description of the new processes and present the features
of the relative experimental data we include in our study.
We then turn to a detailed discussion of the way the \texttt{NNPDFpol1.1} parton set 
is obtained via Bayesian reweighting of prior PDF Monte Carlo ensembles, 
followed by unweighting. We also present its main features in
comparison to \texttt{NNPDFpol1.0}. Finally,  we discuss some phenomenological 
implications for the spin content of the proton, 
based on our new polarized parton set.
The analysis discussed in this Chapter has been presented in preliminary form in 
Refs.~\cite{Nocera:2013spa,Nocera:2013yia}.

\item {\textbf{Chapter~\ref{sec:conclusions}. Conclusions and outlook.}}
We will draw our conclusions, highlighting the main results
presented in this Thesis.
We also provide an outlook on future possible developments 
in the determination of polarized parton distributions within the 
NNPDF methodology.

\item {\textbf{Appendix~\ref{sec:appB}. Statistical estimators.}}
We collect the definitions of the statistical estimators
used in the NNPDF analyses presented in 
Chaps.~\ref{sec:chap2}-\ref{sec:chap4}-\ref{sec:chap5}.
Despite they were already described 
in Refs.~\cite{Forte:2002fg,DelDebbio:2004qj,Ball:2010de},
we find it useful to give them for completeness and ease of 
reference here.

\item {\textbf{Appendix~\ref{sec:appC}. 
A \textit{Mathematica} interface to NNPDF parton sets.}}
We present a package for handling 
both unpolarized and polarized NNPDF parton sets
within a \textit{Mathematica} notebook file.
This allows for performig PDF manipulations easily and quickly,
thanks to the powerful features of the \textit{Mathematica} software.
The package was tailored to the users who are not familiar with 
\texttt{FORTRAN} or \texttt{C++} programming codes, on which the standard
available PDF interface, LHAPDF~\cite{Whalley:2005nh,web:LHAPDF}, is based.
However, since our \textit{Mathematica} package includes all the features
available in the LHAPDF interface, any user can benefit from the interactive
usage of PDFs within \textit{Mathematica}.
The \textit{Mathematica} interface to NNPDF parton sets 
appeared as a contribution to conference proceedings
in Ref.~\cite{Hartland:2012ia}. 

\item {\textbf{Appendix~\ref{sec:appA}. The FONLL scheme for $g_1(x,Q^2)$ 
up to $\mathcal{O}(\alpha_s)$.}}
We collect the relevant explicit formulae for the practical
computation of the polarized DIS structure function 
of the proton, $g_1^p(x,Q^2)$, within the
FONLL approach~\cite{Forte:2010ta} up to $\mathcal{O}(\alpha_s)$. 
In particular, we will restrict to the heavy charm quark contribution $g_{1}^{p,c}$
to the polarized proton structure function $g_1$, 
which might be of interest for studies at an
Electron-Ion Collider in the future, as mentioned in Chap.~\ref{sec:chap4}.

\end{list}

\chapter{Polarized Deeply-Inelastic Scattering}
\label{sec:chap1}

This Chapter is devoted to a detailed discussion of
Deeply-Inelastic Scattering (DIS) with both polarized
lepton beams and nuclear targets.
In particular, we will focus on neutral-current DIS,
limited to the kinematic regime in which the exchange
of a virtual photon between the lepton and the nucleon
provides the leading contribution to the process.
In Sec.~\ref{sec:DISth}, we rederive the expression 
for the differential cross-section of polarized DIS  
in terms of polarized structure functions.
We present the naive parton model expectations for spin asymmetries 
in Sec.~\ref{sec:parton-model} and we discuss how they
should be modified in the framework of QCD in
Sec.~\ref{sec:QCDevol}. We complete our theoretical
overview on polarized DIS with a sketch  
of spin sum rules in Sec.~\ref{sec:sumrule}.
Finally, we summarize the relevant phenomenological relations
between structure functions and measured observables in 
Sec.~\ref{sec:phenopol}, and the formalism adopted to 
take into account kinematic target mass corrections
in Sec.~\ref{sec:TMC}.

\section{General formalism}
\label{sec:DISth}

Let us consider the inclusive, neutral-current, 
inelastic scattering of a polarized 
lepton (electron or muon) beam off a polarized nucleon target,
\begin{equation}
l(\ell) + N(P) \to l^\prime(\ell^\prime) + X(P_X)
\,\mbox{,}
\label{eq:DIS}
\end{equation}
where the four-momenta of the incoming (outgoing) lepton $l$ ($l^\prime$),
the nucleon target $N$ and the undetected final hadronic system $X$
are labelled as $\ell$ ($\ell^\prime$), $P$, and $P_X$ respectively.  
If the momentum transfer involved in the reaction is much smaller
than the $Z^0$ boson mass, as it is customary at 
polarized DIS facilities, the only sizable contribution to the 
process is given by the exchange of a virtual photon, see
Fig.~\ref{fig:DISfeyn}.
\begin{figure}[t]
\centering
\begin{fmffile}{simple}
\fmfframe(0,10)(0,20)
{
\begin{fmfgraph*}(200,125)
\fmfleft{li,P1}
\fmfright{lo,X1}
\fmflabel{$l$}{P1}
\fmflabel{$N$}{li}
\fmflabel{$X$}{lo}
\fmflabel{$l^\prime$}{X1}
\fmf{fermion,tension=2,label=$P$,l.side=left}{li,v2}
\fmf{fermion,tension=2,label=$P_X$,l.side=left}{v2,lo}
\fmf{photon,label=$\gamma^{*}(q)$,l.side=right}{v1,v2}
\fmf{fermion,label=$\ell$,l.side=right}{P1,v1}
\fmf{fermion,label=$\ell^\prime$,l.side=right}{v1,X1}
\fmfblob{.15w}{v2}
\fmffreeze
\fmfi{plain,tension=2}{vpath(__li,__v2) shifted (thick*(+0.0,+1.7))}
\fmfi{plain,tension=2}{vpath(__li,__v2) shifted (thick*(+0.5,-1.5))}
\fmfi{plain,tension=2}{vpath(__v2,__lo) shifted (thick*(+0.0,+1.7))}
\fmfi{plain,tension=2}{vpath(__v2,__lo) shifted (thick*(-0.5,-1.5))}
\end{fmfgraph*}
}
\end{fmffile}
\mycaption{Virtual-photon-exchange contribution to neutral-current DIS.}
\label{fig:DISfeyn}
\end{figure}

In order to work out the kinematics, we 
denote the nucleon mass, $M$, the lepton mass, $m_{\ell}$, the 
covariant spin four-vector of the incoming (outgoing) lepton
$s_\ell$ ($s_{\ell^\prime}$) and the spin four-vector of the nucleon, $S$. 
In the target rest frame, we define the four-momenta to be
\begin{equation}
\begin{array}{rcll}
\ell  & = & (E,  \mathbf{\ell})  & \mbox{incoming lepton,}  \\
\ell ' & = & (E', \mathbf{\ell '}) & \mbox{outgoing lepton,}  \\
P  & = & (M,  \mathbf{0})  & \mbox{proton.}
\end{array}
\label{eq:four-momenta}
\end{equation}
The deeply inelastic regime is identified by 
the invariant mass $W$ of the final hadronic system to be much larger than
the nucleon mass, namely
\begin{equation}
W^2=M^2+Q^2\frac{1-x}{x}\gg M^2
\,\mbox{.}
\label{eq:Wmass}
\end{equation}
This allows us to neglect all masses and to use the approximation 
\begin{equation}
\ell ^2 = {\ell '}^2 \approx 0 \mbox{,     } \ \ \  
E \approx |\mathbf{\ell}|      \mbox{,     } \ \ \  
E' \approx |\mathbf{\ell '}|   \mbox{.     } \ \ \  
\label{eq:approximation}
\end{equation}
Based on these assumptions, only two kinematical variables (besides the 
centre of mass energy $s=(\ell + P)^2$ or, alternatively, the lepton beam 
energy $E$) are needed to describe the process in Eq.~(\ref{eq:DIS}). 
They can be chosen among the following invariants:
\begin{eqnarray}
Q^2 & = & -q^2 
= 
(\ell - \ell ')^2 = 2EE'(1-\cos \theta) 
= 
4EE' \sin^2 \left(\frac{\theta}{2}\right) 
\\
& & \mbox{the laboratory-frame photon square momentum,}
\nonumber
\\
\nu & = & E-E' = \frac{P \cdot q}{M}
\\
& & \mbox{the laboratory-frame photon energy,}
\nonumber
\\
x & = & \frac{Q^2}{2P \cdot q} = \frac{Q^2}{2M\nu} 
\\
& & \mbox{the Bjorken scaling variable,}
\nonumber
\\
y & = & \frac{P \cdot q}{P \cdot \ell} = \frac{\nu}{E}
\label{eq:ydef}
\\
& & \mbox{the energy fraction lost by the incoming lepton $\ell$,}
\nonumber
\label{eq:invariants}
\end{eqnarray}
where $\theta$ is the scattering angle between the incoming and the outgoing 
lepton beams. 

The differential cross-section for lepton-nucleon scattering then reads
\begin{equation}
d^3\sigma(\ell N \to \ell ' X)
=
\frac{1}{2s} \frac{d^3 \ell '}{(2\pi)^3 2E'}
\sum_{s_{\ell '}}
\sum_{X} \int d\Pi_X
|\mathcal{M}(\ell N \to \ell ' X)|^2
\,\mbox{,}
\label{eq:cross_section}
\end{equation}
where 
\begin{equation}
\int d \Pi_X= \int \frac{d^3 \mathbf{P}_X}{(2\pi)^3 2E_X } (2\pi)^4
\delta^4(P+q-P_X)
\label{eq:hadronic_phase_space}
\end{equation}
is the phase-space factor for the unmeasured hadronic system and
\begin{eqnarray}
|\mathcal{M}(\ell N \to \ell ' X)|^2
& = &
\frac{e^4}{q^4}
\left [
\bar{u}(\ell,s_{\ell}) \gamma^{\nu} u (\ell ', s_{\ell '}) 
\bar{u}(\ell ', s_{\ell '})\gamma^{\mu} u (\ell, s_\ell)
\right]
\nonumber
\\
& \times & 
\left[
\langle P,S | J_{\nu}^{\dag}(q) | P_X \rangle
\langle P_X | J_{\mu}(q) | P,S \rangle
\right]
\label{eq:squarematrix}
\end{eqnarray}
is the squared amplitude including the Fourier transform of the quark
electromagnetic current $J^\mu(q)$ flowing through the hadronic vertex.
Since we are describing the scattering of polarized leptons on a polarized 
target, with no measurement of the outgoing lepton polarization nor of the 
final hadronic system, in Eq.~(\ref{eq:cross_section}) we must sum over 
the final lepton spin $s_{\ell '}$ and over all final hadrons $X$, 
but must not average over the initial lepton spin, nor sum over 
the nucleon spin.

It is customary to define the leptonic tensor
\begin{equation}
L^{\mu\nu}
=
\sum_{s_{\ell '}}
\left [
\bar{u}(\ell,s_{\ell}) \gamma^{\nu} u (\ell ', s_{\ell '}) 
\bar{u}(\ell ', s_{\ell '})\gamma^{\mu} u (\ell, s_\ell)
\right]
\label{eq:leptonic_tensor}
\end{equation} 
and the hadronic tensor
\begin{equation}
W_{\mu\nu}
=
\frac{1}{2\pi}
\sum_X
\int d\Pi_X
\left[
\langle P,S | J_{\nu}^{\dag}(q) | P_X \rangle
\langle P_X | J_{\mu}(q) | P,S \rangle
\right]
\label{eq:hadronic_tensor}
\end{equation}
in order to rewrite Eq.~(\ref{eq:cross_section}) as
\begin{equation}
d^3\sigma=\frac{1}{2s}
\frac{e^4}{Q^4}
2\pi
L^{\mu\nu}W_{\mu\nu}
\frac{d^3 \mathbf{\ell '}}{(2\pi)^3 2E'}
\label{eq:cross_section1}
\end{equation}
or, in the target rest frame, where $s=2ME$, 
and considering $d^3\ell '=E'^2 dE' d\Omega$, $d\Omega=d\cos\theta d\varphi$,
\begin{equation}
\frac{d^3\sigma}{d\Omega dE'}
=
\frac{\alpha_{em}^2}{2MQ^4} \frac{E'}{E} L^{\mu\nu}W_{\mu\nu}
\,\mbox{.}
\label{eq:cross_section2}
\end{equation} 
This is the differential cross-section for finding the scattered lepton in solid
angle $d\Omega$ with energy $(E^\prime,E^\prime+dE^\prime)$ usually quoted in
the literature (see for example Ref.~\cite{Leader:1985ab}). 
In Eq.~(\ref{eq:cross_section2}), $\alpha_{em}$ 
is the fine-structure electromagnetic constant, while $\varphi$ is the 
azimuthal angle of the outgoing lepton. 
The variables $E'$ and $\theta$ are natural ones, in that they are measured 
in the laboratory frame, by detecting the scattered lepton. 
However, it is more convenient to perform the variable transformation 
$(E',\theta) \to (x,y)$ and to express the differential cross-section 
in terms of the latter quantities as
\begin{equation}
\frac{d^3\sigma}{dx dy d\varphi}
=
\frac{\alpha^2_{em} y}{2 Q^4} L^{\mu\nu}W_{\mu\nu}
\,\mbox{,}
\label{eq:cross_section3}
\end{equation}
since these are gauge-invariant and dimensionless.

\subsection{Leptonic tensor}
\label{sec:leptonict}
In a completely general way, the leptonic tensor $L^{\mu\nu}$ can be decomposed 
into a symmetric and an antisymmetric part under $\mu \leftrightarrow \nu$ 
interchange 
\begin{equation}
L_{\mu\nu}=L_{\mu\nu}^{(S)}(\ell, \ell ') + iL_{\mu\nu}^{(A)}(\ell,s_{\ell}, \ell ')
\,\mbox{.}
\label{eq:lept_tens1}
\end{equation} 
Recalling the identity satisfied by the spinor $u(p,s)$, 
for a fermion with polarization vector $s^{\mu}$, 
\begin{equation}
u(p,s)\bar{u}(p,s)=(\slashed{p} + m) \frac{1}{2}(1+\gamma_5 \slashed{s})
\,\mbox{,}
\label{eq:property}
\end{equation}
and summing only on $s_{\ell '}$, the leptonic tensor reads 
\begin{equation}
L_{\mu\nu}
=
\mbox{Tr}
\left\{
\gamma_{\mu} (\slashed{\ell '}+m_{\ell}) 
\gamma_{\nu} (\slashed{\ell}+m_{\ell})
\frac{1}{2}(1+\gamma_5 \slashed{s}_{\ell})
\right\}
\,\mbox{.}
\label{eq:lept_tens2}
\end{equation}
Trace computation via Dirac algebra finally leads to (retaining lepton masses)
\begin{equation}
L_{\mu\nu}^{(S)}
 = 
2
\left[
\ell_{\mu} \ell '_{\nu} + \ell_{\nu} \ell '_{\mu} - g_{\mu\nu} (\ell \cdot \ell ' - m_{\ell}^2)
\right]
\,\mbox{,}
\label{eq:lept_tens3}
\end{equation}
\begin{equation}
L_{\mu\nu}^{(A)}
 = 
2 m_{\ell} \epsilon_{\mu\nu\rho\sigma} s_{\ell}^{\rho}(\ell - \ell ')^{\sigma}
\,\mbox{.}
\label{eq:lept_tens4}
\end{equation}
If the incoming lepton is longitudinally polarized, 
its spin vector can be expressed as
\begin{equation}
s_{\ell}^{\mu} 
= 
\frac{\lambda_{\ell}}{m_{\ell}} (|\mathbf{\ell}|, \hat{\mathbf{\ell}} E)
, \ \ 
\hat{\mathbf{\ell}}
=
\frac{\mathbf{\ell}}{|\mathbf{\ell}|}
\,\mbox{,}
\label{eq:spin}
\end{equation}
\textit{i.e.} it is parallel ($\lambda_\ell=+1$) or antiparallel 
($\lambda_\ell=-1$) to the direction of motion ($\lambda_\ell=\pm 1$ 
is twice the lepton helicity). Then, Eq. (\ref{eq:lept_tens4}) reads
\begin{equation}
L_{\mu\nu}^{(A)}
 = 
2 \lambda_{\ell} \epsilon_{\mu\nu\rho\sigma} \ell^{\rho}(\ell - \ell ')^{\sigma}
 = 
2\lambda_{\ell} \epsilon_{\mu\nu\rho\sigma} \ell^{\rho} q^{\sigma}
\,\mbox{.}
\label{eq:lept_tens5}
\end{equation}
Notice that the lepton mass $m_{\ell}$ appearing in Eq.~(\ref{eq:lept_tens4}) 
has been cancelled by the denominator in Eq.~(\ref{eq:spin}), which 
refers to a longitudinally polarized lepton. 
In contrast, if it is transversely polarized, that is, 
$s_{\ell}^{\mu}=s_{\ell \perp}^{\mu}$, no such cancellation occurs and the 
corresponding contribution is suppressed by a factor $m_{\ell}/E$. 

\subsection{Hadronic tensor}
\label{sec:hadronict}

The hadronic tensor $W_{\mu\nu}$ allows for a decomposition analogous to
Eq.~(\ref{eq:lept_tens1}), that is
\begin{equation}
W_{\mu\nu}
=
W_{\mu\nu}^{(S)} (q,P) 
+iW_{\mu\nu}^{(A)} (q;P,S)
\,\mbox{,}
\label{eq:had_tens1}
\end{equation}
where the symmetric and antisymmetric parts can be expressed in terms
of two pairs of structure functions, $W_1$, $W_2$ and $G_1$, $G_2$, as
\begin{eqnarray}
\frac{1}{2M} W_{\mu\nu}^{(S)}
& = &
\left(
-g_{\mu\nu} + \frac{q_{\mu} q_{\nu}}{q^2} 
\right)
W_1(P \cdot q, q^2)
\nonumber
\\
& + &
\frac{1}{M^2}
\left(
P_{\mu} - \frac{P \cdot q}{q^2} q_{\mu}
\right)
\left(
P_{\nu} - \frac{P \cdot q}{q^2} q_{\nu}
\right)
W_2(P \cdot q, q^2)
\,\mbox{,}
\label{eq:had_tens_simm}
\end{eqnarray}
\begin{eqnarray}
\frac{1}{2M} W_{\mu\nu}^{(A)}
& = &
\epsilon_{\mu\nu\rho\sigma} q^{\rho}
\Big{\{}
MS^{\sigma}G_1(P \cdot q, q^2)
\nonumber
\\
& + &
\frac{1}{M}
\left[
P \cdot q S^{\sigma} - S \cdot q P^{\sigma}
\right]
G_2(P \cdot q, q^2)
\Big{\}}
\,\mbox{.}
\label{eq:had_tens_asimm}
\end{eqnarray}
It is customary to introduce the dimensionless structure functions
\begin{equation}
F_1(x,Q^2) \equiv MW_1(\nu, Q^2)
\,\mbox{,}
\ \ \ \ \ \ \ \ \ \
F_2(x,Q^2) \equiv \nu W_2 (\nu, Q^2)
\,\mbox{,}
\label{eq:struct_funcF}
\end{equation}
\begin{equation}
g_1(x,Q^2) \equiv M^2 \nu G_1(\nu, Q^2)
\,\mbox{,}
\ \ \ \ \ \ \ \ \ \
g_2(x,Q^2) \equiv M \nu^2 G_2 (\nu, Q^2)
\,\mbox{,}
\label{eq:struct_funcg}
\end{equation}
and to rewrite the symmetric and antisymmetric parts of the hadronic tensor as
\begin{eqnarray}
W_{\mu\nu}^{(S)}
& = & 
2
\left(
-g_{\mu\nu} + \frac{q_{\mu} q_{\nu}}{q^2} 
\right)
F_1(x, Q^2)
\nonumber
\\
& + &
\frac{2}{P \cdot q}
\left(
P_{\mu} - \frac{P \cdot q}{q^2} q_{\mu}
\right)
\left(
P_{\nu} - \frac{P \cdot q}{q^2} q_{\nu}
\right)
F_2(x, Q^2)
\,\mbox{,}
\label{eq:had_tens_sym}
\end{eqnarray} 
\begin{equation}
W_{\mu\nu}^{(A)}
 =  
\frac{2M\epsilon_{\mu\nu\rho\sigma} q^{\rho}}{P \cdot q}
\left\{
S^{\sigma} g_1 (x, Q^2)
+
\left[
S^{\sigma} - \frac{S \cdot q}{P \cdot q} P^{\sigma}
\right]
g_2 (x,Q^2)
\right\}
\,\mbox{.}
\label{eq:had_tens_asym}
\end{equation} 
These expressions give the most general gauge-invariant decompositions
of the hadronic tensor for pure electromagnetic, parity conserving, 
interaction, see \textit{e.g}~\cite{Anselmino:1994gn} and references therein. 
A theoretical description of both neutral- and charged-current DIS 
at energies of the order of the weak boson masses (or higher) must include
parity violating terms in the decomposition of the hadronic tensor.
Because of them, one no longer has the correspondence that
its symmetric part, Eq.~(\ref{eq:had_tens_sym}),
is spin independent and its antisymmetric part, 
Eq.~(\ref{eq:had_tens_asym}), is spin dependent.
Actually, the spin-dependent part of the hadronic tensor becomes 
a superposition of symmetric and antisymmetric pieces.
Four more independent structure functions appear in this case,
usually called $F_3$, $g_3$, $g_4$ and $g_5$: the first multiplies
a term independent from the lepton or nucleon spin four-vector, while the 
three latter do not. 
Such a general decomposition of the hadronic tensor in DIS,
with particular emphasis on the polarized case, can be found
in Ref.~\cite{Anselmino:1993tc}. 

Since experimental data on 
polarized DIS are taken with electron or muon beams at 
momentum transfer values not exceeding $Q^2\sim 100$ GeV$^2$, 
we can safely neglect the contribution from a weak boson exchange
to describe them properly. 
However, the most general decomposition will be needed
in the future to handle neutral-current DIS at high energies, 
as it may be performed at an 
Electron-Ion Collider~\cite{Boer:2011fh,Accardi:2012qut},
or charged-current DIS with neutrino 
beams, as it might be available at a neutrino factory~\cite{Mangano:2001mj}. 

\subsection{Polarized cross-sections differences}
\label{sec:polxsec}

\begin{figure}[t]
\centering
\includegraphics[scale=0.7]{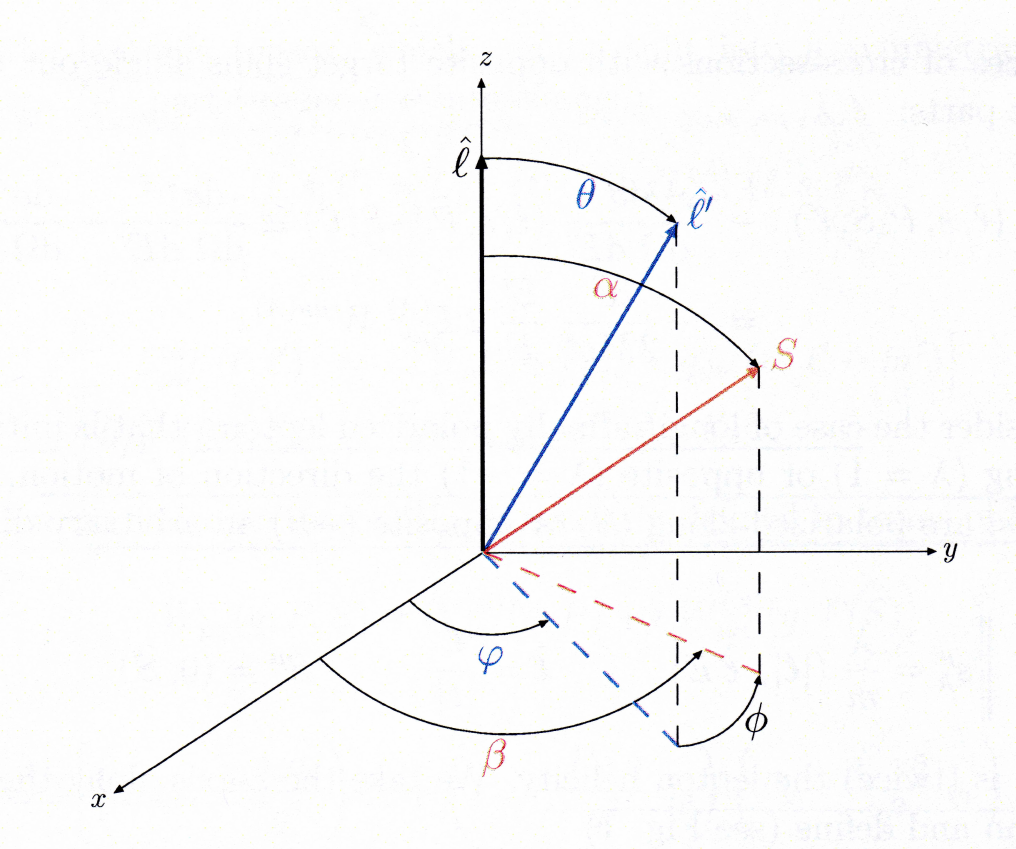} 
\mycaption[Target spin and lepton momenta]{Azimuthal and polar angles of the 
final lepton momentum, $\mathbf{\ell '}$, and the nucleon polarization vector, 
$\mathbf{S}$. The initial lepton moves along the positive $z$-axis. 
Often one defines the $(\hat{\mathbf{\ell}},\hat{\mathbf{\ell '}})$ lepton 
plane as the $\varphi=0$ plane.}
\label{fig:vectors}
\end{figure}

Insertion of Eqs. (\ref{eq:lept_tens1}) and (\ref{eq:had_tens1}) 
into Eq. (\ref{eq:cross_section3}) yields the expression
\begin{equation}
\frac{d^3\sigma^{s_{\ell},S}}{dx dy d \varphi}
=
\frac{\alpha_{em}^2 y}{2Q^4}
\left[
L_{\mu\nu}^{(S)} W^{\mu\nu (S)} - L_{\mu\nu}^{(A)} W^{\mu\nu (A)}
\right]
\,\mbox{,}
\label{eq:cross_section4}
\end{equation} 
and differences of cross-sections with opposite target helicity states 
single out 
the tensor antisymmetric parts
\begin{equation}
\frac{d^3\sigma^{\lambda_{\ell},+S}}{dx dy d\varphi} 
-
\frac{d^3\sigma^{\lambda_{\ell},-S}}{dx dy d\varphi} 
=
- \frac{\alpha_{em}^2 y}{Q^4} L_{\mu\nu}^{(A)} W^{\mu\nu(A)}
\mbox{ .}
\label{eq:asymmetry1}
\end{equation}
In the target rest frame, using the notation of Fig.~\ref{fig:vectors},
we parametrize the nucleon spin four-vector as
\begin{equation}
S^{\mu}=(0,\hat{\mathbf{S}})
=
(0,\sin\alpha \cos\beta, \sin\alpha \sin\beta, \cos\alpha)
\,\mbox{,}
\label{eq:spin_vector}
\end{equation}
where we have assumed $|\mathbf{S}|=1$.
Taking the direction of the incoming lepton to be along the $z$-axis, 
we also have
\begin{eqnarray}
\ell^{\mu}
& =&
E(1,0,0,1),
\\
\ell '^{\mu}
& = &
E'(1, \sin\theta \cos\varphi, \sin\theta \sin\varphi, \cos \varphi)
\,\mbox{.}
\label{eq:lept_vect}
\end{eqnarray}

Supposing now that the incoming lepton is polarized collinearly to its 
direction of motion, \textit{i.e.} $\lambda_{\ell}=+1$ and 
$s_{\ell}^{\mu}=\frac{\ell^{\mu}}{m_{\ell}}$, we have
\begin{eqnarray}
L_{\mu\nu}^{(A)} W^{\mu\nu (A)}
& = &
-\frac{8}{\nu}
\Big{\{}
\left[
(\ell \cdot q) (S \cdot q) - q^2 (S \cdot q)
\right]
g_1 (x, Q^2)
\nonumber
\\
& - & 
q^2
\left[
S \cdot \ell - \frac{(P \cdot \ell) (S \cdot q)}{P \cdot q}
\right]
g_2 (x, Q^2)
\Big{\}}
\,\mbox{.}
\label{eq:antisym_lept_had}
\end{eqnarray}
Note that, owing to currrent conservation, 
we have $L_{\mu\nu} q^{\mu} = L_{\mu\nu} q^{\nu} = 0$ and the terms proportional 
to $q^{\mu}$ and $q^{\nu}$ in Eq. (\ref{eq:had_tens_sym}) 
do not contribute when contracted with the leptonic tensor. 
Explicit computation of four-momentum products in this equation yields to
a new expression for the differential asymmetry~(\ref{eq:asymmetry1})
\begin{eqnarray}
& & 
\frac{d^3\sigma^{+;+S}}{dx dy d\phi}
-
\frac{d^3\sigma^{+;-S}}{dx dy d\phi}
\nonumber
\\
& = &
-\frac{4 \alpha_{em}^2}{Q^2} y
\left\{
\cos\alpha
\left[
\left(
\frac{E}{\nu} + \frac{E'}{\nu} \cos\theta
\right)
g_1 (x, Q^2)
+
\frac{2 E E'}{\nu^2}
(\cos\theta -1)
g_2 (x, Q^2)
\right]
\right.
\nonumber
\\
& + &
\left.
\sin\alpha \cos\phi\left[
\frac{E'}{\nu} \sin\theta g_1 (x, Q^2)
+
\frac{2 E E'}{\nu^2} \sin\theta g_2 (x, Q^2)
\right]
\right\}
\,\mbox{,}
\label{eq:asymmetry2}
\end{eqnarray}
where $\phi=\beta - \varphi$ is the azimuthal angle between the 
lepton plane and the $(\hat{\mathbf{\ell}} ,\hat{\mathbf{S}})$ plane. 
Notice that the r.h.s. of this equation is not expressed in terms 
of the usual invariants $x$ and $y$; to this purpose, 
let us define the ${\mathcal O}(1/Q)$ quantity
\begin{equation}
\gamma \equiv \frac{M x}{Q}
\label{eq:gamma}
\end{equation}
and work out a little algebra to obtain the final expression for the 
differential polarized cross-section difference Eq.~(\ref{eq:asymmetry1}) 
\begin{eqnarray}
& &
\frac{d^3\sigma^{+;+S}}{dx dy d\phi} 
-
\frac{d^3\sigma^{+;-S}}{dx dy d\phi} 
\nonumber
\\
& = &
-\frac{4 \alpha_{em}^2}{Q^2}
\left\{
\left[
\left(
2-y-\frac{\gamma^2 y^2}{4}
\right)
g_1 (x, Q^2)
-
\gamma^2 y g_2 (x, Q^2)
\right]
\cos\alpha
\right.
\nonumber
\\
& + &
\left.
\sqrt{1-y-\frac{\gamma^2 y^2}{4}}
\left[
y g_1 (x, Q^2) + 2 g_2 (x, Q^2)
\right]
\sin\alpha \cos\phi
\right\}
\,\mbox{.}
\label{eq:asymmetry_-}
\end{eqnarray}
Results obtained so far need a few comments.
\begin{enumerate}
\item The terms \textit {longitudinal} and \textit {transverse}, 
when speaking about the nucleon polarization, are somewhat ambiguous, 
insofar as a reference axis is not specified. From an experimental point 
of view, the \textit {longitudinal} or \textit {transverse} nucleon 
polarizations are defined with respect to the lepton beam axis, 
thus \textit {longitudinal} (\textit {transverse}) indicates 
the direction parallel (orthogonal) to this axis. 
We will use the large arrows $\Rightarrow$ ($\Uparrow$) to denote these 
two cases respectively.   
\item Eq.~(\ref{eq:asymmetry_-}) refers to the scattering of longitudinally 
polarized (positive helicity) leptons off a nucleon with positive or negative 
polarization along an arbitrary direction $\hat{\mathbf{S}}$. 
According to Eqs.~(\ref{eq:lept_tens4})-(\ref{eq:asymmetry1}),
the cross-section difference is proportional to $L_{\mu\nu}^{(A)}$, 
which contains a small factor
$m_{\ell}$; as already noticed, this small factor is cancelled by the $1/m_{\ell}$
factor appearing in the lepton-helicity four-vector, Eq.~(\ref{eq:spin}). 
This would not be the case with transversely polarized leptons, 
for which one would have $s^{\mu}=(0,\hat{\mathbf{s}})$, with 
$\hat{\mathbf{s}} \cdot \mathbf{\ell}=0$. 
Then, transversely polarized leptons lead to tiny cross-section 
asymmetries of order of $m_\ell/E$. 
\item Eq.~(\ref{eq:asymmetry_-}) can be specialized to particular 
cases of the nucleon polarization. For longitudinally polarized nucleons, 
that is $\hat{\mathbf{S}} \parallel \mathbf{\ell}$, one has $\alpha=0$ and 
the differential cross-section reads
\begin{equation}
\frac{d^3\sigma^{+;\Rightarrow}}{dx dy d\phi}
-
\frac{d^3\sigma^{+;\Leftarrow}}{dx dy d\phi}
=
-\frac{4\alpha_{em}^2}{Q^2}
\left[
\left(
2-y-\frac{\gamma^2 y^2}{2}
\right)
g_1 (x,Q^2)
-
\gamma^2 y g_2(x,Q^2) 
\right]
\,\mbox{;}
\label{eq:asymmetry_parallel}
\end{equation}
for nucleons polarized transversely to the lepton direction, 
one has $\alpha=\pi/2$ and the differential cross-section is
\begin{equation}
\frac{d^3\sigma^{+;\Uparrow}}{dx dy d\phi}
-
\frac{d^3\sigma^{+;\Downarrow}}{dx dy d\phi}
=
-\frac{4\alpha_{em}^2}{Q^2}
\gamma
\sqrt{1-y-\frac{\gamma^2 y^2}{4}}
\left[
y g_1(x,Q^2) + 2 g_2(x,Q^2)
\right]
\cos\varphi
\,\mbox{.}
\label{eq:asymmetry_perpendicular}
\end{equation}
In general, the term proportional to $g_2$ is suppressed by a factor 
$\gamma$, Eq.~(\ref{eq:gamma}), with respect to the one proportional to $g_1$: 
in the Bjorken limit, 
Eqs.~(\ref{eq:asymmetry_parallel})-(\ref{eq:asymmetry_perpendicular})
decouple and only $g_1$ is asymptotically relevant.
We emphasize that, in the case of transverse polarizations, both the $g_1$ and
$g_2$ structure functions equally contribute, but the whole cross-section
difference is suppressed by the overall factor, Eq.~(\ref{eq:gamma}),
of order $1/Q$.
In the following, we will mostly concentrate on the longitudinally polarized
cross-section difference, Eq.~(\ref{eq:asymmetry_parallel}).
\item From Eq.~(\ref{eq:cross_section4}), it is straightforward to obtain
the unpolarized cross-section for inclusive DIS by averaging over spins 
of the incoming lepton ($s_{\ell}$) and of the nucleon ($S$) 
and by integrating over the azimuthal angle $\varphi$. It reads
\begin{equation}
\frac{d^2 \sigma^{unp}}{dx dy}
=
2\pi
\frac{1}{2} \sum_{s_{\ell}}
\frac{1}{2} \sum_{S}
\frac{d^2\sigma^{s_{\ell},S}}{dx dy}
=
\frac{\alpha_{em}^2 y}{2Q^4}
L_{\mu\nu}^{(S)} W^{\mu\nu (S)}
\,\mbox{.}
\label{eq:unp_cross_section}
\end{equation}
Finally, the unpolarized cross-section, expressed in terms of the usual
unpolarized structure functions $F_1$ and $F_2$, 
when neglecting contributions of order $M^2/Q^2$, is
\begin{equation}
\frac{d^2 \sigma^{unp}}{dx dy}
=
\frac{4 \pi \alpha_{em}^2 s}{Q^4}
\left[
x y^2 F_1(x, Q^2) + (1-y) F_2 (x, Q^2) 
\right]
\,\mbox{.}
\label{eq:unp_cross_section1}
\end{equation}
\end{enumerate}

\section{Factorization of structure functions}
\label{sec:sf-factorization}

In the previous Section, we have parametrized the hadronic tensor,
which describes the coupling of the virtual photon to the composite
nucleon, in terms of four structure functions, namely $F_1$, $F_2$ and
$g_1$, $g_2$, see Eqs.~(\ref{eq:had_tens_sym})-(\ref{eq:had_tens_asym}).
We have then derived the expression for 
the differential cross-section asymmetries of longitudinally 
and transversely polarized nucleons in terms of $g_1$ and $g_2$,
Eqs.~(\ref{eq:asymmetry_parallel})-(\ref{eq:asymmetry_perpendicular}).
In principle, by performing DIS experiments with nucleons polarized both
longitudinally and transversely, one should learn
about the structure functions $g_1$ and $g_2$, as we will discuss in detail 
in Sec.~\ref{sec:phenopol} below.
In this Section, we would like to provide a description of DIS 
in the framework of QCD, and in particular give a \textit{factorized}
expression for the structure function $g_1$.
Actually, even though QCD is asymptotically free, the computation of any  
cross-section does involve non-perturbative contributions, since the 
initial and final states are not the fundamental degrees of freedom of the 
theory, but compound states of quarks and gluons. 
As we shall see, the factorization theorem allows for the separation of
a hard, perturbative and process-dependent part from a low energy,
process-independent contribution. The latter is given by the Parton
Distribution Functions (PDFs), which parametrize our ignorance 
on the inner structure of the proton. 
In order to deal with the factorized expression for the structure function
$g_1$, we will first provide the leading-order (LO) QCD description
of polarized DIS, starting from the naive parton model;
we will then give a heuristic development of the 
next-to-leading order (NLO) perturbative QCD corrections to polarized DIS,
focusing on their effects on the $g_1$ structure function. 

\subsection{Naive parton-model expectations}
\label{sec:parton-model}

The information on the \textit{a priori} unknown structure of a 
polarized nucleon is carried by the structure functions $g_1$ and $g_2$.
As discussed in Sec.~\ref{sec:hadronict}, they can only be functions of
$x$ and $Q^2$. In the naive parton 
model~\cite{Bjorken:1968dy,Feynman:1969ej,Bjorken:1969ja,Callan:1969uq},
they allow for simple expressions, since the cross-section for 
lepton-nucleon scattering is regarded as the incoherent sum of point-like 
interactions between the lepton and a free, massless parton
\begin{equation}
\frac{d^2\sigma}{dx dy}
=
\sum_q e_q^2 f_{q/p}(x) \frac{d\hat{\sigma}}{dy}
\,\mbox{.}
\label{eq:convolutionspm}
\end{equation}
In this expression, $e_q$ is the fractional charge carried by a parton $q$, 
$\frac{d\hat{\sigma}}{dy}$ is the cross-section for the elementary
QED process $\ell q \to \ell^\prime q$, and $f_{q/p}$ is the PDF, 
the probability density distribution for the momentum 
fraction $x$ of any parton $q$ in a nucleon $p$. In the simple picture
provided by the parton model, PDFs do not depend on the
scale $Q^2$ and the structure functions are observed to obey the
scaling law $g_{1,2}(x,Q^2)\to g_{1,2}(x)$~\cite{Bjorken:1968dy}. 
This property is related to the assumption that the transverse
momentum of the partons is small. In the framework of QCD, however, 
the radiation of hard gluon from the quarks violates this assumption
beyond leading order in pertubation theory, as we will discuss below.
Of course, the naive parton model predates QCD, but we find it
of great value for its intuitive nature.

If we specialize Eq.~(\ref{eq:convolutionspm}) to polarized cross-section 
asymmetries, we should write
\begin{equation}
\frac{d^2\sigma^{+;\Rightarrow}}{dx dy}
-
\frac{d^2\sigma^{+; \Leftarrow}}{dx dy}
=
\sum_q e_q^2 \Delta f_{q/p} (x)
\left[
\frac{d\hat{\sigma}^{+;+}}{dy} 
-
\frac{d\hat{\sigma}^{+;-}}{dy} 
\right]
\,\mbox{,}
\label{eq:asymmetry_parallel_pdf}
\end{equation}
where $\frac{d\hat{\sigma}^{\lambda_\ell,\lambda_q}}{dy}$ denotes the elementary 
cross-section retaining helicity states of both the lepton ($\lambda_\ell$) 
and the struck parton ($\lambda_q$). We also introduced helicity-dependent,
or polarized PDFs, $\Delta f_{q/p}$, defined as the momentum densities of partons
with spin aligned parallel or antiparallel to the longitudinally polarized 
parent nucleon:
\begin{equation}
\Delta f_{q/p}(x)\equiv f_{q/p}^{\uparrow}(x) - f_{q/p}^{\downarrow}(x)
\,\mbox{.}
\label{eq:helPDFs}
\end{equation}

Explicit expressions for the elementary cross-sections appearing in the 
r.h.s of Eq.~(\ref{eq:asymmetry_parallel_pdf}) are easily computed 
at the lowest order in Quantum Electrodynamics (QED):
\begin{equation}
\frac{d\hat{\sigma}^{+,+}}{dy}
=
\frac{4 \pi \alpha_{em}^2}{Q^2}
\frac{1}{y}
\,\mbox{,}
\ \ \ \ \
\frac{d\hat{\sigma}^{+,-}}{dy}
=
\frac{4 \pi \alpha_{em}^2}{Q^2}
\frac{(1 - y)^2}{y}
\,\mbox{.}
\label{eq:helicity_amplitudes}
\end{equation} 
Replacing these elementary cross-sections 
in Eq.~(\ref{eq:asymmetry_parallel_pdf})
leads to the expression
\begin{equation}
\frac{d^2\sigma^{+;\Rightarrow}}{dx dy}
-
\frac{d^2\sigma^{+; \Leftarrow}}{dx dy}
=
\frac{4 \pi \alpha_{em}^2}{Q^2}
\left[
\sum_q e_q^2 \Delta q (x)
(2-y)
\right]
\,\mbox{,}
\label{eq:pol_cs}
\end{equation}
which can be directly compared to Eq.~(\ref{eq:asymmetry_parallel}),
provided the latter is integrated over the azimuthal angle $\phi$.
Neglecting terms of order $\mathcal{O}(\gamma^2)$, we finally obtain
the naive parton model relations between structure functions
$g_1(x)$, $g_2(x)$ and the polarized distributions $\Delta f_{q/p}(x)$:
\begin{eqnarray}
g_1 (x)
& = &
\frac{1}{2} \sum_q e_q^2 \Delta f_{q/p}(x)
\label{eq:g_relations1}
\,\mbox{,}
\\
g_2 (x)
& = &
0
\,\mbox{.}
\label{eq:g_relations2}
\end{eqnarray}
These results require a few comments.
\begin{enumerate}
\item The structure function $g_1$, Eq.~(\ref{eq:g_relations1}), 
can be unambigously expressed in terms 
of quark and antiquark polarized parton distributions. 
Assuming the number of flavors to be $n_f=3$, we can define the singlet,
$\Delta\Sigma$, and nonsinglet triplet, $\Delta T_3$, and octet, $\Delta T_8$,
combinations of polarized quark densities
\begin{eqnarray}
\Delta\Sigma
& = &
\Delta u^++\Delta d^++\Delta s^+
\,\mbox{,}
\label{eq:partbasis1}
\\
\Delta T_3
& = & 
\Delta u^+ -\Delta d^+
\,\mbox{,}
\label{eq:partbasis2}
\\
\Delta T_8
& = &
\Delta u^+ + \Delta d^+ -2\Delta s^+
\,\mbox{,}
\label{eq:partbasis3}
\end{eqnarray}
where $\Delta q^+=\Delta q + \Delta \bar{q}$, $q=u,d,s$ 
are the total parton densities. 
Then, the structure function $g_1$, Eq~(\ref{eq:g_relations1}), 
can be cast into the form
\begin{equation}
g_1(x) 
=
\frac{1}{9}\Delta\Sigma(x) + \frac{1}{12}\Delta T_3(x)
+\frac{1}{36} \Delta T_8(x) 
\label{eq:g153}
\,\mbox{.}
\end{equation}
The structure function $g_1$ does not receive any contribution
from gluons, yet we shall see in Sec.~\ref{sec:QCDevol} that it is not true in
the framework of QCD beyond Born approximation.
\item The structure function $g_2$ is zero, Eq.~(\ref{eq:g_relations2}).
However, non-zero values of $g_2$ can be obtained by allowing the quarks to 
have an intrinsic Fermi motion inside the nucleon. In this case, there is
no unambiguous way to calculate $g_2$ in the naive parton model.
We will not further investigate this issue in this Thesis;
a detailed discussion of the problem can be found 
in Ref.~\cite{Anselmino:1994gn}.
\item The Wilson Operator Product Expansion (OPE) can be applied to
the expression of the hadronic tensor $W_{\mu\nu}$ in terms of the 
Fourier transform of the nucleon matrix elements of the elctromagnetic current 
$J_\mu(x)$, Eq.~(\ref{eq:hadronic_tensor}).
this way, one can give the moments of the structure functions
$g_1$ and $g_2$ in terms of hadronic matrix elements of certain operators 
multiplied by perturbatively calculable Wilson coefficient functions.
In particular, it can be shown (see \textit{e.g.}~\cite{Anselmino:1994gn})
that the first moment of the singlet quark density
\begin{equation}
a_0=\int_0^1 dx \Delta\Sigma(x)
\label{eq:naivea0}
\end{equation}
is related to the matrix element of the flavor singlet axial current.
Hence, $a_0$ can be interpreted as the contribution of quarks 
and antiquarks to the proton's spin, intuitively 
twice the expectation value of the sum
of the $z$-components of quark and antiquark spins
\begin{equation}
a_0=2\langle S_z^{\mathrm{quarks+antiquarks}}\rangle 
\,\mbox{.}
\label{eq:naivespin}
\end{equation}
Uncritically, one should expect $a_0\approx 1$, while in a more realistic 
relativistic model one finds $a_0\approx 0.6$~\cite{Jaffe:1989jz}. 
In the late 80s, this expectation was found to be in contrast with the
anomalously small value measured by the European Muon Collaboration at 
CERN~\cite{Ashman:1987hv,Ashman:1989ig}. 
This result could be argued to imply that the sum of the spins carried by
the quarks in a proton, $\langle S_z^{\mathrm{quarks+antiquarks}}\rangle$,
was consistent with zero rather than $1/2$, 
suggesting a \textit{spin crisis} in the parton model~\cite{Leader:1988vd}. 
This led to an intense scrutiny of the basis of the theoretical calculation of 
the structure function $g_1$ and the spin crisis was immediately recognized 
not to be a fundamental problem, but rather an interesting property of spin 
structure functions to be understood in terms of QCD.
We will give a summary of such a description in the following Section.
\end{enumerate}

\subsection{QCD corrections and evolution}
\label{sec:QCDevol}

The parton model predates the formulation of QCD. As soon as QCD is
accepted as the theory of strong interactions, with quark and gluon fields
as the fundamental fields, one should describe the lepton scattering  
off partons in the nucleon perturbatively. At Born level, the interaction
is described by the Feynman diagram in Fig.~\ref{fig:gluecorrections}-$(a)$
as the tree-level scattering of a quark (or antiquark) 
off the virtual photon $\gamma^*$.
In this case, quarks are free partons and one recovers
the parton model expressions for the structure functions, 
Eqs.~(\ref{eq:g_relations1})-(\ref{eq:g_relations2}).
At $\mathcal{O}(\alpha_s)$, several new contributions appear:
the emission of a gluon, Fig.~\ref{fig:gluecorrections}-$(b)$,
the one-loop correction, Fig.~\ref{fig:gluecorrections}-$(c)$,
and the process initiated by a gluon which then splits into a quark-antiquark
pair, the so-called photon-gluon fusion (PGF) process, 
Fig.~\ref{fig:gluecorrections}-$(d)$.
The main impact of the QCD interactions is twofold: first, they introduce 
a mild, calculable, logarithmic $Q^2$ dependence in the parton distributions; 
second, the correction in Fig.~\ref{fig:gluecorrections}-$(d)$ generate 
a contribution to the structure function $g_1$ arising from the polarization
of the gluons in the nucleon. We shall describe both these effects in the
following.
\begin{figure}[t]
\begin{center}
\begin{tabularx}{\textwidth}{XXXX}
 $(a)$ & $(b)$ & $(c)$ & $(d)$ 
\end{tabularx}
\begin{fmffile}{QCDcorr}
\fmfframe(0,0)(0,-70)
{
\begin{fmfgraph*}(80,160)
\fmfleft{li}
\fmfright{lo}
\fmftop{gamma}
\fmf{fermion}{li,v2}
\fmf{fermion}{v2,lo}
\fmf{photon,tension=2}{gamma,v2}
\end{fmfgraph*}
\ \ \ \ \
\begin{fmfgraph*}(80,160)
\fmfleft{li}
\fmfright{lo}
\fmfright{ghost}
\fmftop{gamma}
\fmf{fermion}{li,v3,v2}
\fmf{photon}{gamma,v2}
\fmf{vanilla}{v2,v4}
\fmf{fermion}{v4,lo}
\fmffreeze
\fmf{gluon}{v3,vf}
\fmf{phantom,tension=1}{vf,ghost}
\end{fmfgraph*}
\ \ \ \ \
\begin{fmfgraph*}(80,160)
\fmfleft{li}
\fmfright{lo}
\fmftop{gamma}
\fmf{fermion}{li,v3,v2}
\fmf{fermion}{v2,v4,lo}
\fmf{photon}{gamma,v2}
\fmffreeze
\fmf{gluon}{v3,v4}
\end{fmfgraph*}
\ \ \ \ \
\begin{fmfgraph*}(80,160)
\fmfleft{li}
\fmfright{lo}
\fmfright{ghost}
\fmftop{gamma}
\fmf{gluon}{li,v3}
\fmf{fermion}{v3,v2}
\fmf{photon}{gamma,v2}
\fmf{vanilla}{v2,v4}
\fmf{fermion}{v4,lo}
\fmffreeze
\fmf{fermion}{vf,v3}
\fmf{phantom,tension=1}{vf,ghost}
\end{fmfgraph*}
}
\end{fmffile}
\end{center}
\mycaption{Leading contribution $(a)$ 
and next-to-leading order corrections $(b), (c), (d)$ to DIS.}
\label{fig:gluecorrections}
\end{figure}

\subsubsection{Scale dependence of parton distributions}
When including NLO corrections, Fig.~\ref{fig:gluecorrections}-$(b)$-$(c)$,
problems arise from the so-called collinear singularities linked to 
the effective masslessness of quarks. The factorization theorem
is probed~\cite{Collins:1989gx} to allow for a
separation of the process into a hard and a soft part 
and for the absorption of the infinity
into the soft part (the PDF), which in any case
cannot be calculated and has to be determined from experimental data.
The scale at which the separation is made is called factorization scale $\mu^2$.
Schematically, one finds terms of the form $\alpha_s\ln(Q^2/M)$, 
which one splits as follows
\begin{equation}
\alpha_s\ln \frac{Q^2}{M}
=
\alpha_s\ln\frac{Q^2}{\mu^2}+\alpha_s\ln\frac{\mu^2}{M^2}
\,\mbox{;}
\label{eq:log}
\end{equation}
one then absorbs the first term on the r.h.s. of Eq.~(\ref{eq:log}) 
into the hard part of the process, and the second term into the soft part. 
The factorization scale $\mu^2$ can be
chosen arbitrarily and, in exact calculations, physical results must not 
depend on it. In practice, since we never calculate to all orders in 
perturbation theory, it can make a difference what value we choose, but
it turns out that an optimal choice is $\mu^2=Q^2$. Consequently,
parton distributions no longer obey exact Bjorken scaling, but develop a slow
logarithmic dependence on $Q^2$. Actually, if one keeps only the leading-log
terms (proportional to $\alpha_s\ln(Q^2/\mu^2)$), one finds that the 
parton model expressions, 
Eqs.~(\ref{eq:g_relations1})-(\ref{eq:g_relations2}), still hold, provided
the replacement
\begin{equation}
\Delta f_{q/p}(x) \longrightarrow \Delta f_{q/p}(x,Q^2)
\end{equation}
to the $Q^2$-dependent PDF is made.
We can think of the scale dependence of PDFs within the following picture.
As the scale increases, the photon starts to \textit{see}
evidence for the point-like valence quarks within the proton. 
If the quarks were non-interacting, no further structure would be resolved 
increasing the resolving scale: the Bjorken scaling would set in, 
and the naive parton model would be satisfactory. 
For this reason, we can consider the naive parton model 
as the approximation of QCD to Born level.
However, QCD predicts that on increasing the resolution, 
one should see that each quark is itself surrounded by a cloud of partons. 
The number of resolved partons which share
the proton's momentum increases with the scale.

The perturbative dependence of the polarized PDFs on the scale $Q^2$ is given
by the Altarelli-Parisi evolution equations~\cite{Altarelli:1977zs},
a set of $(2n_f+1)$ coupled integro-differential equations. 
It is customary to write them in the \textit{evolution basis}, \textit{i.e.}
in terms of linear combinations of the individual parton
distributions such that the $(2n_f+1)$ equations maximally decouple from each other. 
To this purpose, we define the polarized gluon distribution $\Delta g(x,Q^2)$ as in 
Eq.~(\ref{eq:helPDFs}) and the singlet and nonsinglet quark PDF combinations as
\begin{eqnarray}
\Delta q_{\mathrm{NS}}(x,Q^2)
& \equiv &
\sum_{i=1}^{n_f}\left(\frac{e^2_i}{\langle e^2 \rangle} - 1\right)
\left[\Delta q_i(x,Q^2)+\Delta \bar q_i(x,Q^2)\right]
\,\mbox{,}
\label{eq:qNS}
\\
\Delta \Sigma(x,Q^2)
& \equiv &
\sum_{i=1}^{n_f}\left[\Delta q_i(x,Q^2)+\Delta \bar q_i(x,Q^2)\right]
\,\mbox{,}
\label{eq:qSigma}
\end{eqnarray}
where $\Delta q_i(x,Q^2)$ and $\Delta\bar{q}_i(x,Q^2)$ are the 
scale-dependent quark and 
antiquark polarized densities of flavor $i$, also defined according to 
Eq.~(\ref{eq:helPDFs}). 
The evolution equations are coupled for the singlet quark-antiquark combination 
and the gluon distribution 
\begin{equation}
\mu^2\frac{\partial}{\partial \mu^2}
\left(
\begin{array}{c}
\Delta\Sigma(x,\mu^2) \\ \Delta g(x,\mu^2)
\end{array}
\right)
=
\frac{\alpha_s(\mu^2)}{2\pi}
\left(
\begin{array}{cc}
\Delta P_{qq}^{\mathrm{S}} & 2n_f\Delta P_{qg}^{\mathrm{S}}\\
\Delta P_{qg}^{\mathrm{S}} & \Delta P_{gg}^{\mathrm{S}}
\end{array}
\right)
\otimes
\left(
\begin{array}{c}
\Delta\Sigma(x,\mu^2) \\ \Delta g(x,\mu^2)
\end{array}
\right)
\,\mbox{,}
\label{eq:evolSigmag}
\end{equation}
while the nonsinglet quark-antiquark combination evolves independently as
\begin{equation}
\mu^2\frac{\partial}{\partial\ln\mu^2}\Delta q_{\mathrm{NS}}(x,\mu^2)
=
\frac{\alpha_s(\mu^2)}{2\pi}\Delta P_{qq}^{\mathrm{NS}}\otimes\Delta q_{\mathrm{NS}}(x,\mu^2)
\,\mbox{.}
\label{eq:evolNS}
\end{equation}
In Eqs.~(\ref{eq:evolSigmag})-(\ref{eq:evolNS}), 
$\Delta P_{ij}^{\mathrm{S/NS}}$, $i,j=q,g$, denotes the singlet/nonsinglet 
spin-dependent splitting functions for quarks and gluons 
and $\otimes$ is the shorthand notation for 
the convolution product with respect to $x$
\begin{equation}
f\otimes g = \int_x^1\frac{dy}{y}f\left(\frac{x}{y} \right) g(y)
\,\mbox{.}
\label{eq:convolution}
\end{equation}
We notice that Eqs.~(\ref{eq:evolSigmag})-(\ref{eq:evolNS}) hold to
all orders in perturbative theory, hence
splitting functions may be expanded in powers of the strong coupling
$\alpha_s$:
\begin{equation}
\Delta P_{ij}^{p}
=
\Delta P_{ij}^{p(0)}(x)+\frac{\alpha_s(\mu^2)}{2\pi}\Delta P_{ij}^{p(1)}(x) 
+\mathcal{O}(\alpha_s^2)
\,\mbox{,}
\label{eq:NLOCandP}
\end{equation}
where $p=$S, NS, and $i,j=q,g$.
The splitting functions for polarized PDFs were computed at LO in 
Ref.~\cite{Altarelli:1977zs} for the first time, the computation was
then extended to NLO in Refs.~\cite{Mertig:1995ny,Vogelsang:1995vh},
while only partial results are available at NNLO 
so far~\cite{Vogt:2008yw}.

The solution of the Altareli-Parisi equations may be written as
\begin{equation}
 \Delta f_i(x,Q^2) 
 =
 \sum_j\Gamma_{ij}(x,\alpha_s,\alpha_s^0)\otimes \Delta f_j(x,Q_0^2)
 \,\mbox{,}
\label{eq:APsolution}
\end{equation}
where: $f_i=\Sigma, \, q_{\mathrm{NS}}, \, g$; $\Delta f_j(x,Q_0^2)$ are the 
corresponding input PDFs, parametrized at an initial scale $Q_0^2$, to be determined from 
experimental data; $\Gamma_{ij}(x,\alpha_s,\alpha_s^0)$ are the 
evolution factors; and we have used the shorthand notation
\begin{equation}
 \alpha_s\equiv\alpha_s(Q^2)
 \ \ \ \ \ \ \ \ \ \ 
 \alpha_s^0\equiv \alpha_s(Q_0^2)
 \,\mbox{.}
 \label{eq:shortand}
\end{equation}
The evolution factors also satisfy evolution equations
\begin{equation}
 \mu^2\frac{\partial}{\partial\mu^2}\Gamma_{ij}(x,\alpha_s,\alpha_s^0)
 =
 \sum_k P_{ik}(x,\alpha_s)\otimes \Gamma_{kj}(x,\alpha_s,\alpha_s^0)
 \,\mbox{,}
 \label{eq:evolfact}
\end{equation}
with boundary conditions $\Gamma_{ij}(x,\alpha_s^0,\alpha_s^0)=\delta_{ij}\delta(1-x)$.
The QCD evolution equations are most easily solved using Mellin moments
since then all convolutions become simple products, and the equations
can be solved in a closed form.
We refer to~\cite{Diemoz:1987xu,Vogt:2004ns,Blumlein:2006be} 
for a comprehensive discussion about details concerning such a technique. 

\subsubsection{The gluon contribution to the $g_1$ structure function}
Another important consequence of QCD corrections is the rise of a contribution
to the $g_1$ structure function from the polarization of gluons in the nucleon,
see for instance Fig.~\ref{fig:gluecorrections}-$(d)$.
For this reason, the factorized leading-twist expression for the 
structure function $g_1$ reads, instead of Eq.~(\ref{eq:g153}),
\begin{equation}
g_1(x,Q^2)
=
\frac{\langle e^2\rangle}{2}
\left[
C_{\mathrm{NS}}\otimes\Delta q_{\mathrm{NS}}
+
C_{\mathrm{S}}\otimes\Delta\Sigma
+
2n_f C_g\otimes \Delta g
\right]
\,\mbox{,}
\label{eq:g1coeff}
\end{equation}
where $\langle e^2\rangle=n_f^{-1}\sum_{i=1}^{n_f} e_i^2$ is the average charge,
with $n_f$ the number of active flavors and $e_i$ their electric charge,
and $\otimes$ denotes the convolution product with respect to $x$,
Eq.~(\ref{eq:convolution}). Besides, $\Delta q_{\mathrm{NS}}$ and $\Delta\Sigma$
are the scale-dependent nonsinglet and singlet quark PDF combinations,
Eqs.~(\ref{eq:qNS})-(\ref{eq:qSigma}) and $\Delta g$ is the gluon PDF.
Finally, $C_{\mathrm{NS}}$, $C_{\mathrm{S}}$ and $C_g$ are the corresponding
coefficient functions related to calculable short-distance cross-sections,
for hard photon-quark and photon-gluon cross-sections respectively.
Coefficient functions are perturbative objects and may be expanded in 
powers of the strong coupling $\alpha_s$
\begin{equation}
C_p(x,\alpha_s)
=
C_p^{(0)}(x)+\frac{\alpha_s(\mu^2)}{2\pi}C_p^{(1)}(x)+\mathcal{O}(\alpha_s)
\,\mbox{,}
\end{equation}
with $p=$S, NS, $g$ and $i,j=q,g$. At the lowest order in $\alpha_s$,
$C_{\mathrm{NS}}^{(0)}=C_{\mathrm{S}}^{(0)}=\delta(1-x)$
and $C_g^{(0)}=0$~\cite{Altarelli:1977zs}, hence
the structure function $g_1$ decouples from the gluon contribution,
see Eq.~(\ref{eq:g1coeff}), and the parton model
prediction, Eq.~(\ref{eq:g153}) is recovered.
Polarized coefficient functions have been computed up to 
$\mathcal{O}(\alpha_s^2)$ so far~\cite{Zijlstra:1993sh}.

Notice incidentally that the substitution of Eq.~(\ref{eq:APsolution})
into Eq.~(\ref{eq:g1coeff}) leads to the relation
\begin{equation}
g_1(x,Q^2)
=
\frac{\langle e^2\rangle}{2}
\sum_{j=\mathrm{NS}, \Sigma, g} K_j(x,\alpha_s,\alpha_s^0)\otimes\Delta f_j(x,Q_0^2)
\,\mbox{,}
 \label{eq:g1fullfactor}
\end{equation}
where the hard kernel defined as
\begin{equation}
K_j(x,\alpha_s,\alpha_s^0)
=
\sum_{k=\mathrm{NS},\mathrm{S},g} C_k(x,\alpha_s)\otimes\Gamma_{ij}(x,\alpha_s,\alpha_s^0)
 \label{eq:hardkernel}
\end{equation}
is completely computable in perturbation theory. 
Hence, in Eq.~(\ref{eq:g1fullfactor}) we have fully
separated the perturbative and the non-perturbative parts
entering the structure function $g_1$. Besides, the hard kernels
in Eq.~(\ref{eq:hardkernel}) are independent of the particular
set of input PDFs, and may thus be computed separately once and for all,
suitably interpolated and stored. This is of uttermost importance 
while performing a fit of PDFs to experimental data, as we will further 
delineate in Sec.~\ref{sec:QCDanalysis}, since this involves
the evaluation of only the one set of convolutions 
Eq.~(\ref{eq:g1fullfactor}), which is amenable to computational
optimization.

The gluonic term in the expression of the $g_1$ structure function, 
see for instance Eq.~(\ref{eq:g1coeff}),
can be shown~\cite{Altarelli:1988nr,Altarelli:1988mu,Carlitz:1988ab} 
to entail an additional 
contribution to the singlet axial charge of the form
\begin{equation}
a_0^{\mathrm{gluons}}=-n_f\frac{\alpha_s(Q^2)}{2\pi}\int_0^1dx\Delta g(x,Q^2)
\,\mbox{.}
\label{eq:gluonanomaly}
\end{equation}
We remind that $n_f$ is the number of light
flavors, $u$, $d$, $s$, and heavy flavors are assumed not to contribute.
Hence, the naive parton
model expectation for the axial current $a_0$, Eq.~(\ref{eq:naivea0}),
should be replaced by 
\begin{equation}
a_0=\int_0^1dx \Delta\Sigma(x,Q^2) + a_0^{\mathrm{gluons}}
\,\mbox{.}
\label{eq:a0anomaly}
\end{equation}

At first sight, one should expect that the gluonic term in 
Eq.~(\ref{eq:a0anomaly}) would not survive at large $Q^2$, since it looks like
an $\alpha_s$ correction which would disappear 
as the running coupling $\alpha_s$ vanishes.
However, the first moment of the gluon contribution 
$\int_0^1dx\Delta g(x,Q^2)$ grows as $[\alpha_s(Q^2)]^{-1}$ for large
values of $Q^2$, as dictated by Altarelli-Parisi evolution equations,
see Eq.~(\ref{eq:evolSigmag}).\footnote{Higher moments are instead decreasing 
functions of $\log Q^2$, falling at a faster rate than for the unpolarized
gluon density.} 
Hence, the gluon does not decouple
from $g_1$ asymptotically and the parton model expression for
the structure function $g_1$, Eq.~(\ref{eq:g_relations1}),
is not recovered in perturbative QCD even in the limit $\alpha_s\to 0$.
The $Q^2$ behavior of the first moment of the polarized
gluon density was originally derived when the QCD evolution equations were 
first written down in $x$ space~\cite{Altarelli:1977zs}. In fact, the first 
moment of the polarized gluon splitting function is finite and proportional 
to the first coefficient of the QCD beta function, which establishes the quoted 
relation with the running coupling $\alpha_s(Q^2)$. This relation between
the $Q^2$ evolution of the first moment of the polarized gluon density and 
the running coupling is induced by the axial anomaly~\cite{Altarelli:1988mu}
corresponding to the QCD version of the anomalous \textit{triangle 
diagram}~\cite{Adler:1969gk,Bell:1969ts}.

As a consequence, the definition of the singlet quark first moment 
becomes totally ambiguous, because two generic definitions differ by terms
of order $\alpha_s(Q^2)\int_0^1 dx\Delta g(x,Q^2)$. For the first moment,
what is formally a NLO correction is potentially of the same size. 
Owing to Eq.~(\ref{eq:a0anomaly}),
the singlet quark first moment, Eq.~(\ref{eq:naivea0}), 
defined directly from the structure function 
$g_1$ and used by the EMC experiment when the 
\textit{spin crisis}~\cite{Leader:1988vd} was announced, does not have 
to coincide with the constituent quark value, \textit{i.e.} the total fraction
of the spin carried by quarks. 
Only for exactly conserved quantities do the corresponding values for
constituent and parton quarks have to coincide. The first moments of the quark
densities are in general only conserved at LO by the QCD evolution, but, due
to the axial anomaly, the singlet quark first moment defined from $g_1$ 
is not conserved in higher orders. We conclude that a definition of the singlet
quark density $\Delta\Sigma(x,Q^2)$ must be carefully specified, as
further discussed in Sec.~\ref{sec:schemedep} below.

The result quoted in Eqs.~(\ref{eq:gluonanomaly})-(\ref{eq:a0anomaly})
was advocated to reconcile the EMC result with the theoretical expectation
for the proton spin content. As explained above, what the EMC experiment
actually observed was the singlet axial charge $a_0$, 
Eq.~(\ref{eq:a0anomaly}). The almost vanishing value 
measured for this quantity can be explained as a cancellation between a 
reasonably large quark spin contribution, \textit{e.g.}
$\Delta\Sigma\simeq 0.6 - 0.7$, as expected intuitively, and the anomalous
gluon contribution. However, in order to accomplish this cancellation, one 
should require a large gluon spin contribution, 
\textit{e.g.} $\int_0^1dx\Delta g(x,Q^2)\simeq 4$
at $\langle Q^2\rangle\simeq 10$ GeV$^2$. 
As we have explained, the latter momentum grows indefinitely as $Q^2$ 
increases, so that in principle such a large value cannot be ruled out.
Hence, we must carefully investigate with which accuracy we are
able to determine each term in Eq.~(\ref{eq:a0anomaly}), particularly
the gluon contribution, by scrutinizing both the available experimental data
and the methodology we use to determine parton distributions from them.
This is exactly the goal of this Thesis: 
we will find that the gluon is still largely uncertain, 
in contrast to somewhat common belief, and that its determination is still a 
challenge in spin physics.

\subsection{Scheme dependence of parton distribution moments}
\label{sec:schemedep}
Beyond leading order, coefficient and splitting functions are no longer 
universal, hence even though the scale dependence of the 
structure function $g_1$ is determined uniquely, at least up to
higher order corrections, its separation into contributions due to
quarks and gluons is scheme dependent (and thus essentially arbitrary).
The NLO coefficient functions may be modified by a change of the 
factorization scheme which is partially compensated by a corresponding change 
in the NLO splitting functions, hence both are required for a 
consistent NLO computation. 
A comprehensive discussion of scheme dependence can be found in 
Ref.~\cite{Leader:2001ab}. Here, we summarize the main features of the 
schemes which are commonly used in the analysis of polarized DIS.
\begin{enumerate}
\item The most popular renormalization scheme is the so-called 
$\overline{\mathrm{MS}}$ scheme~\cite{Mertig:1995ny,Vogelsang:1995vh}, 
in which the first
moment of the gluon coefficient function vanishes. In this scheme, the gluon
density does not contribute to the first moment of the structure function 
$g_1$ and the scale-dependent singlet axial charge is equal to the singlet 
quark first moment:
\begin{equation}
a_0(Q^2)=\left.\int_0^1 dx \Delta\Sigma(x,Q^2)\right|_{\overline{\mathrm{MS}}}
\,\mbox{.}
\label{eq:MSbar1}
\end{equation} 
Also, the first moments of the nonsinglet triplet and octet PDF combinations
\begin{equation}
a_3=\int_0^1dx\Delta T_3(x,Q^2)
\ \ \ \ \ \ \ \ \ \
a_8=\int_0^1dx\Delta T_8(x,Q^2)
\label{eq:a3a8}
\end{equation}
are independent of $Q^2$.
\item An alternatively scheme is the so-called Adler-Bardeen 
scheme~\cite{Ball:1995td}
defined such that the first moment of the singlet PDF combination is
independent of $Q^2$, thus it can be identified with the total
quark helicity.
The polarized gluon density is defined as in the 
$\overline{\mathrm{MS}}$ scheme, but directly contributes to the singlet axial
charge, and consequently to the first moment of $g_1$, which now reads
\begin{equation}
a_0(Q^2)=\left.\int_0^1\Delta\Sigma(x,Q^2)\right|_{\mathrm{AB}}
-n_f\frac{\alpha_s(Q^2)}{2\pi}\int_0^1\Delta g (x,Q^2)
\,\mbox{.}
\label{eq:AB1}
\end{equation}
\end{enumerate}
The relation between the first moments of the singlet quark combination
in AB and $\overline{\mathrm{MS}}$ renormalization schemes
is then simply obtained by comparing Eqs.~(\ref{eq:MSbar1})-(\ref{eq:AB1}).
As discussed above, the difference is proportional to 
$\alpha_s(Q^2)\int_0^1 dx \Delta g(x,Q^2)$ and is due to the anomalous
nonconservation of the singlet axial current. At LO, it is scale-invariant:
this implies that the first moment of the polarized gluon distribution
increases as $1/\alpha_s(Q^2)$ with $Q^2$, hence the gluon contribution in 
Eq.~(\ref{eq:AB1}) is not asymptotically suppressed by powers of $\alpha_s$.
As a consequence, this scheme dependence does not vanish at large $Q^2$, and
the definition of the singlet quark first moment is therefore maximally
ambiguous.

\section{Sum rules}
\label{sec:sumrule}

Moments of structure functions are a powerful tool to study some fundamental 
properties of the nucleon structure, like the total momentum fraction carried 
by quarks or the total contribution of quark spin to the
spin of the nucleon. While a complete description of structure functions 
based on fundamental QCD principles may be unattainable for now, 
moments of structure functions can be directly compared to
rigorous theoretical results, like sum rules, lattice QCD calculations 
and chiral perturbation theory.

The light-cone expansion of the current product in 
Eq.~(\ref{eq:hadronic_tensor}) implies that the $n$-th moments of the
structure functions $g_1$ and $g_2$, at leading twist are given
by~\cite{Leader:2001ab}
\begin{equation}
\int_0^1dx x^{n-1} g_1(x,Q^2)
=
\frac{1}{2}\sum_i\delta_ia_n^iC_{1,i}^{n}(Q^2,\alpha_s)
\ \ \ \ \ \ \ \ \ \ \ \ \ \ \ \ \ \ \ \ \ \ \ \ \ \ \ \ \ \ \ \ \ \ \ \ \ \ \ \ \ \ 
n=1,3,5,\dots
\label{eq:g1sr}
\end{equation}
\begin{equation}
\int_0^1dx x^{n-1}g_2(x,Q^2)
=
\frac{1-n}{2n}\sum_i\delta_i
\left[
a_n^iC_{1,i}^{n}(Q^2,\alpha_s)-d_n^i C_{2,i}^n(Q^2,\alpha_s)
\right]  \ \ \ \ \ \
n=3,5,7,\dots
\label{eq:g2sr}
\end{equation}
where the $\delta_i$ are numerical coefficients, the $C_i^n(Q^2,\alpha_s)$
are the coefficient functions and the $a_n^i$ and $d_n^i$ are related to the
hadronic matrix elements of the local operator. The label $i$ indicates 
what kind of operator is contributing: for flavor-nonsinglet operators,
only quark fields an their covariant derivatives occur.

In the case of the first moment of the $g_1$ structure function, 
Eq.~(\ref{eq:g1sr}) can be recast as
\begin{equation}
\Gamma_1^{p,n}\equiv\int_0^1dx g_1(x,Q^2)
=
\frac{1}{12}
\left[
C_{\mathrm{NS}}(Q^2)\left(\pm a_3+\frac{1}{3}a_8\right)
+\frac{4}{3}C_{\mathrm{S}}(Q^2)a_0
\right]
\,\mbox{,}
\label{eq:gammapn}
\end{equation}
where the plus (minus) sign refers to a proton (neutron) target.
In the above, $a_3$ and $a_8$ are measures of the proton matrix elements
of an $SU(3)$ flavor octet of quark axial-vector currents. 

The octet of axial-vector currents is precisely the set of currents that 
controls the weak $\beta$-decays of the neutron and of the spin-$1/2$ hyperons.
Consequently, $a_3$ and $a_8$ can be expressed in terms of two parameters
$F$ and $D$ measured in hyperon $\beta$ decays~\cite{Nakamura:2010zzi}
\begin{equation}
a_3=F+D=1.2701\pm 0.0025
\,\mbox{,}
\ \ \ \ \ \ \ \ \ \ 
a_8=\frac{1}{\sqrt{3}}(3F-D)=0.585\pm 0.025
\,\mbox{.}
\label{eq:hypdecayconst}
\end{equation}

It follows that a measurement of $\Gamma_1^p(Q^2)$ in polarzed DIS
can be interpreted as a measurement of $a_0(Q^2)$. Indeed, 
Eq.~(\ref{eq:gammapn}) can be rewritten as
\begin{equation}
C_{\mathrm{S}}(Q^2)a_0(Q^2)
=
9\Gamma_1^p(Q^2)-\frac{1}{2}C_{\mathrm{NS}}(Q^2)(3F+D)
\,\mbox{.}
\label{eq:a0fromgammap}
\end{equation}
Since the two terms on the r.h.s. are roughly of the same order, 
the value of $a_0$ arises from a large cancellation between 
them (see \textit{e.g.} Ref.~\cite{Forte:1994dw}).
The present measured value for $a_0$ is still disturbingly small,
as briefly noted at the end of previous Sec.~\ref{sec:QCDevol}.

Finally, in going from the case of a proton to a neutron, $a_0$ and 
$a_8$ in Eq.~(\ref{eq:gammapn}) remain unchanged, whereas $a_3$
reverses its sign. One thus finds the Bjorken sum 
rule~\cite{Bjorken:1966jh,Bjorken:1969mm}
\begin{equation}
\Gamma_1^p(Q^2)-\Gamma_1^n(Q^2)
=
\frac{1}{6}C_{\mathrm{NS}}(Q^2)a_3
\label{eq:bjorken1}
\,\mbox{,}
\end{equation}
which was originally derived from current algebra and isospin asymmetry.
A comparison with experimental data, thus allows for a direct test of
isospin, as well as of the predicted scale dependence. Furthermore,
since the nonsiglet coefficient function $C_{\mathrm{NS}}$ is known
up to three loops~\cite{Larin:1991tj}, Eq.~(\ref{eq:bjorken1}) potentially
provides a theoretically very accurate handle on the strong coupling 
$\alpha_s$~\cite{Altarelli:1998nb}. 
This feature will be discussed in Sec.~\ref{sec:bjorken}
in the framework of a determination of an unbiased parton set
from inclusive polarized DIS data.

Finally, we notice that relations like Eq.~(\ref{eq:gammapn})
can be obtained also for the $g_2$ structure function. 
These include the Burkhardt-Cottingham sum 
rule~\cite{Burkhardt:1970ti} and the Efremov-Leader-Teryaev sum 
rule~\cite{Efremov:1996hd},
for a discussion of which we refer to~\cite{Anselmino:1994gn,Kuhn:2008sy}. 
Here we only notice that
data on $g_2$ are not yet accurate enough for a significant test of them. 

\section{Phenomenology of polarized structure functions}
\label{sec:phenopol}

Experimental information on the structure functions $g_1(x,Q^2)$ and
$g_2(x,Q^2)$ is extracted from measured cross-section asymmetries,
both longitudinal, $A_\parallel$, and transeverse, $A_\perp$. 
These are defined by considering longitudinally polarized leptons 
scattering off a hadronic target, polarized either
longitudinally or transversely with respect to the collision axis, and
read
\begin{equation}
 A_{\parallel}=
\frac{d\sigma^{\rightarrow\Rightarrow}-d\sigma^{\rightarrow\Leftarrow}}
{d\sigma^{\rightarrow\Rightarrow}+d\sigma^{\rightarrow\Leftarrow}}
\,\mbox{;}\quad
 A_{\perp}=
\frac{d\sigma^{\rightarrow\Uparrow}-d\sigma^{\rightarrow\Downarrow}}
{d\sigma^{\rightarrow\Uparrow}+d\sigma^{\rightarrow\Downarrow}}
\,\mbox{.}
\label{eq:xsecasy}
\end{equation}
The numerator of these expressions is given by 
Eqs.~(\ref{eq:asymmetry_parallel})-(\ref{eq:asymmetry_perpendicular}), 
while the denominator is twice the unpolarized cross-section,
Eq.~(\ref{eq:unp_cross_section1}).

Inversion of 
Eqs.~(\ref{eq:asymmetry_parallel})-(\ref{eq:asymmetry_perpendicular}) 
gives the explicit relation between 
the polarized structure functions and the measurable asymmetries
Eq.~(\ref{eq:xsecasy})
\begin{align}
g_1(x,Q^2)&=
\frac{F_1(x,Q^2)}{(1+\gamma^2)(1+\eta\zeta)}
\left[
(1+\gamma\zeta)\frac{A_{\parallel}}{D}-(\eta-\gamma)\frac{A_{\perp}}{d}
\right]
\,\mbox{,}
\label{g1toA}
\\
g_2(x,Q^2)&=
\frac{F_1(x,Q^2)}{(1+\gamma^2)(1+\eta\zeta)}
\left[
\left(\frac{\zeta}{\gamma}-1\right)\frac{A_{\parallel}}{D}
+\left(\eta+\frac{1}{\gamma}\right)\frac{A_{\perp}}{d}
\right]
\,\mbox{,}
\label{g2toA}
\end{align}
where we have defined the kinematic factors
\begin{align}
\label{eq:ddef}
d&=\frac{D\sqrt{1-y-\gamma^2 y^2/4}}{1-y/2}\,\mbox{,}\\
\label{eq:Ddef}
D&=\frac{1-(1-y)\epsilon}{1+\epsilon R(x,Q^2)}\,\mbox{,}\\
\label{eq:etadef}
\eta&=\frac{\epsilon\gamma y}{1-\epsilon(1-y)}\,\mbox{,}\\
\label{eq:zetadef}
\zeta&=\frac{\gamma(1-y/2)}{1+\gamma^2 y/2}\,\mbox{,}\\
\label{eq:epsilondef}
 \epsilon &= \frac{4(1-y) - \gamma^2 y^2}{2 y^2 + 4 (1-y) + \gamma^2 y^2}
\,\mbox{.}
\end{align}
The unpolarized structure function $F_1$ and unpolarized
structure function ratio $R$ which enter the definition
of the asymmetries, Eqs.~(\ref{g1toA})-(\ref{g2toA}),
may be expressed in terms of $F_2$ and $F_L$ by
\begin{eqnarray}\label{eq:fonedef}
F_1(x,Q^2)&\equiv&\frac{F_2(x,Q^2)}{2x\left[1+R(x,Q^2)\right]}
\left(1+\gamma^2\right)
\,\mbox{,}
\\\label{eq:Rdef}
R(x,Q^2)&\equiv&\frac{F_L(x,Q^2)}{F_2(x,Q^2)-F_L(x,Q^2)}
\,\mbox{.}
\end{eqnarray}

The longitudinal and transverse asymmetries are sometimes expressed in terms
of the virtual photo-absorption asymmetries $A_1$ and $A_2$ according
to
\begin{equation}
\label{eq:asyrel}
A_\parallel=D(A_1+\eta A_2)
\mbox{ ,}
\qquad\qquad
A_\perp=d(A_2-\zeta A_1),
\end{equation}
where 
\begin{equation}
\label{eq:gammaasy}
A_1(x,Q^2)
\equiv 
\frac{\sigma^T_{1/2}-\sigma^T_{3/2}}{\sigma^T_{1/2}+\sigma^T_{3/2}}
\mbox{ ,}
\qquad\qquad
A_2(x,Q^2)
\equiv
\frac{2\sigma^{TL}}{\sigma^T_{1/2}+\sigma^T_{3/2}}.
\end{equation}
Recall that $\sigma^T_{1/2}$ and $\sigma^T_{3/2}$
are cross-sections for the scattering of
virtual transversely polarized photons
(corresponding to longitudinal lepton polarization)
with helicity of the photon-nucleon system equal to 1/2 and 3/2
respectively, and $\sigma^{TL}$ denotes the interference term between the
transverse and longitudinal photon-nucleon amplitudes.
In the limit $M^2\ll Q^2$ Eqs.~(\ref{eq:asyrel}) reduce to $D=A_\parallel/A_1$,
$d=A_\perp/A_2$, thereby providing a physical interpretation of
$d$ and $D$ as depolarization factors.

Using Eqs.~(\ref{eq:asyrel}) in Eqs.~(\ref{g1toA})-(\ref{g2toA}) we may
express the structure functions in terms of $A_1$ and $A_2$ instead:
\begin{align}
\label{g1toA1}
g_1(x,Q^2) &= \frac{F_1(x,Q^2)}{1+\gamma^2} \left[ A_1(x,Q^2) 
+ \gamma A_2 (x,Q^2) \right]\,\mbox{,}\\\label{g2toA2}
g_2(x,Q^2)&=
\frac{F_1(x,Q^2)}{1+\gamma^2}
\left[\frac{A_2}{\gamma}- A_1\right]
\,\mbox{.}
\end{align}

We are interested in the structure function $g_1(x,Q^2)$,
whose moments are proportional to nucleon matrix elements of twist-two
longitudinally polarized quark and gluon operators, and therefore can
be expressed in terms of longitudinally polarized quark and gluon distributions.
Using Eqs.~(\ref{g1toA})-(\ref{g2toA}),
we may obtain an expression of it in terms of
the two asymmetries $A_{\parallel}$, $A_{\perp}$, or, using 
Eqs.~(\ref{g1toA1})-(\ref{g2toA2}), in terms of
the two asymmetries $A_1$, $A_2$.
Clearly, up to corrections of
${\mathcal O}\left(M/Q\right)$, $g_1$ is fully determined by
$A_{\parallel}$, which coincides with $A_1$ up to
${\mathcal O}\left(M/Q\right)$ terms, while $g_2$ is
determined by $A_{\perp}$ or $A_2$.
It follows that, even though in principle a measurement of both
asymmetries is necessary for the determination of $g_1$, in practice
most of the information comes from $A_{\parallel}$ or $A_1$, with the
other asymmetry only providing a relatively small correction unless
$Q^2$ is very small. 

It may thus be convenient to express $g_1$ in terms of
$A_{\parallel}$
and $g_2$
\begin{equation}
\label{eq:g1tog2}
g_1(x,Q^2)
=
\frac{F_1(x,Q^2)}{1+\gamma\eta}\frac{A_{\parallel}}{D}
+\frac{\gamma(\gamma-\eta)}{\gamma\eta+1}g_2(x,Q^2)
\,\mbox{,}
\end{equation}
or, equivalently, in terms of  $A_1$ and $g_2$
\begin{equation}
\label{eq:g1tog2p}
g_1(x,Q^2) = A_1(x,Q^2) F_1(x,Q^2) + \gamma^2 g_2(x,Q^2)
\,\mbox{.}
\end{equation}
It is then possible to use Eq.~(\ref{eq:g1tog2}) or
Eq.~(\ref{eq:g1tog2p}) to determine $g_1(x,Q^2)$ from a dedicated
measurement of the longitudinal asymmetry, and an independent
determination of $g_2(x,Q^2)$.

In practice, experimental information on the transverse asymmetry and
structure function $g_2$ is
scarce~\cite{Abe:1998wq,Anthony:2002hy,Airapetian:2011wu}. 
However, the Wilson expansion for polarized DIS implies 
that the structure function $g_2$ can be
written as the sum of a twist-two
and a twist-three contribution~\cite{Wandzura:1977qf}:
\begin{equation}
g_2(x,Q^2)=g_2^{\mathrm{t2}}(x,Q^2)+g_2^{\mathrm{t3}}(x,Q^2).
\end{equation}
The twist-two contribution to $g_2$ is simply related
to $g_1$. One finds
\begin{equation}
g_2^{\mathrm{t2}}(x,Q^2)= -g_1(x,Q^2)+\int_x^1\frac{dy}{y} g_1(y,Q^2)
\label{wweq}
\end{equation}
which in Mellin space becomes
\begin{equation}
g_2^{\mathrm{t2}}(N,Q^2)= -\frac{N-1}{N}g_1(N,Q^2).
\label{wweqN}
\end{equation}
It is important to note that $g_2^{\mathrm{t3}}$ is
not suppressed by a power of $M/Q$ in comparison to
$g_2^{\mathrm{t2}}$, because in the polarized case the availability of the spin
vector allows the construction of an extra scalar
invariant. Nevertheless, experimental evidence 
suggests that $g_2^{\mathrm{t3}}$ is compatible with zero at low scale
$Q^2\sim M^2$. Fits to $g_2^{\mathrm{t3}}$~\cite{Accardi:2009au,Blumlein:2012se}, 
as well as theoretical
estimates of it~\cite{Accardi:2009au,Braun:2011aw} support the
conclusion that
\begin{equation}
g_2(x,Q^2)\approx g_2^{\mathrm{t2}}(x,Q^2)\equiv g_2^{\mathrm{WW}}(x,Q^2)
\,\mbox{,}
\label{eq:wwrel}
\end{equation}
which is known as the  Wandzura-Wilczek~\cite{Wandzura:1977qf}
relation. 
The effect of such an assumption on a determination of parton distributions
from experimental data could be tested to compare to results obtained with
the opposite assumption, \textit{i.e.} $g_2(x,Q^2)=0$.
We will follow this strategy
in our extraction of unbiased polarized PDFs presented in 
Chap.~\ref{sec:chap2}.

\section{Target mass corrections}
\label{sec:TMC}

A large part of experimental data in polarized DIS are taken at relatively 
low values of $Q^2$, typically a few GeV$^2$, and at medium-to-large $x$
values. In this kinematical region, the target-mass factor $\gamma$,
Eq.~(\ref{eq:gamma}), is of order unity with the finite value of the 
nucleon mass $M$, hence contributions which are elsewhere suppressed 
may play a relevant role. These corrections are usually referred to as
kinematic higher-twist terms. Another source of terms suppressed by
inverse powers of $Q^2$ arises from the Wilson expansion of the
hadronic tensor, Eqs.~(\ref{eq:had_tens_sym})-(\ref{eq:had_tens_asym}),
\textit{i.e.} from matrix elements of operators of non-leading twist.
These corrections are referred to as dynamical higher-twist terms.
A detailed analysis of the effect of twist-3 and twist-4 
corrections to $g_1$, as well as a twist-3 correction to the
$g_2$ structure function on the determination of a set of polarized PDFs
from inclusive DIS data has been recently 
presented~\cite{Jimenez-Delgado:2013boa}. They find that 
higher-twist terms could have
a sizable effect, particularly in the high-$x$ and low-$Q^2$ kinematic
region.

In order to include exactly the effect of kinematic target mass corrections,
TMCs henceforth, one has to deal with Eqs.~(\ref{eq:g1tog2})-(\ref{eq:g1tog2p}),
supplemented by some model assumption on $g_2$, as discussed in 
Sec.~\ref{sec:phenopol}. This is required since experimental data on $g_2$
are restricted to a limited range in $(x,Q^2)$ and are affected by large 
uncertainties.
Target mass corrections assume simple expressions 
in Mellin space, where they read~\cite{Piccione:1997zh}
\begin{eqnarray}
\tilde g_1(N,Q^2) 
&=& 
g_1(N,Q^2)+\frac{M^{2}}{Q^2}\frac{N(N+1)}{(N+2)^2}
\nonumber \\
&\times&\left[(N+4)g_1(N+2,Q^2)+4\frac{N+2}{N+1}g_2(N+2,Q^2)\right]
+{\mathcal O}\left(\frac{M^2}{Q^2}\right)^2
\\
\label{g1n1bis}
\tilde g_2(N,Q^2)
&=&
g_2(N,Q^2)+\frac{M^2}{Q^2}\frac{N(N-1)}{(N+2)^2}
\nonumber \\
&\times&\left[N\frac{N+2}{N+1}g_2(N+2,Q^2)-g_1(N+2,Q^2)\right]
+{\mathcal O}\left(\frac{M^2}{Q^2}\right)^2
\,\mbox{.}
\label{g2n1bis}
\end{eqnarray}
Here, we have denoted by $\tilde g_{1,2}(N,Q^2)$ the Mellin space
structure functions with TMCs included, while $g_{1,2}(N,Q^2)$
are the structure functions determined in the  $M=0$ limit.
These expressions can be specialized under assumptions for $g_2$.
In particular, in the Wandzura-Wilczek case, substituting 
Eq.~(\ref{wweqN}) in Eq.~(\ref{g1n1bis}) and taking the inverse 
Mellin transform, we get
\begin{equation}
\tilde g_1(x,Q^2)=
\frac{1}{2\pi i}\int dN\,x^{-N}\left[1
+\frac{M^2x^2}{Q^2}
\frac{(N-2)^2(N-1)}{N^2}\right]g_1(N,Q^2)
\,\mbox{,}
\label{g1tmc1ww}
\end{equation}
where we have shifted $N\to N-2$ in the term proportional to $M^2$.
Inverting the Mellin transform we then obtain
\begin{eqnarray}
\tilde g_1(x,Q^2)
&=&g_1(x,Q^2)+\frac{M^2x^2}{Q^2}
\left[-5g_1(x,Q^2)-x\frac{\partial g_1(x,Q^2)}{\partial x}\right.
\nonumber \\
& & \left.+\int_x^1\frac{dy}{y}\left(8g_1(y,Q^2)
+4g_1(y,Q^2)\log\frac{x}{y}\right)\right]
\,\mbox{.}
\label{g1xWW}
\end{eqnarray}
Conversely, if we simply set $g_2=0$, we have
\begin{equation}
\tilde g_1(x,Q^2)=
\frac{1}{2\pi i}\int dN\,x^{-N}\left[1
+\frac{M^2x^2}{Q^2}\frac{(N^2-4)(N-1)}{N^2}\right]
g_1(N,Q^2)
\,\mbox{,}
\label{g1tmc10}
\end{equation}
whence
\begin{eqnarray}
\tilde g_1(x,Q^2)
&=&g_1(x,Q^2)+\frac{m^2x^2}{Q^2}
\left[-g_1(x,Q^2)-x\frac{\partial g_1(x,Q^2)}{\partial x}\right.
\nonumber \\
& &\left.-\int_x^1\frac{dy}{y}\left(4g_1(y,Q^2)
+4g_1(y,Q^2)\log\frac{x}{y}\right)\right]
\,\mbox{.}
\label{g1x0}
\end{eqnarray}
The usefulness of these realtions will be apparent in Chap.~\ref{sec:chap2},
where we will use them in an unbiased determination of a polarized 
parton set based on inclusive polarized DIS data.

\chapter{Phenomenology of polarized Parton Distributions}
\label{sec:chap3}

In this Chapter, we review how a set of parton distribution functions is
usually determined from a global fit to experimental data. 
In Sec.~\ref{sec:generalstrategy}, we delineate the general strategy
for PDF determination and its main theoretical and methodological
issues, focusing on those which are peculiar to the polarized case.
In Sec.~\ref{sec:NNPDFapproach}, we summarize how some of these 
problems are addressed within the NNPDF methodology, which was 
developed in recent years to provide a statistically sound 
determination of parton distributions and their uncertainties.
In Sec.~\ref{sec:setsummary}, we finally conclude with an overview on 
available polarized PDF sets.

\section{General strategy for \textit{standard} PDF determination}
\label{sec:generalstrategy}

In principle, the general strategy for parton fitting can be simply stated.
Thanks to the factorization theorem, theoretical predictions for the various 
measured observables are expressed as the convolution between 
coefficient functions and parton distributions.
The former are perturbative quantities, computed in field theory 
at desired accuracy, but different for each partonic 
subrocess contributing to the observable.
The latter are non-perturbative objects, but universal, \textit{i.e.}
they do not depend on the observable under investigation. 
In order to determine parton distributions from experimental data, 
they have to be parametrized, usually at an initial energy 
scale $Q_0^2$, and then randomly initialized. They then need to be 
evolved up to the energy scale $Q^2$, relevant for the measurement under 
investigation, by solving Altarelli-Parisi equations.

The agreement between measured observables $\left\{O_i^{\mathrm{(exp)}}\right\}$ and 
corresponding theoretical predictions $\left\{O_i^{\mathrm{(th)}}\right\}$ is quantified 
by a figure of merit, usually chosen as the $\chi^2$ function, 
\begin{equation}
\chi^2=\sum_{i,j}^{N_{\mathrm{dat}}}(O_i^{\mathrm{(exp)}}-O_i^{\mathrm{(th)}})
[\mathrm{cov}_{ij}](O_j^{\mathrm{(exp)}}-O_j^{\mathrm{(th)}})
\,\mbox{,}
\label{eq:erfuncgeneral}
\end{equation}
where $\mathrm{cov}_{ij}$ is the experimental covariance matrix.
If data are provided with no correlated systematics, as the case in most 
polarized measurements, this is the diagonal matrix of the experimental
uncertainties. The best fit is obtained by minimizing 
the figure of merit, Eq.~(\ref{eq:erfuncgeneral}), and the corresponding
best-fit parameters will finally fix the PDF shape.

Despite the apparent simplicity of this strategy, 
the determination of parton distributions from a global set of experimental
data, possibly coming from different processes, is a challenging exercise.
This requires to face several issues, some of which are peculiar to the 
polarized case. We summarize them as follows.

\begin{list}{}{\leftmargin=0pt}
\item {\textbf{Lack of experimental data.}}
Each observable has its own definition in terms of
parton distributions and for this reason a specific observable can
constrain or disentangle some distributions and not all of them. 
In principle, several different observables are needed to determine all the 
$(2n_f+1)$ independent parton components (quark, antiquark and gluon,
with $n_f$ the number of active flavors)
inside the nucleon. If such an information is lacking,
either a subset of parton distributions is determined, or
general assumptions on the unconstrained PDFs have to be made.

For instance, the bulk of experimental data to constrain polarized PDFs
consists of polarized inclusive DIS data. This process only 
allows for the determination of the total quark distributions
$\Delta u^+=\Delta u +\Delta\bar{u}$, $\Delta d^+=\Delta d +\Delta\bar{d}$,
$\Delta s^+=\Delta s +\Delta\bar{s}$, and of the gluon $\Delta g$ (for
details, see Chap.~\ref{sec:chap1}). Information on light sea antiquarks
is provided either by semi-inclusive DIS with identified hadrons 
in the final state (mostly pions or kaons) or $W$ production in 
proton-proton collisions. These reactions actually receive leading 
contributions from partonic subrocesses initiated by $\bar{u}$
and $\bar{d}$ antiquarks, hence they will be able to provide information
on the corresponding spin-dependent 
distributions $\Delta\bar{u}$ and $\Delta\bar{d}$. 

Besides, it is worth noticing that inclusive DIS indirectly 
constrains the polarized gluon, $\Delta g$, through scaling violations.
Unfortunately, these turn out to have a mild effect for its determination, 
due to the small $Q^2$ lever arm of experimental data. 
Observables receiving leading contribution from gluon-initiated partonic
subprocesses will be more suited to probe $\Delta g$ directly. They include
asymmetries for jet and pion production in proton-proton
collisions and for one- or two-hadron and open-charm production in 
fixed-target lepton-nucleon scattering. The theory and phenomenology
of these processes will be discussed in more detail in Chap.~\ref{sec:chap5}.

Finally, we notice that polarized data are less abundant and less accurate 
than their unpolarized counterparts and have a rather limited kinematic
coverage in the $(x,Q^2)$ plane. The various experimental processes 
that provide information on polarized PDFs, together with
the corresponding leading partonic subprocesses, PDFs that are being probed 
and covered kinematic ranges are summarized in Tab.~\ref{tab:polprocesses}.
\begin{table}[t]
\centering
\footnotesize
\begin{tabular}{ccccc}
\toprule
\textsc{reaction} & \textsc{partonic subprocess} & \textsc{probed pdf}
& $x$ & $Q^2$ [GeV$^2$] \\
\midrule
\multirow{2}*{$\ell^\pm\{p,d,n\}\to\ell^\pm X$} &
\multirow{2}*{$\gamma^*q\to q$} &
$\Delta q +\Delta\bar{q}$ &
\multirow{2}*{$0.003\lesssim x \lesssim 0.8$} &
\multirow{2}*{$1\lesssim Q^2 \lesssim 70$}\\
& & $\Delta g$ & 
\\
\multirow{2}*{$\overrightarrow{p}\overrightarrow{p}\to jet(s) X$} &
$gg\to qg$ &
\multirow{2}*{$\Delta g$} &
\multirow{2}*{$0.05\lesssim x \lesssim 0.2$} &
\multirow{2}*{$30\lesssim p_T^2 \lesssim 800$}\\
& $qg\to qg$ & & 
\\
\multirow{2}*{$\overrightarrow{p}p\to W^\pm X$} &
$u_L\bar{d}_R\to W^+$ &
$\Delta u$ $\Delta\bar{u}$ &
\multirow{2}*{$0.05\lesssim x \lesssim 0.4$} &
\multirow{2}*{$\sim M_W^2$}\\
& $d_L\bar{u}_R\to W^-$ & $\Delta d$ $\Delta\bar{d}$ &
\\
\midrule
$\ell^\pm\{p,d\}\to \ell^\pm D X$ &
$\gamma^*g\to c\bar{c}$ &
$\Delta g$ &
$0.06\lesssim x \lesssim 0.2$ &
$0.04 \lesssim p_T^2 \lesssim 4$
\\
\multirow{3}*{$\ell^\pm\{p,d\}\to \ell^\pm h X$} &
\multirow{3}*{$\gamma^*q\to q$} &
$\Delta u$ $\Delta\bar{u}$  &
\multirow{3}*{$0.005\lesssim x \lesssim 0.5$} &
\multirow{3}*{$1\lesssim Q^2\lesssim 60$}
\\
 & & $\Delta d$ $\Delta\bar{d}$ & & \\
 & & $\Delta g$  & & \\
\multirow{2}*{$\overrightarrow{p}\overrightarrow{p}\to \pi X$} &
$gg\to qg$ &
\multirow{2}*{$\Delta g$} &
\multirow{2}*{$0.05\lesssim x \lesssim 0.4$} &
\multirow{2}*{$1\lesssim p_T^2 \lesssim 200$}\\
& $qg\to qg$ & & \\
\bottomrule
\end{tabular}
\mycaption{Summary of polarized processes to determine polarized PDFs. 
For each of them, we show the leading partonic subprocesses, probed polarized
PDFs, and the ranges of $x$ and $Q^2$ that become accessible.
Processes are separated according to the need of using fragmentation functions
for their analysis: processes in the upper part of the Table do not 
involve the fragmentation of the struck quark into an observed hadron, 
while those in the lower part do.}
\label{tab:polprocesses}
\end{table}

\item \textbf{Theoretical issues.}
Many theoretical subtleties may affect the determination
of parton distributions. In particular, we distinguish between issues related
to the QCD analysis and issues related to the methodology adopted
for the fitting of parton distributions. The latter especially include
the choice of the functional form to parametrize PDFs and of the
formalism to propagate uncertainties: they will be addressed separately
in the next paragraphs.

Theoretical details in the QCD analysis concern for instance the
treatment of heavy quark mass effects.
So far, they have been almost completely unaddressed
in the analysis of polarized PDFs. This is because such effects have 
been shown to be relatively small on the scale of present-day
unpolarized PDF uncertainties~\cite{Ball:2011mu}, which are 
rather smaller than those of their polarized counterparts. 
Hence, the effects of heavy quark masses hardly emerge
in polarized experimental data.

Another QCD theoretical issue is related to the treatment of higher-twist
and nuclear corrections. As mentioned in Sec.~\ref{sec:TMC}, 
a large part of experimental data in polarized DIS are taken at relatively 
low values of $Q^2$, typically a few GeV$^2$, and at medium-to-large $x$
values. In this kinematic region, such corrections could play a relevant
role, as demonstrated in Ref.~\cite{Jimenez-Delgado:2013boa}.

Furthermore, assuming exact $SU(3)$ symmetry, 
the first moments of the nonsinglet quark combinations can be 
related to hyperon octet decay constants, see Eqs.~(\ref{eq:hypdecayconst}). 
However, the violation of $SU(3)$ flavor symmetry is debated in the 
literature~\cite{FloresMendieta:1998ii}, 
even though a detailed phenomenological
analysis seems to support it~\cite{Cabibbo:2003cu}.
It should be clear that, due to the lack of experimental information, 
theoretical constraints, such these sum rules,
the positivity of measured cross-sections and the integrability of 
parton distributions,
provide a significant input for determining the shape of PDFs 
in some kinematic regions, as we will discuss in Sec.~\ref{sec:results}.

Finally, we notice that the inclusion of processes involving identified 
hadrons in final states requires the usage of poorly known fragmentation 
functions. Recent work has emphasized the troubles for all available 
fragmentation function sets to describe the most updated inclusive 
charged-particle spectra data at the LHC~\cite{d'Enterria:2013vba}.
Hence, the inclusion of semi-inclusive DIS and collider pion production data
in a global determination of polarized PDFs
is likely to introduce, via fragmentation functions, an uncertainty
which is diffucult to estimate (though it is usually neglected).

\item \textbf{Functional parametrization.}
The choice of the PDF parametrization is a crucial 
point. In principle, since PDFs represent our ignorance of the 
non-perturbative nucleon structure, there should
be complete freedom in choosing their parametric form. 
However, in order to carry out a parton fit, 
one needs to choose a particular functional form for the PDFs at the
initial scale, usually
\begin{equation}
x\Delta f_i(x,Q_0^2)=\eta_iA_ix^{a_i}(1-x)^{b_i}\left(1+\rho_ix^{\frac{1}{2}}+\gamma_ix\right)
\mbox{.}
\label{eq:fixedform}
\end{equation}
Some of the parameters in Eq.~(\ref{eq:fixedform}) can be constrained by 
barion octet decay constants, by Regge interpretation at small-$x$, 
and by constraining to zero parton distributions at $x=1$. 
Arbitrary assumptions are often made on the parameters
to make the fit minimization to converge. Of course, as the number of 
free parameters decreases, the PDF parametrization turns out to be more rigid,
thus introducing a bias in the PDF determination. 

\item \textbf{Error estimates.}
The Hessian formalism~\cite{Martin:2002aw,Pumplin:2001ct} is the most
commonly used method for PDF error determination.
The $\chi^2$ function, Eq.~(\ref{eq:erfuncgeneral}), is quadratically 
expanded about its global minimum
\begin{equation}
\Delta\chi^2=\chi^2-\chi^2_0
=
\sum_{i=1}^{N_{\mathrm{par}}}\sum_{j=1}^{N_{\mathrm{par}}} H_{ij}(a_i-a_i^0)(a_j-a_j^0)
\,\mbox{,}
\label{eq:chihessian}
\end{equation}
with $\chi_0^2=\chi^2(S_0)$ and $\{\mathbf{a}^0\}$ the $\chi^2$ and 
the set of $N_{\mathrm{par}}$ parameters corresponding to the 
best estimate $S_0$ for the PDF set 
$\{\Delta f\}$ respectively; $H_{ij}$ is the Hessian matrix element defined as
\begin{equation}
H_{ij}=\frac{\partial^2\chi^2(\mathbf{\{a\}})}{\partial a_i\partial a_j}
\,\mbox{.}
\label{eq:H}
\end{equation}
The Hessian matrix, Eq.~(\ref{eq:H}), 
has a complete set of $N_{\mathrm{par}}$ orthonormal eigenvectors $v_{ik}$ with 
eigenvalues $\epsilon_k$ defined by
\begin{equation}
\sum_{j=1}^{N_{\mathrm{par}}}H_{ij}\{\mathbf{a}^0\}v_{jk}=\epsilon_k v_{ik}
\,\mbox{,}
\ \ \ \ \ \ \ \ \ \ \
\sum_{i=1}^{N_{\mathrm{par}}} v_{il}v_{jk}=\delta_{lk}
\,\mbox{.}
\label{eq:eigenvectors}
\end{equation}
Moving the parameters around their best value, a shift is observed in 
the $\chi^2$ function, $\Delta\chi^2$.
Each eigenvector determines a direction in the parameter space along which 
the $\chi^2$ variation about the minimum can be expressed in a 
\textit{natural} way. However, the fit quality typically deteriorates 
far more quickly upon variations in some directions than others, hence the 
eigenvalues $\epsilon_k$ are distributed over a wide range that 
covers many orders of magnitude.  
In terms of the diagonalized set of parameters defined with respect to 
the eigenvectors $v_{ij}$
\begin{equation}
z_i=\sqrt{\frac{\epsilon_i}{2}}\sum_{j=1}^{N_{\mathrm{par}}}(a_j-a_j^0)v_{ij}
\label{eq:eigenvar}
\end{equation} 
one obtains $\Delta\chi^2=\sum_{i=1}^{N_{\mathrm{par}}}z_i^2$, \textit{i.e.} 
the region of acceptable fits around the global minimum is 
contained inside a hypersphere of radius $\sqrt{\Delta\chi^2}$.
The scheme for the diagonalization procedure is shown in 
Fig.~\ref{fig:hessemethod}.
\begin{figure}[t]
\centering
\epsfig{width=0.9\textwidth,figure=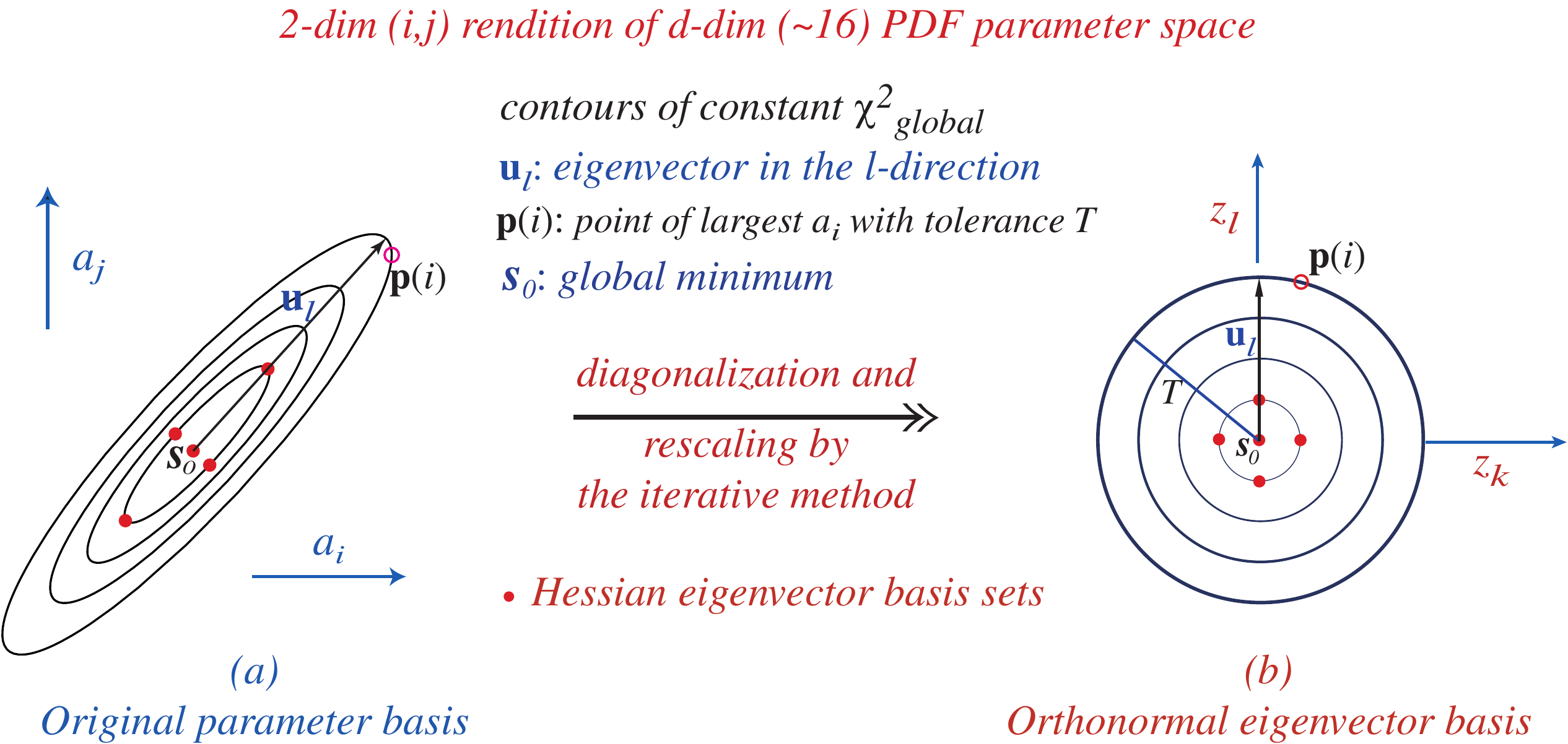}
\mycaption{A schematic representation of the transformation from the 
PDF parameter basis to the orthonormal eigenvector basis as defined by 
Eq.~(\ref{eq:eigenvar}). The figure is taken from Ref.~\cite{Campbell:2006wx}.}
\label{fig:hessemethod}
\end{figure}

An observable $O$, depending on the set of parameters $\{\mathbf{a}\}$
via the PDF parametrization, is assumed to be rather well approximated 
in the neighbourhood of the global minimum by the first term of its
Taylor-series expansion. The deviation of $O$ from its best estimate $O_0$
is then given by $\Delta O = O - O_0\approx \sum_{i=1}^{N_{\mathrm{par}}}O_i z_i$, 
with $O_i\equiv\partial O/\partial z_i |_{\mathbf{z}=\mathbf{0}}$. For a given 
variation $\Delta\chi^2$ of the $\chi^2$, the error estimate on the 
observable $O$ is then evaluated as
\begin{equation}
\Delta O = \left\{\sum_{i=1}^{N_{\mathrm{par}}}\left[O(S_i^+)-O(S_i^-) \right]^2\right\}^{\frac{1}{2}}
\,\mbox{,}
\label{eq:obshessian}
\end{equation}
where $S_i^\pm$ are $2\,N_\mathrm{par}$ sets of PDFs computed at the two 
points defined by
\begin{equation}
z_i^\pm=\pm\frac{\sqrt{\Delta\chi^2}}{2}
\label{eq:zhessian}
\end{equation}
on the edge of the $N_{\mathrm{par}}$-dimensional hypersphere in the $z$ 
parameter space. Together with $S_0$ they form $2N_{\mathrm{par}}+1$ sets
of PDF, that are the ones needed to compute PDFs
errors on $S_0$, and errors 
on the observable $O$ from Eq.~(\ref{eq:obshessian}).

It is worth noticing that the propagation of PDF uncertainties in the Hessian 
method has been derived under the assumption that a first order, linear
approximation is adequate. Of course, due to the complicate nature of a 
global fit, deviations, also from the simple quadratic behavior,
Eq.~(\ref{eq:chihessian}), are inevitable.
Textbook statistics implies that, if
the measurements belong to experimental data sets which are compatible with 
each others, and linear error propagation holds, 
one should have $\Delta\chi^2=1$. 
However, it has been argued that somewhat larger \textit{tolerance} value
$T=\Delta\chi^2$~\cite{Pumplin:2002vw} should be adopted in order for the
distribution of $\chi^2$ values between different experiments in a 
global fit to be reasonable. 
The reasonable value of the tolerance $T$ can be
determined, for instance, by estimating the range of overall 
$\chi^2$ along each of the 
eigenvector directions within which a fit to all data sets can be 
obtained and then averaging the ranges over the $N_{\mathrm{par}}$ 
eigenvector directions. This method has been investigated in detail
in Ref.~\cite{Collins:2001es} by demanding that
indeed $90\%$ of experiments approximately fall within the $90\%$ 
confidence level.
More refined methods involve the determination of a different 
tolerance~\cite{Martin:2009iq} along each Hessian eigenvector 
(the so-called \textit{dynamical} tolerance).
Nevertheless, we should notice that 
the use of $T>1$ is somewhat controversial, 
given that there is no rigorous statistical proof for the criterions 
adopted to estimate it. For all these reasons, error
estimates based on the Hessian method are not necessarily
always accurate.

A way to estimate PDF uncertainties avoiding quadratic approximation 
of $\chi^2$ is provided by the Lagrange multiplier 
method~\cite{Pumplin:2000vx,Stump:2001gu}. This is implemented by 
minimizing a function
\begin{equation}
\Psi(\{a_i\},\{\lambda_i\})=\chi^2(\{a_i\})+\sum_j\lambda_jO_j(\{a_i\})
\label{eq:lagrmultiplier}
\end{equation}
with respect to the set of PDF parameters $\{a_i\}$ for fixed values of the 
Lagrange multipliers $\{\lambda_i\}$. Each multiplier is related to one specific
observable $O_j$, and the choice $\lambda_j=0$ corresponds to the best fit 
$S_0$. By repeating this minimization procedure for many values of $\lambda_j$,
one can map out precisely how the fit to data deteriorates as the expectation
for the observable $O_j$ is forced to change. Unlike the Hessian method, the
Lagrange multiplier technique does not rely on any assumption regarding the
dependence of the $\chi^2$ on the parameters of the fit $\{a_i\}$.

\end{list}

\section{The NNPDF approach to parton fitting}
\label{sec:NNPDFapproach}

In recent years, several polarized PDF sets with uncertainties have been 
released. They slightly differ in the choice of data sets, the form of PDF 
parametrization, and in several details of the QCD analysis, like the
treatment of higher-twist corrections, as will be reviewed with some more
detail in Sec.~\ref{sec:setsummary} below. However, they are all based on the 
\textit{standard} methodology for PDF fitting, based on a fixed functional
parametrization of PDFs and Hessian error estimate. This methodology is 
known~\cite{Forte:2010dt} to run into difficulties especially when
information is scarce, because of the intrinsic bias of fixed parton 
parametrization. This is likely to be particularly the
case for polarized PDFs, which rely on data both less
abundant and less accurate than their unpolarized counterparts.

In order to overcome these difficulties, the NNPDF collaboration
has proposed and developed a new methodology for PDF 
determination~\cite{DelDebbio:2004qj,DelDebbio:2007ee,
Ball:2008by,Ball:2009qv,Ball:2009mk,
Ball:2010de,Ball:2010gb,Ball:2011mu,Ball:2011uy,
Ball:2011eq, Ball:2011gg,Ball:2012cx,Ball:2013hta}.
So far, the NNPDF methodology has been successfully
adopted to determine unpolarized
PDF sets with increasing accuracy, which are now routinely used by the 
LHC collaborations in their data analysis and for data-theory comparisons.
The method is based on a Monte Carlo approach, with neural
networks used as unbiased interpolants.
Monte Carlo sampling allows one to evaluate all quantities, 
such as the uncertainty
or the correlation of PDFs, in a statistically sound way, while
the use of neural networks provides a robust
and flexible parametrization of the parton distributions at the initial scale.
In the following, we will describe in detail the main features of the
NNPDF methodology.

\subsection{Monte Carlo sampling of the probability density distribution}
\label{sec:MCsampling}
Given a PDF or an observable depending on (polarized) PDFs, 
$O(\{\Delta f_i\})$, 
its average is given by the integration - in the functional space 
$V(\{\Delta f_i\})$ 
spanned by the parton distributions and weighted by a suitably defined 
probability measure - of all possible functions describing 
PDFs at a reference scale:
\begin{equation}
\langle O[\Delta f] \rangle =\int_V \mathcal{D}\Delta f\mathcal{P}[\Delta f]O[\Delta f]
\,\mbox{.}
\label{eq:probdensity}
\end{equation}
In the NNPDF approach, the probability measure is represented by a Monte Carlo
sample in the space of PDFs.
An ensemble of replicas of the original data set is generated, 
such that it reproduces the statistical distribution
of the experimental data, followed by its projection into the space of 
PDFs through the fitting procedure. 
Notice that all theoretical assumptions represent a prior for the
determination of such probability measure.
The ensemble in the space of data has to contain all the available 
experimental information. In practice, most data are given with 
multi-Gaussian probability distributions of
statistical and systematic errors, described by a covariance matrix.
In such cases this is the distribution that will be used to 
generate the pseudodata. However, any other probability distribution 
can be used if and when required by the experimental data.
Each replica of the experimental data is a member of the Monte Carlo ensemble 
and contains as many data points as are originally available. 
Whether the given ensemble has the desired statistical features can be verified 
by means of statistical standard tests by
comparing quantities calculated from it with the original 
properties of the data.
Such tests, together with pseudodata generation, will be explicitly discussed
in Sec.~\ref{sec:expdata} for the case of polarized DIS.

Here we should notice that the sampling of the underlying probability
density distribution of data allows one to circumvent the non-trivial
issues related to Hessian propagation of errors. 
Indeed, an ensemble of parton distributions is fitted to pseudodata: 
this means that, at the end of the fitting procedure, one obtains as many 
PDFs as the number of replicas $N_{\mathrm{rep}}$ of the data that were generated.
The experimental values in each replica will fluctuate according to 
their distribution in the Monte Carlo ensemble and the best fit PDFs 
will fluctuate accordingly for each replica. 
Even though individual PDF replicas might fluctuate significantly,
averaged quantities like central value and one-sigma error bands are 
smooth inasmuch as the size of the ensemble increases.
The advantages of the Monte Carlo methodology are then apparent.
First, the expectation value for any observable depending on the PDFs 
or the PDFs themselves can be easily computed as the Monte Carlo average
over the PDF ensemble: Eq.~(\ref{eq:probdensity}) is then replaced by 
\begin{equation}
\langle O[\Delta f]\rangle=\frac{1}{N_{\mathrm{rep}}}\sum_{k=1}^{N_{\mathrm{rep}}}O[\Delta f_k]
\,\mbox{,}
\label{eq:MCmean}
\end{equation}
and similarly uncertainties can be obtained as standard deviations, and so
forth.
Second, non-Gaussian behavior of uncertainties 
can be tested either at the level of 
experimental data, by sampling them according to an arbitrary distribution, 
or at the level of the results in the best fit ensemble of PDFs,
by defining proper confidence levels.
In any case, we do not have to rely on the quadratic assumption, 
Eq.~(\ref{eq:chihessian}), made in the Hessian approach.
Finally, the stability of results upon a change of parametrization 
can be verified by standard statistical tools, for instance by 
computing the distance between results
in units of their standard deviation.  
Likewise, it is possible to verify that fits performed by removing 
data from the set have wider error bands but remain compatible within 
these enlarged uncertainties, or to address how results change within different 
theoretical assumptions. 
The reliability of the results can thus be assessed directly.

\subsection{Neural network parametrization}
\label{sec:NNparametrization}
The Monte Carlo technique adopted to propagate the experimental error 
into the space of PDFs is completely independent of the method used 
to parametrize parton distributions; it might well be used along with 
standard parametrization, Eq.~(\ref{eq:fixedform})~\cite{Signori:2013mda}. 
On the other hand, in order to get a faithful determination of 
parton distributions, one ought to make sure that the chosen functional form 
is redundant enough not to introduce a theoretical bias which would 
artificially reduce parton uncertainty in regions where data
do not constrain enough PDFs. 
There are several ways for obtaining such a redundant
parametrization. One may use some clever polynomial basis, or more refined tools
such as self-organising maps~\cite{Askanazi:2013ota}. 

Within the NNPDF methodology, each parton distribution is parametrized by 
a neural network, which provide a redundant and minimally 
biased parametrization. The only theoretical assumption is smoothness,
a presumed feature of PDFs, 
which is ensured by the flexibility and adaptability of
neural networks. In particular, we use feed-forward neural 
networks~\cite{Forte:2002fg}. They are made of
a set of interconnected units, called neurons, eventually 
organized in groups, called layers. The state or activation of a given
neuron $i$ in a given layer $l$, $\xi_i^{(l)}$,
is a real number, determined as a function of the activation
of the neurons connected to it, namely those in the previous, $l-1$, layer. 
Each pair of neurons $(i,j)$ is then connected by a synapsis, characterized
by a real number $\omega_{ij}^{(l-1)}$, called weight. The activation
of each neuron $i$ in a given layer $l$ is a function $g$
of the difference between a weighted average of input from 
neurons in the preceding layer and a threshold $\theta_i^{(l)}$:
\begin{equation}
\xi_i^{(l)}=g
\left( 
\sum_{j=1}^{N_{l-1}}\omega_{ij}^{(l-1)}\xi_j^{(l-1)}-\theta_i^{(l)}
\right)
\,\mbox{,}
\label{eq:NNactivation}
\end{equation}
where $N_{l-1}$ is the number of neurons in the $(l-1)^{\mathrm{th}}$ layer.
The input and output vectors are labeled as $\mathbf{\xi}^{(1)}$ and 
$\mathbf{\xi}^{(L)}$ respectively, with $L$ the number of layers in the 
network.

The activation function $g$ is in general non-linear. 
The simplest example of activation function $g(x)$ is the step function
$g(x) = \Theta(x)$, which produces binary activation only. However,
it turns out to be advantageous to use an activation function
with two distinct regimes, linear
and non-linear, such as the sigmoid
\begin{equation}
g(x)=\frac{1}{1-e^{-\beta x}}
\,\mbox{.}
\label{eq:sigmoid}
\end{equation}
This function approaches the step function at large
$\beta$; without loss of generality, in the NNPDF methodology we 
usually take $\beta=1$. The sigmoid activation function has a linear 
response when $x\approx 0$, and it saturates
for large positive or negative arguments. 
If weights and thresholds are such that the sigmoids
work on the crossover between linear and saturation regimes, 
the neural network behaves in a
non-linear way. Thanks to this non-linear behavior, the 
neural network is able to reproduce nontrivial functions.

Basically, multilayer feed-forward neural networks provide a non-linear map
between some input $\mathbf{\xi}^{(1)}_i$ and output $\mathbf{\xi}^{(L)}_j$ 
variables, parametrized by weights, thresholds and activation function,
\begin{equation}
\mathbf{\xi}^{(L)}=F
\left[
\mathbf{\xi}^{(1)};\{\omega_{ij}^{(l)}\},\{\theta_i^{(l)}\};g
\right]
\,\mbox{.}
\label{eq:NNmap}
\end{equation}
For given activation function, the parameters can be tuned in such 
a way that the neural
network reproduces any continuous function. The behavior
of a neural network is determined
by the joint behavior of all its connections and thresholds, 
and it can thus be built to be
redundant, in the sense that modifying, adding or removing a
neuron has little impact on the final output. 
Because of these reasons, neural networks can be considered to be robust,
unbiased universal approximants.

In Sec.~\ref{sec:minim}, we will explicitly discuss how polarized PDFs 
can be parametrized in terms of neural networks. 
In particular, we will discuss their architecture
and preprocessing of data to enforce the asymptotic small- and large-$x$ 
behavior of PDFs.

\subsection{Minimization and stopping}
\label{sec:minstop}
Once each independent PDF is parametrized in terms of neural networks at
an inital reference scale $Q_0^2$, physical observables are computed by 
convolving hard kernels with PDFs evolved to the scale of the experimental 
measurements by Altarelli-Parisi evolution.
The best-fit set of parton distributions is determined by comparing 
the theoretical computation of the observable 
with its experimental value, for each Monte Carlo replica.
This is performed by evaluating
a suitable figure of merit, \textit{e.g.} Eq.~(\ref{eq:erfuncgeneral}).

Both the minimization and the determination of the best-fit
in the wide, non-local, space of parameters spanned by the neural network 
parameters are delicate issues.  
To minimize the error function, a genetic algorithm is used.
The main advantage of such an algorithm 
is that it works on a population of solutions, rather than tracing
the progress of one point through parameter space. 
Thus, many regions of parameter space are explored simultaneously, 
thereby lowering the possibility of getting trapped
in local minima.

The basic idea underlying the genetic algorithm is the following.
Starting from a randomly chosen set of parameters, a
pool of possible new sets is generated by mutation of one or more parameters 
at a time. Each new set that has undergone mutation is a mutant. 
A value of the figure of merit that is being minimized corresponds
to each set of parameters, so those configurations 
that fall far away from the minimum can be discarded. 
This procedure is iterated over a sufficiently large number of generations.
Of course, one has to be sure that the final set of parameters corresponds
to acceptable PDFs, \textit{i.e.} they must satisfy theoretical constraints
like positivity of cross-sections and sum rules. Usually, this requirement
is fulfilled by penalizing unacceptable replicas during the minimization
by arbitrarily increasing their figure of merit. In Sec.~\ref{sec:minim},
we will discuss technical details related to how this issue is faced 
in the determination of polarized distributions.

The redundancy of the parametrization also implies another subtle problem,
known as \textit{overlearning}, which happens when neural networks
start to fit statistical fluctuations of data, rather than their
underlying physical law.
The solution to this problem is achieved using a cross-validation method 
to determine a criterion for the fit to stop before entering 
the overlearning regime. Technical details about the specific implemetation of
stopping in the polarized case will be extensively discussed in 
Sec.~\ref{sec:minim}.

\begin{figure}[t]
\centering
\epsfig{width=0.7\textwidth,figure=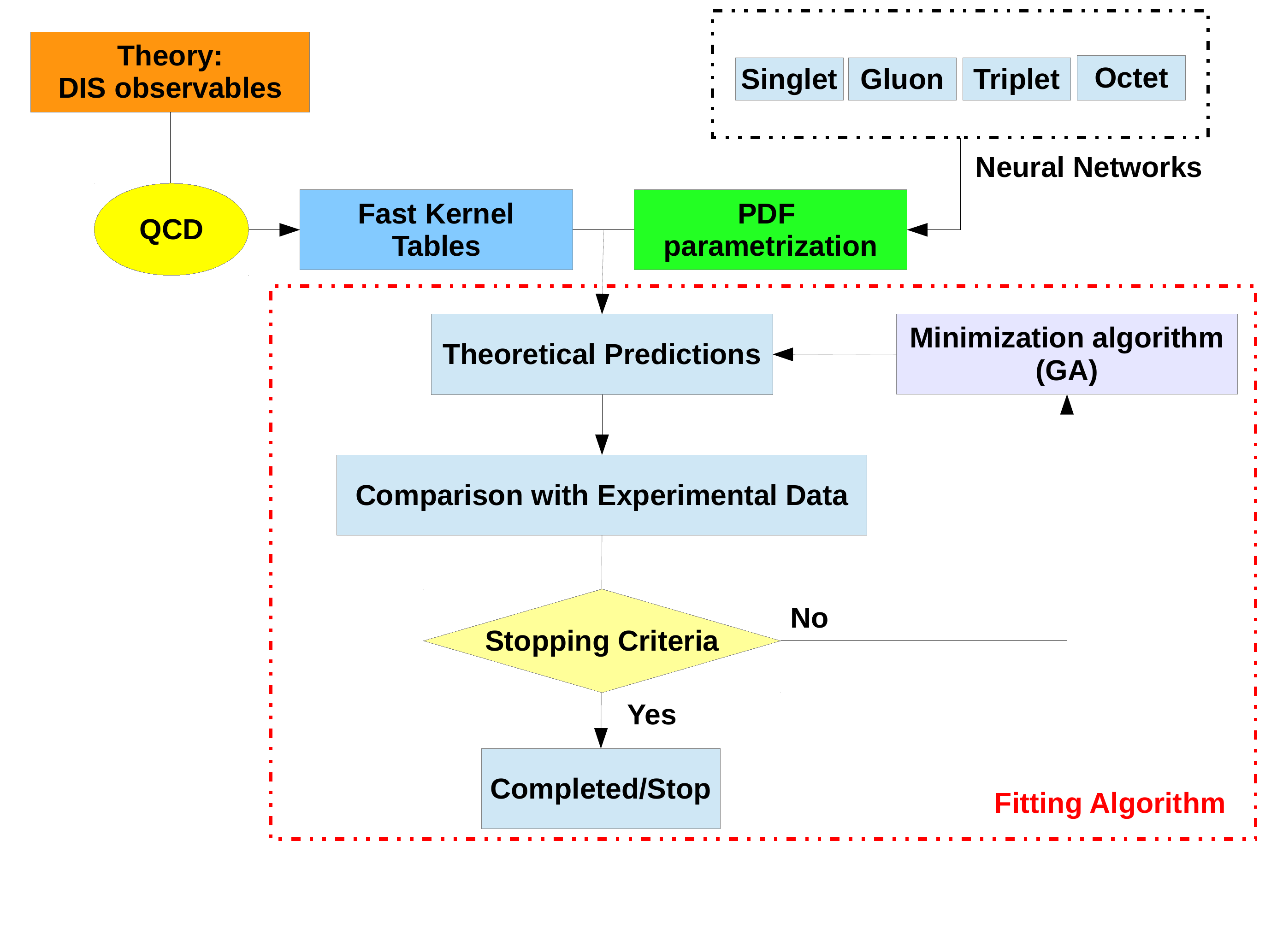}
\mycaption{Scheme of the NNPDF methodology for parton fitting.}
\label{fig:scheme} 
\end{figure}

The main ingredients of the NNPDF methodology described above,
namely Monte Carlo sampling of data distribution, neural network 
parametrization of PDFs, minimization and stopping, are finally sketched in 
Fig.~\ref{fig:scheme}.

\subsection{Bayesian reweighting of Monte Carlo PDF ensembles}
\label{sec:breweighting}
Monte Carlo sampling of the underlying probability density distribution of 
data used in the NNPDF methodology allows for applying
standard statistical tools to the resulting PDF ensembles.
Most importantly, Bayesian inference can be exploited to determine the 
impact on PDFs of new data sets that were not included in a fit. 
The methodology, referred to as reweighting, was presented in detail in
Refs.~\cite{Ball:2010gb,Ball:2011gg}. In short, the main idea underlying 
reweighting is to assign to each replica in the PDF ensemble a weight
which assesses the probability that this replica agrees with new data.
The expectation value for an observable $O$, taking into account the new data,
is then given by the weighted average
\begin{equation}
\langle O[\Delta f]\rangle_{\mathrm{new}} =\int_V \mathcal{D}\Delta f
\mathcal{P}_{\mathrm{new}}[\Delta f]O[\Delta f]
=
\frac{1}{N_{\mathrm{rep}}}\sum_{k=1}^{N_{\mathrm{rep}}}w_k O[\Delta f_k]
\,\mbox{.}
\label{eq:obsrw}
\end{equation}
The weights $w_k$ are computed from the $\chi^2$ of the new data to the 
prediction obtained using a given replica, according to the formula
\begin{equation}
w_k=\frac{(\chi_k^2)^{\frac{1}{2}(n-1)}e^{-\frac{1}{2}\chi_k^2}}
{\frac{1}{N_{\mathrm{rep}}}\sum_{k=1}^{N_{\mathrm{rep}}}(\chi_k^2)^{\frac{1}{2}(n-1)}e^{-\frac{1}{2}\chi_k^2}}
\,\mbox{,}
\label{eq:weightformula}
\end{equation}
where $n$ is the number of new data. The formula, Eq.~(\ref{eq:weightformula}),
is derived under the assumption that new data have Gaussian errors and that
they are statistically independent of the old data. By the law of multiplication
of probabilities it then follows that
\begin{equation}
\mathcal{P}_{\mathrm{new}}
=
\mathcal{N}_\chi\mathcal{P}(\chi|\Delta f)\mathcal{P}_{\mathrm{old}}(\Delta f)
\,\mbox{,}
\label{eq:multprob}
\end{equation}
where $\mathcal{N}_\chi$ is a normalization factor chosen such that 
$\sum_{k=1}^{N_{\mathrm{rep}}}w_k=N_{\mathrm{rep}}$ and
\begin{equation}
w_k=\mathcal{N}_\chi\mathcal{P}(\chi|\Delta f_k)
\propto
(\chi_k^2)^{\frac{1}{2}(n-1)}e^{-\frac{1}{2}\chi_k^2}
\,\mbox{.}
\label{eq:weightb}
\end{equation}

Notice that, after reweighting a given PDF ensemble of $N_{\mathrm{rep}}$ 
replicas, 
the efficiency in describing the distribution of PDFs is no longer the same. 
In fact, the weights give the relative importance of the different
replicas, and the replicas with very small weights will become almost 
irrelevant in ensemble averages.
The reweighted replicas will thus no longer be as efficient as the old:
for a given $N_{\mathrm{rep}}$, the accuracy of the representation of 
the underlying distribution 
$\mathcal{P}_{\mathrm{new}}(\Delta f)$ will be less than it would be 
in a new fit. This loss of efficiency
can be quantified using the Shannon entropy to determine the effective number 
of replicas left after reweighting:
\begin{equation}
N_{\mathrm{eff}}
=
exp
\left\{
\frac{1}{N_{\mathrm{rep}}}\sum_{k=1}^{N_{\mathrm{rep}}}w_k\ln\left(\frac{N_{\mathrm{rep}}}{w_k}\right) 
\right\}
\label{eq:shannon}
\,\mbox{.}
\end{equation}
Clearly $0<N_{\mathrm{eff}}<N_{\mathrm{rep}}$: the reweighted fit has the same 
accuracy as a refit with $N_{\mathrm{eff}}$ replicas. Hence, if 
$N_{\mathrm{eff}}$ becomes too low, the reweighting procedure will no longer
be reliable, either because the new data contain a lot of information
on the PDFs, neccessitating a full refitting, or
because the new data are inconsistent with the old.

Bayesian reweighting allows for the inclusion of new pieces of 
experimental information in an ensemble of PDFs without performing a new fit. 
This is desirable, 
particularly when dealing with observables for the computation of
which no fast code is available. Also notice that, from a conceptual point 
of view, a determination of a PDF set might be performed by including all
data through reweighting of a first reasonable guess, 
called \textit{prior}, as suggested in
Refs.~\cite{Giele:1998gw,Giele:2001mr}. 
Typically, this is the result of a previous PDF fit to 
data (other than those with which 
the parton set will be reweighted) or a model
based on some theoretical assumptions. The reweighting method  
works fine provided the prior is unbiased,
so that one can check that the final results do not depend on the
choice of the initial guess. 
We will devote Chap.~\ref{sec:chap5}
to the discussion of this aspect
in a particular case of physical interest: 
we will reweight a DIS-based parton set with jet and $W$ 
production data from polarized proton-proton collisions.

Results obtained via reweighting of existing Monte Carlo 
PDF sets were shown to be
statistically equivalent to those obtained via global 
refitting: in particular, the method was validated in the unpolarized case
by considering inclusive jet and LHC $W$ lepton asymmetry 
data~\cite{Ball:2010gb,Ball:2011gg}.
As a further illustration of possible
applications of the method, reweighting was used to quantify the impact 
on unpolarized PDFs
of direct photon data~\cite{d'Enterria:2012yj} and of top quark production
data~\cite{Czakon:2013tha}, and to study the constraints on nuclear PDFs
from LHC pPb data~\cite{Armesto:2013kqa}.
Also, notice that the same technique can be straightforwardly 
applied in the case of Hessian PDF sets~\cite{Watt:2012tq}.

Once a reweighted PDF set has been determined, it would be interesting 
to be able to produce a new PDF ensemble with the same probability 
distribution as a reweighted set, but without the need to include the 
weight information. A method of unweighting has therefore been developed, 
whereby the new set is constructed by deterministically sampling
with replacement the weighted probability 
distribution~\cite{Ball:2011gg}. 
This means that replicas with a
very small weight will no longer appear in the final unweighted set 
while replicas with large weight will occur repeatedly.

If the probability for each replica and the probability cumulants 
are defined as
\begin{equation}
p_k=\frac{w_k}{N_{\mathrm{rep}}}
\ \ \ \ \ \ \ \ \ \
p_k\equiv p_{k-1}+p_k=\sum_{j=0}^k p_j
\,\mbox{,}
\label{eq:cumulants}
\end{equation}
it is possible to quantitatively describe the unweighting procedure. 
Starting with $N_{\mathrm{rep}}$ replicas with weights $w_k$, $N_{\mathrm{rep}}$ 
new weights $w_k^\prime$ are determined:
\begin{equation}
w_k^\prime=\sum_{j=1}^{N_{\mathrm{rep}}^\prime}
\theta
\left( 
\frac{j}{N_{\mathrm{rep}}^\prime}-p_{k-1}
\right)
\theta
\left(
p_k-\frac{j}{N_{\mathrm{rep}}^\prime}
\right)
\,\mbox{.}
\label{eq:unweighting}
\end{equation}
These weights are therefore either zero or a positive integer. 
By construction they satisfy
\begin{equation}
N_{\mathrm{rep}}^\prime\equiv\sum_{k=1}^{N_{\mathrm{rep}}}w_k^\prime
\,\mbox{,}
\end{equation}
\textit{i.e.} the new unweighted set consists of $N_{\mathrm{rep}}$ replicas, 
simply constructed by taking $w_k$ copies of the $k$-th replica, 
for all $k=1, \dots, N_{\mathrm{rep}}$. 

In Chap.~\ref{sec:chap5}, Bayesian reweighting of polarized parton 
distributions, determined within the
NNPDF methodology from inclusive DIS data, will be used to assess the impact
of open-charm production data in fixed-target DIS as well as 
$W$ and jet production data in proton-proton collisions. 
A new polarized parton set will be then determined via unweighting.

\section{Overview of available polarized parton sets}
\label{sec:setsummary}

First studies of the polarized structure of the nucleon were aimed at
an accurate determination of polarized first moments, including detailed
uncertainty estimates~\cite{Ball:1995td,Altarelli:1996nm,Altarelli:1998nb},
but did not attempt a determination of a full PDF set. This was first
proposed in Ref.~\cite{Gehrmann:1995ag}, but without uncertainty estimation.
Several polarized parton sets have been delivered in recent years, all
at NLO accuracy, usually
updated with the inclusion of new data, theoretical or statistical features.
In this Section, we review the main features of the presently available 
polarized PDF sets; the discussion about recent developments in the 
determination of a polarized parton set based on the NNPDF methodology will be 
addressed in Chaps.~\ref{sec:chap2}-\ref{sec:chap4}-\ref{sec:chap5}.

\subsection{DIS-based fits}
The bulk of experimental information on longitudinally polarized 
nucleon structure consists of inclusive, photon induced, deep-inelastic
scattering with both polarized charged lepton beams and nucleon targets.
Actually, only data sets coming from this process are included in most
of the polarized PDF determinations, as follows.

\begin{list}{}{\leftmargin=0pt}
\item {\textbf{BB10.}}
The \texttt{BB10} parton set~\cite{Blumlein:2010rn} includes 
world-available data on the asymmetry $A_1$ or the structure function $g_1$
from polarized inclusive DIS. Four independent polarized PDFs are determined
there, namely the valence combinations 
$\Delta u^-=\Delta u-\Delta\bar{u}$ and $\Delta d^-=\Delta d-\Delta\bar{d}$,
the gluon, $\Delta g$ and the quark sea, $\Delta\bar{q}$, assuming symmetric
sea, $\Delta\bar{u}=\Delta\bar{d}=\Delta\bar{s}=\Delta s$. Each PDF is 
parametrized in terms of a fixed functional form like that of 
Eq.~(\ref{eq:fixedform}),
but only a subset of them are actually taken to be free, depending on the
PDF. Also the QCD scale $\Lambda_{\mathrm{QCD}}$ is a parameter to be determined 
in the fit. Errors are determined within the standard Hessian approach 
without any tolerance criterion. The analysis is supplemented by the 
inclusion of experimental systematic uncertainties from different sources,
namely data, NMC parametrization of $F_2$~\cite{Arneodo:1995cq}
and $R$ parametrization~\cite{Whitlow:1990gk} both entering
Eq.~(\ref{g1toA}). 
Also, theoretical systematic uncertainties were estimated from a
variation of the factorization and renormalization scale and input parametrization 
scale. Finally, higher twist contributions from heavy flavor Wilson
coefficients in fixed-flavor number (FFN) scheme were included in the
theoretical QCD analysis.

The \texttt{BB10} parton set is publicly available together with a 
\texttt{FORTRAN} program which allows for computing PDF central values
and errors.

\item {\textbf{JAM13.}}
The \texttt{JAM13} parton set~\cite{Jimenez-Delgado:2013boa}, instead of
the asymmetry $A_1$ or the structure function $g_1$, directly fits the 
measured longitudinal and transverse asymmetries $A_\parallel$ and 
$A_\perp$, Eqs.~(\ref{eq:xsecasy}). This makes a differencee with 
most of other analyses, which usually include the information from the
$g_1$ structure function extracted from observed asymmetries by each 
experimental collaboration within different assumptions. Besides, 
a large number of preliminary inclusive DIS data from JLAB are included
in the \texttt{JAM13} study. Since these data lie in the large-$x$ and 
small-$Q^2$ kinematic region, they are expected to be particularly sensitive 
to target mass and higher-twist effects. For this reason, both twist-$3$ 
and twist-$4$ corrections to
$g_1$, as well as a twist-$3$ correction to the $g_2$ structure function
are taken into account. Moreover, they consistently apply the nuclear
smearing corrections to both the
$g_1$ and $g_2$ structure functions for both deuterium and
$^3\mathrm{He}$, within the framework of the weak binding 
approximation~\cite{Ethier:2013hna,Kulagin:2007ph,Kulagin:2008fm}.

Six independent polarized PDFs are determined, 
namely the total quark combinations,
$\Delta u^+=\Delta u+\Delta\bar{u}$ and $\Delta d^+=\Delta d+\Delta\bar{d}$,
the gluon, $\Delta g$, and the quark sea, $\Delta\bar{q}$ ($q=u, d, s$)
and assuming a symmetric strangeness $\Delta s =\Delta\bar{s}$.
These are parametrized with the functional form  Eq.~(\ref{eq:fixedform});
however, since the experimental piece of information does not allow
for a complete determination of all six PDFs above, several additional
contraints are adopted. In particular, the $\Delta\bar{u}$ and $\Delta\bar{d}$
parton distributions, which do not contribute directly to the description
of inclusive DIS data included in the analysis, are fixed by requiring
\begin{equation}
\lim_{x\to 0}\Delta\bar{q}(x,Q_0^2)=\frac{1}{2}\lim_{x\to 0}\Delta q^+(x,Q_0^2)
\,\mbox{,}
\label{eq:seafixing}
\end{equation}
with $q=u,d$. In addition, in order to avoid unphysical results and provide 
\textit{reasonable} values for all distributions, the following 
constraint is imposed 
\begin{equation}
\frac{1}{2}
\left(
\left|\frac{\Delta\bar{q}^{(2)}}{\Delta\bar{s}^{(2)}} \right|
+
\left|\frac{\Delta\bar{s}^{(2)}}{\Delta\bar{q}^{(2)}} \right|
\right)
=
1\pm 0.25
\,\mbox{.}
\label{eq:seafixing1}
\end{equation}
Both Eq.~(\ref{eq:seafixing}) and Eq.~(\ref{eq:seafixing1}) entail
the decrease in the number of free parameters whose values are actually fitted
to data. The choice of limiting the flexibility of the parametrization
given by Eq.~(\ref{eq:fixedform}) is also made for the gluon, for which 
the values of the exponents $a_g$ and $b_g$ are somewhat arbitrarily fixed. 
The propagation of uncertainties is performed within the Hessian approach
without tolerance criterion.

\item {\textbf{ABFR98.}}
The \texttt{ABFR98} parton set~\cite{Altarelli:1998nb} 
is based on a less updated set of inclusive
DIS experimental data and provides four polarized PDFs, both the 
total combinations $\Delta q^+$ and the gluon $\Delta g$, within
fixed functional parametrization and Hessian error estimate.
Despite the \texttt{ABFR98} is less recent than 
other analyses discussed here, it should be worth mentioning it at least for
two reasons. First, this provides polarized PDFs that are fitted in the
AB renormalization scheme (see Sec.~\ref{sec:QCDevol}) instead of 
$\overline{\mathrm{MS}}$ used in all other PDF determinations. Second,
it includes a detailed discussion of theoretical uncertainties
originated by neglected higher orders, higher twists, 
position of heavy quark thresholds, value of
the strong coupling, violation of SU(3) flavor symetry 
and finally uncertainties related to the choice of the functional form.

\end{list}

\subsection{Global PDF fits}
In recent years, the knowledge on longitudinally polarized
nucleon structure has been supplemented by data coming from processes
different from polarized inclusive DIS. As outlined 
in Sec.~\ref{sec:generalstrategy},
these include fixed-target SIDIS and hadron or jet production in 
polarized proton-proton collisions. Including these pieces of experimental
information in a global NLO analysis 
is a challenging task, both theoretical and computational, because of the 
need to deal with hadronic observables. They may also depend on the 
fragmentation of quarks in the final measured hadrons and hence their
analysis require the usage of poorly known fragmentation functions.
Several such global parton sets have been determined in very recent years,
as summarized as follows.

\begin{list}{}{\leftmargin=0pt}

\item {\textbf{AAC08.}}
Besides polarized inclusive DIS data, the \texttt{AAC08} 
analysis~\cite{Hirai:2008aj} 
also includes $\pi^0$ production data at RHIC, via a $K$-factor 
approximation for the NLO corrections. Because these data only 
provide constraints on the gluon polarization, only four idependent
PDFs are determined: they are the same as in the \texttt{BB10} analysis
(symmetric sea is also assumed), but the functional form they use reads
\begin{equation}
\Delta f(x,Q_0^2) = \left[ \delta x^\nu - \kappa (x^\nu - x^\mu)\right]
f(x,Q_0^2)
\,\mbox{,}
\label{eq:ACCfuncform}
\end{equation}
where $\delta$, $\mu$, $\nu$, $\kappa$ are free parameters to be determined 
in the fit and $f(x,Q_0^2)$ is the corresponding unpolarized PDF
which was taken from Ref.~\cite{Gluck:1998xa}. For the description 
of the fragmentation into a pion, the \texttt{HKNS07} set~\cite{Hirai:2007cx}
is used.
Error estimates are handled via standard Hessian approach, but a tolerance
$\Delta\chi^2=12.95$ is assumed in order for the
distribution of $\chi^2$ values between different experiments 
in the global fit to be reasonable.

The \texttt{AAC08} parton set is publicly available together with a 
\texttt{FORTRAN} program which allows for computing PDF central values
and uncertainties.

\item {\textbf{LSS10.}}
The \texttt{LSS10} parton set~\cite{Leader:2010rb} is based on 
world-available data from incluisve and semi-inclusive DIS. These allow
for a determination of light antiquarks. Six polarized PDFs are parametrized
according to Eq.~(\ref{eq:fixedform}), namely the total PDF combinations
$\Delta u^+$ and $\Delta d^+$, the antiquarks $\Delta\bar{u}$, 
$\Delta\bar{d}$ and $\Delta\bar{s}$ and the gluon $\Delta g$
(for the four latter $\rho_i=0$ in Eq.~(\ref{eq:fixedform})).
The assumption $\Delta s=\Delta\bar{s}$ is also made. Hessian error 
propagation is performed assuming $\Delta\chi^2=1$. Fragmentations 
functions are taken from the \texttt{DSS07} 
analysis~\cite{deFlorian:2007aj,deFlorian:2007hc}.
Similarly to the \texttt{JAM13} fit, the theoretical QCD analysis 
takes into account the $1/Q^2$ terms, arising from kinematic target mass 
corrections and dynamic higher twist corrections, in the expression of the
nucleon spin structure function $g_1$.

The \texttt{LSS10} parton set is publicly available together with a 
\texttt{FORTRAN} program which allows for computing PDF central values.

\item {\textbf{DSSV.}}
Different PDF sets belong to the \texttt{DSSV} family, due to the
remarkable effort put by this collaboration in updating their 
polarized parton sets with new available data. The first global
analysis to include polarized collider measurements at RHIC, 
besides inclusive and semi-inclusive polarized DIS data,
was \texttt{DSSV08}~\cite{deFlorian:2008mr,deFlorian:2009vb}.
Semi-inclusive pion and single-inclusive jet production data
were considered in their study. This was recently updated by
the \texttt{DSSV+} and \texttt{DSSV++} 
fits~\cite{deFlorian:2011ia,Aschenauer:2013woa}; in particular, the latter
includes the most recent jet production data at RHIC, which were found
to constrain the gluon shape in the mid-to-large $x$ region with
unprecedented accuracy.

Six independent polarized PDFs are determined from data, 
as in the \texttt{LSS10} analysis,
and the \texttt{DSS07} fragmentation functions
are used. The uncertainty estimates are provided 
through the Lagrange multiplier method described 
in Sec.~\ref{sec:generalstrategy}
with the conservative assumption $\Delta\chi^2/\chi^2=2\%$,
even though the standard Hessian approach, with $\Delta\chi^2=1$,
is also used for comparison. The \texttt{DSSV08} set is publicly available
as $38$ hessian eigenvector sets (a set for each minimized 
parameter for each direction of variation) 
plus a central value set.

\end{list}
\begin{table}[t]
\footnotesize
\centering
\begin{tabular}{lccccc}
\toprule
Fit & 
Ref. & 
Data set(s) &
Scheme &
Parton Distributions &
Uncertainties\\
\midrule
\texttt{BB10} & \cite{Blumlein:2010rn} & DIS & $\overline{\mathrm{MS}}$
& $\Delta u^-$, $\Delta d^-$, $\Delta\bar{q}$, $\Delta g$
& Hessian $\Delta\chi^2=1$\\
\texttt{JAM13} & \cite{Jimenez-Delgado:2013boa} & DIS & $\overline{\mathrm{MS}}$
& $\Delta u^+$, $\Delta d^+$, $\Delta\bar{u}$, 
$\Delta\bar{d}$, $\Delta\bar{s}$, $\Delta g$
& Hessian $\Delta\chi^2=1$\\
\texttt{ABFR98} & \cite{Altarelli:1998nb} & DIS & AB
& $\Delta u^+$, $\Delta d^+$, $\Delta s^+$, $\Delta g$
& Hessian $\Delta\chi^2=1$\\
\texttt{AAC08} & \cite{Hirai:2008aj} & DIS, $\pi^0$ & 
$\overline{\mathrm{MS}}$
& $\Delta u^+$, $\Delta d^+$, $\Delta s^+$, $\Delta g$
& Hessian $\Delta\chi^2=12.95$\\
\texttt{LSS10} & \cite{Hirai:2008aj} & DIS, SIDIS & 
$\overline{\mathrm{MS}}$
& $\Delta u^+$, $\Delta d^+$, $\Delta\bar{u}$, $\Delta\bar{d}$, 
$\Delta\bar{s}$, $\Delta g$
& Hessian $\Delta\chi^2=1$\\
\multirow{2}{*}{\texttt{DSSV08}} & 
\multirow{2}{*}{\cite{deFlorian:2009vb}} & 
DIS, SIDIS, & 
\multirow{2}{*}{$\overline{\mathrm{MS}}$} 
& \multirow{2}{*}{$\Delta u^+$, $\Delta d^+$, $\Delta\bar{u}$, $\Delta\bar{d}$, 
$\Delta\bar{s}$, $\Delta g$}
& Hessian $\Delta\chi^2=1$\\
& & $\pi^0$, Jets & & &
Lagr. mult. $\Delta\chi^2/\chi^2=2\%$\\
\bottomrule
\end{tabular}
\mycaption{Main features of the available polarized PDF fits.}
\label{tab:polPDFsets}
\end{table}

In Tab.~\ref{tab:polPDFsets}, 
we summarize the features of the available polarized 
parton sets; a detailed comparison of polarized PDF between them and 
the NNPDF determination will be discussed in Chap.~\ref{sec:chap2}.

\chapter{Unbiased polarized PDFs from inclusive DIS}
\label{sec:chap2}

In this Chapter, we present the first determination of 
polarized parton distributions 
based on the NNPDF methodology, \texttt{NNPDFpol1.0}. 
This analysis includes all available data from inclusive,
neutral-current, polarized DIS and aims at an unbiased extraction
of total quark-antiquark and gluon distributions at NLO accuracy.
In Sec.~\ref{sec:expdata} we present the data sets used in the
present analysis, and we discuss how their statistical distribution 
is sampled with Monte Carlo generation of pseudodata.
We provide details of the QCD analysis in Sec.~\ref{sec:QCDanalysis},
then we discuss the PDF parametrization in terms of neural networks
in Sec.~\ref{sec:minim}; we give particular emphasis on the
minimization strategy and its peculiarities in the polarized case.
The \texttt{NNPDFpol1.0} parton set is presented in 
Sec.~\ref{sec:results}, where we illustrate its statistical features, 
and its stability upon the variation of several theoretical and 
methodological assumptions. We also compare our results to 
other recent polarized PDF sets reviewed in Sec.~\ref{sec:setsummary}.
Finally, we discuss phenomenological implications for the spin 
content of the proton and the test of the Bjorken sum rule 
in Sec.~\ref{sec:phenoimplications}.
The analysis presented in this Chapter mostly reproduces 
Ref.~\cite{Ball:2013lla}. 

\section{Experimental input}
\label{sec:expdata}

We present the features of the experimental data
sets included in the \texttt{NNPDFpol1.0} analysis and we discuss
in detail which piece of information they provide on the polarized
structure functions. Then, we summarize the construction and the 
validation of the Monte Carlo pseudodata sample from the input 
experimental data.

\subsection{The data set: 
observables, kinematic cuts, uncertainties and correlations} 
\label{sec:dataset}

\begin{figure}[t]
\begin{center}
\epsfig{width=0.5\textwidth,figure=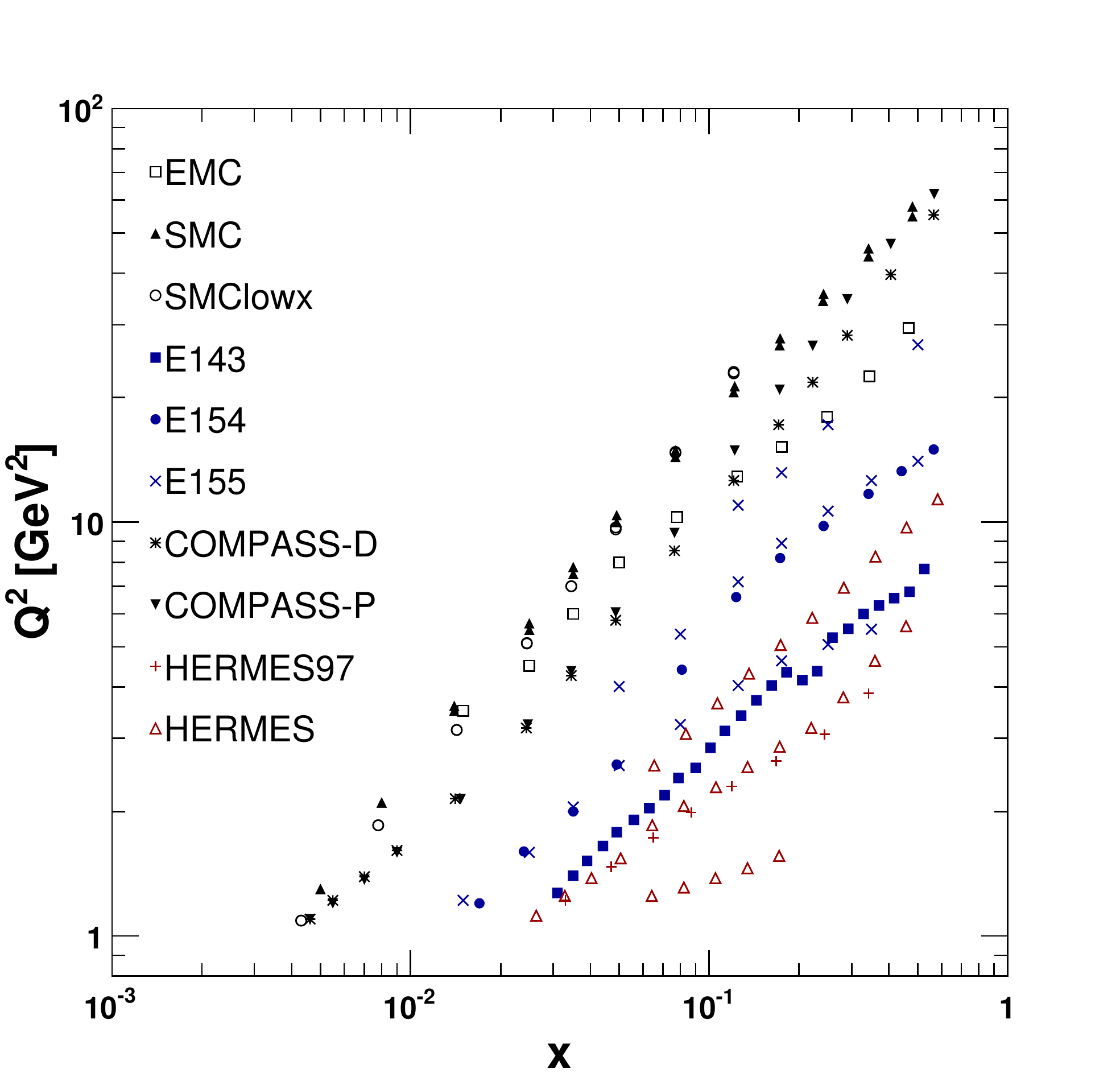}  
\mycaption{Experimental data in the $(x,Q^2)$ plane 
(after kinematic cuts): black points are from CERN experiments~\cite{Ashman:1989ig,Adeva:1998vv,Adeva:1999pa,Alexakhin:2006oza,Alekseev:2010hc}, blue 
from SLAC~\cite{Abe:1998wq,Abe:1997cx,Anthony:2000fn} and red from 
DESY~\cite{Ackerstaff:1997ws,Airapetian:2007mh}.}
\label{fig:dataplot}
\end{center}
\end{figure}

We consider inclusive, neutral current, lepton-nucleon DIS data
coming from-all-over-the-world experiments performed at 
CERN~\cite{Ashman:1989ig,Adeva:1998vv,Adeva:1999pa,Alexakhin:2006oza,Alekseev:2010hc}, 
SLAC~\cite{Abe:1998wq,Abe:1997cx,Anthony:2000fn}
and DESY~\cite{Ackerstaff:1997ws,Airapetian:2007mh}.
These experiments use either electron or muon beams, and
either proton or neutron (deuteron or $^3$He) targets.
The main features of the data sets included in our analysis are
summarized in Tab.~\ref{tab:exps-sets}, where we show, for each
of them, the number of available data points, the covered kinematic range,
and the published observable we use to reconstruct the 
$g_1$ structure function. Their kinematic coverage in the $(x,Q^2)$ plane
is also shown in Fig.~\ref{fig:dataplot}.
\begin{sidewaystable}[p]
 \centering
 \footnotesize
 \begin{tabular}{llccccccc}
  \toprule
  Experiment & Set                             
             & $N_{\mathrm{dat}}$ & $x_{\mathrm{min}}$ &  $x_{\mathrm{max}}$    
             & $Q^2_{\mathrm{min}}$ [GeV$^2$] & $Q^2_{\mathrm{max}}$  [GeV$^2$] 
             & $F$ & Ref.\\ 
 \midrule
 \multicolumn{2}{l}{EMC} & & & & & & & \\
             &  EMC-A1P & 10  & .0150 
             & .4660 & 3.5 & 29.5   
             & $A_{\parallel}^p/D$ & \cite{Ashman:1989ig}\\
 \midrule
 \multicolumn{2}{l}{SMC} & & & & & & & \\
             &  SMC-A1P & 12 & .0050 
             & .4800 & 1.3 & 58.0    
             & $A_{\parallel}^p/D$ & \cite{Adeva:1998vv}\\
             &  SMC-A1D  & 12  & .0050 
             & .4790 & 1.3 & 54.8   
             & $A_{\parallel}^d/D$ & \cite{Adeva:1998vv}\\
 \midrule
 \multicolumn{2}{l}{SMClowx} & & & & & & & \\
             &  SMClx-A1P & 15 (8) & .0001 (.0043)
             & .1210 &  0.02 (1.09) & 23.1    
             & $A_{\parallel}^p/D$ & \cite{Adeva:1999pa}\\
             &  SMClx-A1D & 15 (8) & .0001 (.0043) 
             & .1210 & 0.02 (1.09) & 22.9   
             & $A_{\parallel}^d/D$ & \cite{Adeva:1999pa}\\
 \midrule
 \multicolumn{2}{l}{E143} & & & & & & & \\
             &  E143-A1P & 28 (25) & .0310 
             & .5260 & 1.27 & 9.52 (7.72)    
             & $A_1^p$ & \cite{Abe:1998wq}\\
             &  E143-A1D & 28 (25) & .0310 
             & .5260 & 1.27 & 9.52 (7.72)    
             & $A_1^d$ & \cite{Abe:1998wq}\\
 \midrule
 \multicolumn{2}{l}{E154} & & & & & & & \\
             &  E154-A1N & 11 & .0170 
             & .5640 & 1.2 & 15.0    
             & $A_1^n$ & \cite{Abe:1997cx}\\
 \midrule
 \multicolumn{2}{l}{E155} & & & & & & & \\
             &  E155-G1P & 22 (20) & .0150 
             & .7500 (.5000) & 1.22 & 34.72 (26.86)   
             & $g_1^p/F_1^p$ & \cite{Anthony:2000fn}\\
             &  E155-G1N & 22 (20) & .0150 
             & .7500 (.5000) & 1.22 & 34.72 (26.86)  
             & $g_1^n/F_1^n$ & \cite{Anthony:2000fn}\\
 \midrule
 \multicolumn{2}{l}{COMPASS-D} & & & & & & & \\
             &  CMP07-A1D & 15 & .0046 
             & .5660 & 1.10 & 55.3   
             & $A_{\parallel}^d/D$ & \cite{Alexakhin:2006oza}\\
 \midrule
 \multicolumn{2}{l}{COMPASS-P} & & & & & & & \\
             &  CMP10-A1P & 15 & .0046 
             & .5680 & 1.10 & 62.1   
             & $A_{\parallel}^p/D$ & \cite{Alekseev:2010hc}\\
 \midrule
 \multicolumn{2}{l}{HERMES97} & & & & & & & \\
             &  HER97-A1N &  9 (8) & .0330 
             & .4640 (.3420) &  1.22 & 5.25 (3.86)    
             & $A_{\parallel}^n/D$ & \cite{Ackerstaff:1997ws}\\
 \midrule
 \multicolumn{2}{l}{HERMES} & & & & & & & \\
             &  HER-A1P & 38 (28) & .0264 
             & .7311 (.5823) &  1.12 & 14.29 (11.36)    
             & $A_{\parallel}^p$ & \cite{Airapetian:2007mh}\\
             &  HER-A1D & 38 (28) & .0264 
             & .7311 (.5823) & 1.12 & 14.29 (11.36)    
             & $A_{\parallel}^d$ & \cite{Airapetian:2007mh}\\
 \midrule
 \multicolumn{2}{l}{Total} &  290 (245) & \multicolumn{6}{c}{}\\
 \bottomrule
 \end{tabular}
\mycaption{Experimental data sets included in the \texttt{NNPDFpol1.0}
analysis. For each experiment we show the number of data points before and
after (in parenthesis) applying kinematic cuts, the covered kinematic range
and the measured observable.}
\label{tab:exps-sets}
\end{sidewaystable}
This quantity differs experiment by experiment, since the primary observable
can be one of the asymmetries or structure functions discussed 
in Sec.~\ref{sec:phenopol}. In the following, we summarize how 
we reconstruct the $g_1$ structure function from the published 
experimental observables for individual experiments
(labelled as in Tab.~\ref{tab:exps-sets}).

\begin{list}{}{\leftmargin=0pt}

\item {\textbf{EMC, SMC, SMClowx, COMPASS, HERMES97.}}
All these experiments have performed a measurement of
$A_\parallel$, then they determined $A_1$ using
Eq.~(\ref{eq:asyrel}), under the assumption $\eta\approx0$. Therefore,
the observable published by these experiments actually 
corresponds to a measurement of $\frac{A_\parallel}{D}$. 
We determine $g_1$ from $\frac{A_\parallel}{D}$
using Eq.~(\ref{eq:g1tog2}). This is possible because $D$
is completely fixed by Eq.~(\ref{eq:Ddef}) in terms of the unpolarized
structure function ratio Eq.~(\ref{eq:Rdef}) and of the kinematics. We
determine the unpolarized structure function ratio using as primary
inputs $F_2$, for which we use the parametrization of
Ref.~\cite{Forte:2002fg,DelDebbio:2004qj}, and $F_L$, which we
determine from its
expression in terms of parton distributions, using the NLO \texttt{NNPDF2.1}
parton set~\cite{Ball:2011uy}.

\item {\textbf{HERMES.}}
This experiment has performed a measurement of  
$A_\parallel$, and it published both $A_\parallel$ and
$A_1$ (which is determined using Eq.~(\ref{eq:asyrel}) and a
parametrization of $A_2$). We use the published values of
$A_\parallel$, which are closer to the experimentally measured
quantity, to determine $g_1$ through Eq.~(\ref{eq:g1tog2}).

\item {\textbf{E143.}}
This experiment has taken data with three different
beam energies, $E_1=29.1$ GeV, $E_2=16.2$ GeV, $E_3=9.7$ GeV. 
For the highest energy both $A_\parallel$ and $A_\perp$ are independently 
measured and $A_1$ is extracted from them using Eq.~(\ref{eq:asyrel}); 
for the two lowest energies only $A_\parallel$ is measured and 
$A_1$ is extracted from it using Eqs.~(\ref{g1toA1}-\ref{g2toA2}),
while assuming the form Eq.~(\ref{eq:wwrel}) for $g_2$. 
The values of $A_1$ obtained with the three beam energies are combined into a
single determination of $A_1$; radiative corrections are applied at
this combination stage. Because of this, we should use this combined
value of $A_1$, from which we then determine $g_1$
using Eq.~(\ref{eq:g1tog2p}). In order to determine $y$,
Eq.~(\ref{eq:ydef}), which depends on the beam energy, we use the
mean of the three energies.

\item {\textbf{E154.}}
This experiment measures $A_\parallel$ and $A_\perp$
independently, and then extracts a determination of $A_1$. 
We use these values of $A_1$ to determine $g_1$ by means of 
Eq.~(\ref{eq:g1tog2p}).

\item {\textbf{E155.}}
This experiment only measures the longitudinal asymmetry $A_\parallel$, from
which the ratio $g_1/F_1$ is extracted using Eq.~(\ref{g1toA1}) with
the Wandzura-Wilczek form of $g_2$, Eq.~(\ref{eq:wwrel}). In this
case, we use these values of 
$g_1/F_1$, and we extract $g_1$ using
Eq.~(\ref{eq:fonedef}) for $F_1$, together with
the parametrization of
Ref.~\cite{Forte:2002fg,DelDebbio:2004qj} for $F_2$ and
the expression in terms 
of parton distributions and the NLO \texttt{NNPDF2.1}
parton set~\cite{Ball:2011uy} for $F_L$. 
\end{list}

All these experiments also provide an extraction of the $g_1$ 
structure function in their specific framework, based on different 
assumptions on $g_2$. For this reason, we preferred to use the experimental
asymmetries, instead of the corresponding structure functions, which 
instead we reconstructed on our own. We checked that they are consistent 
with those provided by the experimental collaborations themselves within
their uncertainties.

We have excluded from our analysis all data points with
$Q^2 \le Q^2_{\mathrm{cut}}=1$ GeV$^2$, since below such energy scale
perturbative QCD cannot be considered reliable. A similar choice
of cut is usually made in all polarized analyses, specifically in
Refs.~\cite{Ball:1995td,Altarelli:1996nm,Altarelli:1998nb,
deFlorian:2009vb,Blumlein:2010rn,Hirai:2008aj}.
Notice that this value of $Q^2_{\mathrm{cut}}$
is somewhat lower than that adopted in unpolarized
fits, where $Q^2_{\mathrm{cut}}=2$ GeV$^2$.
This difference arises not to exclude 
a large piece of experimental information
which, in the polarized case, lies in the small-$Q^2$ region
(see Fig.~\ref{fig:dataplot}).

We further impose a cut on the squared invariant mass of the 
hadronic final state $W^2=Q^2(1-x)/x$
in order to remove points which may be affected by sizable dynamical
higher-twist corrections. The cut is chosen based on
a study presented in Ref.~\cite{Simolo:2006iw}, 
where higher twist terms were added to the
observables, with a coefficient fitted to the data: it was shown there
that the higher twist contributions become compatible with zero if one
imposes the cut $W^2 \ge W^2_{\mathrm{cut}}=6.25$ GeV$^2$. 
We will follow  this choice, which
excludes data points with large Bjorken-$x$ at moderate values of
the squared momentum transfer $Q^2$, roughly corresponding to the
bottom-right corner of the $(x,Q^2)$-plane, see
Fig.~\ref{fig:dataplot}: in particular, it excludes
all available JLAB data~\cite{Zheng:2004ce,Fatemi:2003yh,Dharmawardane:2006zd}.
The number of data points surviving the kinematic cuts for each data set
is given in parenthesis in Tab.~\ref{tab:exps-sets}.

As can be seen from Fig.~\ref{fig:dataplot}, the
region of the $(x,Q^2)$-plane where data are available  
after kinematic cuts is
roughly restricted to $4\cdot 10^{-3}\lesssim x\lesssim 0.6$ and 
$1$~GeV$^2\leq Q^2\lesssim 60$~GeV$^2$. In recent years,
the coverage of the low-$x$ region has been
improved by a complementary set of SMC data~\cite{Adeva:1999pa} and by
the more recent COMPASS
data~\cite{Alexakhin:2006oza,Alekseev:2010hc},
both included in the present analysis. In the
large-$x$ region, information is provided at rather high $Q^2$ by the
same COMPASS data and at lower energy by the latest HERMES
measurements~\cite{Airapetian:2007mh}. 
In the near future, additional polarized inclusive DIS measurements
are expected from an update of COMPASS data~\cite{slides:DIS2013}
and from JLAB spin program~\cite{Jimenez-Delgado:2013boa}. 
However, the latter will cover the large-$x$
and small $Q^2$ corner, hence their inclusion in a global fit will require
a careful treatment of higher-twist corrections, as
performed in Ref.~\cite{Jimenez-Delgado:2013boa}.
The data set used in this paper is the same as that of 
Ref.~\cite{Blumlein:2010rn}, and also the same as the DIS data of the
fit of Ref.~\cite{deFlorian:2009vb}, which however has a wider data
set beyond inclusive DIS.

Each experimental collaboration provides uncertainties on the
measured quantities listed in the next-to-last column of
Tab.~\ref{tab:exps-sets}.
Correlated systematics are only provided by EMC and
E143, which give the values of the
systematics due to the uncertainty in the beam and target 
polarizations, while all other experiments do not provide any piece of
information on the covariance matrix. For each experiment, we determine 
the uncorrelated uncertainty on $g_1$ by combining the uncertainty on
the experimental observable with that of the unpolarized structure
function using standard error propagation. 
For EMC and E143 experiments, we also include all available correlated
systematics. These are provided by both experimental collaborations as
a percentage correction to $g_1$ (or, alternatively, to the asymmetry
$A_1$): we apply the percentage uncertainty on $g_1$ to the structure
function determined by us as discussed in Sec.~\ref{sec:phenopol}.

We then construct the covariance matrix
\begin{equation}
\mathrm{cov}_{pq}
=
\left(
\sum_i\sigma^{(c)}_{i,p}\sigma^{(c)}_{i,q} + \delta_{pq} \sigma^{(u)}_{p}\sigma^{(u)}_{q}
\right)
g_{1,p}g_{1,q}
\,\mbox{,}
\label{eq:covmat}
\end{equation}
where $p$ and $q$ run over the experimental data points, 
$g_{1,p}\equiv g_1(x_p,Q_p^2)$ ($g_{1,q}\equiv g_1(x_q,Q_q^2)$),  
$\sigma^{(c)}_{i,p}$ are the various sources of correlated uncertainties,  
and $\sigma^{(u)}_{p}$ are the uncorrelated uncertainties, 
which are in turn found as a sum in
quadrature of all uncorrelated sources of statistical
$\sigma^{\mathrm{(stat)}}_{i,p}$ and systematic $\sigma^{\mathrm{(syst)}}_{i,p}$
uncertainties on each point:
\begin{equation}
\left(
\sigma^{(u)}_{p}\right)^2
=
\sum_i \left(\sigma^{\mathrm{(stat)}}_{i,p}\right)^2 +\sum_j
\left(\sigma^{\mathrm{(syst)}}_{j,p}\right)^2
\,\mbox{.}
\label{eq:uncsum}
\end{equation}
The correlation matrix is defined as
\begin{equation}
\rho_{pq} 
= 
\frac{{\mathrm{cov}}_{pq}}{\sigma^{\mathrm{(tot)}}_{p}\sigma^{\mathrm{(tot)}}_{q}g_{1,p}g_{1,q}}
\,\mbox{,}
\label{eq:cormatr}
\end{equation}
where the total uncertainty $\sigma^{\mathrm{(tot)}}_{p}$ on the $p$-th data 
point is
\begin{equation}
\left(\sigma^{\mathrm{(tot)}}_{p}\right)^2 =
(\sigma^{(u)}_{p})^2+\sum_i\left(\sigma^{(c)}_{i,p}\right)^2 
\,\mbox{.}
\label{eq:sigmatot}
\end{equation}

In Tab.~\ref{tab:exps-err}, we show the
average experimental uncertainties for each data set, with
uncertainties separated into statistical and correlated systematics. 
All values are given as absolute uncertainties and refer to the
structure function $g_1$, which has been reconstructed for each experiment
as discussed above.
As in the case of Tab.~\ref{tab:exps-sets}, we provide
the values before and after kinematic cuts (if different).
\begin{table}[t]
\centering
\footnotesize
 \begin{tabular}{llccc}
 \toprule
 Experiment & Set
 & $\langle\delta{g_1}_{\mathrm{s}}\rangle$
 & $\langle\delta{g_1}_{\mathrm{c}}\rangle$
 & $\langle\delta{g_1}_{\mathrm{tot}}\rangle$\\
 \midrule
 \multicolumn{2}{l}{EMC       } & & & \\
 & EMC-A1P    &        0.144         &        0.037 &        0.150\\
 \midrule
 \multicolumn{2}{l}{SMC       } & & & \\
 & SMC-A1P    &        0.098         &        --      &        0.098\\
 & SMC-A1D    &        0.116         &        --      &        0.116\\
 \midrule
 \multicolumn{2}{l}{SMClowx   } & & & \\
 & SMClx-A1P  &       18.379 (0.291) &        -- (--) &       18.379 (0.291)\\
 & SMClx-A1D  &       22.536 (0.649) &        -- (--) &       22.536 (0.649)\\
 \midrule
 \multicolumn{2}{l}{E143      } & & & \\
 & E143-A1P   &        0.042 (0.046) &  0.009 (0.009) &        0.043 (0.047)\\
 & E143-A1D   &        0.053 (0.058) &  0.004 (0.005) &        0.054 (0.059)\\
 \midrule
 \multicolumn{2}{l}{E154      } & & & \\
 & E154-A1N   &        0.044         &        --      &        0.044\\
 \midrule
 \multicolumn{2}{l}{E155      } & & & \\
 & E155-G1P   &        0.040 (0.043) &        -- (--) &        0.040 (0.043)\\
 & E155-G1N   &        0.124 (0.135) &        -- (--) &        0.124 (0.135)\\
 \midrule
 \multicolumn{2}{l}{COMPASS-D } & & & \\
 & CMP07-A1D  &        0.061 &                --      &        0.061\\
 \midrule
 \multicolumn{2}{l}{COMPASS-P } & & & \\
 & CMP10-A1P  &        0.101 &                --      &        0.101\\
 \midrule
 \multicolumn{2}{l}{HERMES97  } & & & \\
 & HER97-A1N  &        0.087 (0.093) &        -- (--) &        0.087 (0.093)\\
 \midrule
 \multicolumn{2}{l}{HERMES    } & & & \\
 & HER-A1P    &        0.067 (0.062) &        -- (--) &        0.067 (0.062)\\
 & HER-A1D    &        0.040 (0.034) &        -- (--) &        0.040 (0.034)\\
 \bottomrule
 \end{tabular}
\mycaption{Averaged statistical, correlated systematic and total
uncertainties before and after (in parenthesis) kinematic cuts for each of 
the experimental sets included in the present analysis. Uncorrelated systematic
uncertainties are considered as part of the statistical uncertainty
and they are added in quadrature. 
All values are absolute uncertainties and refer to the structure function $g_1$,
which has been reconstructed for each experiment as discussed in the text.
Details on the number of points and
the kinematics of each data set are provided in Tab.~\ref{tab:exps-sets}.
\label{tab:exps-err}}
\end{table}

Finally, notice that in both Tabs.~\ref{tab:exps-sets}-\ref{tab:exps-err}
we distinguish between 
experiments, defined as groups of data which cannot be correlated to
each other, and data sets within a given experiment, which could in
principle be correlated with each other, as they
correspond to measurements of
different observables in the same experiment, or measurements of the
same observable in different years. Even though, in practice, only two
experiments provide such correlated systematics (see Tab.~\ref{tab:exps-err}), 
this distinction will be useful in the 
minimization strategy, see Sec.~\ref{sec:minim} below. 

\subsection{Monte-Carlo generation of the pseudo-data sample}
\label{sec:MCgeneration}

Error propagation from experimental data to the fit is handled by a
Monte Carlo sampling of the probability distribution defined by
data, as discussed in Sec.~\ref{sec:NNPDFapproach}. 
The statistical sample is obtained by generating
$N_{\mathrm{rep}}$ pseudodata replicas, according to a multi-Gaussian
distribution centered at the data points and with a covariance equal
to that of the original data. Explicitly, given an
experimental data point $g_{1,p}^{(\mathrm{exp})}\equiv g_1(x_p,Q_p^2)$, 
we generate $k=1,\dots,N_\mathrm{rep}$ artificial points 
$g_{1,p}^{(\mathrm{art}),(k)}$ according to 
\begin{equation}
g_{1,p}^{(\mathrm{art}),(k)} (x,Q^2)
= 
\left[
1+\sum_i r_{(c),p}^{(k)}\sigma_{i,p}^{(c)} + r_{(u),p}^{(k)}\sigma_p^{(u)}
\right]
g_{1,p}^{(\mathrm{exp})} (x,Q^2)
\,\mbox{,}
\label{eq:MCgeneration}
\end{equation}
where $r_{(c),p}^{(k)}$, $r_{(u),p}^{(k)}$ are univariate Gaussianly distributed 
random numbers, and $\sigma_{i,p}^{(c)}$ and $\sigma_{p}^{(u)}$ are respectively
the relative correlated systematic and
statistical uncertainty. Unlike in the unpolarized case, 
Eq.~(\ref{eq:MCgeneration}) receives no contribution from 
normalization uncertainties, given that
all polarized observables are obtained as cross-section asymmetries.

\begin{figure}[p]
\centering
\epsfig{width=0.40\textwidth,figure=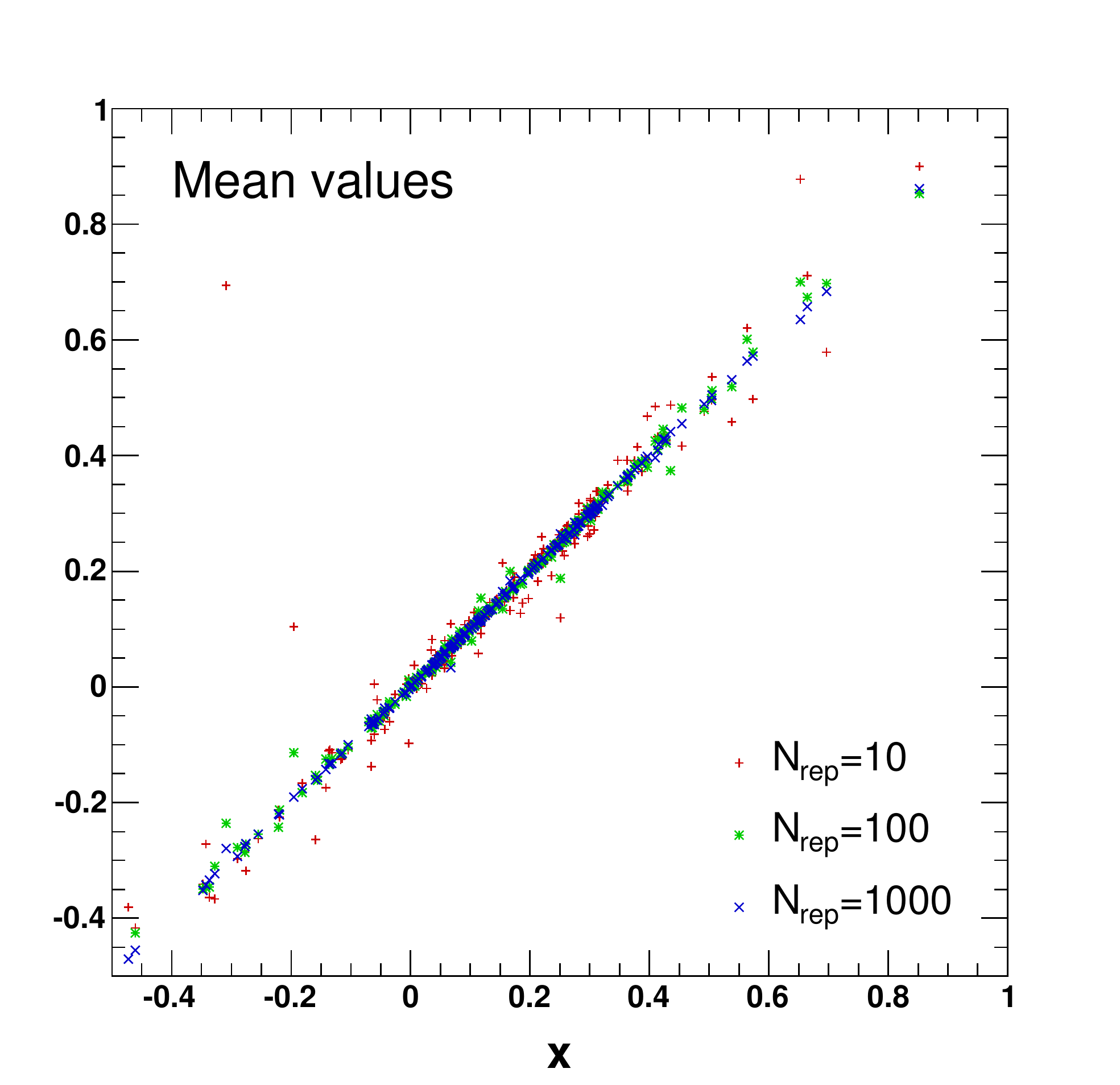}  
\epsfig{width=0.40\textwidth,figure=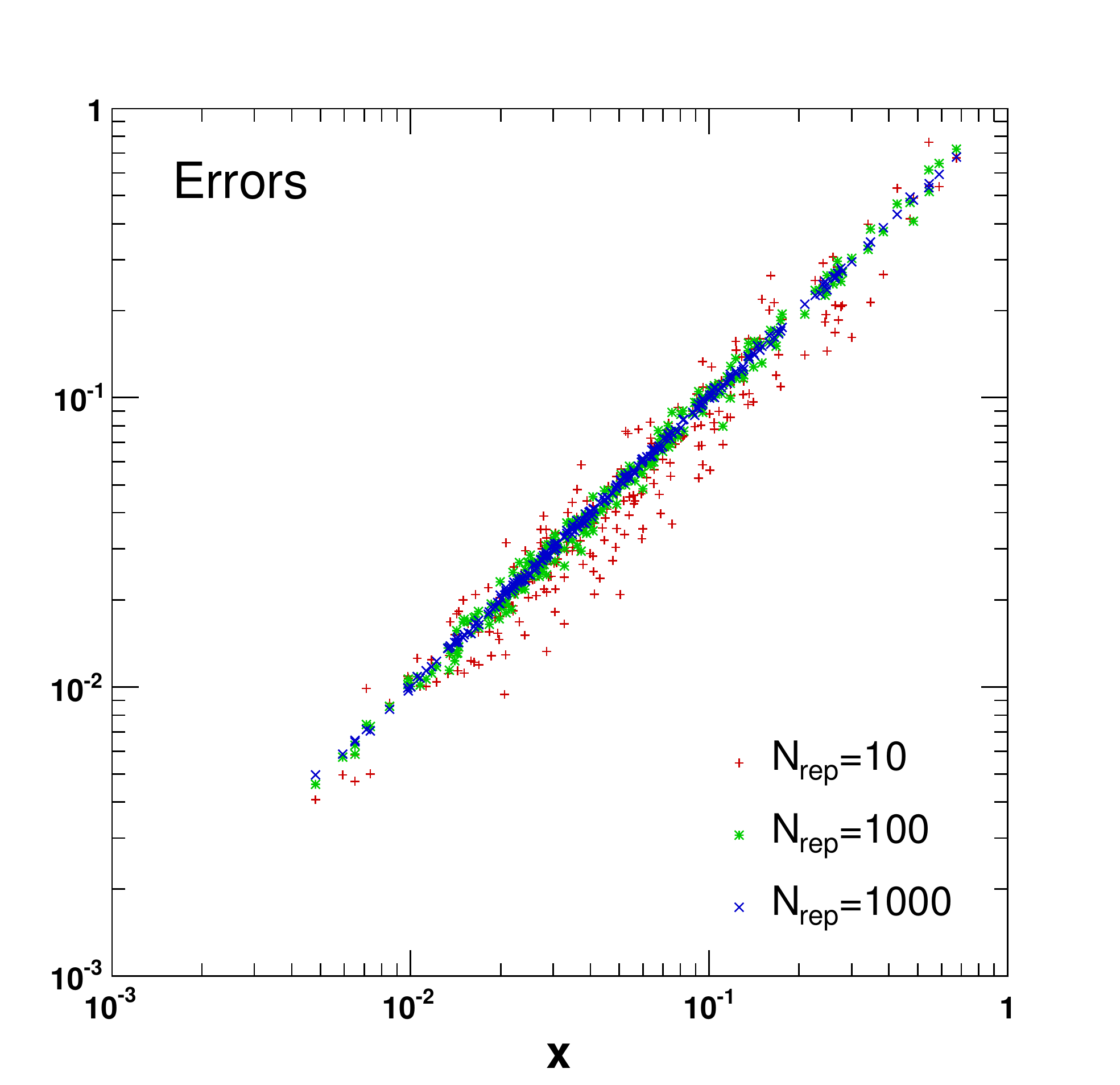} 
\mycaption{Scatter plot of experimental versus artificial Monte Carlo mean central values and absolute uncertainties of polarized
structure functions computed from 
ensembles made of $N_{\mbox{\scriptsize{rep}}}=10,100,1000$ replicas.}
\label{fig:splots}
\end{figure}

The number of Monte Carlo replicas of the data is determined by
requiring that the central values, uncertainties and  correlations of the
original experimental data can be reproduced to a given accuracy by
taking  averages, variances and
covariances over the replica sample. 
A comparison between expectation values and variances of the Monte
Carlo set and the corresponding input experimental values as a
function of the number of replicas is shown in Fig.~\ref{fig:splots},
where we display scatter plots of the central values and errors for
samples of $N_{\mbox{\scriptsize{rep}}}=10,100$ and $1000$ replicas.
A more quantitative comparison can be performed by defining suitable
statistical estimators (see, for example, 
Appendix B of Ref.~\cite{DelDebbio:2004qj}).

In Tabs.~\ref{tab:est1gen}-\ref{tab:est2gen} we show the percentage
error and the scatter correlation $r$ for central values and errors 
respectively, whose definition is recalled in appendix~\ref{sec:appB}. 
The scatter correlation $r$ is, crudely speaking, the
correlation between the input value and the value computed from the
replica sample.
We do not compute values for correlations, as these
are available for a small number of data points from only two experiments,
see Tab.~\ref{tab:exps-err}. 
Some large values of the percentage uncertainty are due to the fact
that, for some experiments, $g_1$ can take values which are very close
to zero. It is clear from both the tables and the plots that a Monte Carlo 
sample of pseudodata with $N_\mathrm{rep}=100$ is
sufficient to reproduce the mean values and the errors of experimental
data to an accuracy which is better than $5\%$, while the improvement in
going up to $N_\mathrm{rep}=1000$ is moderate. Therefore, we will
henceforth use a $N_\mathrm{rep}=100$ replica sample as a default 
for our reference fit.
\begin{table}[p]
\centering
\footnotesize
\begin{tabular}{clcccccc}
\toprule
& Estimator
& \multicolumn{3}{c}{$\left\langle \mbox{PE} \left[\langle g_1^{\mbox{\scriptsize{(art)}}}\rangle \right] \right\rangle$ [\%]}
& \multicolumn{3}{c}{$r\left[ g_1^{\mbox{\scriptsize{(art)}}}\right]$} \\
\midrule
& $N_{\mbox{\scriptsize{rep}}}$ & 10 & 100 & 1000 & 10 & 100 & 1000 \\
\midrule
\multirow{10}{*}{\begin{sideways}Experiment\end{sideways}}
& EMC          
 &   23.7 &    3.5 & 2.9 
 & .76037   & .99547   & .99712 \\
& SMC         
 &   19.4 & 5.6    & 1.2   
 & .94789   & .99908   & .99993 \\
& SMClowx     
 &  183 & 25.8  & 15.4 
 & .80370   & .99239   & .99960 \\
& E143        
 &   18.5 & 5.7    & 2.1 
 & .99159   & .99860   & .99984 \\
& E154        
 &   239 & 44.0   & 21.9
 & .99635   & .99981   & .99994 \\
& E155        
 &   37.3 & 13.4   & 4.3   
 & .99798   & .99993   & .99998 \\
& COMPASS-D   
 &   26.4 & 8.6    & 3.2
 & .96016   & .98774   & .99917 \\
& COMPASS-P   
 & 16.4   & 1.9    & 1.5  
 & .91942   & .99829   & .99902 \\
& HERMES97    
 & 22.5   & 6.2    & 2.2 
 & .96168   & .99762   & .99979 \\
& HERMES      
 & 10.5   & 5.8    & 1.2 
 & .98564   & .99916   & .99973 \\
\bottomrule
\end{tabular}
\mycaption{Table of statistical estimators for the mean value computed from
the Monte Carlo sample with $N_\mathrm{rep}=10,100,1000$ replicas.
Estimators refer to individual experiments and are defined in Appendix B of Ref.~\cite{DelDebbio:2004qj}.}
\label{tab:est1gen}
\end{table}
\begin{table}[p]
\centering
\footnotesize
\begin{tabular}{clcccccc}
\toprule
& Estimator
& \multicolumn{3}{c}{$\left\langle \mbox{PE} \left[\langle \delta g_1^{\mbox{\scriptsize{(art)}}}\rangle \right] \right\rangle$ [\%]}
& \multicolumn{3}{c}{$r\left[ \delta g_1^{\mbox{\scriptsize{(art)}}}\right]$} \\
\midrule
& $N_{\mbox{\scriptsize{rep}}}$ & 10 & 100 & 1000 & 10 & 100 & 1000 \\
\midrule
\multirow{10}{*}{\begin{sideways}Experiment\end{sideways}}
& EMC          
 &   12.8 &    4.9 & 2.0 
 & .97397   & .99521   & .99876 \\
& SMC         
 &   22.4 & 5.4    & 1.7    
 & .96585   & .99489   & .99980 \\
& SMClowx     
 &  16.9 & 6.2  & 2.1 
 & .97959   & .99490   & .99905 \\
& E143        
 &   16.0 & 7.4    & 2.0 
 & .95646   & .98684   & .99946 \\
& E154        
 &   19.1 & 3.7   & 1.3
 & .99410   & .99871   & .99992 \\
& E155        
 &   21.2 & 5.6   & 1.8   
 & .99428   & .99971   & .99997 \\
& COMPASS-D   
 &   15.5 & 5.2    & 1.6
 & .99375   & .99687   & .99993 \\
& COMPASS-P   
 & 18.4   & 7.4    & 1.5  
 & .99499   & .99005   & .99988 \\
& HERMES97    
 & 17.9   & 6.4    & 1.6 
 & .89065   & .97318   & .99894 \\
& HERMES      
 & 19.5   & 6.0    & 1.6 
 & .91523   & .99237   & .99942 \\
\bottomrule
\end{tabular}
\mycaption{Same as Tab.~\ref{tab:est1gen}, but for errors.}
\label{tab:est2gen}
\end{table}

\section{Details of the QCD analysis}
\label{sec:QCDanalysis}

We will now briefly outline some details of the QCD analysis of polarized 
structure functions. The observable with which we fit experimental data
is the $g_1$ structure function, Eq.~(\ref{eq:g1coeff}), expressed in terms 
of the PDF combinations in Eqs.~(\ref{eq:qNS})-(\ref{eq:qSigma}).
From these relations, supplemented with
Eqs.~(\ref{eq:g_relations1})-(\ref{eq:g153}), it is clear that
neutral-current $g_1$ data only allow for a direct determination of the 
four polarized PDF combinations $\Delta\Sigma$, $\Delta T_3$, $\Delta T_8$ 
and $\Delta g$: these will form the basis of polarized PDFs to be determined
in our analysis.  
In principle, an intrinsic polarized component could also be present
for each heavy flavor, as observed in Sec.~\ref{sec:generalstrategy}. 
However, we will neglect it here and assume 
that heavy quark PDFs are dynamically generated above
threshold by (massless) Altarelli-Parisi evolution, in a zero-mass
variable-flavor number (ZM-VFN) scheme. In such a scheme all heavy
quark mass effects are neglected. While they can be introduced for instance
through the \texttt{FONLL} method~\cite{Forte:2010ta}, these effects have been 
shown to be relatively small already on the scale of present-day unpolarized 
PDF uncertainties, and thus are most likely negligible in the polarized case 
where uncertainties are rather larger.
We will further comment on this issue in Sec.~\ref{sec:g1charm}
and Appendix~\ref{sec:appA}, where we will sketch how to handle 
intrinsic charm contribution via the \texttt{FONLL} scheme.

The proton and neutron PDFs are related to each other by isospin,
which we will assume to be exact, thus yielding
\begin{equation}
\Delta u^p=\Delta d^n,\quad \Delta d^p=\Delta u^n, \quad \Delta s^p=\Delta s^n
\,\mbox{,}
\label{eq:kernels}
\end{equation}
and likewise for the polarized anti-quarks. In the following we will
always assume that PDFs refer to the proton.

As discussed at length in Sec.~\ref{sec:QCDevol}, beyond leading order in 
QCD the first moment of all non-singlet combinations of quark and antiquark 
distributions are scale independent due to axial current conservation.
Besides, we enforce $SU(2)$ and $SU(3)$ flavor asymmetry by requiring the
first moments of the non-singlet, $C$-even, combinations, 
Eqs.~(\ref{eq:a3a8}), to be fixed by the experimental values of baryon
octet decay constants, Eqs.~(\ref{eq:hypdecayconst}). 
Actually, a much larger uncertainty on the octet axial charge,  
up to about 30\%, is found if SU(3) symmetry is 
violated~\cite{FloresMendieta:1998ii} with respect to that 
quoted in Eqs.~(\ref{eq:hypdecayconst}). Even though a detailed 
phenomenological analysis does not 
seem to support this conclusion~\cite{Cabibbo:2003cu}, we will 
take as default this more conservative 
uncertainty estimation
\begin{equation}
\label{eq:a8p}
a_8 = 0.585 \pm 0.176 
\,\mbox{.}
\end{equation}
The impact of replacing this with the more aggressive determination 
given in Eq.~(\ref{eq:hypdecayconst}) will be 
studied in Sec.~\ref{sec:srres}.

Structure functions will be computed in terms of polarized parton
distributions using the so-called NNPDF \texttt{FastKernel} method,
introduced in Ref.~\cite{Ball:2010de}. In short, in this method the
PDFs at scale $Q^2$ are obtained by convoluting the parton
distributions at the parametrization scale $Q_0^2$ with a set of Green's
functions, which are in turn obtained by solving the QCD evolution
equations in Mellin space (below also denoted as $N$-space). 
These Green's functions are then convoluted
with coefficient functions, so that the structure function can be
directly expressed in terms of the PDFs at the parametrization scale
through suitable kernels $K$. In terms of the polarized PDFs at the
input scale (labelled with the subscript $0$) we have
\begin{equation}
g_1^p=\left\{
 K_{g_1, \Delta\Sigma}\otimes \Delta \Sigma_0 
+K_{g_1, \Delta g} \otimes \Delta g_0 
+K_{g_1, +} \otimes  \left(\Delta T_{3,0}
+ \frac{1}{3}\Delta T_{8,0} \right)\right\}
\,\mbox{,}
\label{eq:Kg1p}
\end{equation}
where the kernels $K_{g_1, \Delta\Sigma}$, $K_{g_1, \Delta g}$,
$K_{g_1, +}$ take into account both the coefficient functions and
$Q^2$ evolution. This way of expressing structure
functions is amenable to numerical optimization, because all kernels
can then be precomputed and stored, and convolutions may be reduced to
matrix multiplications by projecting onto a set of suitable basis
functions. 

The neutron polarized structure function $g_1^n$ is given
in terms of the proton and deuteron ones as
\begin{equation}
g_1^n = 2\frac{g_1^d}{1-1.5\omega_D}-g_1^p
\,\mbox{,}
\label{eq:g1n}
\end{equation}
with $\omega_D=0.05$ the probability that the deuteron is found in a D state.
Under the assumption of exact isospin symmetry, the expression
of $g_1^n$ in terms of parton densities is obtained from Eq.~\eqref{eq:kernels}
by interchanging the up and down quark PDFs, which amounts
to changing the sign of $\Delta T_3$. 
We will
assume the values $\alpha_s(M_Z^2)=0.119$ for the strong coupling
constant and $m_c=1.4$ GeV and $m_b=4.75$ GeV for the charm and bottom
quark masses respectively.

We have benchmarked our implementation of the evolution of polarized
parton densities up to NLO by cross-checking against the
Les Houches polarized PDF evolution benchmark
tables~\cite{Dittmar:2005ed}.\footnote{Note that in 
Ref.~\cite{Dittmar:2005ed}
the  polarized sea PDFs are given incorrectly, and should be 
$x\Delta \bar{u}=-0.045 x^{0.3} (1-x)^7$ and
$x\Delta \bar{d}=-0.055 x^{0.3} (1-x)^7$.}  
These tables were obtained from a
comparison of the \texttt{HOPPET}~\cite{Salam:2008qg} and 
\texttt{PEGASUS}~\cite{Vogt:2008yw} evolution codes, which are $x-$space and
$N-$space codes respectively. In order to perform a meaningful
comparison, we use the so-called iterated solution of the $N-$space evolution
equations and use the same initial PDFs and running coupling as
in~\cite{Dittmar:2005ed}. 
The relative difference $\epsilon_{\mathrm{rel}}$ 
between our PDF evolution and the benchmark tables of
Refs.~\cite{Dittmar:2005ed} at NLO in the ZM-VFNS
are tabulated in Tab.~\ref{tab:lhacc} for various combinations of
polarized PDFs: the accuracy of our code is $\mathcal{O}(10^{-5})$  
for all relevant values of $x$, which is the nominal accuracy of the 
agreement between \texttt{HOPPET} and \texttt{PEGASUS}.
\begin{table}[t]
\footnotesize
\begin{center}
\begin{tabular}{ccccc}
\toprule
$x$  & $\epsilon_{\mathrm{rel}}(\Delta u^-)$ 
     & $\epsilon_{\mathrm{rel}}(\Delta d^-)$ 
     & $\epsilon_{\mathrm{rel}}(\Delta \Sigma)$ 
     & $\epsilon_{\mathrm{rel}}(\Delta g)$ \\
\midrule
$10^{-3}$ & $1.1\,10^{-4}$ & $9.2\,10^{-5}$ & $9.9\,10^{-5}$& $1.1\,10^{-4}$\\
$10^{-2}$  & $1.4\,10^{-4}$ & $1.9\,10^{-4}$ & $3.5\,10^{-4}$& $9.3\,10^{-5}$\\
$0.1$  & $1.2\,10^{-4}$ & $1.6\,10^{-4}$ & $5.4\,10^{-6}$& $1.7\,10^{-4}$\\
$0.3$  & $2.3\,10^{-6}$ & $1.1\,10^{-5}$ & $7.5\,10^{-6}$& $1.7\,10^{-5}$\\
$0.5$  & $5.6\,10^{-6}$ & $9.6\,10^{-6}$ & $1.6\,10^{-5}$& $2.5\,10^{-5}$\\
$0.7$  & $1.2\,10^{-4}$ & $9.2\,10^{-7}$ & $1.6\,10^{-4}$& $7.8\,10^{-5}$\\
$0.9$  & $3.5\,10^{-3}$ & $1.1\,10^{-2}$ & $4.1\,10^{-3}$& $7.8\,10^{-3}$\\
\bottomrule
\end{tabular}
\end{center}
\mycaption{Percentage difference between FastKernel perturbative
evolution of  polarized PDFs and  the Les Houches benchmark 
tables~\cite{Dittmar:2005ed}
for different
polarized PDF combinations at NLO in the ZM-VFNS.} 
\label{tab:lhacc}
\end{table}
Therefore, we can conclude that the accuracy of the polarized
PDF evolution in the \texttt{FastKernel} framework is satisfactory
for precision phenomenology.

Finally, we include exactly all kinematic target mass corrections within the
formalism presented in Sec.~\ref{sec:TMC}. However, we notice that the 
numerical implementation of Eqs.~(\ref{g1xWW}) or
Eq.~(\ref{g1x0}) is difficult, because of the presence
of the first derivative of $g_1$ in the correction term.
For this reason, we do not factorize TMCs into the hard kernels,
as it was done in the unpolarized case, where  
the first derivative of $F_1$ does not appear~\cite{Ball:2008by}.
Rather, we will include target mass
effects in an iterative way: 
we start by performing a fit in which we set $M=0$ and
at each iteration of the minimization procedure 
the target mass corrected $g_1$ structure function
is computed by means of Eqs.~(\ref{g1xWW}-\ref{g1x0}) using the   
$g_1$ obtained in the previous minimization step.
We found that this strategy allows for convergence in a few minimization
steps, hence TMCs are properly included when the fit is stopped.

\section{Fitting strategy}
\label{sec:minim}

We will now discuss some details of the fitting strategy adopted in the 
\texttt{NNPDFpol1.0} analysis. In particular, 
we describe how PDFs are
parametrized in terms of neural networks and how they are trained to
experimental data to obtain the optimal fit. 
The main steps of the minimization strategy have been 
summarized in Sec.~\ref{sec:NNPDFapproach}, and can be found in 
Ref.~\cite{Ball:2010de}. 
Here, we emphasize some specific features which were introduced 
to deal with issues peculiar to the polarized case, in particular with
respect to the implementation of theoretical constraints.

\subsection{Neural network parametrization}

The four independent polarized PDF flavor combinations in the evolution
basis, $\Delta\Sigma$, $\Delta T_3$ and $\Delta T_8$, and the gluon 
$\Delta g$ are separately parametrized using a multi-layer feed-forward
neural network~\cite{Ball:2011eq}. All neural networks have the same
architecture, namely 2-5-3-1, which corresponds to 37 free parameters
for each PDF, and thus a total of 148 free parameters. This is to be
compared to about 10-15 free parameters for all other available
determinations of polarized PDFs within the \textit{standard} methodology,
see Sec.~\ref{sec:generalstrategy}.
This parametrization has been
explicitly shown to be redundant in the unpolarized case, in that
results are unchanged when a smaller neural network architecture is
adopted: this ensures that results do not depend on the
architecture~\cite{Ball:2011eq}. Given that polarized data are much
less abundant and affected by much larger uncertainties than their
unpolarized counterparts, this architecture is surely adequate in the
polarized case too.

The neural network parametrization is supplemented with a
preprocessing function. In principle, large enough neural networks
can reproduce any functional form given sufficient training
time. However, the training can be made more efficient by adding a
preprocessing step, \textit{i.e.} by multiplying the output of the neural
networks by a fixed function. The neural network then only fits the
deviation from this function, which improves the speed of the
minimization procedure if the preprocessing function is suitably
chosen. We thus write the input PDF basis in terms of preprocessing
functions and neural networks ${\mathrm{NN}}_{\mathrm{\Delta pdf}}$ as follows
\begin{eqnarray}
\Delta \Sigma(x,Q_0^2)
&=&
{(1-x)^{m_1}}{x^{-n_1}}{\mathrm{NN}}_{\Delta\Sigma}(x)
\,\mbox{,}
\nonumber\\
\Delta T_3(x,Q_0^2)
&=&
A_3{(1-x)^{m_3}}{x^{-n_3}}{\mathrm{NN}}_{ \Delta T_3}(x)  
\,\mbox{,}
\nonumber\\
\Delta T_8(x,Q_0^2)
&=&
A_8{(1-x)^{m_8}}{x^{-n_{ \Delta T_8}}}{\mathrm{NN}}_{ \Delta T_3}(x)  
\,\mbox{,}\\
\Delta g(x,Q_0^2)
&=&
{(1-x)^{m_g}}{x^{-n_g}}{\mathrm{NN}}_{\Delta g}(x)
\,\mbox{.}
\nonumber
\label{eq:PDFbasisnets}
\end{eqnarray}

Of course, one should check that no bias is introduced in the choice
of preprocessing functions. To this purpose, we first select a
reasonable range of values for the large and small-$x$
preprocessing exponents $m$ and $n$, and produce a PDF determination
by choosing for each replica a value of the exponents at random with
uniform distribution within this range. We then determine effective 
exponents for each replica, defined as 
\begin{equation}
m_{\mathrm{eff}}(Q^2)
\equiv
\lim_{x\to1}\frac{\ln \Delta f(x,Q^2) }{\ln(1-x)}
\,\mbox{,}
\label{eq:effexp2}
\end{equation}
\begin{equation}
n_{\mathrm{eff}}(Q^2)
\equiv
\lim_{x\to0} \frac{\ln \Delta f(x,Q^2)}{\ln\frac{1}{x}}
\,\mbox{,}
\label{eq:effexp1}
\end{equation}
where $\Delta f = \Delta\Sigma\mbox{, }\Delta T_3\mbox{, }\Delta T_8\mbox{, }\Delta g$.
Finally, we check that the range of variation of the preprocessing
exponents is wider than the range of effective exponents for each PDF.
If it is not, we enlarge the range of variation of preprocessing, then
repeat the PDF determination, and iterate
until the condition is satisfied. 
This ensures that the range of effective large- and
small-$x$ exponents found in the fit is not biased, and in particular not
restricted, by the range of preprocessing exponents. Our final values
for the preprocessing exponents are summarized in
Tab.~\ref{tab:prepexps}, while the effective exponents obtained in
our fit will be discussed in Sec.~\ref{sec:prepexp}.
\begin{table}[t]
\centering
\footnotesize
\begin{tabular}{ccc}
\toprule 
PDF & $m$ & $n$ \\
\midrule
$\Delta\Sigma(x,Q_0^2)$  & $[1.5,3.5]$ & $[0.2,0.7]$ \\
\midrule
$\Delta g(x,Q_0^2)$  & $[2.5,5.0]$ & $[0.4,0.9]$ \\
$\Delta T_3(x,Q_0^2)$  & $[1.5,3.5]$ & $[0.4,0.7]$ \\
$\Delta T_8(x,Q_0^2)$  & $[1.5,3.0]$ & $[0.1,0.6]$ \\
\bottomrule
\end{tabular}
\mycaption{Ranges for the small and large $x$
preprocessing exponents Eq.~(\ref{eq:PDFbasisnets}).}
\label{tab:prepexps}
\end{table}
It is apparent from Tab.~\ref{tab:prepexps} that
the allowed range  of preprocessing exponents is rather 
wider than in the unpolarized case, as a
consequence of the limited amount of experimental information. 

The nonsinglet triplet and octet PDF combinations in the parametrization basis,
Eq.~(\ref{eq:PDFbasisnets}), $\Delta T_3$ and $\Delta T_8$, are supplemented by
a prefactor. This is because these PDFs 
must satisfy the sum rules Eqs.~(\ref{eq:a3a8}),
which are enforced by letting
\begin{eqnarray}
A_3
&=&
\frac{a_3}{\int_0^1 dx\,(1-x)^{m_3}x^{-n_3} {\mathrm{NN}}_{\Delta T_3}(x)}
\,\mbox{,}
\nonumber\\ 
A_8
&=&
\frac{a_8}{\int_0^1 dx\, (1-x)^{m_8}x^{-n_8} {\mathrm{NN}}_{\Delta T_8}(x)  } 
\,\mbox{.} 
\label{eq:sumrules1}
\end{eqnarray}
The integrals are computed numerically each time the parameters of 
the PDF set are modified. The values of $a_3$ and $a_8$ are chosen
for each replica as Gaussianly distributed numbers, with central value
and width given by the corresponding experimental values,
Eqs.~(\ref{eq:hypdecayconst})-(\ref{eq:a8p}).

\subsection{Genetic algorithm minimization}
\label{sec:genmin}
As discussed at length in Ref.~\cite{Ball:2008by}
and summarized in Sec.~\ref{sec:NNPDFapproach}, minimization
with a neural network parametrization of PDFs must be performed through an
algorithm which explores the very wide functional space efficiently.
This is done by means of a genetic algorithm, which is
used to minimize a suitably defined figure
of merit, namely the error function~\cite{Ball:2008by},
\begin{equation}
E^{(k)}
=
\frac{1}{N_{\mathrm{dat}}}\sum_{I,J=1}^{N_{\mathrm{dat}}}
\left(g_I^{(\mathrm{art})(k)}-g_I^{(\mathrm{net})(k)}\right)
\left(\left({\mathrm{cov}}\right)^{-1}\right)_{IJ}
\left(g_J^{(\mathrm{art})(k)}-g_J^{(\mathrm{net})(k)}\right) 
\,\mbox{.}
\label{eq:errfun}
\end{equation}
Here $g_I^{\mathrm{(art)}(k)}$ is the value of the observable $g_I$ at the
kinematical point $I$ corresponding to the Monte Carlo replica $k$,
and $g_I^{\mathrm{(net)}(k)}$ is the same observable computed from the neural
network PDFs; the covariance matrix 
$\left({\mathrm{cov}}\right)_{IJ}$ is defined in Eq.~(\ref{eq:covmat}).  

The minimization procedure we adopt follows
closely that of Ref.~\cite{DelDebbio:2007ee}, to which we refer for a more
general discussion. Minimization is perfomed by means of a genetic
algorithm, which  minimizes the
figure of merit, Eq.~(\ref{eq:errfun}) by generating, at each
minimization step, a pool of new
neural nets, obtained by randomly mutating the parameters of the
starting set, and retaining the configuration which corresponds to the
lowest value of the figure of merit.

The parameters which characterize the behavior of the genetic
algorithm are tuned in order to optimize the efficiency of
the minimization procedure. We essentially rely on previous experience of
the development of unpolarized NNPDF sets: in particular, the
algorithm is characterized by a mutation rate, which 
decreases as a function of the number of the algorithm iterations 
$N_{\mathrm{ite}}$ according to the law~\cite{Ball:2008by}
\begin{equation}
 \eta_{i,j}=\eta_{i,j}^{(0)}/N_{\mathrm{ite}}^{r_\eta}
\,\mbox{.}
\label{eq:etarate}
\end{equation}
This way, in the early stages of the training large mutations are
allowed, while they become less likely as one approaches the
minimum. The starting mutation rates are chosen to be larger for PDFs
which contain more information. We perform two mutations per PDF at
each step, with the starting rates given in Tab.~\ref{tab:etapars}.
The exponent $r_\eta$ has been
introduced in order to optimally span the whole range of possible
beneficial mutations and it is randomized between $0$ and $1$ at each
iteration of the genetic algorithm, as in Ref.~\cite{Ball:2010de}.
\begin{table}[t]
\centering
\footnotesize
\begin{tabular}{cccc}
\toprule
$\eta^{(0)}_{i,\Delta\Sigma}$ &
$\eta^{(0)}_{i,\Delta g}$ & 
$\eta^{(0)}_{i,\Delta T_3}$ & 
$\eta^{(0)}_{i,\Delta T_8}$ \\
\midrule
$5, 0.5$ & $5, 0.5$ & $2, 0.2$ & $2, 0.2$\\
\bottomrule 
\end{tabular}
\mycaption{The initial values of the mutation rates for the two 
mutations of each PDF.}
\label{tab:etapars}
\end{table}

Furthermore, following Ref.~\cite{Ball:2010de}, we let the number of new
candidate solutions depend on the stage of the minimization.
At earlier stages of the minimization, when 
the number of generations is smaller than $N^{\mathrm{mut}}$, we
use a large population of mutants, $N_{\mathrm{mut}}^{a}\gg 1$, so a larger
space of mutations is being explored. At later stages of the
minimization, as the minimum is approached, a smaller
number of mutations $N_{\mathrm{mut}}^{b}\ll N_{\mathrm{mut}}^{a}$ is used. 
The values of the parameters 
$N_{\mathrm{gen}}^{\mathrm{mut}}$, $N_{\mathrm{mut}}^{a}$ and 
$N_{\mathrm{mut}}^{b}$ are collected in Tab.~\ref{tab:mutpars}.
\begin{table}[t]
\centering
\footnotesize
\begin{tabular}{ccccc}
\toprule
$N_{\mathrm{gen}}^{\mathrm{mut}}$ & $N^a_{\mathrm{mut}}$ & 
$N^b_{\mathrm{mut}}$ & $N_{\mathrm{gen}}^{\mathrm{wt}}$ & $E^{\mathrm{sw}}$ \\
\midrule
200 & 50 & 10 &  5000 & 2.5\\
\bottomrule
\end{tabular}
\mycaption{Values of the parameters of the genetic algorithm.}
\label{tab:mutpars}
\end{table}

Because the minimization procedure stops the fit to all experiments at
once, we must make sure that the quality of the fit to different
experiments is approximately the same. This is nontrivial, because of
the variety of experiments and data sets included in the
fit. Therefore, the figure of merit per data point for a given set is 
not necessarily a reliable indicator of the quality of the fit to that set,
because some experiments may have systematically underestimated or
overestimated uncertainties. Furthermore, unlike for unpolarized PDF
fits, information on the experimental covariance matrix is only
available for a small subset of experiments, so for most experiments
statistical and systematic errors must be added in quadrature, thereby
leading to an overestimate of uncertainties: this leads to a wide
spread of values of the figure of merit, whose value depends on the
size of the correlated uncertainties which are being treated as
uncorrelated.

A methodology to deal with this situation was developed in
Ref.~\cite{Ball:2010de}. The idea is to first determine the optimal value of
the figure of merit for each experiment, \textit{i.e.} 
a  set of target values $E_{i}^{\mathrm{targ}}$ for each of the $i$ experiments, 
then during the fit  give more weight to experiments for which the figure
of merit is further away from its target value, and stop to train
experiments which have already reached the target value. This is done by
minimizing, instead of the figure of merit Eq.~(\ref{eq:errfun}), the
weighted figure of merit
\begin{equation}
E_{\mathrm{wt}}^{(k)}=\frac{1}{N_{\mathrm{dat}}}
\sum_{j=1}^{N_{\mathrm{sets}}}p_j^{(k)} N_{\mathrm{dat},j}E_j^{(k)}
\,\mbox{,}
\label{eq:weight_errfun}
\end{equation}
where $E_j^{(k)}$ is the error function for the $j$-th data set with
$N_{{\mathrm{dat}},j}$  points, and the weights  $p_j^{(k)}$ are given by 
\begin{enumerate}
\item If $E_{i}^{(k)} \ge E_{i}^{\mathrm{targ}}$, then 
$p_i^{(k)}=\left( E_{i}^{(k)}/E_{i}^{\mathrm{targ}}\right)^n$
\,\mbox{,} 
\item If $E_{i}^{(k)} < E_{i}^{\mathrm{targ}}$, then $p_i^{(k)}=0$
\,\mbox{,}
\end{enumerate}
with $n$ a free parameter which essentially determines the amount of
weighting. In the unpolarized fits of
Refs.~\cite{Ball:2010de,Ball:2011mu,Ball:2011uy,Ball:2012cx} the value
$n=2$ was used. Here instead we will choose $n=3$. This larger value,
determined by trial and error, is justified by the wider spread of
figures of merit in the polarized case, which in turn is related 
to the absence of correlated systematics for most
experiments.

The target values $E_{i}^{\mathrm{targ}}$ are determined through an
iterative procedure: they are set to one at first, then a very long
fixed-length fit is run, and the values of $E_{i}$ are taken as
targets for a new fit, which is performed until stopping
(according to the criterion to be discussed in the following 
Section). The values of $E_{i}$ at the end of this fit are then taken
as new targets until convergence is reached, usually after a couple
iterations. 

Weighted training stops after the first $N_{\mathrm{gen}}^{\mathrm{wt}}$ 
generations, unless the total error function
Eq.~(\ref{eq:errfun}) is above some threshold $E^{(k)}\geq E^{\mathrm{sw}}$. 
If it is, weighted training continues until $E^{(k)}$ falls
below the threshold value. Afterwards, the error function is just the
unweighted error function Eq.~(\ref{eq:errfun}) computed on
experiments. This ensures that the figure of merit behaves smoothly in
the last stages of training. The values for the parameters 
$N_{\mathrm{gen}}^{\mathrm{wt}}$ and $E^{\mathrm{sw}}$ are also given in
Tab.~\ref{tab:mutpars}. 

\subsection{Determination of the optimal fit}

Because the neural network parametrization is extremely redundant, it may
be able to fit not only the underlying behavior of the PDFs, but also
the statistical noise in the data. Therefore, the best fit does not 
necessarily coincide with the absolute minimum of the figure of merit 
Eq.~(\ref{eq:errfun}). We thus determine the best fit, 
as in Refs.~\cite{DelDebbio:2007ee,Ball:2008by}, using a 
cross-validation method~\cite{Bishop:1995}:
for each replica, the data are randomly divided in two sets, training
and validation, which include a fraction $f_{\mathrm{tr}}^{(j)}$ and 
$f_{\mathrm{val}}^{(j)}=1-f_{\mathrm{tr}}^{(j)}$ of the data points respectively.  
The figure of merit Eq.~(\ref{eq:errfun}) is then computed for both sets.
The training figure of merit function is minimized through the genetic
algorithm, while the validation figure of merit is monitored: when the
latter starts increasing while the former still decreases, the fit is
stopped. This means that the fit is stopped as soon as the neural
network is starting to learn the statistical fluctuations of the
points, which are different in the training and validation sets,
rather than the underlying law which they share.

In the unpolarized fits of
Refs.~\cite{DelDebbio:2007ee,Ball:2008by,Ball:2010de,Ball:2011mu,Ball:2011uy,Ball:2012cx} 
equal training and validation fractions were uniformly chosen,
$f_{\mathrm{tr}}^{(j)}=f_{\mathrm{val}}^{(j)}=1/2$.
However, in this case we have to face the problem that the number of
data points is quite small: most experiments include about ten
data points (see Tab.~\ref{tab:exps-sets}). Hence, it is difficult to
achieve a stable minimization if only half of them
are actually used for minimization, as we have explicitly verified. Therefore,  
we have chosen to include 80\% of the data in the training set,
\textit{i.e.}  $f_{\mathrm{tr}}^{(j)}=0.8$ and 
$f_{\mathrm{val}}^{(j)}=0.2$. We have explicitly
verified that the fit quality which is obtained in this case is
comparable to the one achieved when including all data in the training set
(\textit{i.e.} with $f_{\mathrm{tr}}^{(j)}=1.0$ and 
$f_{\mathrm{val}}^{(j)}=0.0$), but the presence of a
nonzero validation set allows for a satisfactory stopping, as we have
checked by explicit inspection of the profiles of the figure of merit
as a function of training time.

In practice, in order to implement cross-validation we must determine
a stopping criterion, namely, give conditions which must be satisfied
in order for the minimization to stop.
First, we require that the weighted training stage
has been completed, \textit{i.e.}, that the genetic algorithm has been run for
at least  $N_{\mathrm{gen}}^{\mathrm{wt}}$ minimization steps. Furthermore,
we check that all experiments have reached a value of the figure of merit
below a minimal threshold $E_{\mathrm{thr}}$. Note that because stopping
can occur only after weighted training has been switched off, and this
in turn only happens when the figure of merit falls below the value
$E^{\mathrm{sw}}$, the total figure of merit must be below this value in
order for stopping to be possible.

We then compute moving averages
\begin{equation}
\langle E_{\mathrm{tr,val}}(i)\rangle\equiv
\frac{1}{N_{\mathrm{smear}}}
\sum_{l=i-N_{\mathrm{smear}}+1}^iE_{\mathrm{wt;\,tr,val}}(l)
\,\mbox{,}
\label{eq:smearing}
\end{equation}
of the figure of merit Eq.~(\ref{eq:weight_errfun}) for either the
training or the validation set at the  $l$-th genetic minimzation step.
The fit is then stopped if 
\begin{equation}
r_{\mathrm{tr}} < 1-\delta_{\mathrm{tr}}\quad{\mathrm{and}}
\quad r_{\mathrm{val}} > 1+\delta_{\mathrm{val}}
\,\mbox{,}
\label{eq:trratcond}
\end{equation}
where
\begin{equation}
r_{\mathrm{tr}}\equiv\frac{\langle E_{\mathrm{tr}}(i)\rangle}
{\langle E_{\mathrm{tr}}(i-\Delta_{\mathrm{smear}})\rangle} 
\,\mbox{,}
\label{eq:dec-train}
\end{equation}
\begin{equation}
r_{\mathrm{val}}\equiv\frac{\langle E_{\mathrm{val}}(i)\rangle}
{\langle E_{\mathrm{val}}(i-\Delta_{\mathrm{smear}})\rangle} 
\,\mbox{.}
\label{eq:dec-valid}
\end{equation}

The parameter $N_{\mathrm{smear}}$ determines the width of the moving
average; the parameter $\Delta_{\mathrm{smear}}$ determines the distance
between the two points along the minimization path which are compared
in order to determine whether the figure of merit is increasing or
decreasing; and the parameters $\delta_{\mathrm{tr}}$, $\delta_{\mathrm{val}}$ are
the threshold values for the decrease of the training and increase of
the validation figure of merit to be deemed significant.
The optimal value of these parameters should be chosen in such a way
that the fit does not stop on a statistical fluctuation, yet it does
stop before the fit starts overlearning (\textit{i.e.} learning statistical
fluctuation). As explained in Ref.~\cite{Ball:2010de}, this is done 
studying the profiles of the error functions for individual data set
and for individual replicas. 
In order to avoid unacceptably long fits, training is stopped anyway
when a maximum number of iterations $N_{\mathrm{gen}}^{\mathrm{max}}$ 
is reached, even though the stopping
conditions Eqs.~(\ref{eq:trratcond}) are not satisfied.  This leads to
a small loss of accuracy of the corresponding fits:  this is
acceptable provided it only happens for a small enough fraction of
replicas. If a fit stops at $N_{\mathrm{gen}}^{\mathrm{max}}$ without the
stopping criterion having been satisfied, we also check that the
total figure of merit is below the value $E^{\mathrm{sw}}$ at which
weighted training is switched off. If it hasn't, we conclude that the
specific fit has not converged, and we retrain the same replica, \textit{i.e.},
we perform a new fit to the same data starting with a different random
seed. This only occurs in about one or two percent of cases.
The full set of parameters which determine the stopping criterion is given in
Tab.~\ref{tab:stopping_pars}.
\begin{table}[t]
\centering
\footnotesize
\begin{tabular}{cccccc}
\toprule
$N_{\mathrm{gen}}^{\mathrm{max}}$ & 
$E_{\mathrm{thr}}$ & $N_{\mathrm{smear}}$ & 
$\Delta_{\mathrm{smear}}$ & $\delta_{\mathrm{tr}}$ & $\delta_{\mathrm{val}}$ \\
\midrule
$20000$ & $8$ & $100$ & $100$ & $5 \cdot 10^{-4}$ & $5 \cdot 10^{-4}$ \\
\bottomrule
\end{tabular}
\mycaption{Parameters for the stopping criterion.}
\label{tab:stopping_pars}
\end{table}

\begin{figure}[t]
 \centering
 \epsfig{width=0.40\textwidth,figure=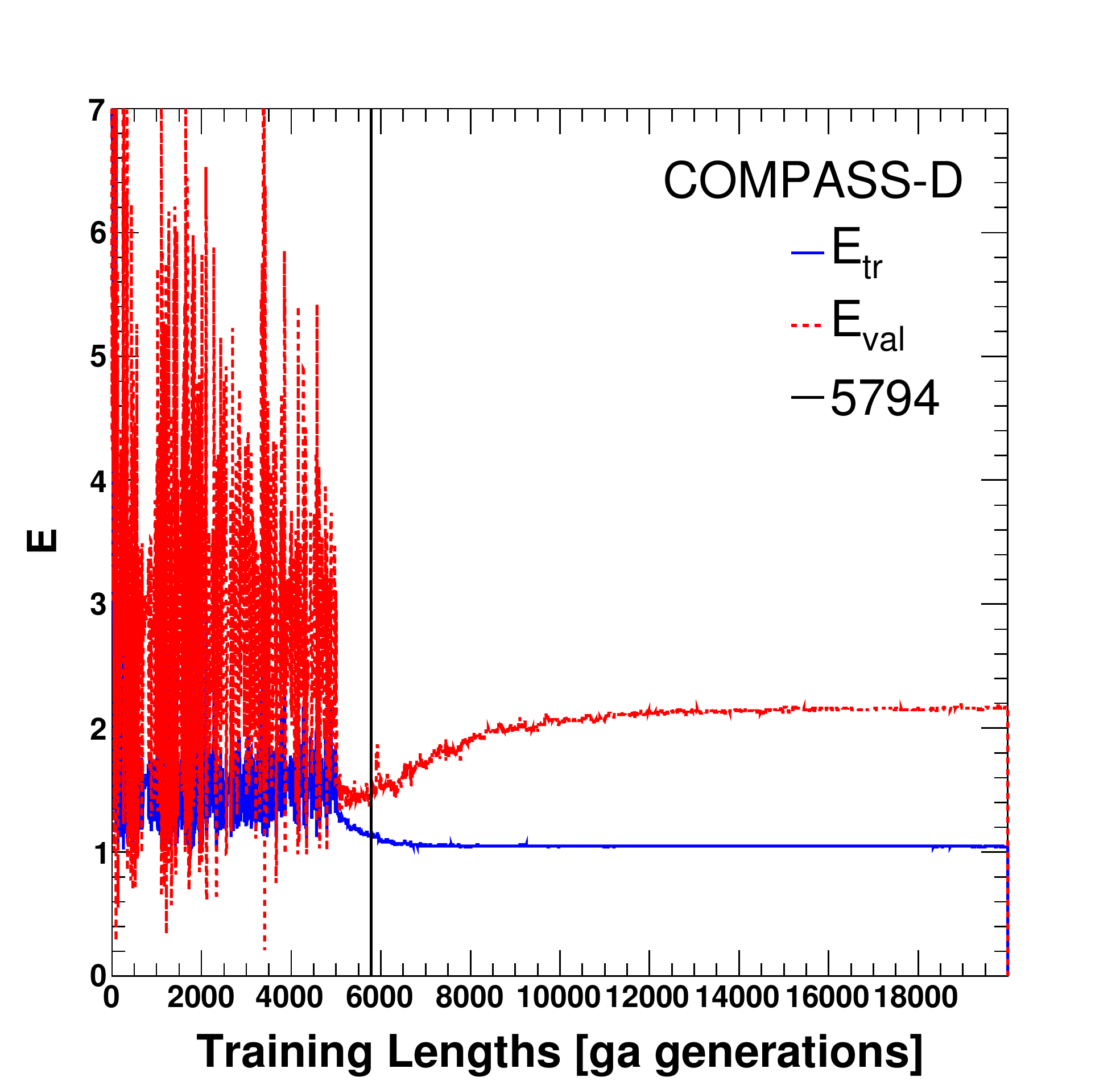}
 \epsfig{width=0.40\textwidth,figure=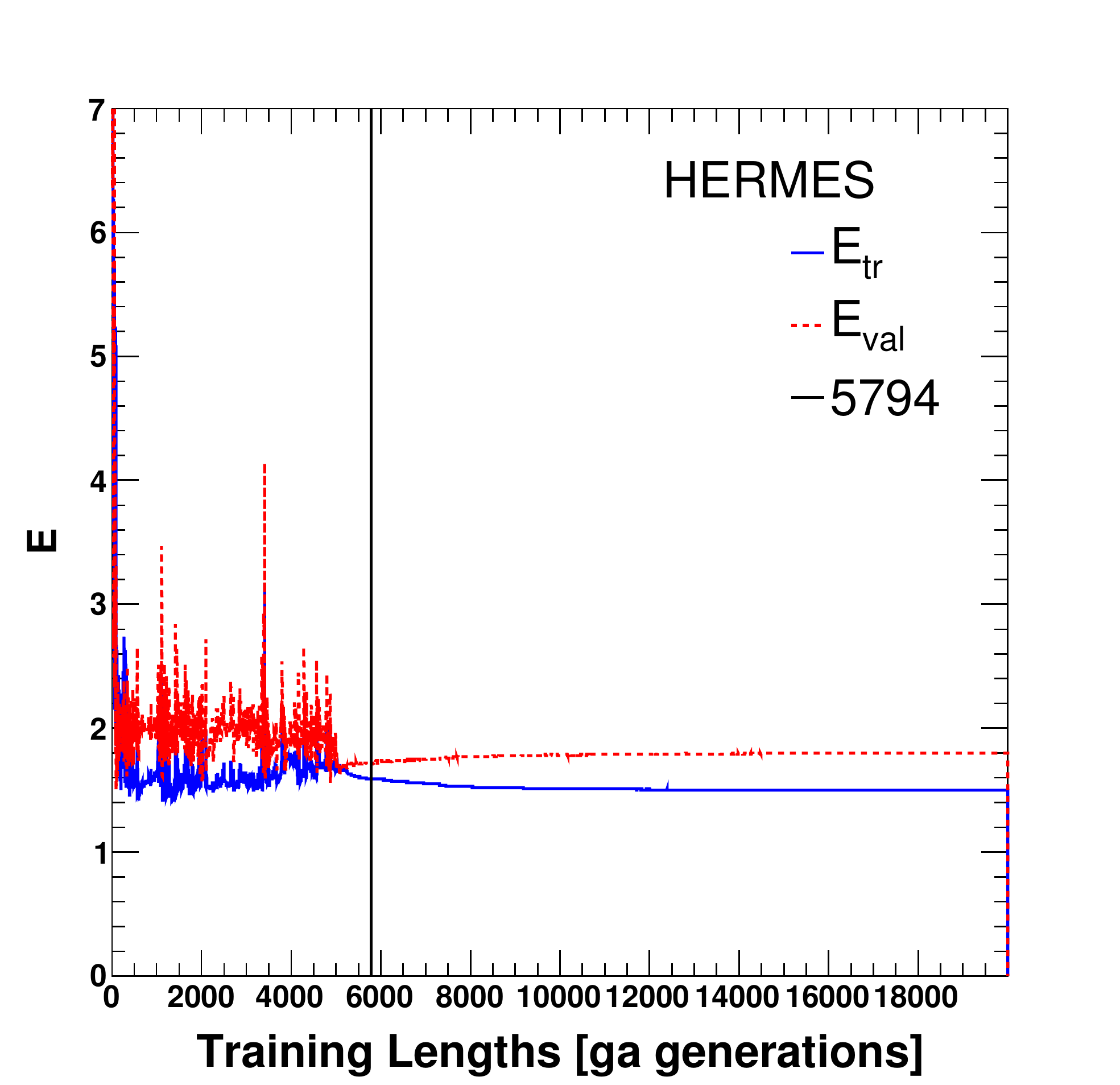}
 \mycaption{Behaviour of the moving average Eq.~(\ref{eq:smearing})
of the training and validation figure of merit for two different
data sets included in a global fit (\texttt{COMPASS-P} and
\texttt{HERMES}) as a function of training length. The straight vertical line
indicates the point at which the fit stops with the stopping parameters
of Tab.~\ref{tab:stopping_pars}. The weighted
training is switched off at $N_{\mathrm{gen}}^{\mathrm{wt}}=5000$.}
\label{fig:stop}
\end{figure}

An example of how the stopping criterion works in practice is
shown in Fig.~\ref{fig:stop}. We display the moving averages
Eq.~(\ref{eq:smearing}) of the
training and validation error functions 
$\langle E_{\mathrm{tr,val}}^{(k)}\rangle$,
computed with the parameter settings of Tab.~\ref{tab:stopping_pars}, and
plotted as a function of the number of iterations of the
genetic algorithm, for a particular replica and for two
of the experiments included in the fit. The wide fluctuations which
are observed in the first part of training, up to the 
$N_{\mathrm{gen}}^{\mathrm{wt}}$-th generation, are due to the fact that the weights
which enter the definition of the figure of merit 
Eq.~(\ref{eq:weight_errfun}) are frequently adjusted. Nevertheless,
the downwards trend of the figure of merit is clearly visible.
Once the weighted training is switched off,
minimization proceeds smoothly. The vertical line denotes the point at
which the stopping criterion is satisfied. Here, we have let the
minimization go on beyond this point, and we clearly see that the
minimization has entered  an overlearning regime, in which
the validation error function $E_{\mathrm{val}}^{(k)}$ is rising while the training
$E_{\mathrm{tr}}^{(k)}$ is still decreasing. Note that the stopping point,
which in this particular case occurs at
$N_{\mathrm{gen}}^{\mathrm{stop}}=5794$, is determined by verifying that the
stopping criteria are satisfied by the total 
figure of merit, not that of individual experiments shown here. 
The fact that the two different experiments
considered here both start overlearning at the same point shows that
the weighted training has been effective in synchronizing the fit
quality for different experiments.

\subsection{Theoretical constraints}
\label{sec:thconstraints}

Polarized PDFs are only loosely constrained by data, which are 
scarce and  not very accurate. Theoretical constraints are
thus especially important in reducing the uncertainty on the PDFs. We
consider in particular positivity and integrability. 

Positivity of the individual cross-sections
which enter the polarized asymmetries Eq.~(\ref{eq:xsecasy}) implies
that, up to power-suppressed corrections, longitudinal polarized
structure functions are bounded by their unpolarized counterparts,
\textit{i.e.}  
\begin{equation}
|g_1(x,Q^2)| \le F_1(x,Q^2) 
\,\mbox{.}
\label{eq:pos}
\end{equation}
At leading order, structure functions are proportional to parton
distributions, so imposing Eq.~(\ref{eq:pos}) for any process (and a
similar condition on an asymmetry which is sensitive to polarized
gluons~\cite{Altarelli:1998gn}), would imply 
\begin{equation}
|\Delta  f_i(x,Q^2)|\le f_i(x,Q^2)
\label{eq:pospdf}
\end{equation}
for any pair of unpolarized and polarized PDFs $f$ and $\Delta f$, for
all quark flavors and gluon $i$, for all $x$, and for all $Q^2$.
Beyond leading order, the condition
Eq.~(\ref{eq:pos}) must still hold, but it does not necessarily imply 
Eq.~(\ref{eq:pospdf}). Rather, one should then impose at least a number of
conditions of the form of Eq.~(\ref{eq:pos}) on physically measurable
cross-sections which is equal to the number of independent polarized
PDFs. For example, in principle one may require that the condition
Eq.~(\ref{eq:pos}) is separately satisfied for each flavor, \textit{i.e.}
when only contributions from the
$i$-th flavor are included in the polarized and unpolarized structure
function: this corresponds to requiring positivity of semi-inclusive
structure functions which could in principle be measured 
(and that fragmentation effects cancel in the ratio).
A condition on the gluon can be obtained by imposing
positivity of the polarized and unpolarized cross-sections for 
inclusive Higgs production in gluon-proton
scattering~\cite{Altarelli:1998gn}, again measurable in principle if
not in practice.

Because $g_1/F_1\sim x$ as $x\to0$~\cite{Ball:1995ye}, the positivity
bound Eq.~(\ref{eq:pos}) is only significant at large enough
$x\gtrsim10^{-2}$. On the other hand, at very large $x$ the NLO
corrections to the LO positivity bound become
negligible~\cite{Altarelli:1998gn,Forte:1998kd}. Therefore, the NLO
positivity bound in practice only differs from its LO counterpart
Eq.~(\ref{eq:pospdf}) in a small region $10^{-2}\lesssim x\lesssim0.3$, and
even there by an amount of rather less that
10\%~\cite{Altarelli:1998gn}, which is negligible in comparison to the size
of PDF uncertainties, as we shall see explicitly  in
Sec.~\ref{sec:results}.

Therefore, we will impose the leading-order positivity bound
Eq.~(\ref{eq:pospdf}) on each flavor combination $\Delta q_i+\Delta
\bar q_i$ and on the gluon $\Delta g$ (denoted as $\Delta f_i$ below).
We do this by requiring 
\begin{equation}
|\Delta  f_i(x,Q^2)| \le  f_i(x,Q^2) + \sigma_i(x,Q^2)  
\,\mbox{,}
\label{eq:possigma}
\end{equation}
where $\sigma_i(x,Q^2)$ is the uncertainty on the corresponding unpolarized PDF
combination $f_i(x,Q^2)$ at the kinematic point $(x,Q^2)$. This choice is
motivated by two considerations. First, it is clearly meaningless to
impose positivity of the polarized PDF 
to an accuracy which is greater than that with which
the unpolarized PDF has been determined. Second, because the
unpolarized PDFs satisfy NLO positivity, they can become negative and
thus they may have nodes. As a consequence, the LO bound
Eq.~(\ref{eq:pospdf}) would imply that the polarized PDF must vanish
at the same point, which would be clearly meaningless.

As in Ref.~\cite{Ball:2010de} positivity is imposed during the
minimization procedure, thereby guaranteeing that the genetic
algorithm only explores the subspace of acceptable physical
solutions. This is done through a Lagrange multiplier 
$\lambda_{\mathrm{pos}}$, \textit{i.e.}  by computing the polarized PDF at 
$N_{\mathrm{dat,pos}}$ fixed kinematic points $(x_p,Q_0^2)$ and then adding 
to the error function Eq.~(\ref{eq:errfun}) a contribution 
\begin{eqnarray}
&&E_{\mathrm{pos}}^{(k)}={\lambda_{\mathrm{pos}}}\sum_{p=1}^{N_{\mathrm{dat,pos}}}
\Bigg\{
\sum_{j=u+\bar{u},d+\bar{d},s+\bar{s},g} \Theta\left[
\left|\Delta f_j^{\mathrm{(net)}(k)}(x_p,Q_0^2)\right| -\left(f_j + \sigma_j\right) 
(x_p,Q_0^2) \right] \nonumber\\
&&\qquad\qquad \times
\left[\left|\Delta f_j^{\mathrm{(net)}(k)}(x_p,Q_0^2)\right| - 
\left(f_j + \sigma_j\right)(x_p,Q_0^2) \right]  
\Bigg\} 
\,\mbox{.}
\label{eq:lagrmult}
\end{eqnarray}
This provides a penalty, proportional to the violation of positivity,
which enforces  Eq.~(\ref{eq:possigma}) separately for all
the non-zero quark-antiquark combinations. The values of the
unpolarized PDF combination $f_j(x,Q^2)$ and its uncertainty 
$\sigma_j(x,Q^2)$ are computed using the
\texttt{NNPDF2.1} PDF set at NLO~\cite{Ball:2011mu}, while $\Delta
f_j^{\mathrm{(net)}(k)}$ is the corresponding polarized PDF computed from
the neural network parametrization for the $k$-th replica. The
polarized and unpolarized PDFs are evaluated at $N_{\mathrm{dat,pos}}=20$
points with $x$ equally spaced in the interval
\begin{equation} 
x \in [10^{-2},0.9] 
\,\mbox{.}  
\end{equation}
Positivity is imposed at the initial scale $Q_0^2=1$ GeV$^2$ since once 
positivity is enforced at low scales, it is automatically
satisfied at larger scales~\cite{Altarelli:1998gn,Forte:1998kd}.
After stopping, we finally test the positivity condition
Eq.~(\ref{eq:possigma}) is satisfied on a grid of $N_{\mathrm{dat,pos}}=40$ 
points in the same intervals. Replicas for which positivity is violated in one
or more points are discarded and retrained.

In the unpolarized case, in which positivity only played a minor role
in constraining PDFs, a fixed value of the Lagrange multiplier
$\lambda_{\mathrm{pos}}$ was chosen. In the polarized case it turns out to
be necessary to vary the Lagrange multiplier along the minimization.
Specifically, we let
\begin{equation}
\left\{
\begin{array}{rcll}
\lambda_{\mathrm{pos}} 
& = & 
\lambda_{\mathrm{max}}^{(N_{\mathrm{gen}}-1)/(N_{\lambda_{\mathrm{max}}}-1)} & 
N_{\mathrm{gen}} < N_{\lambda_{\mathrm{max}}}\\
\lambda_{\mathrm{pos}} 
& = & 
\lambda_{\mathrm{max}} & N_{\mathrm{gen}} \geq N_{\lambda_{\mathrm{max}}}.
\end{array}
\right.
\label{eq:lagrmult2}
\end{equation}
This means that the Lagrange multiplier increases as
the minimization proceeds, starting from $\lambda_{\mathrm{pos}}=1$, at the
first minimization step, $N_{\mathrm{gen}}=1$, up to 
$\lambda_{\mathrm{pos}} = \lambda_{\mathrm{max}}\gg 1$ when 
$N_{\mathrm{gen}} = N_{\lambda_{\mathrm{max}}}$. After
$N_{\lambda_{\mathrm{max}}}$ generations $\lambda_{\mathrm{pos}}$ is then kept
constant to $\lambda_{\mathrm{max}}$. The rationale behind this choice is
that the genetic algorithm can thus learn experimental data and
positivity at different stages of minimization.  During the early
stages, the contribution coming from the modified error function
Eq.~(\ref{eq:lagrmult}) is negligible, due to the moderate value of
the Lagrange multiplier; hence, the genetic algorithm will mostly
learn the basic shape of the PDF driven by experimental data. As soon
as the minimization proceeds, the contribution coming from the
Lagrange multiplier increases, thus ensuring the proper learning of
positivity: at this stage, most of the replicas which will not fulfill
the positivity bound will be discarded.

The final values of $N_{\lambda_{\mathrm{max}}}=2000$ and $\lambda_{\mathrm{max}}=10$
have been  determined as follows. 
First of all, we have performed a fit without
any positivity constraint and we have observed that data were mostly
learnt in about $2000$ generations: hence we have taken this value for
$N_{\lambda_{\mathrm{max}}}$. Then we have tried different values for
$\lambda_{\mathrm{max}}$ until we managed to reproduce the same $\chi^2$
obtained in the previous, positivity unconstrained, fit.  This ensures
that positivity is not learnt to the detriment of the global fit
quality.

Notice that the value of $\lambda_{\mathrm{max}}$ is rather small if
compared to the analogous Lagrange multiplier used in the unpolarized
case~\cite{Ball:2011mu}. This depends on the fact that, in this latter
case, positivity is learnt at the early stages of minimization, when
the error function can be much larger than its asymptotic value:
a large Lagrange multiplier is then needed to select the best
replicas. Also, unpolarized PDFs are quite well constrained by data
and positivity is almost automatically fulfilled, except in some
restricted kinematic regions; only a few replicas 
violate positivity and need to be penalized. This means that the behavior
of the error function Eq.~(\ref{eq:errfun}), which governs the fitting
procedure, is essentially dominated by data instead of positivity.  

In the polarized case, instead, positivity starts to be effectively
implemented only after some minimizaton steps, when the error function
has already decreased to a value of a few units. Furthermore, we have
checked that, at this stage, most of replicas slightly violate the
positivity condition Eq.~(\ref{eq:possigma}): thus, a too large value
of the Lagrange multiplier on the one hand would penalize replicas which
are good in reproducing experimental data and only slightly worse in
reproducing positivity; on the other, it would promote replicas which
fulfill positivity but whose fit to data is quite bad. As a
consequence of this behavior, the convergence of the minimization
algorithm would be harder to reach.  We also verified that, using a value 
for the Lagrange multiplier up to $\lambda_{\mathrm{pos}}=100$ leads to no 
significant improvement neither in the fulfillment of positivity requirement 
nor in the fit quality.  
We will show in detail the effects of the positivity 
bound Eq.~(\ref{eq:possigma}) 
on the fitted replicas and on polarized PDFs in Sec.~\ref{sec:results}.

Finally, we impose that PDFs are integrable, \textit{i.e.} that they have
finite first moments. This corresponds to the assumption that the nucleon
matrix element of the axial current for the $i$-th flavor is finite.
The integrability condition is imposed by computing at each
minimization step the integral of each of the polarized PDFs
in a given interval, 
\begin{equation} 
I(x_1,x_2)=\int_{x_1}^{x_2} dx~\Delta q_i(x,Q_0^2) \,
\qquad \Delta q_i=\Delta\Sigma, \Delta g, \Delta T_3, \Delta T_8 
\end{equation} 
with $x_1$ and $x_2$ chosen in the small $x$ region, well below the
data points, and verifying that in this region the growth of the
integral as $x_1$ decreases for fixed $x_2$ is less than logarithmic. 
In practice, we test for the condition
\begin{equation}
\frac{I(x_1,x_2)}{I(x_1^\prime,x_2)} < 
\frac{\ln\frac{x_2}{x_1}}{\ln\frac{x_2}{x_1\prime}}, 
\end{equation}
with $x_1<x_1^\prime$. 
Mutations which do not satisfy the 
condition are rejected during the minimization procedure. In our default fit,
we chose $x_1=10^{-5}$,  $x_1^\prime=2\cdot 10^{-5}$ and $x_2=10^{-4}$.

\section{Results}
\label{sec:results}

In this Section, we present the first determination of a polarized PDF
set based on the NNPDF methodology, \texttt{NNPDFpol1.0}.
We will first illustrate the statistical features of our PDF fit, then
compare the PDFs in our set to those from other recent determinations
introduced in Sec.~\ref{sec:setsummary}. 
We will also 
discuss the stability of our results upon the variation of several
theoretical and methodological assumptions, namely the treatment of 
target mass corrections, the use of sum rules to fix the nonsinglet
axial charges, the effect of positivity constraints on polarized PDFs,
and impact of preprocessing of neural networks on small- and large-$x$
PDF behavior.

\subsection{Statistical features}
\label{sec:stat_features}

The statistical features of the \texttt{NNPDFpol1.0} analysis are summarized in 
Tabs.~\ref{tab:chi2tab1}-\ref{tab:chi2tab2}, for the full
data set and for individual experiments and sets respectively. 
\begin{table}[t]
\begin{center}
\footnotesize
\begin{tabular}{cc}
\toprule
\multicolumn{2}{c}{\texttt{NNPDFpol1.0}} \\
\midrule
$\chi^{2}_{\mathrm{tot}}$ &      0.77 \\
$\langle E \rangle \pm \sigma_{E} $   &       1.82 $\pm$       0.18    \\
$\langle E_{\mathrm{tr}} \rangle \pm \sigma_{E_{\mathrm{tr}}}$ &  1.66 $\pm$ 0.49 \\
$\langle E_{\mathrm{val}} \rangle \pm \sigma_{E_{\mathrm{val}}}$&  1.88 $\pm$ 0.67 \\
$\langle{\mathrm{TL}} \rangle \pm \sigma_{\mathrm{TL}}$   &  6927 $\pm$ 3839 \\
\midrule
$\langle \chi^{2(k)} \rangle \pm \sigma_{\chi^{2}} $  & 0.91 $\pm$       0.12    \\
\bottomrule
\end{tabular}
\end{center}
\mycaption{Statistical estimators for \texttt{NNPDFpol1.0} with 
$N_\mathrm{rep}=100$ replicas.}
\label{tab:chi2tab1}
\end{table}
\begin{table}[t]
\footnotesize
\centering
\begin{tabular}{llcc}
\toprule 
Experiment  & Set  & $\chi^{2}_{\mathrm{tot}}$  
            & $\langle E\rangle \pm \sigma_{E}$ \\
\midrule
EMC       & &  0.44&  1.54 $ \pm $  0.64 \\
&EMC-A1P   &  0.44&  1.54 $ \pm$  0.64 \\
\midrule
SMC       & &  0.93&  1.93 $ \pm $  0.51 \\
&SMC-A1P   &  0.40&  1.44 $ \pm$  0.54 \\
&SMC-A1D   &  1.46&  2.42 $ \pm$  0.82 \\
\midrule
SMClowx   & &  0.97&  1.90 $ \pm $  0.67 \\
&SMClx-A1P &  1.40&  2.32 $ \pm$  1.13 \\
&SMClx-A1D &  0.53&  1.48 $ \pm$  0.69 \\
\midrule
E143      & &  0.64&  1.68 $ \pm $  0.29 \\
&E143-A1P  &  0.43&  1.49 $ \pm$  0.34 \\
&E143-A1D  &  0.88&  1.90 $ \pm$  0.45 \\
\midrule
E154      & &  0.40&  1.69 $ \pm $  0.61 \\
&E154-A1N  &  0.40&  1.69 $ \pm$  0.61 \\
\midrule
E155      & &  0.89&  1.96 $ \pm $  0.36 \\
&E155-G1P  &  0.89&  2.00 $ \pm$  0.51 \\
&E155-G1N  &  0.88&  1.93 $ \pm$  0.47 \\
\midrule
COMPASS-D & &  0.65&  1.72 $ \pm $  0.53 \\
&CMP07-A1D &  0.65&  1.72 $ \pm$  0.53 \\
\midrule
COMPASS-P & &  1.31&  2.38 $ \pm $  0.72 \\
&CMP10-A1P &  1.31&  2.38 $ \pm$  0.72 \\
\midrule
HERMES97  & &  0.34&  1.37 $ \pm $  0.69 \\
&HER97-A1N &  0.34&  1.37 $ \pm$  0.69 \\
\midrule
HERMES    & &  0.79&  1.79 $ \pm $  0.30 \\
&HER-A1P   &  0.44&  1.49 $ \pm$  0.39 \\
&HER-A1D   &  1.13&  2.09 $ \pm$  0.50 \\
\bottomrule
\end{tabular}
\mycaption{Same as Tab.~\ref{tab:chi2tab1}, but for individual experiments.}
\label{tab:chi2tab2}
\end{table}
The mean value of the error function, Eq.~(\ref{eq:errfun}), $\langle E\rangle$,
shown in the tables both for the total, training and validation data sets
is the figure of merit for the quality of the fit of each PDF replica
to the corresponding data replica. The quantity which is actually
minimized during the neural network training is this figure of merit
for the training set, supplemented by weighting in the early stages of
training according to Eq.~(\ref{eq:weight_errfun}) and by a Lagrange
multiplier to enforce positivity according to Eq.~(\ref{eq:lagrmult}).
In the table we also show the average over all replicas, 
$\langle\chi_{\mathrm{tot}}^{2(k)}\rangle$, of $\chi_{\mathrm{tot}}^{2(k)}$
computed for the $k$-th replica, which coincides with the figure of
merit Eq.~(\ref{eq:weight_errfun}), but with the data replica 
$g_I^{\mathrm{(art)}(k)}$
replaced by the experimental data $g_I^{\mathrm{(dat)}}$. We finally show 
$\chi^2_{\mathrm{tot}}$, which  coincides with the figure of
merit Eq.~(\ref{eq:weight_errfun}), but again with  
$g_I^{\mathrm{(art)}(k)}$ replaced by $g_I^{\mathrm{(dat)}}$, and also with
$g_I^{(\mathrm{net})(k)}$ replaced by 
$\langle g_I^{(\mathrm{net})(k)}\rangle$, 
\textit{i.e.} the average of the observable over
replicas, which provides our best prediction.
The average number of iterations of the genetic algorithm at stopping, 
$\langle \mathrm{TL}\rangle$, is also given in this table. 

\begin{figure}[t]
\begin{center}
\epsfig{width=0.40\textwidth,figure=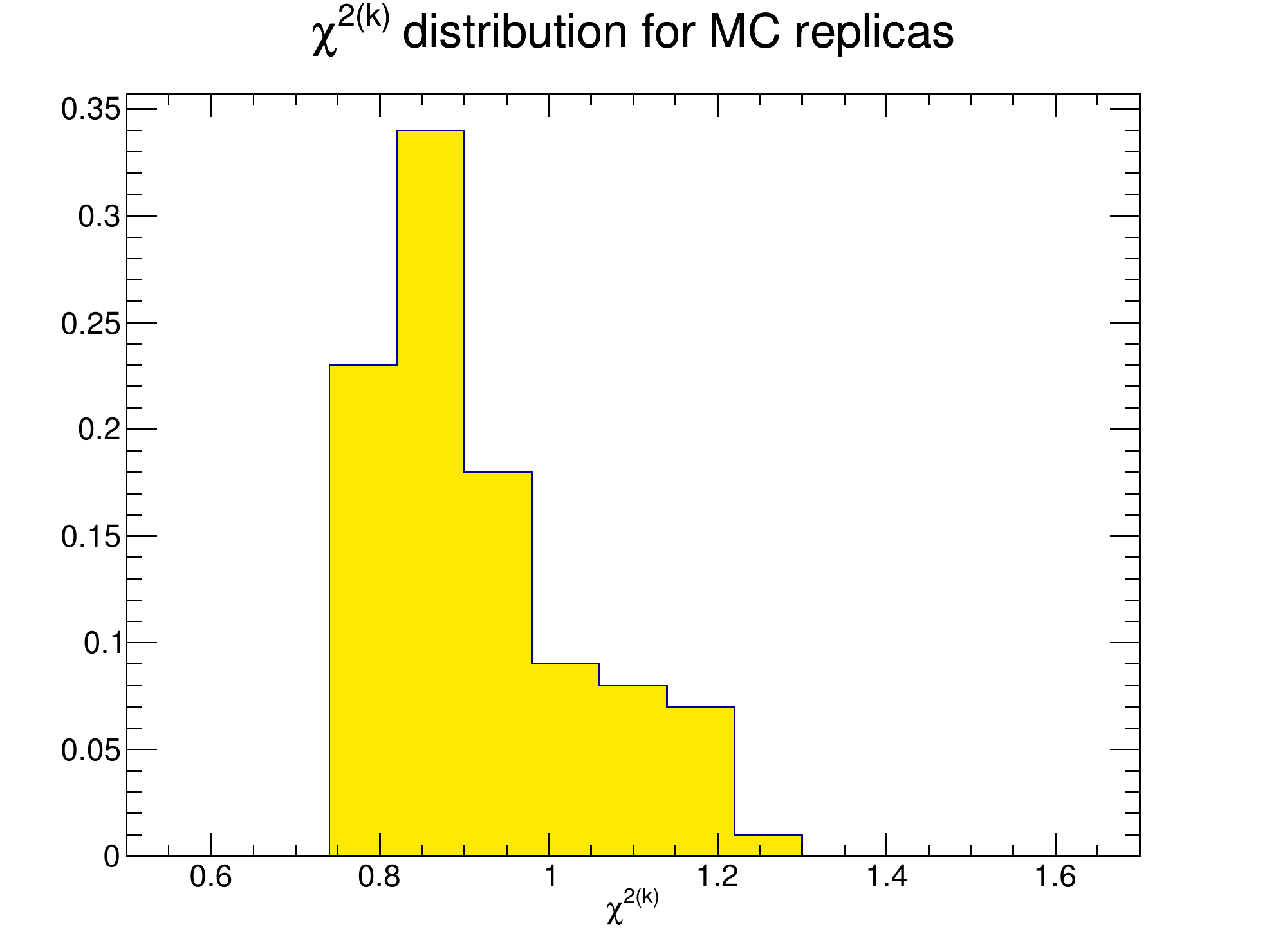}
\epsfig{width=0.40\textwidth,figure=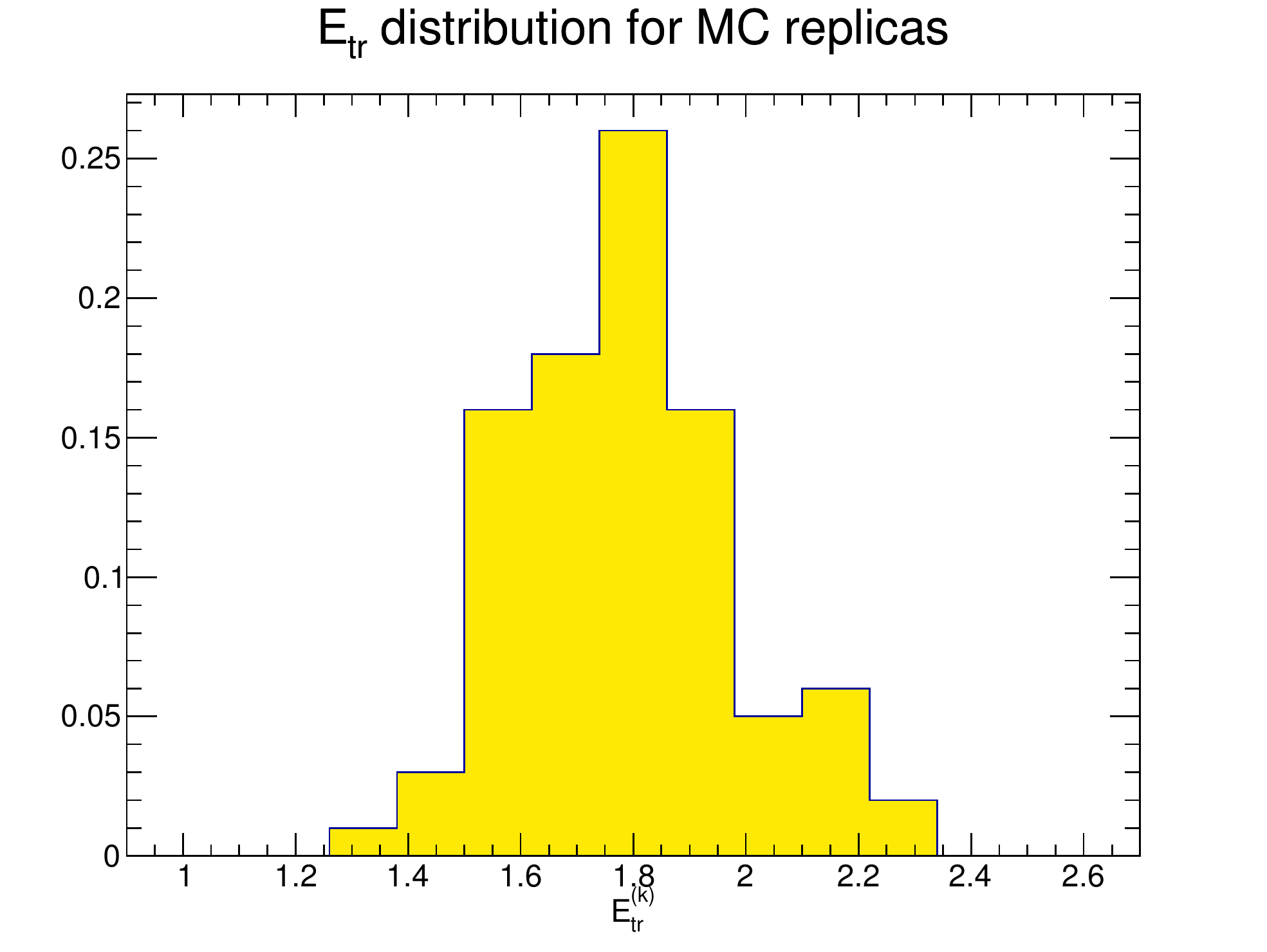}
\mycaption{Distribution of $\chi^{2(k)}$ and $E_{\mathrm{tr}}^{(k)}$ over 
the sample of $N_{\mathrm{rep}}=100$ replicas.} 
\label{fig:chi2-Etr-distr}
\end{center}
\end{figure}

\begin{figure}[t]
\begin{center}
\epsfig{width=0.40\textwidth,figure=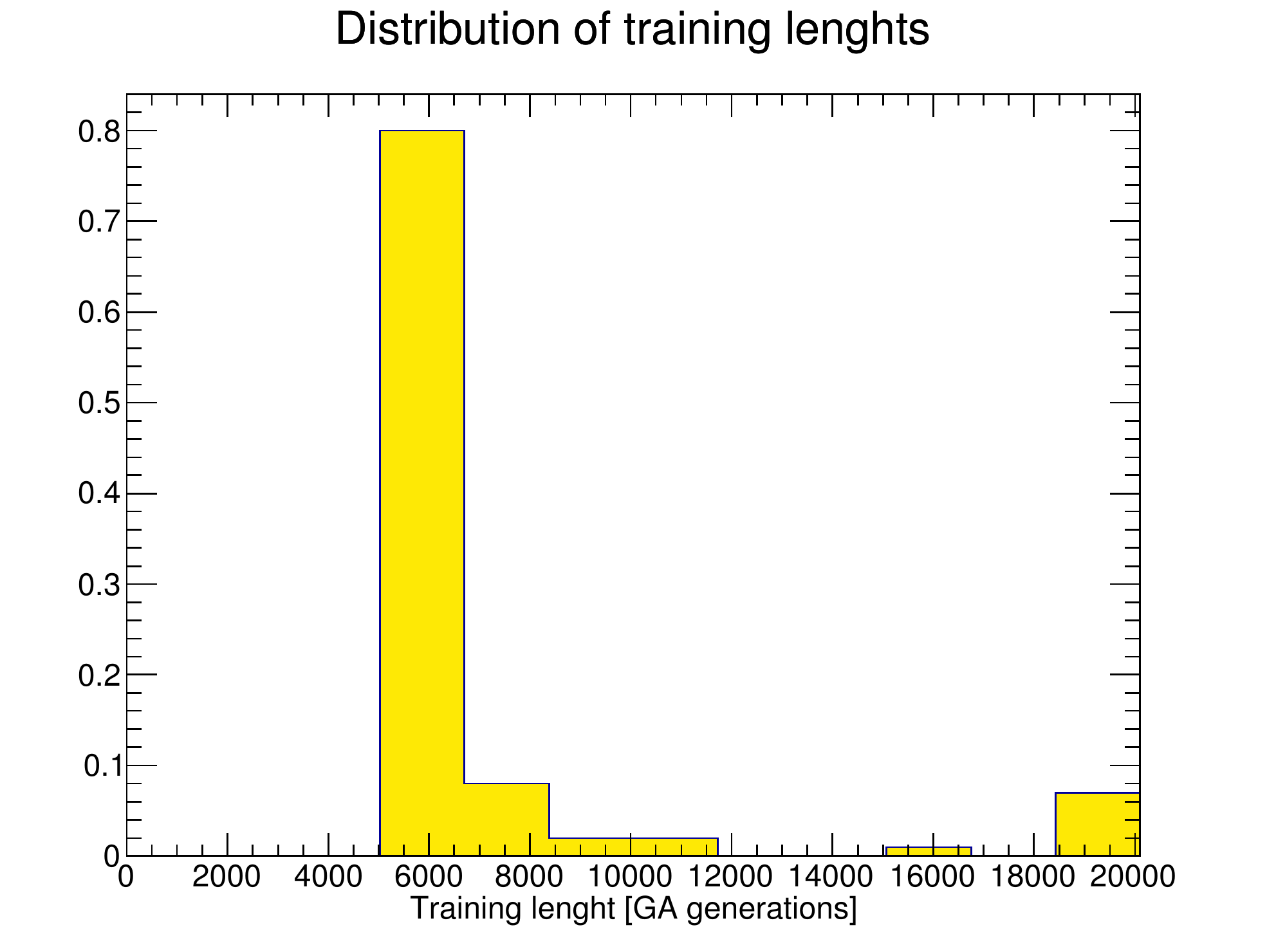}
\mycaption{\small Distribution of training lengths over the sample of 
$N_{\mathrm{rep}}=100$ replicas.} 
\label{fig:TL-distr}
\end{center}
\end{figure}

The distribution of $\chi^{2(k)}$, $E_{\mathrm{tr}}^{(k)}$, and
training lengths among the $N_{\mathrm{rep}}=100$ replicas are shown
in Fig.~\ref{fig:chi2-Etr-distr} and Fig.~\ref{fig:TL-distr}
respectively. Note that the latter has a long
tail which causes an accumulation of points at the maximum
training length, $N_{\mathrm{gen}}^{\mathrm{max}}$. This means that
there is a fraction of replicas that do not 
fulfill the stopping criterion. 
This may cause a loss in accuracy in outlier fits,
which however make up fewer than $10\%$ of the total sample. 

\begin{figure}[t]
\begin{center}
\epsfig{width=0.7\textwidth,figure=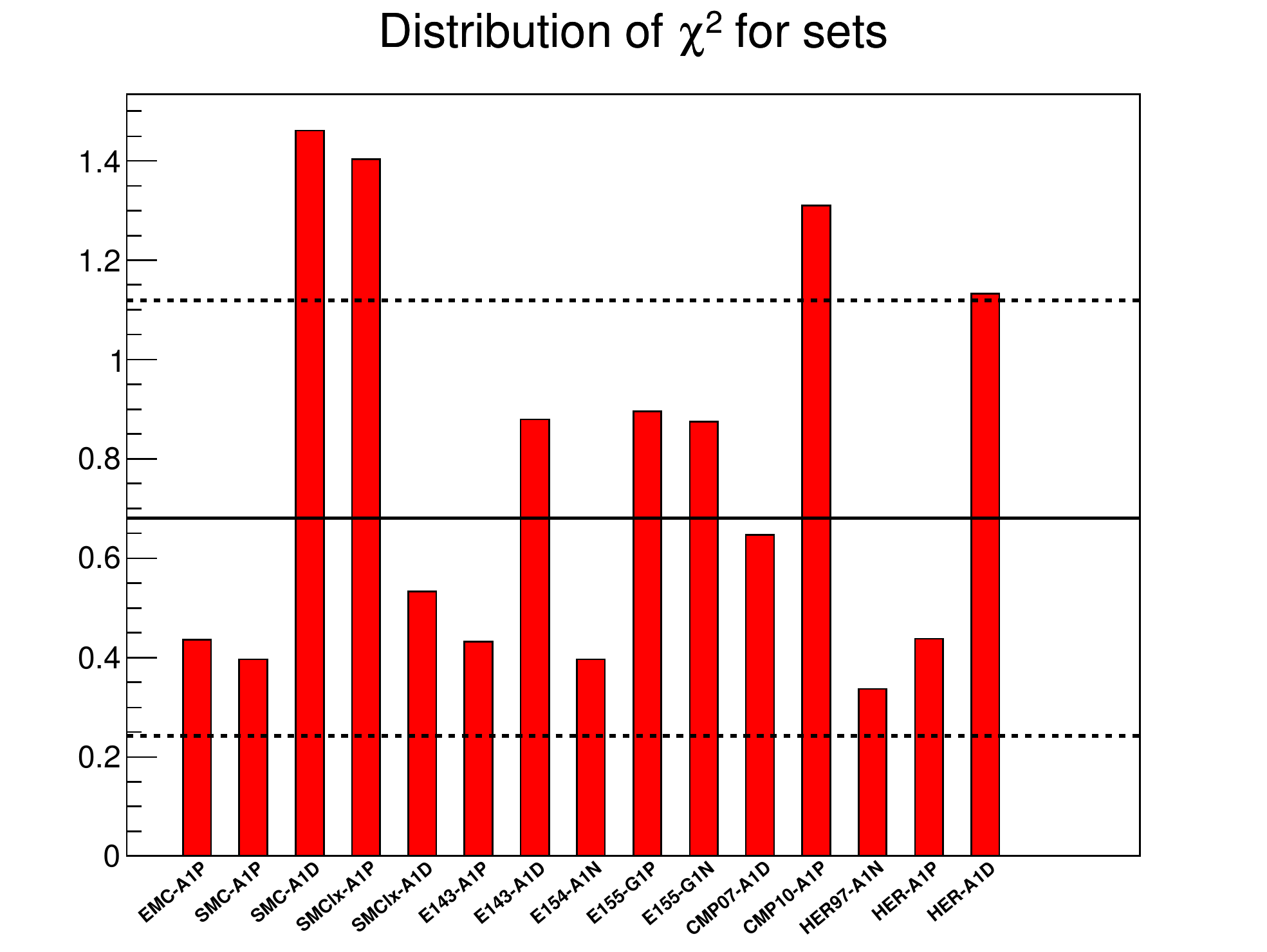}
\mycaption{Value of the $\chi^2$ per data
point for the data sets
included in the \texttt{NNPDFpol1.0} reference fit, 
listed in Tab.~\ref{tab:chi2tab2}. 
The horizontal line is 
the unweighted average of these $\chi^2$ over the data sets and 
the black dashed lines give  the one-sigma interval about it.} 
\label{fig:chi2-dist}
\end{center}
\end{figure}

The features of the fit can be summarized as follows:
\begin{itemize}
\item The quality of the central fit, as measured by its
  $\chi_{\mathrm{tot}}^{2}=0.77$, is good. However, this value should
  be taken with care in view of the fact that uncertainties for all
  experiments but two are overestimated because  the
  covariance matrix is not available and thus correlations between
  systematics cannot be properly accounted for. This explains the
  value lower than one for this quantity, which would be very unlikely
  if it had included correlations.
\item The values of $\chi_{\mathrm{tot}}^{2}$ and $\langle E \rangle$ differ
  by approximately one unit. This is due to the fact that replicas
  fluctuate within their uncertainty about the experimental data, which in
  turn are Gaussianly distributed about a true
  value~\cite{Forte:2002fg}: it shows that the neural network is correctly
  reproducing the underlying law thus being closer to the true
  value. This is confirmed by the fact that $\langle \chi^{2(k)}\rangle$ is of
  order one.
\item The distribution of $\chi^2$ for different experiments (also
  shown as a histogram in Fig.~\ref{fig:chi2-dist}) shows sizable
  differences, and indeed the standard deviation (shown as a dashed
  line in the plot) about the mean (shown as a solid line) is very
  large. This can be understood as a consequence of the lack of
  information on the covariance matrix: experiments where
  large correlated uncertainties are treated as uncorrelated will
  necessarily have a smaller value of the  $\chi^2$.
\end{itemize}

\begin{figure}[p]
\begin{center}
\epsfig{width=0.35\textwidth,figure=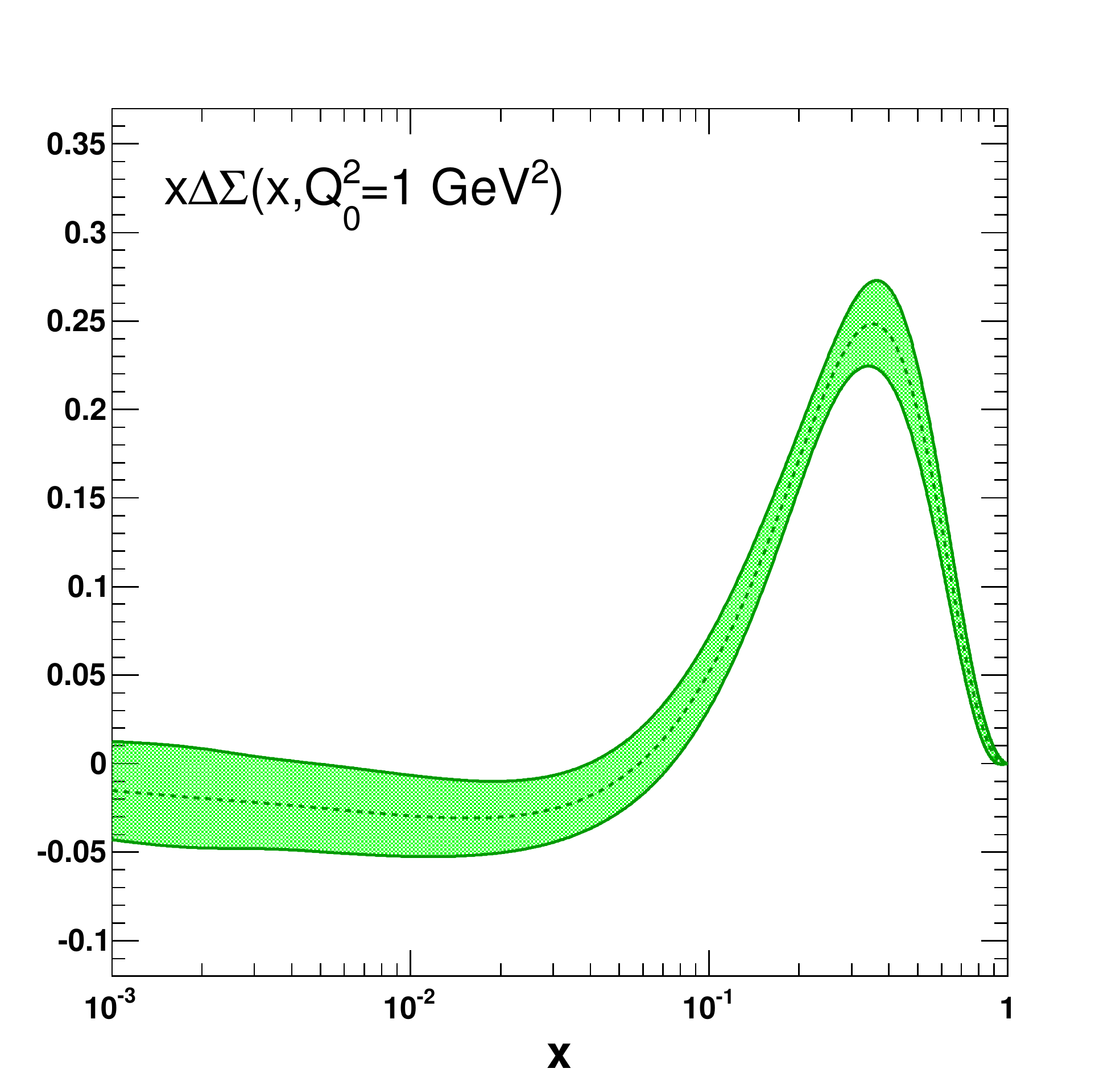}
\epsfig{width=0.35\textwidth,figure=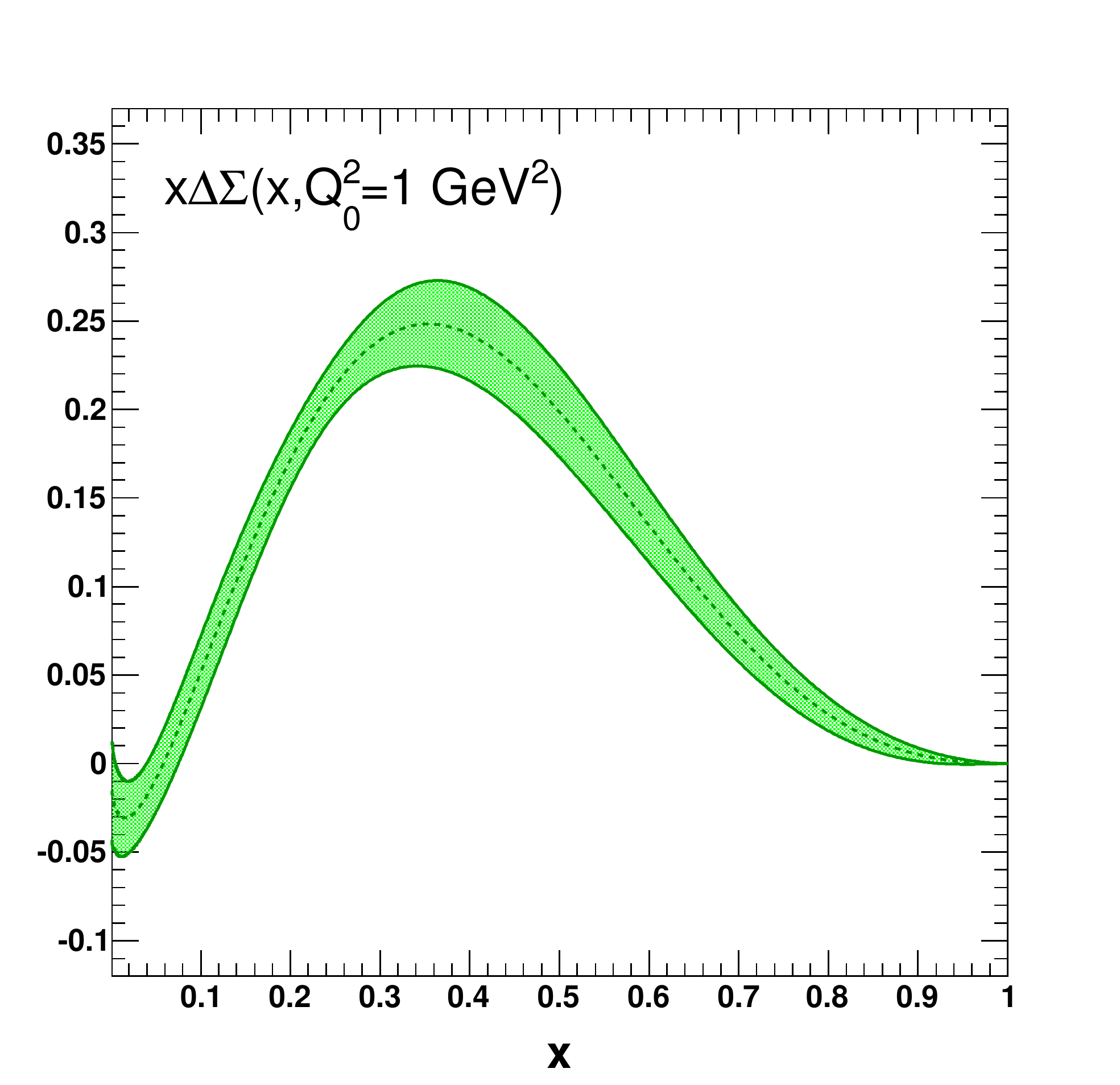}
\epsfig{width=0.35\textwidth,figure=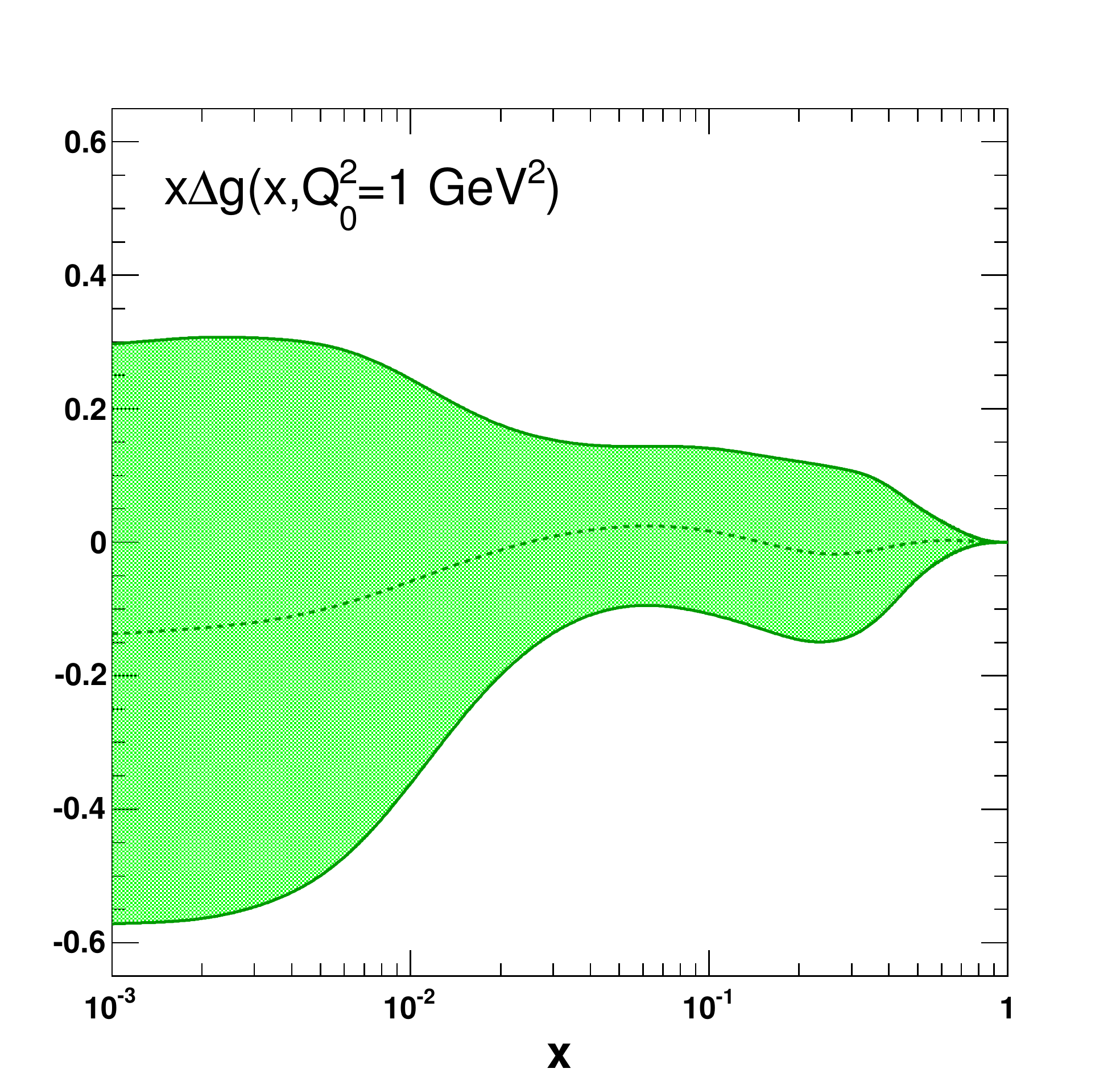}
\epsfig{width=0.35\textwidth,figure=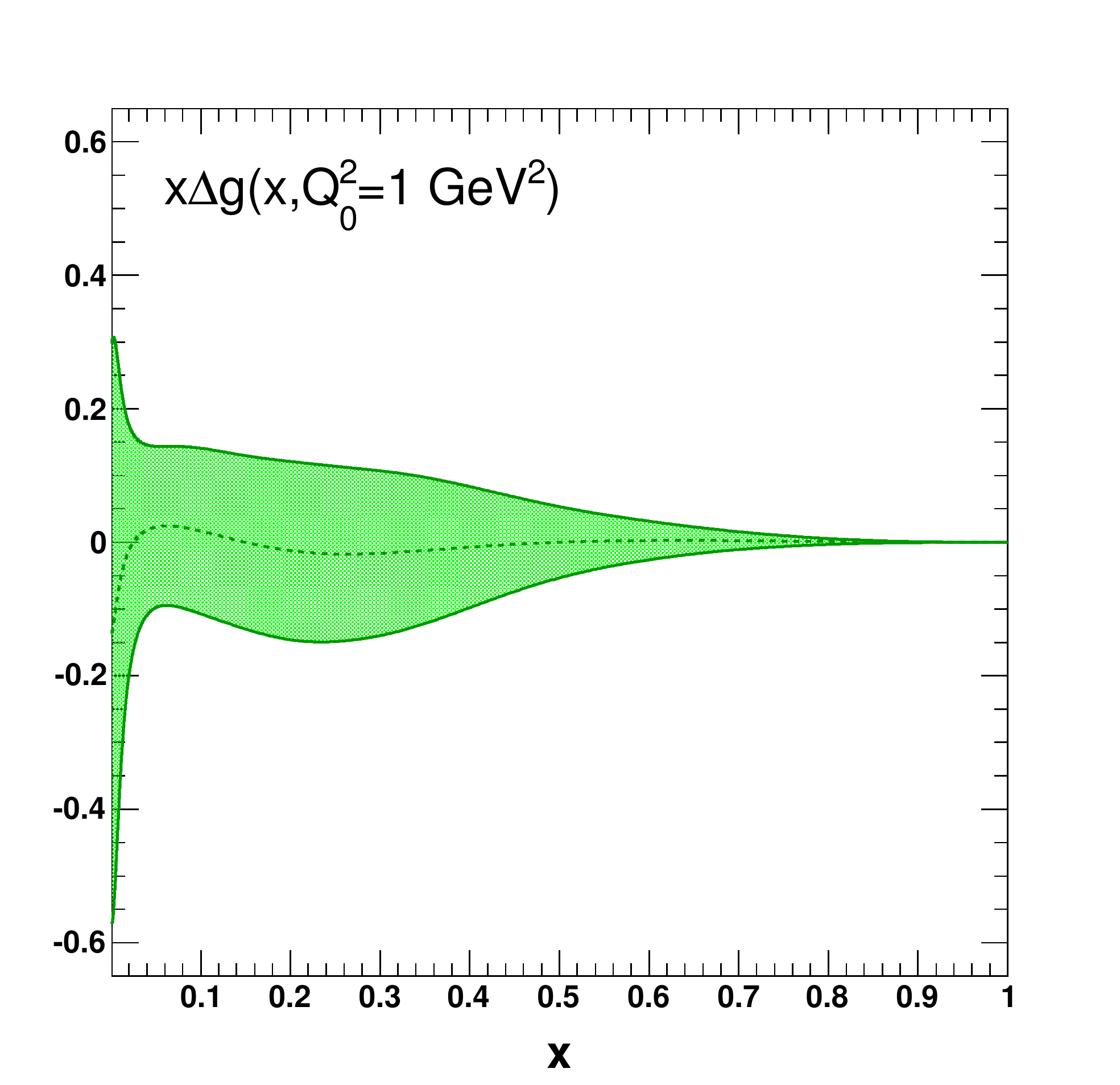}
\epsfig{width=0.35\textwidth,figure=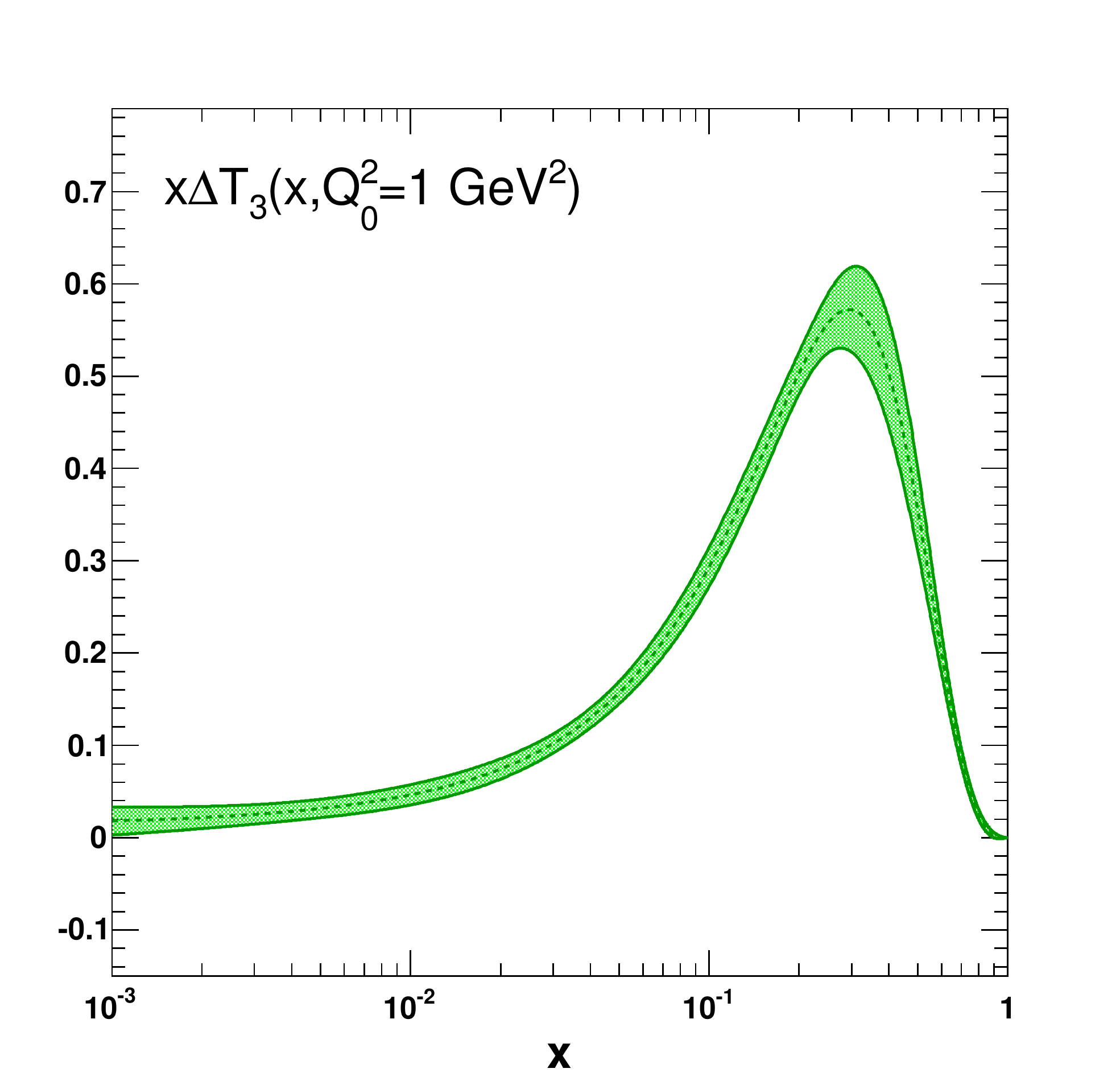}
\epsfig{width=0.35\textwidth,figure=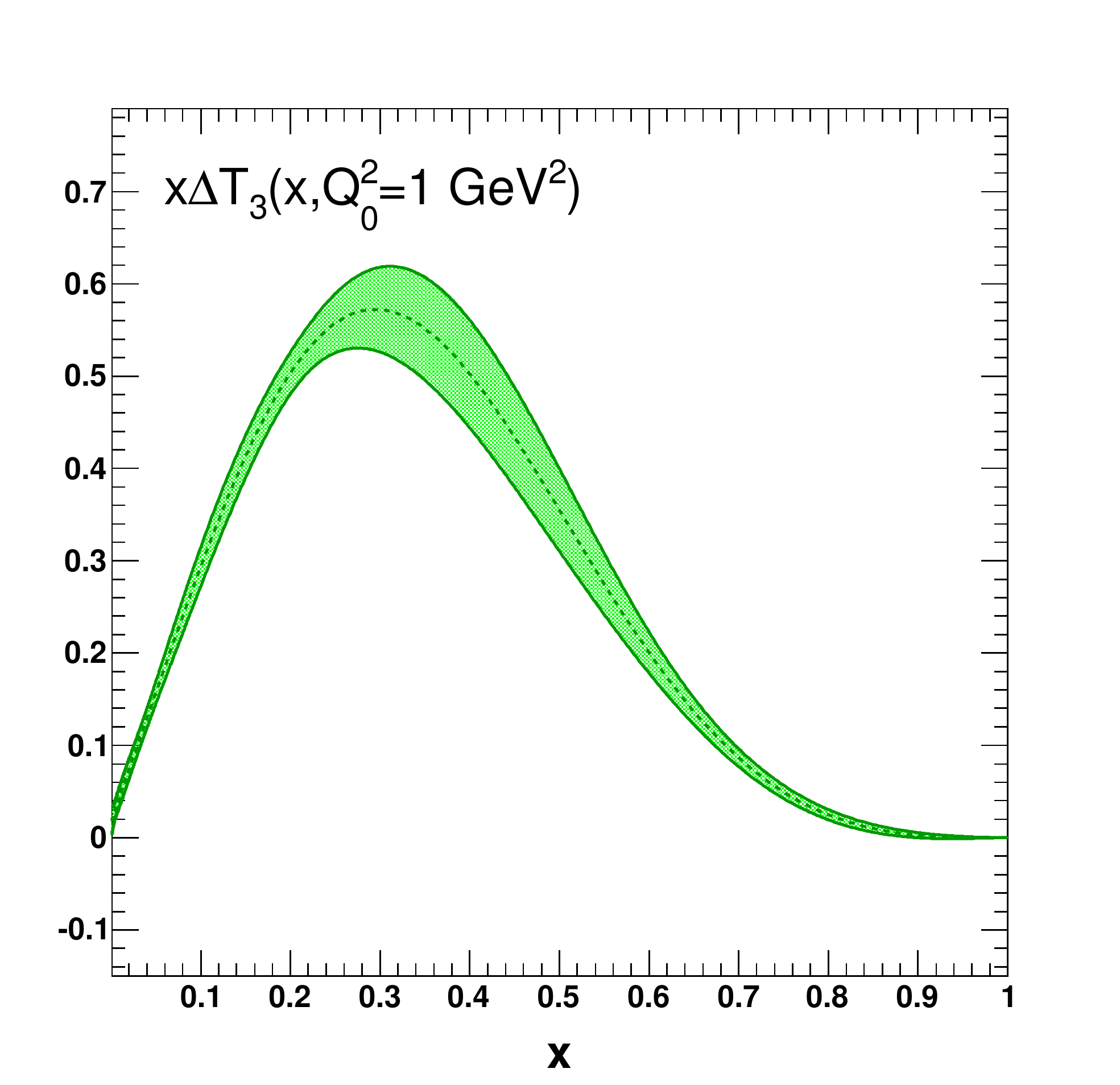}
\epsfig{width=0.35\textwidth,figure=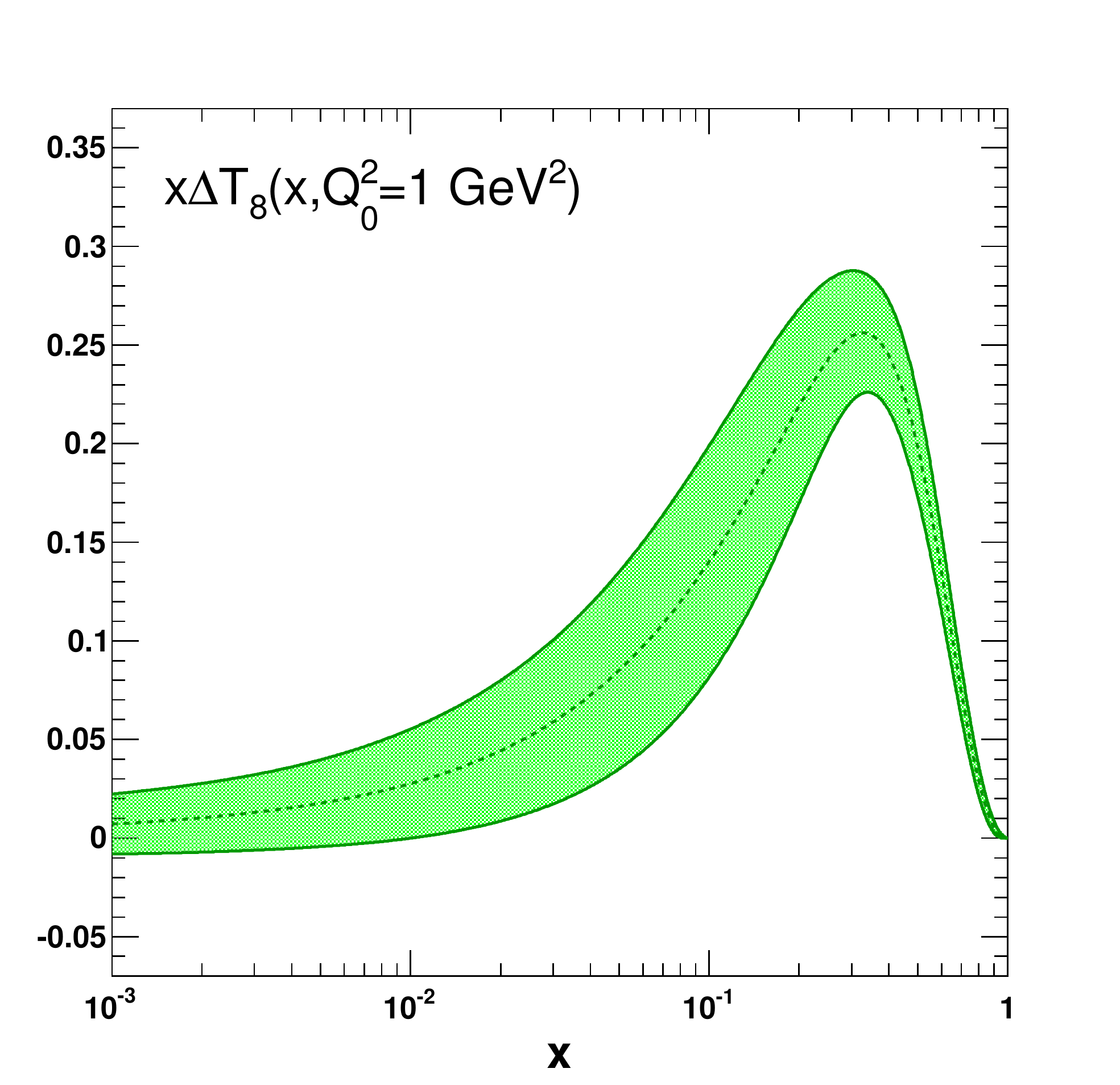}
\epsfig{width=0.35\textwidth,figure=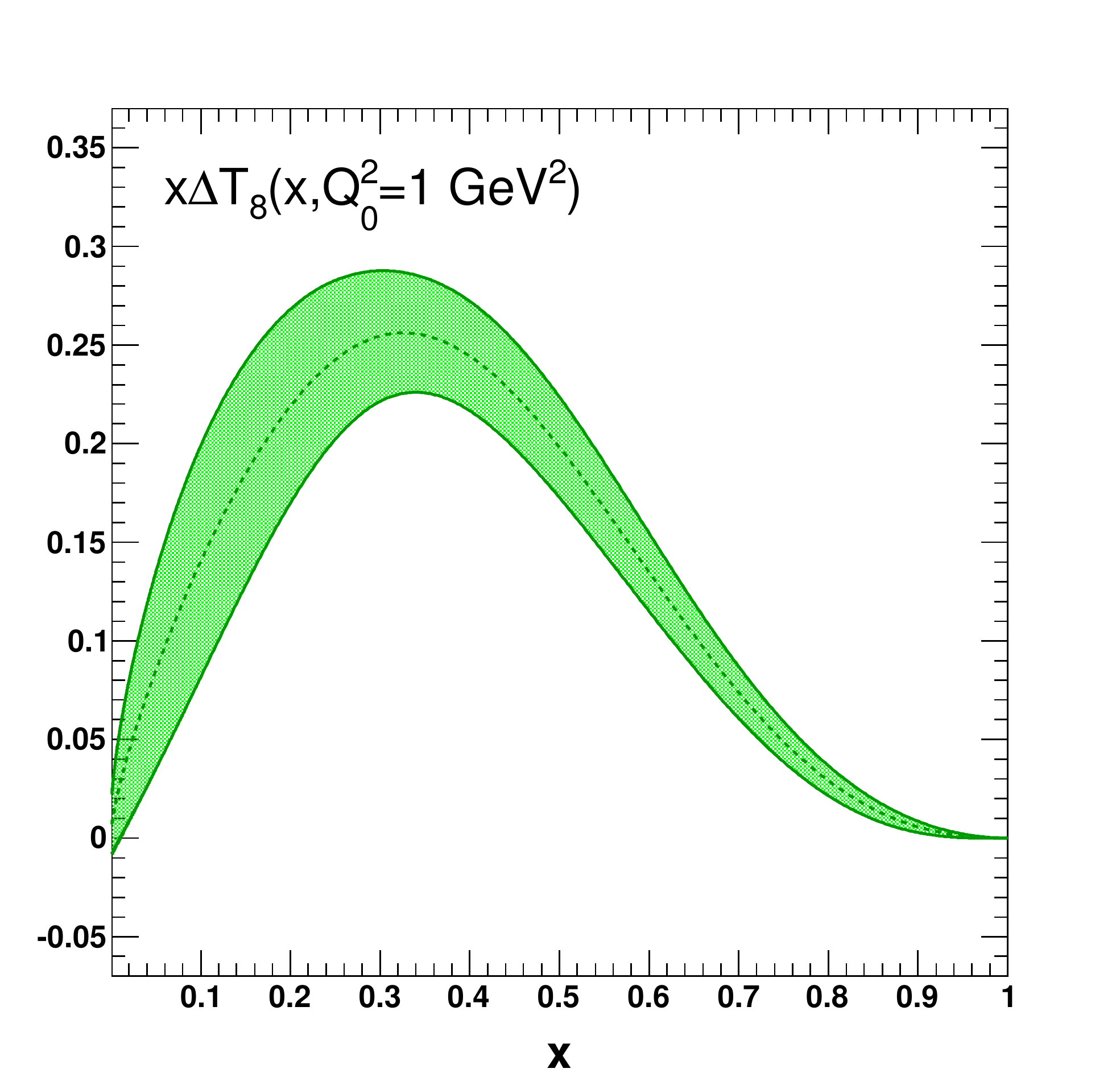}
\mycaption{The \texttt{NNPDFpol1.0} parton distributions at 
$Q_0^2=1$ GeV$^2$ in the parametrization basis plotted as a function of $x$,
on a logarithmic (left) and linear (right) scale.}
\label{fig:ppdfs-100}
\end{center}
\end{figure}

\begin{figure}[p]
\begin{center}
\epsfig{width=0.35\textwidth,figure=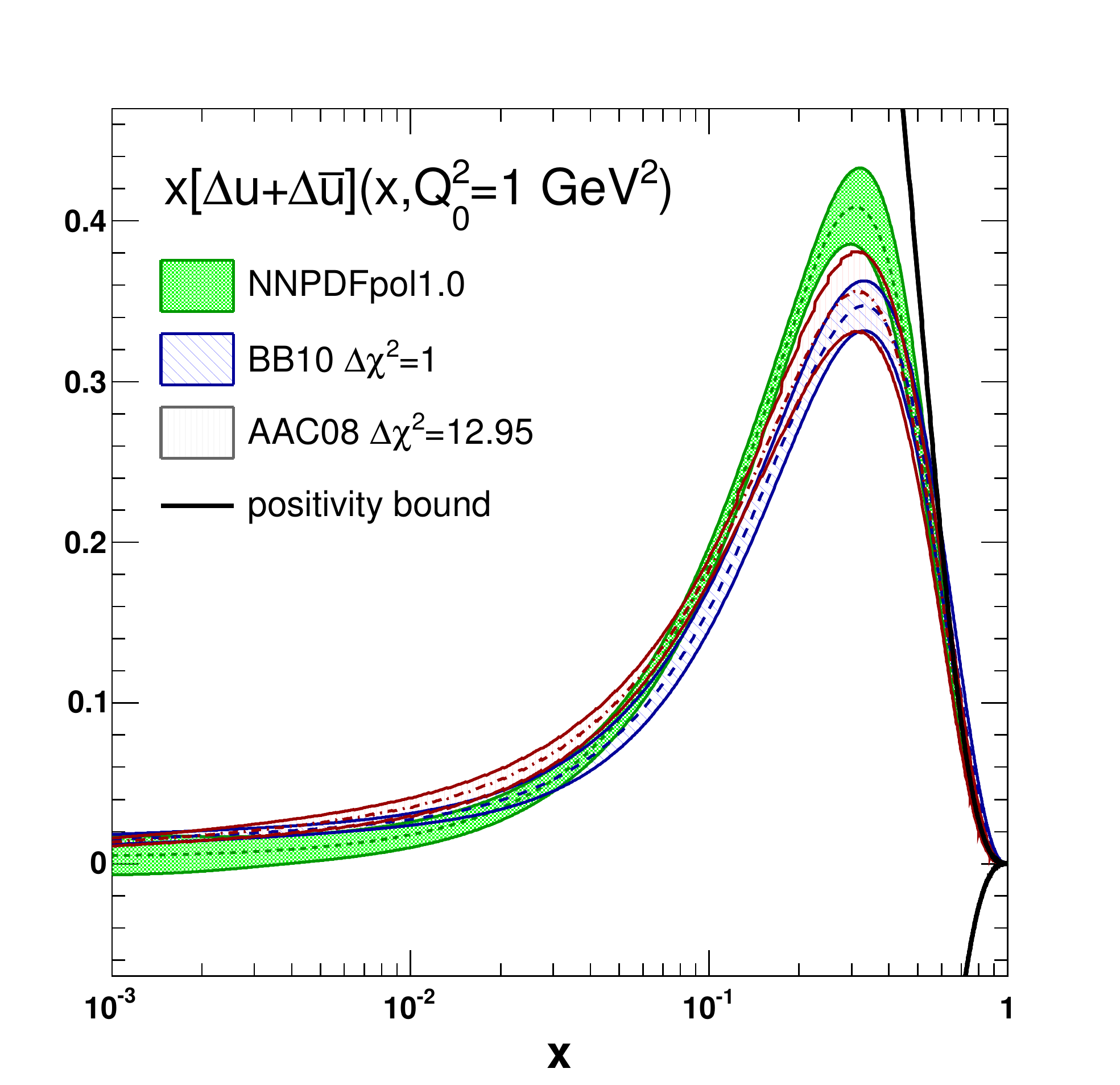}
\epsfig{width=0.35\textwidth,figure=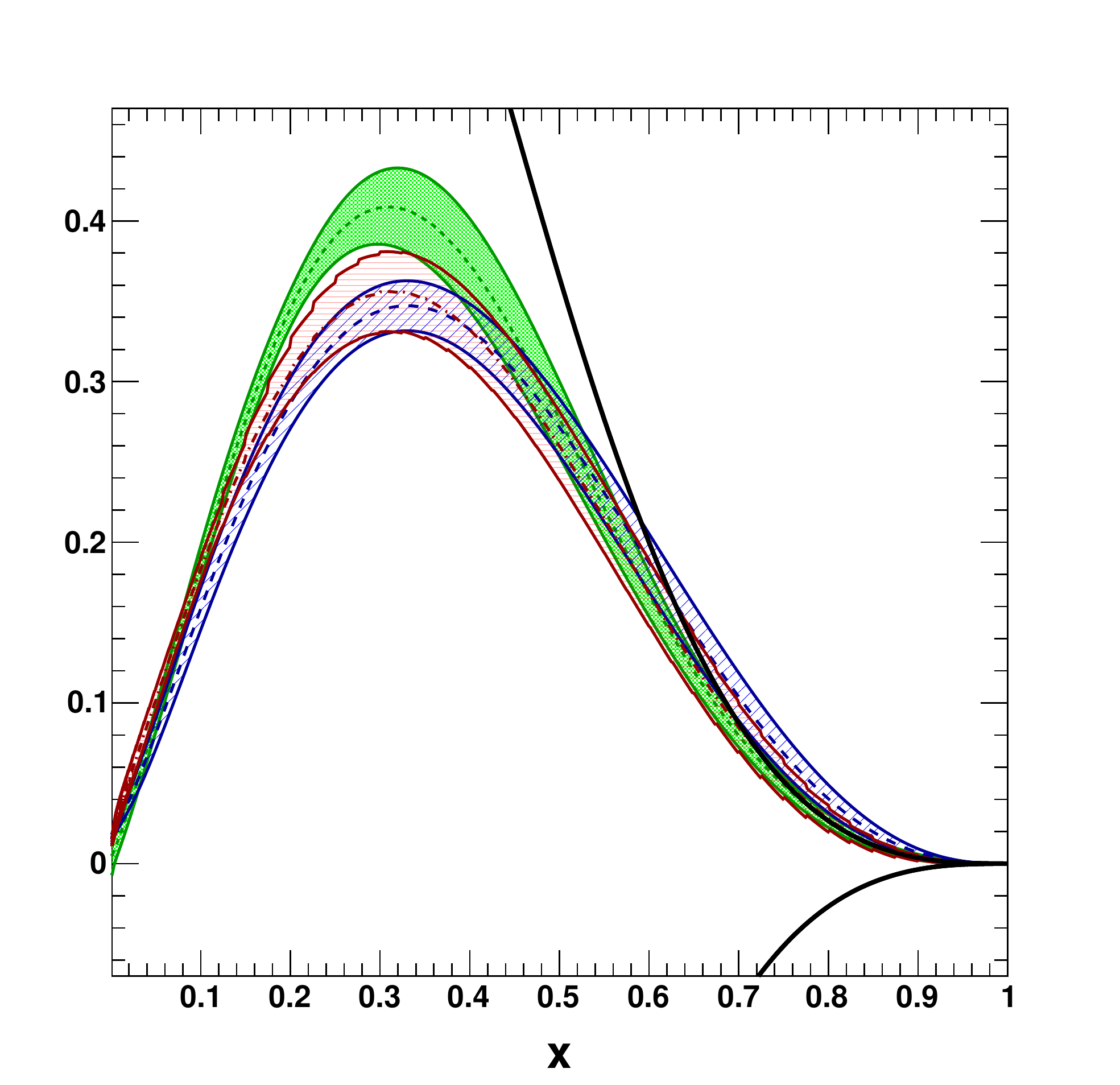}
\epsfig{width=0.35\textwidth,figure=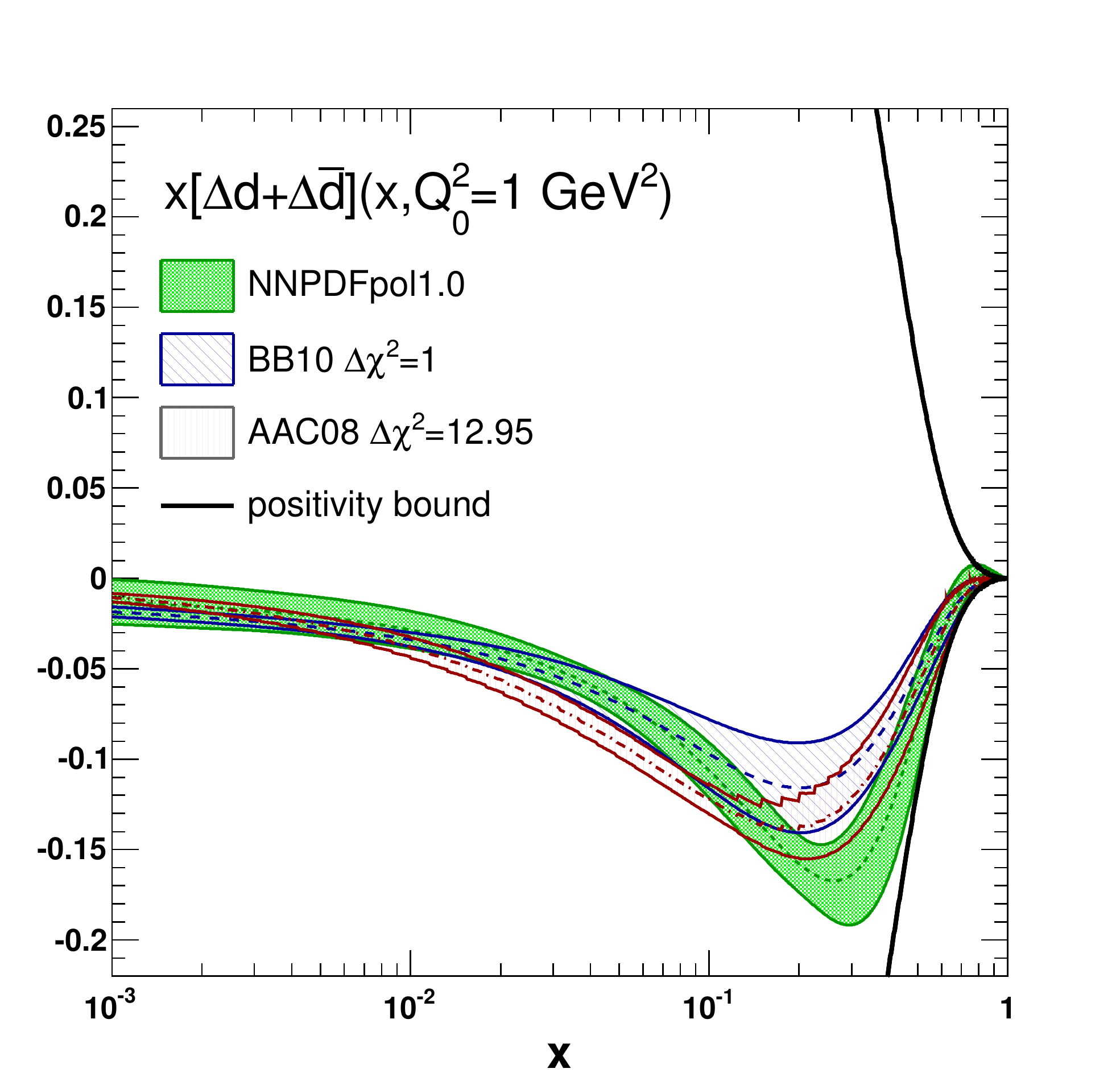}
\epsfig{width=0.35\textwidth,figure=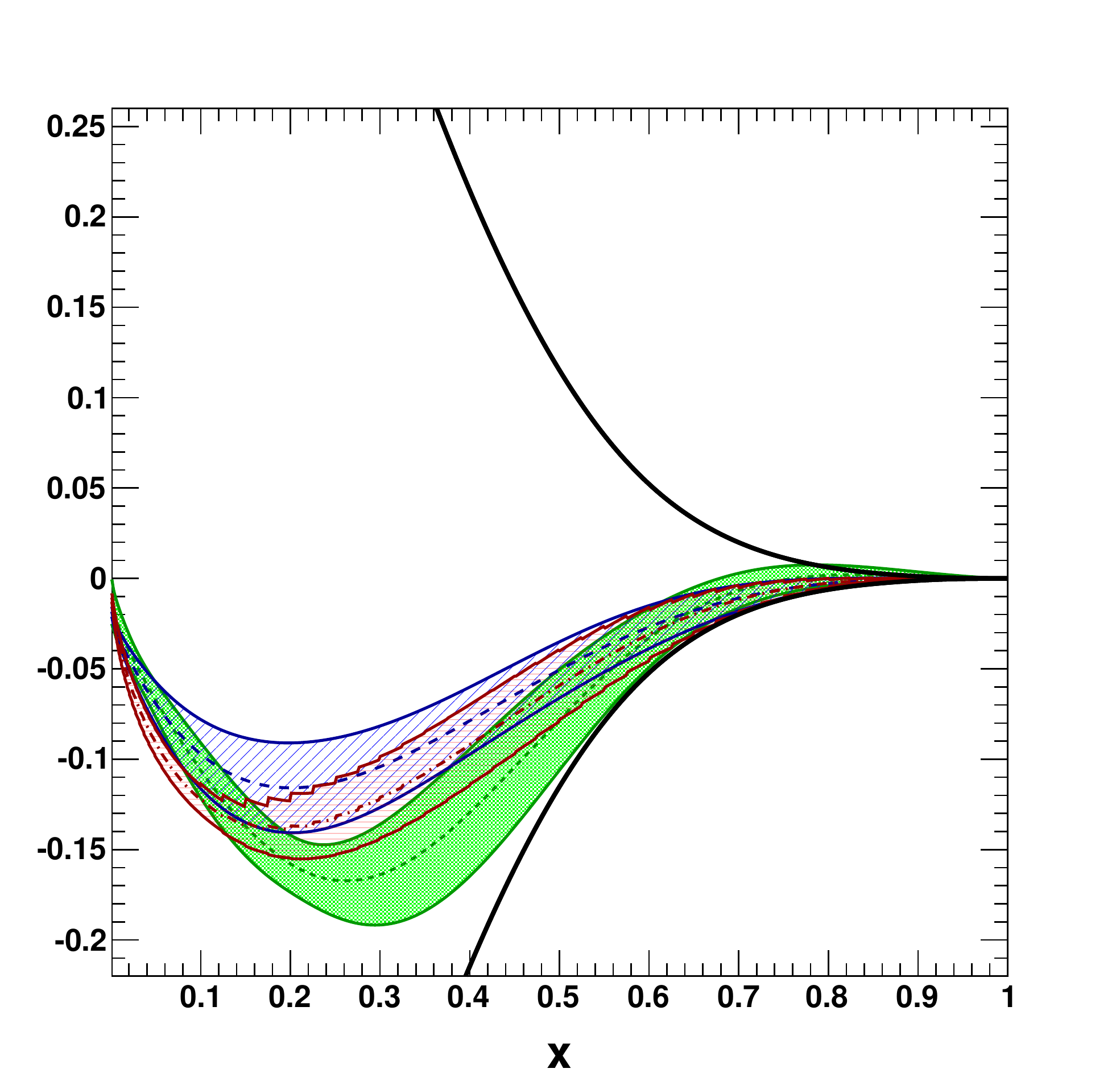}
\epsfig{width=0.35\textwidth,figure=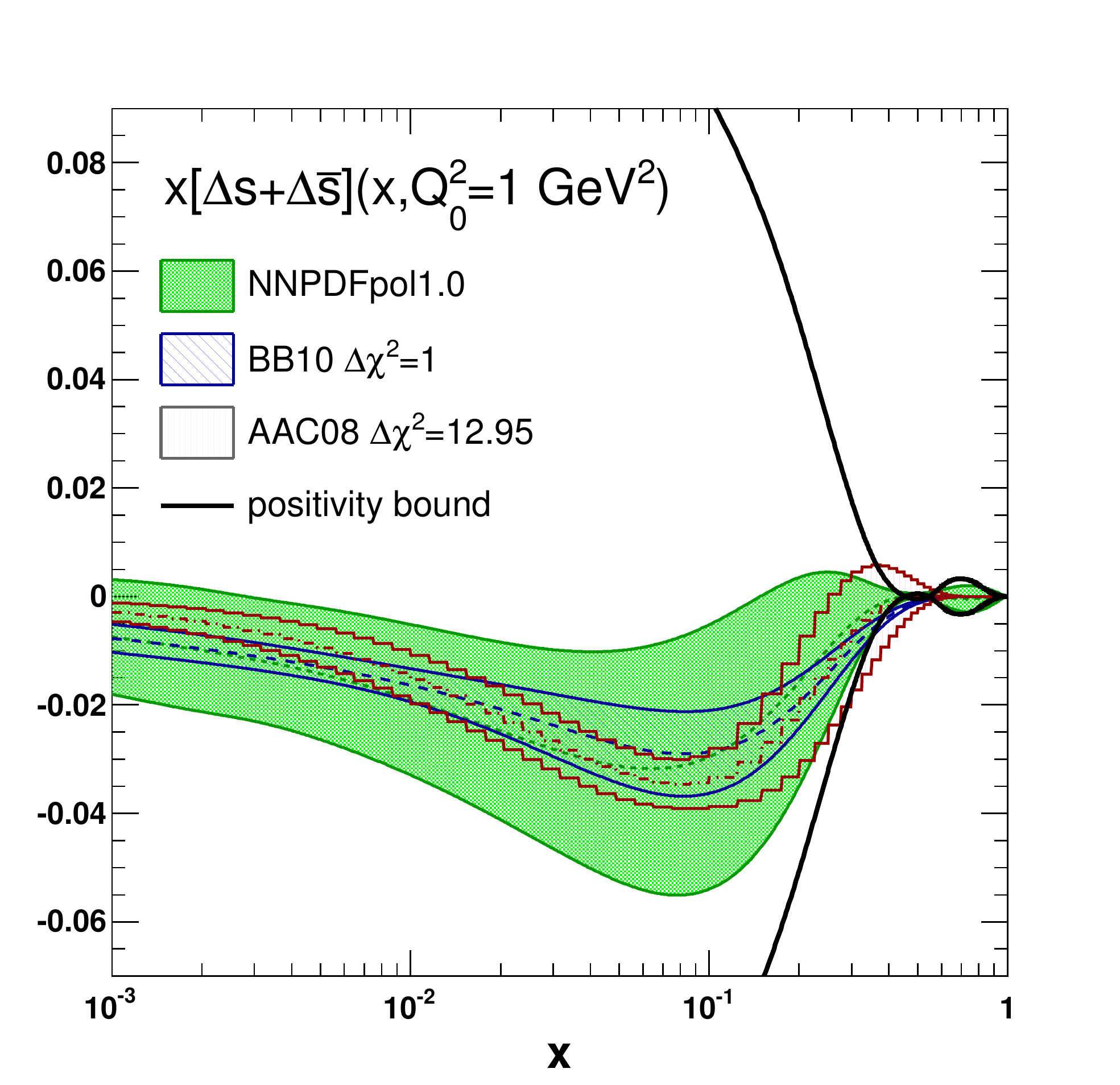}
\epsfig{width=0.35\textwidth,figure=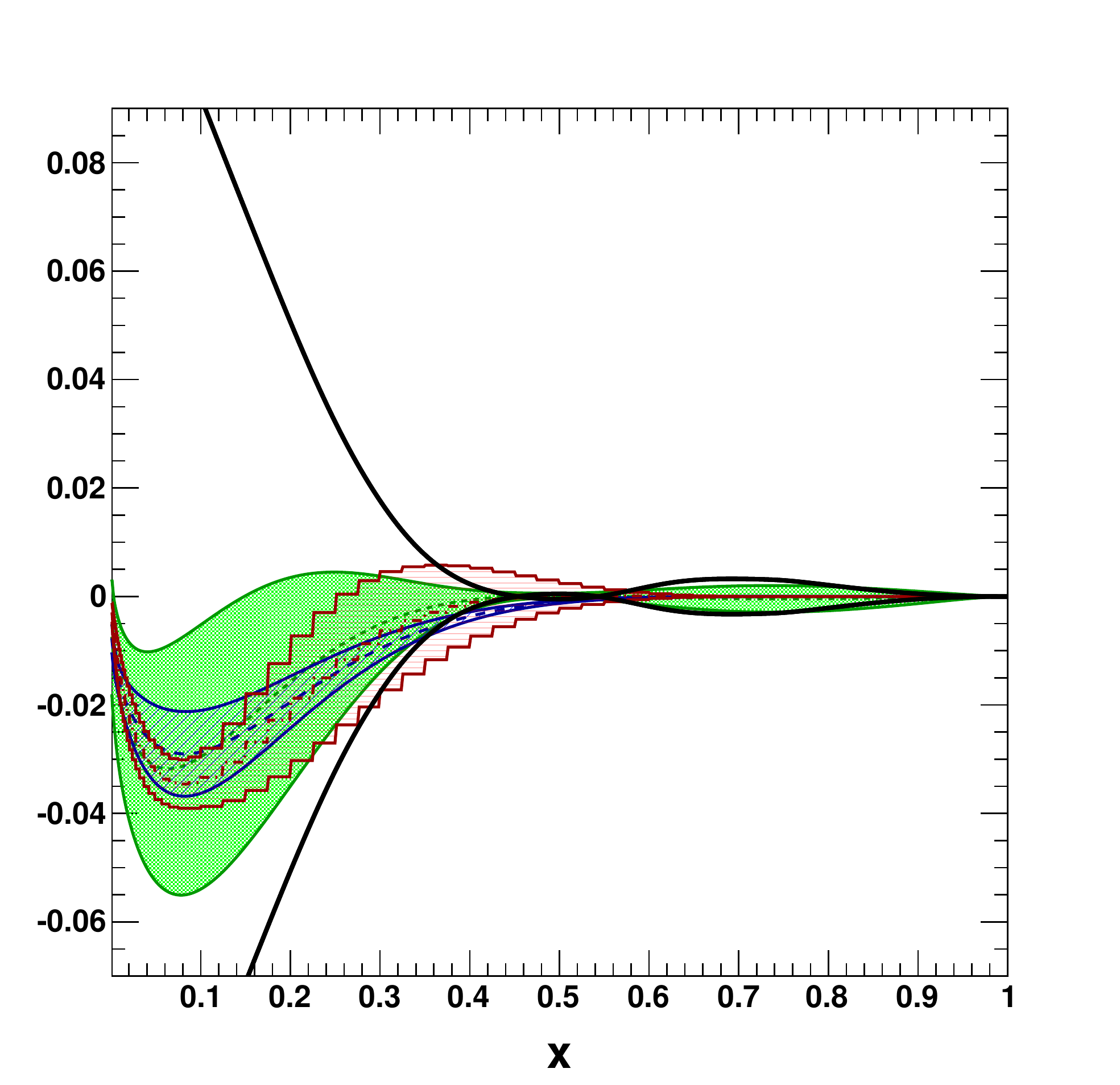}
\epsfig{width=0.35\textwidth,figure=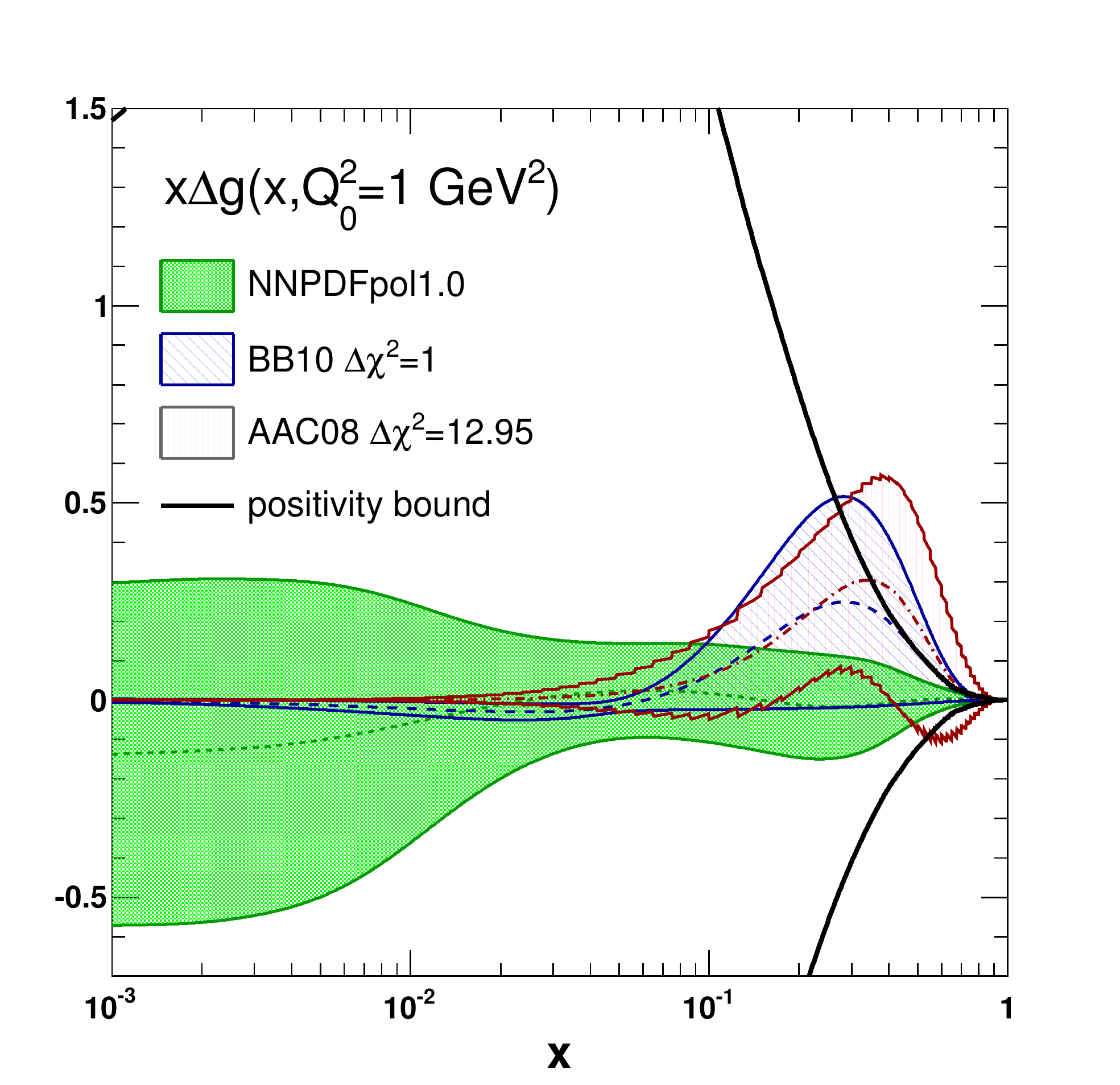}
\epsfig{width=0.35\textwidth,figure=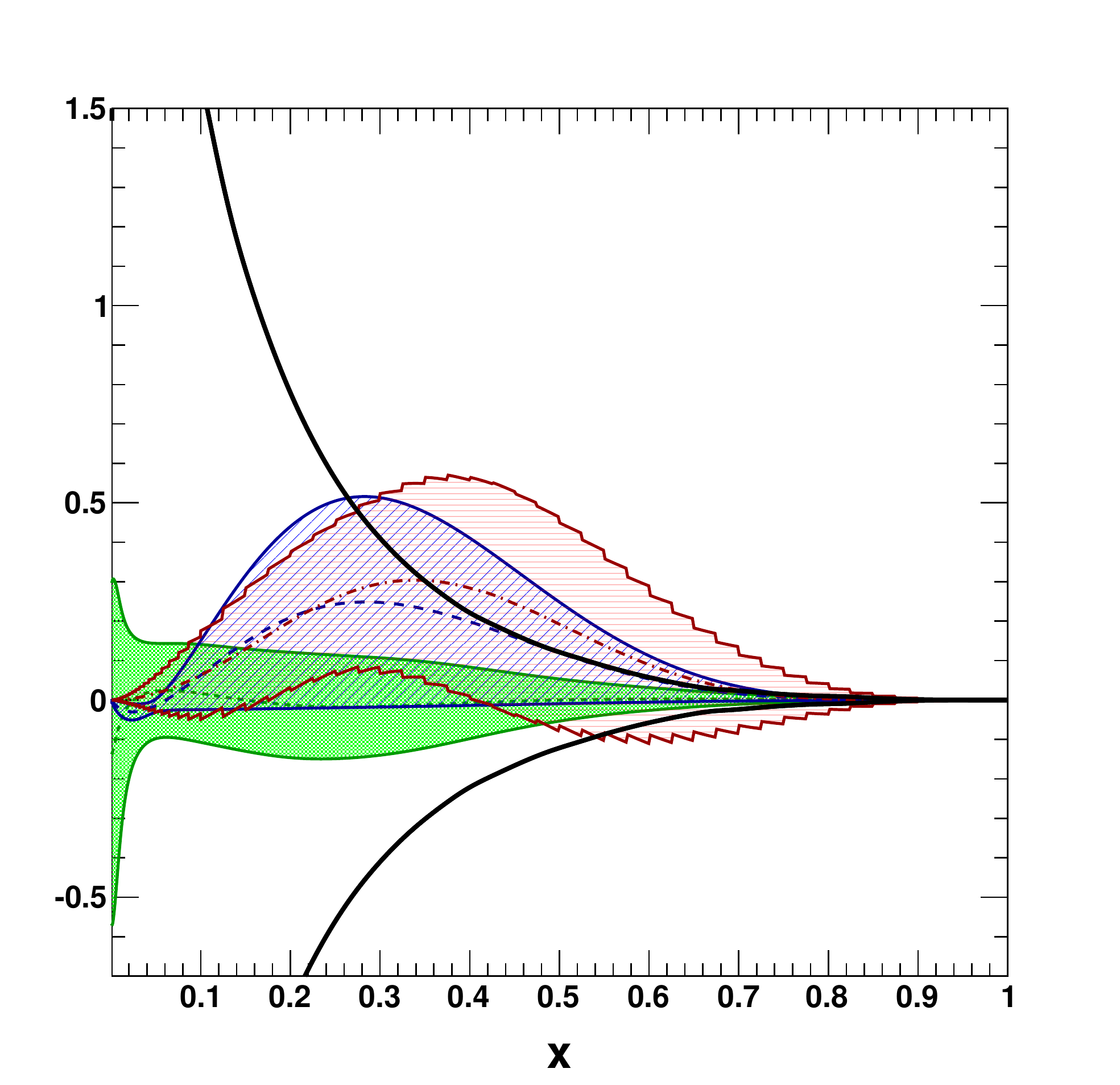}
\mycaption{The \texttt{NNPDFpol1.0} parton distributions at 
$Q_0^2=1$ GeV$^2$ in the flavor basis plotted as a function of $x$,
on a logarithmic (left) and linear (right) scale and compared to
\texttt{BB10} and \texttt{AAC08} parton sets.}
\label{fig:ppdfs3}
\end{center}
\end{figure}

\begin{figure}[p]
\begin{center}
\epsfig{width=0.35\textwidth,figure=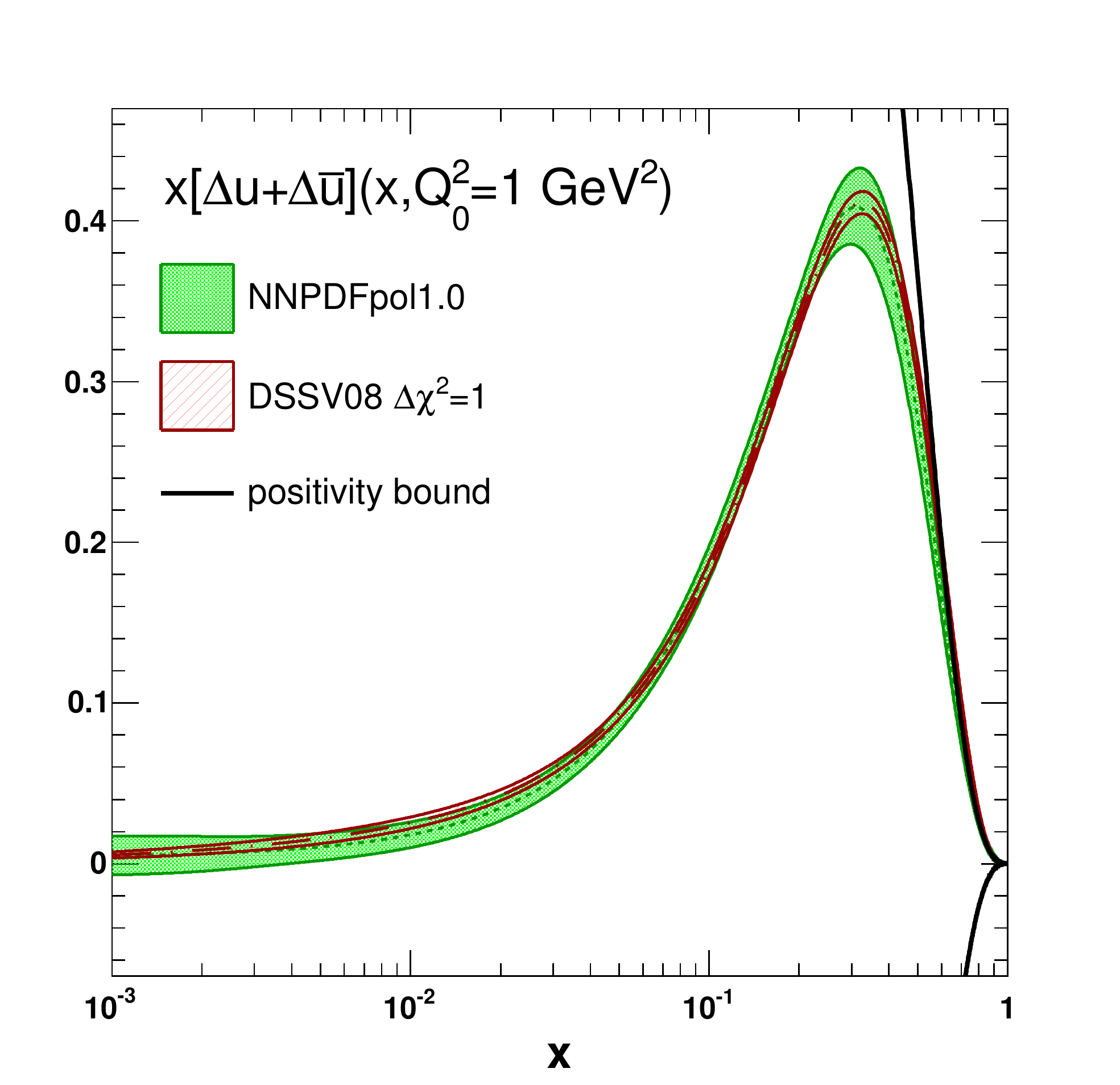}
\epsfig{width=0.35\textwidth,figure=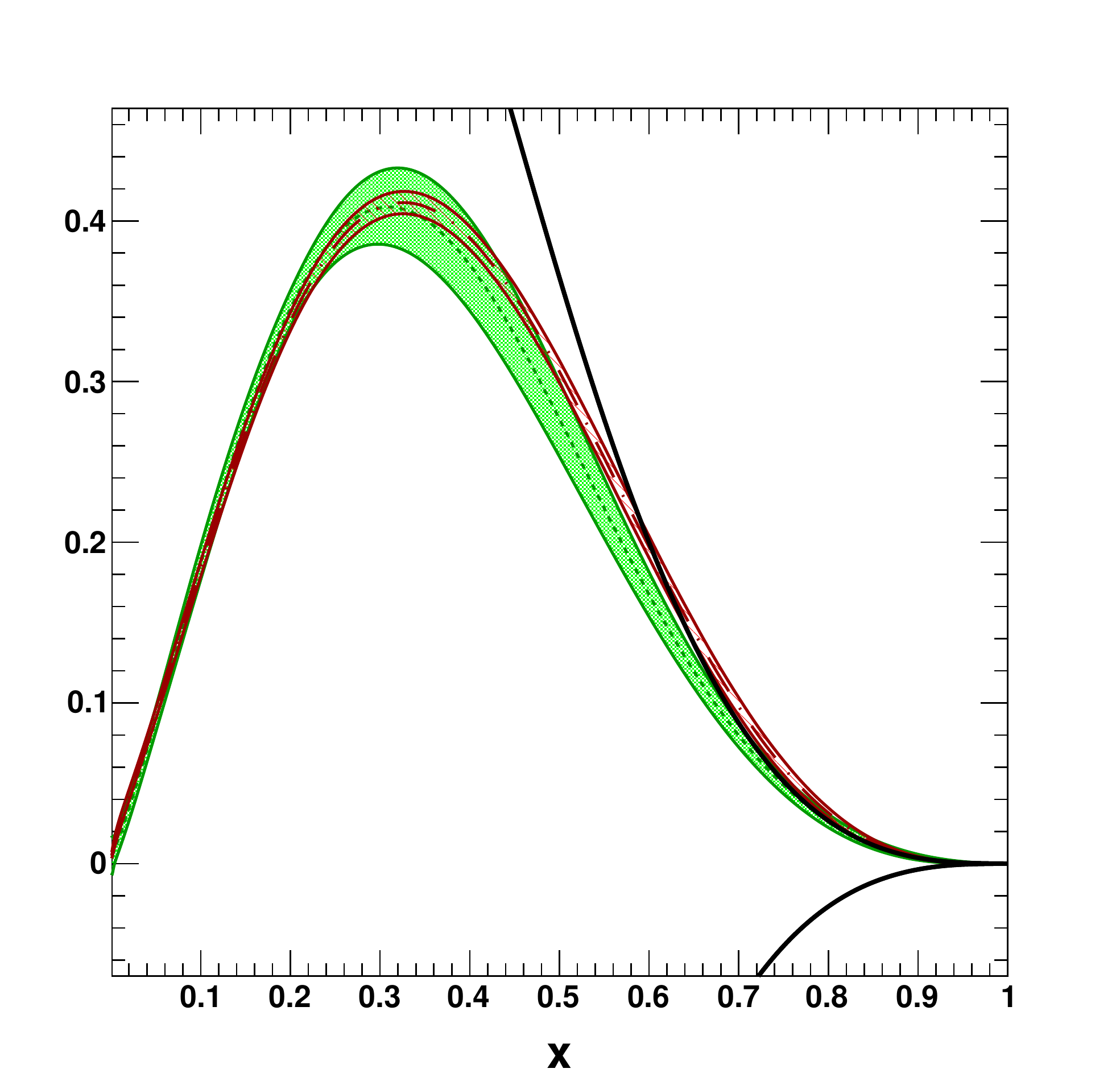}
\epsfig{width=0.35\textwidth,figure=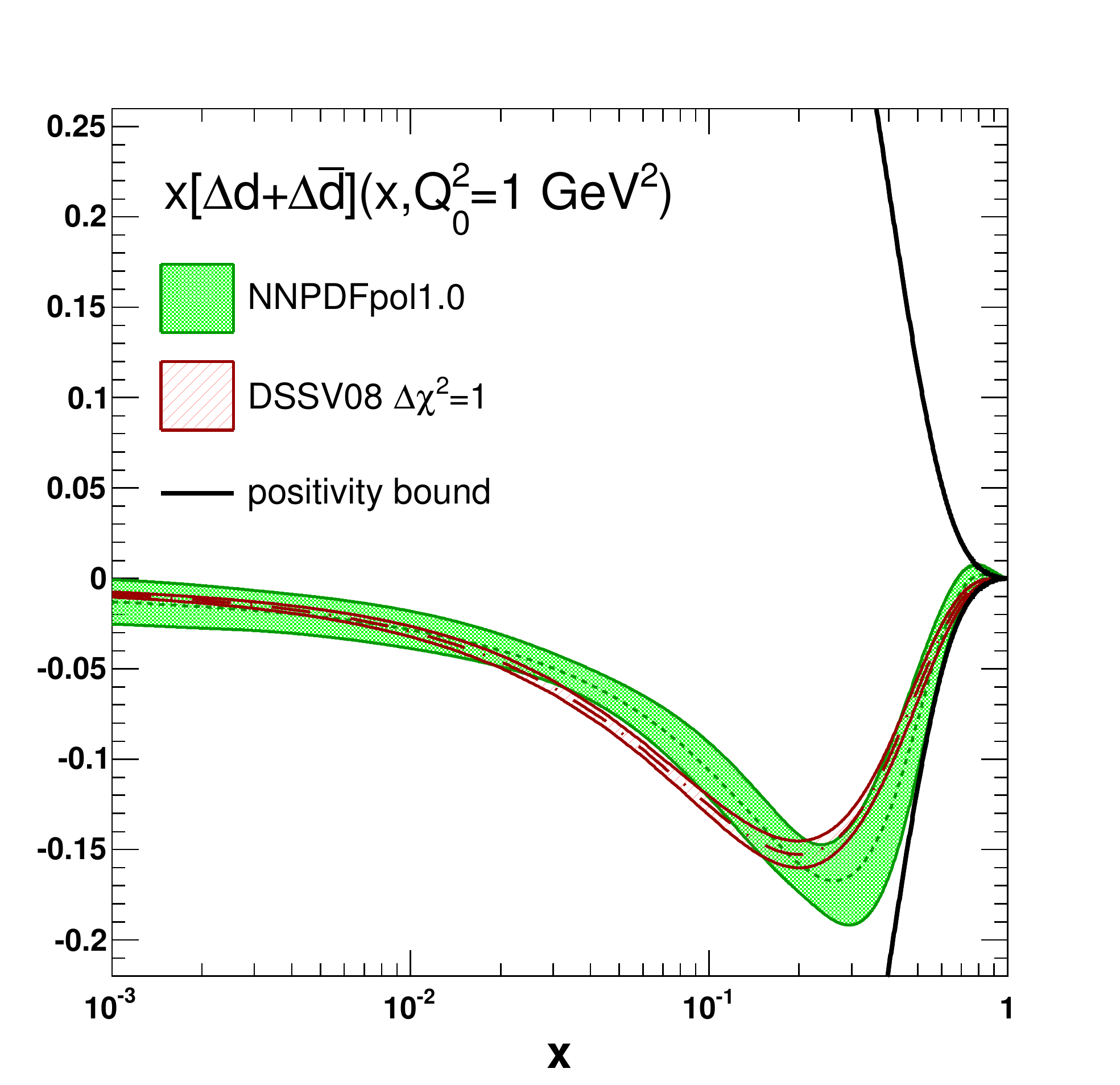}
\epsfig{width=0.35\textwidth,figure=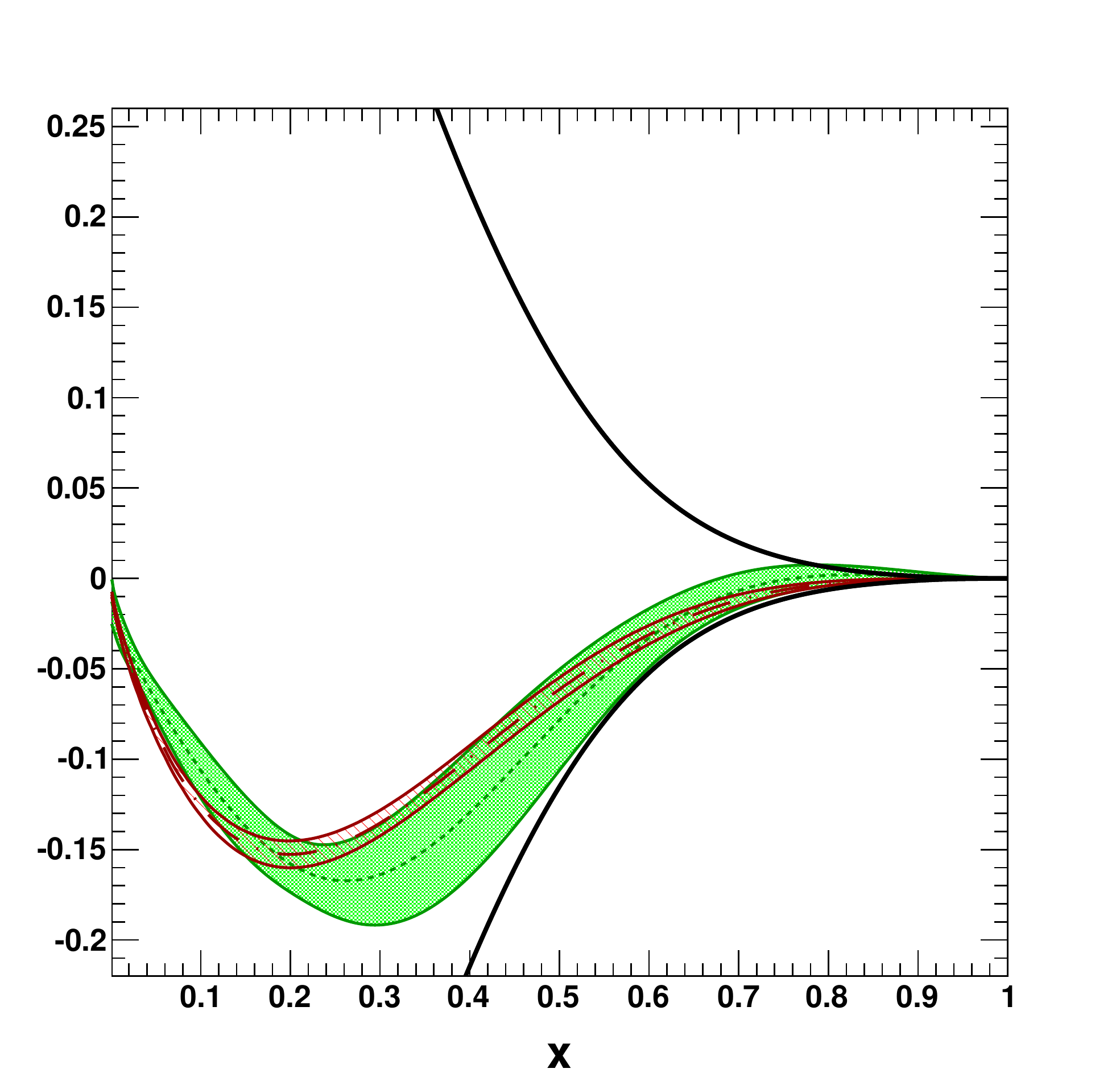}
\epsfig{width=0.35\textwidth,figure=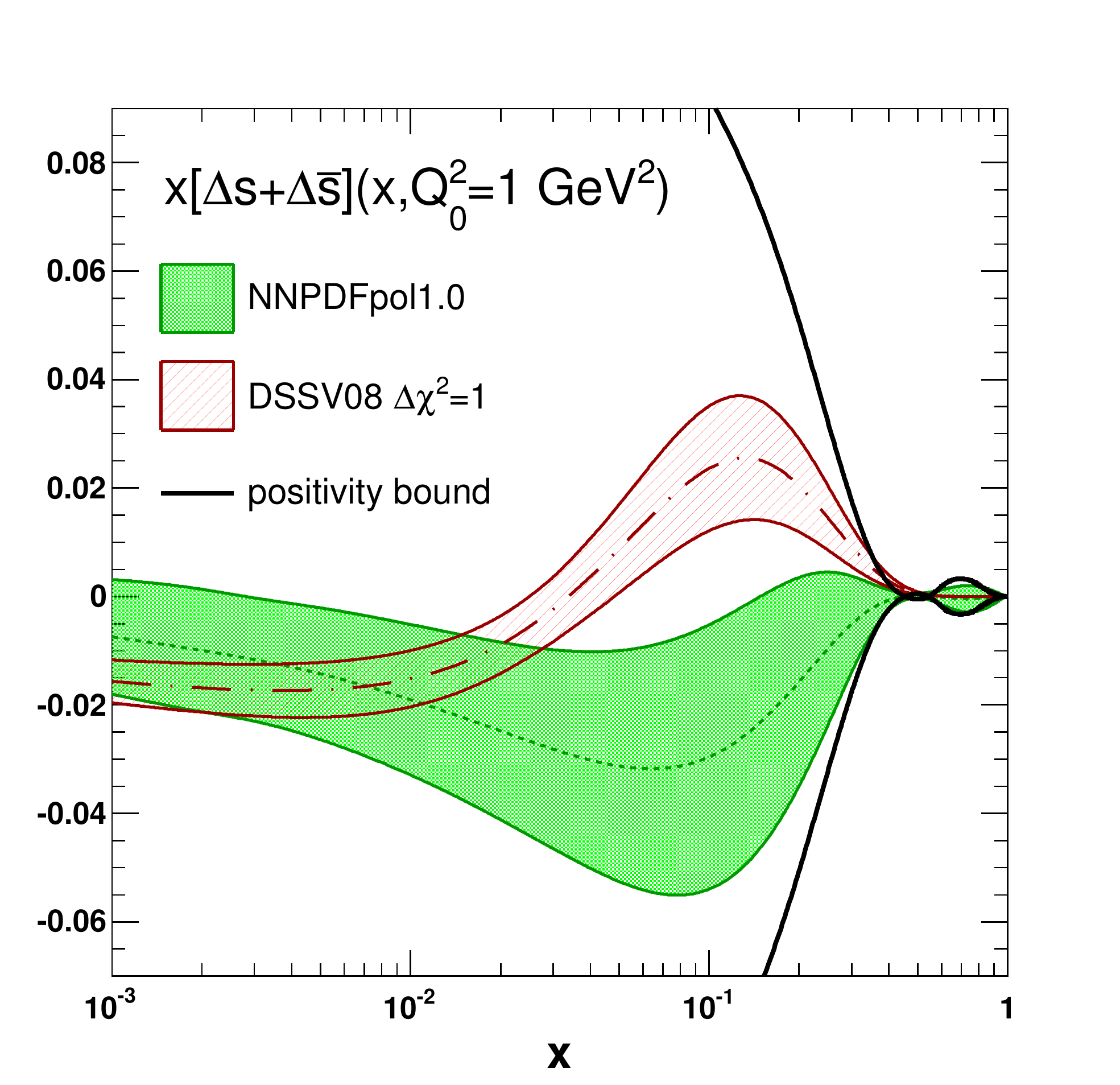}
\epsfig{width=0.35\textwidth,figure=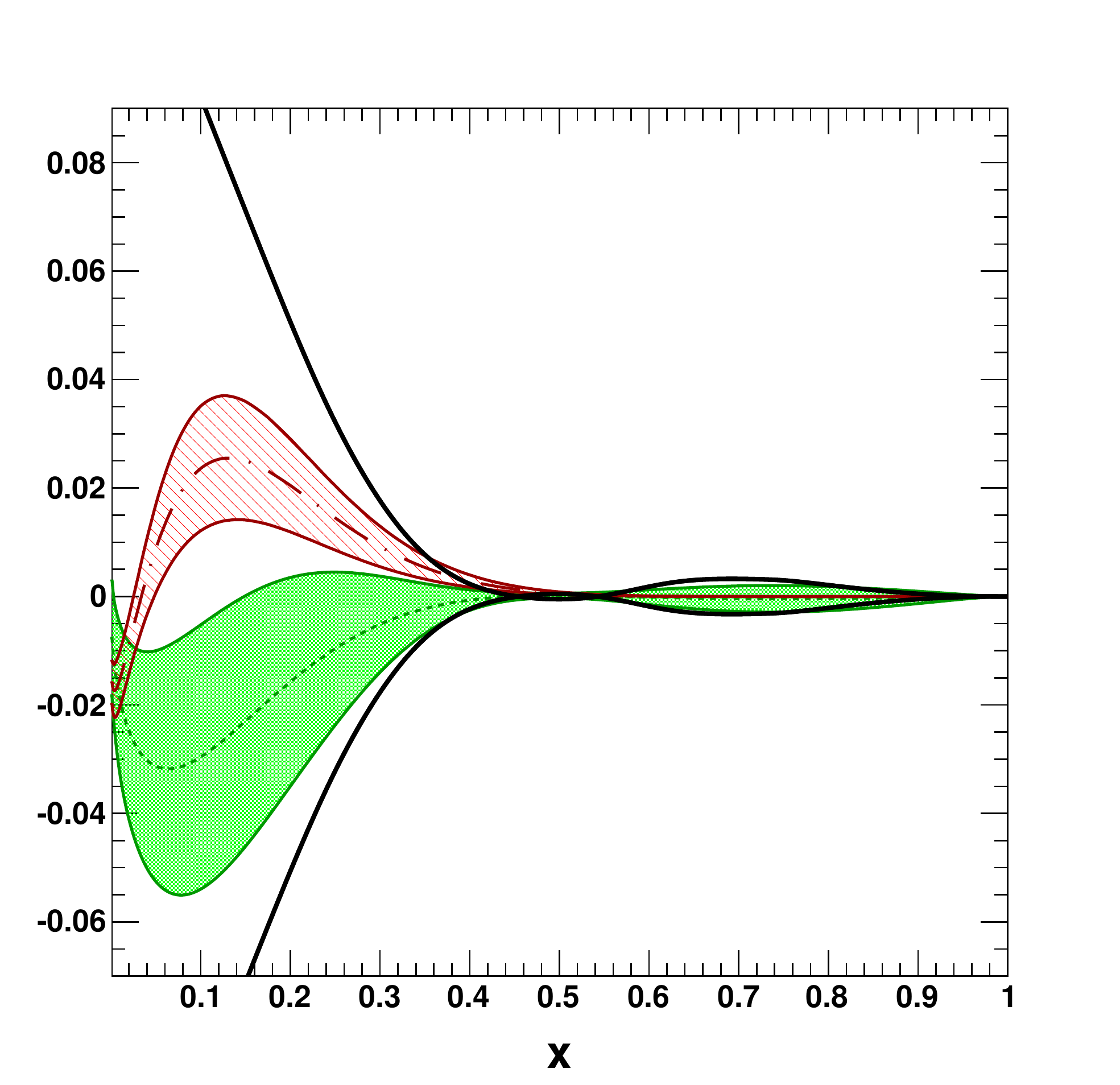}
\epsfig{width=0.35\textwidth,figure=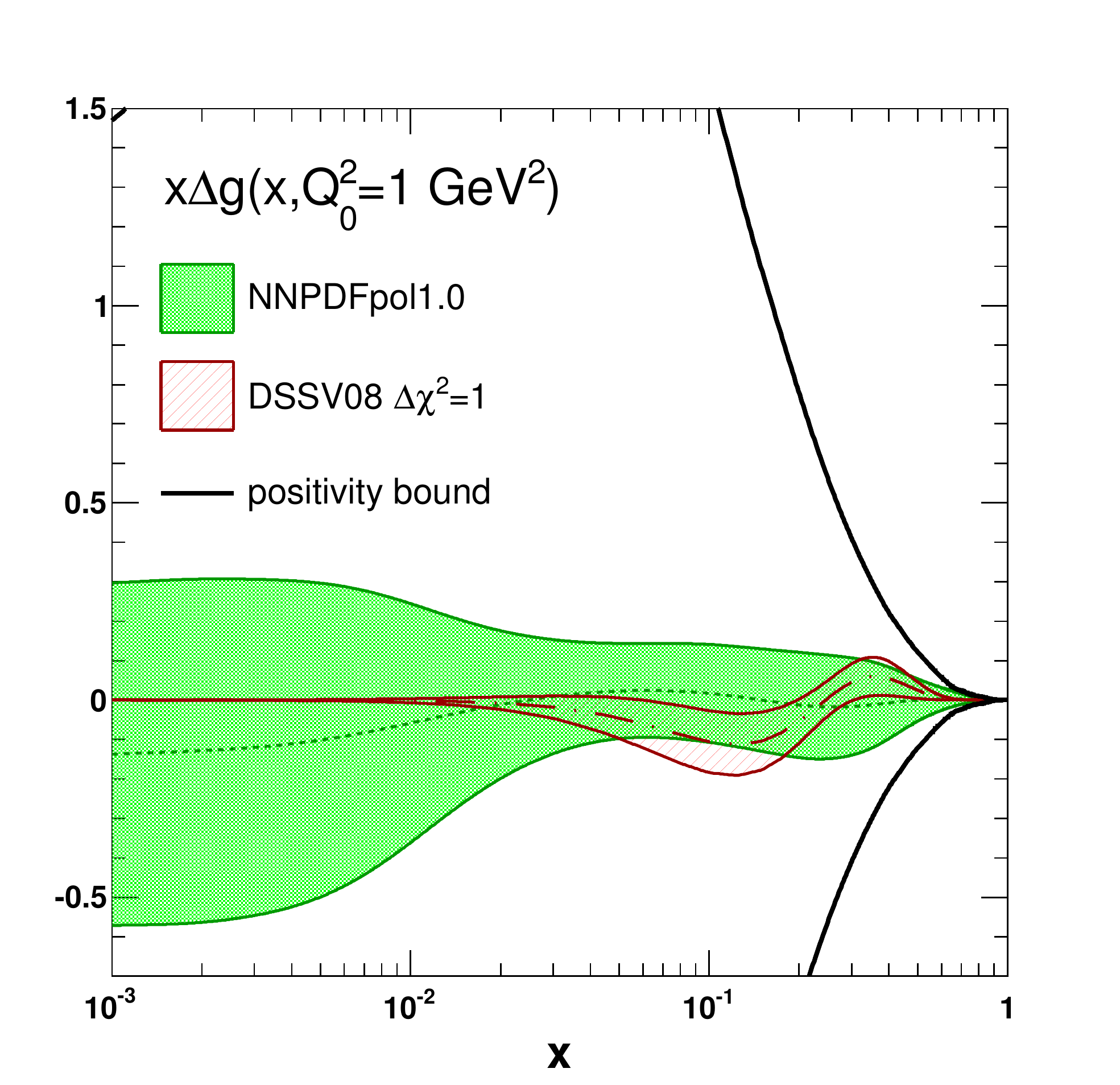}
\epsfig{width=0.35\textwidth,figure=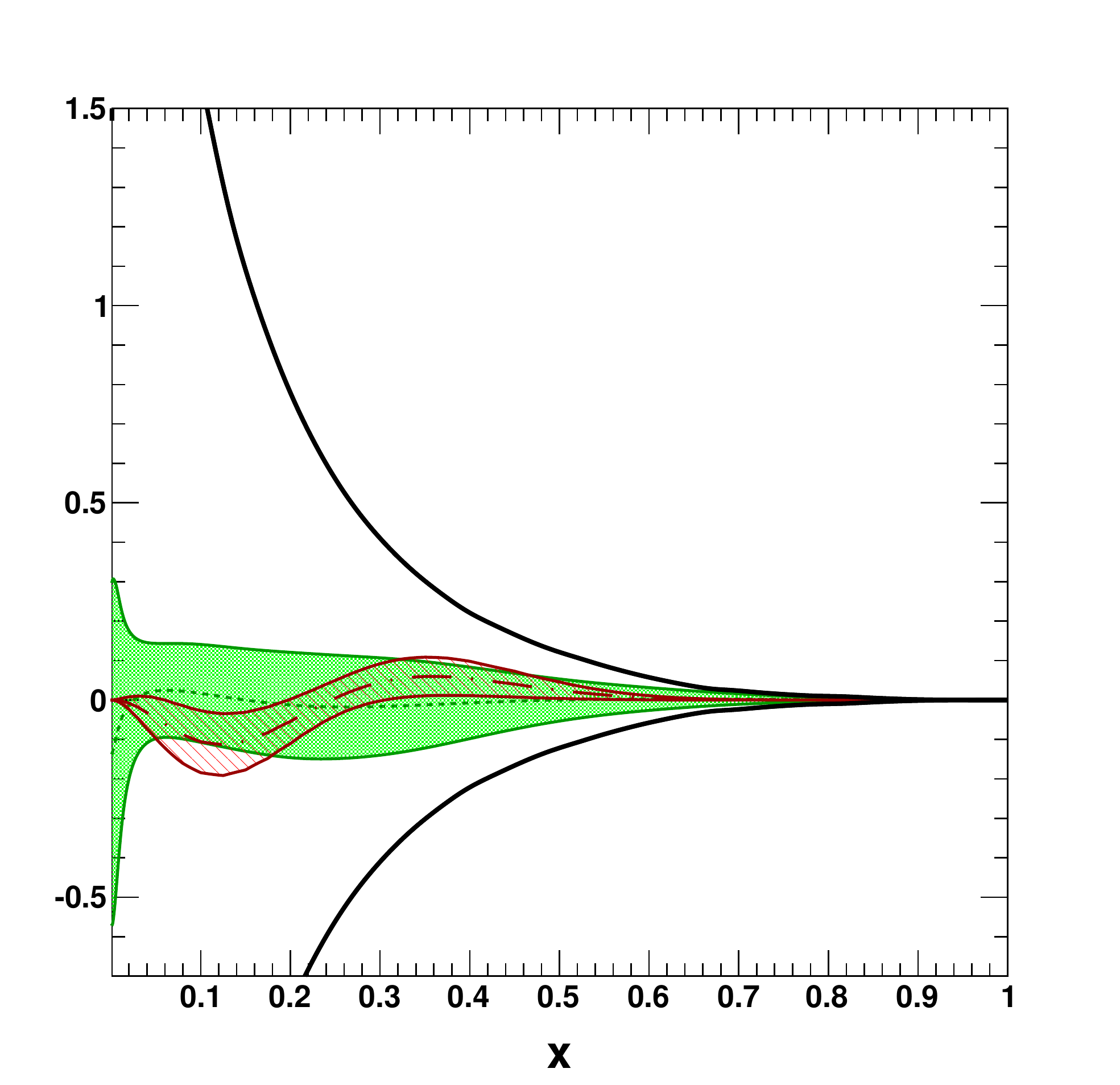}
\mycaption{Same as Fig.~\ref{fig:ppdfs3}, but compared to \texttt{DSSV08} parton set.} 
\label{fig:ppdfs2}
\end{center}
\end{figure}

The \texttt{NNPDFpol1.0} parton distributions, computed from a set of
$N_{\mbox{\scriptsize{rep}}}=100$ replicas, are displayed in
Fig.~\ref{fig:ppdfs-100} at the input scale $Q_0^2=1$ GeV$^2$, in the
PDF parametrization basis as a function of
$x$ both on a logarithmic and linear scale.  
In Figs.~\ref{fig:ppdfs3}-\ref{fig:ppdfs2} the same PDFs are plotted in
the flavor basis, and compared to other available NLO PDF sets:
\texttt{BB10}~\cite{Blumlein:2010rn} and \texttt{AAC08}~\cite{Hirai:2008aj} in
Fig.~\ref{fig:ppdfs3}, and \texttt{DSSV08}~\cite{deFlorian:2009vb} in
Fig.~\ref{fig:ppdfs2}. We do not show a direct comparison to the 
\texttt{LSS10}~\cite{Leader:2010rb} nor 
\texttt{JAM13}~\cite{Jimenez-Delgado:2013boa} PDF sets
because they are not publicly available.
We remind from Sec.~\ref{sec:setsummary}
that all these parton determinations are based on somewhat different data sets.
For instance, \texttt{BB10} contains purely DIS data and 
\texttt{AAC08} contains DIS data supplemented by a few high-$p_T$-$\pi^0$ 
production data from RHIC: hence they are closely comparable to our PDF 
determination. Instead, the \texttt{DSSV08} determination includes, 
on top of DIS data, polarized jet production data, 
and, more importantly, a large amount of
semi-inclusive DIS data which in particular allow for
quark-antiquark separation and a more direct handle on
strangeness. In these plots, NNPDF uncertainties correspond to 
the nominal one-sigma error bands, while for other PDF sets they are 
Hessian uncertainties corresponding to the default value assumed in 
each analysis. We remind from Sec.~\ref{sec:setsummary} that it is assumed
to be $T=12.95$ for \texttt{AAC08}, while $T=1$ for all other PDF sets. 

The main conclusions of this comparison can be summarized as follows.
\begin{itemize}
\item The central values of the $\Delta u + \Delta\bar{u}$ and the
$\Delta d + \Delta\bar{d}$  PDF combinations are in
reasonable agreement with those of other parton sets. 
The \texttt{NNPDFpol1.0} results are in best agreement with 
\texttt{DSSV08}, in slightly worse agreement with \texttt{AAC08}, 
and in worst agreement with \texttt{BB10}. 
Uncertainties on these PDFs are generally slightly larger for
\texttt{NNPDFpol1.0} than for other sets, 
especially \texttt{DSSV08}, which however is based on a much wider data set. 
\item The \texttt{NNPDFpol1.0} determination of $\Delta s +\Delta\bar{s}$ 
is affected by a much larger uncertainty than \texttt{BB10}
and \texttt{AAC08}, for almost all values of $x$.
Overall, the \texttt{AAC08} and \texttt{BB10} total strange distributions 
fall well within the \texttt{NNPDFpol1.0} uncertainty band.
\item The \texttt{NNPDFpol1.0} determination of total strangeness,
$\Delta s +\Delta\bar{s}$ is inconsistent at the two sigma level 
in the medium-to-small $x\sim0.1$ region with \texttt{DSSV08}, which is also
rather more accurate. 
However, we notice that total strangeness is constrained in the two 
analyses by rather different experimental information. In \texttt{NNPDFpol1.0},
it is determined through its $Q^2$ evolution at different scales together with 
fixing the first moments of the nonsinglet PDF combinations 
to the baryonic octet decay constants. Conversely, in \texttt{DSSV08} 
the total strangeness is mostly determined from semi-inclusive data with strange
hadrons in the final states. Hence, the flavor combination $\Delta s+\Delta\bar{s}$
is also sensitive to the corresponding 
fragmentations functions. Since these are poorly known, especially for strange 
hadrons (namely kaons), the result obtained in the \texttt{DSSV08} analysis
is likely to be biased by the form assumed for the fragmentation funcions.
\item The gluon PDF is affected by a large uncertainty,
rather larger than any other set, especially at small $x$. In
particular, the \texttt{NNPDFpol1.0} polarized gluon distribution 
is compatible with zero for all values of $x$. At $0.04\lesssim x\lesssim 0.2$,
the gluon determination in the \texttt{DSSV08} parton set 
benefits from sensitivity to pion and jet production data, which are not
included in the other determinations.
\item Uncertainties on the PDFs in the regions where no data are available
tend to be larger than those of other sets. At very large
values of $x$ the PDF uncertainty band is largely determined by the
positivity constraint, while at small values of $x$ it is prevented to blow up
arbitrarily by demanding the integrability of its first moment.
\end{itemize}

\begin{figure}[t]
\centering
\epsfig{width=0.45\textwidth,figure=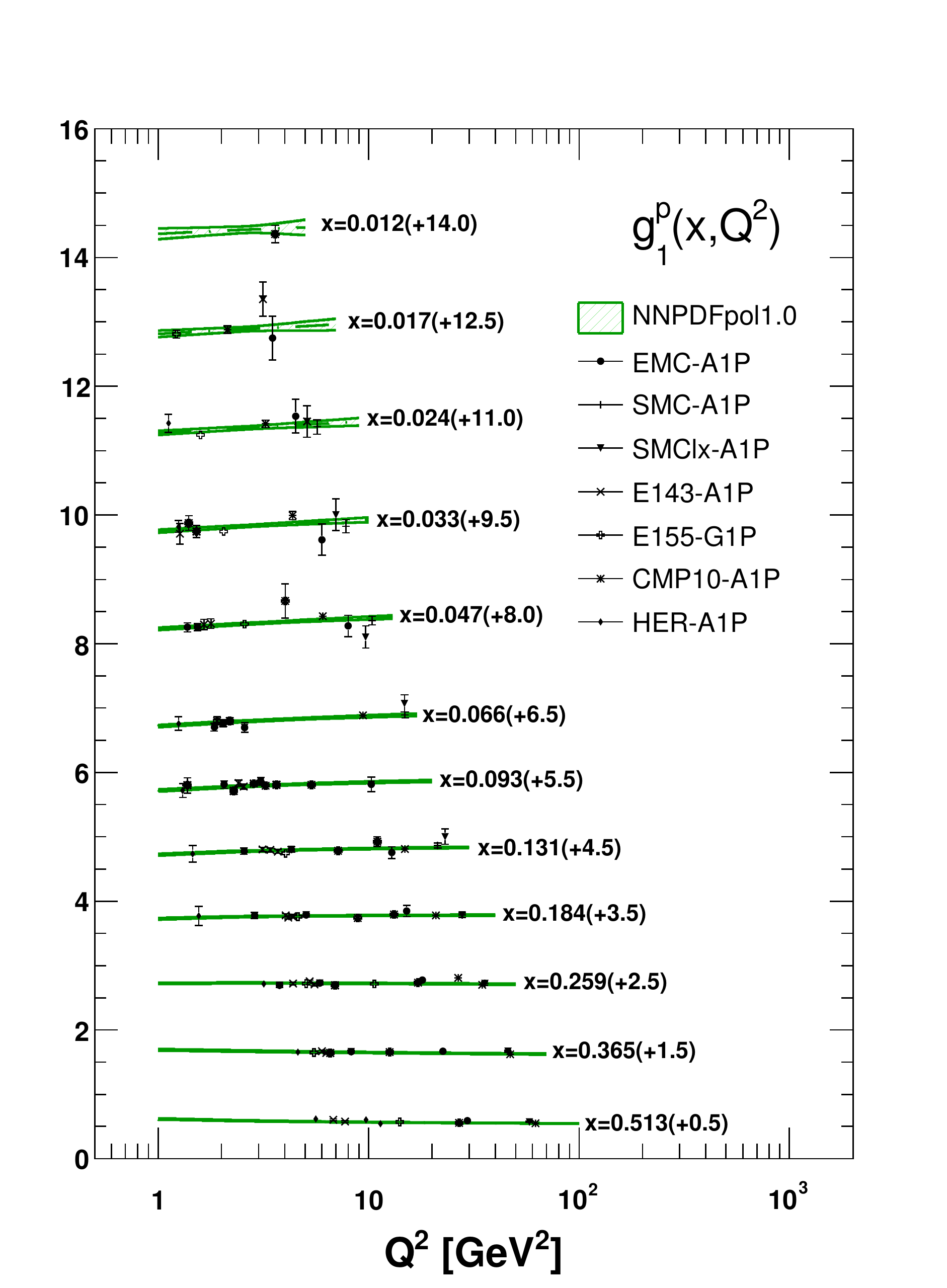}
\epsfig{width=0.45\textwidth,figure=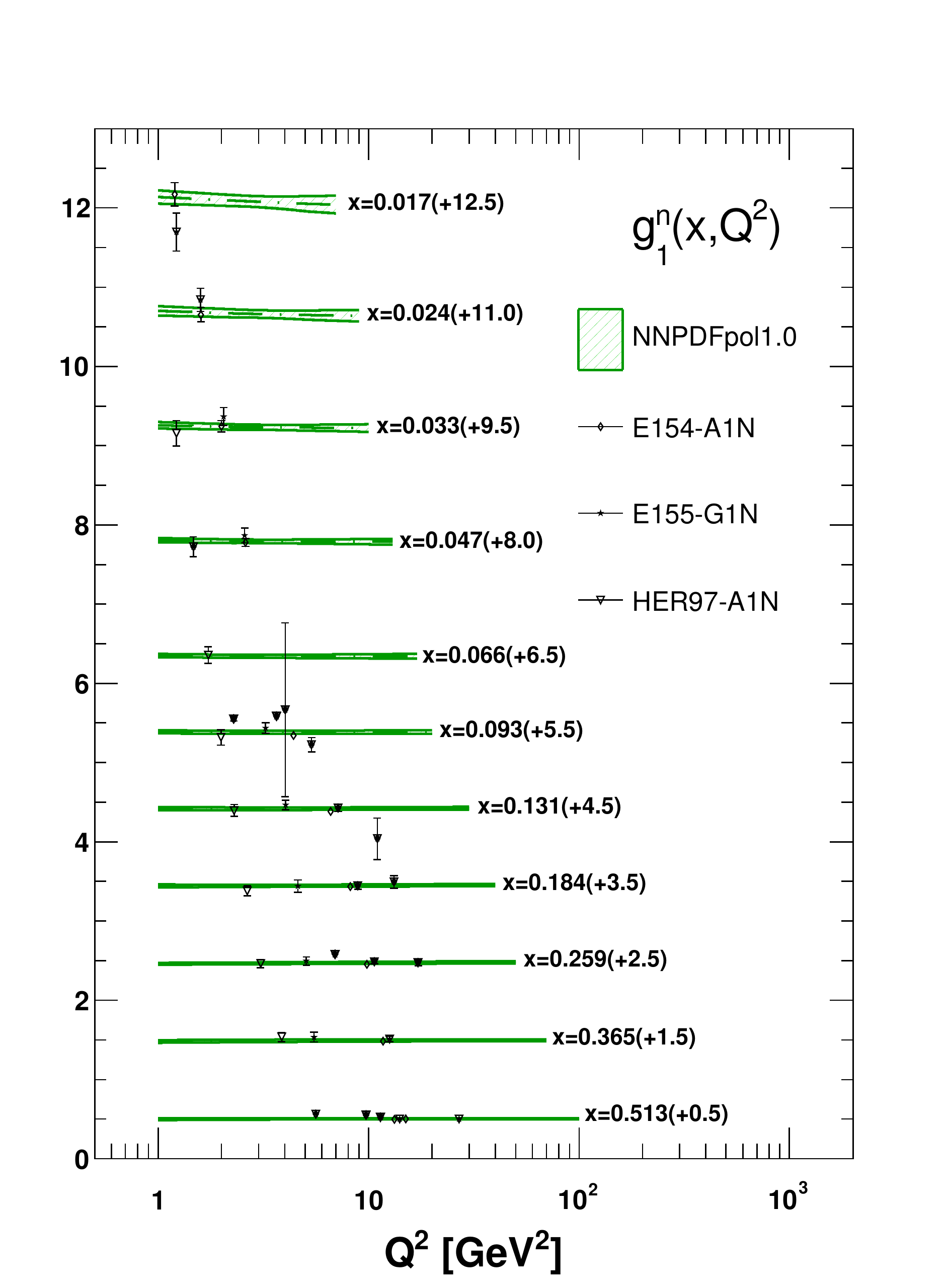}
\mycaption{The proton and neutron 
structure function $g_1(x,Q^2)$ displayed as a function of $Q^2$ in 
different bins of $x$ compared to experimental data. 
Experimental data are grouped in bins of $x$, while
\texttt{NNPDFpol1.0} results are given at the center of each bin,
whose value is given next to each curve. In order to improve
legibility,
the values of $g_1(x,Q^2)$ 
have been shifted by the
amount given next to each curve.}
\label{fig:g1}
\end{figure}

In Fig.~\ref{fig:g1} we compare the 
structure function $g_1(x,Q^2)$  for proton and neutron, 
computed using \texttt{NNPDFpol1.0} (with
its one-sigma uncertainty band) to the experimental data included in
the fit. Experimental data are grouped in bins of $x$
with a logarithmic spacing, while the theoretical
prediction and its uncertainty are computed at the central value of
each bin.
The uncertainty band in the \texttt{NNPDFpol1.0} result is typically
smaller than the experimental errors, except at small-$x$ where a much
more restricted data set is available; in that region, the
uncertainties are comparable. Scaling violations of the polarized
structure functions are clearly visible, especially for $g_1^p$,
despite the limited range in $Q^2$.

\subsection{Stability of the results}
\label{sec:stab}

Our results have been obtained with a number of theoretical
and methodological assumptions, discussed in
Secs.~\ref{sec:QCDanalysis}-\ref{sec:minim}. 
We will now test their stability
upon variation of these assumptions.

\subsubsection{Target-mass corrections and $g_2$.}
\label{sec:tmcres}

We have consistently included in our determination of $g_1$
corrections suppressed by powers of the nucleon mass which are of
kinematic origin. Thus in particular, we have included
target-mass corrections (TMCs) up to first order in
${M^2}/{Q^2}$. Furthermore, both TMCs and the relation between the
measured asymmetries and the structure function $g_1$ involve
contributions to the structure
function $g_2$ proportional to powers of ${M^2}/{Q^2}$ which we
include according to Eq.~(\ref{eq:g1tog2}) or Eq.~(\ref{eq:g1tog2p})
(see the discussion in Sec.~\ref{sec:TMC}). 
Our default PDF set is obtained
assuming that $g_2$ is given by the Wandzura-Wilczek relation,
Eq.~(\ref{eq:wwrel}).

In order to assess the impact of these assumptions on our results, we
have performed two more PDF determinations. In the first, we set $M=0$
consistently everywhere, both in the extraction of the structure functions 
from the
asymmetry data and in our computation of structure functions. 
This thus removes TMCs, and also contributions
proportional to $g_2$. In the second, we retain
mass effects, but we  assume $g_2=0$.

The statistical estimators for each of these three fits over the 
full data set are shown in Tab.~\ref{tab:tmc_estimators}. Clearly, all fits
are of comparable quality.
\begin{table}[t]
\centering
\footnotesize
\begin{tabular}{cccc}
\toprule
Fit & \texttt{NNPDFpol1.0} $g_2=g_2^{\mbox{\tiny WW}}$ 
    & \texttt{NNPDFpol1.0} $M=0$ 
    & \texttt{NNPDFpol1.0} $g_2=0$ \\
\midrule
$\chi^{2}_{\mathrm{tot}}$ & 0.77 & 0.78 & 0.75 \\
$\langle E \rangle \pm \sigma_{E}$                  
& 1.82 $\pm$ 0.18 & 1.81 $\pm$ 0.16 & 1.83 $\pm$ 0.15  \\
$\langle E_{\mathrm{tr}} \rangle \pm \sigma_{E_{\mathrm{tr}}}$ & 1.66 $\pm$ 0.49 
& 1.62 $\pm$ 0.50  & 1.70 $\pm$ 0.38\\
$\langle E_{\mathrm{val}} \rangle \pm \sigma_{E_{\mathrm{val}}}$ 
& 1.88 $\pm$ 0.67   & 1.84 $\pm$ 0.70 & 1.96 $\pm$ 0.56\\
\midrule
$\langle \chi^{2(k)} \rangle \pm \sigma_{\chi^{2}}$  
& 0.91 $\pm$ 0.12 & 0.90 $\pm$ 0.09 & 0.86 $\pm$ 0.09 \\
\midrule
\end{tabular}
\mycaption{The statistical estimators of Tab.~\ref{tab:chi2tab1}
(obtained assuming $g_2=g_2^{\mbox{\tiny WW}}$) compared to a fit with $M=0$ or
with $g_2=0$.}
\label{tab:tmc_estimators}
\end{table}

\begin{figure}[t]
\begin{center}
\epsfig{width=0.40\textwidth,figure=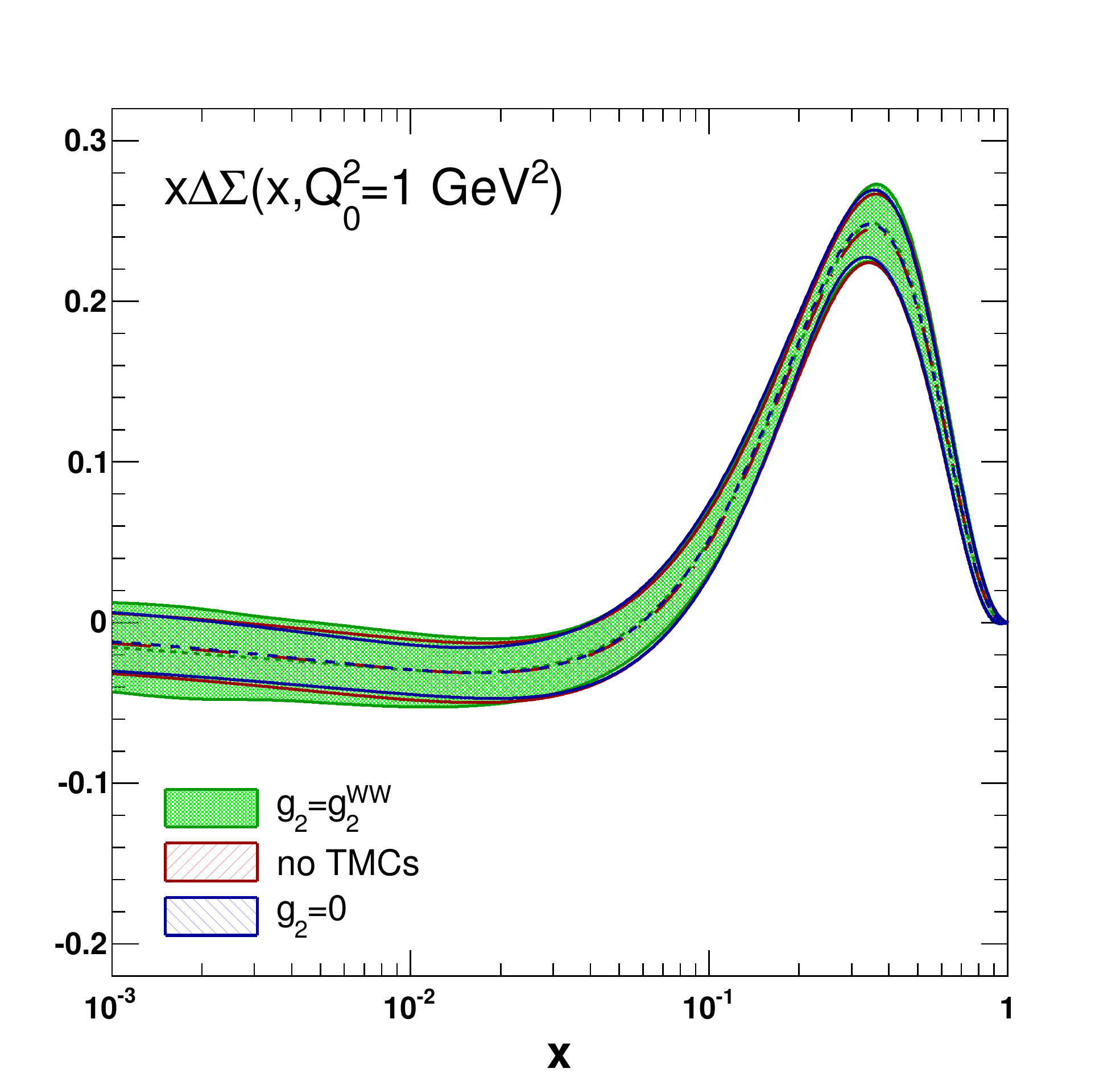}
\epsfig{width=0.40\textwidth,figure=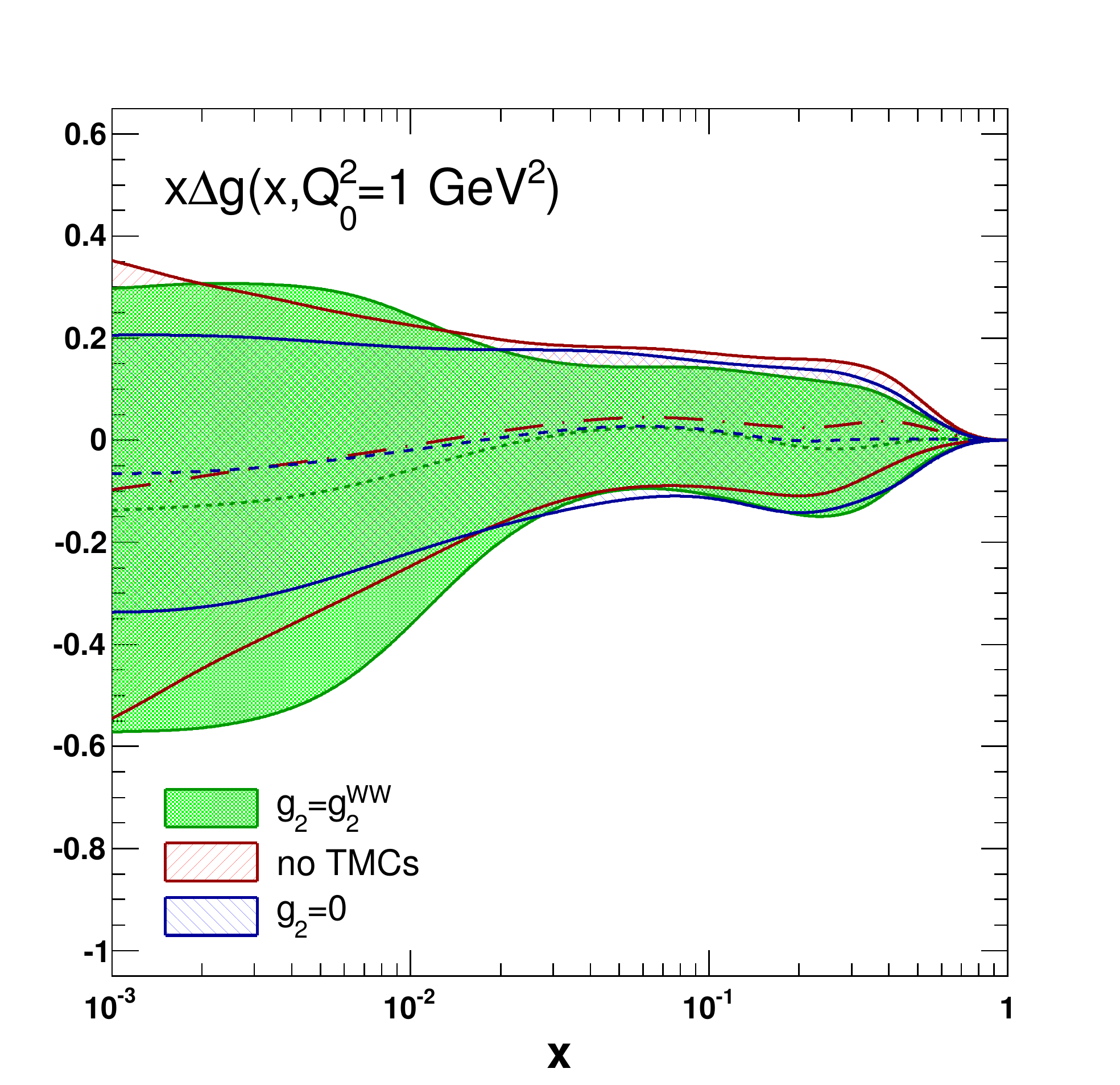}
\epsfig{width=0.40\textwidth,figure=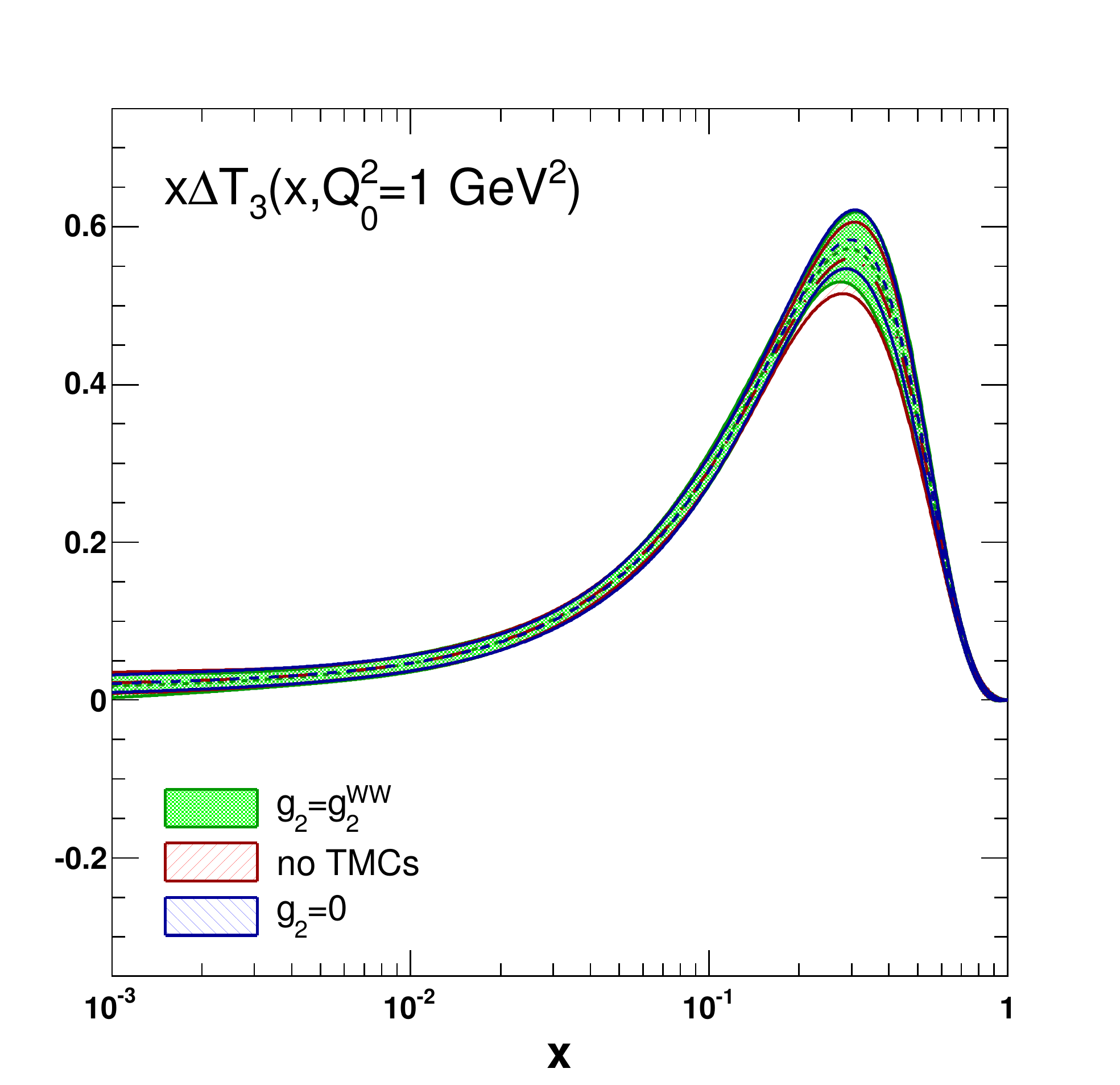}
\epsfig{width=0.40\textwidth,figure=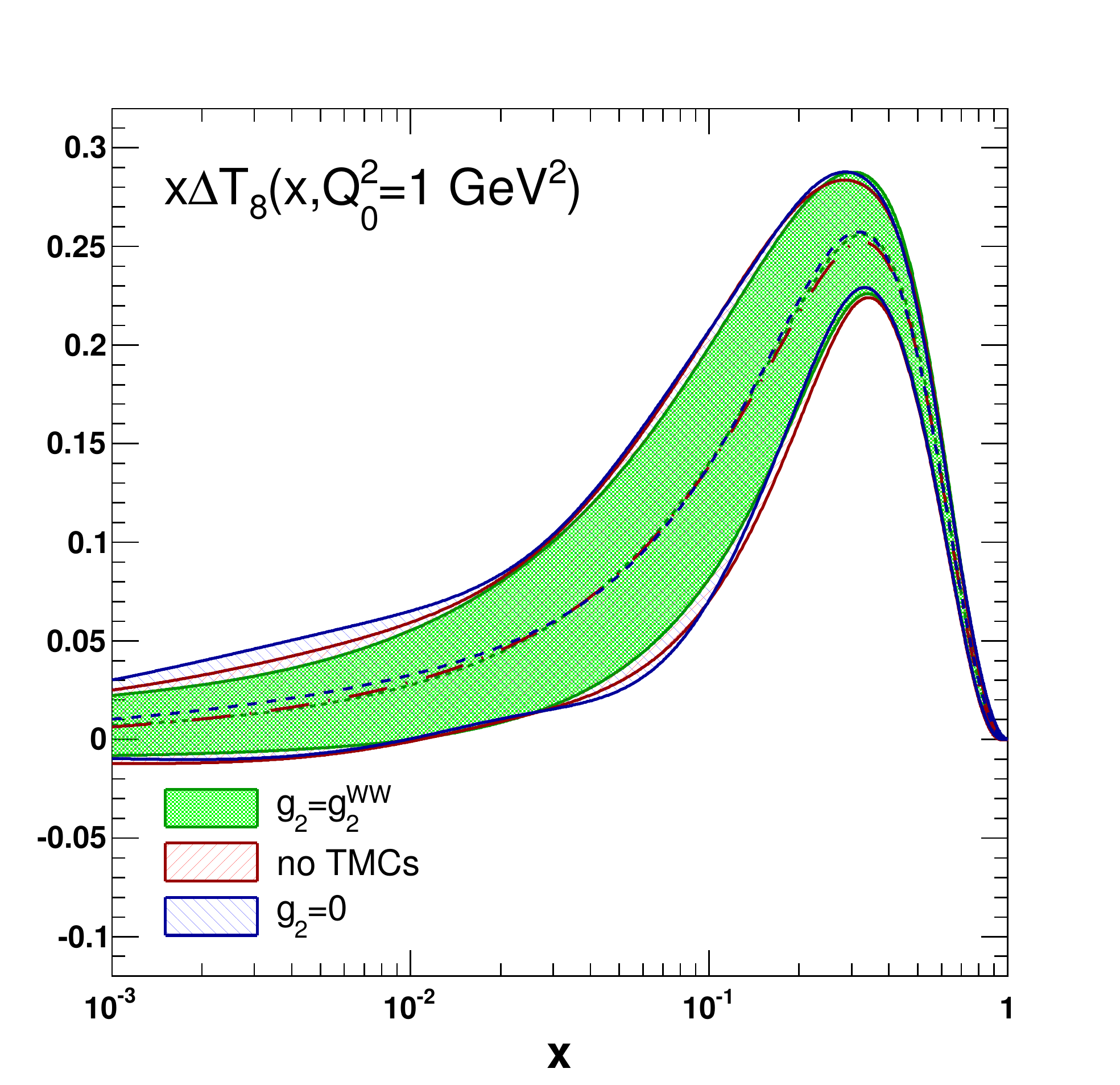}
\mycaption{Comparison between the default \texttt{NNPDFpol1.0}
PDFs (labeled as $g_2=g_2^{\mbox{\tiny WW}}$ in the plot), 
PDFs with $M=0$ (labeled as noTMCs in the plot) and PDFs with $g_2=0$;
each corresponds to the statistical estimators of 
Tab.~\ref{tab:tmc_estimators}.}
\label{fig:TMC_comparison}
\end{center}
\end{figure}

Furthermore, in
Fig.~\ref{fig:TMC_comparison} we compare the PDFs at
the initial scale $Q_0^2$ determined in these fits to our default set:
differences are hardly visible.
This comparison can be made more quantitative by using the distance
$d(x,Q^2)$ between different fits, as defined in Appendix~\ref{sec:appB} 
(see also Appendix~A of Ref.~\cite{Ball:2010de}).  
The distance is defined in such a way that if we
compare two different samples of $N_{\mathrm{rep}}$ replicas each extracted
from the same distribution, then on average $d=1$, while if the two
samples are extracted from two distributions which differ by one
standard deviation, then on average $d=\sqrt{N_{\mathrm{rep}}}$ (the
difference being due to the fact that the standard deviation of the
mean scales as $1/\sqrt{N_{\mathrm{rep}}}$).

\begin{figure}[t]
\begin{center}
\epsfig{width=0.90\textwidth,figure=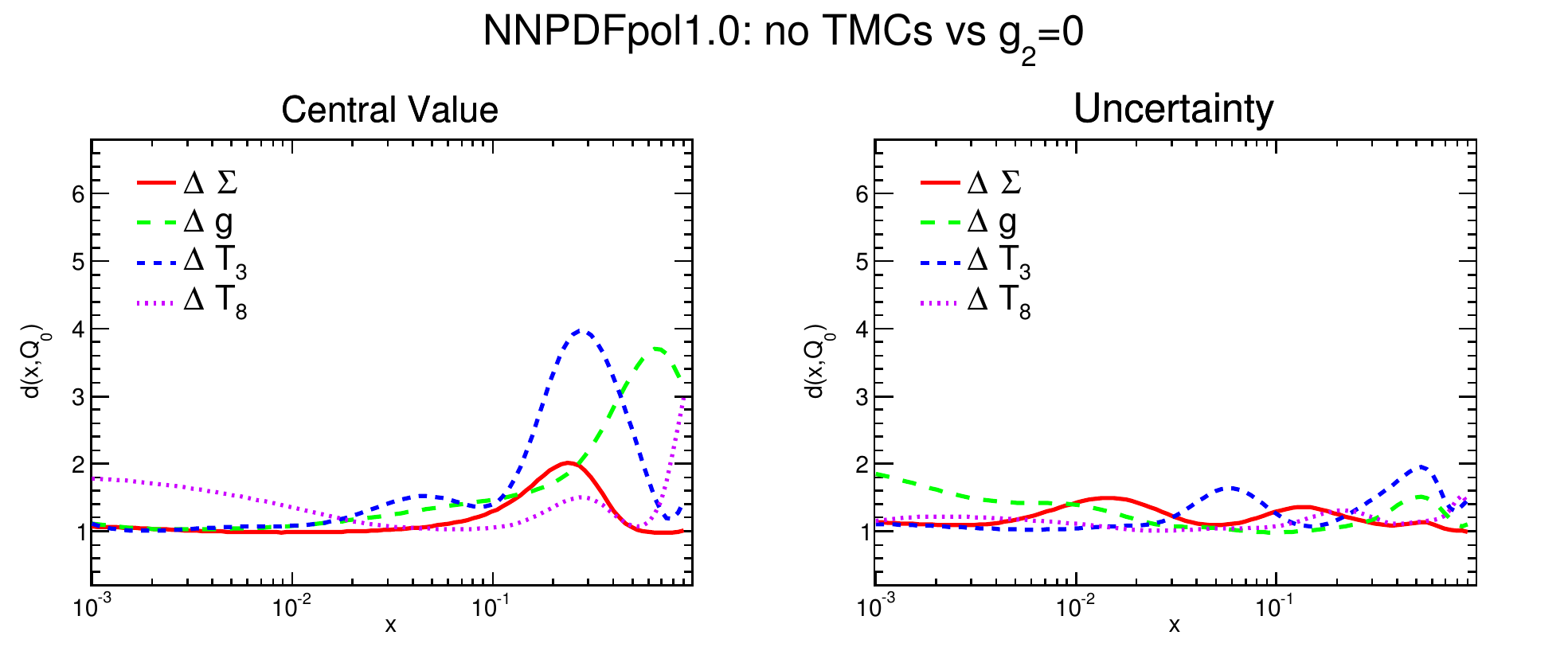}\\
\epsfig{width=0.90\textwidth,figure=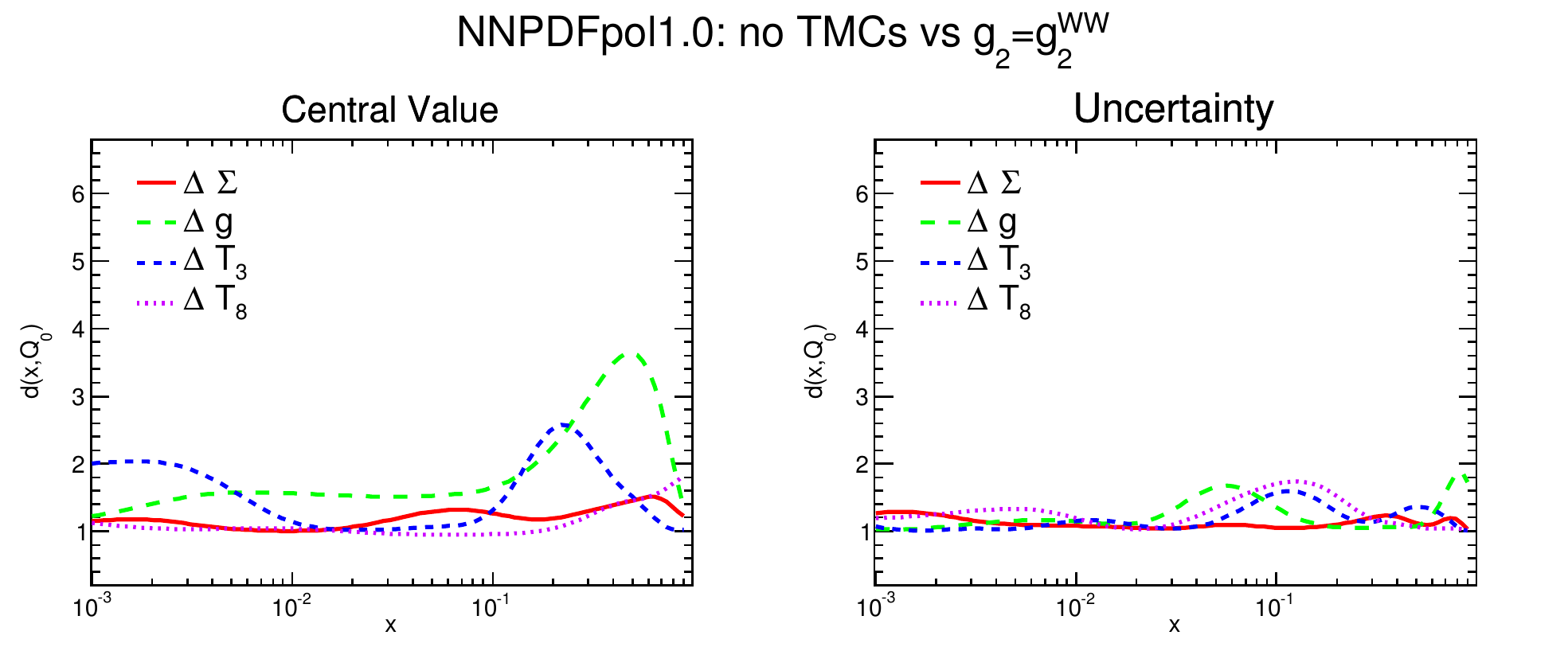}\\
\epsfig{width=0.90\textwidth,figure=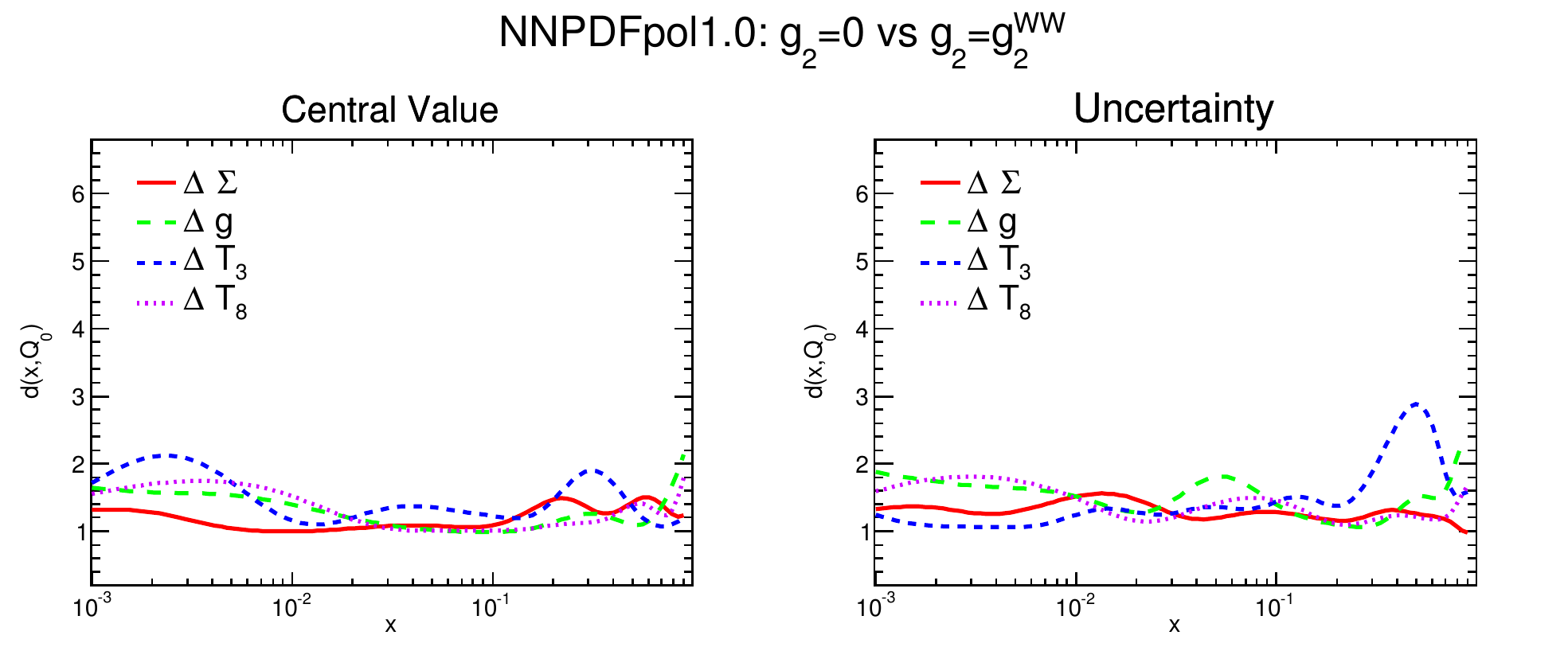}
\mycaption{Distances between each pair of the three sets  of PDFs
shown in Fig.~\ref{fig:TMC_comparison}.}
\label{fig:distances_noTMCs}
\end{center}
\end{figure}

The distances $d(x,Q^2)$
between central values and uncertainties of the three fits
of Tab.~\ref{tab:tmc_estimators} are shown in
Fig.~\ref{fig:distances_noTMCs}. They never exceed $d=4$, which means less
than half a standard deviation for $N_{\mathrm{rep}}=100$.
It is interesting to observe that distances tend to be larger in the
large-$x$ region, where the expansion in powers of $M^2/Q^2$ is less
accurate, and the effects of dynamical higher twists can become relevant.
It is reassuring that even in this region the distances are
reasonably small.

We conclude that inclusive DIS data, with our kinematic cuts, do not show 
sensitivity to finite nucleon mass effects, neither in terms of fit quality, 
nor in terms of the effect on PDFs.

\subsubsection{Sum rules}
\label{sec:srres}

Our default PDF fit is obtained by assuming that
the triplet axial charge $a_3$ is fixed to its value extracted from
$\beta$ decay, 
Eq.~(\ref{eq:hypdecayconst}), and that the octet axial charge $a_8$ is fixed to 
the value of $a_8$ determined from baryon octet
decays, but with an inflated uncertainty in order to allow for $SU(3)$ 
violation, Eq.~(\ref{eq:a8p}). As discussed after Eq.~(\ref{eq:sumrules1})
uncertainties on them are included by randomizing their values among replicas.

In order to test the impact of these assumptions, we have produced two
more PDF determinations. In the first, we have not imposed the triplet
sum rule, so in particular $a_3$ is free and
determined by the data, 
instead of being fixed  to the value Eq.~(\ref{eq:hypdecayconst}). In the
second, we have assumed that the uncertainty on $a_8$ is given by
the much smaller value of Eq.~(\ref{eq:hypdecayconst}). 

\begin{table}[t]
\centering
\footnotesize
\begin{tabular}{ccc}
\toprule
Fit & free $a_3$  & $a_8$ Eq.~(\ref{eq:hypdecayconst}) \\
\midrule
$\chi^{2}_{\mathrm{tot}}$ & 0.79 & 0.77 \\
$\langle E \rangle \pm \sigma_{E}$ 
& 1.84 $\pm$ 0.19 & 1.86 $\pm$ 0.19  \\
$\langle E_{\mathrm {tr}} \rangle \pm \sigma_{E_{\mathrm {tr}}}$ 
& 1.73 $\pm$ 0.41 & 1.66 $\pm$ 0.53  \\
$\langle E_{\mathrm{val}} \rangle \pm \sigma_{E_{\mathrm{val}}}$ 
& 1.93 $\pm$ 0.58 & 1.87 $\pm$ 0.71 \\
\midrule
$\langle \chi^{2(k)} \rangle \pm \sigma_{\chi^{2}}$ 
& 0.93 $\pm$ 0.12 & 0.92 $\pm$ 0.15 \\
\bottomrule
\end{tabular}
\mycaption{The statistical estimators of Tab.~\ref{tab:chi2tab1}, but for
fits in which  the triplet sum rule is not imposed (free $a_3$) or
in which the octet sum rule is imposed with the smaller uncertainty
Eq.~(\ref{eq:hypdecayconst}).}
\label{tab:sr_estimators}
\end{table}

\begin{figure}[t]
\begin{center}
\epsfig{width=0.90\textwidth,figure=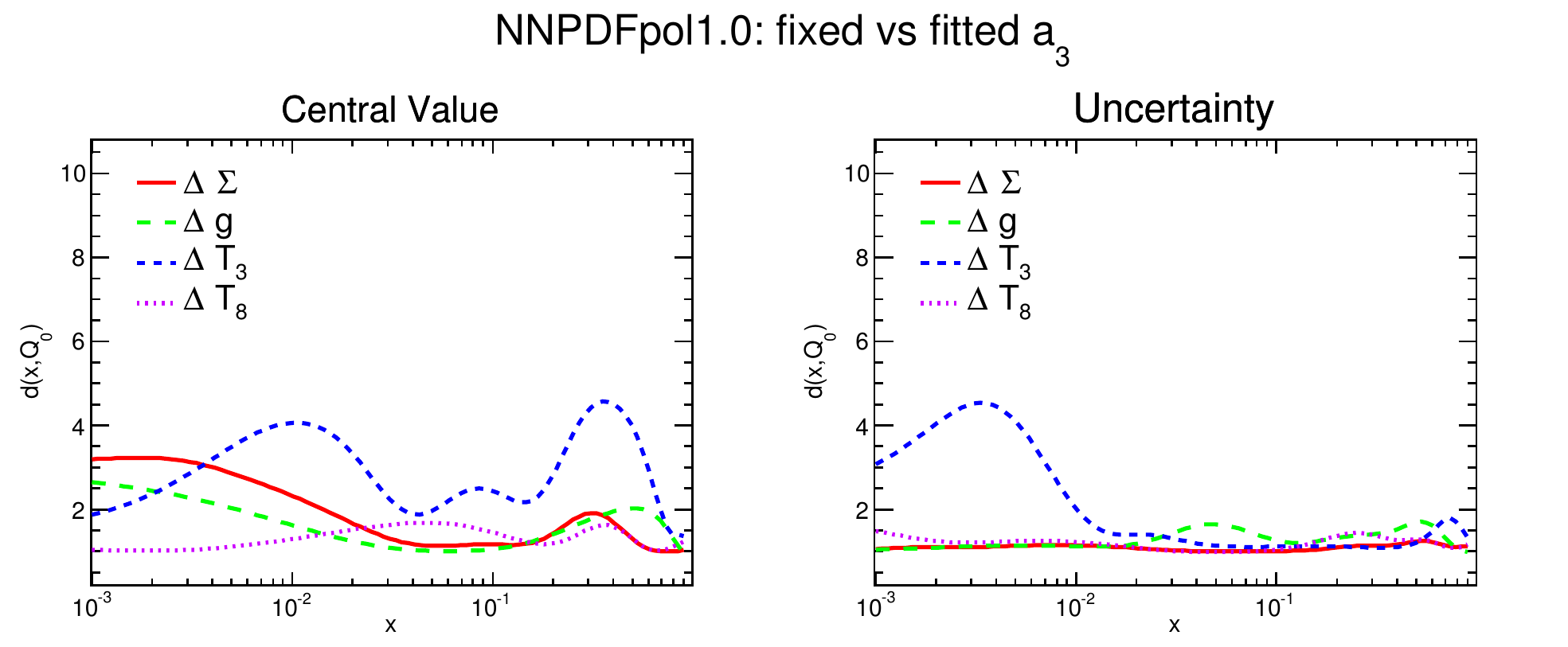}
\mycaption{Distances between PDFs (central values and
uncertainties) for the
default fit, with $a_3$ fixed,  and the fit with free $a_3$,
computed using  $N_{\mbox{\tiny rep}}=100$
replicas from each set.}
\label{fig:distances_a3}
\end{center}
\end{figure}

\begin{figure}[t]
\begin{center}
\epsfig{width=0.40\textwidth,figure=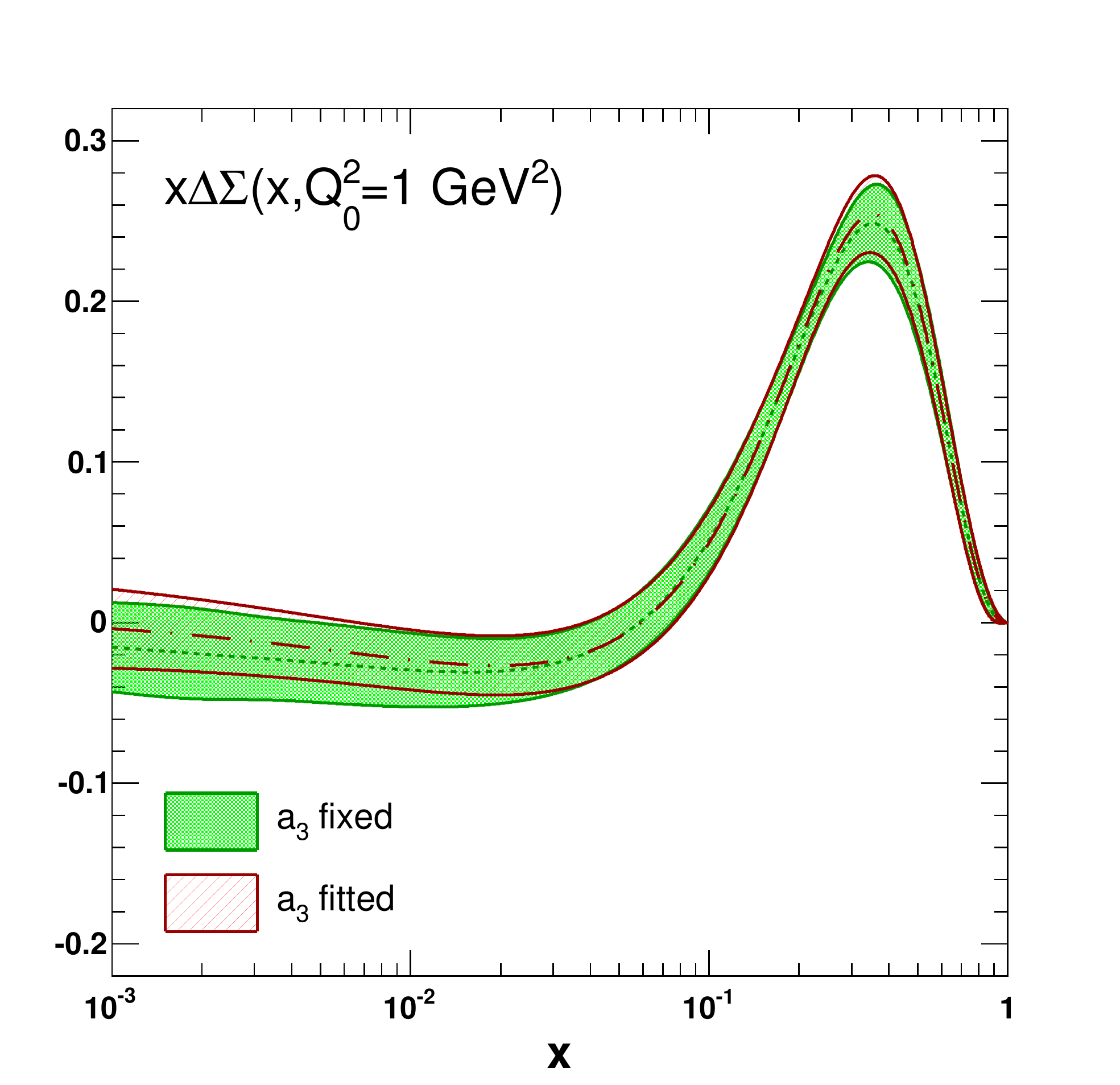}
\epsfig{width=0.40\textwidth,figure=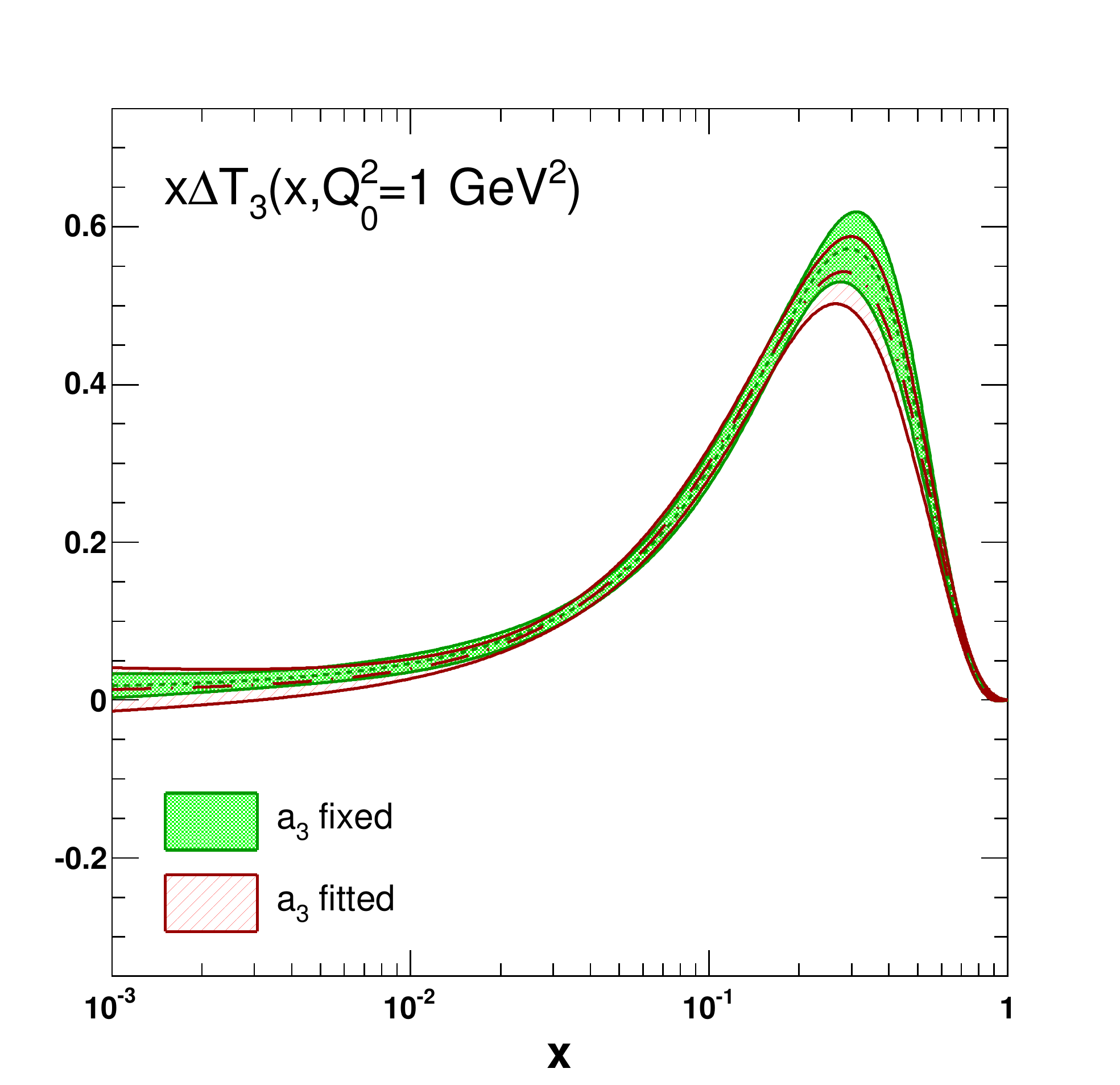}
\mycaption{Comparison of the singlet and triplet PDFs for the
default fit, with $a_3$ fixed, and the fit with free $a_3$.}
\label{fig:fit_a3}
\end{center}
\end{figure}

The statistical estimators for the total data set  for each of these fits
are shown in Tab.~\ref{tab:sr_estimators}. Here too, there is no
significant difference in fit quality between these fits and the default.
The distances between PDFs in the default and the free $a_3$
fits are displayed in Fig.~\ref{fig:distances_a3}. As one may expect,
only the triplet is affected significantly:
the central value is shifted by about $d \sim 5$, 
\textit{i.e.} about half-$\sigma$, in the region $x\sim 0.3$, 
where $x\Delta T_3$ has a maximum, and also around $x\sim 0.01$. 
The uncertainties on the PDFs are very similar in both cases for
all PDFs, except $\Delta T_3$ at small-$x$: in this case, removing the
$a_3$ sum rule results in a moderate increase of the uncertainties;
the effect of removing $a_3$ is otherwise negligible.
The singlet and triplet PDFs for these two fits are compared
in Fig.~\ref{fig:fit_a3}.

\begin{figure}[t]
\begin{center}
\epsfig{width=0.90\textwidth,figure=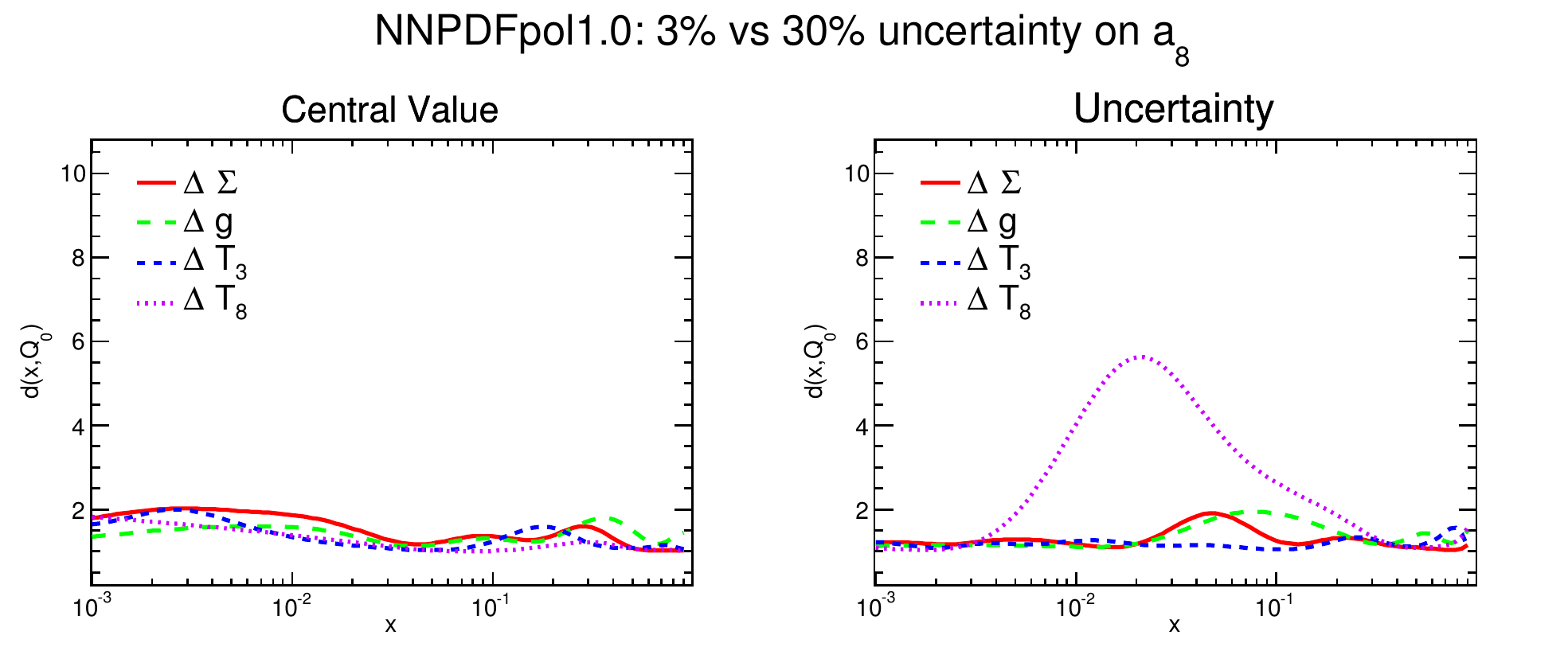}
\mycaption{Distances between PDFs (central values and
uncertainties) for the default fit, with $a_8$ Eq.~(\ref{eq:a8p}),
and the fit with the value of $a_8$ with smaller uncertainty,
Eq.~(\ref{eq:hypdecayconst}).}
\label{fig:distances_a8}
\end{center}
\end{figure}

\begin{figure}[t]
\begin{center}
\epsfig{width=0.40\textwidth,figure=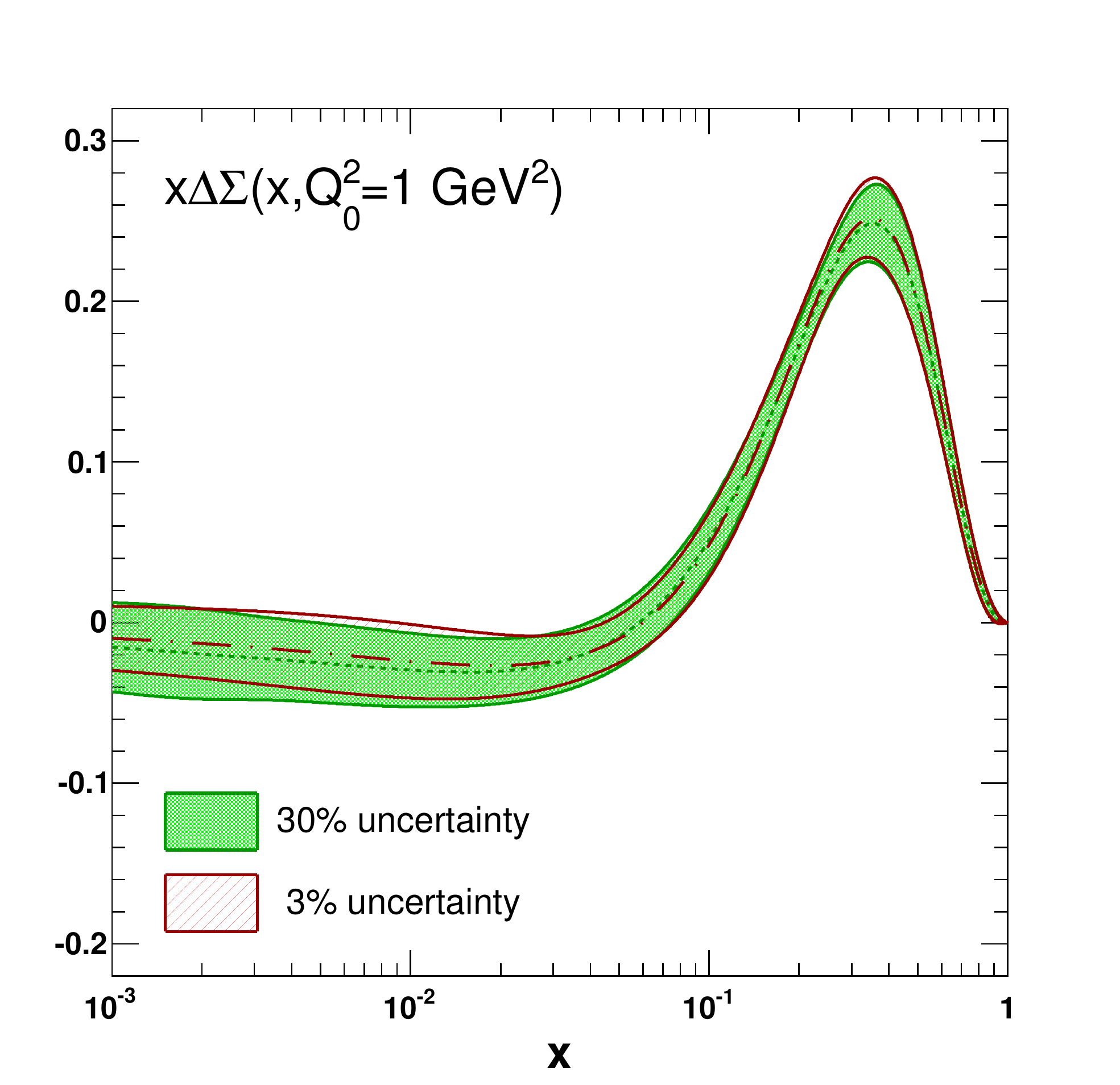}
\epsfig{width=0.40\textwidth,figure=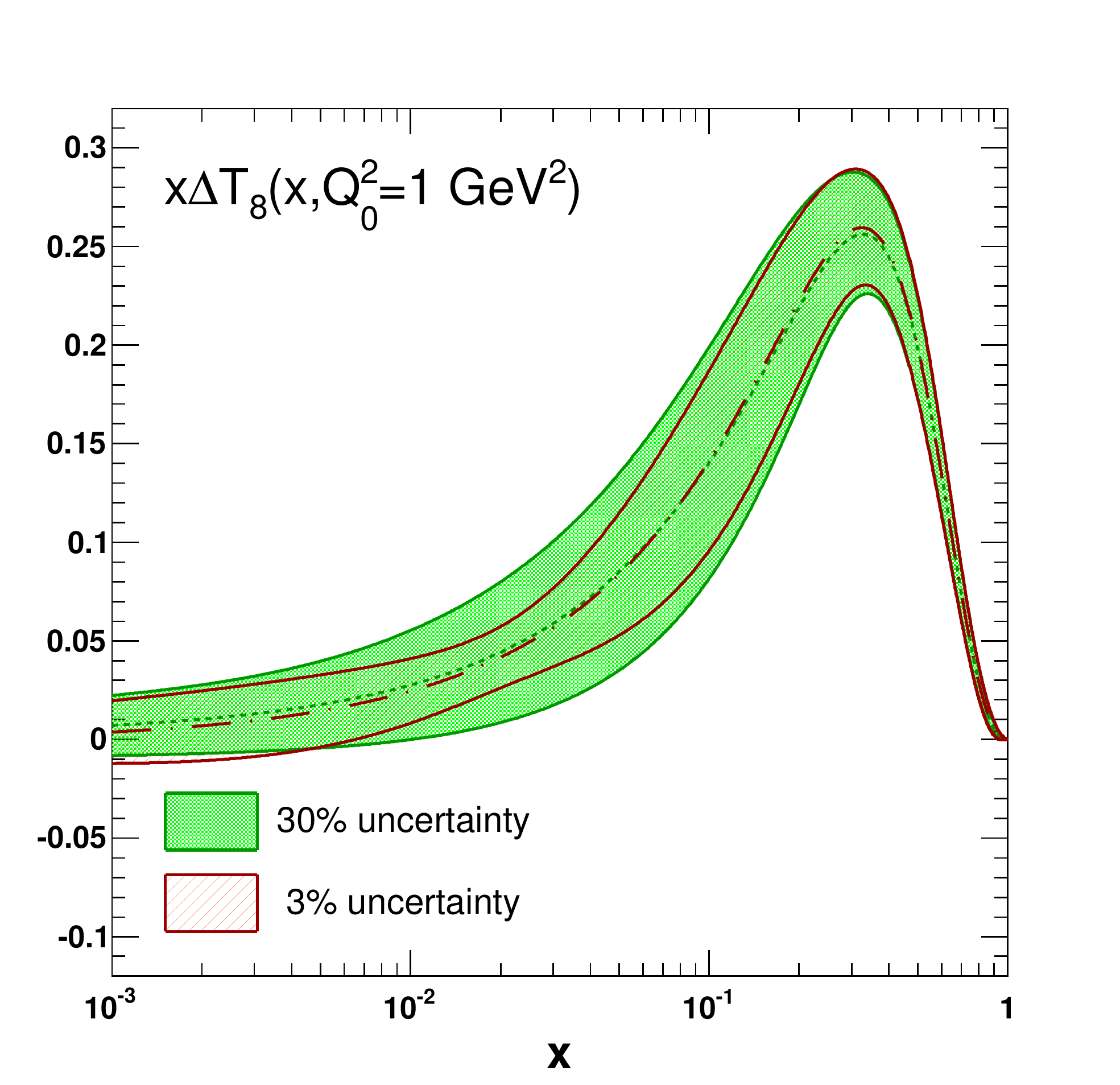}
\mycaption{Comparison of the singlet and octet PDFs for the
default fit, with $a_8$ Eq.~(\ref{eq:a8p}), and the fit with the
value of $a_8$ with smaller uncertainty, Eq.~(\ref{eq:hypdecayconst}).}
\label{fig:fit_a8}
\end{center}
\end{figure}

The distances between the default and the fit with the smaller uncertainty
on $a_8$
are shown in Fig.~\ref{fig:distances_a8}. In this case, again as
expected, the only effect is on the $\Delta T_8$ uncertainty, which changes
in the region $10^{-2}\lesssim x \lesssim 10^{-1}$
by up to $d\sim 6$ (about half a standard deviation): if a more
accurate value of $a_8$ is assumed, the determined $\Delta T_8$
is correspondingly more accurate. Central values are unaffected.
The singlet and octet PDFs for this fit are compared to the default 
in Fig.~\ref{fig:fit_a8}. We conclude that the size of the uncertainty
on  $\Delta T_8$ has a moderate effect on our fit; on the other hand
it is clear that if the octet sum rule were not imposed at all, the
uncertainty on the octet and thus on strangeness would increase very
significantly, as we have checked explicitly.
We conclude that our fit results are quite stable upon variations 
of our treatment of both the triplet and the octet sum rules.

\subsection{Positivity}
As discussed in Sec.~\ref{sec:minim}, positivity of the individual
cross-sections entering the polarized asymmetries
Eq.~(\ref{eq:xsecasy}) has been imposed at leading order according to
Eq.~(\ref{eq:possigma}), using the NLO \texttt{NNPDF2.1} 
PDF set~\cite{Ball:2011mu}, separately for the lightest polarized quark
PDF combinations $\Delta u + \Delta\bar{u}$, $\Delta d +
\Delta\bar{d}$, $\Delta s + \Delta\bar{s}$ and for the polarized gluon
PDF, by means of a Lagrange multiplier Eq.~(\ref{eq:lagrmult}). After
stopping, positivity is checked a posteriori and replicas which do not
satisfy it are discarded and retrained. 

\begin{figure}[t]
\begin{center}
\epsfig{width=0.40\textwidth,figure=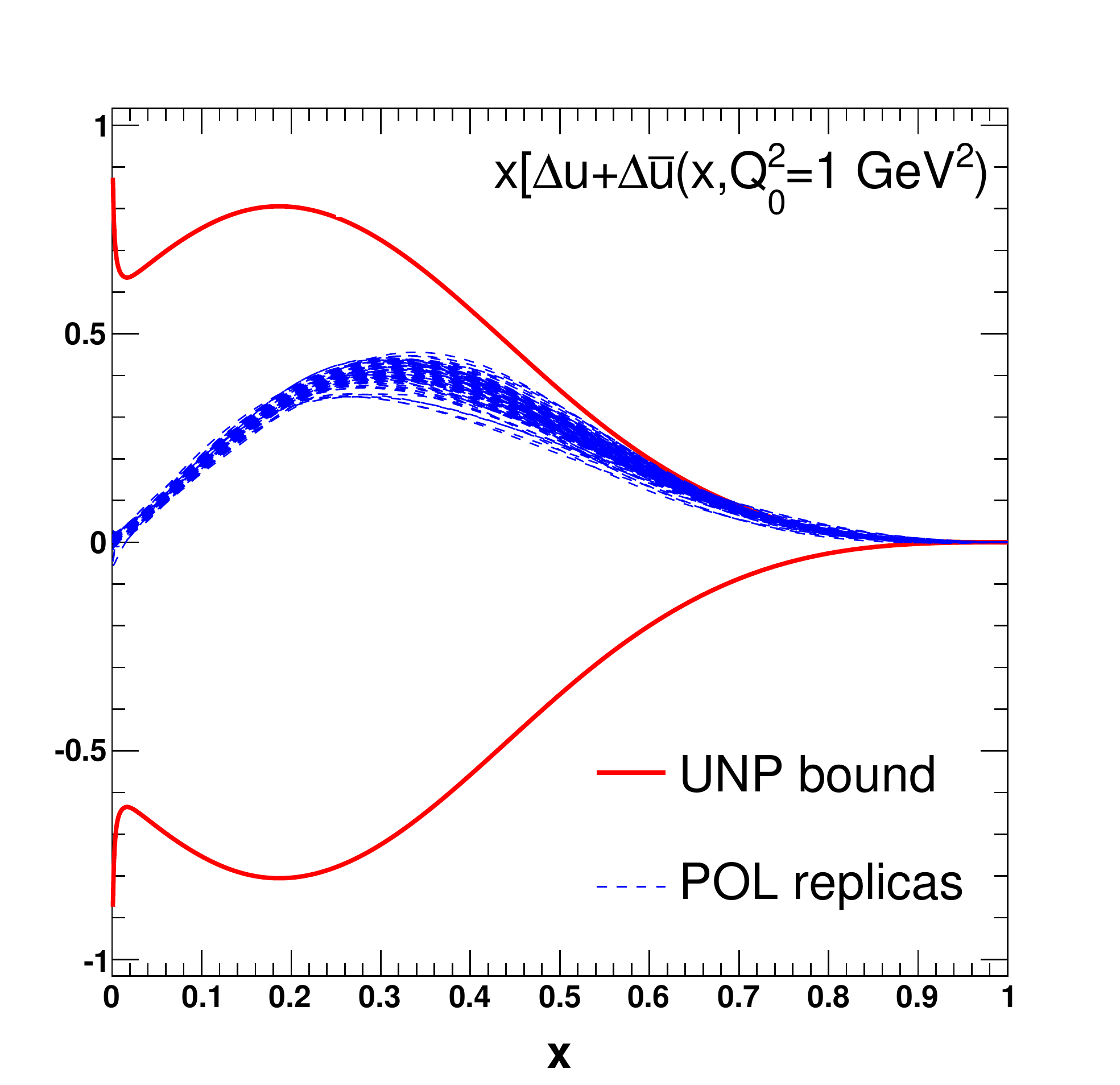}
\epsfig{width=0.40\textwidth,figure=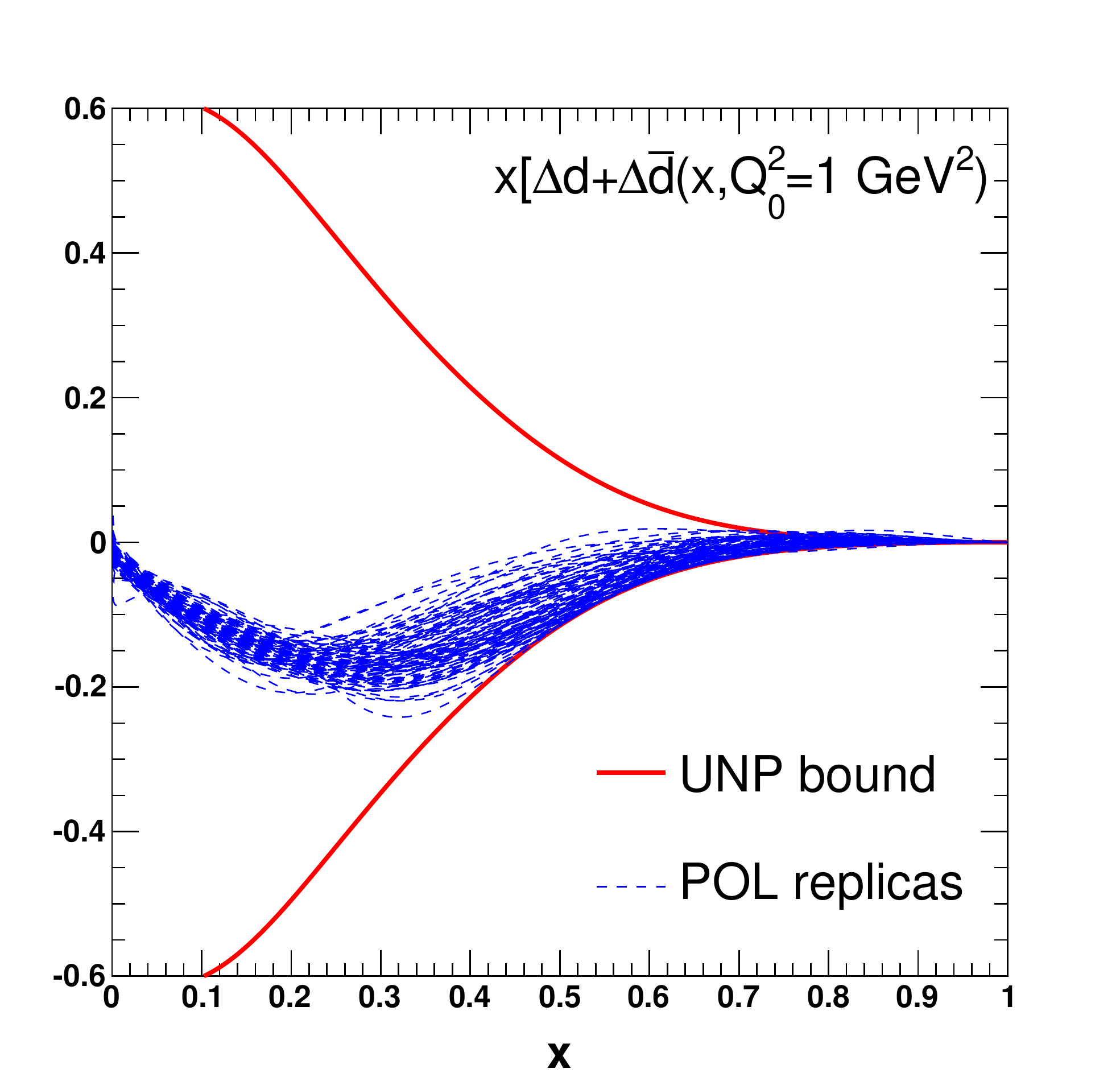}\\
\epsfig{width=0.40\textwidth,figure=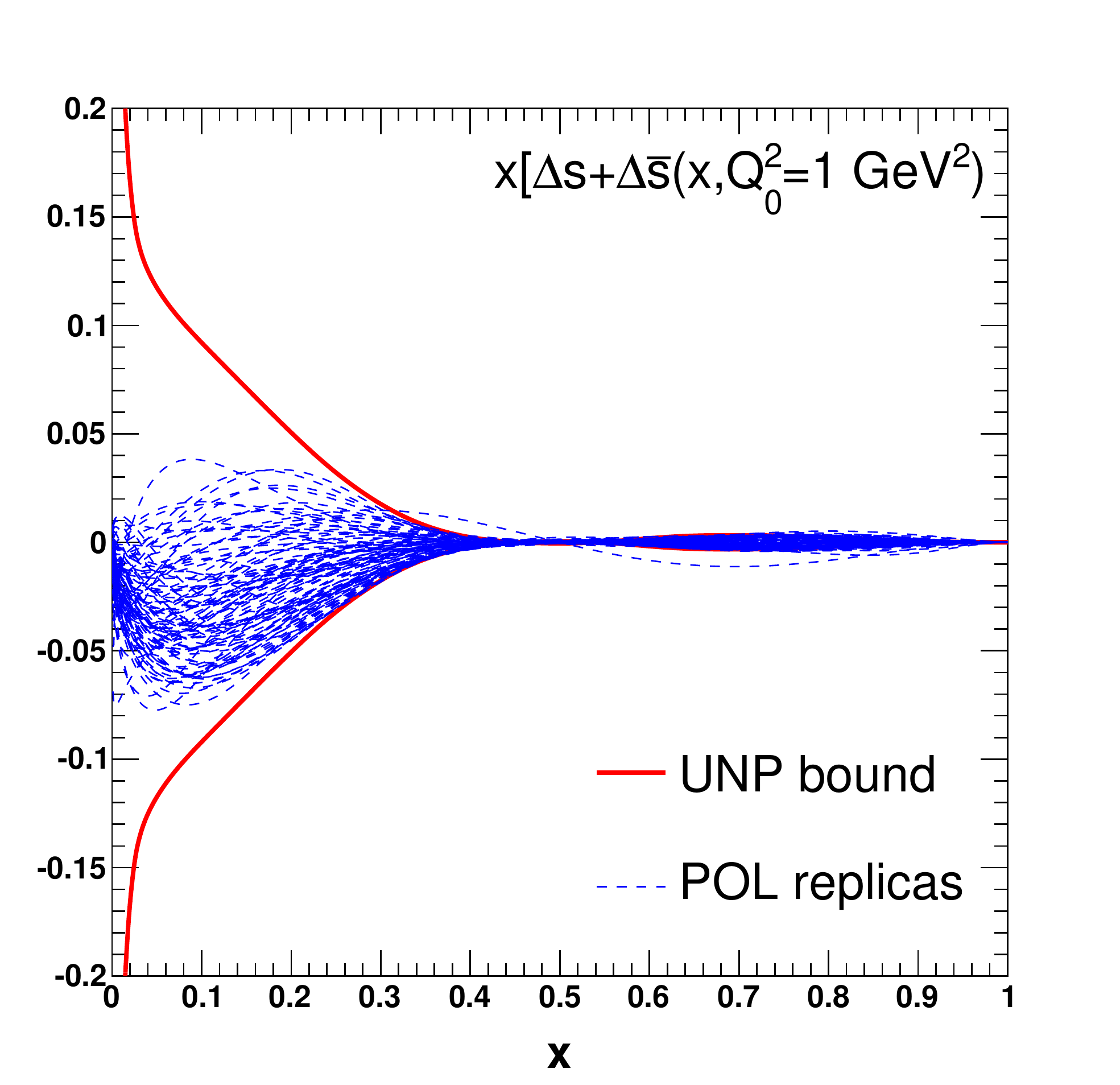}
\epsfig{width=0.40\textwidth,figure=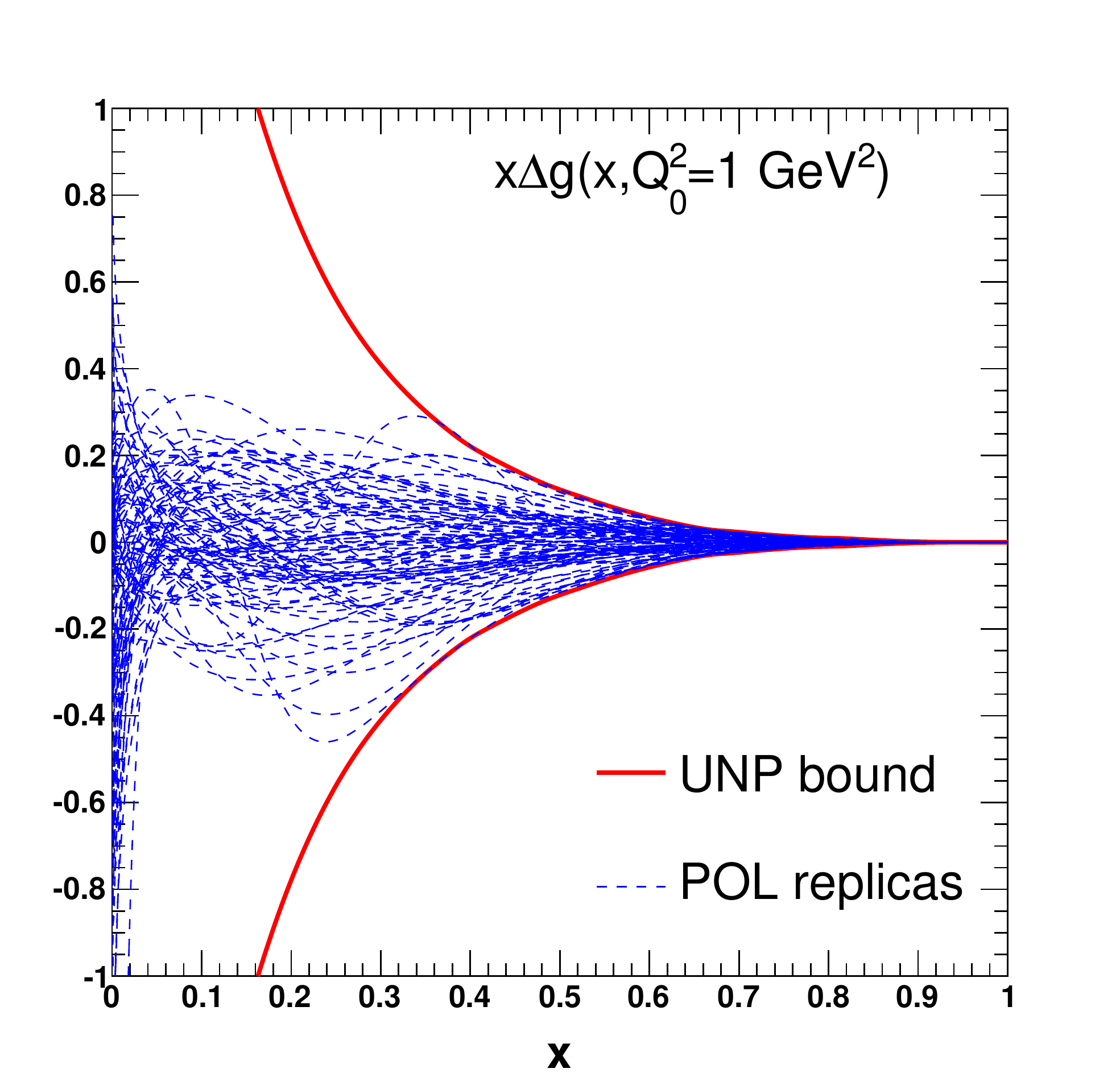}
\mycaption{The positivity bound Eq.~(\ref{eq:possigma}), compared
to a set of $N_{\mbox{\tiny rep}}=100$  replicas  (dashed
lines).}
\label{fig:pdfposconstr}
\end{center}
\end{figure}

In Fig.~\ref{fig:pdfposconstr} we compare  to the positivity bound for
the up, down, strange PDF combinations and
gluon PDF a set of $N_{\mbox{\tiny rep}}=100$  replicas obtained  
by enforcing positivity through a
Lagrange multiplier, but before the final, \textit{a posteriori} check. 
Almost all replicas satisfy the constraint, but at least one replica
which clearly violates it for the total
strangeness combination (and thus will be discarded) is seen.

In order to assess the effect of the positivity
constraints, we have performed a
fit without imposing positivity. Because positivity significantly
affects PDFs in the region where no data are available, and 
thus in particular their large-$x$ behavior, preprocessing exponents for this
PDF determination had to be determined again using the procedure
described in Sec.~\ref{sec:minim}. The values of the large $x$
preprocessing exponents used in the fit without positivity are shown
in Tab.~\ref{tab:prepexpsnopos}. The small $x$
exponents are the same as in the
baseline fit, Tab.~\ref{tab:prepexps}.
\begin{table}[t]
\centering
\footnotesize
\begin{tabular}{cc}
\hline PDF & $m$ \\
\toprule
$\Delta\Sigma(x, Q_0^2)$ & $[0.5, 5.0]$ \\
\midrule
$\Delta g(x, Q_0^2)$ & $[0.5, 5.0]$\\
\midrule
$\Delta T_3(x, Q_0^2)$ & $[0.5, 4.0]$\\
\midrule
$\Delta T_8(x, Q_0^2)$ & $[0.5, 6.0]$ \\
\midrule
\end{tabular}
\mycaption{Ranges for the large-$x$
preprocessing exponents Eq.~(\ref{eq:PDFbasisnets})
for the fit in which no positivity
is imposed. The small-$x$ exponents are the same as in the
baseline fit Tab.~\ref{tab:prepexps}.}
\label{tab:prepexpsnopos}
\end{table}

\begin{figure}[t]
\begin{center}
\epsfig{width=0.40\textwidth,figure=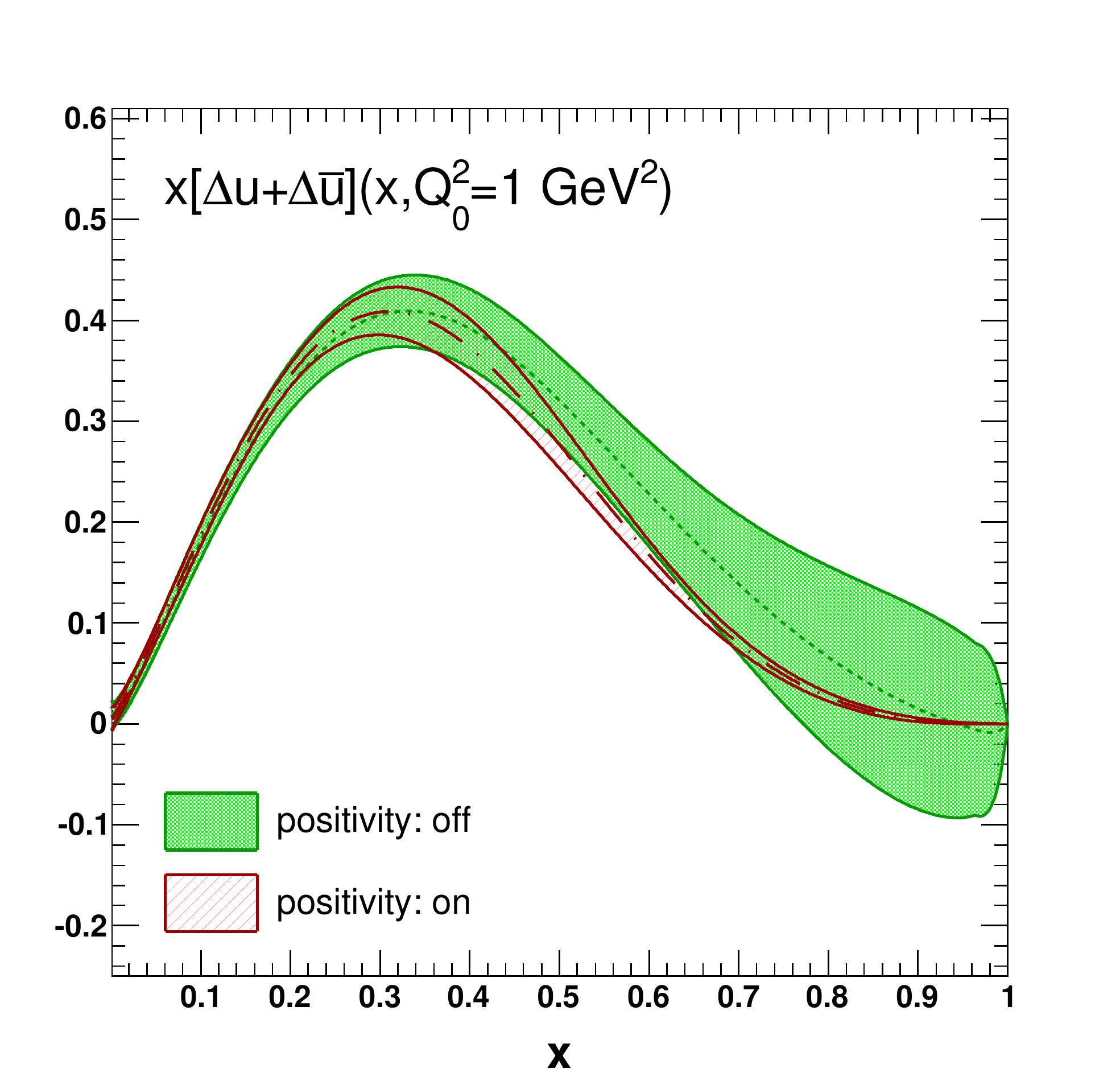}
\epsfig{width=0.40\textwidth,figure=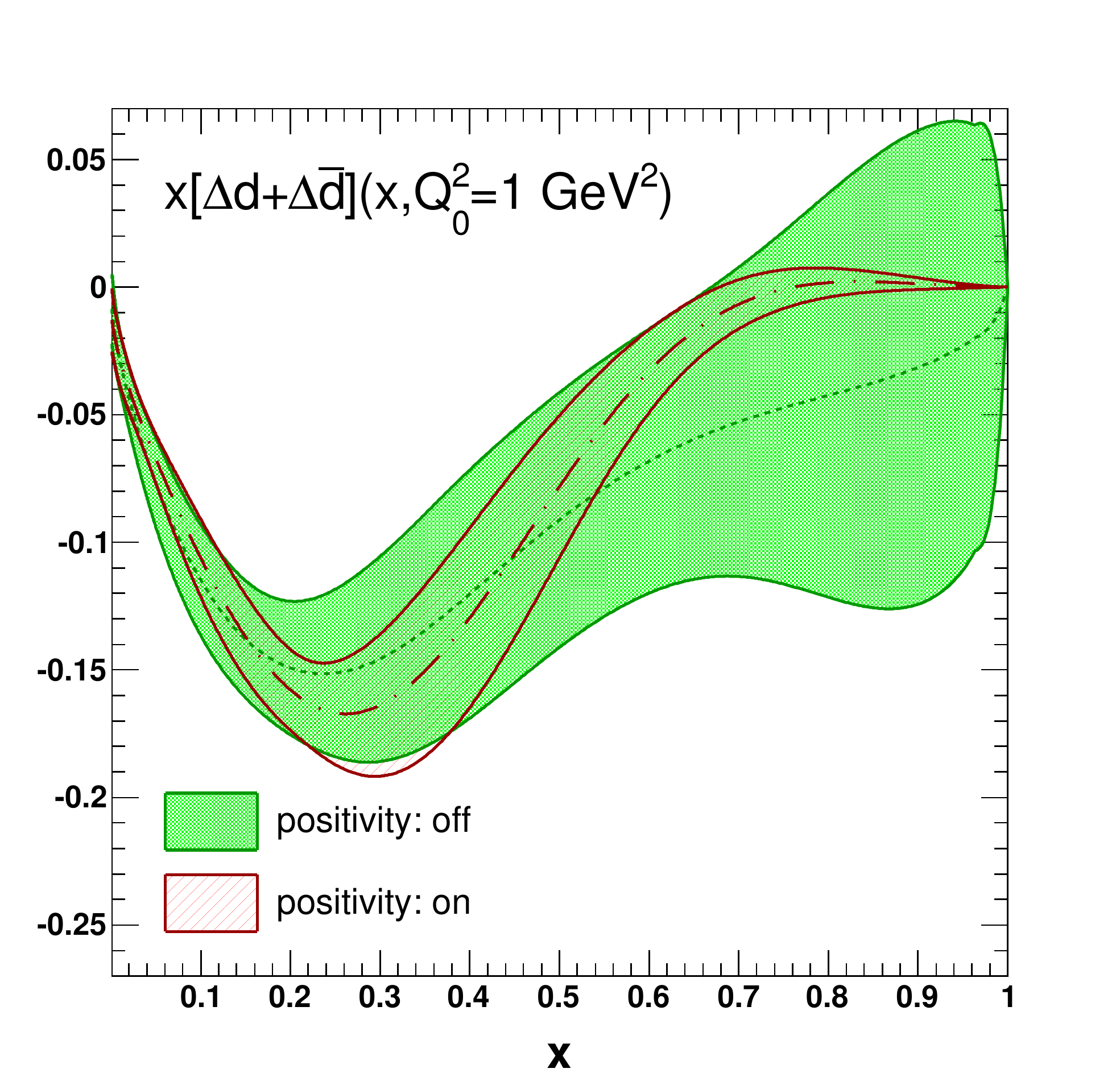}\\
\epsfig{width=0.40\textwidth,figure=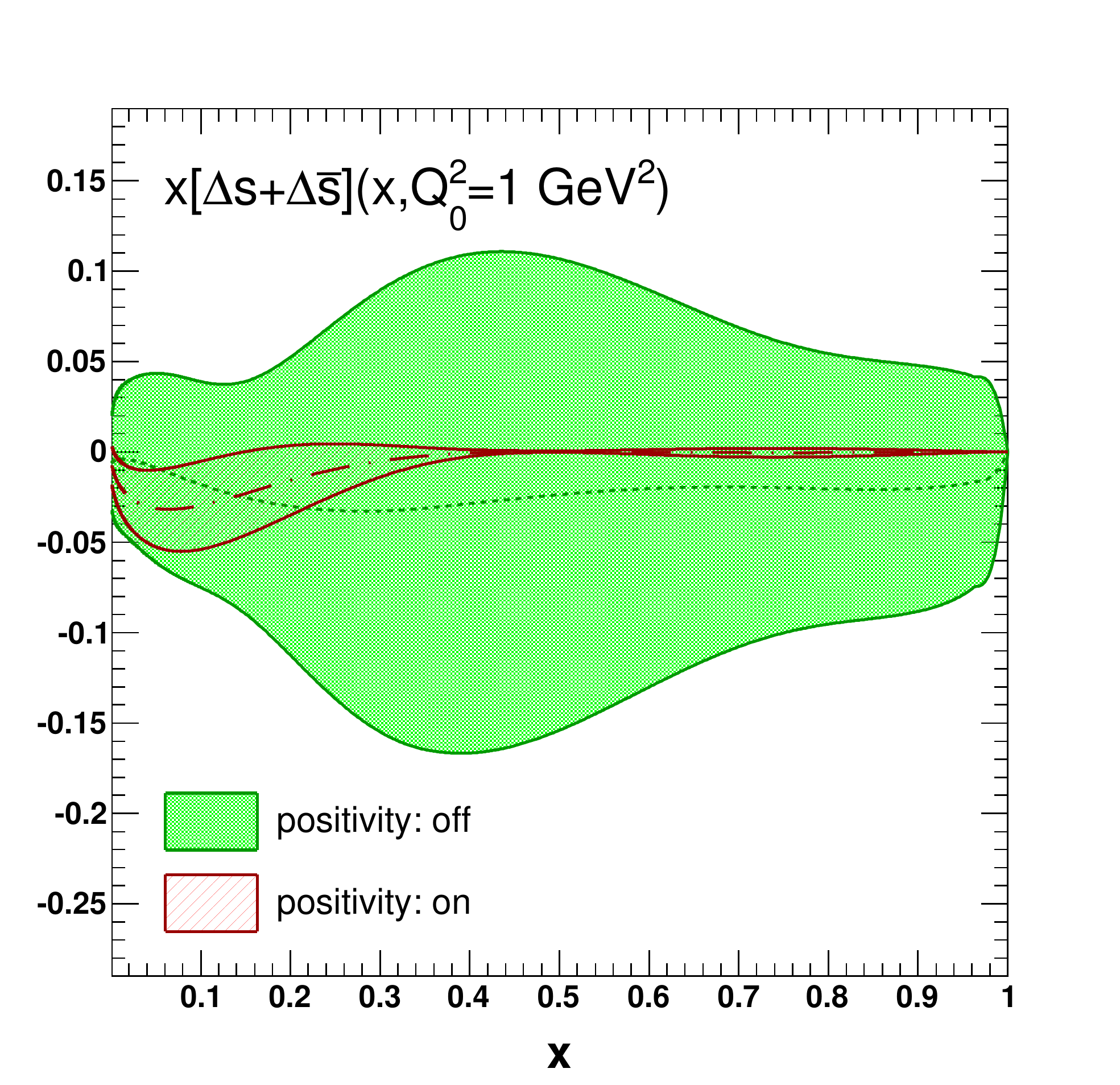}
\epsfig{width=0.40\textwidth,figure=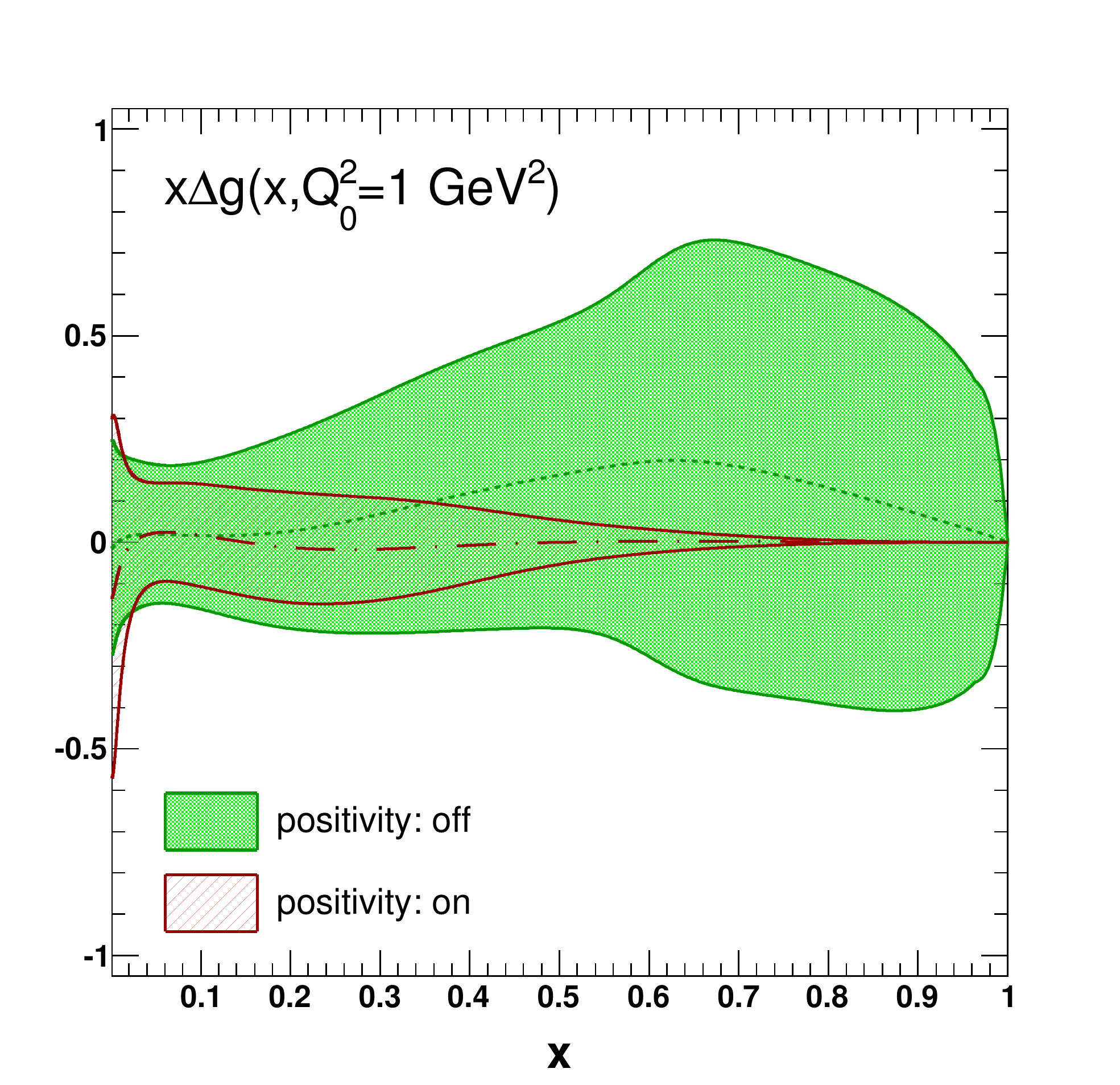}
\mycaption{The \texttt{NNPDFpol1.0} PDFs with and without
positivity constraints compared at the initial parametrization scale
$Q_0^2=1$ GeV$^2$ in the flavor basis. \label{fig:pdfposbench}}
\end{center}
\end{figure}

The corresponding estimators are
shown in Tab.~\ref{tab:pos_estimators}. Also in this case, we see no
significant change in fit quality, with only a slight improvement in 
$\chi^2_{\mathrm{tot}}$ when the constraint is removed. This shows that
our PDF parametrization is flexible enough to easily accommodate positivity.
On the other hand, clearly the positivity bound has a significant
impact on PDFs, especially in the large-$x$ region, as shown in 
Fig.~\ref{fig:pdfposbench}, where PDFs obtained from this fit are
compared to the baseline.
At small $x$, instead, the impact of positivity is moderate, because
$g_1/F_1\sim x$ as
$x\to0$~\cite{Ball:1995ye} so there is no constraint in the
limit. This in particular implies that there is no significant loss of
accuracy in imposing the LO positivity bound, because 
in the small $x\lesssim10^{-2}$  region, where the LO and NLO 
positivity bounds differ
significantly~\cite{Forte:1998kd} the bound is not significant.
\begin{table}[t]
\centering
\footnotesize
\begin{tabular}{cc}
\toprule
Fit & \texttt{NNPDFpol1.0} no positivity  \\
\midrule
$\chi^{2}_{\mathrm{tot}}$ &  0.72\\
$\langle E \rangle \pm \sigma_{E}$ &  1.84 $\pm$ 0.22\\
$\langle E_{\mathrm{tr}} \rangle \pm \sigma_{E_{\mathrm{tr}}}$ &  1.60 $\pm$ 0.20\\
$\langle E_{\mathrm{val}} \rangle \pm \sigma_{E_{\mathrm{val}}}$ &  2.07 $\pm$ 0.39\\
\midrule
$\langle \chi^{2(k)} \rangle \pm \sigma_{\chi^{2}}$ &  0.95 $\pm$ 0.16 \\
\bottomrule
\end{tabular}
\mycaption{The statistical estimators of Tab.~\ref{tab:chi2tab1}
for a fit without positivity constraints.}
\label{tab:pos_estimators}
\end{table}

\subsection{Small- and large-\texorpdfstring{$x$}{x} behavior and preprocessing}
\label{sec:prepexp}

The asymptotic behavior of both polarized and unpolarized
PDFs for $x$ close to 0 or 1 is not controlled by perturbation theory,
because powers of $\ln\frac{1}{x}$ and $\ln(1-x)$
respectively appear in the perturbative coefficients, thereby spoiling
the reliability of the perturbative expansion close to the endpoints. 
Non-perturbative effects are also expected to set in eventually (see
\textit{e.g.}~\cite{Ball:1995ye}). For this reason, 
our fitting procedure makes no assumptions on the large- and small-$x$
behaviors of PDFs, apart from the positivity and integrability constraints
discussed in the previous Section.

\begin{figure}[p]
\begin{center}
\epsfig{width=0.34\textwidth,figure=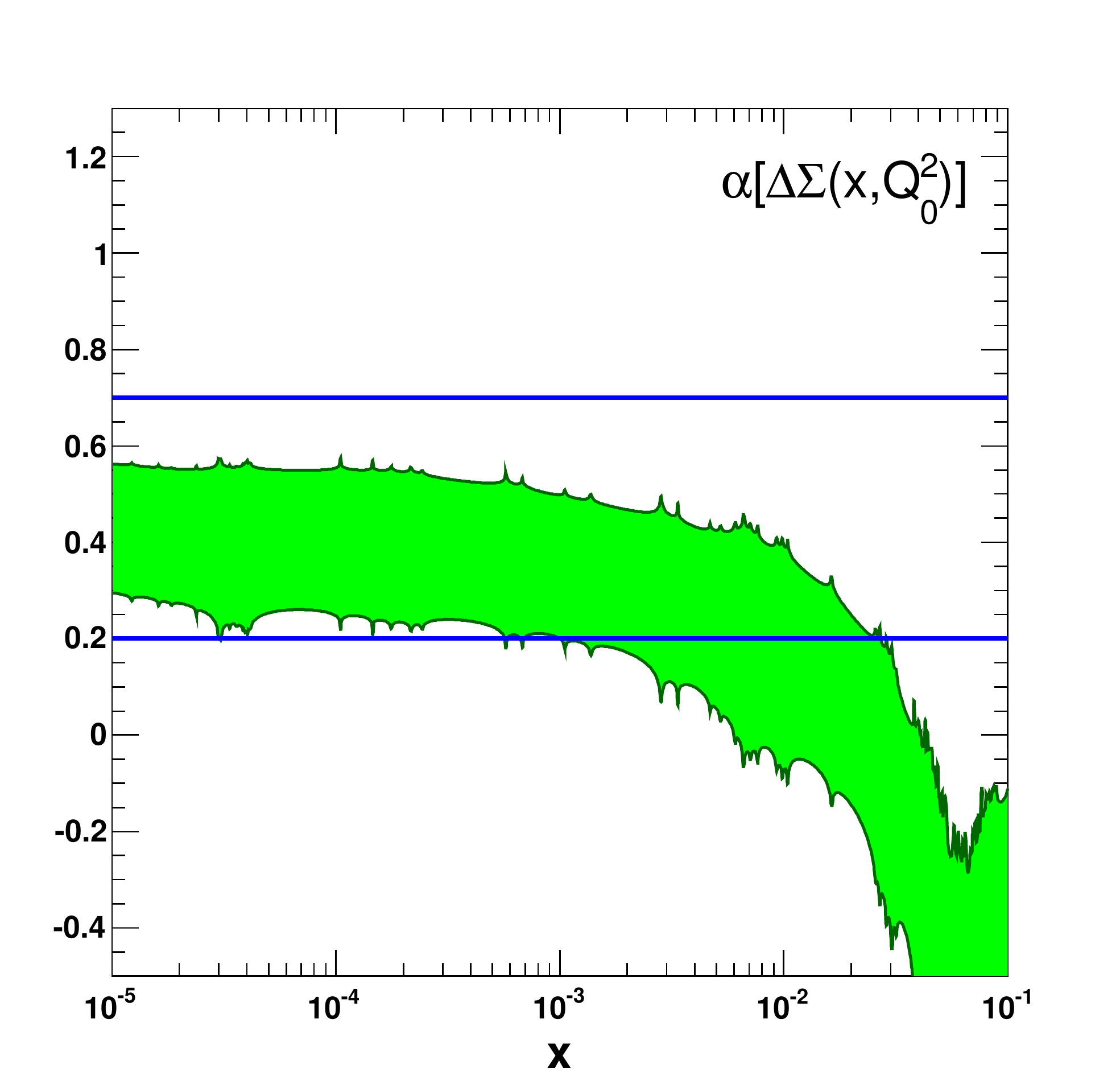}
\epsfig{width=0.34\textwidth,figure=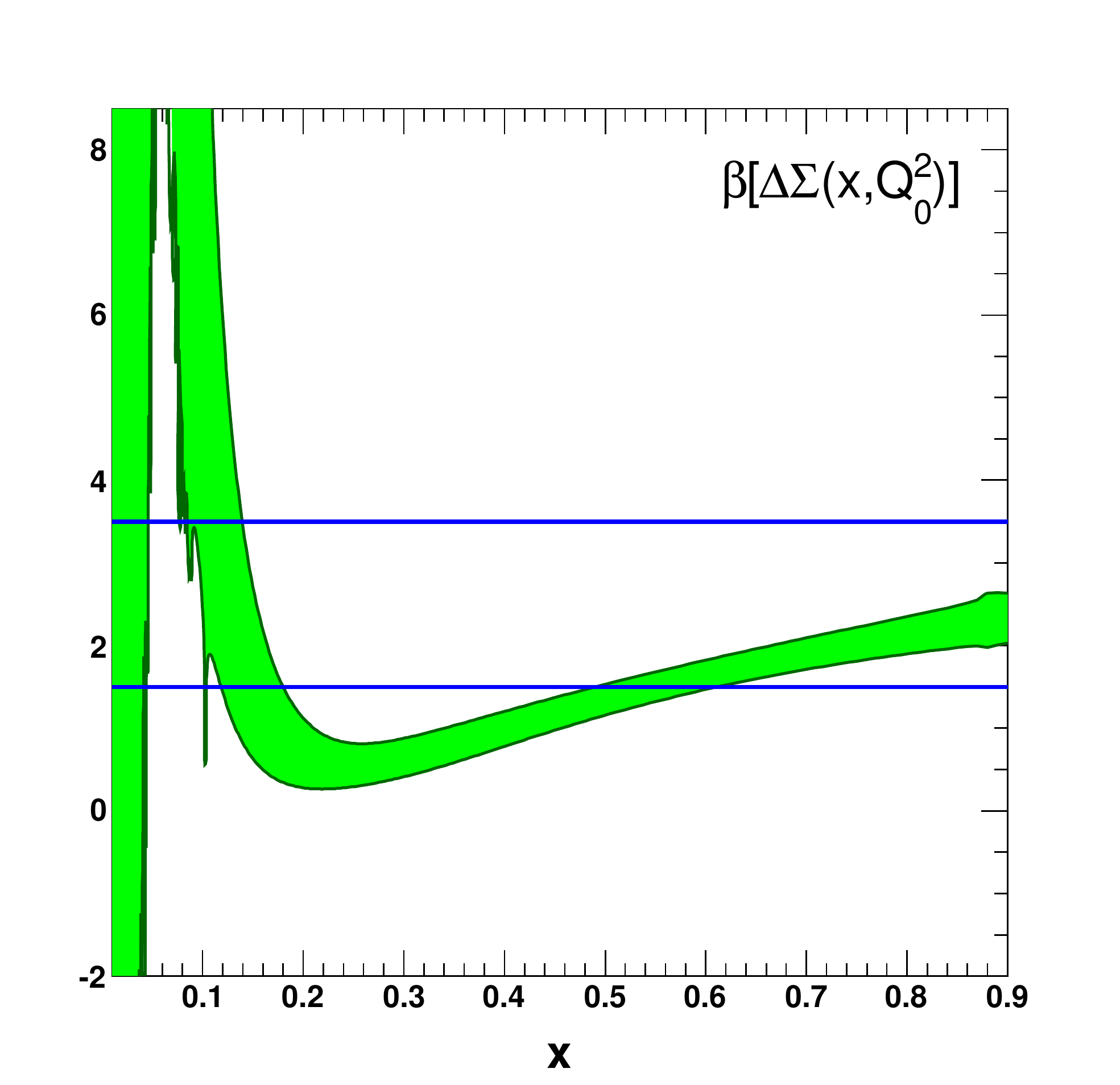}\\
\epsfig{width=0.34\textwidth,figure=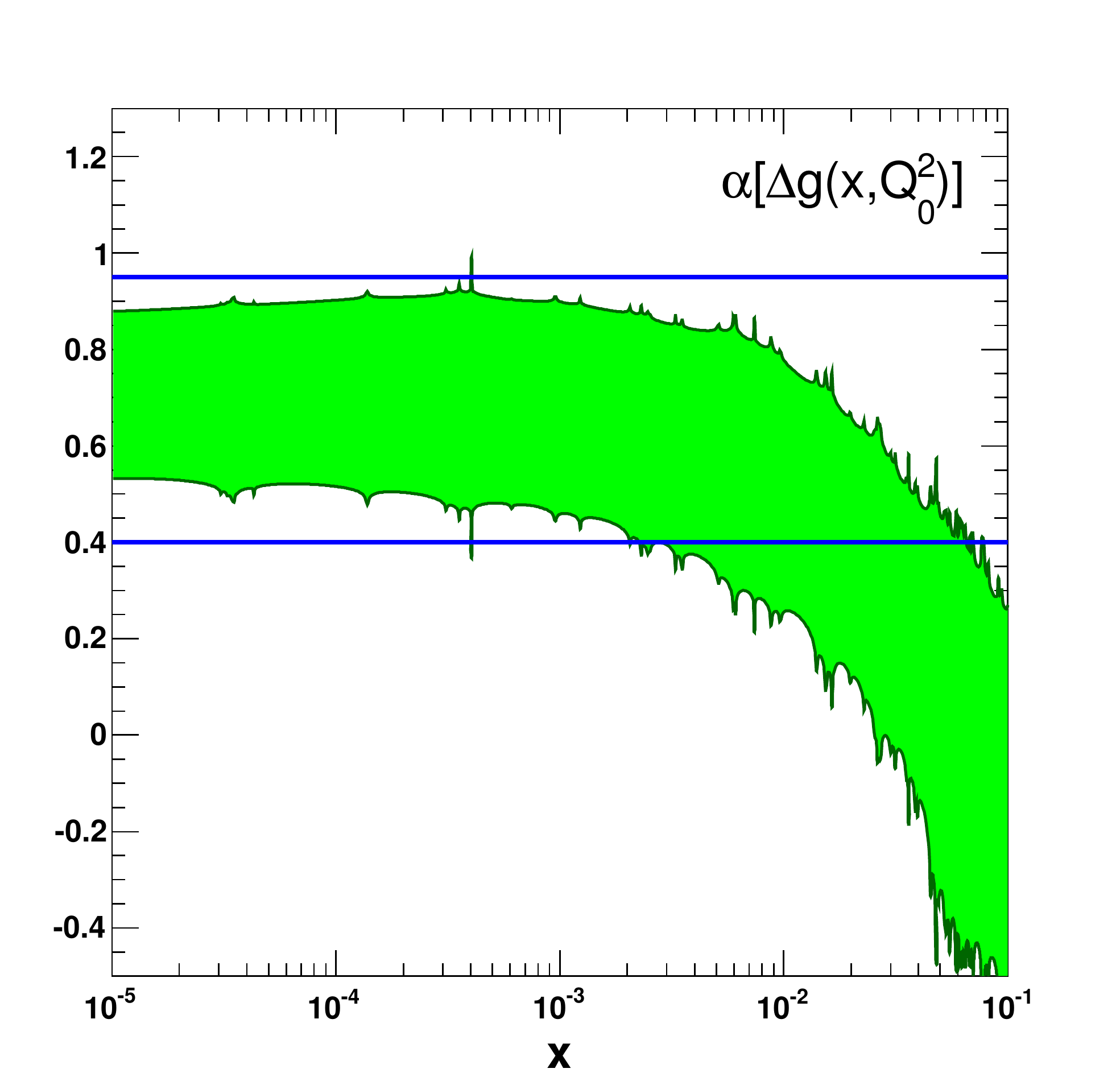}
\epsfig{width=0.34\textwidth,figure=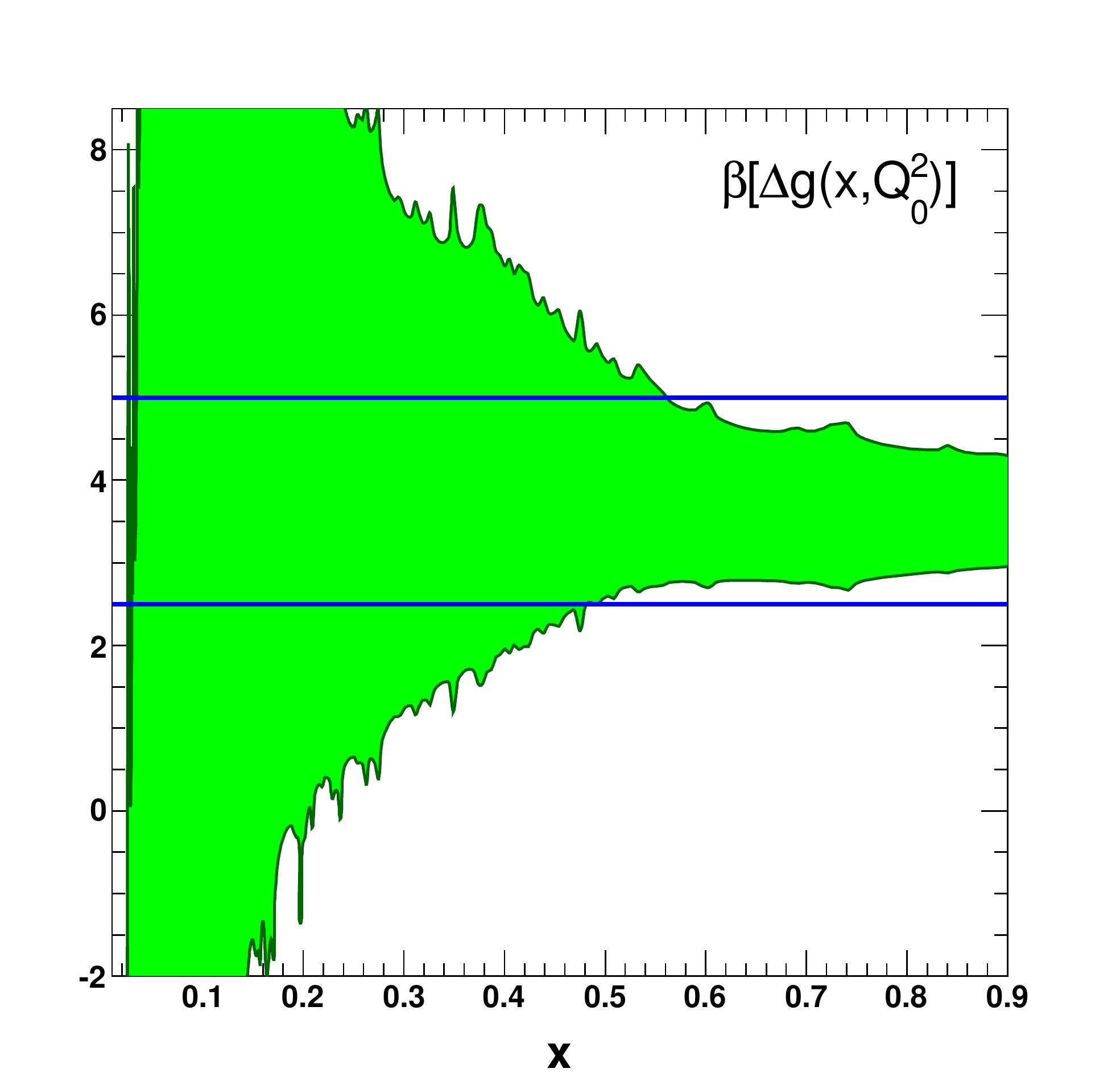}\\
\epsfig{width=0.34\textwidth,figure=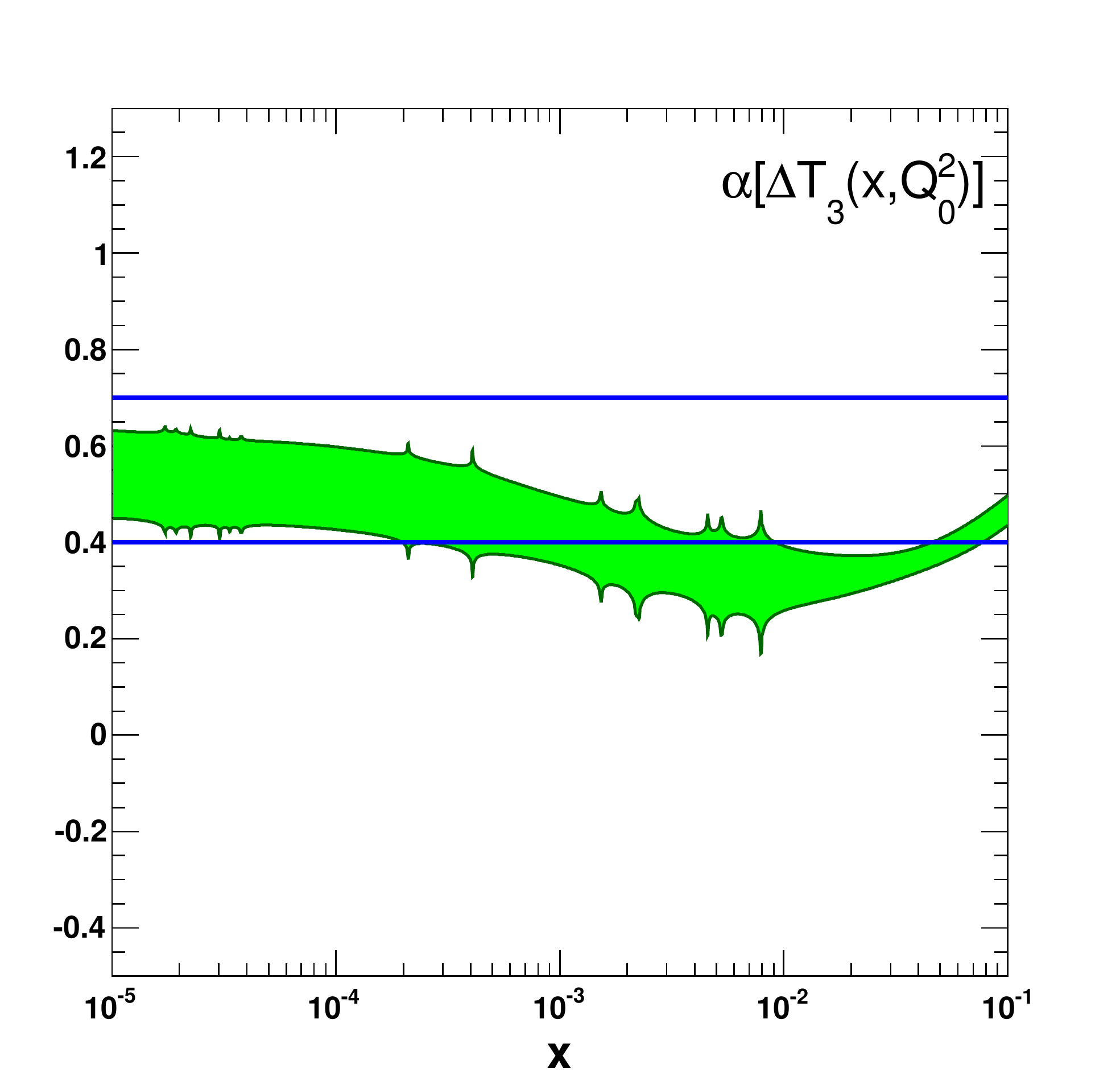}
\epsfig{width=0.34\textwidth,figure=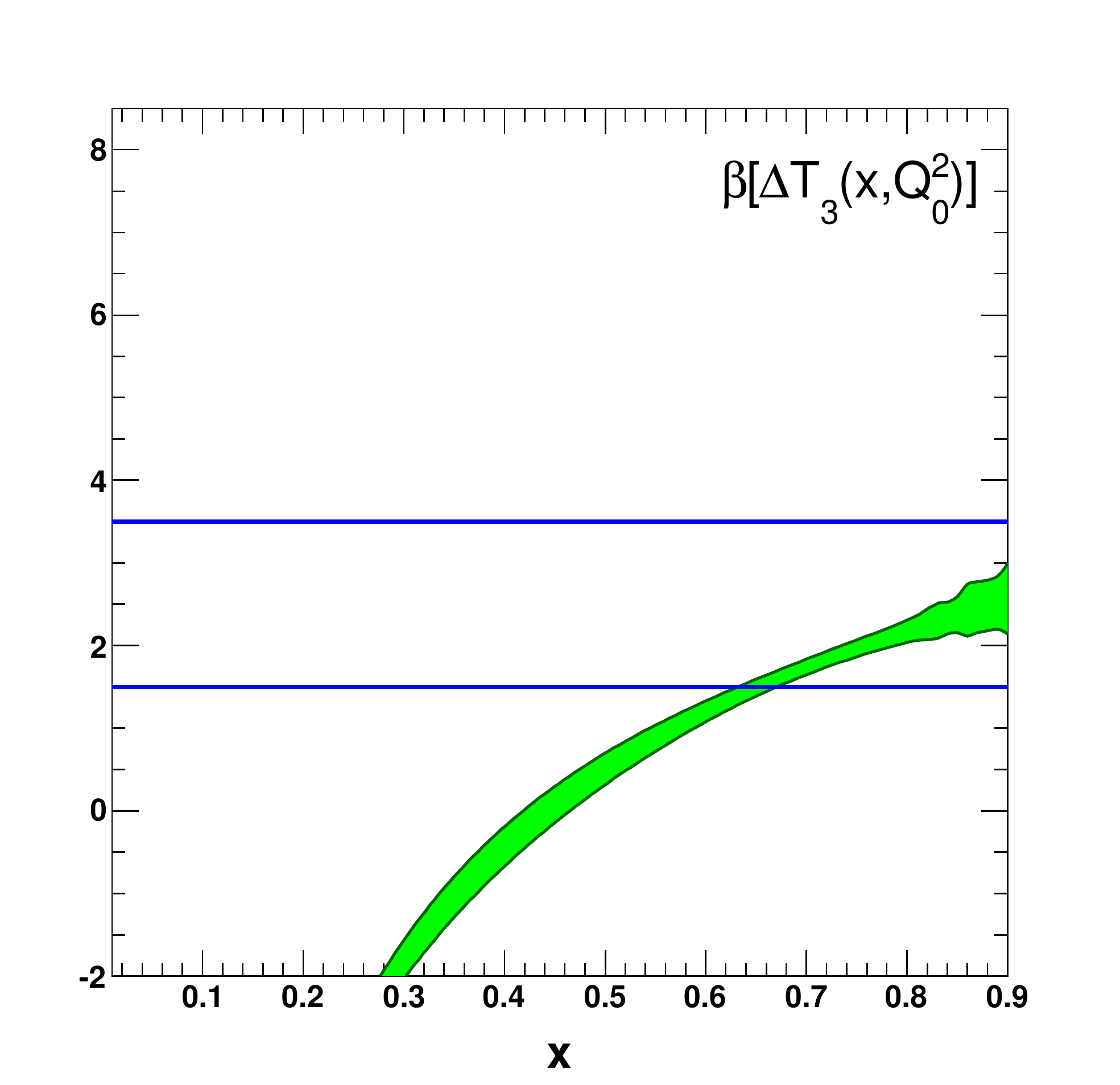}\\
\epsfig{width=0.34\textwidth,figure=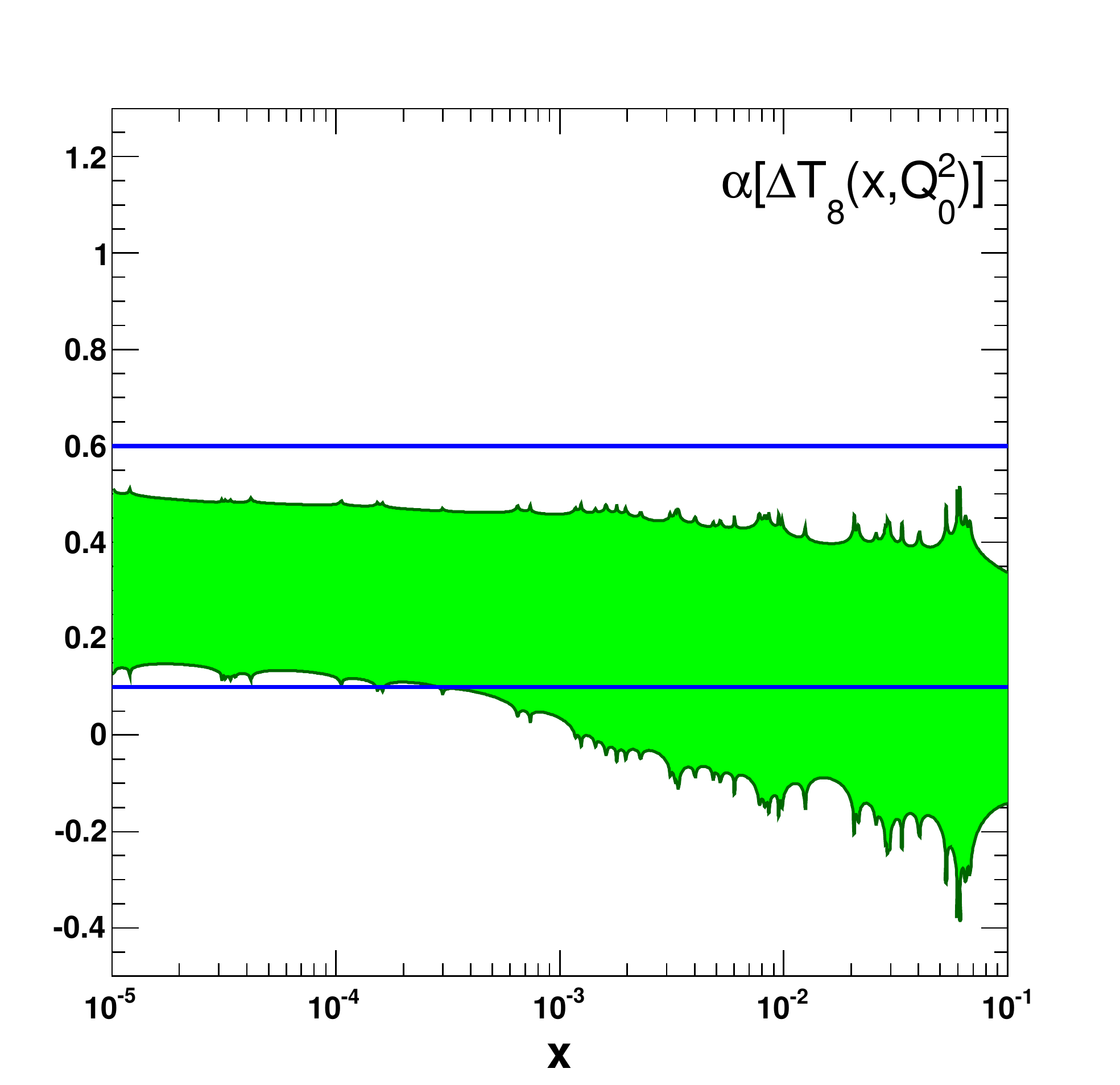}
\epsfig{width=0.34\textwidth,figure=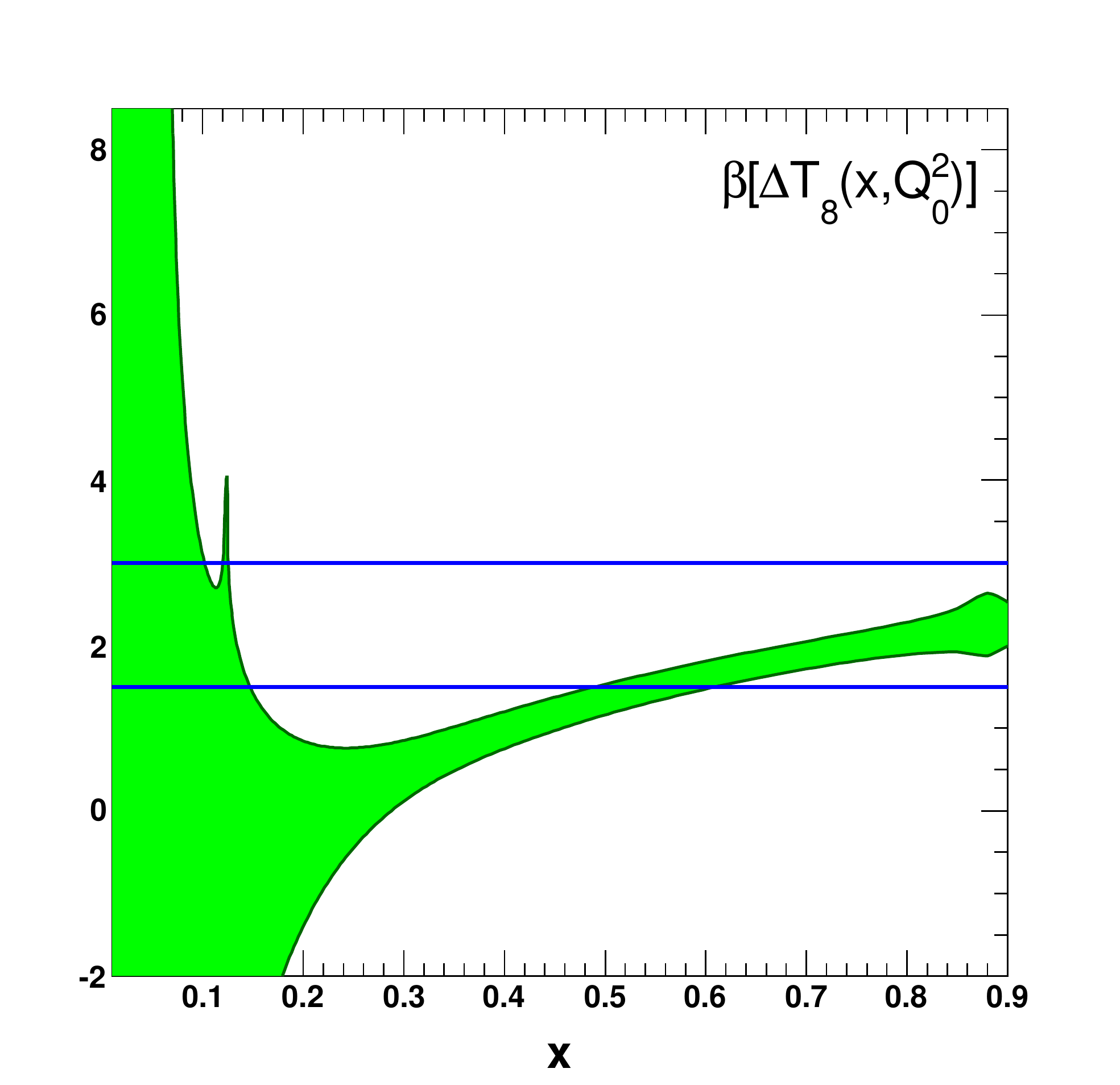}\\
\mycaption{The 68\% confidence level of the distribution of
effective small- and large-$x$ exponents
Eqs.~(\ref{eq:exp2})-(\ref{eq:exp1}) for the default $N_{\mathrm{rep}}=100$ 
replica \texttt{NNPDFpol1.0} set at $Q_0^2=1$ GeV$^2$, plotted as
a functions of $x$. The range of variation of the preprocessing
exponents of Tab.~\ref{tab:prepexps} is also shown in each case
(solid lines).}
\label{fig:prep}
\end{center}
\end{figure}

It is however necessary to check that no bias is introduced by the
preprocessing. We do this following the iterative method
described in Sec.~\ref{sec:minim}. The outcome of the procedure
is the set of exponents Eq.~(\ref{eq:PDFbasisnets}), listed in
Tab.~\ref{tab:prepexps}. The lack of bias with these choices is
explicitly demonstrated  in  Fig.~\ref{fig:prep}, where we plot the 68\%
confidence level of the distribution of 
\begin{align}
&\alpha[\Delta q(x,Q^2)]=\frac{\ln \Delta q(x,Q^2)}{\ln\frac{1}{x}}\mbox{ ,}
\label{eq:exp2}
\\
&\beta[\Delta q(x,Q^2)]=
\frac{ \ln \Delta q(x,Q^2) }{\ln(1-x)}\mbox{ ,}
\label{eq:exp1}
\end{align}
$\Delta q=\Delta\Sigma\mbox{, }\Delta g\mbox{, }
\Delta T_3\mbox{, }\Delta T_8$, for the default 
\texttt{NNPDFpol1.0} $N_{\mathrm{rep}}=100$ replica
set, at $Q^2=Q_0^2=1$ GeV$^2$, and compare them to the ranges of
Tab.~\ref{tab:prepexps}.
It is apparent that  as the endpoints 
$x=0$ and $x=1$ are approached, the uncertainties on both
the small-$x$ and the large-$x$ exponents lie well within the range
of the preprocessing exponents for all PDFs, thus confirming
that the latter do not introduce any bias.

\section{Polarized nucleon structure}
\label{sec:phenoimplications}

We use the \texttt{NNPDFpol1.0} parton set to compute the the first moments 
of the polarized PDFs. These are the quantities of greatest physical interest, 
in that they are directly related to the spin structure of the nucleon,
as discussed in Chap.~\ref{sec:chap1}.
We also assess whether the isotriplet 
first moment determined within our parton set could provide an 
unbiased handle on the strong coupling $\alpha_s$, via the Bjorken sum rule.

\subsection{First moments}
\label{sec:spinmom}

\begin{figure}[t]
\begin{center}
\epsfig{width=0.40\textwidth,figure=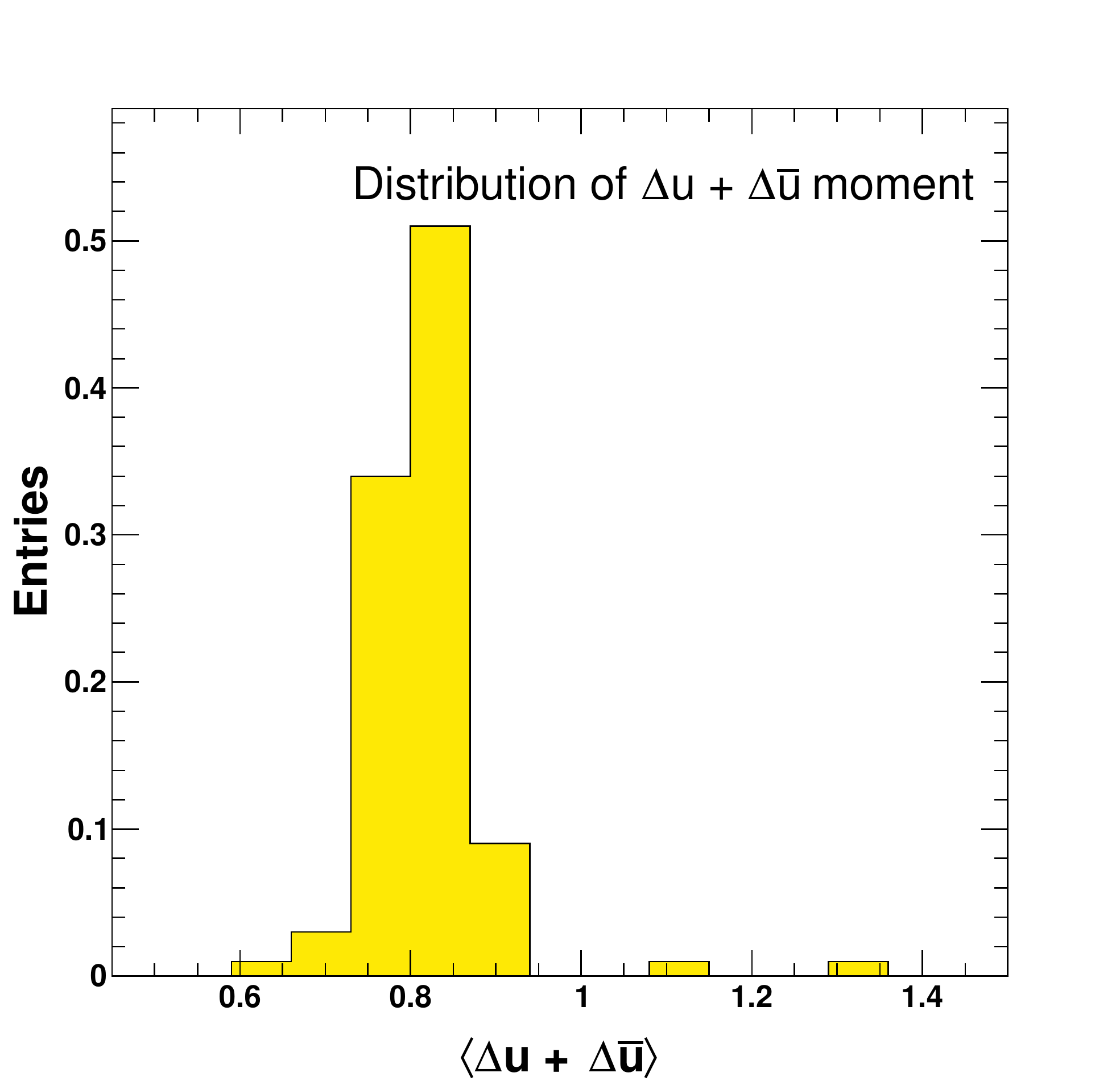}
\epsfig{width=0.40\textwidth,figure=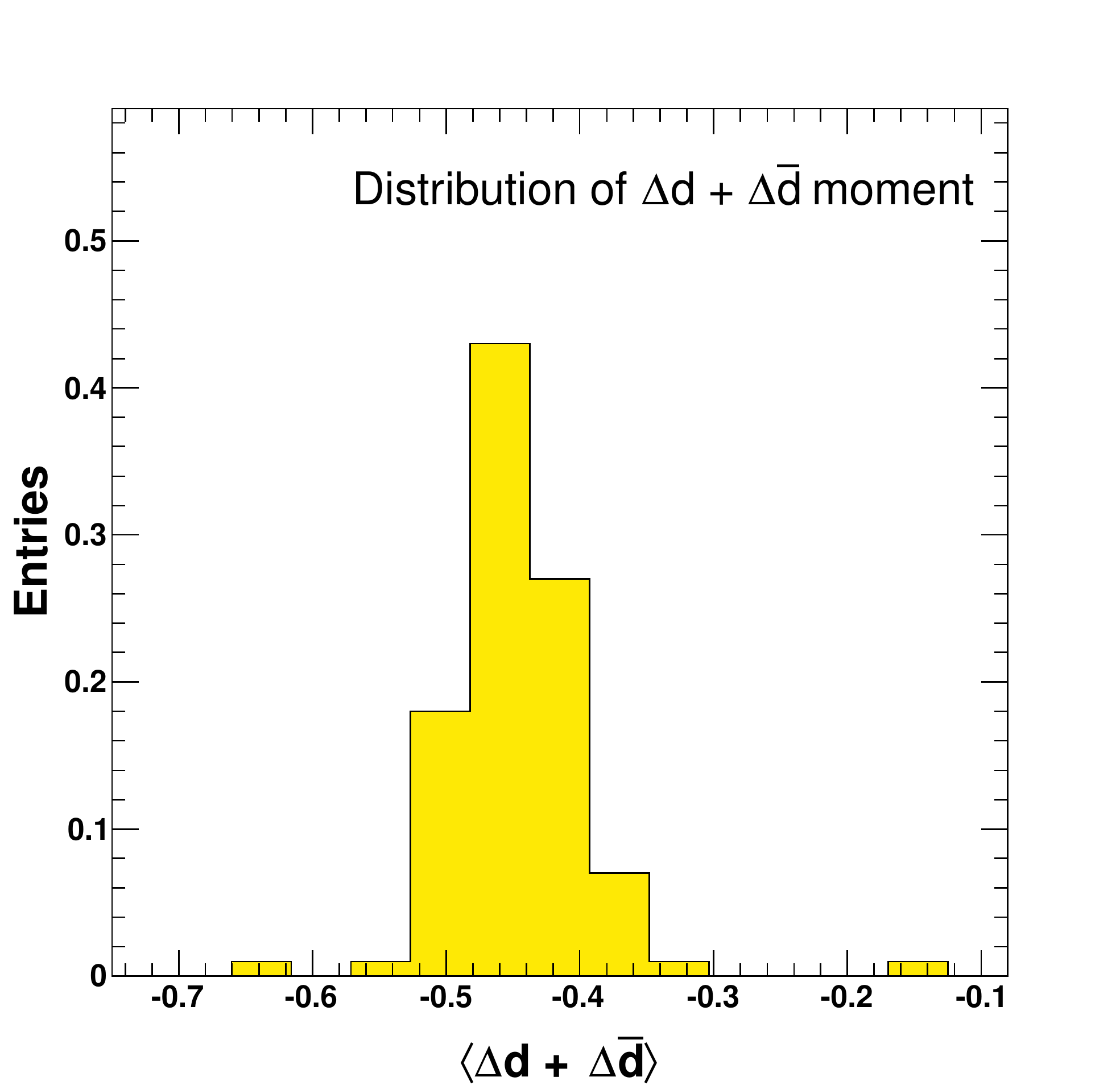}\\
\epsfig{width=0.40\textwidth,figure=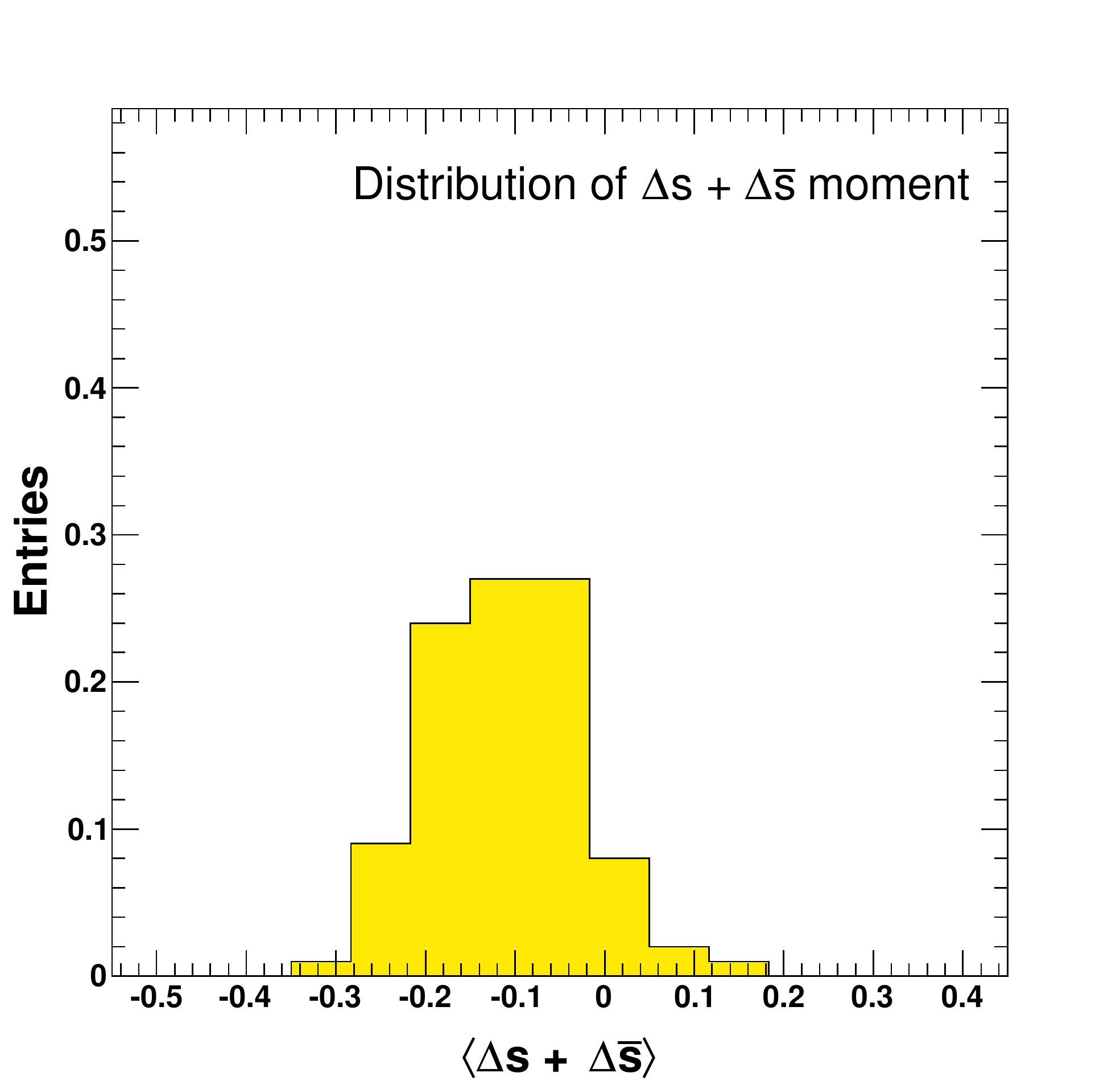}
\epsfig{width=0.40\textwidth,figure=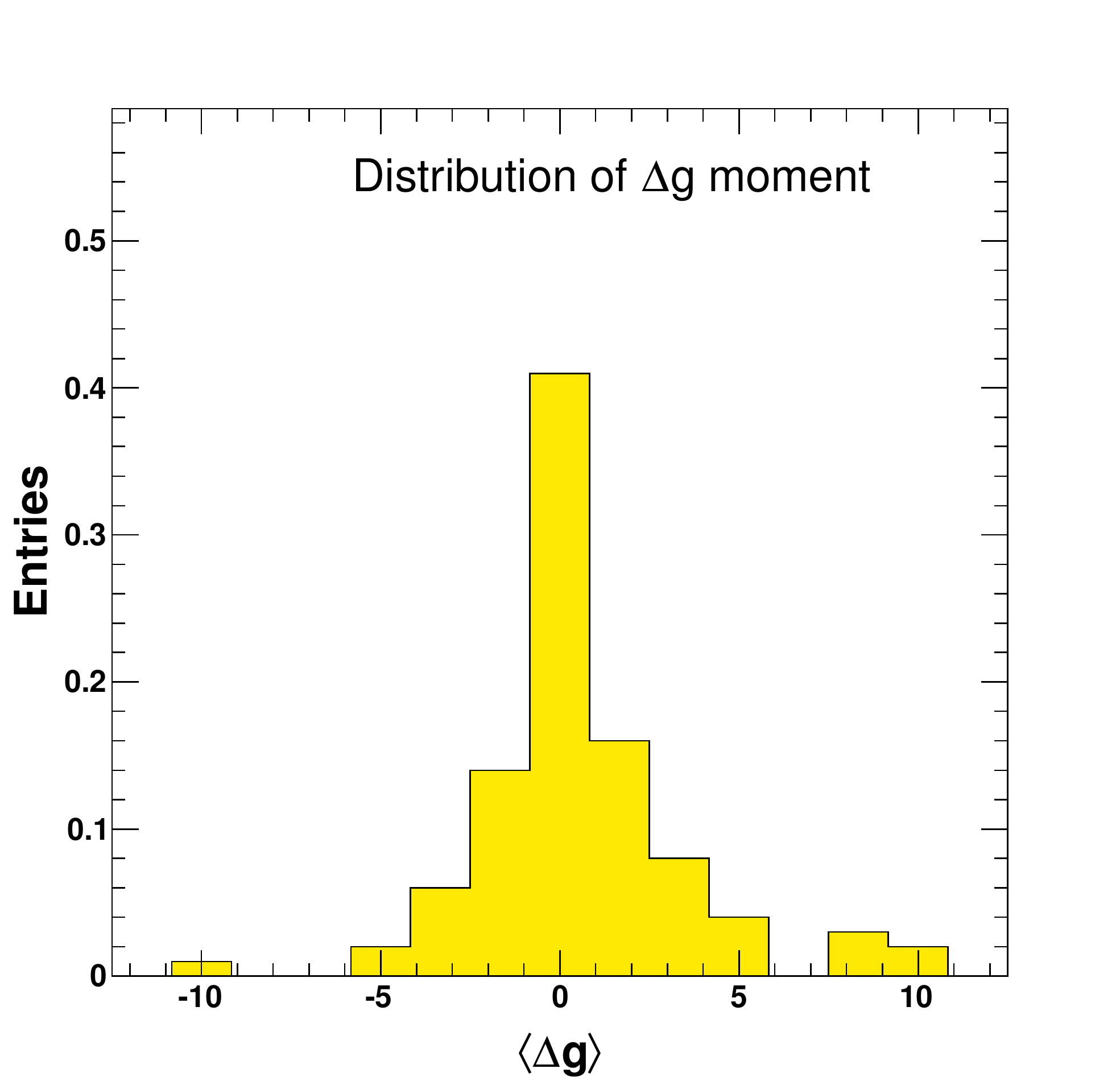}\\
\mycaption{Distribution of the first moments of 
$\Delta u + \Delta\bar{u}$ (top left), $\Delta d + \Delta\bar{d}$ (top right),
$\Delta s + \Delta\bar{s}$ (bottom left) and $\Delta g$ (bottom right)
over a set of $N_\mathrm{rep}=100$  \texttt{NNPDFpol1.0} PDF replicas.}
\label{fig:mom_distr}
\end{center}
\end{figure}

We have computed the first moments
\begin{equation}
\langle \Delta f(Q^2) \rangle 
\equiv
\int_0^1 dx \, \Delta f(x,Q^2)
\label{eq:moments}
\end{equation}
of each light polarized quark-antiquark, $\Delta u +\Delta\bar{u}$,
 $\Delta d +\Delta\bar{d}$, $\Delta s +\Delta\bar{s}$,
and gluon, $\Delta g$, distribution using a sample of
$N_\mathrm{rep}=100$  \texttt{NNPDFpol1.0} PDF replicas.
The histogram  of the distribution of first moments over the replica
sample at $Q_0^2=1$ GeV$^2$ are displayed in Fig.~\ref{fig:mom_distr}: 
they appear to be reasonably approximated by a Gaussian.

The central value and one-sigma uncertainties of the quark-antiquark
combination first moments are listed in Tab.~\ref{tab:spin2}, 
while those of the singlet quark combination
and the gluon are given in Tab.~\ref{tab:spin1}.
Results are compared to those from other
parton sets, namely \texttt{ABFR98}~\cite{Altarelli:1998nb},
\texttt{DSSV08}~\cite{deFlorian:2009vb}, \texttt{AAC08}~\cite{Hirai:2008aj},
\texttt{BB10}~\cite{Blumlein:2010rn} and
\texttt{LSS10}~\cite{Leader:2010rb}. Results from other PDF sets are not
available for all combinations and scales, because public codes only
allow for the computation of first moments in a limited $x$ range, in
particular down to a minimum value of $x$: hence we must rely on
published values for the first moments. In particular, the
\texttt{DSSV08} and \texttt{AAC08} results are shown at 
$Q_0^2=1$ GeV$^2$, while the \texttt{BB10} and \texttt{LSS10} 
results are shown at $Q^2=4$ GeV$^2$. For ease of reference, 
in Tab.~\ref{tab:spin1} the \texttt{NNPDFpol1.0} values 
for both scales are shown.
\begin{table}[t]
\begin {center}
\footnotesize
\begin{tabular}{p{1.5cm}cccccccccccc}
\toprule
         & \multicolumn{4}{c}{$\langle \Delta u +\Delta \bar u \rangle$} 
         & \multicolumn{4}{c}{$\langle \Delta d +\Delta \bar d \rangle$} 
         & \multicolumn{4}{c}{$\langle \Delta s +\Delta \bar s \rangle$}\\
\midrule
         & \textsl{cv} & \textsl{exp} 
         & \textsl{th} & \textsl{tot}
         & \textsl{cv} & \textsl{exp} 
         & \textsl{th} & \textsl{tot}
         & \textsl{cv} & \textsl{exp} 
         & \textsl{th} & \textsl{tot}\\
\midrule
\texttt{NNPDFpol1.0}
         &  0.80 & 0.08 & -- & 0.08
         & -0.46 & 0.08 & -- & 0.08
         & -0.13 & 0.09 & -- & 0.09 \\
\texttt{DSSV08}~\cite{deFlorian:2008mr}
         &  0.82 & 0.01 & 0.01 & 0.02
         & -0.45 & 0.01 & 0.04 & 0.04
         & -0.11 & 0.02 & 0.10 & 0.10 \\
\bottomrule
\end{tabular}
\end {center}
\mycaption{First moments of the polarized quark distributions at $Q_0^2=1$ 
GeV$^2$; \textit{cv} denotes the central value, while \textit{exp} and 
\textit {th} denote uncertainties (see text) whose sum in quadrature 
is given by \textit{tot}.}
\label{tab:spin2}
\end{table}
\begin{table}[t]
\begin{center}
\footnotesize
\begin{tabular}{lcr@{.}lr@{.}lr@{.}lr@{.}lr@{.}lr@{.}lr@{.}lr@{.}l}
\toprule
        & & \multicolumn{8}{c}{$\langle\Delta\Sigma\rangle$} 
          & \multicolumn{8}{c}{$\langle\Delta g\rangle$} \\
\midrule
        & & \multicolumn{2}{c}{\textsl{cv}} & \multicolumn{2}{c}{\textsl{exp}} 
          & \multicolumn{2}{c}{\textsl{th}} & \multicolumn{2}{c}{\textsl{tot}} 
          & \multicolumn{2}{c}{\textsl{cv}} & \multicolumn{2}{c}{\textsl{exp}} 
          & \multicolumn{2}{c}{\textsl{th}} & \multicolumn{2}{c}{\textsl{tot}} \\
\midrule
\multirow{2}*{\texttt{NNPDFpol1.0}} & (1GeV$^2$)     
            &  0 & 22    & 0 & 20  & \multicolumn{2}{c}{---} & 0 & 20 
            & -1 & 2     & 4 & 2   & \multicolumn{2}{c}{---} & 4 & 2 \\
                                    & (4GeV$^2$)     
            &  0 & 18    & 0 & 20  & \multicolumn{2}{c}{---} & 0 & 20 
            & -0 & 9     & 3 & 9   & \multicolumn{2}{c}{---} & 4 & 2 \\
\midrule
\texttt{ABFR98}~\cite{Altarelli:1998nb} 
          & &  0 & 12    & 0 & 05  & \multicolumn{2}{c}{$^{+0.19}_{-0.12}$} 
            & \multicolumn{2}{c}{$^{+0.19}_{-0.13}$} 
            &  1 & 6     & 0 & 4   & 0 & 8                   & 0 & 9 \\
\texttt{DSSV08}~\cite{deFlorian:2008mr} 
          & &  0 & 26    & 0 & 02  & 0 & 13   & 0 & 13
            & -0 & 12    & 0 & 12  & 0 & 06   & 0 & 13 \\
\multirow{2}*{\texttt{AAC08}~\cite{Hirai:2008aj}}  & (\textsl{positive})      
            &  0 & 26    & 0 & 06  & \multicolumn{2}{c}{---} & 0 & 06 
            &  0 & 40    & 0 & 28  & \multicolumn{2}{c}{---} & 0 & 28 \\
                                                   & (\textsl{node})           
            &  0 & 25    & 0 & 07  & \multicolumn{2}{c}{---} & 0 & 07 
            & -0 & 12    & 1 & 78  & \multicolumn{2}{c}{---} & 1 & 78  \\
\texttt{BB10}~\cite{Blumlein:2010rn}    
          & &  0 & 19    & 0 & 08  & 0 & 23   & 0 & 24 
            &  0 & 46    & 0 & 43  & 0 & 004  & 0 & 43 \\
\multirow{2}*{\texttt{LSS10}~\cite{Leader:2010rb}}  & (\textsl{positive})      
            &  0 & 21    & 0 & 03  & \multicolumn{2}{c}{---}  & 0 & 03 
            &  0 & 32    & 0 & 19  & \multicolumn{2}{c}{---}  & 0 & 19 \\
                                                    & (\textsl{node})          
            &  0 & 25    & 0 & 04  & \multicolumn{2}{c}{---}  & 0 & 04 
            & -0 & 34    & 0 & 46  & \multicolumn{2}{c}{---}  & 0 & 46 \\
\bottomrule
\end{tabular}
\end{center}
\mycaption{Same as Tab.~\ref{tab:spin2}, but for the total
singlet quark distribution and the gluon distribution. The \texttt{NNPDFpol1.0}
results are shown both at $Q_0^2=1$ GeV$^2$ and $Q^2=4$ GeV$^2$,  
the \texttt{ABFR98}, \texttt{DSSV08} and \texttt{AAC08} results are 
shown at $Q_0^2=1$ GeV$^2$, and the \texttt{BB10} and \texttt{LSS10} 
are shown at $Q^2=4$ GeV$^2$.}
\label{tab:spin1}
\end{table}

In order to compare the results for first moments shown in
Tabs.~\ref{tab:spin2}-\ref{tab:spin1}, it should be understood that
the uncertainties shown, and sometimes also the central values, 
have somewhat different meanings.

\begin{list}{}{\leftmargin=0pt}

\item {\textbf{NNPDFpol1.0}.}
The \textit{exp} uncertainty, determined as 
the standard deviation of the replica sample, is a pure PDF
uncertainty: it  includes the propagation of the experimental data 
uncertainties and the uncertainty due to the interpolation and extrapolation.

\item {\textbf{ABFR98}.}
The central values were obtained in the AB factorization scheme discussed
in Sec.~\ref{sec:QCDevol}. 
In this scheme, the first moment of the gluon coincides 
with that in the $\overline{\mathrm{MS}}$ scheme used in all other PDF fits
presented here, and thus the corresponding value from 
Ref.~\cite{Altarelli:1998nb} is shown in Tab.~\ref{tab:spin1}. 
Conversely, the singlet first moments in the two schemes are different,
but are related by the simple relation Eq.~(\ref{eq:AB1}).
In Ref.~\cite{Altarelli:1998nb} a value of the singlet axial charge
$a_0$ in the limit of infinite $Q^2$ was also given. 
In the $\overline{\mathrm{MS}}$, the singlet axial charge 
and the first moment of $\Delta\Sigma$
coincide (see Sec.~\ref{sec:QCDevol}), hence we have determined 
$\langle\Delta\Sigma\rangle$ for \texttt{ABFR98} by evolving down to 
$Q^2=1$ GeV$^2$ the value of $a_0(\infty)$ given in 
Ref.~\cite{Altarelli:1998nb}, at NLO and with
$\alpha_s(M_z)=0.118$~\cite{Beringer:1900zz} (the impact of the
$\alpha_s$ uncertainty is negligible). We have checked that
the same result is obtained if $a_0$ is computed as the appropriate
linear combination of $\langle\Delta\Sigma\rangle$ in the AB scheme and the
first moment of $\Delta g$, Eq.~(\ref{eq:gluonanomaly}).
In the \texttt{ABFR98} study, the \textit{exp} uncertainty is the 
Hessian uncertainty on the best fit, and it thus includes the propagated 
data uncertainty. The \textit{th} uncertainty includes the uncertainty 
originated by neglected higher orders (estimated by renormalization and
factorization scale variations), higher twists, position of heavy
quark thresholds, value of the strong coupling, violation of SU(3), and finally
uncertainties related to the choice of functional form, estimated by
varying the functional form. This latter source of theoretical
uncertainty corresponds to interpolation and extrapolation
uncertainties which are included in the \textit{exp} for \texttt{NNPDFpol1.0}. 

\item {\textbf{DSSV08}, \textbf{BB10}.}
The central value is obtained by
computing the first moment integral of the best-fit with a fixed
functional form restricted to the data region, and then
supplementing it with a contribution due to the extrapolation in the
unmeasured, small-$x$, region. The \textit{exp} uncertainty in the
table is the Hessian uncertainty given by \texttt{DSSV08} or \texttt{BB10} 
on the moment in the measured region, and it thus includes the propagated
data uncertainty. In both cases, we have determined the \textit{th}
uncertainty shown in the table as the difference between the full first
moment quoted by \texttt{DSSV08} or \texttt{BB10}, and the first moment in the
measured region. It is thus the contribution from the extrapolation
region, which we assume to be $100\%$ uncertain. In both cases,
we have computed  the truncated first moment in the measured region
using publicly available codes, and checked that it coincides with
the values quoted by \texttt{DSSV08} and \texttt{BB10}.

\item {\textbf{AAC08}.}
The central value is obtained by computing the first moment
integral of the best-fit with a fixed functional form, and the
\textit{exp} uncertainty is the Hessian uncertainty on it. However,
\texttt{AAC08} uses the tolerance~\cite{Pumplin:2002vw} criterion for
the determination of Hessian uncertainties, which rescales the
$\Delta\chi^2=1$ region by a suitable factor, in order to
effectively keep into account also interpolation errors. Hence, the
\textit{exp} uncertainties include propagated data uncertainties, as
well as uncertainties on the PDF shape.

\item{\textbf{LSS10}.}
The central value is obtained by computing the first
moment of the
best fit with a fixed functional form, and the \textit{exp} uncertainty is
the Hessian uncertainty on it. Hence it includes the
propagated data uncertainty.

\end{list}

In all cases, the total uncertainty is computed as the sum in quadrature of the
\textit{exp} and \textit{th} uncertainties. Roughly speaking, for
\texttt{LSS10} this includes only the data uncertainties; 
for \texttt{DSSV08}, and \texttt{BB10}
it also includes extrapolation uncertainties; 
for \texttt{AAC08} interpolation uncertainties;
for \texttt{NNPDFpol1.0} both extrapolation and
interpolation uncertainties; and for \texttt{ABFR98} all of the
above, but also theoretical (QCD) uncertainties. 
For \texttt{LSS10} and \texttt{AAC08},
we quote the results obtained from two different fits, both
assuming positive- or node-gluon PDF: their spread gives a feeling for
the missing uncertainty due to the choice of functional form. Remind
that the \texttt{AAC08} results correspond to their Set B which includes,
besides DIS data, also RHIC $\pi^0$ production data; the \texttt{DSSV08} fit
also includes, on top of these, RHIC jet data and semi-inclusive DIS
data; \texttt{LSS10} includes, beside DIS, also semi-inclusive DIS data. 
All other sets are based on DIS data only.

Coming now to a comparison of results, we see that for the 
singlet first moment  $\langle\Delta\Sigma \rangle$ the \texttt{NNPDFpol1.0} 
result is consistent within uncertainties with that of other groups. 
The uncertainty on the \texttt{NNPDFpol1.0} result is
comparable, if somewhat larger, to that found whenever the
extrapolation uncertainty has been included. 
For individual quark flavors we find excellent agreement in the central
values obtained between \texttt{NNPDFpol1.0} and \texttt{DSSV08}, see
Tab.~\ref{tab:spin2}; the \texttt{NNPDFpol1.0}
uncertainties are rather larger, but this could also be due to the
fact that the data set included in 
\texttt{DSSV08} is sensitive to quark-antiquark separation. 

For the gluon first moment $\langle\Delta g\rangle$, 
the \texttt{NNPDFpol1.0} result is characterized by an
uncertainty which is much larger than that of any other determination:
a factor of three or four larger than \texttt{ABFR98} and \texttt{AAC08}, 
ten times larger than \texttt{BB10}, and twenty times larger than 
\texttt{DSSV08} and \texttt{LSS10}. It is
compatible with zero within this large uncertainty.
We have seen that for the quark singlet, the \texttt{NNPDFpol1.0}
uncertainty is similar to that of groups which include an estimate of
extrapolation uncertainties. In order to assess the impact
of the extrapolation uncertainty for the  gluon, we have
computed the gluon first moment truncated in the region
$x\in[10^{-3},1]$:
\begin{equation}
\int_{10^{-3}}^1dx\, \Delta g(x, Q^2=1 \mathrm{GeV}^2) = -0.26 \pm 1.19 
\,\mbox{,}
\end{equation}
to be compared with the result of Tab.~\ref{tab:spin1}, which is
larger by almost a factor four. 

We must conclude that the experimental status of the gluon first
moment is still completely uncertain, unless one is willing to make
strong theoretical assumptions on the behavior of the polarized gluon
at small $x$, and that previous different conclusions were affected by
a significant underestimate of the impact of the bias in the choice
of functional form, in the data and especially in the extrapolation
region. Because of the large uncertainty related to the extrapolation
region, only low-$x$ data can improve this situation, such as those
which could be collected at a high
energy Electron-Ion Collider~\cite{Deshpande:2005wd,Boer:2011fh},
as we will show in Chap.~\ref{sec:chap4}.

\subsection{The Bjorken sum rule}
\label{sec:bjorken}

The Bjorken sum rule presented in Sec.~\ref{sec:sumrule},
Eq.~(\ref{eq:bjorken1})
\begin{equation}
\Gamma_1^{\mathrm{NS}}\equiv\Gamma_1^p\left(Q^2\right) - \Gamma_1^n\left(Q^2\right) 
= 
\frac{1}{6}
\Delta C_{\mathrm{NS}} (\alpha_s(Q^2)) a_3 
\label{eq:bjorken}
\end{equation}
potentially provides a theoretically very accurate handle on the strong coupling
$\alpha_s$. We recall that we have defined the first moment of the proton
(neutron) structure function $\Gamma_1^{p,n}(Q^2)$ in
Eq.~(\ref{eq:gammapn}), 
the first moment of the nonsinglet triplet PDF combination
in the first relation of Eqs.~(\ref{eq:a3a8})
and $\Delta C_{\mathrm{NS}}(\alpha_s(Q^2))$ is 
the first moment of the corresponding coefficient function,
which is known up to three loops.
In principle, the truncated isotriplet first moment
\begin{equation}
\Gamma_1^{\mathrm{NS}}\left(Q^2,x_{\mathrm{min}}\right) 
\equiv
\int_{x_{\mathrm{min}}}^1 dx 
\left[g_1^p\left( x,Q^2 \right) - g_1^n\left( x,Q^2 \right) \right]
\label{eq:bjorken-cut}
\end{equation}
can be extracted from data without any theoretical assumption. Given a 
measurement of $\Gamma_1^{\mathrm{NS}}\left(Q^2,0\right)$ at a certain scale, the 
strong coupling can then be
extracted from Eq.~(\ref{eq:bjorken}) using the value of $a_3$ from
$\beta$ decays, while given a measurement of 
$\Gamma_1^{\mathrm{NS}}\left(Q^2,0\right)$ at two scales, both $a_3$ and the value
of $\alpha_s$ can be extracted simultaneously.

In Ref.~\cite{DelDebbio:2009sq}, $a_3$ and $\alpha_s$ 
where simultaneously determined
from a set of nonsinglet truncated moments, both the first and higher
moments, by exploiting the scale
dependence of the latter~\cite{Forte:1998nw}, with the result
$a_3=1.04\pm0.13$ and $\alpha_s(M_z)=0.126^{+0.006}_{-0.014}$, where
the uncertainty is dominated by the data, interpolation and
extrapolation, but also includes theoretical QCD
uncertainties. In this reference, truncated moments were determined
from a neural network interpolation of existing data, sufficient for 
a computation of moments at any scale. However, because the small-$x$
behavior of the structure function is only weakly constrained by
data, the $x\to0$ extrapolation was done by assuming a powerlike
Regge behavior~\cite{Close:1994he}.

The situation within \texttt{NNPDFpol1.0} can be understood by exploiting
the PDF determination in which $a_3$ is not fixed by the triplet sum
rule, discussed in Sec.~\ref{sec:srres}. Using the results of this
determination, we find 
\begin{equation}
a_3=\int_0^1dx\,\Delta T_3 (x, Q^2) = 1.19 \pm 0.22
\,\mbox{.}
\label{eq:a3exp}
\end{equation}
The uncertainty is about twice that of the determination of
Ref.~\cite{DelDebbio:2009sq}. As mentioned, the latter was obtained 
from a neural network parametrization of the data with no theoretical
assumptions, and based on a methodology which is quite close to that
of the \texttt{NNPDFpol1.0} PDF determination discussed here, the only
difference being the assumption of Regge behavior in order to perform
the small-$x$ extrapolation. This strongly suggests that, as in the
case of the gluon distribution discussed above, the uncertainty on the
value Eq.~(\ref{eq:a3exp}) is dominated by the small-$x$ extrapolation.

\begin{figure}[t]
\begin{center}
\epsfig{width=0.40\textwidth,figure=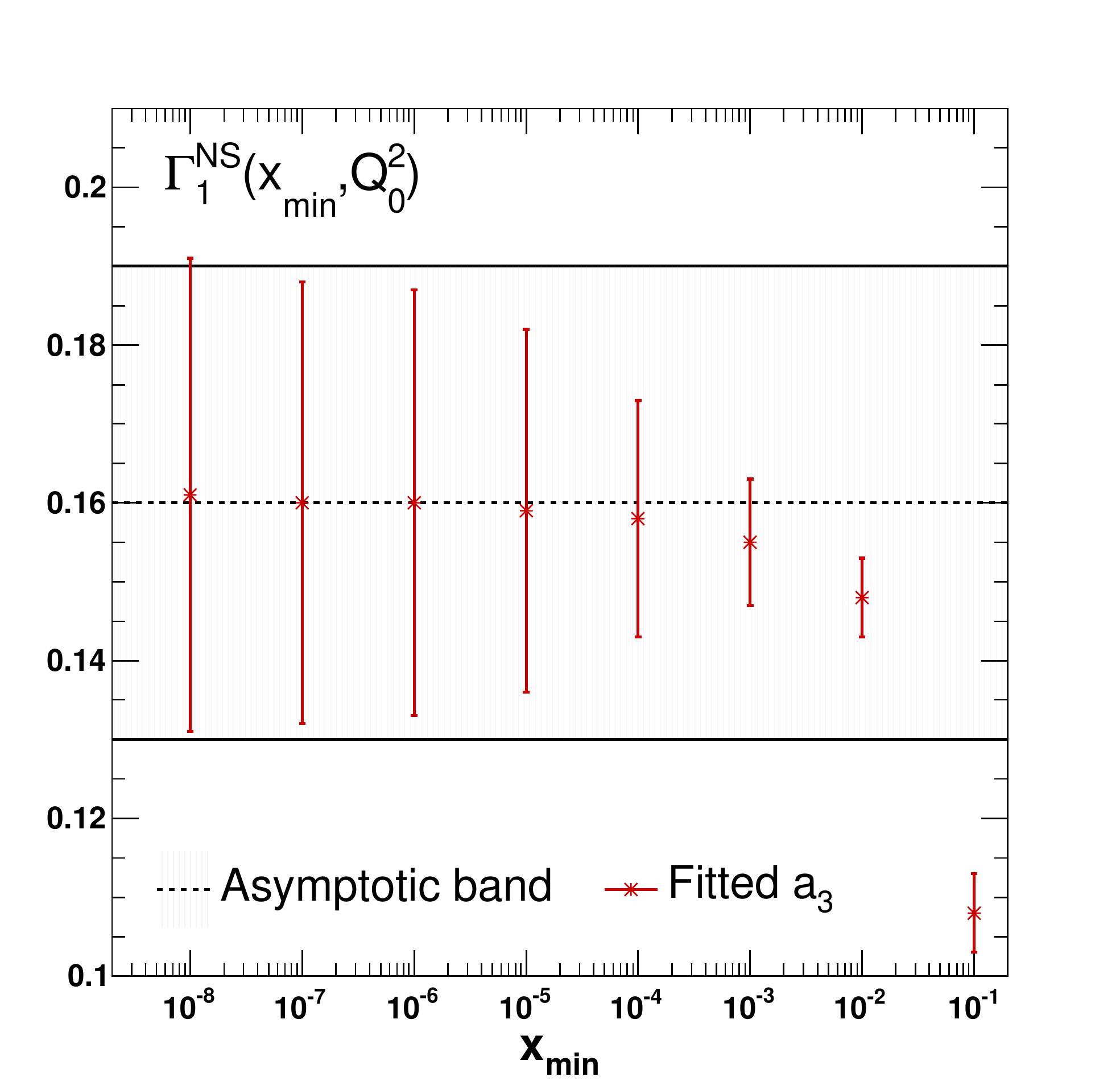}
\epsfig{width=0.40\textwidth,figure=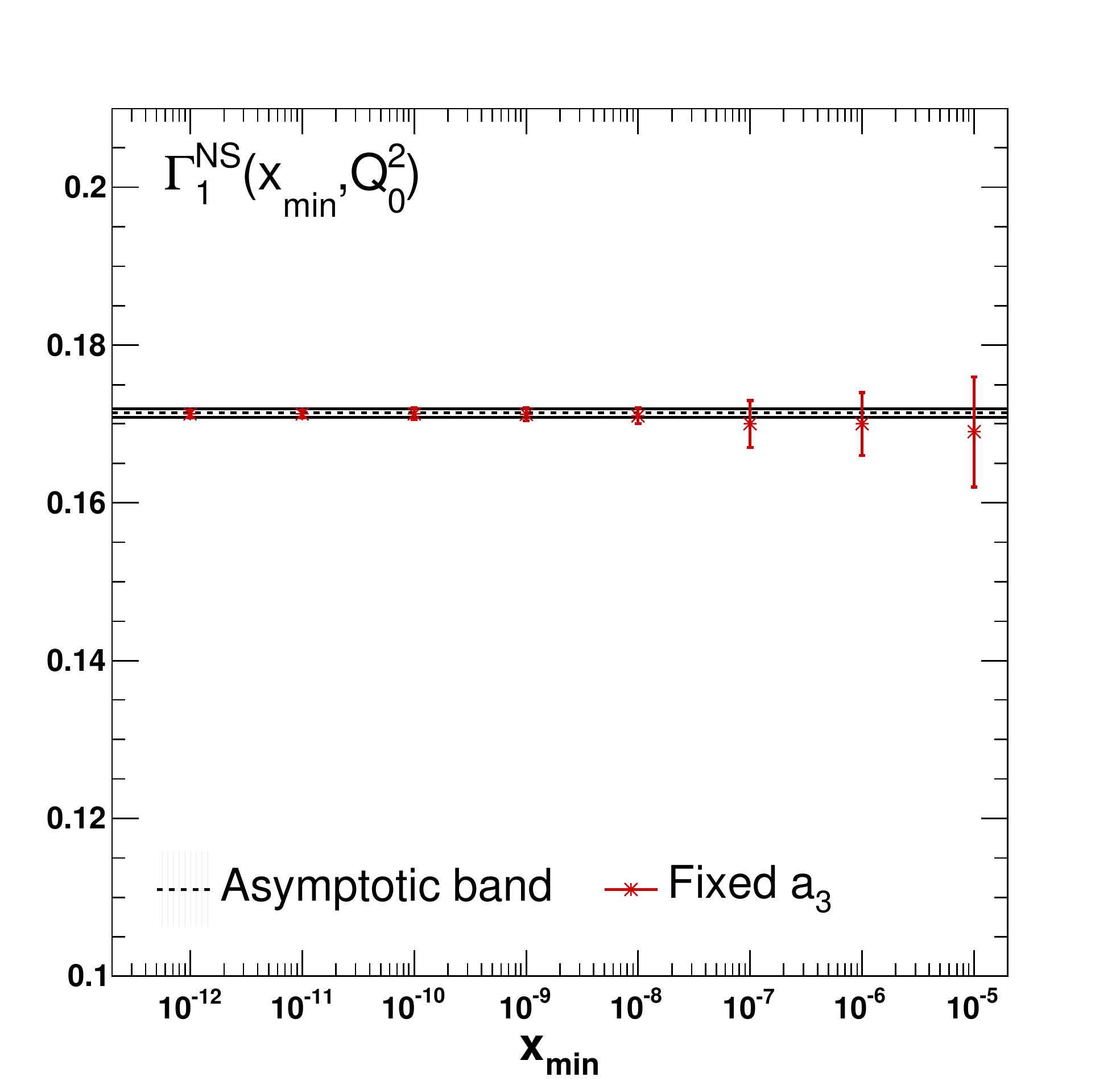}
\mycaption{The truncated 
Bjorken sum rule $\Gamma_1^{\mathrm{NS}}\left(Q^2,x\right)$ 
Eq.~(\ref{eq:bjorken-cut}) plotted as a function of $x$ for $Q^2=1$ GeV$^2$,
for the fit with free $a_3$ (left) and for the reference fit with $a_3$ fixed 
to the value Eq.~(\ref{eq:hypdecayconst}) (right). In the left plot,
the shaded band corresponds to the asymptotic value of the truncated sum rule, 
Eq.~(\ref{eq:extriplet}), while in the right plot it corresponds to the
experimental value Eq.~(\ref{eq:hypdecayconst}).}
\label{fig:bjsr}
\end{center}
\end{figure}

To study this effect, in Fig.~\ref{fig:bjsr} 
we plot the value of the truncated Bjorken sum
rule $\Gamma_1^{\mathrm{NS}}\left(Q^2,x_{\mathrm{min}}\right)$
Eq.~(\ref{eq:bjorken-cut}) as a function 
of the lower limit of integration $x_{\mathrm{min}}$ at $Q_0^2=1$ GeV$^2$, 
along with the asymptotic value 
\begin{equation}
\Gamma_1^{\mathrm{NS}}\left(1 \mathrm{GeV}^2,0\right)= 0.16 \pm 0.03
\label{eq:extriplet}
\end{equation}
which at NLO corresponds to the value of $a_3$ given by Eq.~(\ref{eq:a3exp}).
As a consistency check, we also show the same plot for our baseline fit, 
in which $a_3$ is fixed by the sum rule to the value
Eq.~(\ref{eq:hypdecayconst}). It is clear that indeed the uncertainty is
completely dominated by the small $x$ extrapolation.

We conclude that a determination of $\alpha_s$ from the Bjorken sum
rule is not competitive unless one is willing to make assumptions on
the small $x$ behavior of the nonsinglet structure function in the
unmeasured region. Indeed, it is clear that a determination based on
\texttt{NNPDFpol1.0} would be affected by an uncertainty which is
necessarily larger than that found in Ref.~\cite{DelDebbio:2009sq},
which is already not competitive. The fact that a determination
of $\alpha_s$ from the Bjorken sum rule is not competitive due to
small $x$ extrapolation ambiguities was already pointed out in
Ref.~\cite{Altarelli:1998nb}, where values of $a_3$ and $\alpha_s$
similar to those of Ref.~\cite{DelDebbio:2009sq} were obtained.

\chapter{Polarized PDFs at an Electron-Ion Collider}
\label{sec:chap4}

In this Chapter, we investigate the potential impact of inclusive DIS data
from a future Electron-Ion Collider (EIC) on the determination of polarized 
parton distributions. After briefly motivating our study in 
Sec.~\ref{sec:motivation}, we illustrate in Sec.~\ref{sec:pseudodataEIC} 
which EIC pseudodata sets we use in our analysis and 
in Sec.~\ref{sec:fitoptEIC} how the fitting procedure 
described in Sec.~\ref{sec:minim} needs to be optimized.
Resulting PDFs are presented in Sec.~\ref{sec:resultsEIC}, and they are
compared to \texttt{NNPDFpol1.0} throughout.
Finally, in Sec.~\ref{sec:phenoEIC} we reassess the computation 
of their first moments and we 
give an estimate of the charm contribution to the $g_1$ structure function.
The analysis presented in this Chapter is mostly based on 
Ref.~\cite{Ball:2013tyh}.

\section{Motivation}
\label{sec:motivation}

As already noticed several times in this Thesis,
the bulk of experimental information on longitudinally polarized proton 
structure comes from inclusive, neutral-current DIS,
which allows one to obtain information on the light quark-antiquark combinations
$\Delta q^+\equiv\Delta q +\Delta\bar{q}$, $q=u,d,s$ 
and on the gluon distribution $\Delta g$.
However, presently available DIS data cover only a small 
kinematic region of momentum fractions and energies $(x,Q^2)$, as shown in
fig.~\ref{fig:dataplot}.
On the one hand, the lack of experimental information for 
$x\lesssim10^{-3}$ prevents a reliable determination of polarized PDFs 
at small-$x$. Hence, their first moments
will strongly depend on the functional form one assumes
for PDF extrapolation to the unmeasured $x$ region.
On the other hand, the gluon PDF, which is determined 
by scaling violations, is only 
weakly constrained, due to the small lever-arm in $Q^2$
of the experimental data. Both these limitations were emphasized in 
Chap.~\ref{sec:chap2}, when we have presented the first unbiased 
set of polarized PDFs, \texttt{NNPDFpol1.0}.

For these reasons, despite many efforts, both experimental and theoretical, 
the size of the polarized  gluon contribution to the nucleon spin 
is still largely uncertain, as demonstrated in Sec.~\ref{sec:phenoimplications}
and in Ref.~\cite{deFlorian:2011ia}.
In Sec.~\ref{sec:generalstrategy}, we mentioned that other processes,
receiving leading partonic contributions from gluon-initiated suprocesses,
may provide direct information on the polarized gluon PDF. They include
open-charm photoproduction data from COMPASS~\cite{Adolph:2012ca} and
polarized hadron collider measurements from 
RHIC~\cite{Adare:2008qb,Adamczyk:2012qj,Adamczyk:2013yvv,Adare:2008aa,Adare:2010cc}, 
specifically semi-inclusive particle and jet production data. 
We explicitly assess their impact on the determination of $\Delta g$ in 
Chap.~\ref{sec:chap5}, but we note here that all these data are 
restricted to the medium- and large-$x$ region.

An EIC~\cite{Deshpande:2005wd,Boer:2011fh,Accardi:2012hwp}, with
polarized lepton and hadron beams, would allow for a widening of the
kinematic region comparable to the one achieved in the unpolarized case with 
the DESY-HERA experiments H1 and ZEUS~\cite{Aaron:2009aa}.
Note that a Large Hadron-electron Collider 
(LHeC)~\cite{AbelleiraFernandez:2012cc} would not have the option of
polarizing the hadron beam.
The potential impact of the EIC on the knowledge of the nucleon 
longitudinal spin structure
has been quantitatively assessed
in a recent study~\cite{Aschenauer:2012ve}, in which 
projected neutral-current inclusive DIS and semi-inclusive DIS (SIDIS) 
artificial data were added to the \texttt{DSSV+} polarized PDF
determination~\cite{deFlorian:2011ia}; this study was then extended 
by also providing an estimate of the impact of charged-current 
inclusive DIS pseudo-data on the 
polarized quark-antiquark separation in Ref.~\cite{Aschenauer:2013iia}.
In view of the fact that a substantially larger gluon uncertainty 
is found in \texttt{NNPDFpol1.0} in comparison to previous PDF
determinations~\cite{deFlorian:2009vb,Leader:2010rb,Hirai:2008aj,Blumlein:2010rn}, 
it is worth repeating the study of the impact of EIC
data, but now using NNPDF methodology. 
This is the goal of the study in the present Chapter.

\section{Inclusive DIS pseudodata from an Electron-Ion Collider}
\label{sec:pseudodataEIC}

The realization of an EIC has been proposed for two independent designs so far:
the electron Relativistic Heavy Ion Collider (eRHIC) at  
Brookhaven National Laboratory (BNL)~\cite{BNL:eRHIC} 
and the Electron Light Ion Collider (ELIC)
at Jefferson Laboratory (JLab)~\cite{JLAB:ELIC}.
In both cases, a staged upgrade of the existing facilities
has been planned~\cite{Deshpande:2005wd,Boer:2011fh,Accardi:2012hwp}, so that
an increased center-of-mass energy would be available at each stage.
Concerning the eRHIC option of an EIC~\cite{BNL:eRHIC},
first measurements would be taken by colliding the 
present RHIC proton beam of energy $E_p=100-250$~GeV
with an electron beam of energy $E_e=5$ GeV,
while a later stage envisages electron beams with energy up to $E_e=20$ GeV.

In order to quantitatively assess the impact of future EIC measurements
on the determination of polarized PDFs, 
we have supplemented our QCD analysis presented in Chap.~\ref{sec:chap2} 
and Ref.~\cite{Ball:2013lla}
with  DIS pseudodata from Ref.~\cite{Aschenauer:2012ve}. 
They consist of three sets of data points at different possible
eRHIC electron and proton beam energies, as discussed above. 
These pseudodata were produced by running
the \texttt{PEPSI} Monte Carlo (MC) generator~\cite{Mankiewicz:1991dp}, assuming
momentum transfer $Q^2>1$ GeV$^2$, squared invariant mass 
of the virtual photon-proton system $W^2>10$ GeV$^2$ and fractional energy
of the virtual photon $0.01\leq y \leq 0.95$; they are provided
in five (four)  bins per logarithmic decade in $x$ ($Q^2$). 

For each data set,
the $Q^2$ range spans the values from $Q^2_{\mathrm{min}}=1.39$ GeV$^2$ to 
$Q^2_{\mathrm{max}}=781.2$ GeV$^2$, while the accessible values of momentum
fraction $x=Q^2/(sy)$ depend on the available center-of-mass energy, 
$\sqrt{s}$.
In Tab.~\ref{tab:kintab}, we summarize, for each data set, the number of 
pseudodata $N_{\mathrm{dat}}$, the electron and proton beam 
energies $E_e$, $E_p$, the corresponding center-of-mass energies $\sqrt{s}$,
and the smallest and largest accessible values in the momentum fraction range,
$x_{\mathrm{min}}$  and $x_{\mathrm{max}}$ respectively.
\begin{table}[t]
\footnotesize
\centering
\begin{tabular}{llcccccc}
\toprule
Experiment& Set& $N_{\mathrm{dat}}$& 
$E_e\times E_p$ [GeV] & $\sqrt{s}$ [GeV]&
$x_{\mathrm{min}}$ & $x_{\mathrm{max}}$ & $\langle \delta g_1 \rangle$\\ 
\midrule
EIC 
& EIC-G1P-1 & 56 & $5\times 100$ & $44.7$ & 
$8.2\times 10^{-4}$ & $0.51$& $0.010$\\
& EIC-G1P-2 & 63 & $5\times 250$ & $70.7$ & 
$3.2\times 10^{-4}$ & $0.51$ & $0.032$ \\
& EIC-G1P-3 & 61 & $20\times 250$ & $141$ & 
$8.2\times 10^{-5}$ & $0.32$ & $0.042$\\
\bottomrule
\end{tabular}
\mycaption{The three  EIC pseudodata sets~\cite{Aschenauer:2012ve}. 
For each set we show the number of points
$N_{\mathrm{dat}}$, the electron and proton beam energies $E_e$  and $E_p$,
the center-of-mass energy $\sqrt{s}$, the kinematic coverage
in the momentum fraction $x$, and  the average absolute statistical
uncertainty $\langle \delta g_1 \rangle$.}
\label{tab:kintab}
\end{table} 

\begin{figure}[t]
\begin{center}
\epsfig{width=0.5\textwidth,figure=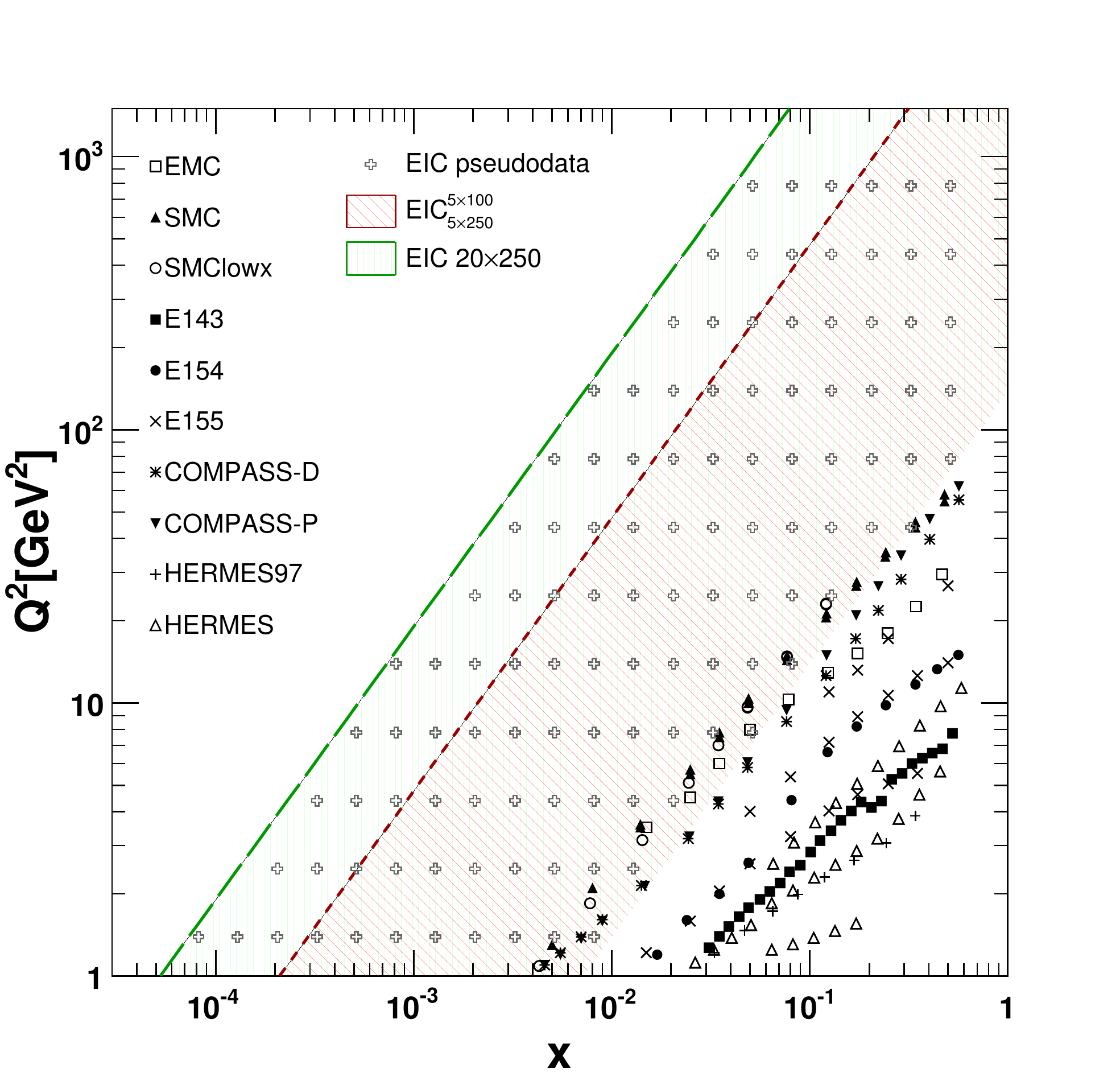}
\mycaption{Kinematic coverage in the $(x,Q^2)$ plane
for the fixed-target experimental data included in the \texttt{NNPDFpol1.0} 
polarized parton fit
and the  EIC pseudodata from~\cite{Aschenauer:2012ve}.
The shaded bands show the expected kinematic reach of
each of the two EIC scenarios discussed in the text.}
\label{fig:kinplotEIC}
\end{center}
\end{figure}

The kinematic coverage of the EIC pseudodata
is displayed in Fig.~\ref{fig:kinplotEIC} together with
the fixed-target DIS data points dscussed in Chap.~\ref{sec:chap2}. 
The dashed regions show the overall kinematic reach of the EIC data
with the two electron beam energies $E_e=5$ GeV or
$E_e=20$ GeV, corresponding to each of the two stages at eRHIC.
It is apparent from Fig.~\ref{fig:kinplotEIC} that EIC data
will extend the kinematic coverage significantly, even for the lowest
center-of-mass energy. 
In particular, hitherto unreachable small-$x$ values, down to $10^{-4}$,
will be attained, thereby leading to a significant reduction of the 
uncertainty in the low-$x$ extrapolation region.
Furthermore, the increased lever-arm
in $Q^2$, for almost all values of $x$ should allow for much more stringent
constraints on $\Delta g(x,Q^2)$ from scaling violations.  

The ratio $g_1(x,Q^2)/F_1(x,Q^2)$ is provided 
in Ref.~\cite{Aschenauer:2012ve} as the inclusive DIS observable,
whose relation with the experimentally measured asymmetries 
was discussed in Sec.~\ref{sec:phenopol}.
The generation of pseudodata
assumes a \textit{true} underlying set of parton distributions: in
Ref.~\cite{Aschenauer:2012ve} these are taken to be
\texttt{DSSV+}~\cite{deFlorian:2011ia} and \texttt{MRST}~\cite{Martin:2002aw} 
polarized and unpolarized PDFs respectively.
Uncertainties are then determined assuming an integrated
luminosity of $10$ fb$^{-1}$, which 
corresponds to a few months operations for the anticipated
luminosities for eRHIC~\cite{BNL:eRHIC}, and a $70\%$ beam
polarization.
Because the \texttt{DSSV+} polarized gluon has rather more structure than that
of \texttt{NNPDFpol1.0}, which is largely compatible with zero,
assuming this input shape will allow us to test whether the EIC data
are sufficiently accurate to determine the shape of the gluon
distribution.

We reconstruct the $g_1$ polarized structure function from the pseudodata
following the same procedure used in Sec.~\ref{sec:expdata} for the
E155 experiment. We provide its average statistical uncertainty in
the last column of Tab.~\ref{tab:kintab}. A comparison 
of these values with the analogous quantities for 
fixed-target experiments (see Tab.~\ref{tab:exps-err} in Sec.~\ref{sec:expdata})
clearly shows that EIC data are expected to be far more precise, 
with uncertainties reduced up to one order of magnitude.
No information on the expected systematic uncertainties is available,
hence we will ignore them in our present analysis.
However, we notice that the projected statistical uncertainties set the scale
at which one needs to control systematics, which arise from luminosity and
polarization measurements, detector acceptance and resolution, and QED
radiative corrections.

We will perform two different fits, corresponding to the two stages envisaged
for the eRHIC option of an EIC~\cite{BNL:eRHIC} discussed above, 
which will be referred to as \texttt{NNPDFpolEIC-A}
and \texttt{NNPDFpolEIC-B}. 
The former includes the first two sets of pseudodata listed in 
Tab.~\ref{tab:kintab}, while the latter also includes the third set.

\section{Fit optimization}
\label{sec:fitoptEIC}

The methodology for the determination of polarized PDFs, including their 
parametrization in terms of neural networks and their minimization through a
genetic algorithm, follows the one discussed in detail in Sec.~\ref{sec:minim}.
However, due to the accuracy and the kinematic coverage of EIC pseudodata, 
which are respectively higher and wider in comparison to their 
fixed-target counterparts, the parameters entering the genetic algorithm and
determining its stopping had to be re-tuned. 

In particular, in order to allow 
the genetic algorithm to explore the space of parameters more efficiently,
we have used a large population of mutants and increased the number of weighted
training generations to ensure that all data sets are learnt with comparable
accuracy. The target values of the figure of merit used in the weighted training
formula, Eq.~(\ref{eq:weight_errfun}), were consistently determined for each
of the three EIC pseudodata sets, following the iterative procedure 
discussed in Sec.~\ref{sec:minim}. As for the stopping criterion, we have 
modified the values of the width of the moving average $N_{\mathrm{smear}}$ and
the smearing parameter $\Delta_{\mathrm{smear}}$; for the \texttt{NNPDFpolEIC-B}
fit, we have also increased the maximum number of genetic algorithm generations
at which the minimization stops if the stopping criterion is not fulfilled.
Also, equal training and validation fractions were chosen for pseudodata sets,
unlike their fixed-target DIS counterparts. Indeed, we checked that the 
EIC pseudodata set size (about fifty points per set) is large enough to 
ensure fit stability.
The values of the minimization and stopping parameters used in the
\texttt{NNPDFpolEIC-A} and \texttt{NNPDFpolEIC-B} determinations are 
collected in Tab.~\ref{tab:minimEIC}: they can be straightforwardly 
compared to those used in the \texttt{NNPDFpol1.0} analysis, see 
Tabs.~\ref{tab:mutpars}-\ref{tab:stopping_pars}.
\begin{table}[t]
\centering
\footnotesize
\begin{tabular}{lccccccc}
\toprule
Fit & $N_{\mathrm{gen}}^{\mathrm{max}}$ & $N_{\mathrm{mut}}^{\mathrm{gen}}$
    & $N_{\mathrm{mut}}^{\mathrm{a}}$ & $N_{\mathrm{mut}}^{\mathrm{b}}$
    & $N_{\mathrm{gen}}^{\mathrm{wt}}$ & $N_{\mathrm{smear}}$ & $\Delta_{\mathrm{smear}}$\\
\midrule
\texttt{NNPDFpolEC-A} & 20000 & 2000 & 80 & 30 &  5000 & 200 & 200\\
\texttt{NNPDFpolEC-B} & 50000 & 2000 & 80 & 30 & 10000 & 200 & 200\\
\bottomrule
\end{tabular}
\mycaption{Values of the minimization and stopping parameters entering the
fitting algorithm. The corresponding values used in the \texttt{NNPDFpol1.0}
analysis are quoted in Tabs.~\ref{tab:mutpars}-\ref{tab:stopping_pars}.}
\label{tab:minimEIC}
\end{table}

Furthermore, we have redetermined the range in which preprocessing 
exponents are randomized, since the new information from EIC pseudodata 
may modify the large- and small-$x$ PDF behavior.
In Tab.~\ref{tab:preprocessing}, we show the values we use for the present fit, 
which can be compared to \texttt{NNPDFpol1.0} from Tab.~\ref{tab:prepexps}.
We have checked that our choice of preprocessing exponents does not bias 
our fit, according to the procedure discussed in 
Sec.~\ref{sec:minim}.
\begin{table}[t]
\footnotesize
\centering
\begin{tabular}{lcc}
\toprule
PDF & $m$ & $n$\\
\midrule
$\Delta\Sigma(x,Q_0^2)$ & $[1.5, 3.5]$ & $[0.1, 0.7]$ \\
$\Delta g(x,Q_0^2)$     & $[2.0, 4.0]$ & $[0.1, 0.8]$ \\
$\Delta T_3(x,Q_0^2)$   & $[1.5, 3.0]$ & $[0.1, 0.6]$ \\
$\Delta T_8(x,Q_0^2)$   & $[1.5, 3.0]$ & $[0.1, 0.6]$ \\
\bottomrule
\end{tabular}
\mycaption{Ranges for the small- and large-$x$ preprocessing exponents.}
\label{tab:preprocessing}
\end{table}

\section{Results}
\label{sec:resultsEIC}

We now present our polarized parton sets based on inclusive DIS pseudodata at 
an EIC discussed in Sec.~\ref{sec:pseudodataEIC}, \texttt{NNPDFpolEIC-A}
and \texttt{NNPDFpolEIC-B}. First, we discuss their statistical features, 
then we show the corresponding parton distributions, compared to 
\texttt{NNPDFpol1.0}. All results presented in this section are obtained
out of PDF ensembles of $N_{\mathrm{rep}}=100$ replicas.

\subsection{Statistical features}
\begin{figure}[p]
\centering
Distribution of $\chi^{2(k)}$\\
\epsfig{width=0.40\textwidth, figure=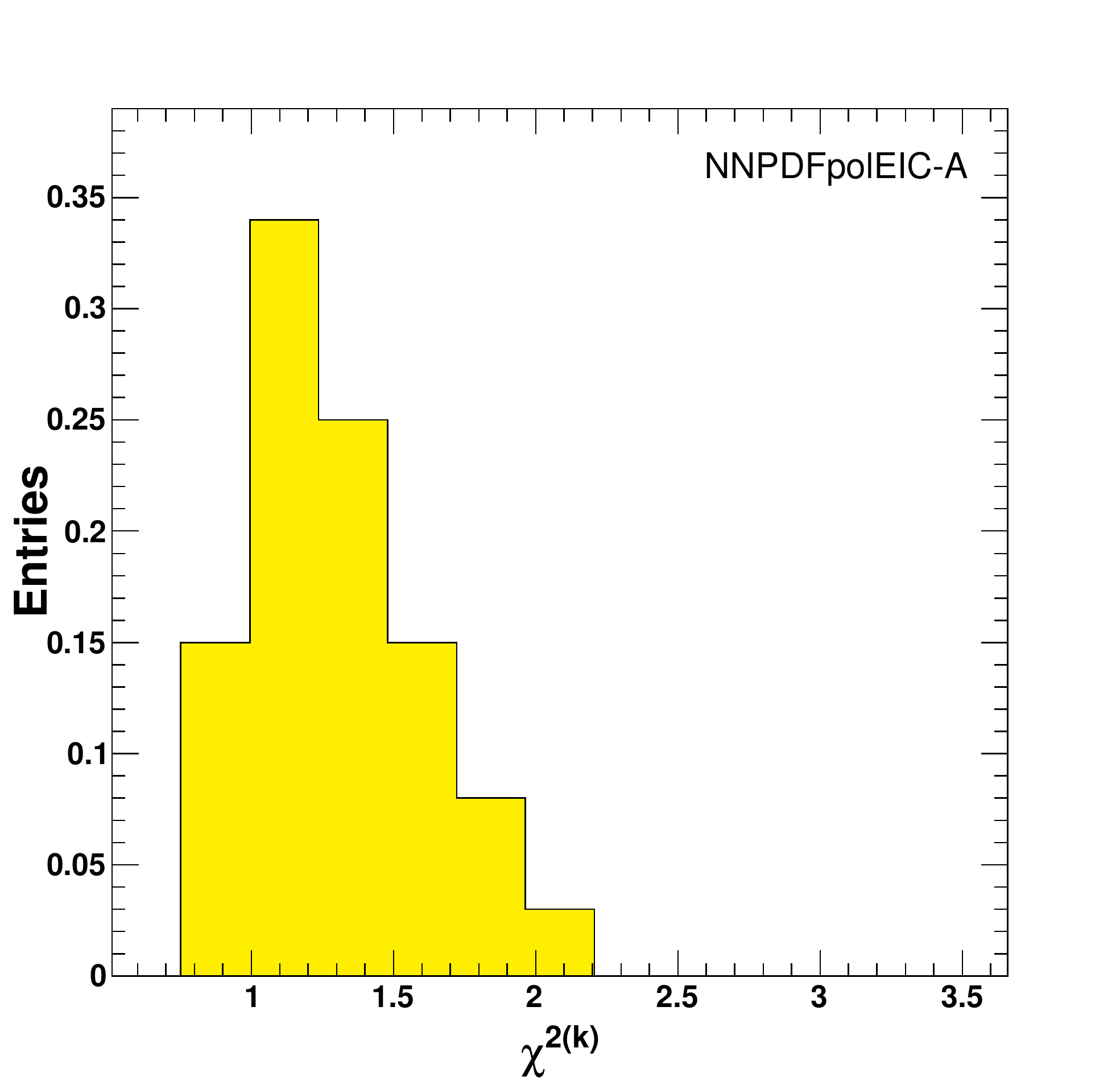}
\epsfig{width=0.40\textwidth, figure=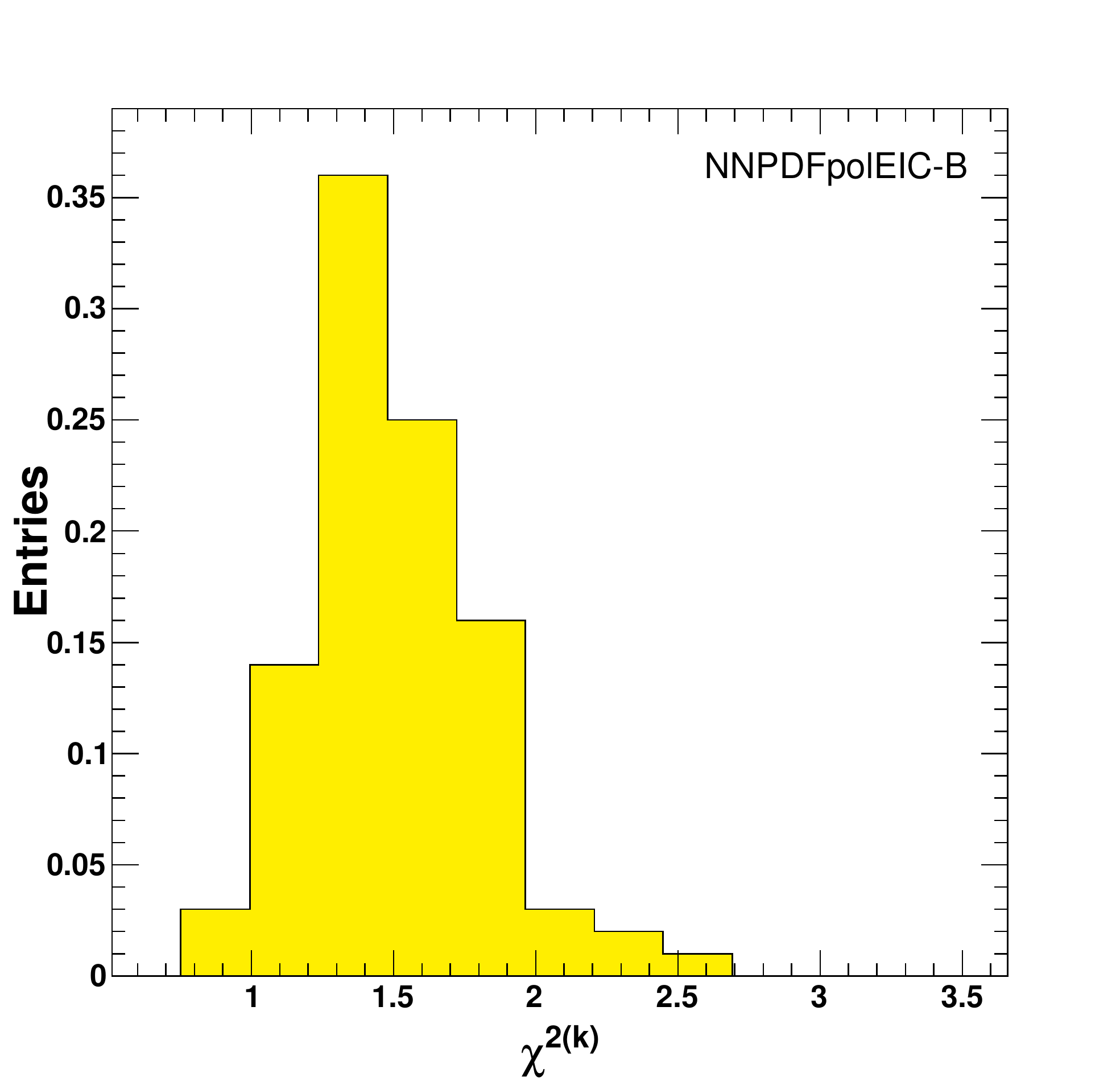}\\
Distribution of $E_{\mathrm{tr}}^{(k)}$\\
\epsfig{width=0.40\textwidth, figure=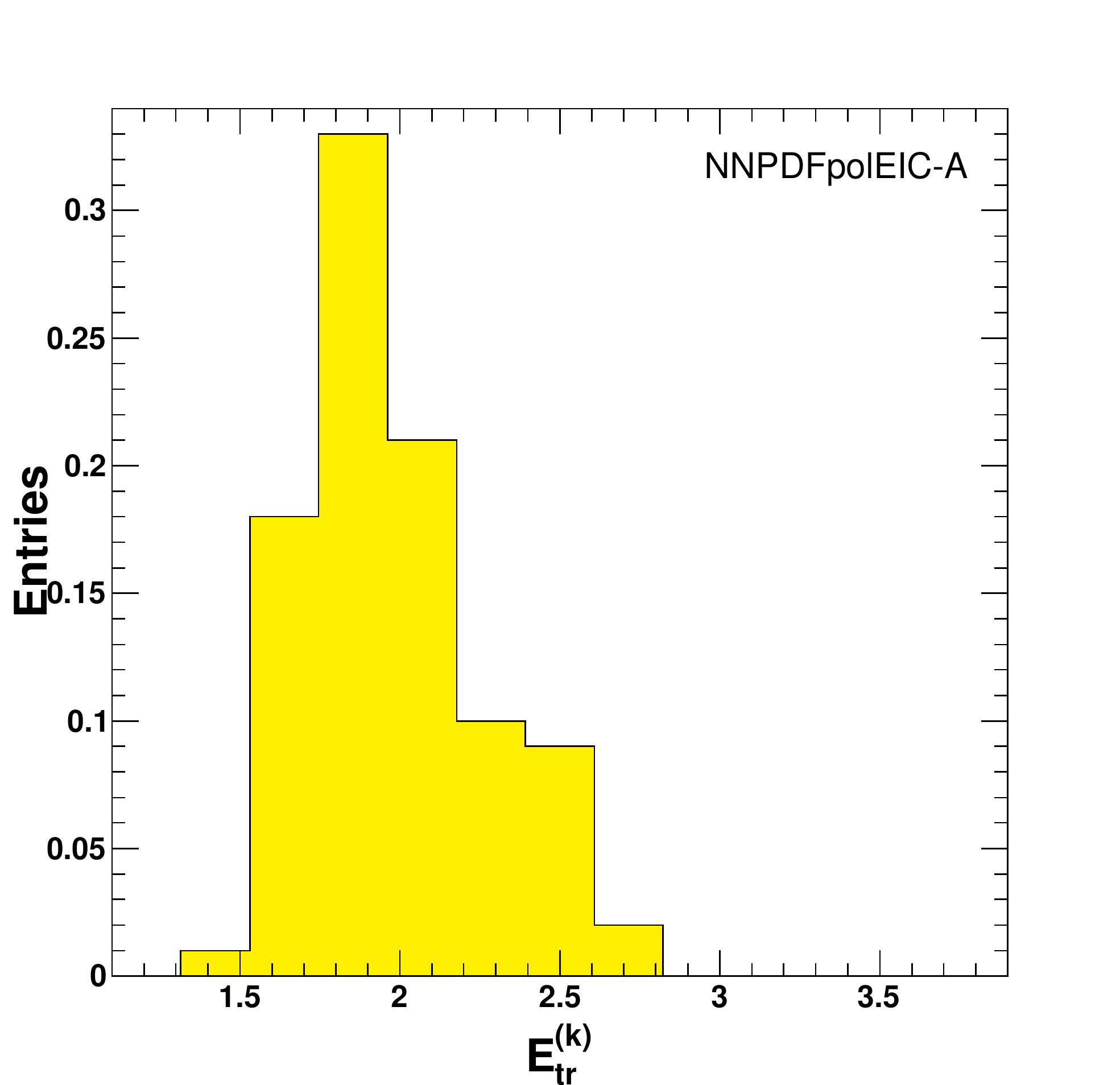}
\epsfig{width=0.40\textwidth, figure=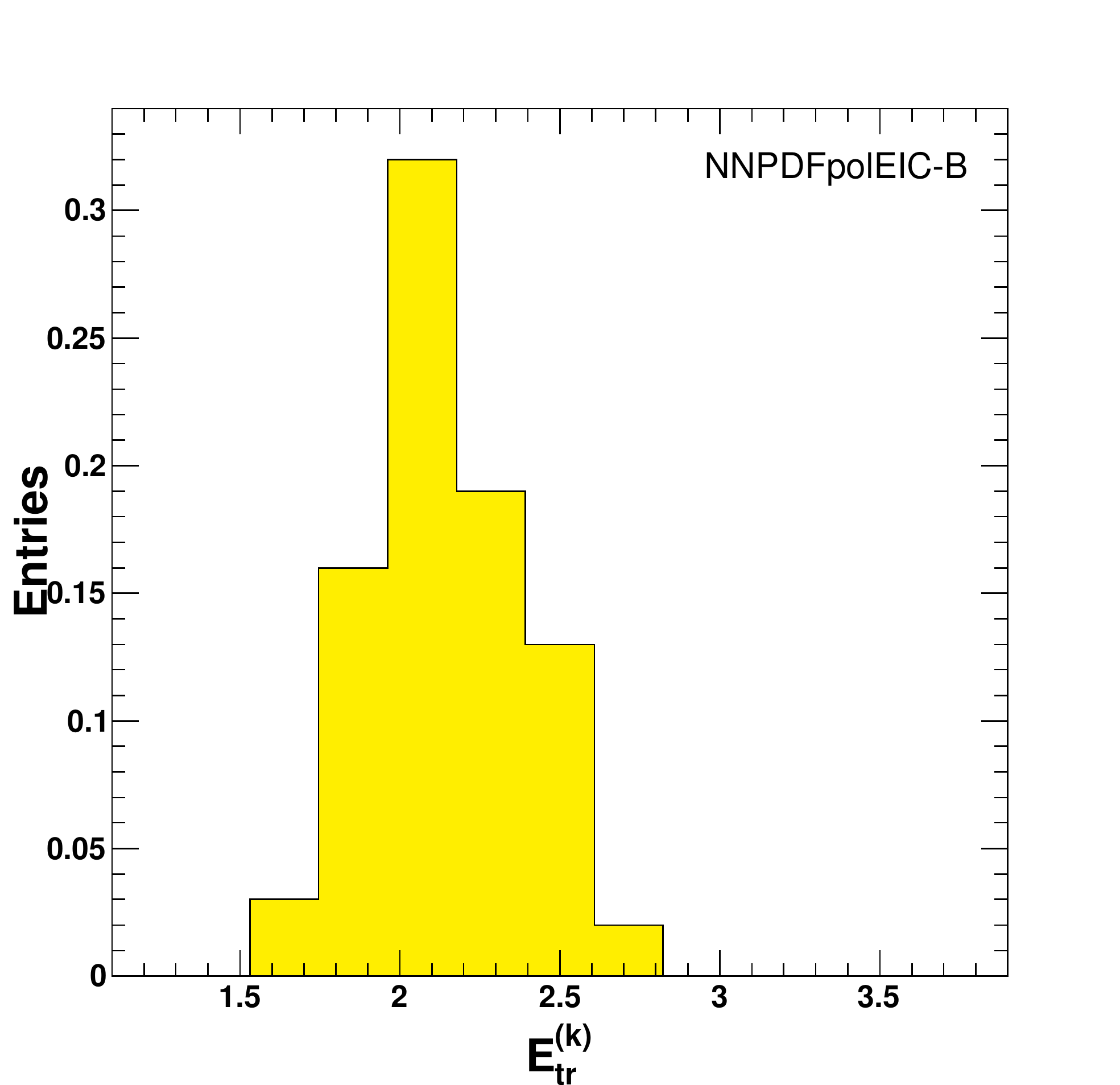}\\
\mycaption{Distribution of $\chi^{2(k)}$ (upper plots) and $E_{\mathrm{tr}}^{(k)}$
(lower plots) from a sample of $N_{\mathrm{rep}}=100$ replicas, for the
\texttt{NNPDFpolEIC-A} (left plots) and \texttt{NNPDFpolEIC-B} (right
plots) parton determinations.}
\label{fig:EICchi2}
\end{figure}

\begin{figure}[p]
\centering
Distribution of training lenghts\\
\epsfig{width=0.40\textwidth, figure=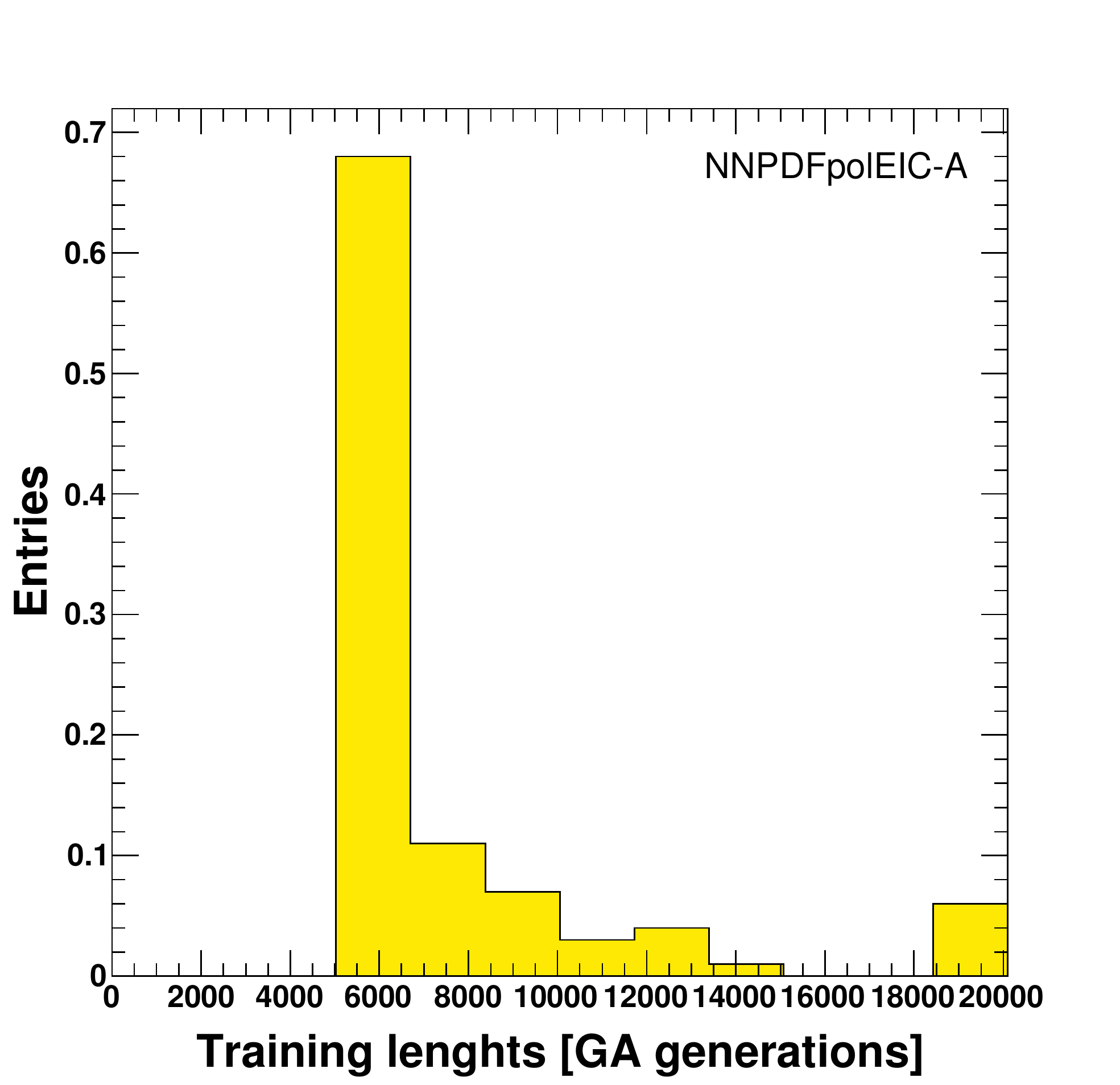}
\epsfig{width=0.40\textwidth, figure=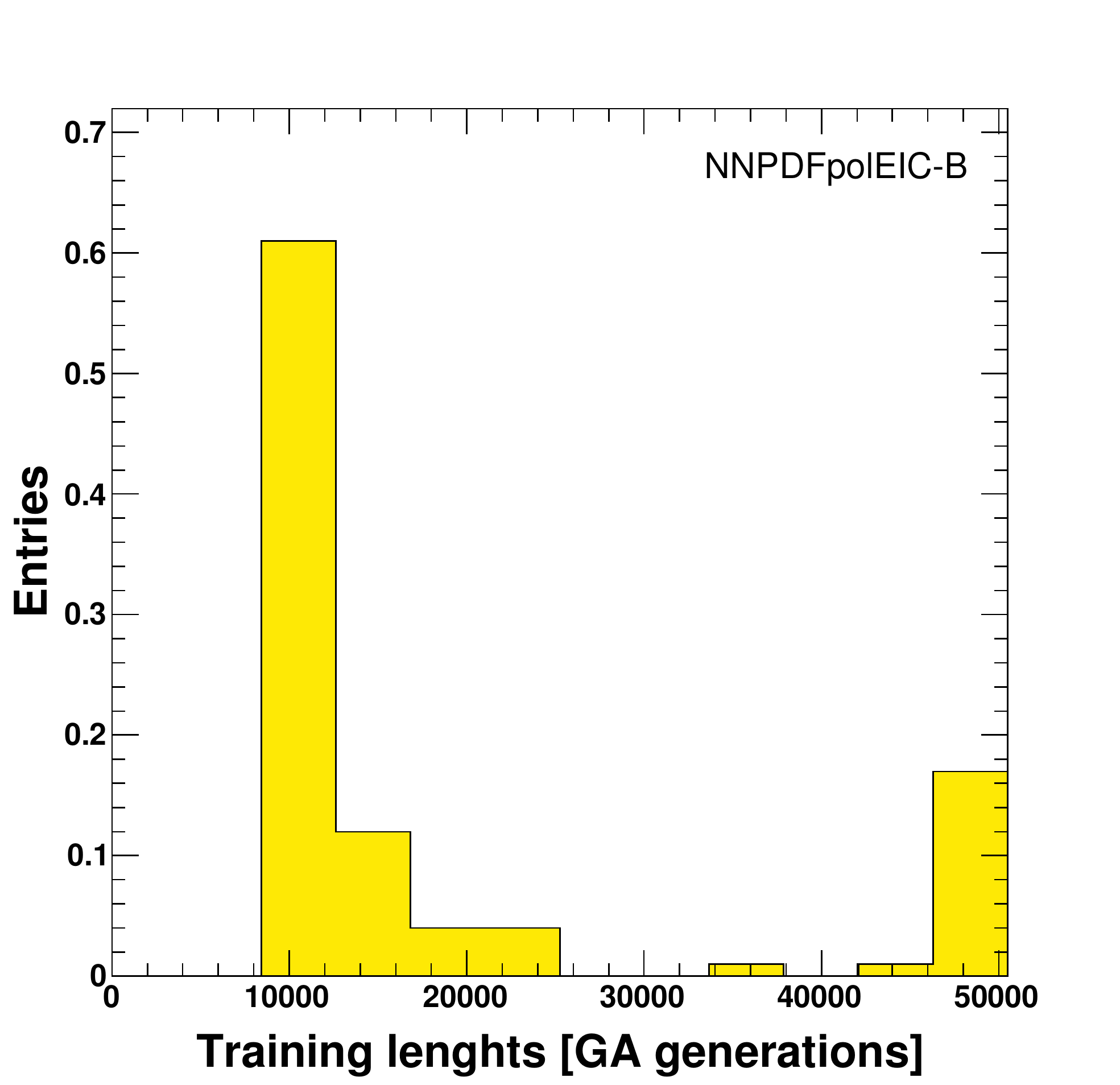}\\
\mycaption{Distribution of training lenghts from a sample 
of $N_{\mathrm{rep}}=100$ replicas, for the
\texttt{NNPDFpolEIC-A} (left plot) and \texttt{NNPDFpolEIC-B} (right
plot) parton determinations.}
\label{fig:EICtl}
\end{figure}

Various general features of the \texttt{NNPDFpolEIC-A} 
and \texttt{NNPDFpolEIC-B} PDF determinations are summarized
in Tab.~\ref{tab:esttot}, and can be straightforwardly compared to 
\texttt{NNPDFpol1.0}, see Tab.~\ref{tab:chi2tab1}. 
These include: the $\chi^2$ per data point of the final best-fit PDF set
compared to data (denoted as $\chi^2_{\mathrm{tot}}$); the average and standard
deviation over the replica sample of the same figure of merit for each
replica when compared to the corresponding data replica (denoted
as $\langle E\rangle \pm\sigma_E$) computed for the total, training
and validation sets; the average and standard deviation of the
$\chi^2$ of each replica when compared to data 
(denoted as $\langle\chi^{2(k)}\rangle$); 
and the average number of iterations of the genetic algorithm at stopping
$\langle \mathrm{TL}\rangle$ and its standard deviation over the
replica sample. All these estimators were introduced in 
Sec.~\ref{sec:stat_features} and are discussed in detail 
in Refs.~\cite{Ball:2008by,Ball:2010de}.
The distributions of $\chi^{2(k)}$, $E_{\mathrm{tr}}^{(k)}$ and training lenghts 
among the $N_{\mathrm{rep}}=100$ replicas are shown in 
Fig.~\ref{fig:EICchi2} and Fig.~\ref{fig:EICtl}.
respectively. As for the training lenghts, notice the different
scale on the horizontal axis for the \texttt{NNPDFpolEIC-A} and 
\texttt{NNPDFpolEIC-B} fits, consistent with the different 
allowed maximum number of
training generations $N_{\mathrm{gen}}^{\mathrm{max}}$ 
(see Tab.~\ref{tab:minimEIC}).
\begin{table}[t]
\footnotesize
\centering
\begin{tabular}{ccc}
\toprule
& \texttt{NNPDFpolEIC-A} 
& \texttt{NNPDFpolEIC-B}\\
\midrule
$\chi^2_{\mathrm{tot}}$  
& $0.79$
& $0.86$\\
$\langle E \rangle \pm \sigma_E$ 
& $2.24\pm 0.34$
& $2.44\pm 0.31$\\
$\langle E_{\mathrm{tr}} \rangle \pm \sigma_{E_{\mathrm{tr}}}$ 
& $1.87\pm 0.54$
& $1.81\pm 0.79$\\
$\langle E_{\mathrm{val}} \rangle \pm \sigma_{E_{\mathrm{val}}}$ 
& $2.61\pm 1.05$
& $2.47\pm 1.17$\\
$\langle\chi^{2(k)}\rangle \pm \sigma_{\chi^2}$
& $1.30\pm 0.31$
& $1.50\pm 0.30$\\
\midrule
$\langle \mathrm{TL} \rangle \pm \sigma_{\mathrm{TL}}$
& $7467\pm 3678$
& $19320\pm 14625$\\
\bottomrule
\end{tabular}
\mycaption{Statistical estimators and average training lengths for the 
two fits to EIC pseudodata described in the text,
\texttt{NNPDFpolEIC-A} and
\texttt{NNPDFpolEIC-B}, with 
$N_{\mathrm{rep}}=100$ replicas. The corresponding estimators for 
\texttt{NNPDFpol1.0} are quoted in Tab.~\ref{tab:chi2tab1}.}
\label{tab:esttot}
\end{table}

\begin{figure}[t]
\begin{center}
\large{Distribution of $\chi^{2(k)}_{\mathrm{tot}}$ for individual sets}\\
\epsfig{width=0.4\textwidth, figure=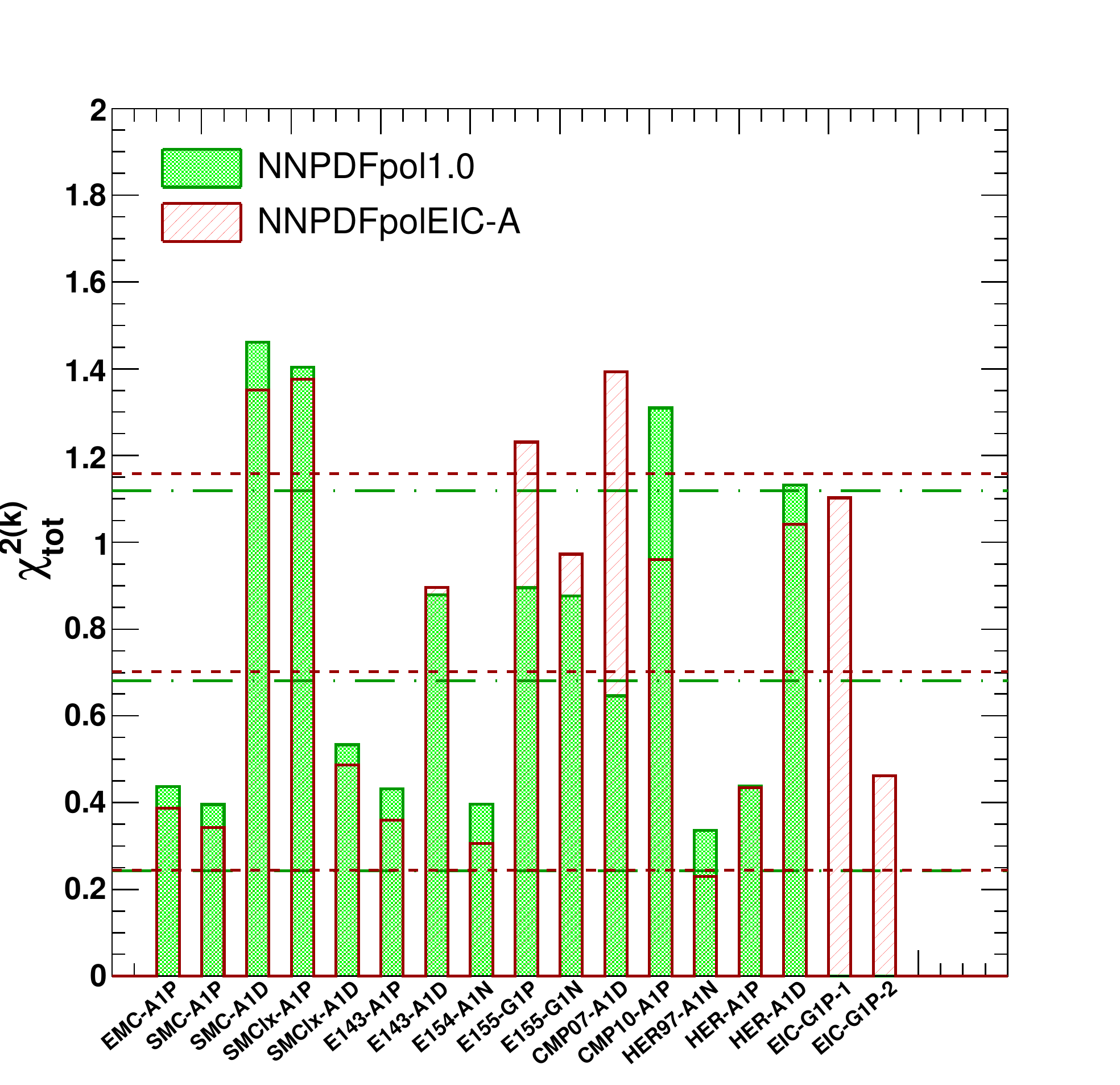}
\epsfig{width=0.4\textwidth, figure=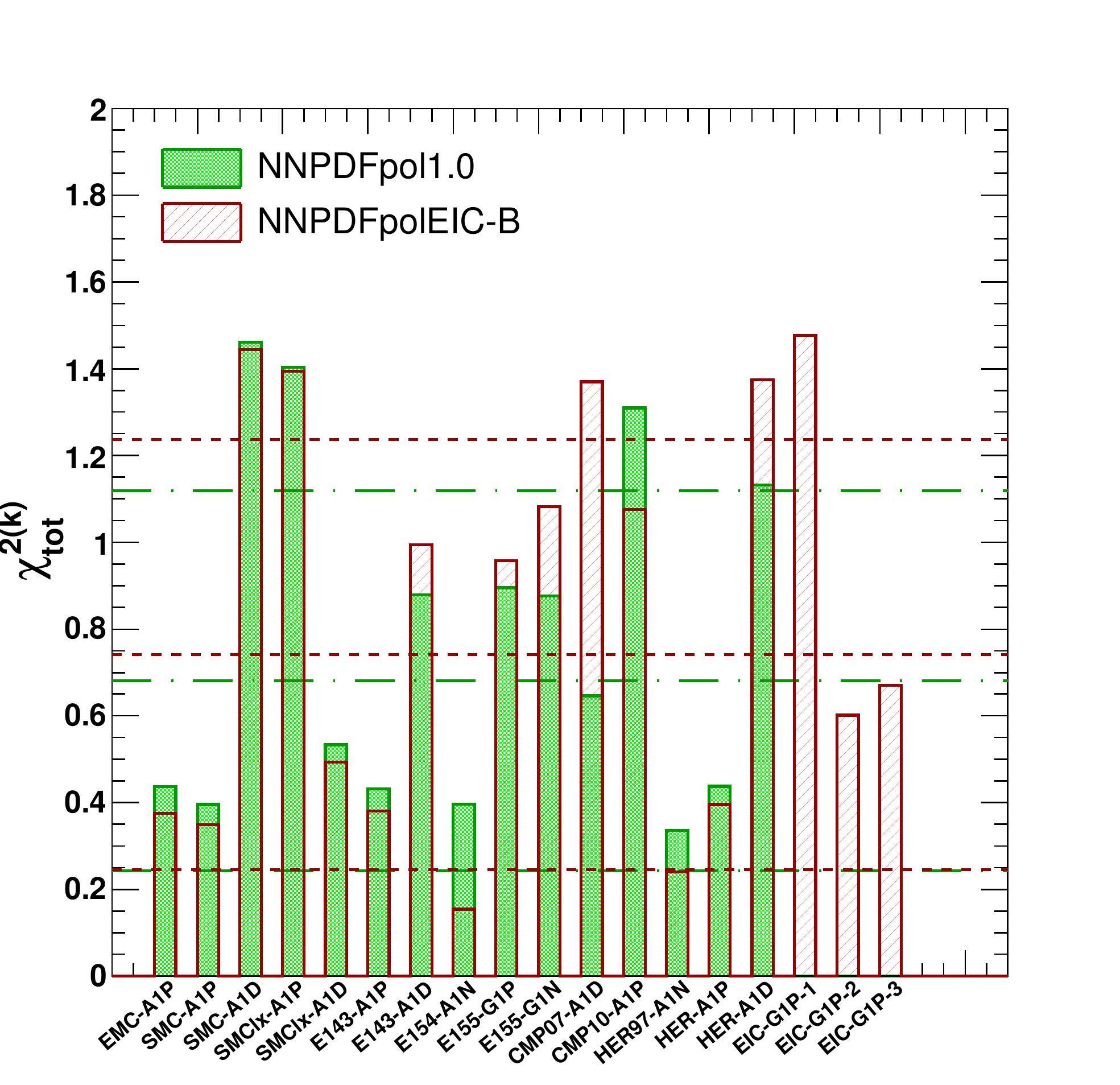}\\
\end{center}
\mycaption{Value of the $\chi^2$ per data point for the data sets 
included in the \texttt{NNPDFpolEIC-A} (left)
and in the \texttt{NNPDFpolEIC-B} (right) fits, 
compared to \texttt{NNPDFpol1.0}~\cite{Ball:2013lla}.
The horizontal lines correspond to the unweighted average of the 
$\chi^2$ values shown, and the one-sigma interval about it.
The dashed lines refer to \texttt{NNPDFpolEIC-A} 
(left plot) or \texttt{NNPDFpolEIC-B} (right plot) fits,
while the dot-dashed lines refer to \texttt{NNPDFpol1.0}.}
\label{fig:chi2sets}
\end{figure}

The  fit quality, as measured by $\chi^2_{\mathrm{tot}}$,
is comparable to that of \texttt{NNPDFpol1.0} 
($\chi^2_{\mathrm{tot}}=0.77$) for both the \texttt{NNPDFpolEIC-A} 
($\chi^2_{\mathrm{tot}}=0.79$) and the \texttt{NNPDFpolEIC-B}
($\chi^2_{\mathrm{tot}}=0.86$) fits. This shows that
our fitting procedure can easily accommodate EIC pseudodata.
The histogram of $\chi^2$ values for each data set included in our fits
is shown in Fig.~\ref{fig:chi2sets}, together with the
\texttt{NNPDFpol1.0} result; the unweighted average 
$\langle\chi^2\rangle_{\mathrm{set}}
\equiv\frac{1}{N_{\mathrm{set}}}\sum_{j=1}^{N_{\mathrm{set}}}\chi^2_{\mathrm{set,j}}$
and standard deviation over data sets are also shown.
As already noticed in Sec.~\ref{sec:results}, $\chi^2$ values
significantly below one are found as a consequence of the fact that
information on correlated systematics is not available for most
experiments, and thus statistical and systematic errors are added in
quadrature. Note that this is not the case for the EIC pseudodata, for
which, as mentioned, no systematic uncertainty was included; this may
explain the somewhat larger (closer to one) value of the $\chi^2$ per
data point which is found when the pseudodata are included.

We notice that EIC pseudodata, which 
are expected to be rather more precise
than fixed-target DIS experimental data, require more training 
to be properly learned by the neural network. This is apparent in the
increase in $\langle TL\rangle$ in Tab.~\ref{tab:esttot}
when going from \texttt{NNPDFpol1.0} to \texttt{NNPDFpolEIC-A} and
then \texttt{NNPDFpolEIC-B}. 
We checked that the statistical features discussed above
do not improve if we run very long fits, 
up to $N_{\mathrm{gen}}^{\mathrm{max}}=50000$ generations,
without dynamical stopping. In particular, we do not observe a 
decrease of the $\chi^2$ for those experiments whose value exceeds the average
by more than one sigma. This ensures that these deviations are not due to 
underlearning, \textit{i.e.} insufficiently long minimization.

\subsection{Parton Distributions}
\begin{figure}[t]
\begin{center}
\epsfig{width=0.40\textwidth, figure=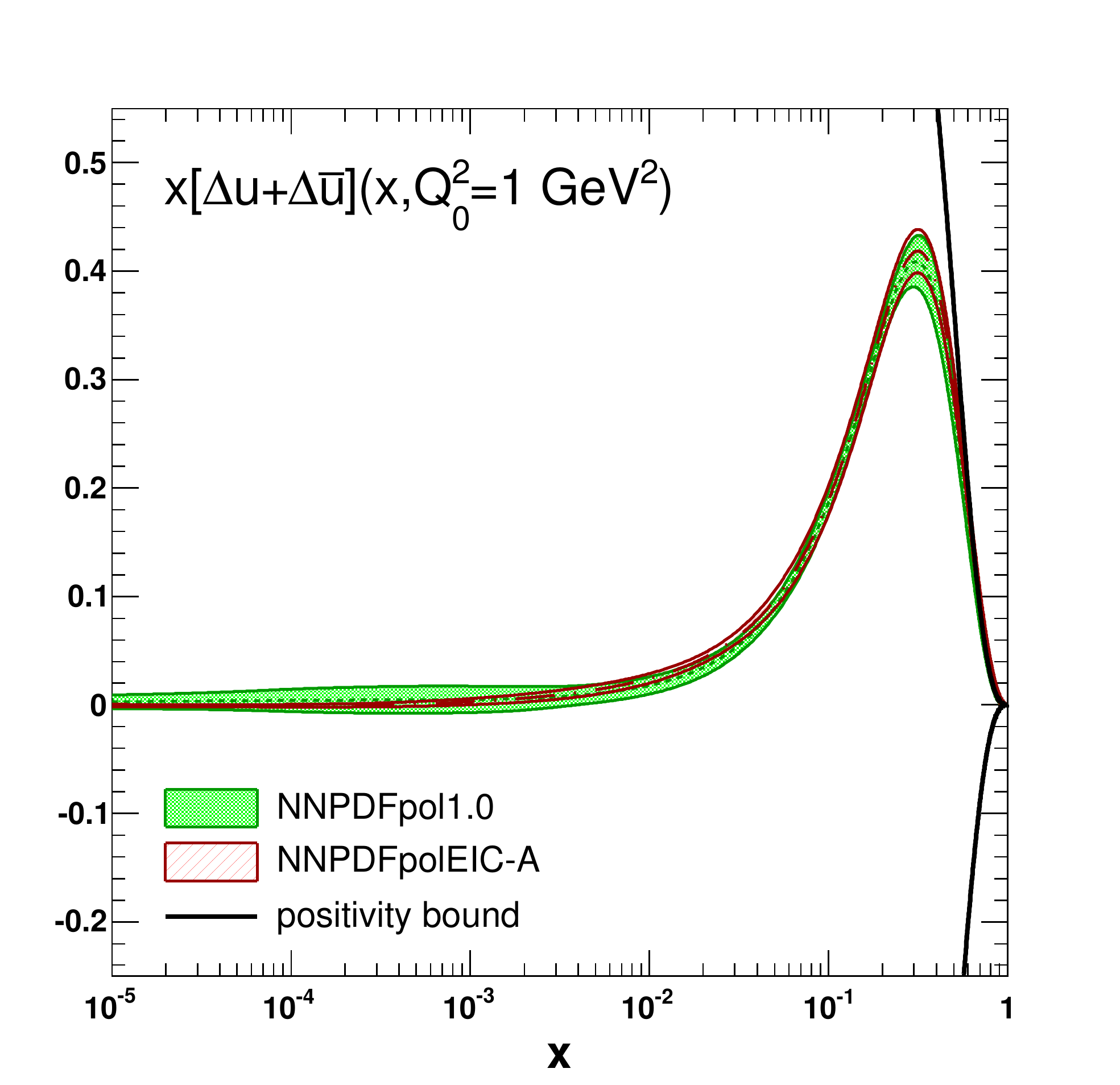}
\epsfig{width=0.40\textwidth, figure=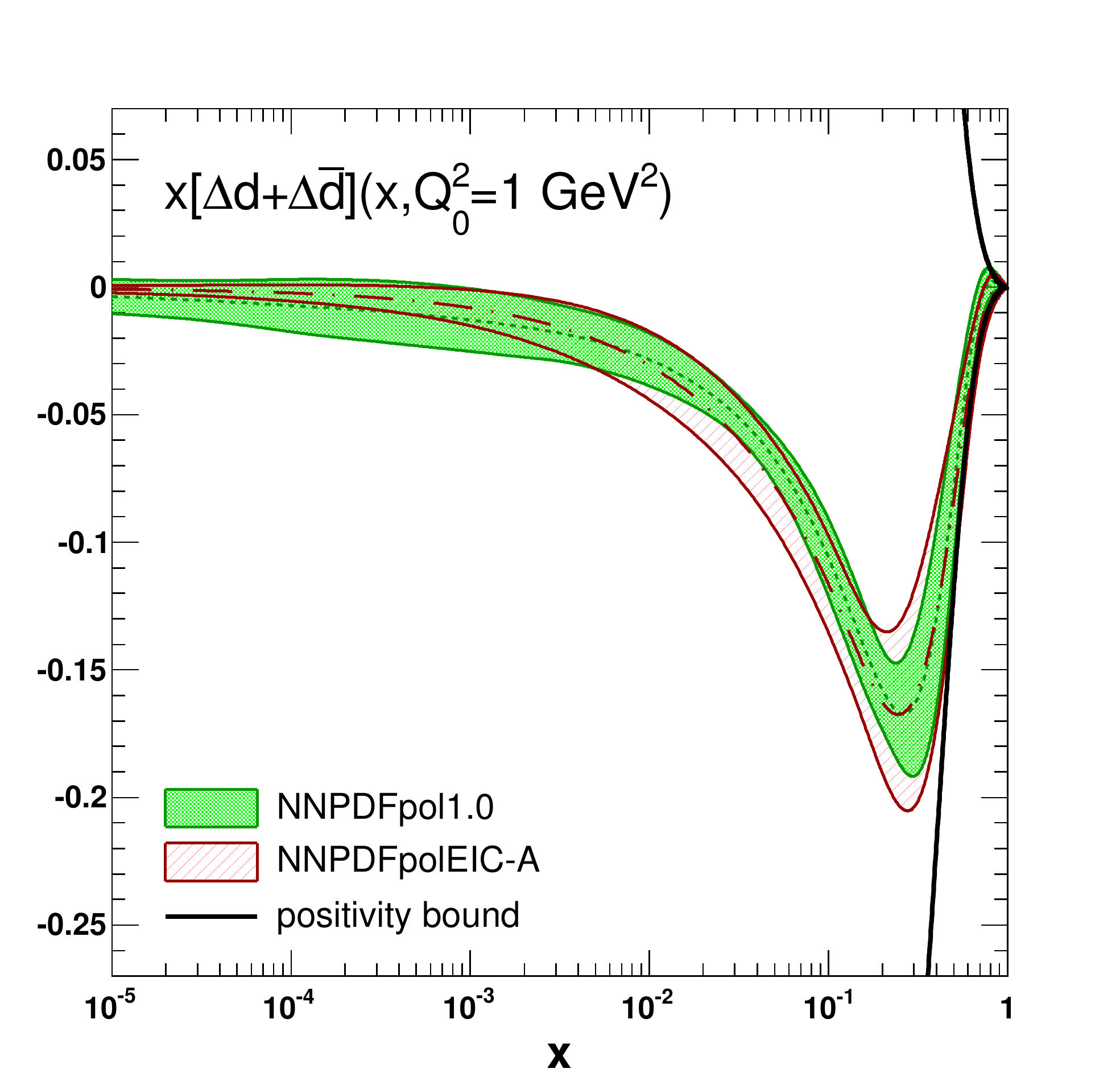}\\
\epsfig{width=0.40\textwidth, figure=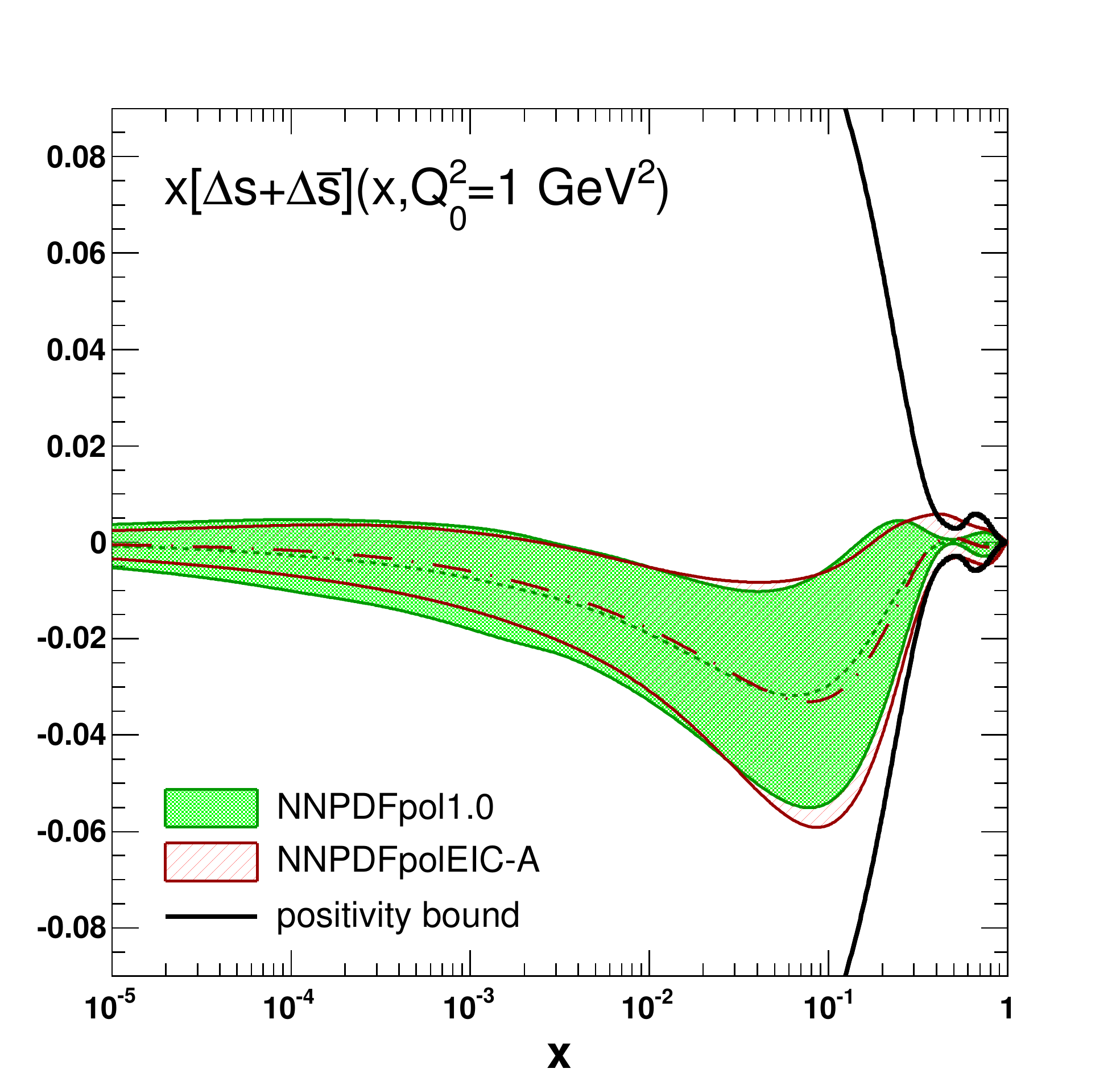}
\epsfig{width=0.40\textwidth, figure=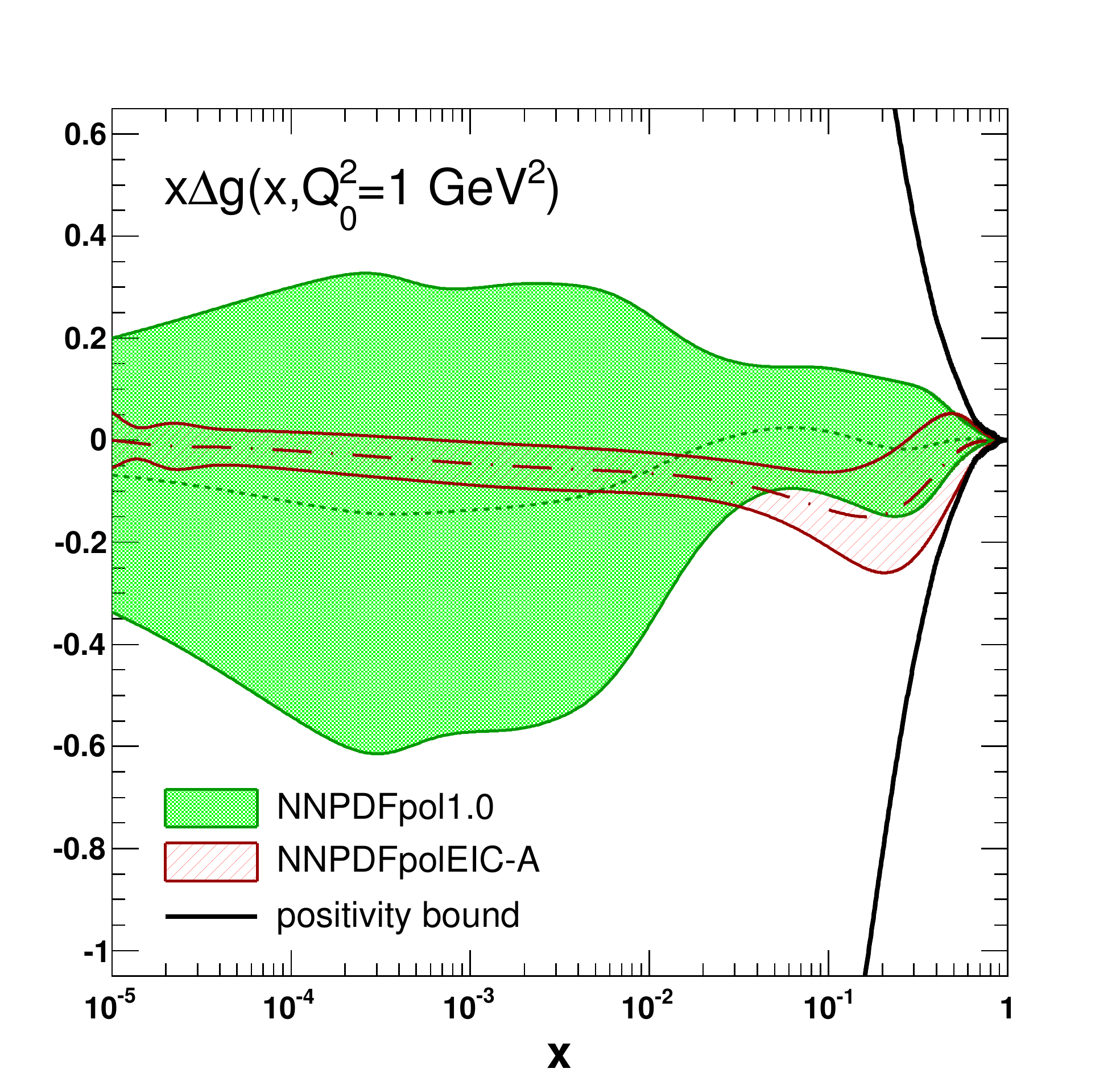}\\
\end{center}
\mycaption{The \texttt{NNPDFpolEIC-A} 
parton distributions at $Q_0^2=1$ GeV$^2$ plotted as a function of $x$ 
on a logarithmic scale, compared to \texttt{NNPDFpol1.0}.}
\label{fig:PDFsEIC1}
\end{figure}

\begin{figure}[t]
\begin{center}
\epsfig{width=0.40\textwidth, figure=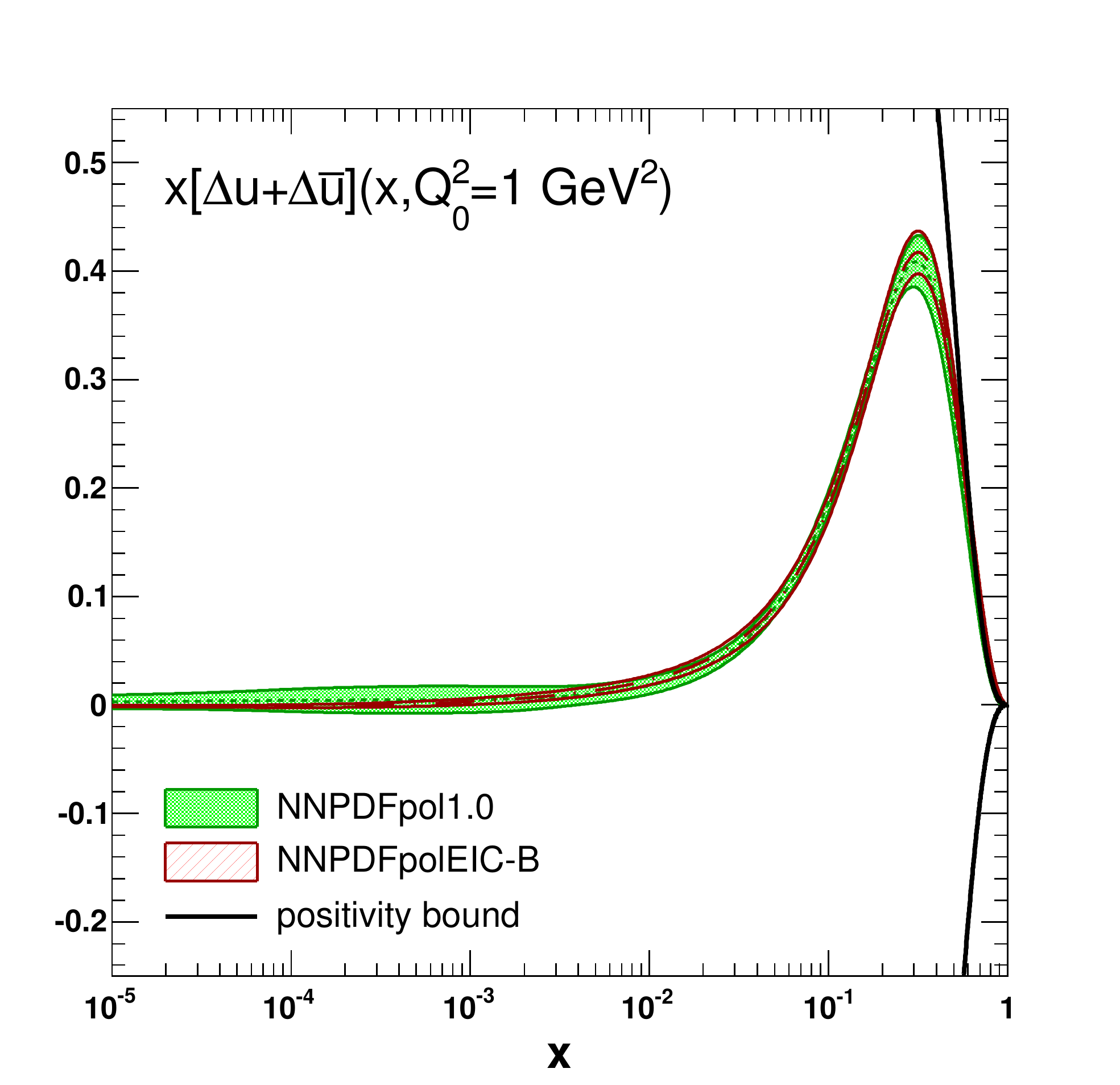}
\epsfig{width=0.40\textwidth, figure=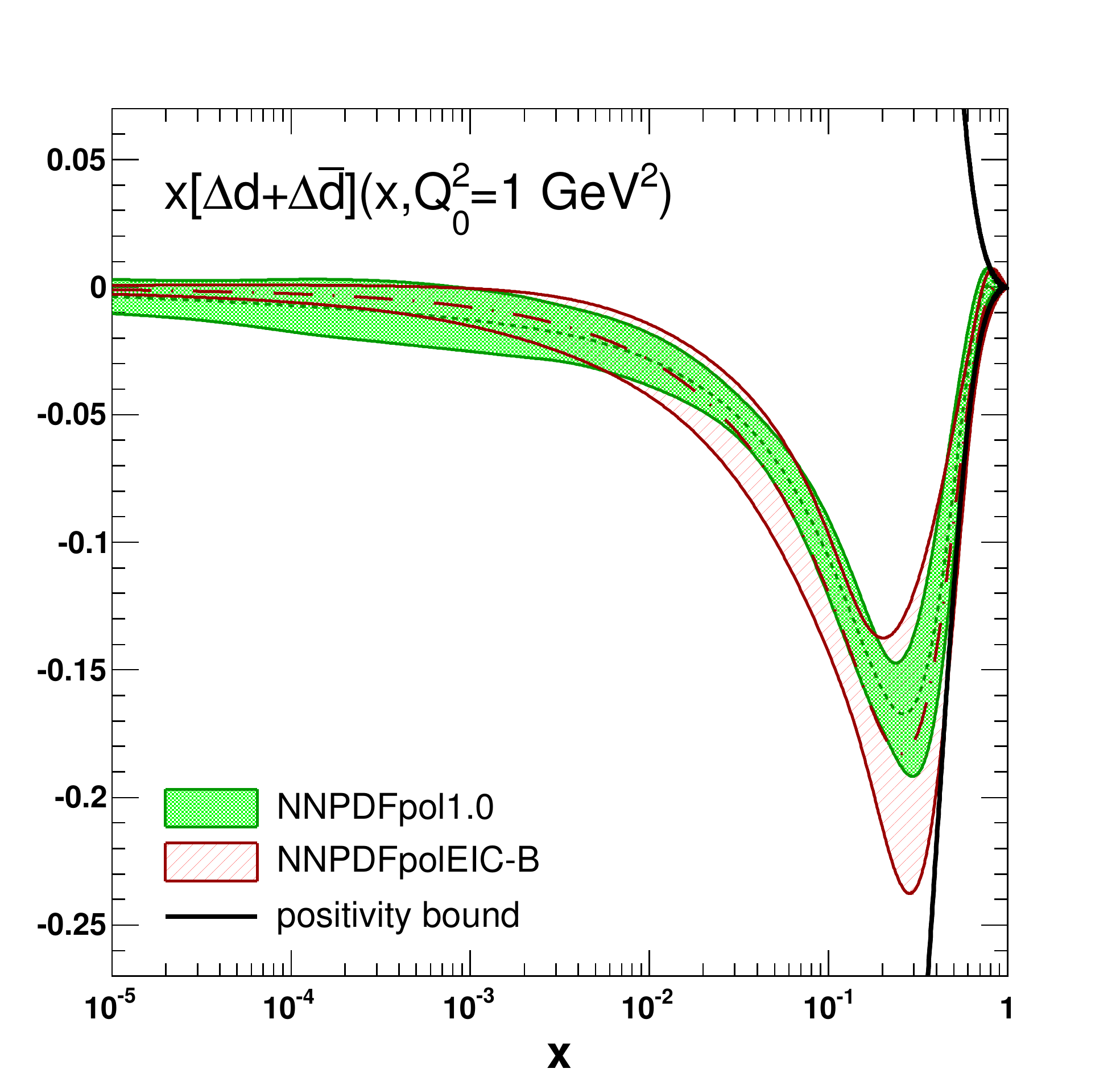}\\
\epsfig{width=0.40\textwidth, figure=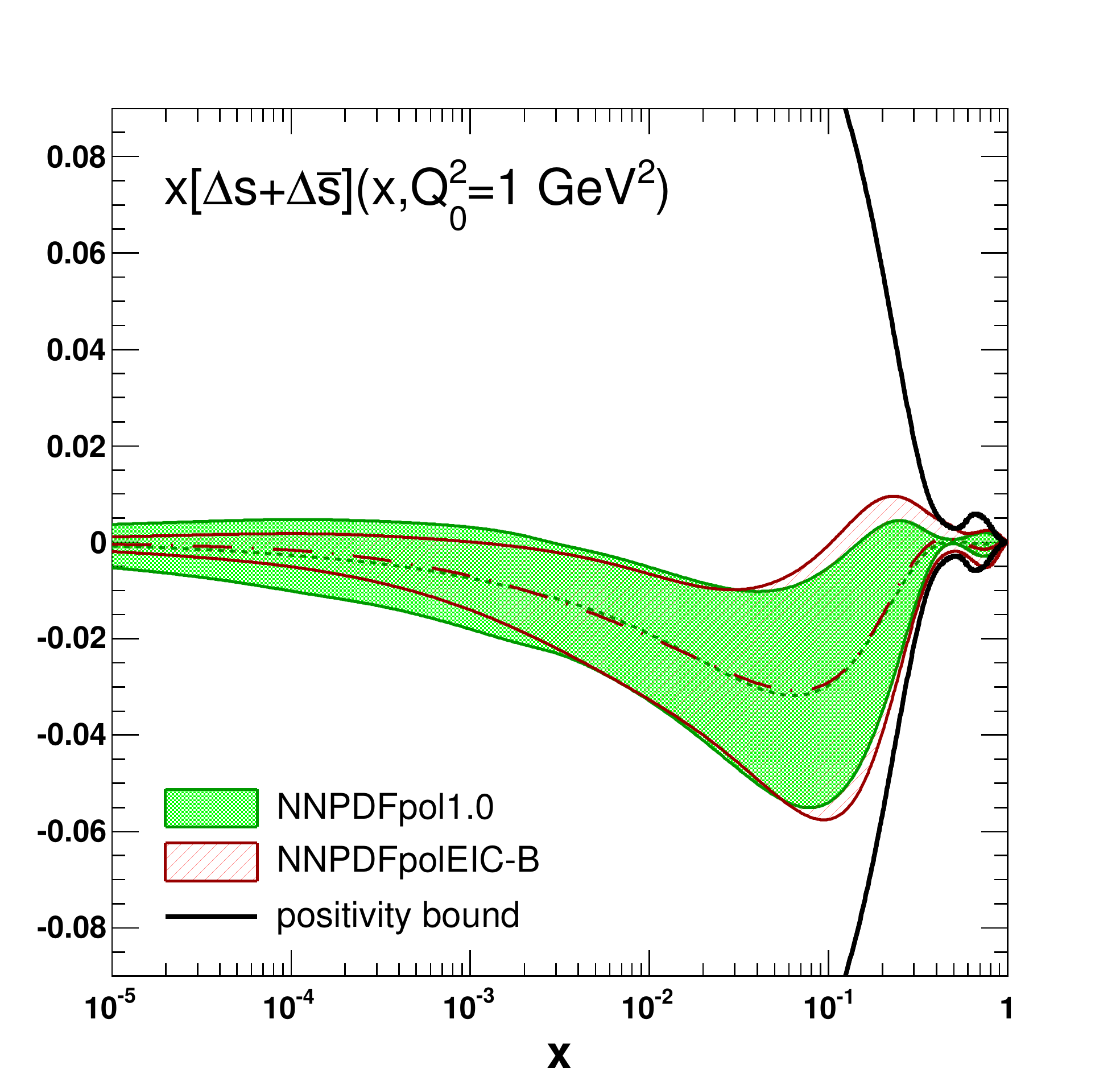}
\epsfig{width=0.40\textwidth, figure=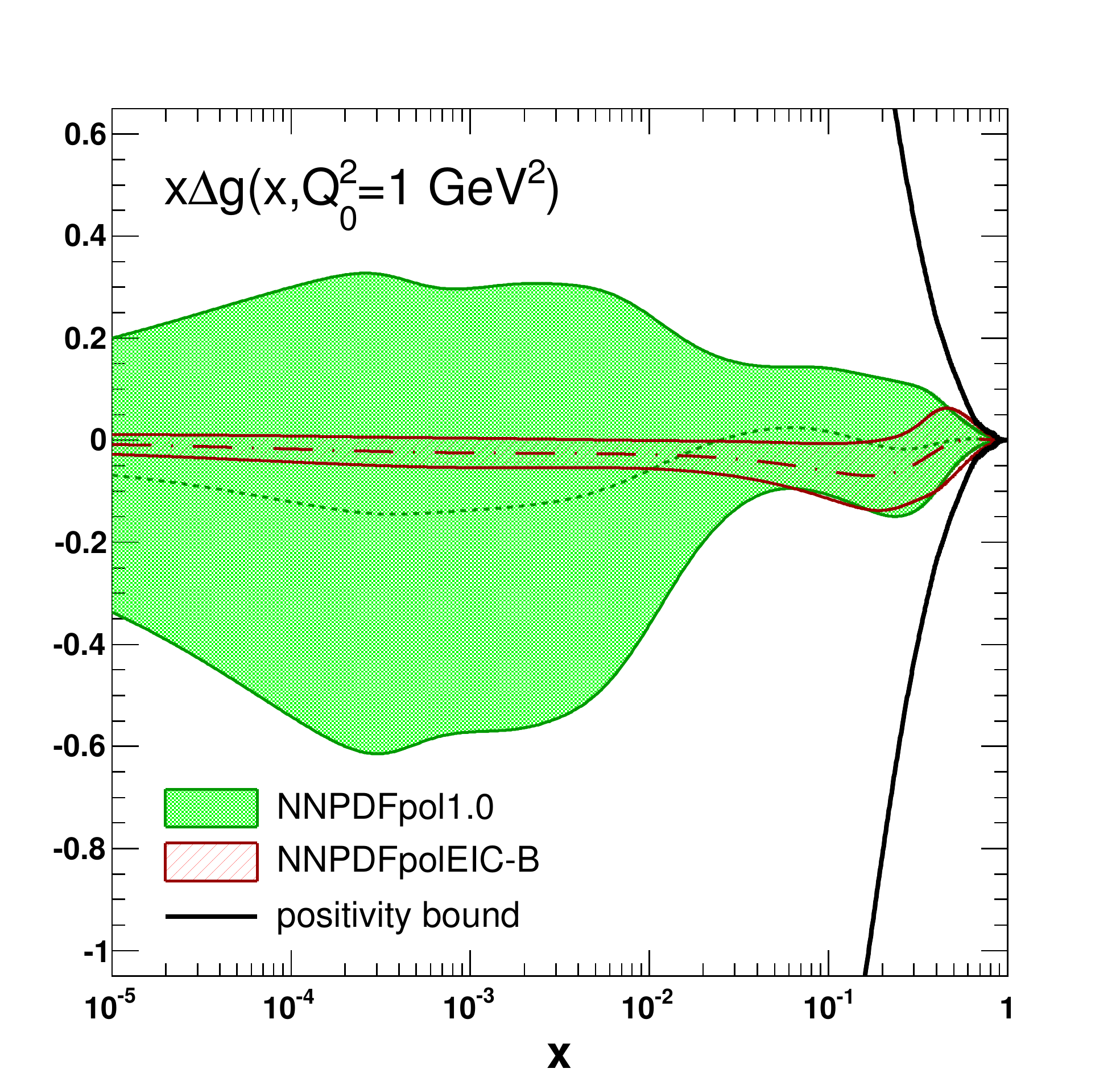}\\
\end{center}
\mycaption{Same as Fig.~\ref{fig:PDFsEIC1}, but for 
\texttt{NNPDFpolEIC-B}, compared to \texttt{NNPDFpol1.0}.}
\label{fig:PDFsEIC2}
\end{figure}

Parton distributions from the
\texttt{NNPDFpolEIC-A} and \texttt{NNPDFpolEIC-B} 
fits are compared to \texttt{NNPDFpol1.0} in 
Figs.~\ref{fig:PDFsEIC1}-\ref{fig:PDFsEIC2} respectively.
In these plots, PDFs are displayed at 
$Q_0^2=1$ GeV$^2$ as a function of $x$ on a logarithmic scale; all
uncertainties shown here are one-sigma bands. The 
positivity bound, obtained from the \texttt{NNPDF2.3} NLO
unpolarized set~\cite{Ball:2012cx} as discussed in 
Sec.~\ref{sec:minim}, is also drawn.

The most visible  impact of inclusive EIC pseudodata in
both our fits is the reduction of PDF uncertainties in the low-$x$ region
($x\lesssim 10^{-3}$) for light flavors and the gluon. 
The size of the effects is  different for different  PDFs. 
As expected, the most dramatic improvement is seen for the gluon,
while uncertainties on light quarks are only reduced by a significant
factor in the small-$x$ region. The uncertainty on the strange
distribution is essentially unaffected:
unlike in Ref.~\cite{Aschenauer:2012ve}, we find no
improvement on strangeness, due to the fact that we do not include
semi-inclusive kaon production data, contrary to what was done there.
When moving from \texttt{NNPDFpolEIC-A} to 
\texttt{NNPDFpolEIC-B} the gluon uncertainty decreases further, while
other PDF uncertainties are basically unchanged. 

\begin{figure}[t]
\begin{center}
\epsfig{width=0.40\textwidth, figure=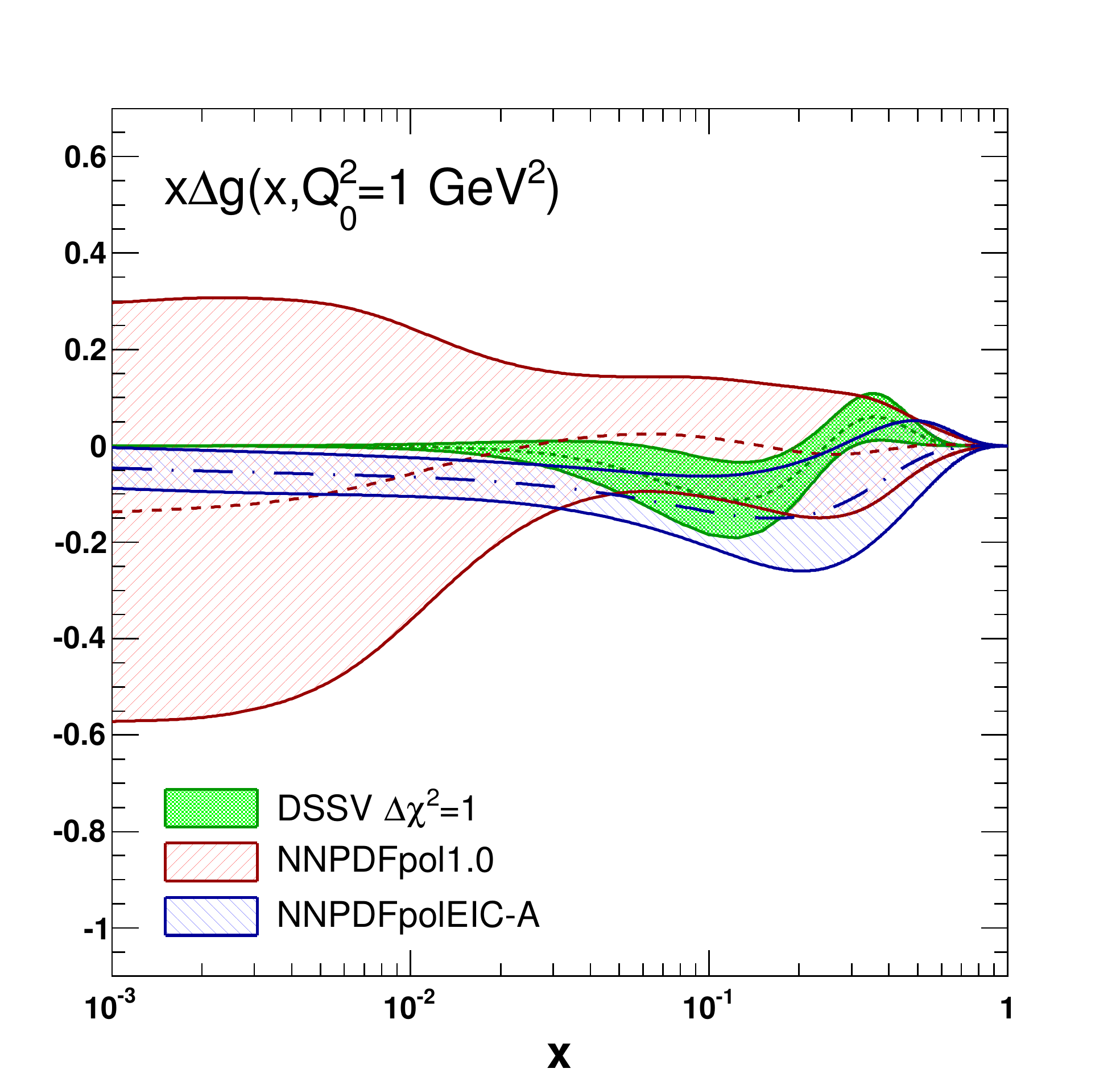}
\epsfig{width=0.40\textwidth, figure=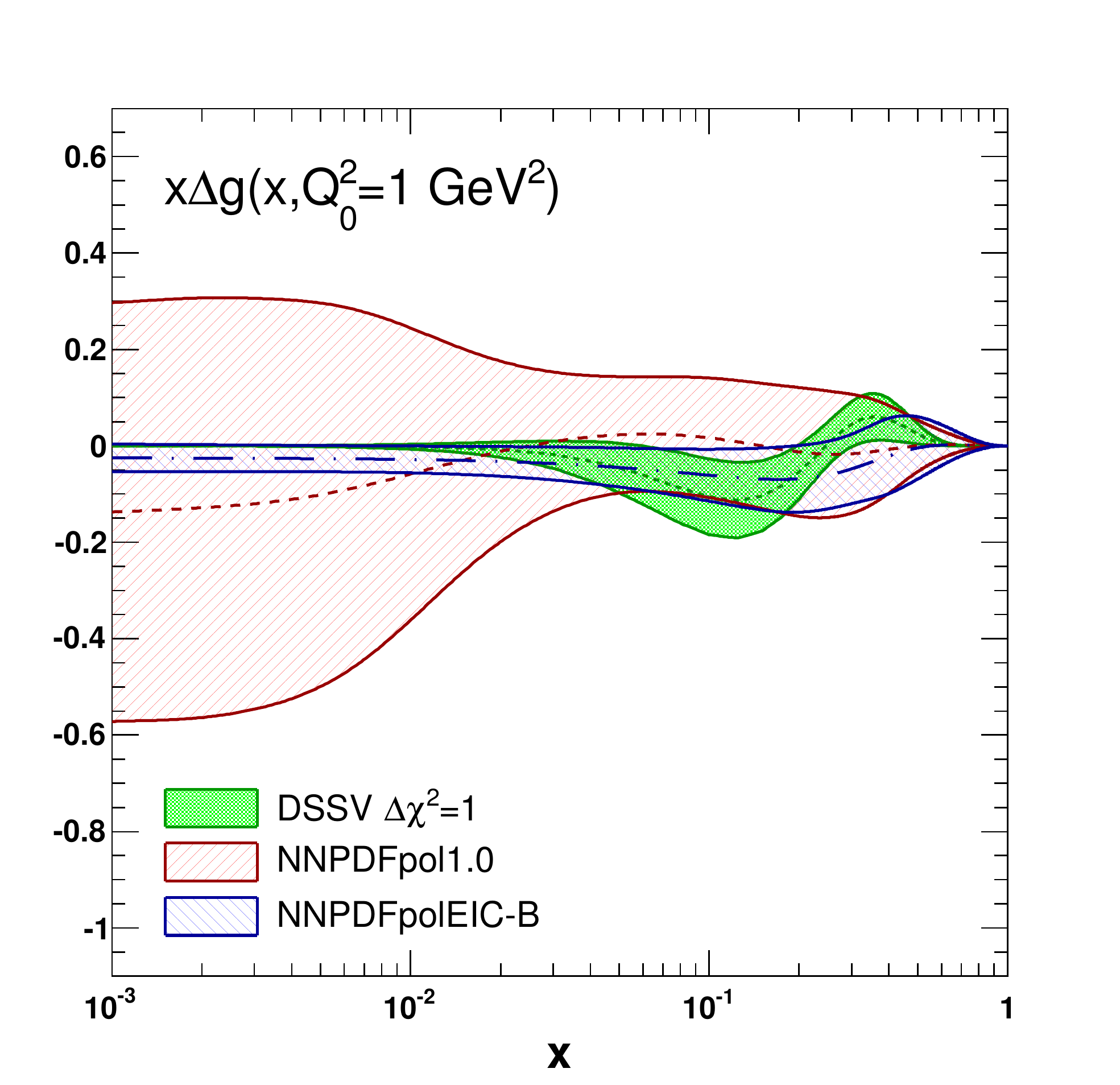}\\
\epsfig{width=0.40\textwidth, figure=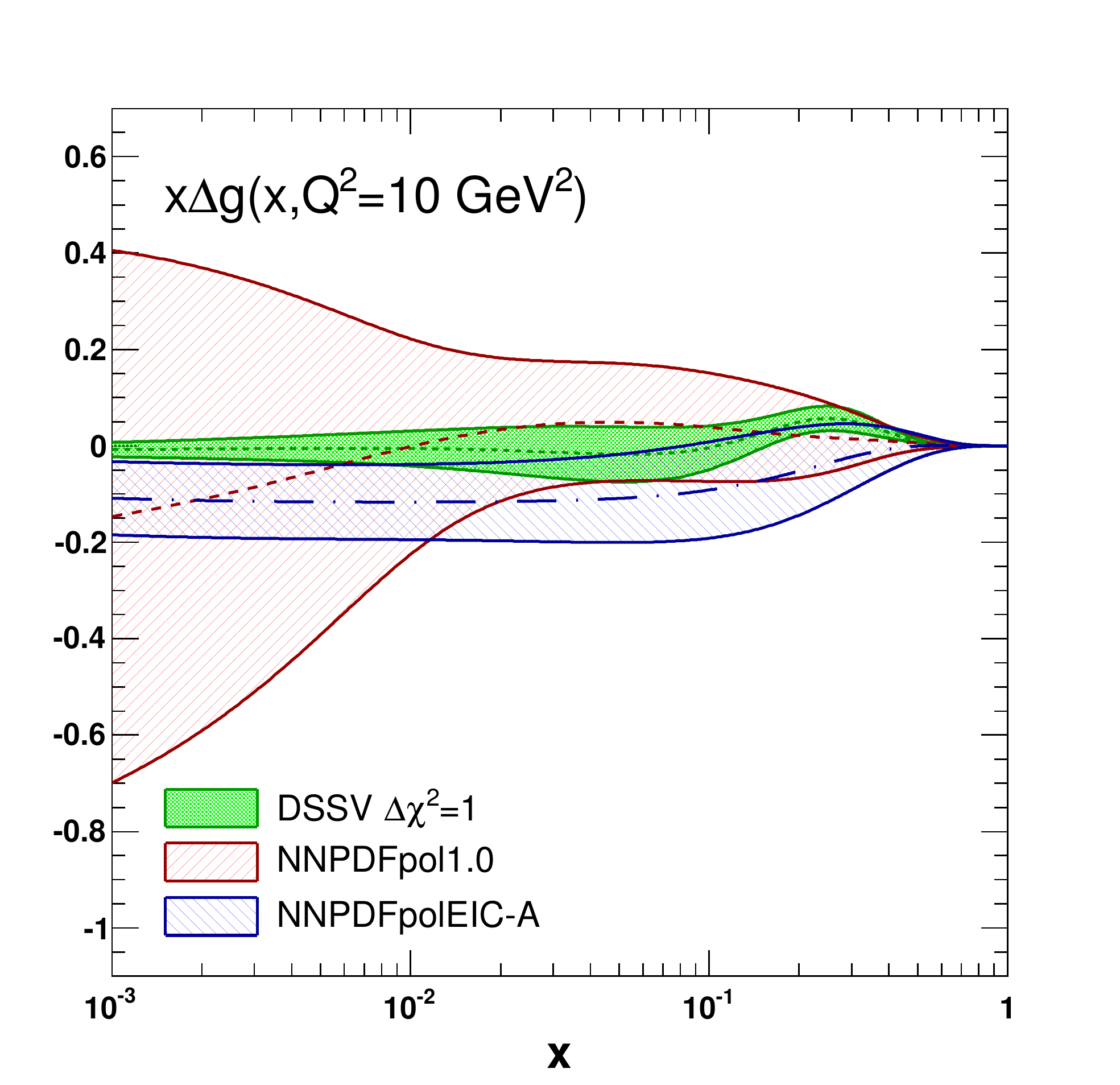}
\epsfig{width=0.40\textwidth, figure=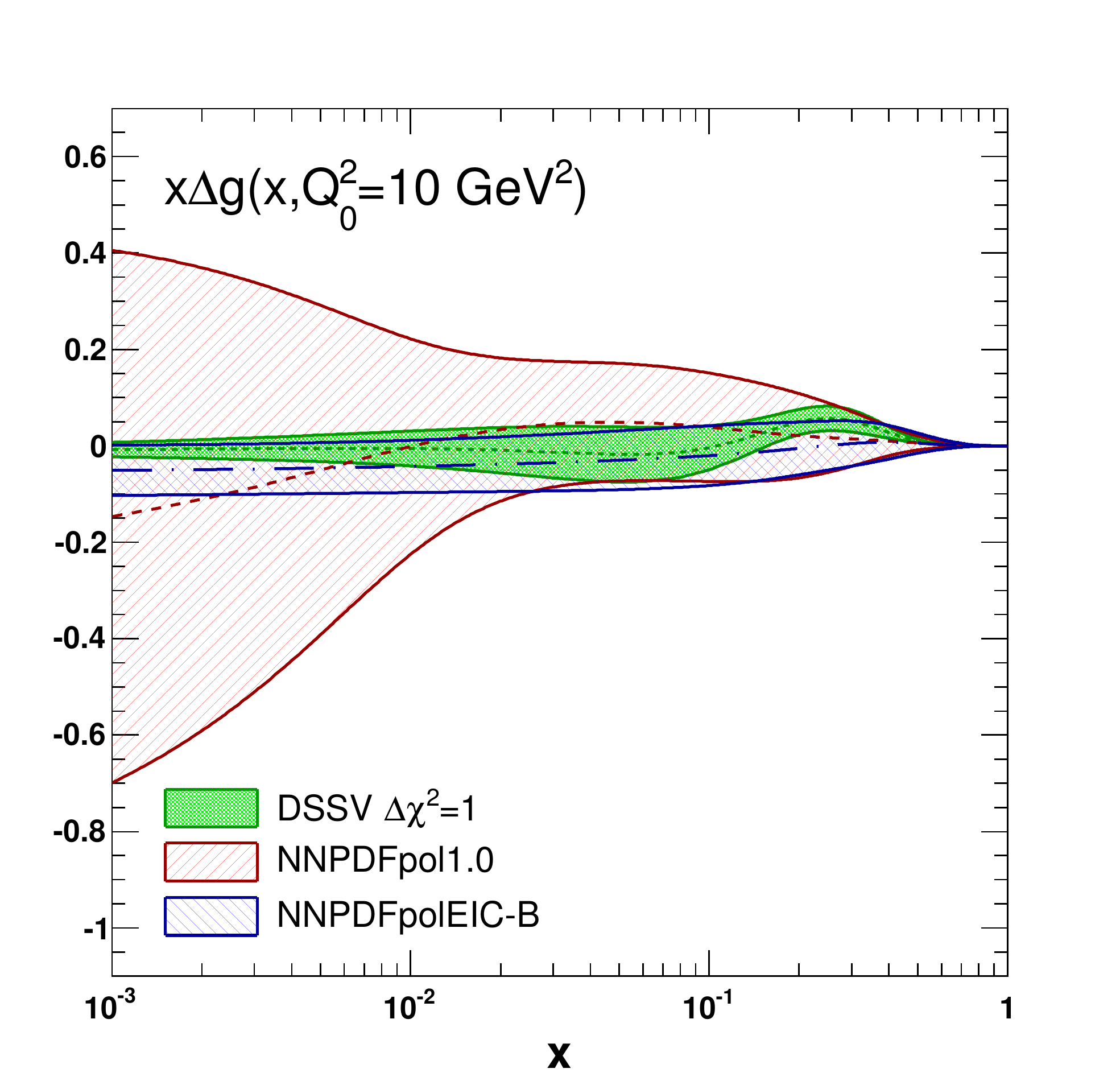}\\
\end{center}
\mycaption{The polarized gluon PDF $\Delta g(x,Q_0^2)$, at $Q_0^2=1$
GeV$^2$ (upper panels) and at $Q^2=10$ GeV$^2$ (lower panels),
in the \texttt{NNPDFpolEIC} PDF sets,
compared to DSSV~\cite{deFlorian:2009vb} and to
\texttt{NNPDFpol1.0}.}
\label{fig:gluon}
\end{figure}

In Fig.~\ref{fig:gluon}, we compare the polarized gluon PDF 
in our EIC fits to the \texttt{DSSV08}~\cite{deFlorian:2009vb} and 
\texttt{NNPDFpol1.0} parton determinations, 
both at $Q_0^2=1$~GeV$^2$ and $Q^2=10$~GeV$^2$.
The \texttt{DSSV08} uncertainty is the Hessian uncertainty computed assuming 
$\Delta\chi^2=1$, which corresponds to the default uncertainty estimate in
Ref.~\cite{deFlorian:2009vb}. This choice may lead to somewhat
underestimated uncertainties, as discussed at length in 
Sec.~\ref{sec:generalstrategy}.

It is clear that 
the gluon PDF from our fits including EIC pseudodata 
is approaching the \texttt{DSSV08} PDF shape, especially at a lower scale where
the corresponding gluon does have some structure, despite the fact that at
higher scales, where much of the data is located, perturbative
evolution tends to wash out this shape. 
Also, this  is more pronounced as more EIC pseudodata
are included in our fit,
\textit{i.e.} moving from \texttt{NNPDFpolEIC-A} to
\texttt{NNPDFpolEIC-B}. 
This means that EIC data would be sufficiently accurate to reveal 
the polarized gluon structure, if any. 

\section{Phenomenological implications of EIC pseudodata}
\label{sec:phenoEIC}

In this Section, we use our \texttt{NNPDFpolEIC-A} and \texttt{NNPDFpolEIC-B}
parton determinations to reassess the spin content of the proton
in the light of future EIC data. 
We also determine the expected contribution of the charm quark to the 
polarized structure function $g_1$, focusing on its potential to further
pin down the uncertainty of the gluon distribution.

\subsection{The spin content of the proton}
\label{sec:spinEIC}
It is particularly interesting to examine how the EIC data affect the
determination of the first moments 
of the polarized PDFs $\Delta f(x,Q^2)$, Eq.~(\ref{eq:moments}),
as they are directly related to the nucleon spin structure.
We have computed  the first moments, Eq.~(\ref{eq:moments}),
of the singlet, lightest quark-antiquark combinations and gluon for
the \texttt{NNPDFpolEIC-A} and \texttt{NNPDFpolEIC-B} PDF sets.
The corresponding central values  and one-sigma uncertainties
at $Q_0^2=1$ GeV$^2$ are shown in Tab.~\ref{tab:fullmom}, compared to 
\texttt{NNPDFpol1.0}. 
\begin{table}[t]
\footnotesize
\centering
\begin{tabular}{lccccc}
\toprule
Fit & $\langle\Delta\Sigma\rangle$ 
& $\langle\Delta u +\Delta\bar{u}\rangle$ 
& $\langle\Delta d +\Delta\bar{d}\rangle$ 
& $\langle\Delta s +\Delta\bar{s}\rangle$ 
& $\langle\Delta g \rangle$\\
\midrule
\texttt{NNPDFpolEIC-A}
&  $0.24\pm 0.08$
&  $0.82\pm 0.02$
& $-0.45\pm 0.02$
& $-0.13\pm 0.07$ 
& $-0.59 \pm 0.86$\\
\texttt{NNPDFpolEIC-B}
&  $0.21\pm 0.06$
&  $0.81\pm 0.02$
& $-0.47\pm 0.02$
& $-0.12\pm 0.07$
& $-0.33 \pm 0.43$\\
\bottomrule
\end{tabular}
\mycaption{First moments of the polarized quark distributions at 
$Q_0^2=1$ GeV$^2$ for the fits in the present analysis.
The corresponding values for \texttt{NNPDFpol1.0} are quoted
in Tabs.~\ref{tab:spin2}-\ref{tab:spin1}.}
\label{tab:fullmom}
\end{table}

It is clear that EIC pseudodata 
reduce all uncertainties significantly.
Note that
moving from \texttt{NNPDFpolEIC-A} to 
\texttt{NNPDFpolEIC-B} does not improve significantly
the uncertainty on quark-antiquark first moments, but it reduces the
uncertainty on the gluon first moment by a factor two.
However, it is worth noticing that, despite a reduction of the
uncertainty on  the 
gluon first moment, even for the most accurate
\texttt{NNPDFpolEIC-B} fit, the value remains compatible with zero
even though the central value is sizable (and negative).

In order to assess the residual extrapolation uncertainty 
on the singlet and gluon first moments, we determine the contribution
to them from the data range $x\in[10^{-3},1]$, \textit{i.e.} 
\begin{equation}
\langle\Delta\Sigma(Q^2)\rangle_{\mathrm{TR}}
\equiv
\int_{10^{-3}}^{1}dx\,\Delta\Sigma(x,Q^2)
\mbox{ ,}
\ \ \ \ \ \ \ \ \ \ 
\langle\Delta g(Q^2)\rangle_{\mathrm{TR}}
\equiv
\int_{10^{-3}}^{1}dx\,\Delta g(x,Q^2)
\mbox{ .}
\label{eq:trmoments}
\end{equation} 
The first moments, Eq.~(\ref{eq:trmoments}), are given in
Tab.~\ref{tab:trmomenta} at $Q_0^2=1$ GeV$^2$ 
and $Q^2=10$ GeV$^2$, where results for central values, uncertainties, and
correlation coefficients between the gluon and quark are collected. 
\begin{table}[t]
\footnotesize
\centering
\begin{tabular}{lccccc}
\toprule
& \multicolumn{2}{c}{$Q^2=1$ GeV$^2$}
& \multicolumn{3}{c}{$Q^2=10$ GeV$^2$}\\
& $\langle\Delta\Sigma(Q^2)\rangle_{\mathrm{TR}}$ 
& $\langle\Delta g(Q^2)\rangle_{\mathrm{TR}}$
& $\langle\Delta\Sigma(Q^2)\rangle_{\mathrm{TR}}$ 
& $\langle\Delta g(Q^2)\rangle_{\mathrm{TR}}$
& $\rho(Q^2)$ \\
\midrule
\texttt{NNPDFpol1.0}
&  $0.25\pm 0.09$
& $-0.26\pm 1.19$ 
&  $0.23\pm 0.16$ 
& $-0.06\pm 1.12$ 
& $+0.861$\\ 
\midrule
\texttt{NNPDFpolEIC-A}
&  $0.27\pm 0.06$
& $-0.53\pm 0.37$ 
&  $0.23\pm 0.05$ 
& $-0.59\pm 0.50$ 
& $-0.186$\\ 
\texttt{NNPDFpolEIC-B}
&  $0.24\pm 0.05$
& $-0.23\pm 0.25$ 
&  $0.22\pm 0.04$ 
& $-0.19\pm 0.32$ 
& $-0.103$\\
\bottomrule
\end{tabular}
\mycaption{The singlet and gluon truncated first moments and
their one-sigma uncertainties at $Q^2=1$ GeV$^2$ and $Q^2=10$ GeV$^2$ 
for the \texttt{NNPDFpolEIC-A} (left) 
and \texttt{NNPDFpolEIC-B} (right) PDF sets,
compared to \texttt{NNPDFpol1.0}.
The correlation coefficient $\rho$ at $Q^2=10$ GeV$^2$ 
is also provided.}
\label{tab:trmomenta}
\end{table}

Comparing the results at $Q^2=1$ GeV$^2$ of Tab.~\ref{tab:fullmom}
and Tab.~\ref{tab:trmomenta} with those 
in Tabs.~\ref{tab:spin2}-\ref{tab:spin1},
we see that in the \texttt{NNPDFpol1.0}
PDF determination for the quark singlet combination the uncertainty on 
the full first moment is about twice as large as that from the measured 
region, and for the gluon it is about four times as large. 
The difference is due to the extra uncertainty coming from the extrapolation. 
In \texttt{NNPDFpolEIC-B} the corresponding increases are by 20\% for the 
quark and 30\% for the gluon, which shows that thanks to EIC data 
the extrapolation uncertainties would be largely under control. 
The correlation coefficient $\rho$ significantly decreases upon 
inclusion of the EIC data: this means that the extra information 
contained in these data allows for an 
independent determination of the quark and gluon first moments.

\begin{figure}[t]
\begin{center}
\epsfig{width=0.40\textwidth, figure=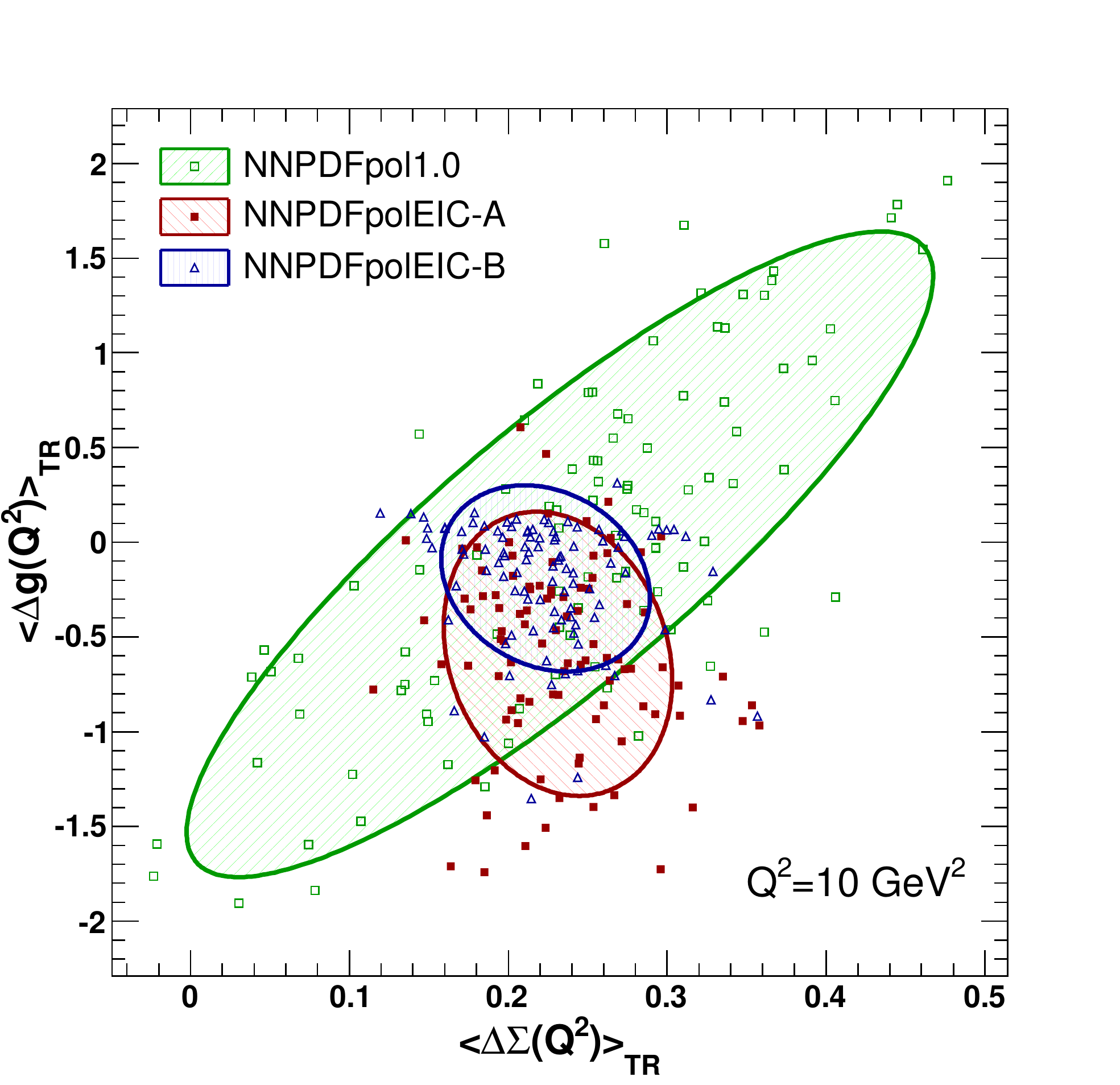}
\end{center}
\mycaption{One-sigma confidence region for the quark singlet and
gluon first moments in the measured region,
Eq.~(\ref{eq:trmoments}). The values for individual replicas are
also shown.}
\label{fig:momcor}
\end{figure}

\begin{figure}[t]
\centering
\epsfig{width=0.40\textwidth, figure=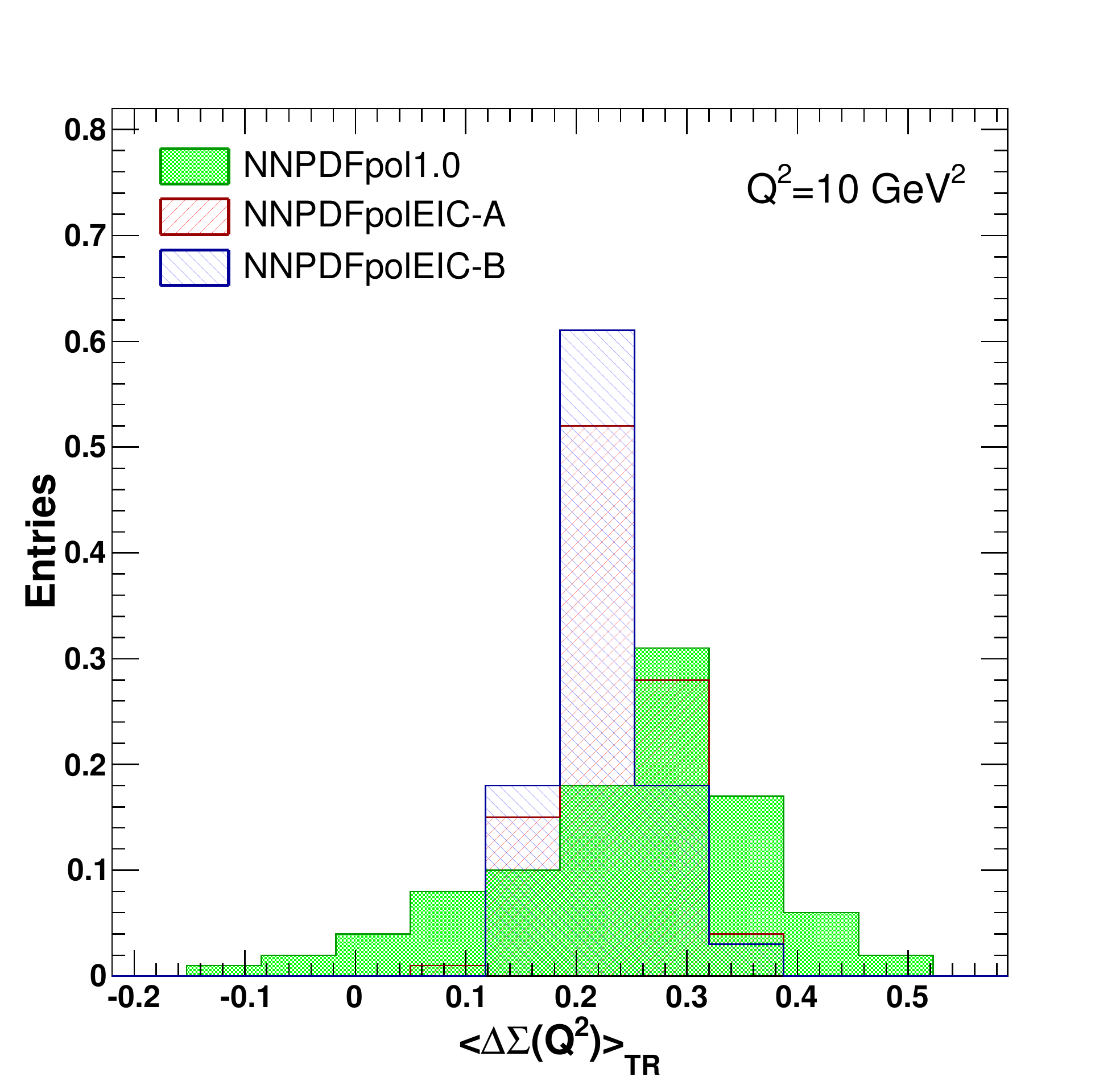}
\epsfig{width=0.40\textwidth, figure=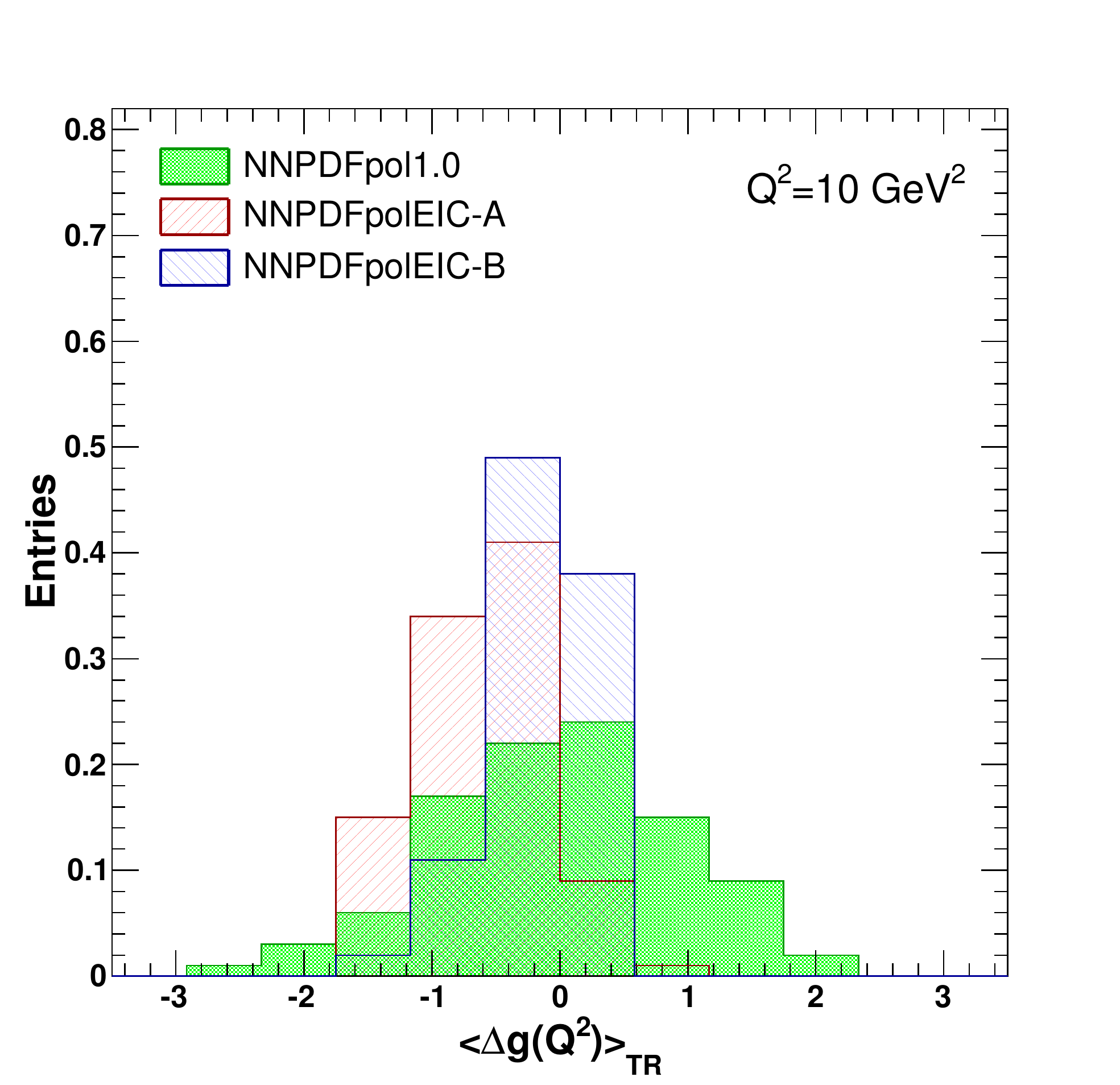}
\mycaption{Distributions of the singlet (left) and gluon (right) truncated 
first moments at $Q^2=10$ GeV$^2$ from a set of $N_{\mathrm{rep}}=100$
replicas in the \texttt{NNPDFpol1.0}, \texttt{NNPDFpolEIC-A} and 
\texttt{NNPDFpolEIC-B} parton ensembles.}
\label{fig:momdistEIC}
\end{figure}

In Fig.~\ref{fig:momcor}, we plot the one-sigma 
confidence region in the 
$(\langle\Delta\Sigma(Q^2)\rangle_{\mathrm{TR}},
\langle\Delta g(Q^2)\rangle_{\mathrm{TR}})$ plane
at $Q^2=10$ GeV$^2$, for 
\texttt{NNPDFpolEIC-A}, \texttt{NNPDFpolEIC-B} and \texttt{NNPDFpol1.0}.
Confidence regions are elliptical, since we have assumed that the truncated
moments are Gaussianly distributed among the $N_{\mathrm{rep}}=100$ replicas
in the PDF ensemble. This is a reasonable assumption for all the three 
parton sets we are considering here, as shown in Fig.~\ref{fig:momdistEIC}.
The main result of our analysis, Fig.~\ref{fig:momcor}, can be
directly compared to Fig.~8 of Ref.~\cite{Aschenauer:2012ve}, which was
based on the DSSV framework and is comparable to our \texttt{NNPDFpolEIC-B}
results.  
In both analyses EIC pseudodata determine the
singlet  first moment in the measured region with an uncertainty of 
about $\pm 0.05$. 

On the other hand, in Ref.~\cite{Aschenauer:2012ve} the uncertainty
on the gluon  was found to be about $\pm 0.02$, while we get a much larger
result of $\pm 0.30$.  One may wonder whether this difference may be
due at least in part to the fact that the DSSV fit on which the result of
Ref.~\cite{Aschenauer:2012ve} is based also includes jet production
and pion production data from RHIC, which may reduce the gluon
uncertainty. To answer this, we have computed the contribution to the
gluon first moment (again at  $Q^2=10$ GeV$^2$) from
the reduced region $0.05\le x\le 0.2$, where the RHIC data are
located. We find that the uncertainty on the contribution to the
gluon first moment in this restricted range is $\pm 0.083$ using
\texttt{NNPDFpolEIC-B}, while it is  $\pm 0.147$  with 
\texttt{NNPDFpol1.0} and ${}^{+0.129}_{-0.164}$ with
\texttt{DSSV+}~\cite{Aschenauer:2013woa}.  
We conclude that before the EIC data are added, the uncertainties
in \texttt{NNPDFpol1.0} and \texttt{DSSV+} are quite
similar despite the fact that \texttt{DSSV+} also includes RHIC data.
Hence, the larger gluon uncertainty we find for the
\texttt{NNPDFpolEIC-B} fit in comparison to
Ref.~\cite{Aschenauer:2012ve} is likely to be due to our more 
flexible PDF parametrization, though some difference might also come from
the fact that the SIDIS pseudodata included in 
Ref.~\cite{Aschenauer:2012ve} provide additional information 
on the gluon through scaling violations of the fragmentation 
structure function $g_1^h$. Of course this also introduces an uncertainty
related to the fragmentation functions which is difficult to quantify.

\subsection{Charm contribution to the \texorpdfstring{$g_1$}{g1} structure function}
\label{sec:g1charm}

In the QCD analysis of presently available DIS data, the contribution of heavy 
quarks to the polarized structure function $g_1$ is usually neglected. 
In both \texttt{NNPDFpol1.0} and \texttt{NNPDFpolEIC} polarized
PDF determinations, heavy quarks are dynamically generated above threshold 
by (massless) Altarelli-Parisi evolution in the ZM-VFN scheme (see 
Sec.~\ref{sec:QCDanalysis}). An exception to this treatment of heavy quark
masses is provided in Ref.~\cite{Blumlein:2010rn}, where charm quark production in 
the photon-gluon fusion process is treated at LO~\cite{Watson:1981ce} 
in the fixed-flavor number (FFN) scheme.

Intrinsic heavy quark effects are neglected in most analyses of polarized PDFs
since they have been shown to be relatively small already on the scale of
present-day unpolarized PDF uncertainties~\cite{Ball:2011mu},
which are rather smaller than their polarized counterparts.
However, as EIC data are expected to be far more accurate than those 
available so far, effects of finite heavy quark masses could be
at least non-negligible.
The treatment of these effects requires a proper 
quark mass scheme, for instance the \texttt{FONLL}
scheme, firstly introduced in Ref.~\cite{Cacciari:1998it}
and explicitly extended to DIS in Ref.~\cite{Forte:2010ta}.  
The method is based upon the idea of looking
at both the massless and massive scheme calculations as power expansions in
the strong coupling constant, and replacing the coefficient of the expansion in
the former with their exact massive counterpart in the latter, when available.
In order to suppress higher order contributions arising in the subtraction 
term near the threshold region, two prescriptions are proposed
in Ref.~\cite{Forte:2010ta}. One consists in damping
the subtraction term by a threshold factor which differs from unity by 
power-suppressed terms; the other consists in using a rescaling variable.
Both prescriptions introduce terms which are formally subleading with 
respect to the order of the calculation, therefore they do
not change its nominal accuracy, but they may in practice improve the 
perturbative stability and smoothness of the results.
\begin{figure}[t]
\centering
\epsfig{width=0.40\textwidth, figure=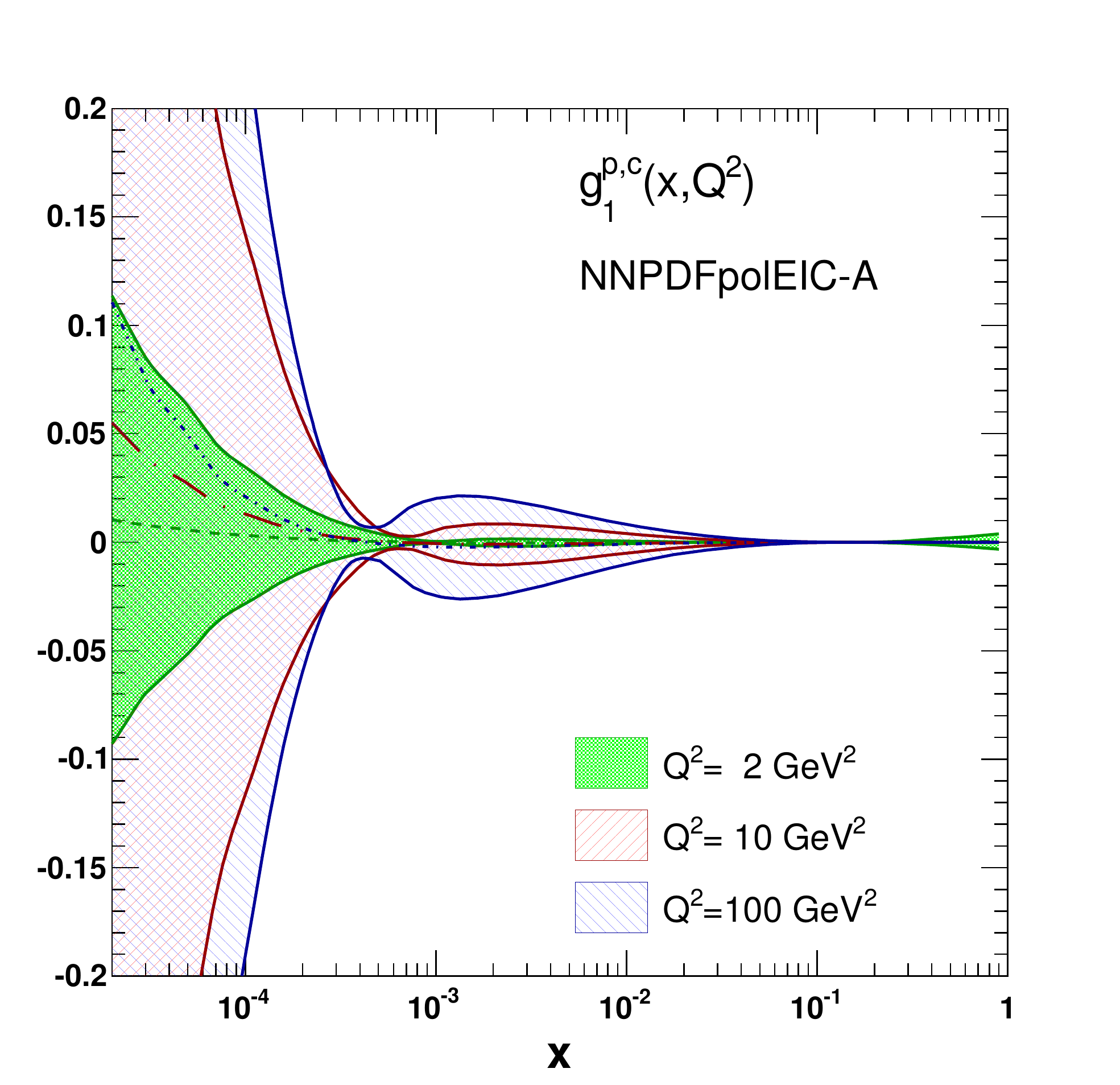}
\epsfig{width=0.40\textwidth, figure=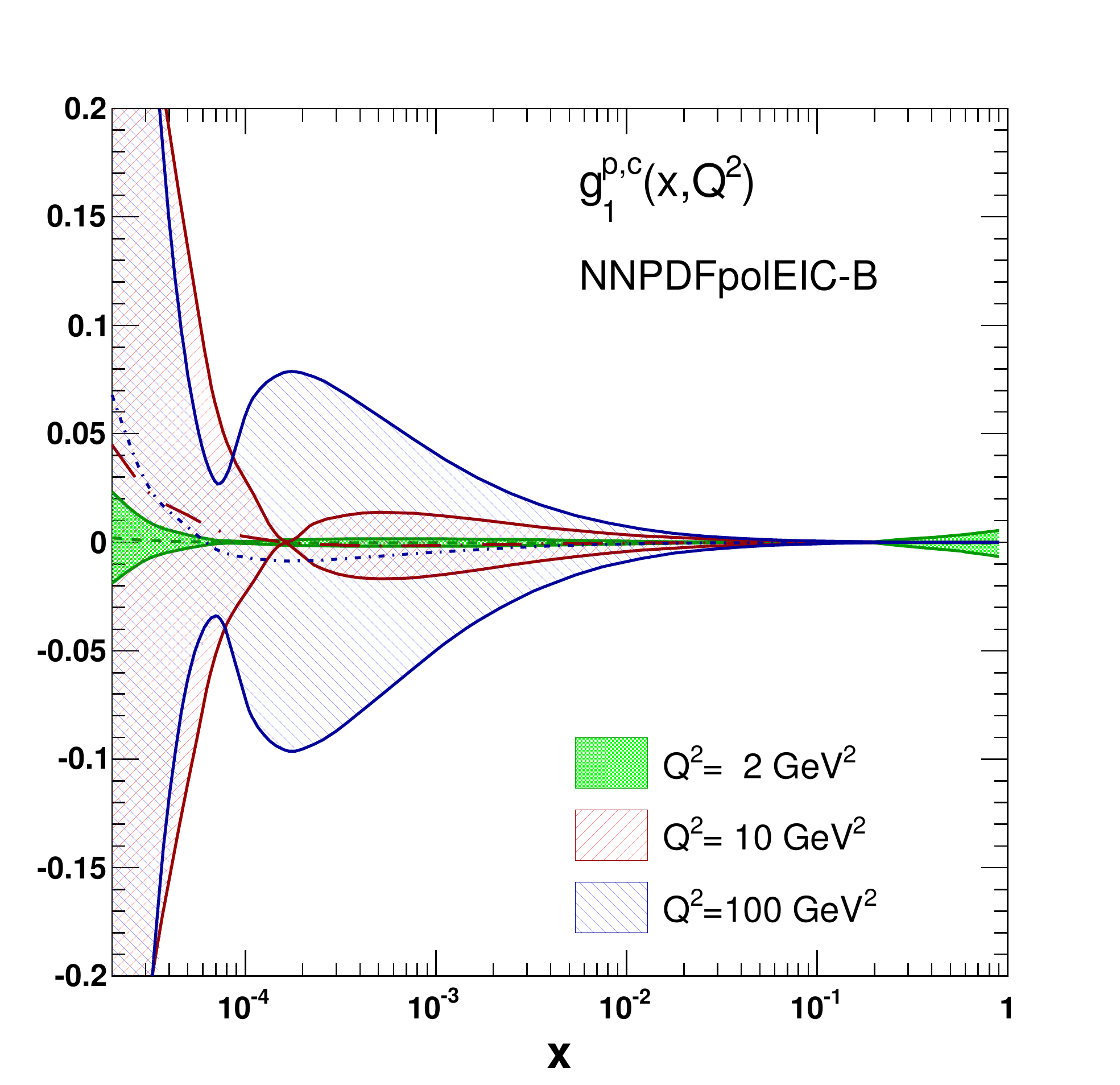}\\
\mycaption{The charm contribution $g_1^{p,c}(x,Q^2)$ to the DIS proton 
polarized structure function $g_1^p$ as a function of $x$ at 
three different energy scales $Q^2$. 
Results are shown for both the \texttt{NNPDFpolEIC-A} (left)
and \texttt{NNPDFpolEIC-B} (right) parton determinations.}
\label{fig:g1cplot}
\end{figure}
\begin{figure}[t]
\centering
\texttt{NNPDFpolEIC-A} 
\ \ \ \ \ \ \ \ \ \ \ \ \ \ \ \ \ \ \ \ \ \ \ \ \ \ \ \ \ \ \ \ 
\texttt{NNPDFpolEIC-B}\\
\epsfig{width=0.40\textwidth, figure=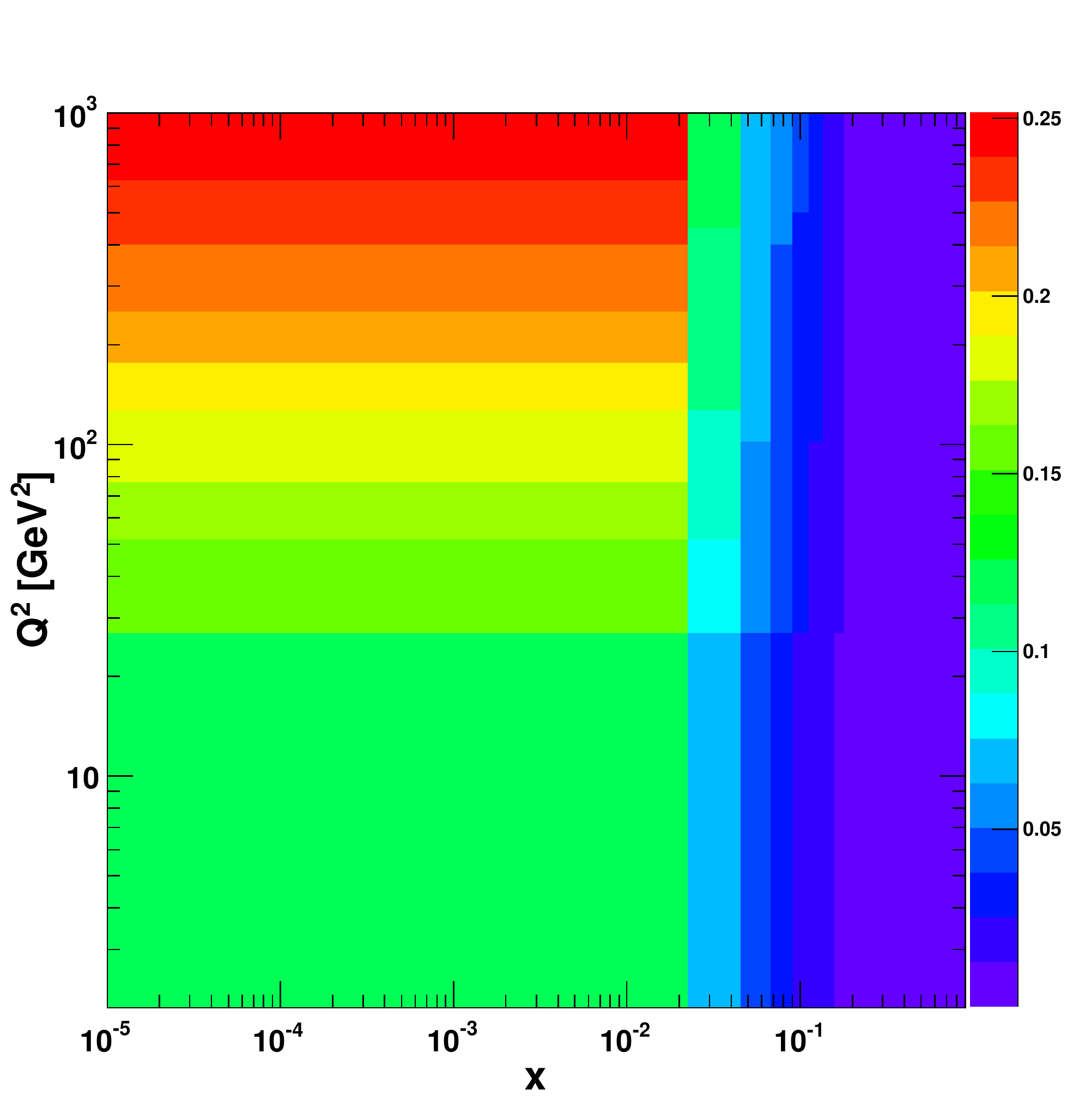}
\epsfig{width=0.40\textwidth, figure=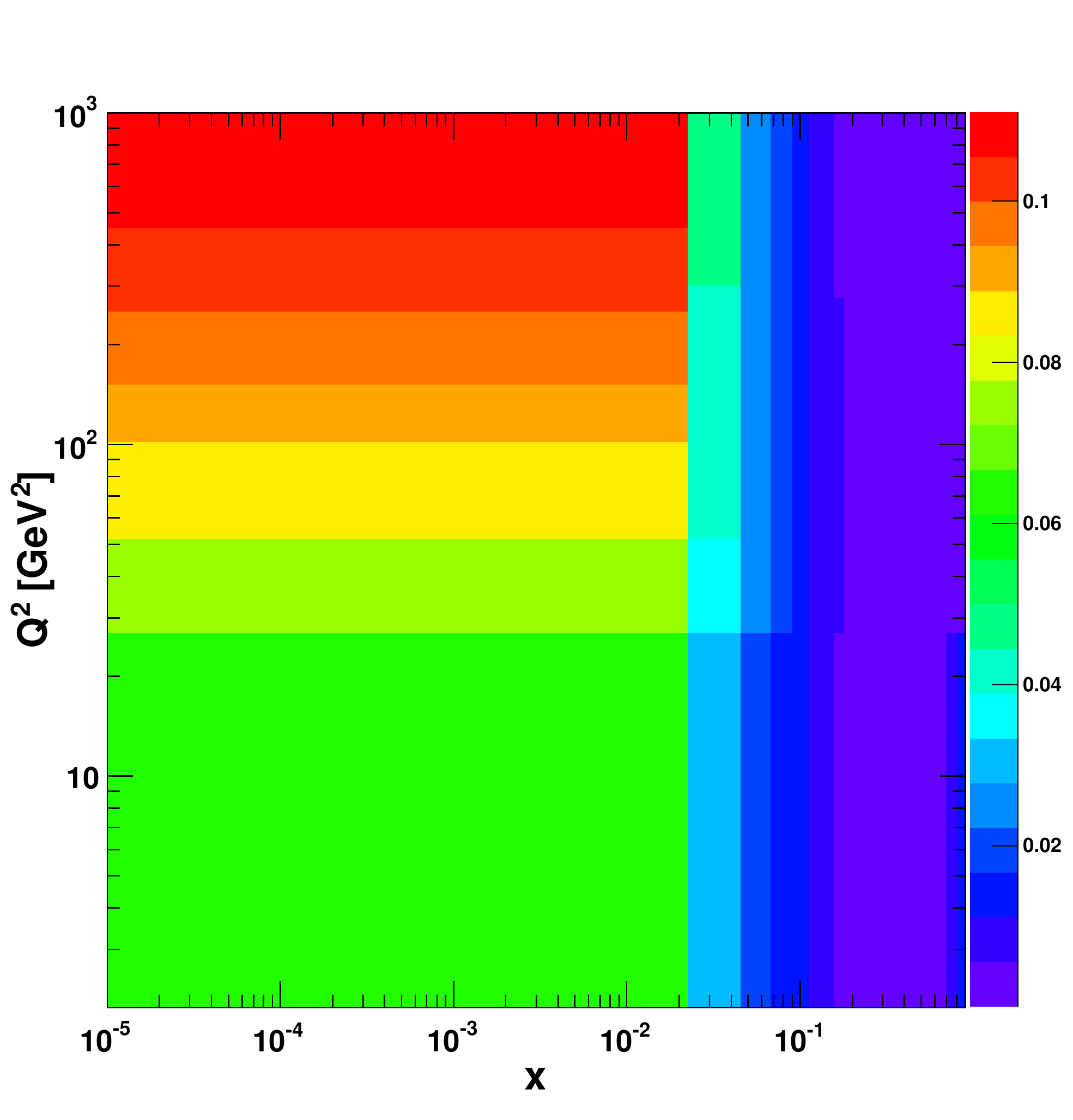}\\
\mycaption{The contour plots for the ratio $g_1^{p,c}(x,Q^2)/g_1^p(x,Q^2)$ 
from the fits to EIC pseudodata, \texttt{NNPDFpolEIC-A} (left)
and \texttt{NNPDFpolEIC-B} (right).}
\label{fig:g1cratio}
\end{figure}

Thanks to its simplicity, the \texttt{FONLL} scheme has been 
implemented in the \texttt{FastKernel} framework for the determination 
of unpolarized PDFs based on the NNPDF methodology~\cite{Ball:2011mu}.
The generalization of the \texttt{FONLL} scheme to the polarized 
structure function $g_1$ is quite simple: details are given
in Appendix~\ref{sec:appA}.
For spin-dependent DIS the charm contribution to the structure function $g_1$,
generated through the photon-gluon fusion process, $\gamma g\to c\bar{c}$, will
very much depend on the currently unknown size of $\Delta g(x,Q^2)$ 
at small $x$.
For instance, we plot in Fig.~\ref{fig:g1cplot} the expectations for the charm 
contribution $g_1^{p,c}$ to the proton structure function $g_1^p$, 
computed from the \texttt{NNPDFpolEIC-A} and \texttt{NNPDFpolEIC-B}
parton determinations. Results are displayed at three different energy scales,
namely $Q^2=2,10,100$ GeV$^2$. In Fig.~\ref{fig:g1cratio}, we also show the 
contour plot for the ratio $g_1^{p,c}(x,Q^2)/g_1^p(x,Q^2)$ for both these fits.
We conclude that the charm contribution to the proton structure function 
$g_1^{p,c}$, though being small, could be as much larger as $10$-$20\%$ of the
total $g_1^p$ in the kinematic region probed by an EIC. 
Hence, in order to further pin down the gluon uncertainty from
intrinsic charm effects, one should be able to measure its corresponding
contribution to the $g_1$ structure function within this accuracy.
Our result is consistent with that obtained in the \texttt{DSSV}
framework presented in Refs.~\cite{Boer:2011fh}.

In summary, the EIC data would entail a considerable reduction in the 
uncertainty on the polarized gluon PDF, they would provide first evidence for
its possible nontrivial $x$ shape and for its possible large contribution to
the nucleon spin. However, this goal would be reached with sizable residual
uncertainty: hence, the measurement of the charm contribution to the proton
structure function $g_1^{p,c}$, which is directly sensitive to the gluon, 
might provide more information on the $\Delta g$ distribution.

\chapter{Global determination of unbiased polarized PDFs}
\label{sec:chap5}

In this Chapter, we present a first global determination of polarized
parton distributions based on the NNPDF methodology: \texttt{NNPDFpol1.1}.
Compared to \texttt{NNPDFpol1.0}, the parton set determined in Chap.~\ref{sec:chap2},
\texttt{NNPDFpol1.1} is obtained using, on top of inclusive DIS data, also data from 
recent measurements of open-charm production in fixed-target DIS,
and of jet and $W$ production in proton-proton collisions. After
motivating our analysis in Sec.~\ref{sec:motivation11}, we will review
the theoretical description of these processes in Sec.~\ref{sec:thoverview}.
The features of the experimental data included in our analysis are then
presented in Sec.~\ref{sec:expinput11}. In Sec.~\ref{sec:reweighting11},
we discuss how the \texttt{NNPDFpol1.1} parton set is obtained via
Bayesian reweighting of prior PDF Monte Carlo ensembles, followed by
unweighting, as outlined in Sec.~\ref{sec:breweighting}.
We also present its main features in comparison to
\texttt{NNPDFpol1.0} and \texttt{DSSV08}. Finally, in Sec.~\ref{sec:pheno11},
we discuss some phenomenological implications of our new polarized parton set
with respect to the spin content of the proton.
Some of the results presented in this Chapter have appeared in 
preliminary form in Refs.~\cite{Nocera:2013spa,Nocera:2013yia}.

\section{Motivation}
\label{sec:motivation11}

The \texttt{NNPDFpol1.0} parton set presented in Chap.~\ref{sec:chap2}
is the first determination of polarized parton distributions
based on the NNPDF methodology. 
However, this is based on inclusive, neutral-current, DIS data only, 
which have two major drawbacks, as pointed out several times in this Thesis.
First, they do not allow for quark-antiquark separation; 
second, their kinematic coverage is rather
limited, both at small-$x$ and high-$Q^2$ values.
For this reason, all PDFs are affected by large uncertainties where
experimental data are not available. Besides, the polarized gluon 
PDF, determined through scaling violations in DIS, is almost
unconstrained because of the rather short $Q^2$ lever arm provided by data.

As discussed in Sec.~\ref{sec:generalstrategy}, one has to resort to
processes other than inclusive DIS to obtain further knowledge 
of polarized parton distributions. 
An impressive set of experimental data
have become available in the last years:
these include semi-inclusive DIS (SIDIS) data in fixed-target 
experiments~\cite{Ackerstaff:1999ey,Adeva:1997qz,Airapetian:2004zf,Alekseev:2007vi,Alekseev:2009ab},
one- or two-hadron and open-charm production data in lepton-nucleon 
scattering~\cite{Airapetian:1999ib,Airapetian:2010ac,Adeva:2004dh,Adolph:2012vj,Adolph:2012ca}, and semi-inclusive particle 
production~\cite{Adare:2008qb,Adare:2008aa,Adamczyk:2013yvv},
high-$p_T$ jet production~\cite{Adare:2010cc,Adamczyk:2012qj} and 
parity-violating $W^\pm$ boson production~\cite{Aggarwal:2010vc,Adare:2010xa}
data in polarized proton-proton collisions at RHIC.
As already summarized in Tab.~\ref{tab:polprocesses}, all these data 
are expected to probe different aspects of polarized PDFs:
semi-inclusive DIS and $W^\pm$ production data allow one
to determine the light quark-antiquark separation,
while jet and pion production data in polarized proton-proton
collisions, as well as hadron or open-charm electroproduction 
data in fixed-target experiments, give a handle on the size
and the shape of the polarized gluon distribution. A theoretical
description of the processes corresponding to these data
will be given in Sec.~\ref{sec:thoverview} below.

Nevertheless, all these measurements fall in the $x$ region already covered by
DIS data. Given that the uncertainties on the first moments of polarized PDFs, 
which eventually determine the contribution of each parton
to the total proton spin, are
already limited by the extrapolation into the unconstrained small-$x$ region 
(see the discussion in Sec.~\ref{sec:spinmom}), 
it is clear that only moderate improvements in these are expected from the
addition of new data other than DIS data.
Only a future high-energy polarized 
Electron-Ion Collider (EIC)~\cite{Boer:2011fh,Accardi:2012hwp,Deshpande:2005wd} 
would be likely to probe the small-$x$ regime of PDFs, and thus 
improve our knowledge of the polarized PDF first 
moments, as we demonstrated in Chap.~\ref{sec:chap4}.

Since no further progress is expected by the time an EIC
will start to operate, some effort has been devoted to perform 
\textit{global} determinations of polarized parton sets, including all
available experimental data. Presently, only two such sets are available:
\texttt{LSS10}~\cite{Leader:2010rb}, which also includes SIDIS 
beside inclusive DIS data, and those from the DSSV 
family~\cite{deFlorian:2009vb,deFlorian:2011ia,Aschenauer:2013woa}, 
which include, on top of these, also inclusive jet and
hadron production measurements 
from polarized proton-proton collisions at RHIC.
The goal of the analysis of this Chapter is to incorporate in the NNPDF
determination of polarized parton distributions
the new experimental information provided by 
some of the processes mentioned above, and thus release the first 
global polarized PDF set based on the NNPDF methodology: \texttt{NNPDFpol1.1}.

\section{Theoretical overview of polarized processes other than DIS}
\label{sec:thoverview}

Before addressing our global determination of a set of polarized 
parton distributions, we provide a theoretical overview of polarized 
processes other than DIS, available from both fixed-target and 
collider experiments.

\subsection{Semi-inclusive lepton-nucleon scattering 
in fixed-target experiments}
\label{sec:SIDISprobes}

The role of fixed-target lepton-nucleon scattering in determining the spin
structure of the nucleon is not restricted to inclusive reactions,
as those we have considered so far. Actually, more exclusive processes,
in which one measures one or more outgoing final state particles, can be 
used to constrain polarized sea quark distributions and to gain some
more knowledge on the polarized gluon distribution. 
In the following, we discuss in turn the potential of 
each of these processes, namely semi-inclusive DIS and
heavy flavor hadron production.

\begin{list}{}{\leftmargin=0pt}

\item {\textbf{Semi-inclusive DIS.}}
Semi-inclusive DIS (SIDIS) is a DIS process in which a hadron $h$,
originated by the fragmentation of the struck quark, 
is detected in the final state:
\begin{equation}
l(\ell)+N(P)\to l^\prime(\ell^\prime)+h(P_h)+X(P_X)
\,\mbox{,}
\label{eq:SIDISgeneral}
\end{equation}
where we use the same notation adopted in Eq.~(\ref{eq:DIS}), supplemented with
the four-momentum of the final hadron, $P_h$.
Due to the statistical correlation between the flavor of the struck quark and 
the type of the hadron formed in the fragmentation process,
semi-inclusive DIS with identified pions or kaons 
may provide a handle on the $\Delta\bar{u}$, $\Delta\bar{d}$ and 
$\Delta\bar{s}$ parton distributions, respectively~\cite{Airapetian:2004zf}. 
For example, roughly speaking the presence of a $\pi^+$ in the final state 
indicates that it is likely that a $u$-quark or a $\bar{d}$-antiquark was struck in the
scattering, because the $\pi^+$ is a $(u\bar{d})$ bound state.

The theoretical description of SIDIS, with longitudinally polarized
lepton beams, closely follows that of
inclusive DIS given in Chap.~\ref{sec:chap1}. In analogy to
Eq.~(\ref{eq:cross_section3}), the SIDIS differential 
cross-section can be written as
\begin{equation}
\frac{d^6\sigma}{dxdyd\phi dzdp_T^2d\Phi}
\propto
L_{\mu\nu}W_h^{\mu\nu}
\,\mbox{;}
\label{eq:SIDISxsec}
\end{equation}
here the leptonic tensor, $L_{\mu\nu}$, has exactly the form given in 
Eq.~(\ref{eq:lept_tens1}) with the symmetric and antisymmetric parts of
Eqs.~(\ref{eq:lept_tens3})-(\ref{eq:lept_tens5}), while the hadronic
tensor, $W_h^{\mu\nu}$, now contains additional degrees of freedom
corresponding to the fractional energy $z$ of the final-state hadron,
the $p_T$ component of the final hadron three momentum, transverse to that
of the virtual photon, and the azimuthal angle $\Phi$ of the hadron
production plane relative to the lepton scattering plane.
Integration over $\Phi$ and $p_T^2$ produces the cross-section relevant for
the experimental observables. These are the longitudinal and transverse 
asymmetries $A_\parallel^h$ and $A_\perp^h$, defined in analogy to
Eq.~(\ref{eq:xsecasy}).

Thanks to factorization, hadronic and leptonic degrees of freedom are
separated, hence kinematic factors depending only on $x$ and $y$ (or $Q^2$)
are carried over directly from inclusive scattering in relating the
measured asymmetries to their virtual photo-absorption counterparts,
Eq.~(\ref{eq:asyrel}). In particular, $A_1$ and $A_2$, 
Eqs.~(\ref{eq:gammaasy}), should now read $A_1^h$ and $A_2^h$ in terms of
the SIDIS cross-sections $\sigma_{1/2,3/2}^h$ of produced hadrons of type $h$.
In particular, we obtain
\begin{equation}
A_1^h(x,Q^2,z)
=
\frac{\sum_q e^2_q\Delta q(x,Q^2) \otimes D_q^h(z,Q^2)}
{\sum_q e^2_{q^\prime}q^\prime(x,Q^2) \otimes D_{q^\prime}^h(z,Q^2)}
\,\mbox{,}
\label{eq:A1SIDIS}
\end{equation}
where $e_q$ are the quark electric charges,
$\Delta q$ ($q$) are the helicity-dependent (-averaged) PDFs,
$D_q^h$ is the fragmentation function for the quark $q$ to
fragment into a hadron $h$, and $\otimes$ denotes the 
convolution product, Eq.~(\ref{eq:convolution}).

Measurements of spin-dependent asymmetries in SIDIS have been performed 
by several experimental collaborations, namely SMC~\cite{Adeva:1997qz},
HERMES~\cite{Airapetian:2004zf}
and COMPASS~\cite{Alekseev:2007vi,Alekseev:2009ac,Alekseev:2010ub},
and have been included in some global determinations
of polarized parton distributions~\cite{Leader:2010rb,deFlorian:2009vb}.
We notice that the analysis of these data requires the
usage of fragmentation functions, which have to be determined in turn
from experimental data. Usually, they are extracted, independently of PDFs,
from electron-positron annihilation, proton-proton collisions 
and possibly SIDIS data (see Ref.~\cite{Albino:2008gy} for a review).
Despite the considerable experimental and theoretical effort
which has gone into the determination of sets of fragmentation 
functions~\cite{Kretzer:2000yf,Kniehl:2000fe,Bourhis:2000gs,Hirai:2007cx,
Albino:2005me,deFlorian:2007aj,deFlorian:2007hc,Albino:2008fy},
their knowledge is still rather poor.
In particular, recent work~\cite{d'Enterria:2013vba} has emphasized
the failure of all available sets of fragmentation functions in describing
the most updated inclusive charged-particle spectra data at the LHC.
For these reasons, when used in a global determination of polarized
PDFs including SIDIS data, they are likely to introduce an uncertainty 
which is difficult to quantify.

\item {\textbf{Heavy flavor hadron production.}}
Heavy flavor hadron production is clearly sensitive to the 
shape and the size of the spin-dependent gluon 
distribution~\cite{Frixione:1996ym,Stratmann:1996xy}. 
In this case, the gluon polarization can be 
accessed via the photon-gluon fusion (PGF) mechanism,
which results in the production of a quark-antiquark pair,
see Fig.~\ref{fig:gluecorrections}-$(d)$.
Experimental signatures to tag this partonic subprocess are the production
of one or two hadrons with high $p_T$ in the final state, 
and open-charm events: in this case, 
the $q\bar{q}$ pair is required to be a $c\bar{c}$ pair and an outgoing 
charmed meson is reconstructed. At LO, the virtual photon-nucleon asymmetry
for open-charm production, $A_{LL}^{\gamma N\to D^0X}$, is expressed as
\begin{equation}
A_{LL}^{\gamma N \to D^0 X}\equiv\frac{\Delta\sigma_{\gamma N}}{\sigma_{\gamma N}}
      =\frac{\Delta\hat{\sigma}_{\gamma g}\otimes\Delta g\otimes D_c^{D^0}}
            {\hat{\sigma}_{\gamma g}\otimes g\otimes D_c^{D^0}}
\,\mbox{,}
\label{eq:ALLgammaN}
\end{equation} 
where $\Delta\hat{\sigma}_{\gamma g}$ ($\hat{\sigma}_{\gamma g}$) is 
the spin-dependent (-averaged) partonic cross-section
for PGF, $\gamma^* g \to c\bar{c}$, 
$\Delta g$ ($g$) is the polarized 
(unpolarized) gluon PDF, 
and $D_{c}^{D^0}$ is the non-perturbative fragmentation
function of a produced charm quark into the observed $D^0$ meson, 
which is assumed to be spin independent. 
In principle, a measurement of the asymmetry in Eq.~(\ref{eq:ALLgammaN}) 
can then provide a direct handle on $\Delta g$.

Measurements of longitudinal spin asymmetries in high-$p_T$ hadron production
were performed by HERMES~\cite{Airapetian:1999ib,Airapetian:2010ac} at DESY 
and by SMC~\cite{Adeva:2004dh} and COMPASS~\cite{Adolph:2012vj} at CERN.
However, in such a reaction, the measured asymmetries receive contributions 
not only from pure PGF events, 
but also from a significant fraction of 
background events, mainly due to the
two competing processes of gluon radiation by QCD Compton scattering
($\gamma^*q\to qg$) and photon absorption at the lowest order of DIS 
($\gamma^*q\to q$). 
At variance with hadron-pair production, open-charm production is free of
background, since the PGF subprocess is the main mechanism for producing
charm quarks in polarized DIS.

The proper theoretical description of these processes depends on the 
virtuality of the probing photon. In case of
photoproduction, where a quasi-real photon is exchanged, on top of direct 
contributions~\cite{Bojak:1998bd,Bojak:1998zm,Merebashvili:2000ya,Contogouris:2000en},
one has to include also \textit{resolved} 
contributions~\cite{Riedl:2009ye,Bojak:2001fx}, 
where the photon fluctuates into a vector meson of the same quantum numbers 
before the hard scattering with partons in
the proton takes place. If the virtuality $Q$ of the photon is
of $\mathcal{O}$(1~GeV) or higher, resolved processes are sufficiently
suppressed but the additional momentum scale $Q$ greatly
complicates the calculations of phase-space and loop integrals.

Only few calculations are available for 
heavy flavor hadron production
in case of polarized beams and targets at NLO 
accuracy~\cite{Bojak:1998bd,Bojak:1998zm,Frixione:1996ym,Stratmann:1996xy}.
A complete phenomenological study of charm quark photoproduction 
in longitudinally polarized lepton-hadron collisions at NLO accuracy  
has been presented only very recently~\cite{Riedl:2012qc}.
For the first time, both direct and resolved photon contributions have been included 
there to compute the relevant cross-sections 
for spin-dependent heavy flavor hadroproduction.
For this reason,  
the available data on hadron and open-charm 
production~\cite{Airapetian:1999ib,Airapetian:2010ac,Adeva:2004dh,Adolph:2012vj,Adolph:2012ca} 
have not been included in global
QCD analyses of polarized parton distributions so far.
Experimental collaborations have analyzed their data only in terms of the
gluon polarization, $\Delta g(x, Q^2)/g(x, Q^2)$, under certain simplifying 
assumptions and based on leading order matrix elements. 
Nonetheless, the results of these exercises,
illustrated in Fig.~\ref{fig:ratiog_NNPDF}, 
are in fairly good agreement with the 
NLO prediction obtained from our \texttt{NNPDFpol1.0} analysis.
We will explicitly show to which extent COMPASS open-charm
data can further pin down the polarized gluon uncertainty
in Sec.~\ref{sec:reweighting-separate}.
\end{list}
\begin{figure}[t]
\begin{center}
\epsfig{width=0.50\textwidth,figure=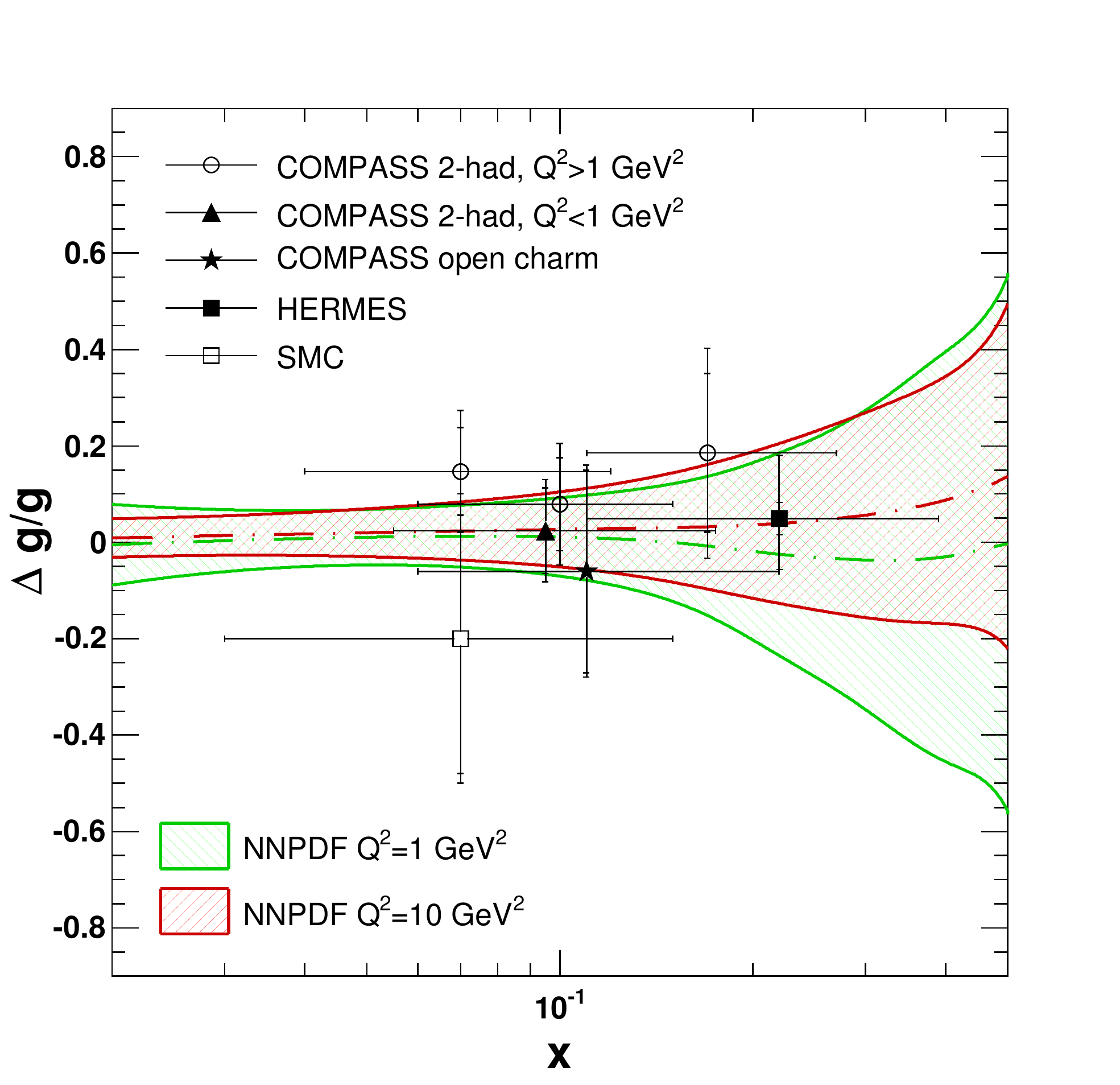}
\mycaption{The theoretical prediction for the ratio $\Delta g(x,Q^2)/g(x,Q^2)$
computed from the polarized (unpolarized) \texttt{NNPDFpol1.0} 
(\texttt{NNPDF2.3}) parton sets at NLO, compared to LO determinations from
one- or two-hadron and open-charm production data in fixed-target DIS 
experiments~\cite{Airapetian:1999ib,Airapetian:2010ac,Adeva:2004dh,Adolph:2012vj,Adolph:2012ca}.}
\label{fig:ratiog_NNPDF}
\end{center}
\end{figure}

\subsection{Spin asymmetries in proton-proton collisions}
\label{sec:PPprobes}

High-energy collisions from longitudinally polarized proton beams, 
as available at the Relativistic Heavy-Ion Collider (RHIC),
provide a unique way to probe proton spin structure
and dynamics~\cite{Bourrely:1993dd,Bunce:2000uv}.
Typically, the quantities measured at RHIC are spin asymmetries.
As an example, for collisions of longitudinally polarized proton beams, 
one defines the double-spin asymmetry for a given process as
\begin{equation}
A_{LL}=\frac{\sigma^{++}-\sigma^{+-}}{\sigma^{++}+\sigma^{+-}}
\equiv\frac{\Delta\sigma}{\sigma}
\,\mbox{,}
\label{eq:RHICasy}
\end{equation}
where $\sigma^{++}$ ($\sigma^{+-}$) is the cross-section for the process 
with equal (opposite) proton beam polarizations. 
As for collisions of unpolarized proton beams, spin-dependent  
inelastic cross-sections factorize into convolutions of polarized 
parton distribution functions of the proton and hard-scattering 
cross-sections describing the spin-dependent interactions of partons.  
Similarly to inclusive DIS, one can 
write schematically
\begin{eqnarray}
\sigma&=&\sum_{a,b,(c)=q,\bar{q},g} f_a \otimes  f_b 
(\otimes D_c^H) \otimes \hat{\sigma}^{(c)}_{ab}
\,\mbox{,}
\label{eq:numRHIC}
\\
\Delta\sigma&=&\sum_{a,b,(c)=q,\bar{q},g}\Delta f_a \otimes \Delta f_b 
(\otimes D_c^H) \otimes \Delta\hat{\sigma}^{(c)}_{ab}
\,\mbox{,}
\label{eq:denRHIC}
\end{eqnarray}
for the denominator and the numerator in Eq.~(\ref{eq:RHICasy}) respectively.
Here, the sum is over all contributing partonic channels
$a + b \to c (\to H) + X$ producing the desired high-$p_T$, large-invariant mass 
final state (or detected hadron $H$). 
As usual, $\otimes$ denotes the convolution, Eq.~(\ref{eq:convolution}),
between unpolarized (polarized) parton distributions, $f_{a,b}$ 
($\Delta f_{a,b}$), parton-to-hadron fragmentation function, $D_c^H$ (only
for those processes with identified hadrons in final states), and 
elementary spin-averaged (-dependent) hard partonic cross-section $\hat{\sigma}^{(c)}_{ab}$
($\Delta\hat{\sigma}^{(c)}_{ab}$). In particular, the spin-dependent 
partonic cross-section is defined as
\begin{equation}
\Delta\hat{\sigma}^{(c)}_{ab}
\equiv
\frac{1}{2}\left[\hat{\sigma}_{ab}^{++}-\hat{\sigma}_{ab}^{+-} \right]
\,\mbox{,}
\label{eq:partonicxsecRHIC}
\end{equation}
the signs denoting the helicity states of the initial partons $a$, $b$. 

At RHIC, there are a number of processes
which allow for the measurement of spin asymmetries like that in
Eq.~(\ref{eq:RHICasy}). Depending on the dominant partonic subrocess,
they can probe different aspects of the nucleon spin structure.
For instance, some of them allow for a clean determination of gluon polarizations, 
while others are more sensitive to quark and antiquark helicity states.
We will discuss below the main processes for which 
measurements of spin asymmetries at RHIC are presently available.
We do not give here details about other measurements 
possible at RHIC, but not yet performed, such as the observation of
high-$p_T$ or \textit{prompt} photon production, heavy-flavor production 
and Drell-Yan production of lepton pairs.
We refer to~\cite{Bunce:2000uv} for a theoretical review on them.

\begin{list}{}{\leftmargin=0pt}

\item {\textbf{High-\texorpdfstring{$p_T$}{pT} inclusive jet production double-spin asymmetry.}}
A clean and theoretically robust process to probe
the polarized gluon PDF, $\Delta g$, is inclusive 
jet production, thanks to the 
dominance of the $gg$ and $qg$ initiated subprocesses in the 
accessible kinematic range~\cite{Bourrely:1990pz,Chiappetta:1992cp}
(see also Tab.~\ref{tab:polprocesses}).
The situation is analogous to the unpolarized case, where inclusive
jet production data from the Tevatron and the 
LHC~\cite{Aaltonen:2008eq,Abazov:2008ae,Aad:2011fc,Chatrchyan:2012bja} 
are instrumental in pinning down the medium- and large-$x$ gluon
behavior.

In polarized collisions, the relevant experimental observable in jet production 
is the longitudinal double-spin asymmetry defined along Eq.~(\ref{eq:RHICasy})
\begin{equation}
A_{LL}^{1jet}=\frac{\sigma^{++}-\sigma^{+-}}{\sigma^{++}+\sigma^{+-}}
\,\mbox{.}
\label{eq:ALL-jet}
\end{equation}
For dijet production, 
 at LO, the parton kinematics is given by
\begin{equation}
x_{1}=\frac{p_T}{2\sqrt{s}}\left( e^{\eta_3} + e^{\eta_4} \right) 
\,\mbox{,}
\ \ \ \ \ \ \ \ \ \ 
x_{2}=\frac{p_T}{2\sqrt{s}}\left( e^{-\eta_3} + e^{-\eta_4} \right)
\mbox{ ,}
\label{eq:part-kin-jet}
\end{equation}
where $p_T$ is the transverse jet momentum, $\eta_{3,4}$ are the rapidities of 
the two jets and $\sqrt{s}$ is the center-of-mass energy.
In single-inclusive jet production, the underlying Born kinematics 
is not uniquely determined because the second jet is being integrated out.
For the illustrative purposes of Fig.~\ref{fig:NNPDFpol11-kin}, we will use 
the following expression to characterize the Born kinematics 
\begin{equation}
x_{1,2}=\frac{p_T}{\sqrt{s}} e^{\pm \eta} \, ,
\end{equation}
with $\eta$ the rapidity of the leading jet, which corresponds to
Eq.~(\ref{eq:part-kin-jet}), provided the incoming partons
carry an equal amount of longitudinal momentum and thus $\eta_3=-\eta_4$. 
This is a good approximation at RHIC, due to the limited coverage 
in rapidity as compared to unpolarized hadron colliders.

The calculation of both the numerator and the denominator in 
Eq.~(\ref{eq:ALL-jet}) requires the definition of a suitable jet algorithm.
Also, notice that corrections up to NLO accuracy should be included in the computation of 
jet cross-sections, through the algorithm for jet reconstruction,
since it is only at NLO that the QCD structure of the jet starts 
to play a role in the theoretical description. This would also provide
for the first time the possibility of realistically matching the 
procedures used in experiment to group final-state particles into jets.

At NLO, a large number of infrared divergencies 
are found in the computation of virtual and
real diagram contributions to the jet cross-section,
due to the large number of color-interacting, massless
partons involved in the hard-scattering processes. 
It is then necessary to devise a procedure to perform the calculation 
of the divergent parts and to show their cancellation in the sum 
which defines any infrared-safe physical observable.
Several independent methods to calculate any infrared-safe quantity
in any kind of hard unpolarized collision are available in the 
literature~\cite{Giele:1993dj,Frixione:1995ms,Catani:1996vz}.
In particular, the subtraction method of Ref.~\cite{Frixione:1995ms}
allows for organizing the computation is such a way 
that the singularities are extracted and canceled by hand, while the remainder
may be integrated numerically over phase space.
This approach has the advantage of being
very flexible; it may be used for any infrared-safe observable, 
with any experimental cut.
On the other hand, the numerical integration involved turns out to 
be rather delicate and time-consuming. 
The subtraction method
has been used in Ref.~\cite{Frixione:1997np} to develop a computer code
that generates partonic events and outputs the momenta of the final-state partons
which can be eventually used to define the physical observables
for one ore more jet production in proton-proton collisions.
Such a \textit{parton generator} is not 
equivalent to the usual Monte Carlo parton shower programs, since 
it is the result of a fixed-order QCD calculation.
The subtraction method and the computer code of 
Refs.~\cite{Frixione:1995ms,Frixione:1997np},
supplemented with the proper matrix elements~\cite{Bern:1991aq,Kunszt:1993sd},
was then extended to the 
case of polarized proton-proton collisions in Ref.~\cite{deFlorian:1998qp}.

The spin-dependent (and spin-averaged) cross-section for 
single-inclusive high-$p_T$ jet production is also available from
Ref.~\cite{Jager:2004jh}. In comparison to Ref.~\cite{deFlorian:1998qp},
the approach of Ref.~\cite{Jager:2004jh} uses a largely analytic technique
for deriving the relevant partonic cross-sections, which becomes
possible if one assumes the jet to be a rather narrow object. This 
assumption is equivalent to the approximation that the cone 
opening $R$ of the jet is not too large, and hence was termed
\textit{small-cone approximation} (SCA) in~\cite{Jager:2004jh}.
In the SCA, one systematically expands the partonic cross-sections
around $R=0$. The dependence on $R$ is of the form 
$\mathcal{A}\log R+\mathcal{B}+\mathcal{O}(R^2)$: the coefficients
$\mathcal{A}$ and $\mathcal{B}$ are retained and calculated analytically,
whereas the remaining terms $\mathcal{O}(R^2)$ and beyond are neglected.
The advantage of this procedure is that it leads to much faster and
more efficient computer code, since 
all singularities arising in intermediate steps have 
explicitly canceled and are not subject to delicate numerical treatments.
The SCA was shown~\cite{Jager:2004jh} to produce results
comparable to those of Ref.~\cite{deFlorian:1998qp}
(whose formalism applies to arbitrary cone openings), 
in the kinematic range where RHIC data are available.
For these reasons, the code provided in Ref.~\cite{Jager:2004jh} is better 
suited than that of Ref.~\cite{deFlorian:1998qp} for the intensive computations
required to include jet data in a global QCD fit.

\item {\textbf{Semi-inclusive \texorpdfstring{$\pi^0$}{pi0} production double-spin asymmetry.}}
In order to constrain the polarized gluon distribution, one can look 
for high-$p_T$ leading hadrons such as $\pi^-$, $\pi^0$, $\pi^+$,
whose production proceeds through the same partonic subprocesses
involved in jet production, in particular $qg\to qg$ and $gg\to qg$
(see Tab.~\ref{tab:polprocesses}). However, the hadronization of the 
struck parton into the final, measured, pion is described by a 
non-perturbative fragmentation function, $D_c^\pi$: this enters the 
theoretical description of the corresponding double-spin asymmetry,
as encompassed in Eqs.~(\ref{eq:RHICasy})-(\ref{eq:numRHIC})-(\ref{eq:denRHIC}).

Next-to-leading order QCD corrections to the spin-dependent cross-section 
for single-inclusive hadron production in proton-proton collisions have been 
computed for the first time in Refs.~\cite{deFlorian:2002az,Jager:2002xm}.
As dictated by Eqs.~(\ref{eq:numRHIC})-(\ref{eq:denRHIC}), 
we need to sum over all possible final states in each channel $ab\to cX$,
in compliance with the requirement of single-inclusiveness of the cross-section. 
For instance, in case of $qg\to qX$ one needs, besides the virtual corrections to 
$qg\to qg$, three different $2\to 3$ reactions: $qg\to q(gg)$, $qg\to q(q\bar{q})$, 
$qg\to q(q^\prime\bar{q}^\prime)$ (where brackets indicate the unobserved
parton pair). The combination of all three processes together will allow 
for obtaining a finite result.

The two computations presented in Refs.~\cite{deFlorian:2002az,Jager:2002xm}
differ from each other in the way the integration
over the entire phase space of the two unobserved partons,
in the $2\to 3$ contributions, is carried out. 
In Ref.~\cite{deFlorian:2002az}, this calculation is performed numerically,
by extending the already mentioned
subtraction method of Refs.~\cite{Frixione:1995ms,Frixione:1997np}
to the case of single-hadron production observables. A computer code 
customized to compute any infrared-safe quantity corresponding to 
one-hadron production at NLO accuracy is then presented as a result.
Conversely, in Ref.~\cite{Jager:2002xm}, the phase-space integration 
of the $2\to 3$ contributions is performed analitically.
As already noticed in the case of jet production, this 
has the main advantage to obtain much faster and more efficient
computer code, which is better suited for the intensive 
computations required by a global QCD fit of polarized parton distributions.

\item {\textbf{Small-\texorpdfstring{$p_T$}{pT} single-spin \texorpdfstring{$W^\pm$}{W+/-} production asymmetry.}}
Production of $W^\pm$ bosons in high energy collisions 
from longitudinally polarized proton beams
provides an ideal tool for the study of individual helicity states of
quarks and antiquarks inside the proton, 
complementary to, but independent of, SIDIS~\cite{Bunce:2000uv}.
Within the stadard model, the process $\overrightarrow{p}p\rightarrow W^\pm X$ 
(the arrow denotes the polarized proton beam)
is driven by a purely weak interaction which couples  
left-handed quarks with right-handed antiquarks only
($u_L\bar{d}_R \rightarrow W^+$ and $d_L\bar{u}_R \rightarrow W^-$,
with some contamination from $s$, $c$, $\bar{s}$ and $\bar{c}$, 
mostly through quark mixing),
thus giving rise to a $W$ parity-violating longitudinal single-spin
asymmetry, sensitive to $\Delta q$ and $\Delta\bar{q}$ flavor dependence.
This asymmetry is defined as
\begin{equation}
 A_L
\equiv 
\frac{\sigma^+ - \sigma^-}{\sigma^+ + \sigma^-}
=
\frac{\Delta\sigma}{\sigma}
\,\mbox{, }
\label{eq:Wasy}
\end{equation}
where $\sigma^{+(-)}$ denotes the cross-section for colliding 
of positive (negative)
longitudinally polarized protons off unpolarized protons.
Notice that this definition differs from that provided in 
Eq.~(\ref{eq:RHICasy}), since only one of the two proton beams 
is polarized. 

If we consider the simplest parton-level process $u\bar{d}\to W^+$ 
at LO, Eq.~(\ref{eq:Wasy}) will read~\cite{Bunce:2000uv}
\begin{equation}
A_L^{W^+}\approx\frac{\Delta u(x_1)\bar{d}(x_2) - \Delta\bar{d}(x_1) u(x_2)}
{u(x_1)\bar{d}(x_2) + \bar{d}(x_1)u(x_2)}\mbox{, }
\label{eq:Wasy+pm}
\end{equation}
where $x_1$ and $x_2$ are the momentum fractions,
carried by quarks and antiquarks,
related to $y_W$, the $W$ boson rapidity relative to
the polarized proton, and to $\sqrt{s}$, the hadronic center-of-mass 
energy, by the relation
\begin{equation}
 x_{1,2}=\frac{M_W}{\sqrt{s}}e^{\pm y_W}\mbox{.}
 \label{eq:x1x2}
\end{equation}
The measurement of the rapidity distribution of the $W$ bosons
thus provide a direct handle on the flavor-separated polarized
quark and antiquark distributions. Indeed, at large rapidities, 
$y_W\gg 0$, where sea distributions are suppressed because $x_1$ is in the 
valence region, the asymmetry $A_L^{W^+}$ 
is given by $\Delta u/u$, whereas it approaches $\Delta\bar{d}/\bar{d}$ 
in the opposite limit, $y_W\ll 0$.
The situation is similar for $W^-$ production, now interchanging 
the roles of $u$ and $d$ flavors.

The naive Born-level picture given above needs to be modified to account 
for additional aspects, both theoretical and experimental~\cite{Bunce:2000uv}.
The former include higher-order perturbative corrections
and other allowed  initial states, including Cabibbo-suppressed channels. 
Concerning the latter, $W$ bosons 
are reconstructed through their leptonic decays
$W^\pm \to e^\pm \nu$ at RHIC, therefore the observed process is actually 
$pp\to \ell^\pm X$, with the neutrino escaping undetected. 
One will then measure the rapidity distributions
of the charged leptons rather than of the $W$ bosons themselves.
All these issues have been taken into account in the NLO 
calculation of the cross-section and longitudinal single-spin asymmetry,
Eq.~(\ref{eq:Wasy}), presented in Ref.~\cite{deFlorian:2010aa}
as a computer program: it may be readily used 
to include experimental spin
asymmetry data in a global analysis of polarized parton densities.

\end{list}

\section{Experimental input}
\label{sec:expinput11}

Among the processes described in Sec.~\ref{sec:thoverview}, in the 
present analysis we only consider open-charm production from COMPASS
and single-inclusive high-$p_T$ jet and $W$ production from RHIC.
Actually, the precise knowledge of fragmentation functions plays a 
minor role in the theoretical description of these processes. 
On the one hand, in the kinematic regime accessed by COMPASS, 
open-charm production shows only a slight dependence 
on the fragmentation of the charm 
quark into a $D$ meson; on the other hand, the theoretical description
of jet and $W$ production asymmetries in proton-proton collisions
does not involve fragmentation into identified hadrons in the final state.
We prefer not to include data whose analysis requires the usage of 
fragmentation functions since these are poorly known objects: 
hence, they are likely 
to affect our unbiased PDFs by an uncertainty difficult to quantify,
as discussed in Sec.~\ref{sec:SIDISprobes}.

\begin{figure}[t]
\begin{center}
\epsfig{width=0.50\textwidth,figure=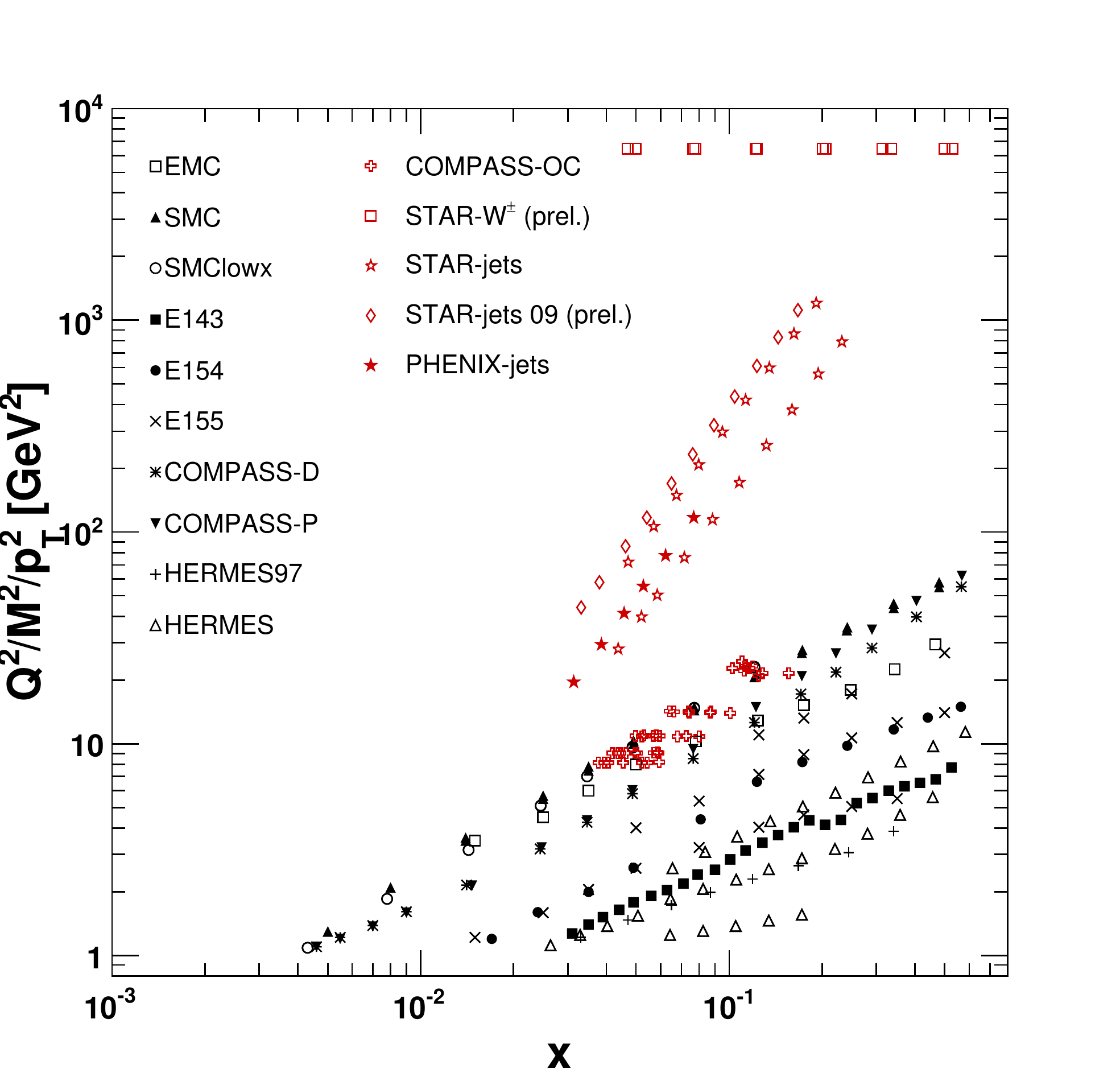}
\mycaption{Kinematic coverage in the $(x,Q^2)$ plane of experimental data
included in the \texttt{NNPDFpol1.1} parton set.
New data are listed in the second column.}
\label{fig:NNPDFpol11-kin}
\end{center}
\end{figure}

In Fig.~\ref{fig:NNPDFpol11-kin}, we plot the new data points considered in 
the present analysis, together with inclusive DIS data already 
included in the fit presented in Chap.~\ref{sec:chap2}. 
More details about leading partonic subprocesses, probed polarized PDFs, 
and the ranges of $x$ and $Q^2$ that become accessible were already 
summarized in Tab.~\ref{tab:polprocesses}. 
We will discuss below the main features of the new data sets, separately 
for each process.
\begin{list}{}{\leftmargin=0pt}

\item {\textbf{Open-charm production at COMPASS.}}
The COMPASS collaboration has recently presented experimental 
results for the photon-nucleon asymmetry
$A_{LL}^{\gamma N \to D^0 X}$, Eq.~(\ref{eq:ALLgammaN}), obtained 
by scattering polarized muons of energy $E_{\mu}=160$ GeV$^2$ 
(center-of-mass energy roughly $\sqrt{s}\sim 18$ GeV$^2$)
off longitudinally polarized 
protons or deuterons from a $^6$LiD or NH$_3$ targets, in the photoproduction 
regime (photon virtuality roughly $Q^2\sim 0$ GeV$^2$)~\cite{Adolph:2012ca}.
A detailed description of the experimental setup can be found in 
Ref.~\cite{Abbon:2007pq}. 
Three different data sets, each one of $N_{\mathrm{dat}}=15$ data points,
were presented there, depending on which
$D^0$ decay mode was assumed to reconstruct the charmed hadron 
from the observed final states: 
$D^0\to K^-\pi^+$, $D^0\to K^-\pi^+\pi^0$ or $D^0\to K^-\pi^+\pi^+\pi^-$.
In the following, they will be referred to as
COMPASS~$K1\pi$, COMPASS~$K2\pi$ and COMPASS~$K3\pi$
respectively. Experimental correlations between systematic uncertainties
are not provided.
Assuming LO kinematics, the experiment may probe the polarized gluon distribution 
at medium momentum fraction values, $0.06 \lesssim x \lesssim 0.22$,
and at energy scale $Q^2 = 4(m_c^2+p_T^2) \sim 13$ GeV$^2$, 
where $m_c$ is the charm quark mass 
and $p_{T}$ is the transverse momentum of the produced charmed hadron,
see Fig.~\ref{fig:NNPDFpol11-kin}.

\item {\textbf{High-$p_T$ jet production at STAR and PHENIX.}}
Both the STAR and PHENIX experiments at RHIC provided 
their measurements of the longitudinal double-spin asymmetry
for inclusive jet production, Eq.~(\ref{eq:ALL-jet}). 
This is obtained by colliding two polarized proton beams at 
center-of-mass energy $\sqrt{s}=200$ GeV. We refer 
to~\cite{Ackermann:2002ad} and references therein for a detailed 
description of the RHIC experimental setup.
Data from STAR are available for the 2005 and 2006 runs and,
in a preliminary form, also for the most recent 2009 run; only one set
of data is available for PHENIX, corresponding to data taken in 2005.
In the following, they will be referred to as STAR 1j-05, STAR 1j-06,
STAR 1j-09 and PHENIX 1j respectively. 
The features of these data sets, 
including the number of data points $N_{\mathrm{dat}}$, jet-finding
algorithm and corresponding cone radius $R$ used for data reconstruction, 
covered range in integrated rapidity $\eta$ 
and intergrated luminosity $\mathcal{L}$,  
are summarized in Tab.~\ref{tab:jet-data}. 
Experimental correlations between systematic uncertainties
are not provided.
Within LO kinematics, jet data roughly cover the range 
$0.05\lesssim x \lesssim 0.2$ and $30\lesssim p_T^2\lesssim 800$ GeV$^2$,
see Fig.~\ref{fig:NNPDFpol11-kin}.
\begin{table}[t]
\footnotesize
\centering
\begin{tabular}{lcccccc}
\toprule
Data set & $N_{\mathrm{dat}}$ & jet-algorithm & $R$  & 
$[\eta_{\mathrm{min}},\eta_{\mathrm{max}}]$ &
$\mathcal{L}$ [pb$^{-1}$] & Ref. \\
\midrule
STAR 1j-05 & $10$ & midpoint-cone & $0.4$ & $[+0.20,+0.80]$ &
 $2.1$ & \cite{Adamczyk:2012qj} \\
STAR 1j-06 & $9$ & midpoint-cone & $0.7$ & $[-0.70,+0.90]$ &
 $5.5$ & \cite{Adamczyk:2012qj} \\
STAR 1j-09 (prel.)  & $11$ & midpoint-cone & $0.7$ & $[-1.00,+1.00]$ &
 $25$ & \cite{Djawotho:2013pga} \\
\midrule
PHENIX 1j & $6$ & seeded-cone & $0.3$ & $[-0.35,+0.35]$ &
 $2.1$ & \cite{Adare:2010cc} \\
\bottomrule
\end{tabular}
\mycaption{Some features of the jet data included in the present analysis:
the number of available data points, $N_{\mathrm{dat}}$, the algortihm used 
for jet reconstruction, the range over which the 
rapidity $\eta$ is integrated and the integrated luminosity, $\mathcal{L}$.}
\label{tab:jet-data}
\end{table}
\item {\textbf{$W$ boson production at STAR.}}
Both the STAR and PHENIX collaborations at RHIC presented first measurements
of the parity-violating spin asymmetry $A_L^{W^\pm}$, Eq.~(\ref{eq:Wasy}), 
based on the 2009 run at 
$\sqrt{s}=500$ GeV.~\cite{Aggarwal:2010vc,Adare:2010xa}.
Unfortunately, due to the low integrated luminosity 
($\mathcal{L}=12$ pb$^{-1}$ and $\mathcal{L}=8.6$ pb$^{-1}$ for
STAR and PHENIX respectively), 
these data will have little impact in determining polarized
antiquark flavors, once included in a polarized parton set.
More interestingly, the STAR collaboration has recently presented preliminary
results for the asymmetry $A_L^{W^\pm}$, based on data collected in 2012
at $\sqrt{s}=510$ GeV and with an integrated luminosity 
$\mathcal{L}=72$ pb$^{-1}$~\cite{Stevens:2013raa}.
Two data sets are provided in six bins of the lepton rapidity $\eta$ and 
at an integrated lepton transverse momentum $25<p_T<50$ GeV,
separately for $W^+$ and $W^-$: they will be referred to
as STAR-$W^+$ and STAR-$W^-$ henceforth.
Given STAR kinematics, these data sets are likely to constrain light 
antiquark PDFs roughly in the momentum fraction interval
$0.05\lesssim x \lesssim 0.4$ and at the energy scale of the $W$ mass
(see Fig.~\ref{fig:NNPDFpol11-kin}).
Besides uncorrelated statistical uncertainties, the measured asymmetries
are provided with a correlated systematic uncertainty related to the
uncertainty in the beam polarization, given as the $3.4\%$
of the measured asymmetry. Other uncorrelated systematics, due to
background and relative luminosity, are estimated to be 
less than $10\%$ of statistical errors in the preliminar STAR
analysis~\cite{Stevens:2013raa}. Even though these uncertainties are
expected to become smaller in the final data released by STAR,
we conservately assume them to be as large as $10\%$.
\end{list}

\section{Determination of parton distributions}
\label{sec:reweighting11}

In this Section, we illustrate how the \texttt{NNPDFpol1.0} parton set
determined in Chap.~\ref{sec:chap2} is supplemented with the piece of
experimental information discussed above. Instead of a global
refit of parton distributions including the new data, we will use 
PDF reweighting, followed by unweighting: this methodology was presented
in Refs.~\cite{Ball:2010gb,Ball:2011gg} and its main features were
summarized in Sec.~\ref{sec:breweighting}.

We recall that parton sets determined trough the NNPDF methodology
are provided as Monte Carlo ensembles made of equally probable PDF replicas,
each fitted to a data replica generated according to the uncertainties 
and the corresponding correlations measured in the experiments.
The number of replicas in a given ensemble, $N_{\mathrm{rep}}$, is determined
by requiring that the central values, uncertainties and correlations of the original 
experimental data can be reproduced to a
given accuracy by taking averages, variances and covariances over the replica sample.
For the case of polarized PDFs, in Sec.~\ref{sec:MCgeneration}
we determined that a Monte Carlo
sample of pseudodata with $N_{\mathrm{rep}}=100$ replicas 
is sufficient to reproduce 
the mean values and the errors of experimental data to 
an accuracy which is better than $5\%$. Only a moderate 
improvement was observed in going up to $N_{\mathrm{rep}}=1000$,
hence our default choice was $N_{\mathrm{rep}}=100$.
The PDF ensemble forms an accurate representation of the underlying
probability distribution of PDFs, conditional on the input data and 
the particular assumptions (such as the details of the QCD analysis)
used in the fit.
Based on statistical inference and Bayes theorem, PDF reweighting cosists 
in assigning to each replica in a Monte Carlo ensemble of PDFs, which 
is referred to as the \textit{prior} ensemble, a weight which assesses
the likelihood that this particular replica agrees with the new data.
We refer to Sec.~\ref{sec:breweighting} for details about
the way the weights are determined from the $\chi^2$ of the new data 
to the prediction obtained using a given replica in the prior.
The theoretical bases of the reweighting methodology were 
carefully checked in Refs.~\cite{Ball:2010gb,Ball:2011gg} and 
it was also shown that results obtained via global refitting 
or reweighting with new data are statistically equivalent between 
each other. 

The main limitation of the reweighting method is that the information brought 
in by new data should be only a moderate correction as compared
to the information already included in the prior PDF ensemble.
This is precisely our present case, 
since we will add to our fit a few dozens of new hadronic data points,
as we discussed in Sec.~\ref{sec:expinput11}.
On the other hand, the reweighting methodology has the main advantage
to avoid global refitting: in particular, for each PDF replica in the prior ensemble, 
the lenghty computation of observables has to be performed only once,
instead of at each minimization step required in the case of refitting.
This is of particular relevance for the polarized case,
since the \texttt{FastKernel} method~\cite{Ball:2010de}, used in 
Sec.~\ref{sec:QCDanalysis} to perform fast evolution and fast computation of 
inclusive DIS observables, has not yet been implemented for the polarized 
hadronic processes considered here. 
Alternatively, exact NLO calculations could be directly used in a global PDF fit
by extending to polarized observables the \texttt{FastNLO}
framework~\cite{Kluge:2006xs} and the general-purpose interface
\texttt{APPLgrid}~\cite{Carli:2010rw}, but this too is 
not yet available.

The main steps of the procedure we follow to determine 
the \texttt{NNPDFpol1.1} parton set are described below.

\subsection{Construction of the prior PDF ensemble}
\label{sec:prior11}

Our goal is to include the piece of new exerimental information 
discussed in Sec.~\ref{sec:expinput11} into 
the determination of polarized parton distributions presented in 
Chap.~\ref{sec:chap2} via Bayesian reweighting. To this purpose, 
we have to compute the theoretical predictions for the measured observables,
\textit{i.e.} the longitudinal spin asymmetries 
Eqs.~(\ref{eq:ALLgammaN})-(\ref{eq:ALL-jet})-(\ref{eq:Wasy}),
based on PDF replicas in \texttt{NNPDFpol1.0}.
The $\chi^2$ of the new data to the prediction obtained 
using a given replica in \texttt{NNPDFpol1.0} is then used to compute the weight
corresponding to this replica according to Eq.~(\ref{eq:weightformula}).
Unfortunately, the \texttt{NNPDFpol1.0} parton set determined in 
Chap.~\ref{sec:chap2} cannot be used to compute the 
theoretical predictions for the asymmetries straightforwardly:
at NLO accuracy, this computation requires the knowledge of quark and 
antiquark distributions separately, which are not provided
by the \texttt{NNPDFpol1.0} parton set, because it was 
determined from a fit to inclusive DIS data only. 

A separation of the polarized quark and antiquark 
distributions can be achieved in a global fit including SIDIS data,
as done in the \texttt{DSSV08}~\cite{deFlorian:2009vb}
and \texttt{LSS10}~\cite{Leader:2010rb} analyses. 
In principle, we could perform a new, global fit to inclusive and
semi-inclusive DIS data and then use the corresponding Monte Carlo
ensemble of parton distributions as the prior for the reweighting
with collider data presented in Sec.~\ref{sec:expinput11}.
However, as already noticed many times in this thesis, the analysis
of SIDIS data requires the usage of poorly known fragmentation 
functions, which may significantly affect the accuracy of our results.
For consistency, we should determine a set of fragmentation functions
based on the NNPDF methodology and then use this set to perform a fit 
including SIDIS data, but this is beyond the scope of the present 
analysis.

Be that as it may, we circumvent the issue related to the 
lack of quark-antiquark 
separation in \texttt{NNPDFpol1.0} using a different approach, which will
be described here for the first time. The idea is to supplement the 
information available in the \texttt{NNPDFpol1.0} parton set for the 
$\Delta u^+$, $\Delta d^+$, $\Delta s^+$ and $\Delta g$ distributions
with some assumptions on $\Delta\bar{u}$ and $\Delta\bar{d}$ in order 
to construct a suitable prior ensemble. 
This is a sensible approach, provided that different ansatz for the 
$\Delta\bar{u}$ and $\Delta\bar{d}$ prior distributions lead to the same
final result after including the new data by reweighting.
Indeed, if new data brings in a sufficient 
amount of information, the final reweighted PDFs will be independent 
of the original choice of prior~\cite{Giele:1998gw,Giele:2001mr}.
Of course, it will be essential to show explicitly that this is 
what happens in our particular situation.

In principle, we could think of making the quark-antiquark separation 
for $u$ and $d$ flavors in the prior arbitrarily: for example, we could assign
to each replica random values for its $\Delta u$, $\Delta d$, $\Delta\bar{u}$ and
$\Delta\bar{d}$ distributions, provided their sum reproduces the 
corresponding total distributions determined in \texttt{NNPDFpol1.0} and they
separately satisfy theoretical constraints (see Sec.~\ref{sec:thconstraints}).
However, such a prior will lead to highly unefficient reweighting, in the 
sense that the new underlying PDF probability distribution after reweighting 
will be sampled by a too small number of effective replicas (for details, see 
Sec.~\ref{sec:breweighting}). In order to avoid this loss of efficiency,
prior ensembles with a huge number of replicas should be produced,
but this would be extremely demanding in terms of computational resources.

Alternatively, the additional information on quark-antiquark separation 
needed to construct suitable prior ensembles
can be obtained from one of the aformentioned fits to SIDIS data.
we can allow for deviations from the corresponding best-fit 
$\Delta\bar{u}$ and $\Delta\bar{d}$ determinations, 
by supplementing them with additional statistical noise 
and uncertainties, until the independence 
of the reweighted results from the prior is achieved. 
The loss of efficiency of these priors will be under control, thus
not requiring to be huge, as we will demonstrate below.

We now discuss in practice how we construct suitable prior ensembles
to be afterwards reweighted with the new data discussed in 
Sec.~\ref{sec:expinput11}. We supplement the \texttt{NNPDFpol1.0}
parton set with the information on $\Delta\bar{u}$ and $\Delta\bar{d}$
distributions from the \texttt{DSSV08} 
parton fit~\cite{deFlorian:2009vb}, which includes all available SIDIS data,
and we construct a collection of independent prior PDF ensembles. 
First of all, we sample the \texttt{DSSV08} $\Delta\bar{u}$ and 
$\Delta\bar{d}$ parton distributions at a fixed reference scale 
$Q_0^2=1$ GeV$^2$. 
We select ten points, half logaritmically and half linearly spaced 
in the interval of momentum fraction $10^{-3}\lesssim x \lesssim 0.4$, 
which roughly corresponds to the range covered by SIDIS experimental 
data relevant for separating quark-antiquark contributions.  
We sample four independent sets of data points assuming the
\texttt{DSSV08} best fit plus one, two, three or four times its nominal    
$\Delta\chi^2=1$ Hessian uncertainty. Separate prior PDF ensembles,
labelled as $1\sigma$, $2\sigma$, $3\sigma$ and $4\sigma$ henceforth,
will then be constructed for each one of these data sets.
Of course, a different ansatz could be made on data:
the rationale we followed was to increase their uncertainty 
until independence of the reweighted results from 
the prior was reached. 
We will explicitly show that this requirement is fulfilled
at least by the $3\sigma$ and $4\sigma$ prior ensembles
at the end of this Section.

Data points sampled from the \texttt{DSSV08} fit are then treated,
separately for $\Delta\bar{u}$ and for $\Delta\bar{d}$,
as sets of \textit{experimental} pseudo-observables.
Henceforth, they will be labelled as \texttt{DSSV08}$_{U}$ and 
\texttt{DSSV08}$_{D}$ respectively.
More precisely, we generate $N_{\mathrm{rep}}=1000$ replicas 
of the original \texttt{DSSV08} pseudodata,
following the procedure described in Sec.~\ref{sec:expdata}, 
and then for each individual replica we perform a neural network 
fit to them: the result gives the $\Delta\bar{u}$ and $\Delta\bar{d}$
distributions with wich we supplement the \texttt{NNPDFpol1.0}
parton set.  
In order to meaningfully fit the $\Delta\bar{u}$
and $\Delta\bar{d}$ pseudodata, we need to supplement
the input PDF basis given in Sec.~\ref{sec:minim},
namely $\Delta \Sigma$, $\Delta T_3$, $\Delta T_8$ and $\Delta g$,  
with two new linearly independent light quark combinations;   
we choose them to be
the total valence, $\Delta V$, and the valence isotriplet, $\Delta V_3$,
\begin{eqnarray}
\Delta V(x,Q_0^2) &=& \Delta u^-(x,Q_0^2)+\Delta d^-(x,Q_0^2)
\,\mbox{,} 
\label{eq:PDFbasis1}
\\
\Delta V_3(x,Q_0^2) &=& \Delta u^-(x,Q_0^2)-\Delta d^-(x,Q_0^2)
\,\mbox{,} 
\label{eq:PDFbasis2}
\end{eqnarray}
where $\Delta q^-=\Delta q-\Delta\bar{q}\mbox{, } q=u,d$.
In addition, 
we have assumed that $\Delta s = \Delta\bar{s}$.
Even though data are presently insufficient to discriminate
between any guess on strage-antistrange distributions and there are
actually no theoretical motivations to support 
a symmetric polarized strangeness,
we adopt the choice $\Delta s =\Delta\bar{s}$ as it is usual in
all polarized PDF analyses. We emphasize that the distribution 
which is physically meaningful is instead the total strange combination
$\Delta s^+$ which was already determined from inclusive DIS data 
in \texttt{NNPDFpol1.0}, see Chap~\ref{sec:chap2}.

Each of the new PDF combinations in 
Eqs.~(\ref{eq:PDFbasis1})-(\ref{eq:PDFbasis2}) 
is  parametrized as usual by means of a neural network supplemented 
with a preprocessing polynomial, 
\begin{eqnarray}
\Delta V(x,Q_0^2) 
&=& 
(1-x)^{m_{\Delta V}} x^{n_{\Delta V}} \textrm{NN}_{\Delta V}(x)
\,\mbox{,}
\label{eq:NNparam1} 
\\
\Delta V_3(x,Q_0^2) 
&=& 
(1-x)^{m_{\Delta V_3}} x^{n_{\Delta V_3}} \textrm{NN}_{\Delta V_3}(x)
\,\mbox{,}
\label{eq:NNparam2}
\end{eqnarray}
where $\textrm{NN}_{\Delta \textrm{pdf}}$, $\textrm{pdf}=V, V_3$, is the output of 
the neural network and the preprocessing exponents $m,n$ are linearly 
randomized for each
Monte Carlo replica within the ranges given in Tab.~\ref{tab:prepexp}.
We have checked that our choice of preprocessing exponents does not bias the
fit, according to the procedure discussed in Sec.~\ref{sec:minim}.
The neural network architecture is the same as in \texttt{NNPDFpol1.0}, namely 
2-5-3-1, see Sec.~\ref{sec:NNparametrization}.
Note that in terms of the  quark PDF input basis, the $\Delta\bar{u}$
and $\Delta\bar{d}$ distributions which here play the role of pseudo-data
are given by the following combinations:
\begin{eqnarray}
\Delta \bar{u}(x,Q_0^2) 
&=& \frac{1}{12}\left( 2\Delta \Sigma  + 3\Delta T_3
+\Delta T_8 -3\Delta V -3\Delta V_3  \right)
(x,Q_0^2)
\,\mbox{,} 
\label{eq:relbasis1}
\\
\Delta \bar{d}(x,Q_0^2) 
&=& \frac{1}{12}\left( 2\Delta \Sigma  - 3\Delta T_3
+\Delta T_8 -3\Delta V +3\Delta V_3  \right)
(x,Q_0^2)
\,\mbox{.}
\label{eq:relbasis2}
\end{eqnarray}
For each pseudodata replica $\Delta \bar{u}^{(k)}$, $\Delta \bar{d}^{(k)}$ 
in Eqs.~(\ref{eq:relbasis1})-(\ref{eq:relbasis2}), $k=1,\ldots,N_{\mathrm{rep}}$, we supplement the
neural networks for $\Delta V$ and $\Delta V_3$ that are being
fitted with random replicas from \texttt{NNPDFpol1.0}.
All the prior PDF ensembles are composed of $N_{\mathrm{rep}}=1000$
replicas; this larger number of PDF members, in comparison to 
that used in the analysis presented in Chap.~\ref{sec:chap2},
where $N_{\mathrm{rep}}=100$, is 
required to ensure that replicas left after reweighting  
still describe the underlying PDF probability distribution
with sufficient accuracy;
$N_{\mathrm{rep}}=1000$ replicas of \texttt{NNPDFpol1.0} 
were generated with this purpose.
\begin{table}[t]
 \centering
 \footnotesize
 \begin{tabular}{ccc}
  \toprule
   PDF & $m$ & $n$ \\
  \midrule
   $\Delta V(x,Q_0^2)$   & [1.5,3.0] & [0.05,0.60]\\
   $\Delta V_3(x,Q_0^2)$ & [1.5,3.0] & [0.01,0.60]\\
  \bottomrule
 \end{tabular}
 \mycaption{Ranges for the small- and large-$x$ preprocessing exponents 
in Eqs.~(\ref{eq:NNparam1})-(\ref{eq:NNparam2}).}
 \label{tab:prepexp}
\end{table}

In these fits to pseudodata, the
minimization is performed by means of a genetic 
algorithm, as discussed in Secs.~\ref{sec:minstop}-\ref{sec:genmin}.
The implementation of theoretical constraints,
both positivity and integrability, also follows consistently the procedure from
Ref.~\cite{Ball:2013lla}
in order to take care of flavor and antiflavor separation.
In particular, Eqs.~(\ref{eq:possigma})-(\ref{eq:lagrmult}) have been enforced
by letting $f=u, \bar{u}, d, \bar{d}$ separately. 
Note that no additional sum rules affect $\Delta V$ and $\Delta V_3$.

\begin{figure}[t]
\begin{center}
\epsfig{width=0.40\textwidth,figure=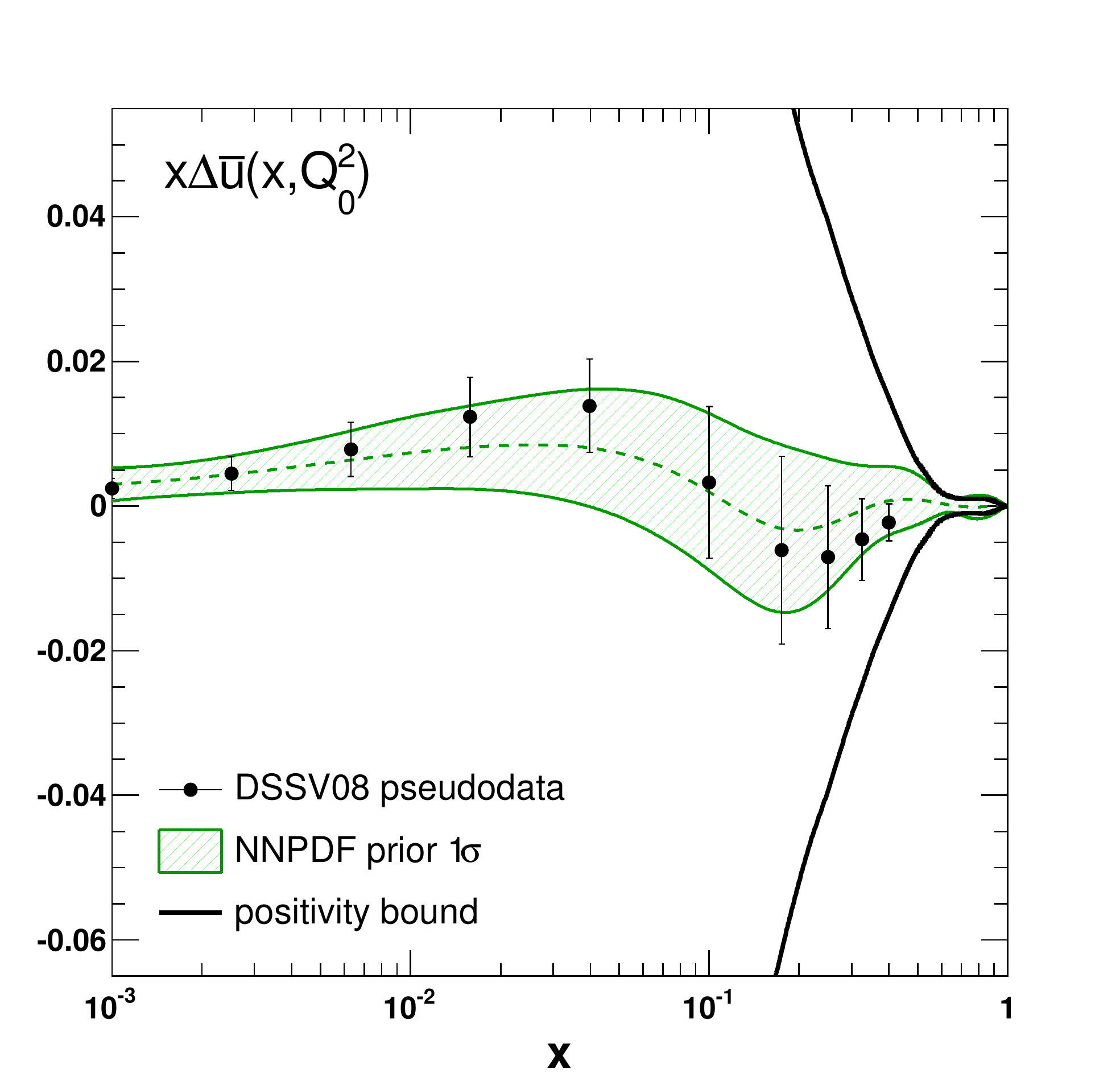}
\epsfig{width=0.40\textwidth,figure=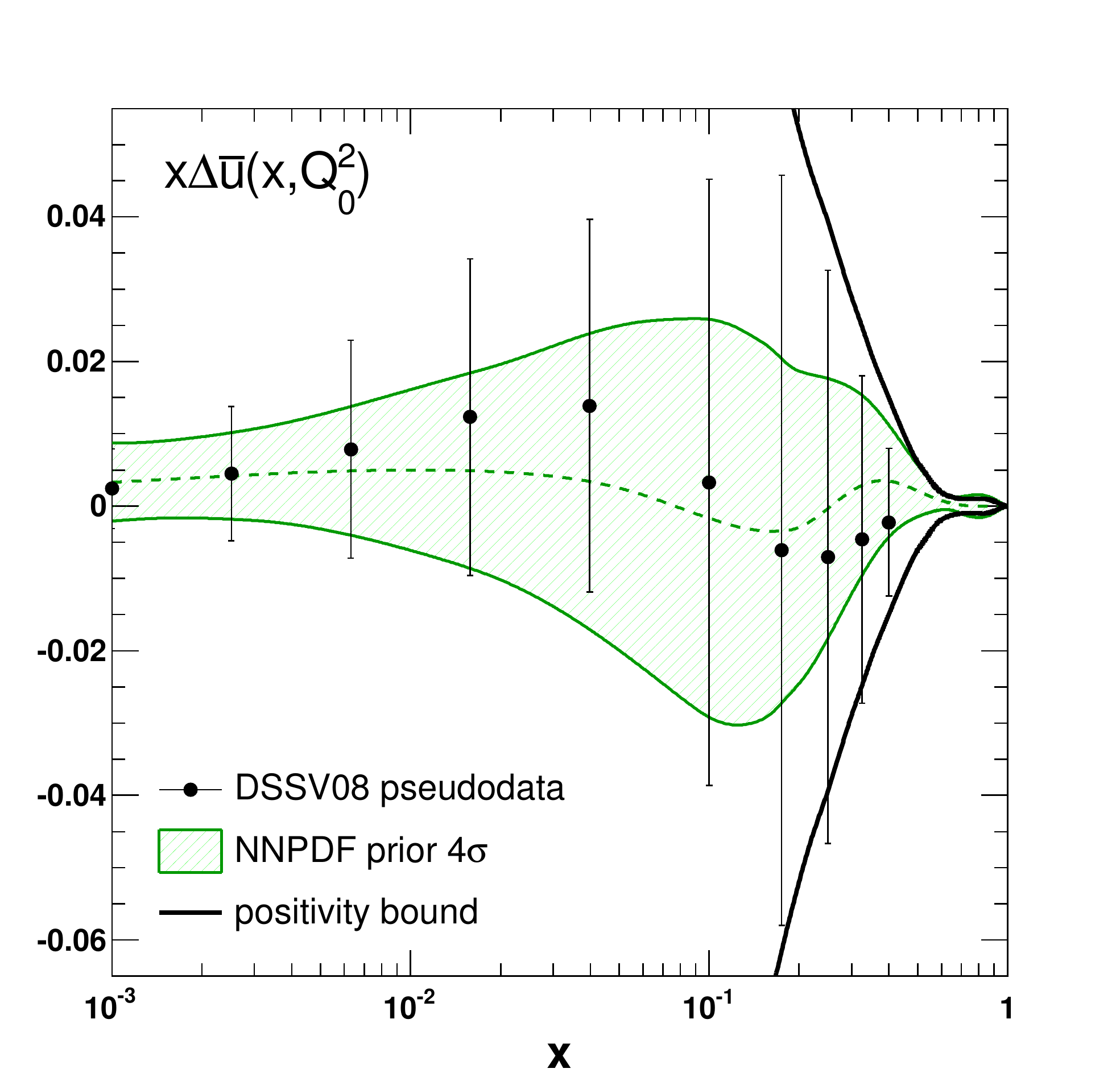}\\
\epsfig{width=0.40\textwidth,figure=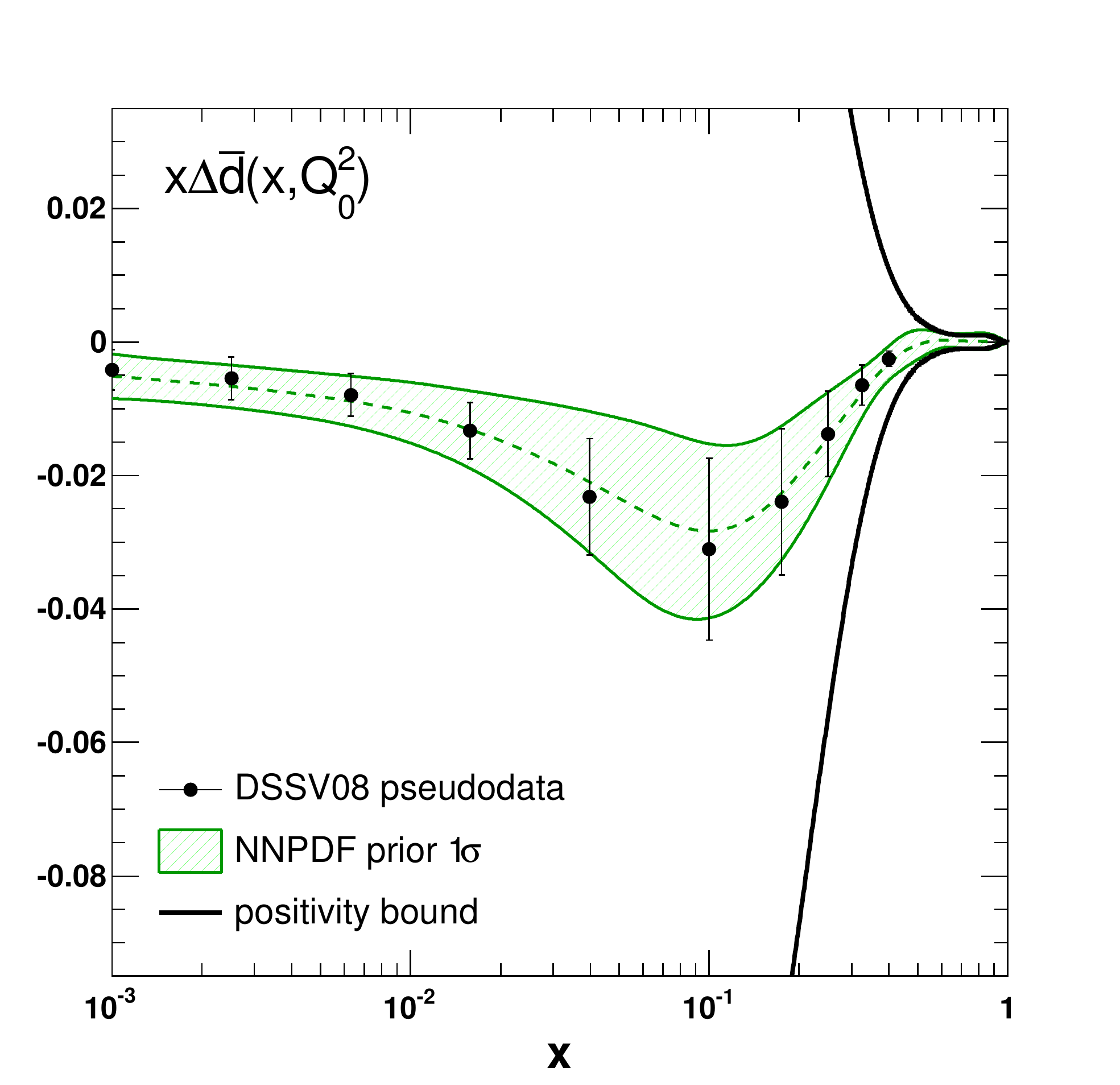}
\epsfig{width=0.40\textwidth,figure=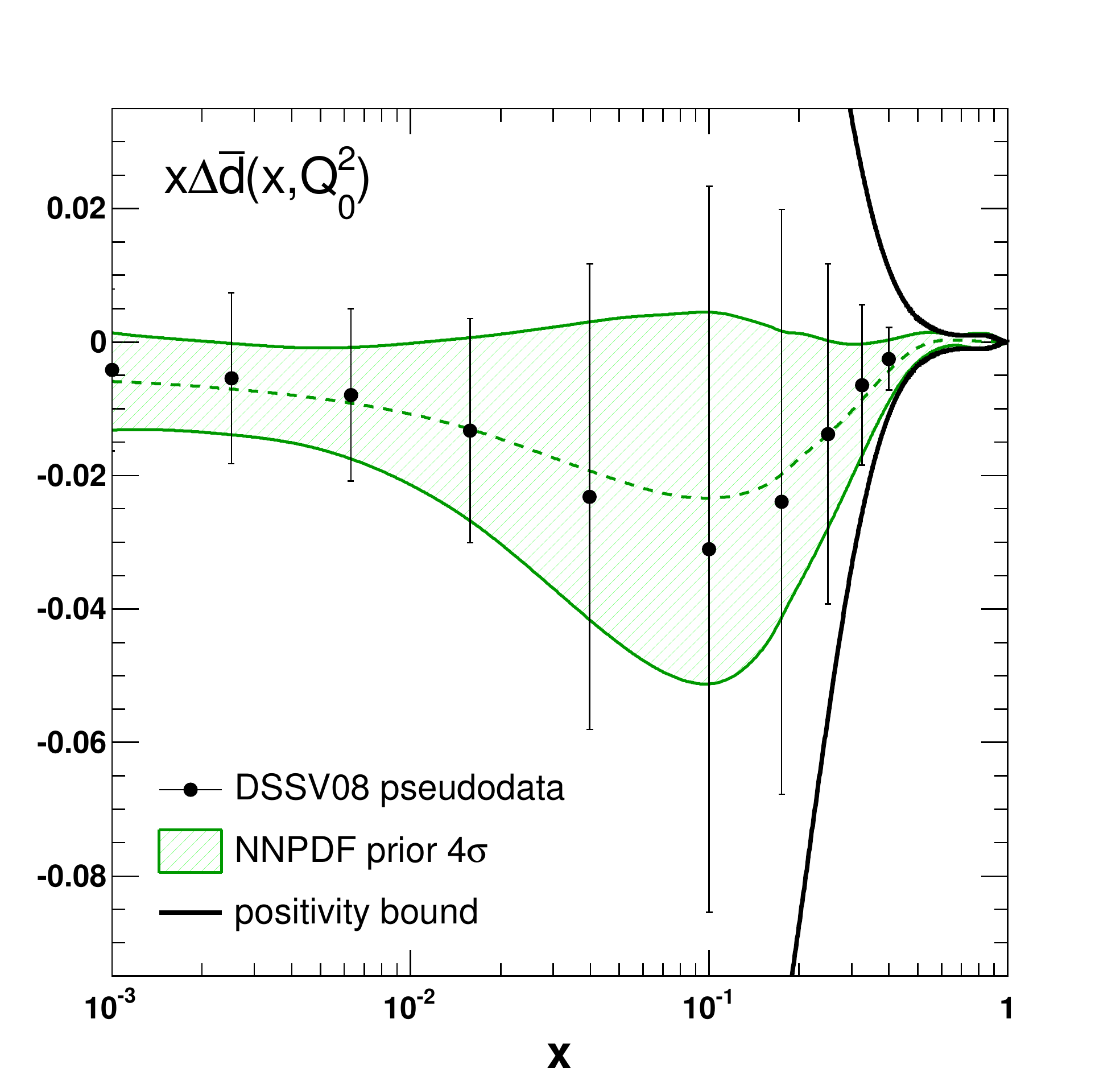}
\mycaption{The polarized sea quark distributions 
$x\Delta\bar{u}(x,Q_0^2)$ (upper plots)
and $x\Delta\bar{d}(x,Q_0^2)$ (lower plots) at the initial energy scale
$Q_0^2=1$ GeV$^2$ from the neural network fit (full band)
to pseudodata sampled from \texttt{DSSV08} parton set
(points with uncertainties).
Results are shown for the $1\sigma$ (left plots) and  
$4\sigma$ (right plots) prior ensembles. 
The positivity bound from the corresponding unpolarized \texttt{NNPDF2.3} 
parton set is also shown.}
\label{fig:prior}
\end{center}
\end{figure}

Following this procedure, we end up with four separate prior PDF ensembles,
labeled as $1\sigma$, $2\sigma$, $3\sigma$
and $4\sigma$, corresponding to the different factors 
by which the \texttt{DSSV08} nominal
PDF uncertainty has been enlarged.
The goodness of the pseudodata fits is quantitatively 
assessed by the $\chi^2$ values
per data point quoted in Tab.~\ref{tab:chi2ud},
which are close to one for both separate and combined 
\texttt{DSSV08}$_{U}$ and \texttt{DSSV08}$_{D}$ 
data sets.
In Fig.~\ref{fig:prior}, we show the 
$x\Delta\bar{u}(x,Q_0^2)$ and $x\Delta\bar{d}(x,Q_0^2)$ PDFs
at the initial energy scale $Q_0^2=1$ GeV$^2$
from the $1\sigma$ and $4\sigma$ prior ensembles we have contructed. 
We have checked that the other priors, 
$2\sigma$ and $3\sigma$, consistently reproduce
intermediate results.
In these plots the positivity bound discussed in Sec.~\ref{sec:minim}
and data points sampled from the \texttt{DSSV08}
parton set~\cite{deFlorian:2009vb} are also shown. 
\begin{table}[t]
\centering
\footnotesize
\begin{tabular}{ll|c|cccc}
\toprule
& & & \multicolumn{4}{c}{$\chi^2_{\mathrm{tot}}$}\\
\midrule
Experiment & Set & $N_{\mathrm{dat}}$ & $1\sigma$ & $2\sigma$ & $3\sigma$ & $4\sigma$ \\
\midrule
 \texttt{DSSV08} &            & $20$ & $1.04$ & $1.10$ & $1.09$ & $0.97$ \\
& \texttt{DSSV08}$_{U}$ & $10$ & $1.13$ & $1.09$ & $1.08$ & $0.97$ \\
& \texttt{DSSV08}$_{D}$ & $10$ & $0.96$ & $1.10$ & $1.09$ & $0.96$ \\
\bottomrule
\end{tabular}
\mycaption{The value of the $\chi^2_{\mathrm{tot}}$ per data point for both 
separate and combined $\Delta\bar{u}$ and $\Delta\bar{d}$ data sets
after the neural network fit to pseudodata sampled from the \texttt{DSSV08}
parton set.}
\label{tab:chi2ud}
\end{table}

Monte Carlo ensembles of polarized parton distributions obtained in this way
are equivalent to \texttt{NNPDFpol1.0}
in the $\Delta q^+=\Delta q +\Delta\bar{q}$ 
($q=u, d, s$) and $\Delta g$ sectors, but
they are supplemented with quark-antiquark separation,
suitable as a starting point for the reweighting procedure.
Notice that they have exactly the same gluon distribution,
since this is not affected by construction.
In the next sections we quantify the impact on these priors of the new
data, and show that results are independent 
of the specific choice of prior starting from the $3\sigma$ case.

\subsection{Reweighting with new data sets}
\label{sec:reweighting-separate}

We would like to reweight the prior PDF ensembles determined 
in Sec.~\ref{sec:prior11} with the data described in 
Sec.~\ref{sec:expinput11}. To this purpose, we should
compute theoretical predictions for the observables 
measured in each process under investigation and then
compare them to experimental data. 
As explained in Sec.~\ref{sec:NNPDFapproach}, we will then assign 
to each replica in the PDF ensembles a weight proportional
to the $\chi^2$ of the new data to the corresponding prediction,
given by Eq.~(\ref{eq:weightformula}).

Before discussing the impact of each new data set in turn, we
summarize a few methodological aspects which apply to all
experiments.
\begin{itemize}
\item For each of the new processess considered in our analysis,
the experimental observable is a longitudinal spin asymmetry,
\textit{i.e.} the ratio between cross-sections depending on 
polarized PDFs (in the numerator) and on unpolarized PDFs (in the
denominator), see Eqs.~(\ref{eq:ALLgammaN})-(\ref{eq:ALL-jet})-(\ref{eq:Wasy}). 
In these expressions, the numerator
will be computed for each polarized replica from the ensembles 
determined in Sec.~\ref{sec:prior11}, while the denominator
will be evaluated only once using the mean value from the unpolarized 
\texttt{NNPDF2.3} parton set~\cite{Ball:2012cx} at NLO.
This strategy accounts for the fact that the uncertainty affecting 
the observed asymmetries is mostly driven by the uncertainty 
of the polarized parton distributions rather than by that of their 
unpolarized counterparts, which is actually negligible.
\item For each data set we will have to check the effectiveness 
of the reweighting procedure. To this purpose, we will look 
at the distribution of $\chi^2$ per data point among replicas
and at its mean value, which is expected to decrease after reweighting.
Also, we will have to keep
under control the loss of accuracy in the description of
the underlying PDF probability distribution, by ensuring that the number of replicas
left after reweighting, $N_{\mathrm{eff}}$, does not become too low.
In particular, we require that $N_{\mathrm{eff}}$ should be 
comparable with the number of replicas, $N_{\mathrm{rep}}$, 
in \texttt{NNPDFpol1.0},
\textit{i.e.} $N_{\mathrm{eff}}\sim N_{\mathrm{rep}}=100$.
Indeed, we determined in Sec.~\ref{sec:MCgeneration} that a Monte Carlo
sample of $N_{\mathrm{rep}}=100$ replicas 
is sufficient to reproduce 
the mean values and the errors of experimental data 
within percent accuracy.
\item We may be interested in determining whether the new 
data are consistent with the old inclusive DIS data.
To this purpose, for each data set we will evaluate the $\mathcal{P}(\alpha)$ 
distribution, defined by Eq.~(12) of Ref.~\cite{Ball:2010gb}.
The parameter $\alpha$ measures the consistency of the data
which are used for reweighting with those included in the prior PDF sets,
by providing the factor by which the uncertainty on the new data must be 
rescaled in order the two sets to be consistent. Hence, if the probability
density for the parameter $\alpha$, $\mathcal{P}(\alpha)$, peaks close to
one, one can conclude that new and old data are consistent with each other.
\end{itemize}

\subsubsection{Open-charm production at COMPASS}
\begin{figure}[p]
\begin{center}
\epsfig{width=0.32\textwidth,figure=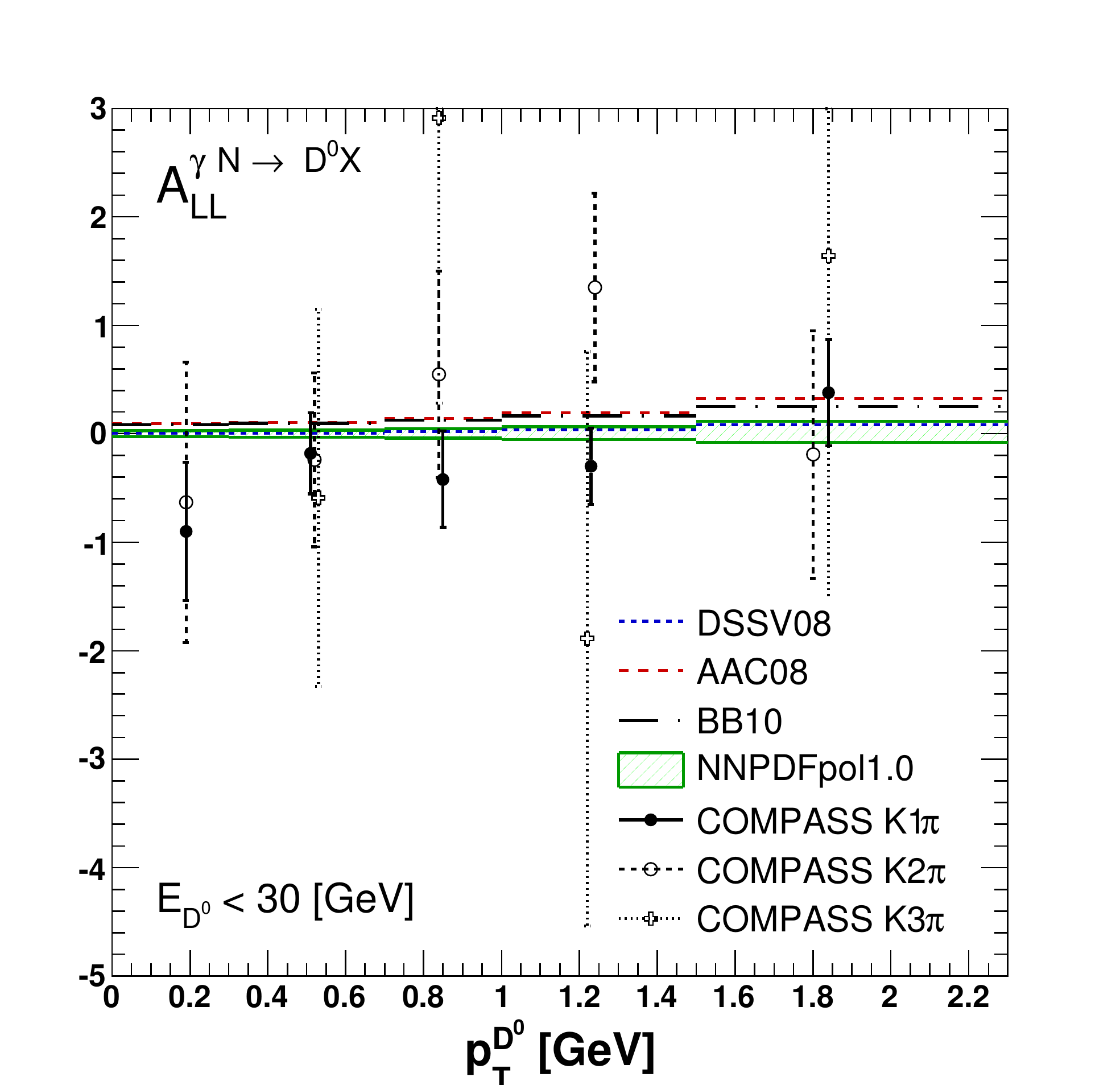}
\epsfig{width=0.32\textwidth,figure=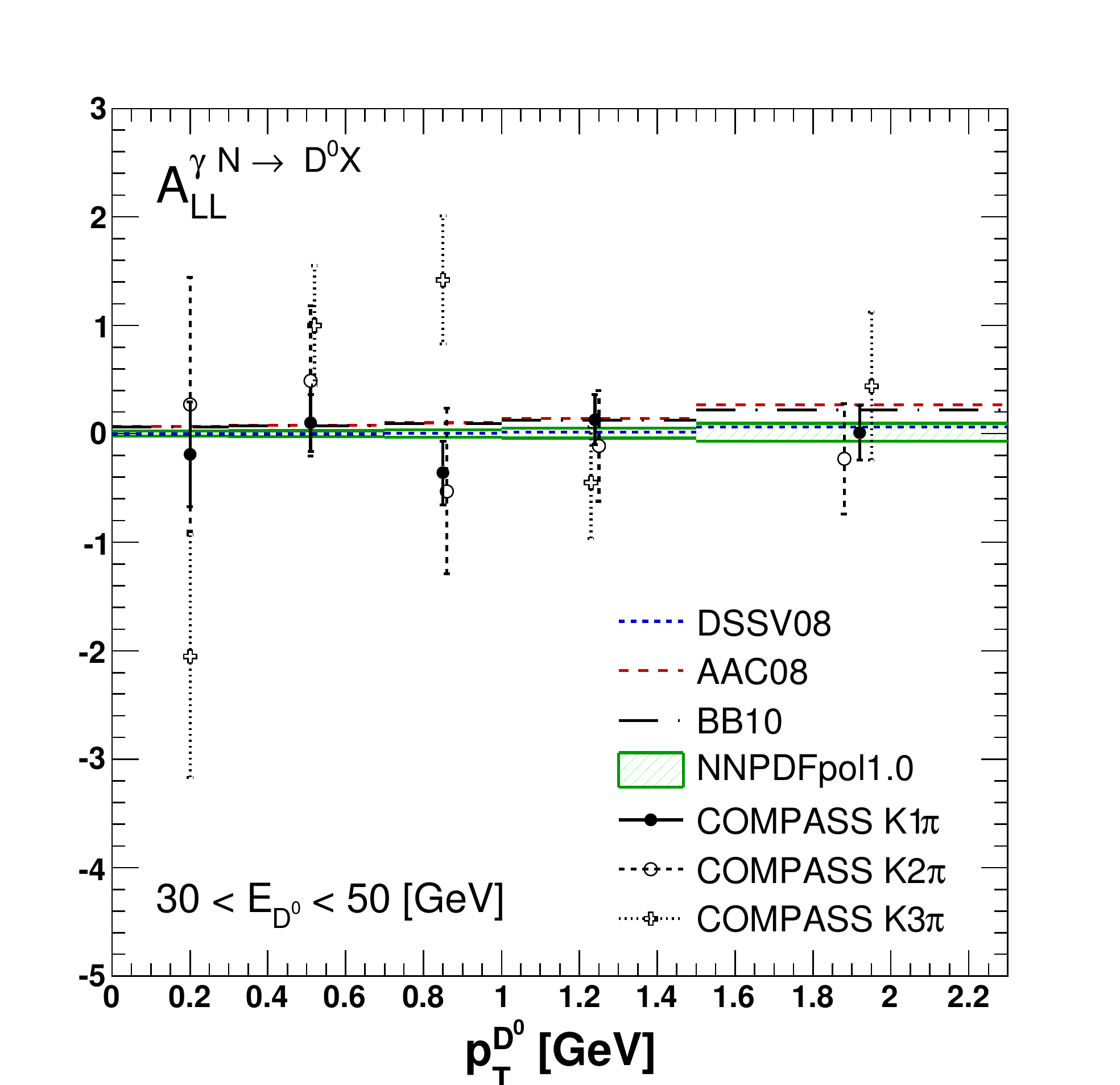}
\epsfig{width=0.32\textwidth,figure=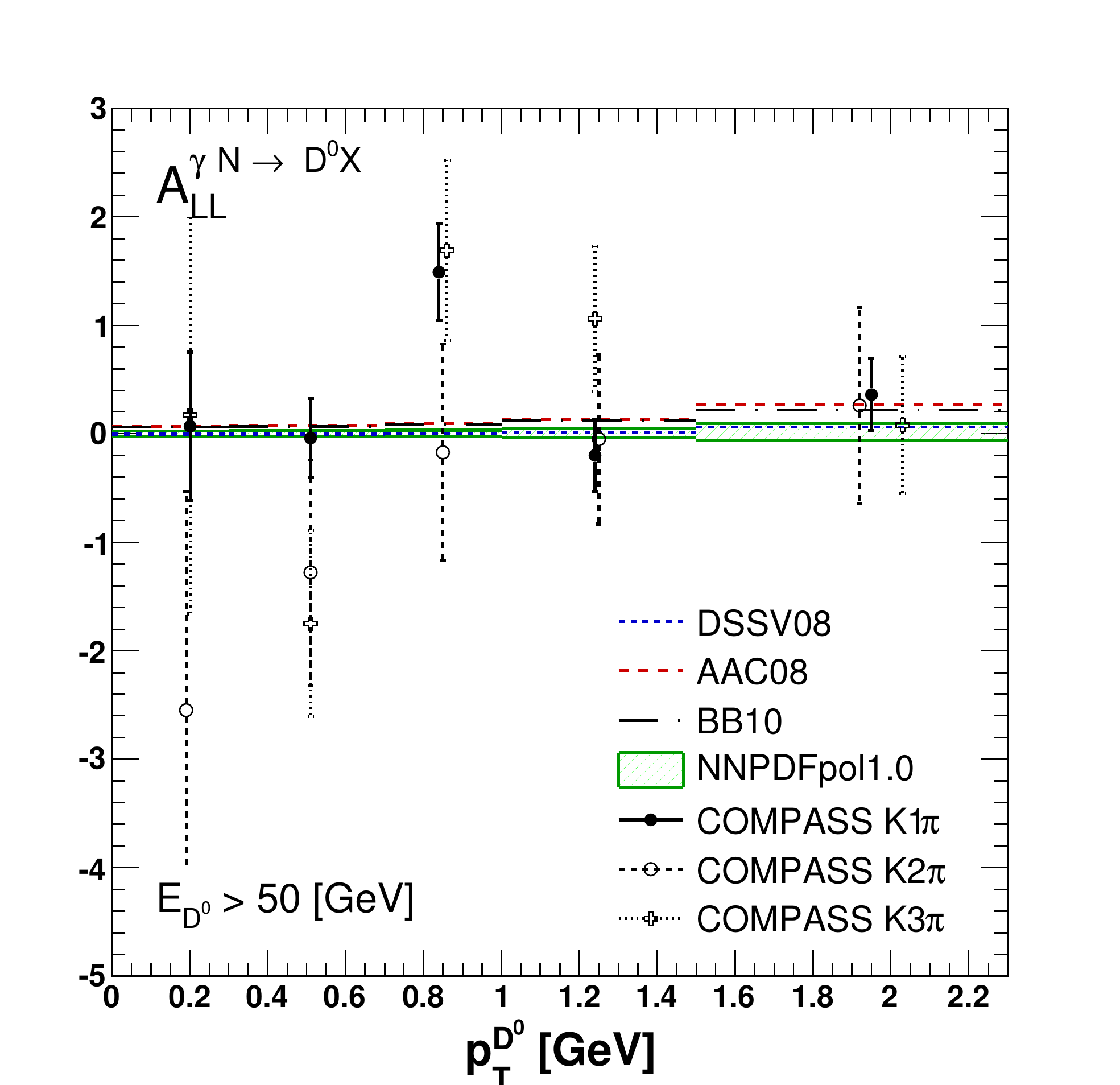}\\
\mycaption{Experimental double-spin asymmetry for $D^0$ meson 
photoproduction $A_{LL}^{\gamma N \to D^0 X}$
measured by COMPASS~\cite{Adolph:2012ca} from three decay channels
compared to its LO prediction, Eq.~(\ref{eq:ALLgammaN}),
computed for different PDF sets in three bins of the charmed hadron energy 
$E_{D^0}$ and in five bins of its transverse momentum $p_{T}^{D_0}$.}
\label{fig:COMPASS_beforerw}
\end{center}
\end{figure}

Predictions for the photon-nucleon asymmetry $A_{LL}^{\gamma N\to D^0 X}$
are computed at LO accuracy, 
using the expressions in Ref.~\cite{Simolo:2006iw}, based in turn on 
Ref.~\cite{Frixione:1996ym}, for both the numerator and the denominator
in Eq.~(\ref{eq:ALLgammaN}).
Results are compared in Fig.~\ref{fig:COMPASS_beforerw} 
to COMPASS experimental data, 
separated into individual decay channels and into three bins of the 
charmed hadron energy $E_{D^0}$. The curves labelled as 
\texttt{DSSV08}, \texttt{AAC08} and \texttt{BB10}
are obtained using the corresponding polarized parton 
sets~\cite{deFlorian:2009vb,Hirai:2008aj,Blumlein:2010rn}
and either the \texttt{CTEQ6}~\cite{Pumplin:2002vw} 
(for \texttt{DSSV08}) or 
the \texttt{MRST2004}~\cite{Martin:2006qz} 
(for \texttt{AAC08} and \texttt{BB10}) unpolarized PDF sets.
The curve labelled as NNPDF is instead computed using the 
\texttt{NNPDFpol1.0} parton set. Notice that we do not need the prior
Monte Carlo ensembles constructed in Sec.~\ref{sec:prior11}, since
we compute the observable at LO and only the polarized gluon appears
in Eq.~(\ref{eq:ALLgammaN}). By construction, this distribution is
exactly the same in all the parton sets discussed in Sec.~\ref{sec:prior11}
and in \texttt{NNPDFpol1.0},
hence they all give the same prediction for the photon-nucleon asymmetry 
$A_{LL}^{\gamma N\to D^0 X}$.
For all the curves shown in Fig.~\ref{fig:COMPASS_beforerw},
we have used the Peterson parametrization of the
fragmentation function $D_c^{D^0}$~\cite{Peterson:1982ak}; 
we checked that results
are unaffected by the choice of other, slightly different, available 
parametrizations~\cite{Colangelo:1992kh},
as noticed in Ref.~\cite{Riedl:2012qc}.

\begin{figure}[p]
\begin{center}
\epsfig{width=0.32\textwidth,figure=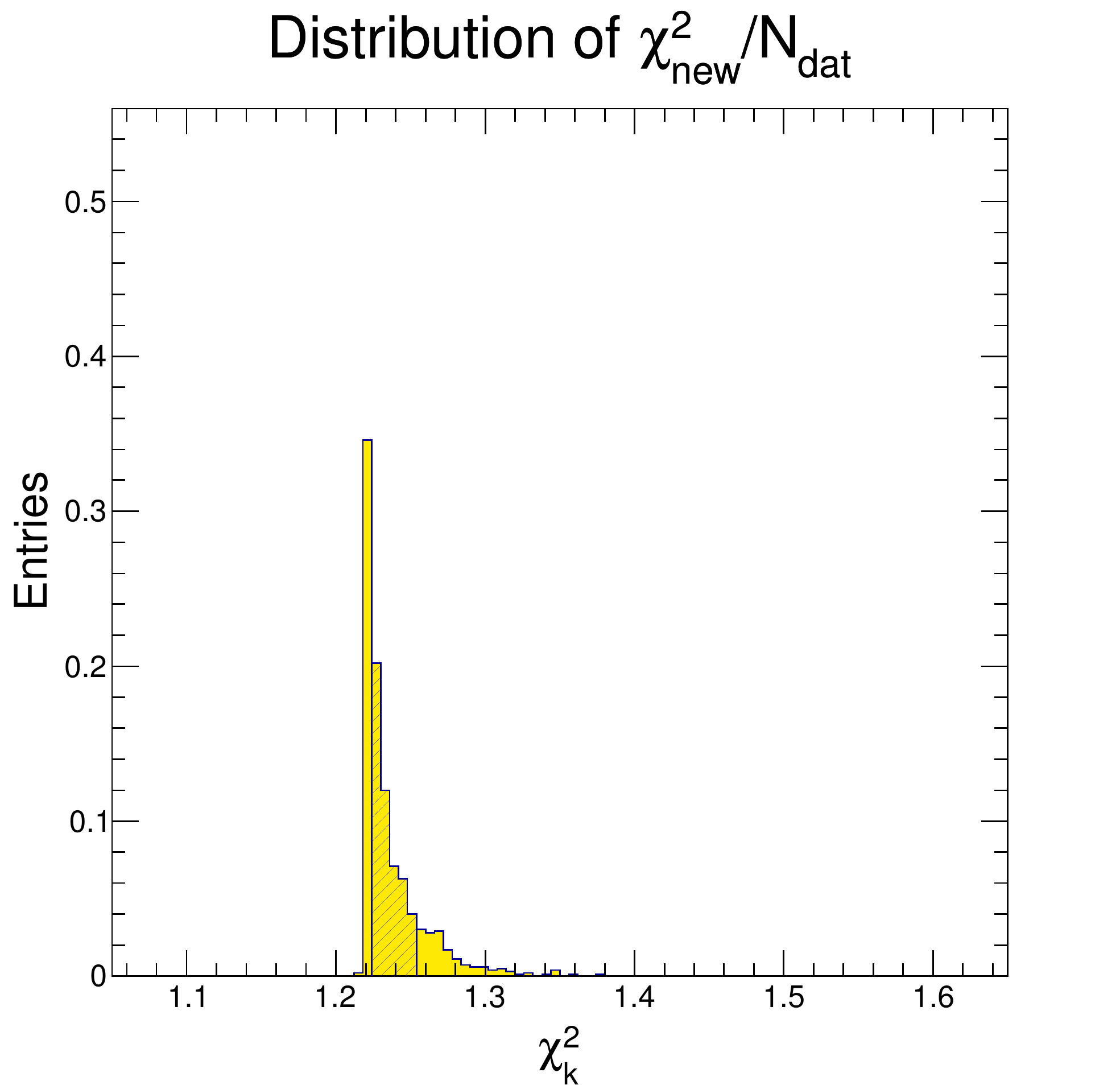}
\epsfig{width=0.32\textwidth,figure=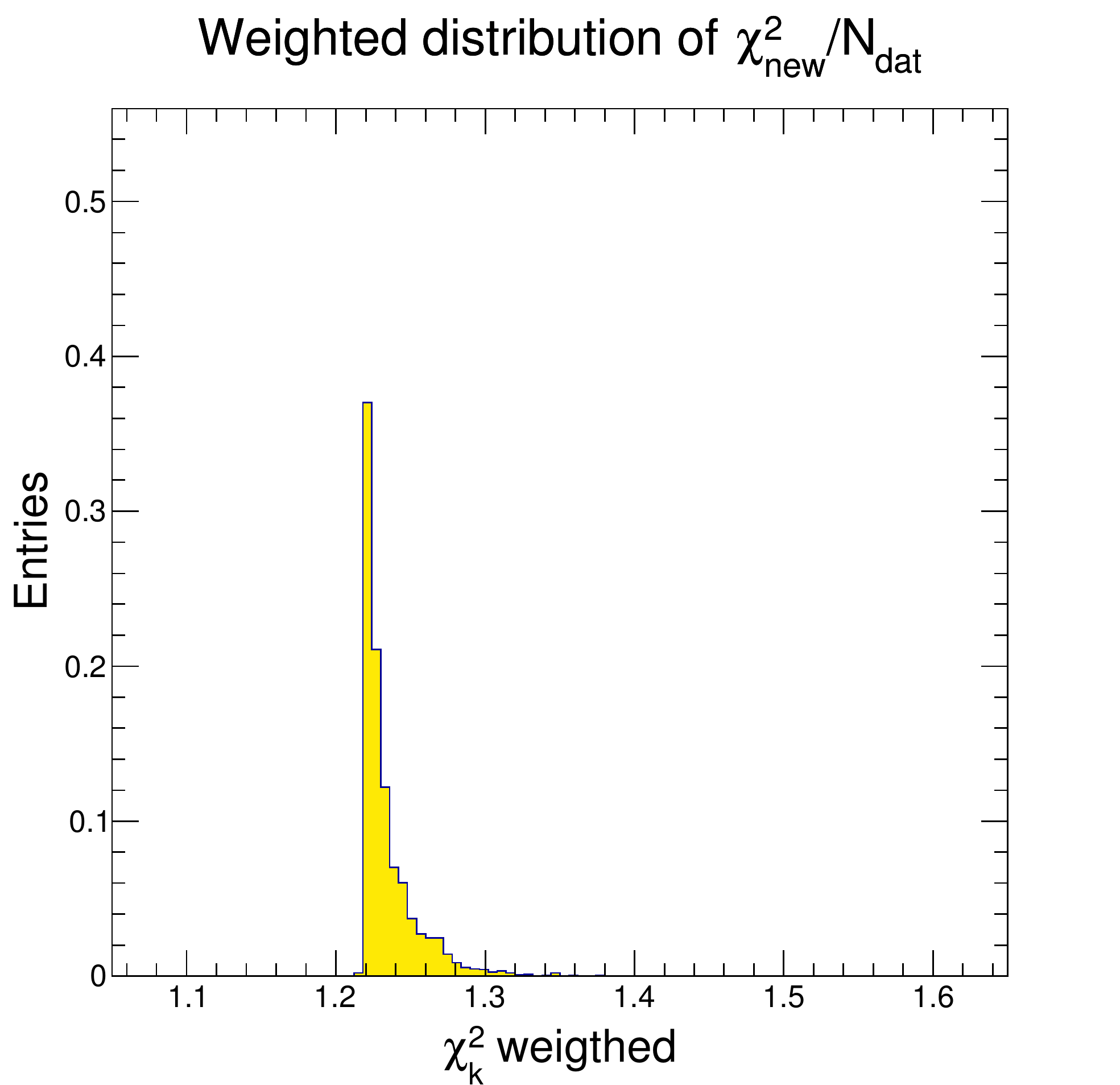}
\epsfig{width=0.32\textwidth,figure=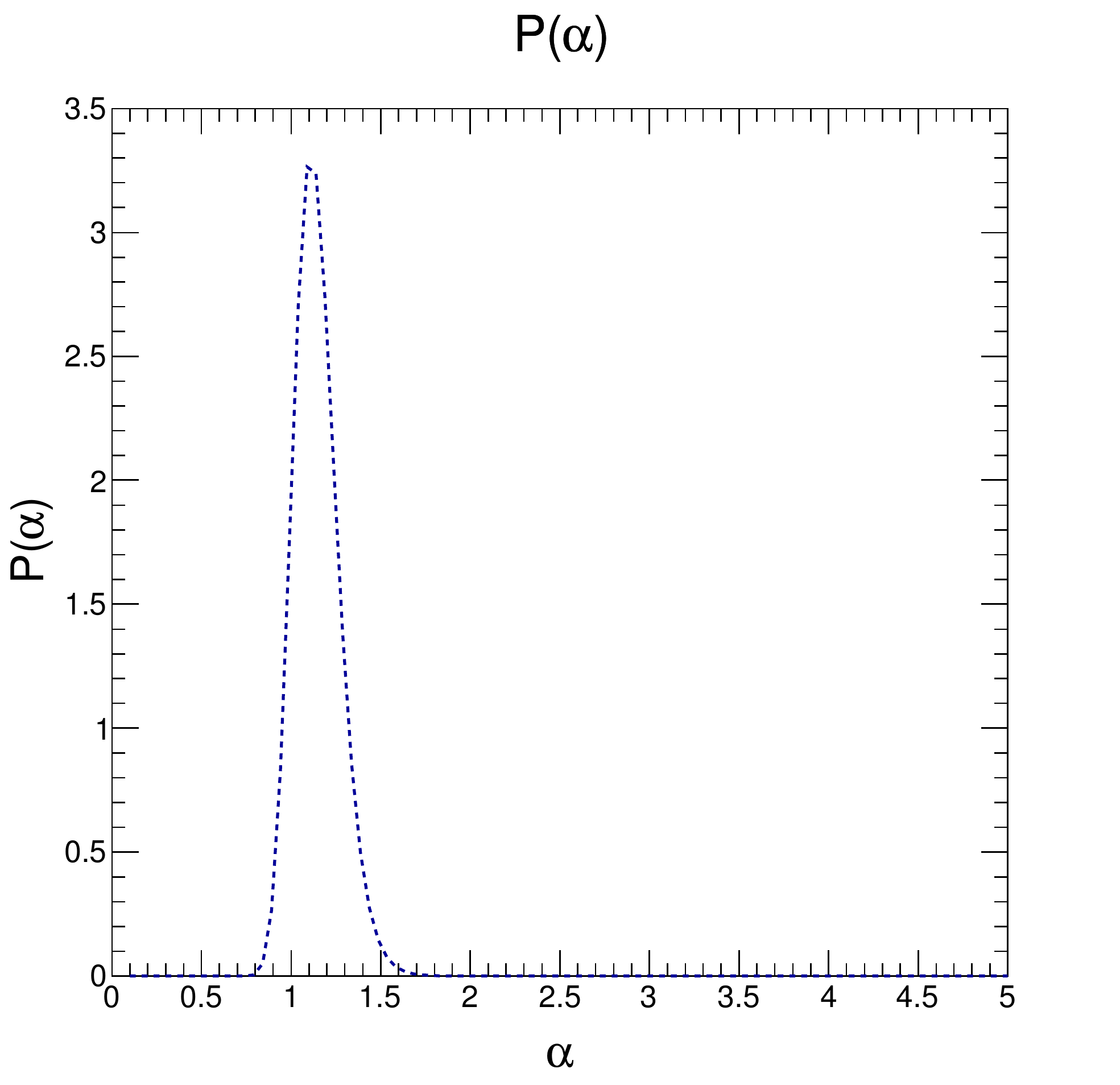}\\
\mycaption{Distribution of $\chi^2/N_{\mathrm dat}$ for individual replicas
before (first panel) and after (second panel) 
reweighting with COMPASS open-charm production data~\cite{Adolph:2012ca}.
The shaded region in the first panel corresponds to the central 
$68\%$ of the distribution. The $\mathcal{P}(\alpha)$ distribution (third panel) 
is also shown.
All plots refer to the three COMPASS data sets combined together.}
\label{fig:COMPASS-rwest1}
\end{center}
\end{figure}

The agreement between predictions from different PDF sets
and experimental data can be quantified by the value of the $\chi^2$ 
per data point, which we compute for both separate
and combined COMPASS data sets, see Tab.~\ref{tab:chi2before}.
The corresponding distribution among NNPDF replicas
is shown in the first panel of Fig.~\ref{fig:COMPASS-rwest1}
only for the combined data set. 
Notice that, lacking the experimental covariance matrix,
the uncertainties which enter the $\chi^2$ definition are taken as the sum 
in quadrature of statistical and systematic uncertainties.
From Tab.~\ref{tab:chi2before}, 
it is clear that, even before reweighting, predictions are 
already in good agreement with 
experimental data, which however are affected by large errors
in comparison to the uncertainty estimate of the asymmetry itself.
Also, all polarized PDF sets provide a similar description of the data.
\begin{table}[t]
 \centering
 \footnotesize
  \begin{tabular}{llccccc}
   \toprule
   \multirow{2}*{Experiment} 
& \multirow{2}*{Set} 
& \multirow{2}*{$N_{\mathrm{dat}}$} & \multicolumn{4}{c}{$\chi^2/N_{\mathrm{dat}}$}\\ 
    &  &  & \texttt{NNPDFpol1.0} & \texttt{DSSV08} & \texttt{AAC08} & \texttt{BB10}\\
   \midrule
     COMPASS &       &  45 & 1.23 & 1.23 & 1.27 & 1.25 \\
   & COMPASS~$K1\pi$ &  15 & 1.27 & 1.27 & 1.43 & 1.38 \\
   & COMPASS~$K2\pi$ &  15 & 0.51 & 0.51 & 0.56 & 0.55 \\
   & COMPASS~$K3\pi$ &  15 & 1.90 & 1.90 & 1.81 & 1.82 \\
   \bottomrule
  \end{tabular}
\mycaption{Values of $\chi^2/N_{\mathrm{dat}}$ before reweighting for different 
polarized parton sets.}
\label{tab:chi2before}
\end{table}

Next, we have quantified the impact of COMPASS open-charm data into 
\texttt{NNPDFpol1.0} using Bayesian reweighting.
In Tab.~\ref{tab:global1}, we quote the $\chi^2$ per data point
after reweighting, $\chi^2_{\mathrm{rw}}/N_{\mathrm{dat}}$, while 
the number of effective replicas left after reweighting, $N_{\mathrm{eff}}$,
and the modal value of the 
$\mathcal{P}(\alpha)$ distribution, $\langle\alpha\rangle$, for both separate and combined data sets
are collected in Tab.~\ref{tab:global2}. 
In Fig.~\ref{fig:COMPASS-rwest1}, 
we also plot the distribution of $\chi^2$ per data point  
before and after reweighting, and of $\mathcal{P}(\alpha)$
for all data sets combined together. 

\begin{figure}[p]
\begin{center}
\epsfig{width=0.32\textwidth,figure=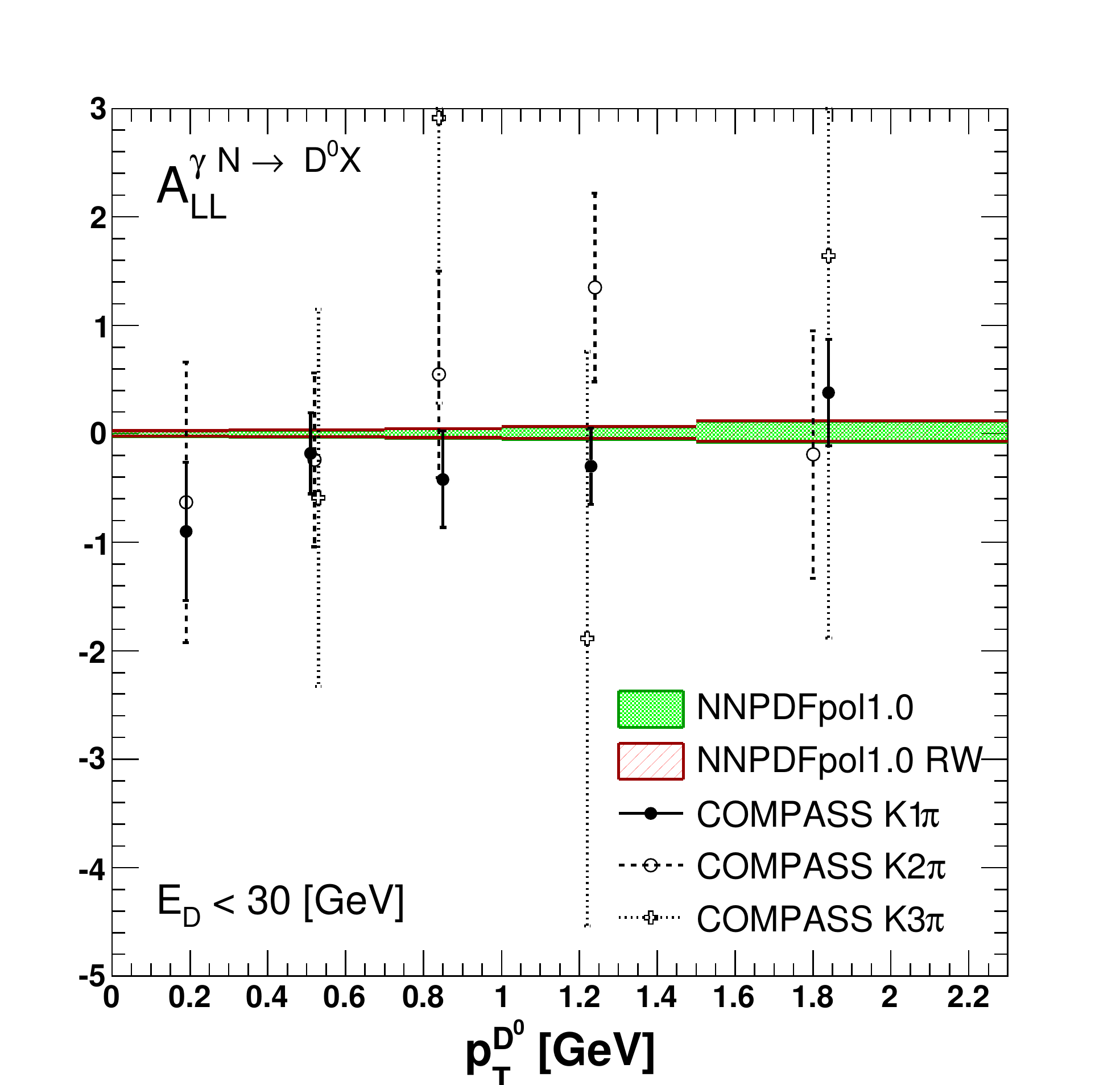}
\epsfig{width=0.32\textwidth,figure=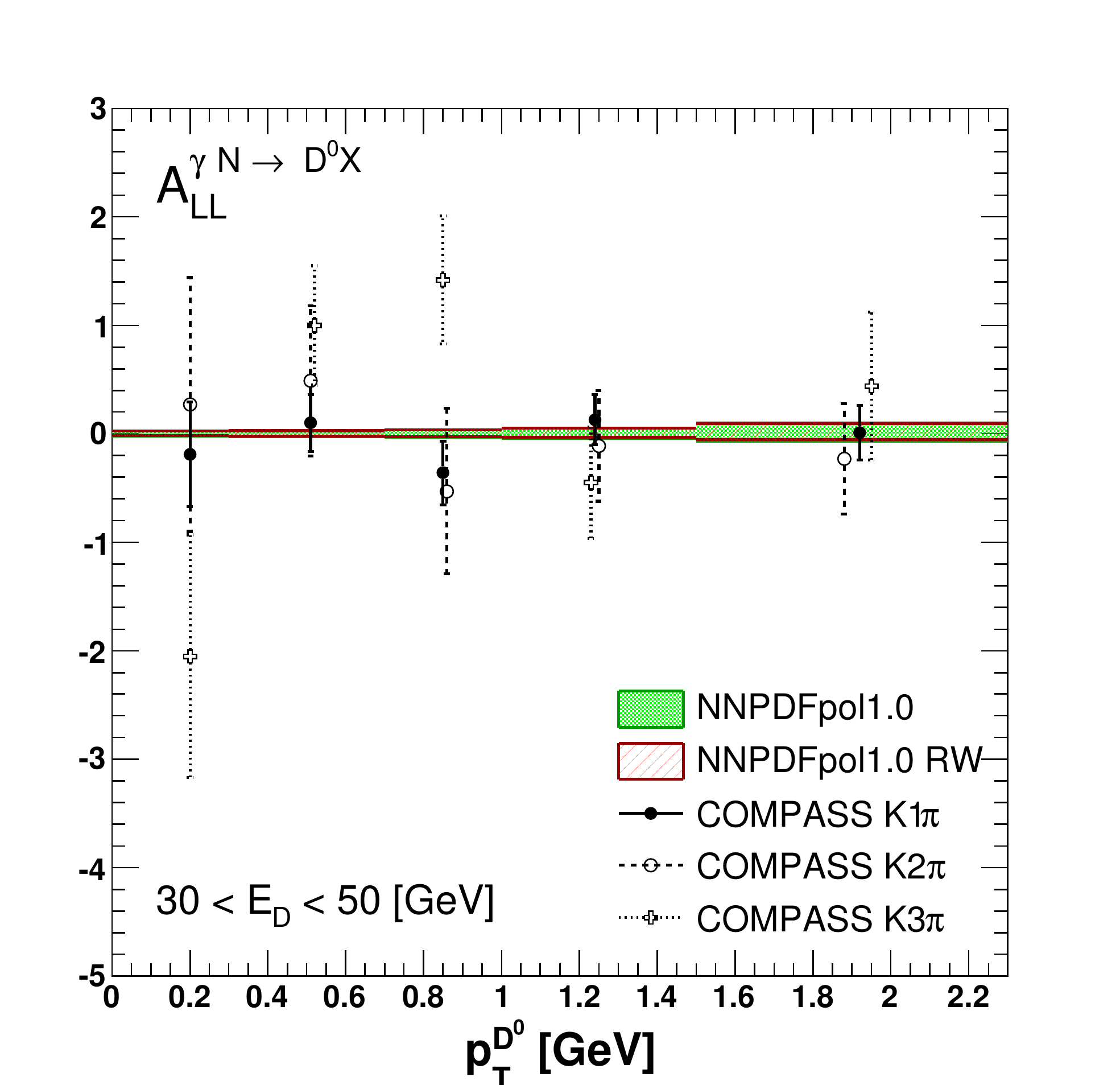}
\epsfig{width=0.32\textwidth,figure=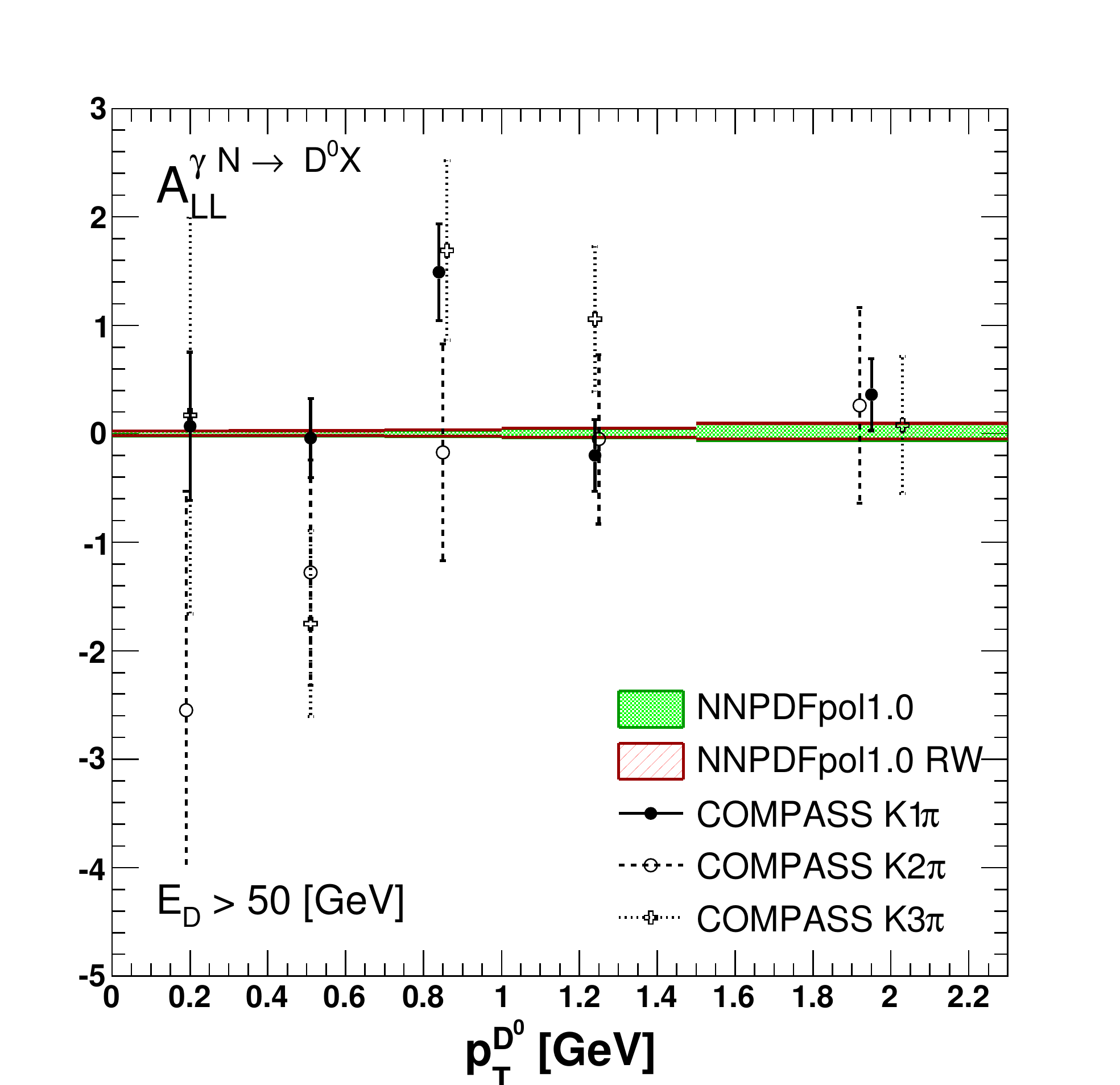}\\
\mycaption{Comparison between the double-spin asymmetry $A_{LL}^{\gamma N \to D^0 X}$,
Eq.~(\ref{eq:ALLgammaN}), before and after reweighting with COMPASS open-charm 
data~\cite{Adolph:2012ca}. Experimental points are also shown.}
\label{fig:COMPASS_afterrw}
\end{center}
\end{figure}

\begin{figure}[t]
\begin{center}
\epsfig{width=0.40\textwidth,figure=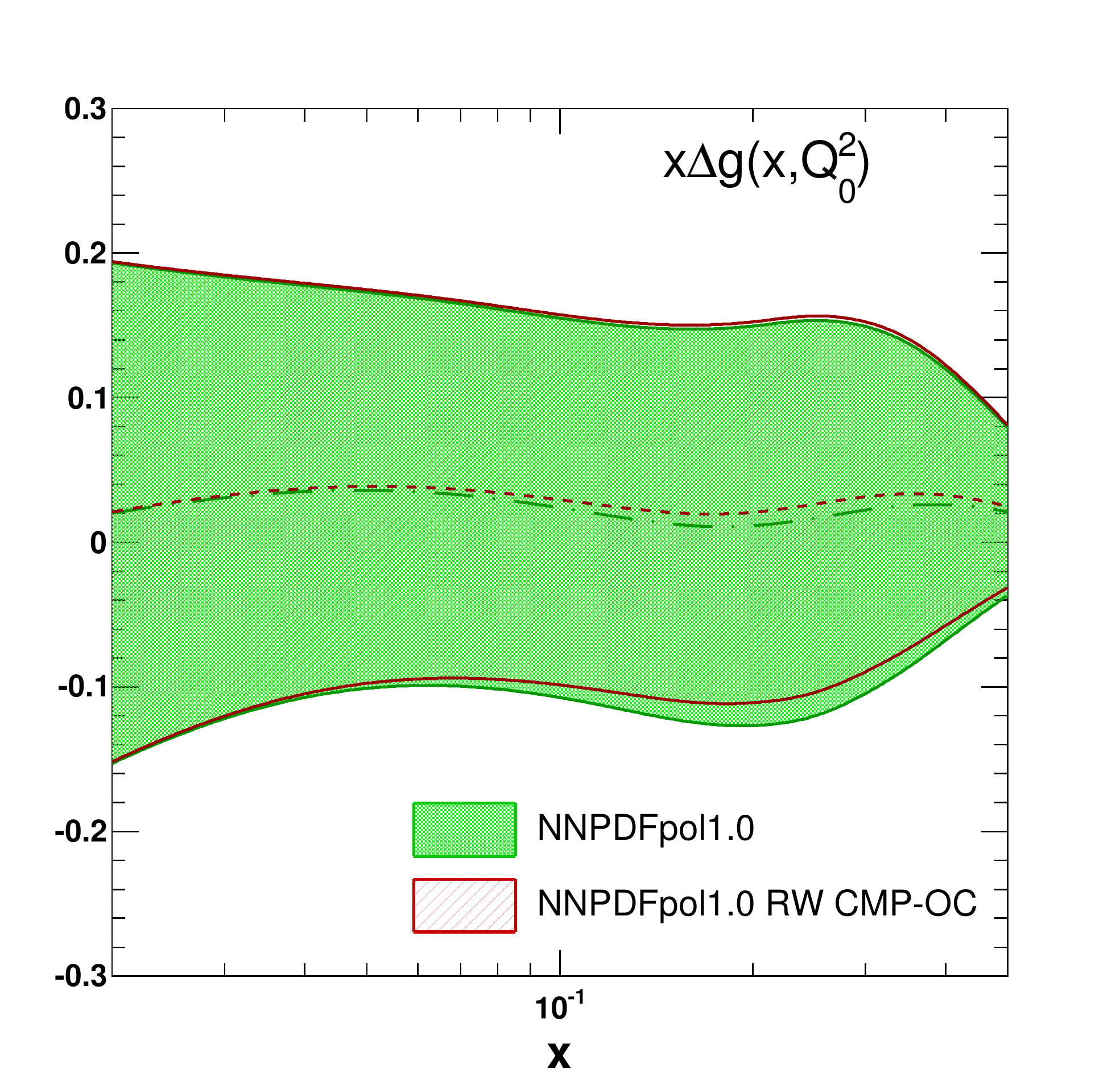}
\epsfig{width=0.40\textwidth,figure=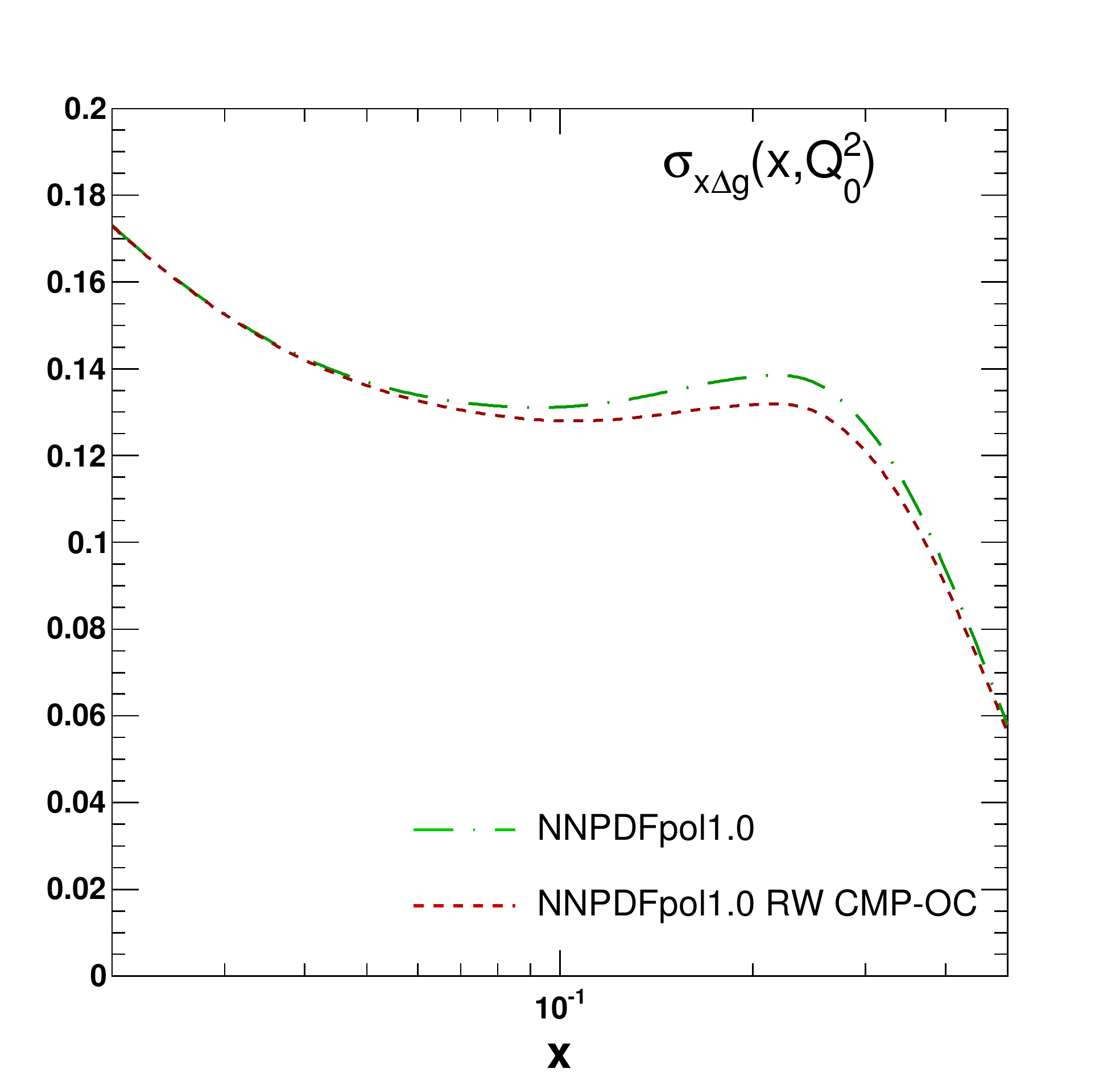}\\
\mycaption{Comparison between the unweighted and the reweighted polarized 
gluon distribution at $Q_0^2=1$ GeV$^2$ (left panel) and the 
improvement in its absolute error (right panel).}
\label{fig:COMPASS-gluon}
\end{center}
\end{figure}

From Tabs.~\ref{tab:global1}-\ref{tab:global2} 
and Fig.\ref{fig:COMPASS-rwest1}, it is evident
that reweighting with COMPASS open-charm data
leaves the prior parton set almost unaffected:
the $\chi^2$ value per data point and its distribution are essentially 
unchanged after reweighting.
Also, almost all replicas in the prior ensemble are preserved: this further
demonstrates the mild constraining power of COMPASS data sets. 
Furthermore, the observable $A_{LL}^{\gamma N \to D^0 X}$, Eq.~(\ref{eq:ALLgammaN}),
is compared before and after reweighting in Fig.~\ref{fig:COMPASS_afterrw},
showing unnoticeable differences. 
Finally, the reweighted polarized gluon PDF is drawn in 
Fig.~\ref{fig:COMPASS-gluon} together with its one-sigma absolute error
and compared to \texttt{NNPDFpol1.0}~\cite{Ball:2013lla}
at $Q_0^2=1$ GeV$^2$. Again, we notice that the differences are mild.

From these results, we can conclude that COMPASS open-charm data 
lead to a moderate improvement 
in our knowledge of the polarized gluon PDF, due
to their large experimental uncertainties.
One may wonder whether this conclusion still holds 
once NLO corrections are taken into account in the 
computation of the photon-nucleon asymmetry, $A_{LL}^{\gamma N \to D^0 X}$.
Such a computation has been recently completed~\cite{Riedl:2012qc}
and it was shown that contributions beyond Born-level 
may significantly affect this asymmetry.
However, COMPASS experimental data are not only affected
by large uncertainties with respect to the corresponding theoretical
predictions (both at LO and NLO), but also do not show a 
clear and stable trend over the covered range of $p_T^{D^0}$,
see Figs.~\ref{fig:COMPASS_beforerw}-\ref{fig:COMPASS_afterrw}.
For this reason, the NLO computation of the photon-nucleon
spin asymmetry will not allow for determining 
the polarized gluon more precisely
than its LO counterpart, though in principle the former contains
more QCD structure than the latter. 
Hence, we expect that the impact of COMPASS open-charm data
on the determination of the polarized gluon PDF 
will be comparable to that found in our LO analysis, 
once they will be included in a global NLO QCD fit of parton distributions,
as also anticipated in Ref.~\cite{Riedl:2012qc}.
\footnote{We were not able to compute the 
photon-nucleon asymmetry, $A_{LL}^{\gamma N \to D^0 X}$,
at NLO accuracy because the code developed to this purpose
in Ref.~\cite{Riedl:2012qc} is not publicly available.
Qualitatively, the impact of the NLO
corrections on this asymmetry can be inferred by comparing our results
in Fig.~\ref{fig:COMPASS_beforerw} (at LO) with those in Fig.~8
of Ref.~\cite{Riedl:2012qc} (at NLO): differences are actually hardly
noticeable.}

\subsubsection{High-\texorpdfstring{$p_T$}{pT} jet production at STAR and PHENIX}
\begin{figure}[t]
\centering
\epsfig{width=0.40\textwidth,figure=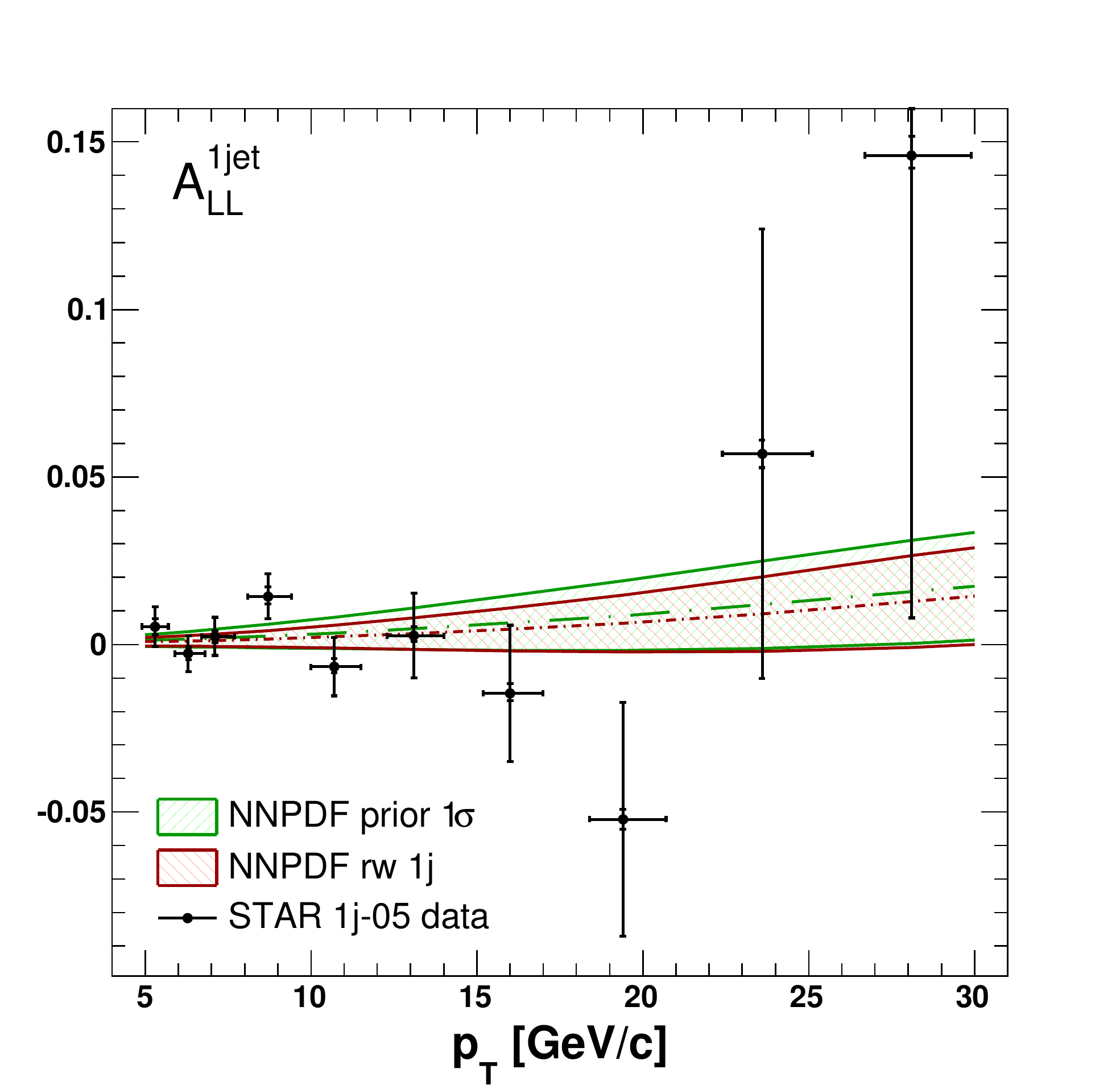}
\epsfig{width=0.40\textwidth,figure=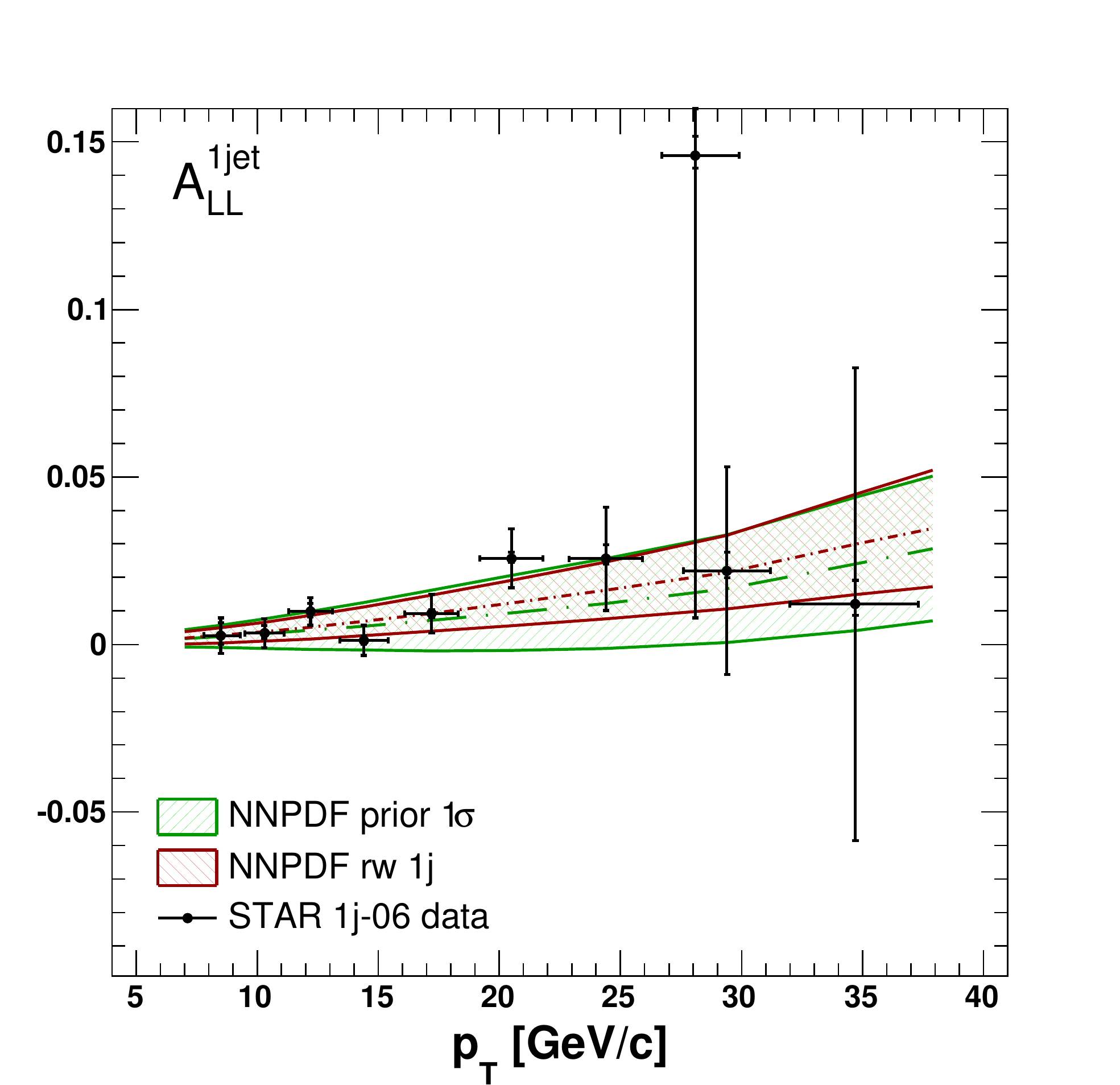}\\
\epsfig{width=0.40\textwidth,figure=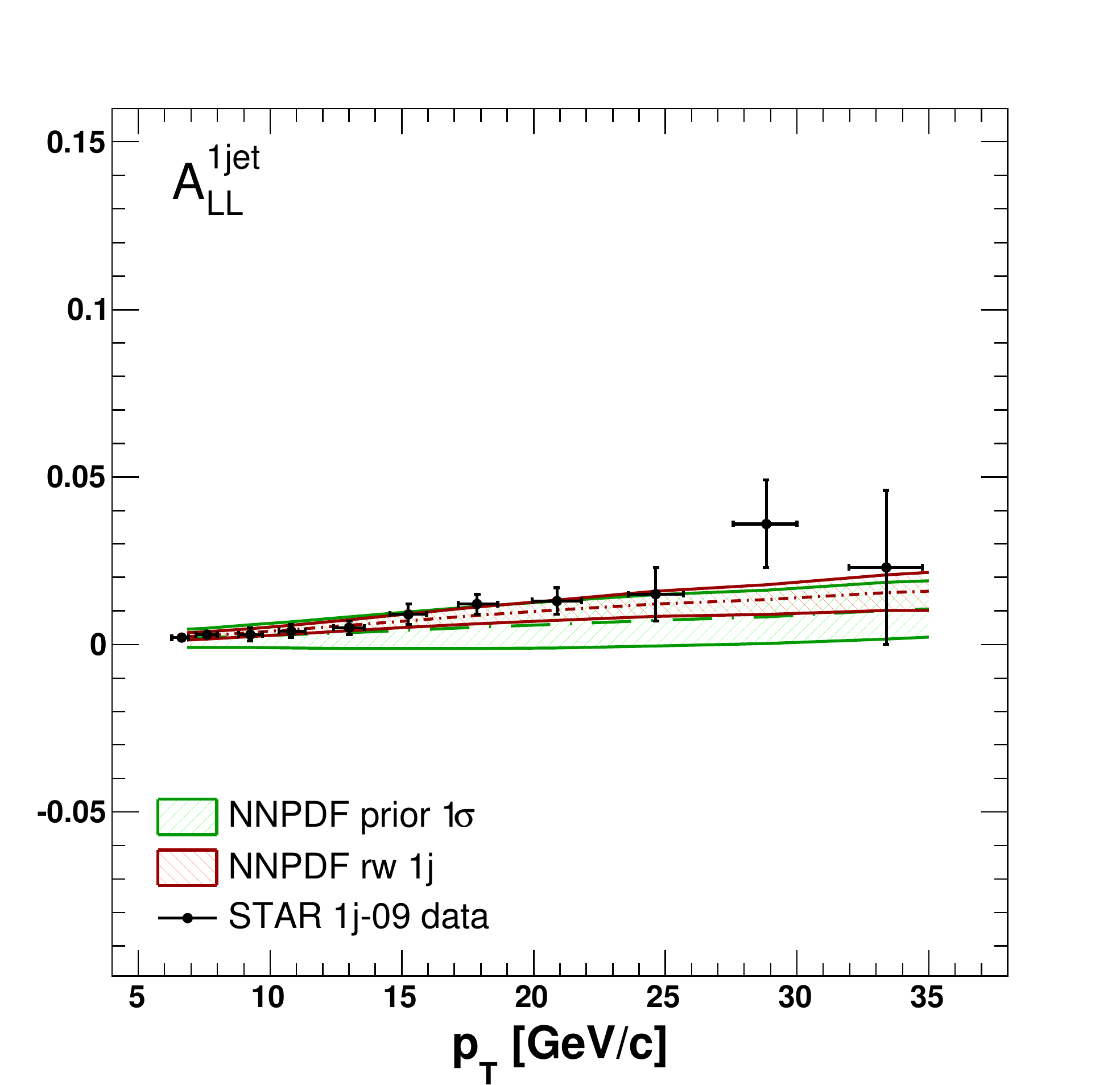}
\epsfig{width=0.40\textwidth,figure=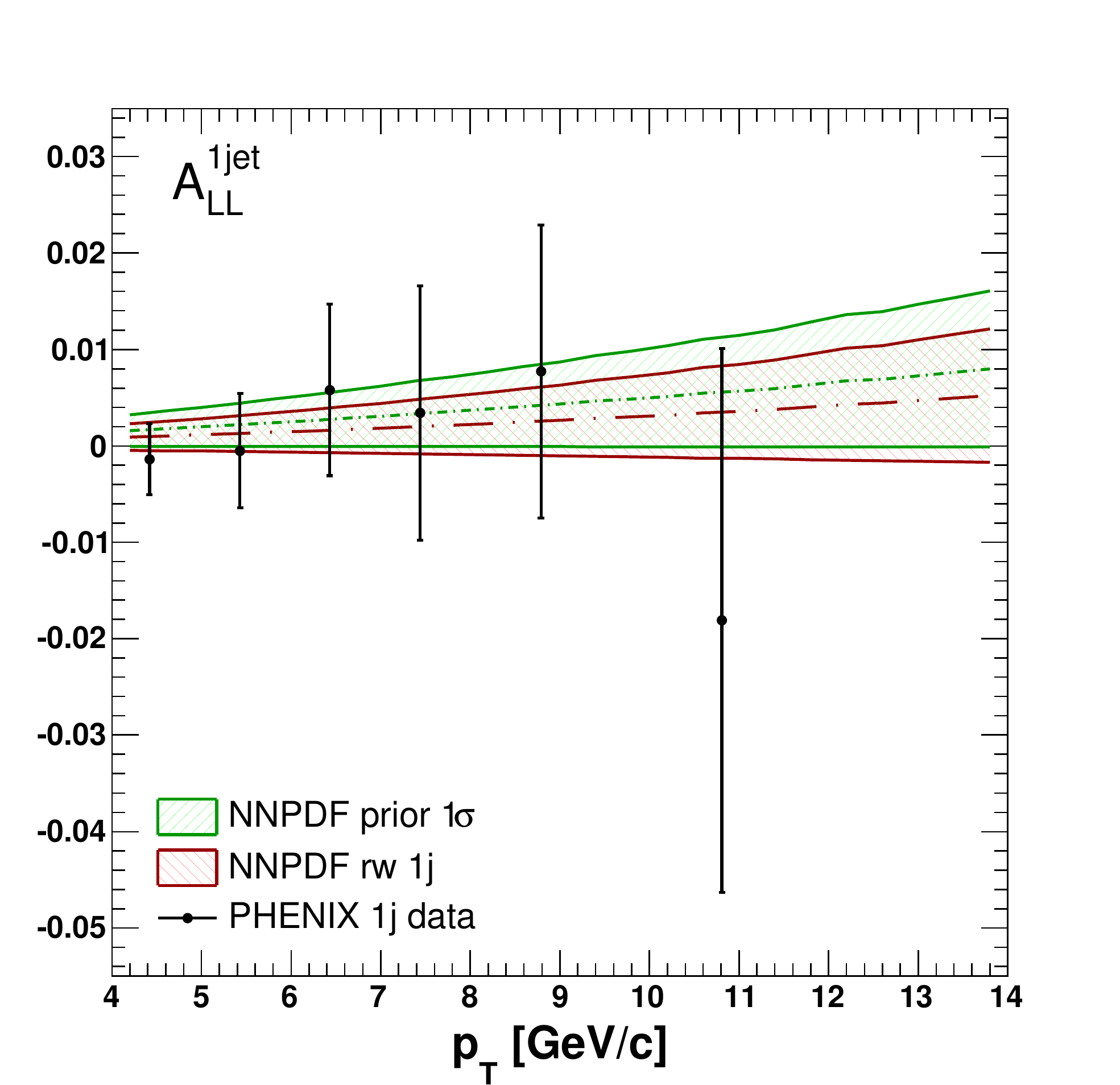}\\
\mycaption{Predictions for the double longitudinal spin asymmetry for 
single-inclusive jet production, $A_{LL}^{1jet}$, before and after reweighting
with RHIC data. Results are obtained from the $1\sigma$ prior PDF ensemble 
discussed in Sec.~\ref{sec:prior11}. Experimental data points are also displayed.
Notice the different scale of vertical axis for PHENIX.}
\label{fig:ALL-plot}
\end{figure}

Predictions for the longitudinal double-spin asymmetry from single-inclusive
jet production, $A_{LL}^{1jet}$, are shown in Fig.~\ref{fig:ALL-plot}.
They are plotted as a function of the transverse jet momentum, $p_T$, 
and compared to each experimental data set in Tab.~\ref{tab:jet-data}.
Predictions are computed at NLO using the code of 
Ref.~\cite{Jager:2004jh}, which was modified 
to handle NNPDF parton sets. We use the same jet algorithm,
cone radius and kinematic cuts
adopted in the experiment (see Tab.~\ref{tab:jet-data}).
In Fig.~\ref{fig:ALL-plot}, results are shown 
only for the $1\sigma$ prior PDF ensemble.
We have explicitly checked their stability upon the choice of any prior PDF
ensemble discussed in Sec.~\ref{sec:prior11}: the asymmetry shows hardly
noticeable differences among different priors, thus proving 
it is not sensitive to quark-antiquark separation,
as expected.

The $\chi^2$ per data point before reweighting for separate
and combined data sets is quoted in Tab.~\ref{tab:global1}.
Neither STAR nor PHENIX jet data are provided with experimental 
covariance matrix,
hence we will assume the systematics to be uncorrelated and then sum in 
quadrature with statistical errors. Also, we have to account for the fact that
data are taken in bins of $p_T$, whereas the corresponding theoretical 
predictions are computed for the center of each bin. We estimate the 
corresponding uncertainty as the maximal variation of the observable 
within each bin and take that value as a further uncorrelated systematic
uncertainty. Since STAR data are provided with asymmetric systematic 
uncertainties, we must take care of symmetrizing them, according to
Eqs.~(7)-(8) in Ref.~\cite{DelDebbio:2004qj}.

We observe that our predictions are in good agreement with experimental data, 
as $\chi^2/N_{\mathrm{dat}} \sim 1$ for all data sets. Nevertheless, notice that,
except for STAR 1j-09 data set,
experimental data points are affected by rather large errors in 
comparison to the uncertainty on the observable due to the polarized PDFs: 
hence, this will limit their potential in constraining the polarized gluon
distribution.
Furthermore, similar results are obtained with different choices 
of the prior PDF ensemble: this strengthens
the conclusion that we are looking at a 
process which is almost insensitive to the quark content of the proton.

\begin{figure}[t]
\centering
\epsfig{width=0.32\textwidth,figure=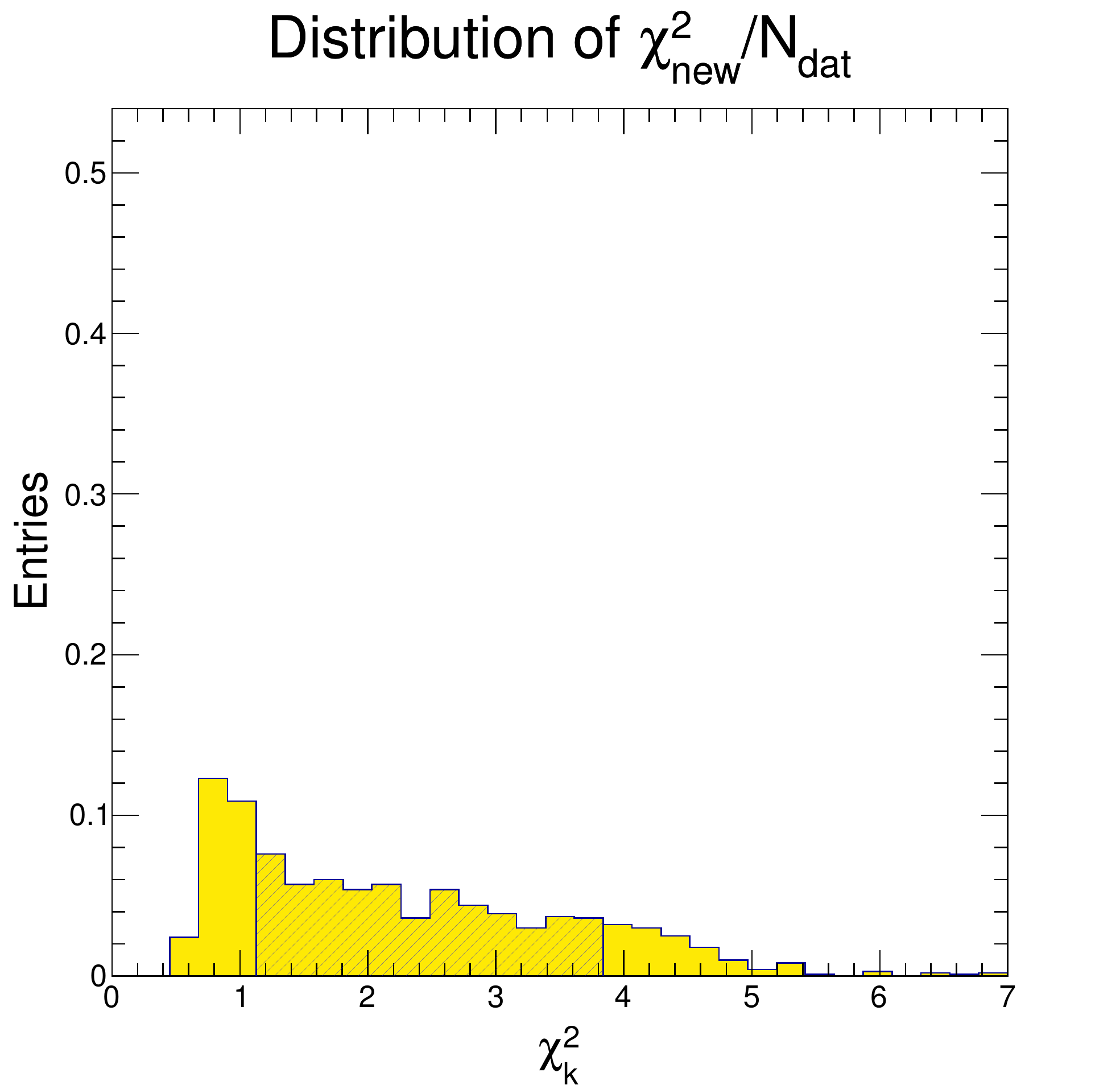}
\epsfig{width=0.32\textwidth,figure=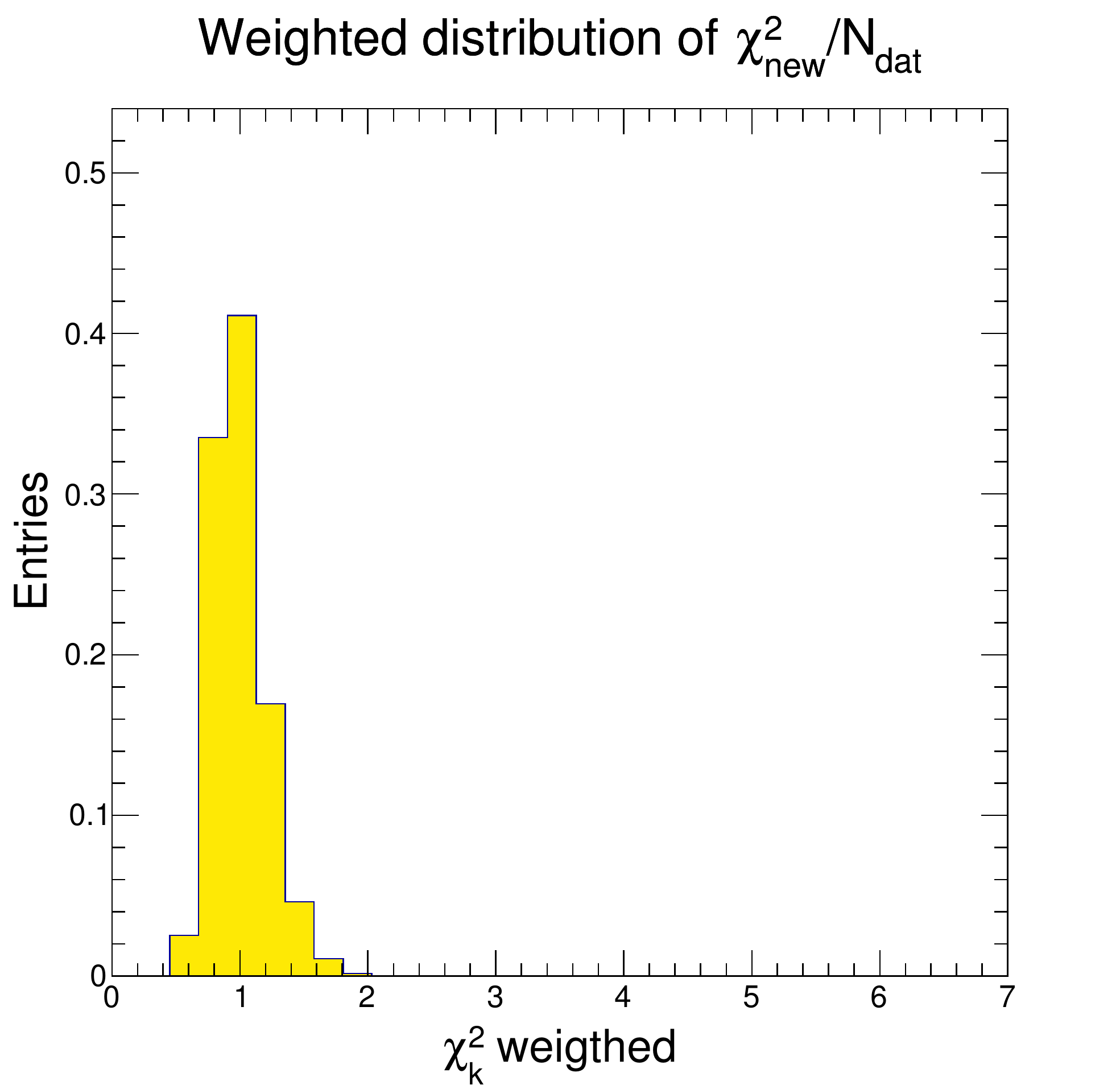}
\epsfig{width=0.32\textwidth,figure=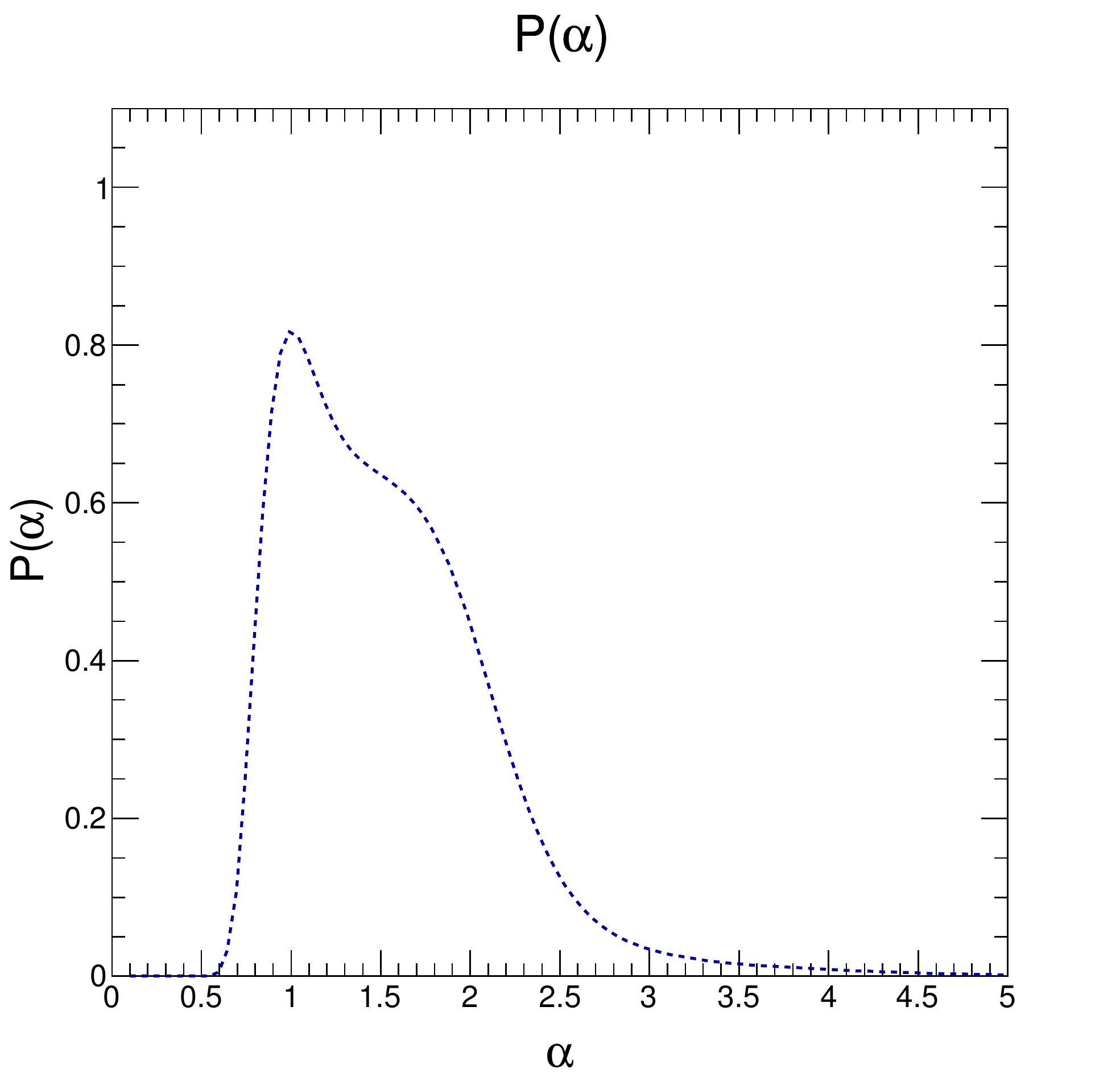}
\mycaption{Same as Fig.~\ref{fig:COMPASS-rwest1}, but for combined RHIC jet data 
and for the $1\sigma$ prior discussed in Sec.~\ref{sec:prior11}.}
\label{fig:est-ALL}
\end{figure}

Next, we reweight the various prior ensembles with both separate and combined
RHIC inclusive jet production data.
In Tab.~\ref{tab:global1}, we quote the values for the $\chi^2$ per data point
after reweighting, $\chi^2_{\mathrm{rw}}/N_{\mathrm{dat}}$,
while the number of effective replicas, $N_{\mathrm{eff}}$, 
and the modal value of the $\mathcal{P}(\alpha)$ distribution,
$\langle\alpha\rangle$, are collected in 
Tab.~\ref{tab:global2}. In Fig.~\ref{fig:est-ALL},
we display, for combined jet data sets, the unweighted distribution of
$\chi^2/N_{\mathrm{dat}}$, the corresponding weighted distribution of 
$\chi^2_{\mathrm{rw}}/N_{\mathrm{dat}}$, and the
$\mathcal{P}(\alpha)$ distribution. In Fig.~\ref{fig:ALL-plot} we finally
compare the asymmetry before and after reweighting.

It is clear from Tab.~\ref{tab:global1} and 
Figs.~\ref{fig:ALL-plot}-\ref{fig:est-ALL} that jet data from 
RHIC carry an important piece of experimental information.
In particular, we observe a substantial improvement in the description of the 
high precision STAR 1j-09 data set, for which the value of the $\chi^2/N_{\mathrm{dat}}$
decreases from 1.69 to 1.02. As we will show below, this improvement
translates into significant constraints on the polarized gluon PDF.
Moreover, the comparison between the $\chi^2$ distributions
before and after reweighting shows that their peak moves close to one, with a slight
narrowing due to the increase in the total number of data points.  
The reweighted observable nicely agrees with experimental data and its
uncertainty is reduced with respect to the prior. 
Also, notice that the PDF set loss of accuracy in describing
the underlying probability distribution is well under control: 
the number of replicas left after reweighting, $N_{\mathrm{eff}}$,
is always larger than $N_{\mathrm{rep}}=100$,
the value roughly required 
for \texttt{NNPDFpol1.0} to reproduce data 
and errors with percent accuracy.
Finally, new data sets are consistent with the experimental information
already included in the prior, as shown by the $\mathcal{P}(\alpha)$
distribution, which is clearly peaked at one (see Fig.~\ref{fig:est-ALL}).

Nevertheless, we notice that different data sets have different power in
improving the fit: the more accurate they are, the more effective they are in
constraining the theoretical prediction for the asymmetry $A_{LL}^{1jet}$
and, in turn, the polarized gluon distribution.
It is clear from Tabs.~\ref{tab:global1}-\ref{tab:global2} that 
a substantial amount of experimental information comes from the STAR 1j-09
data set. We also observe that PHENIX data have a fair impact
in improving the fit, due to their large errors: it is likely that they were
overestimated, since the modal value of the $\alpha$ parameter is far below one
(see Tab.~\ref{tab:global2}).

\begin{figure}[t]
\centering
\epsfig{width=0.32\textwidth,figure=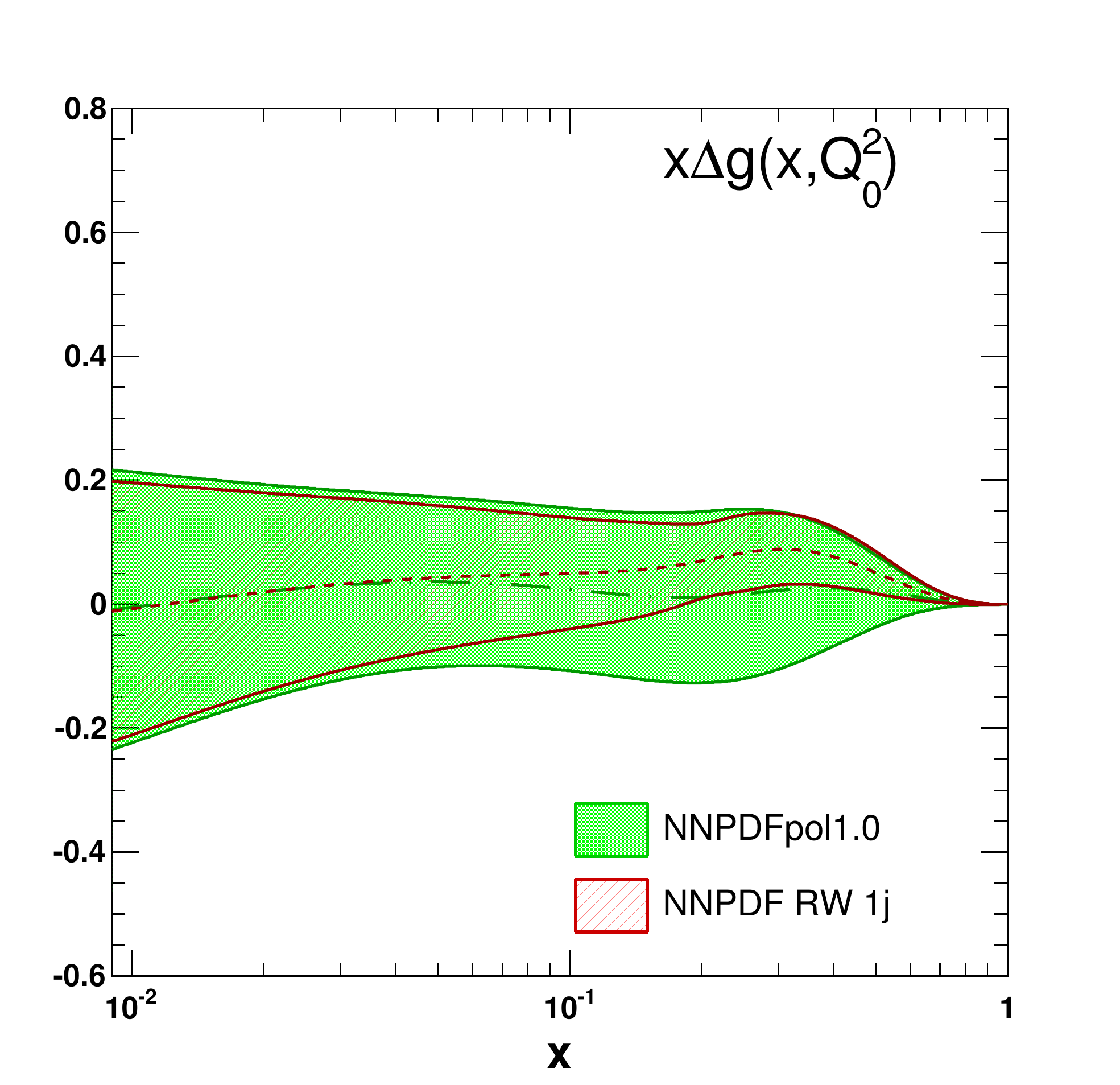}
\epsfig{width=0.32\textwidth,figure=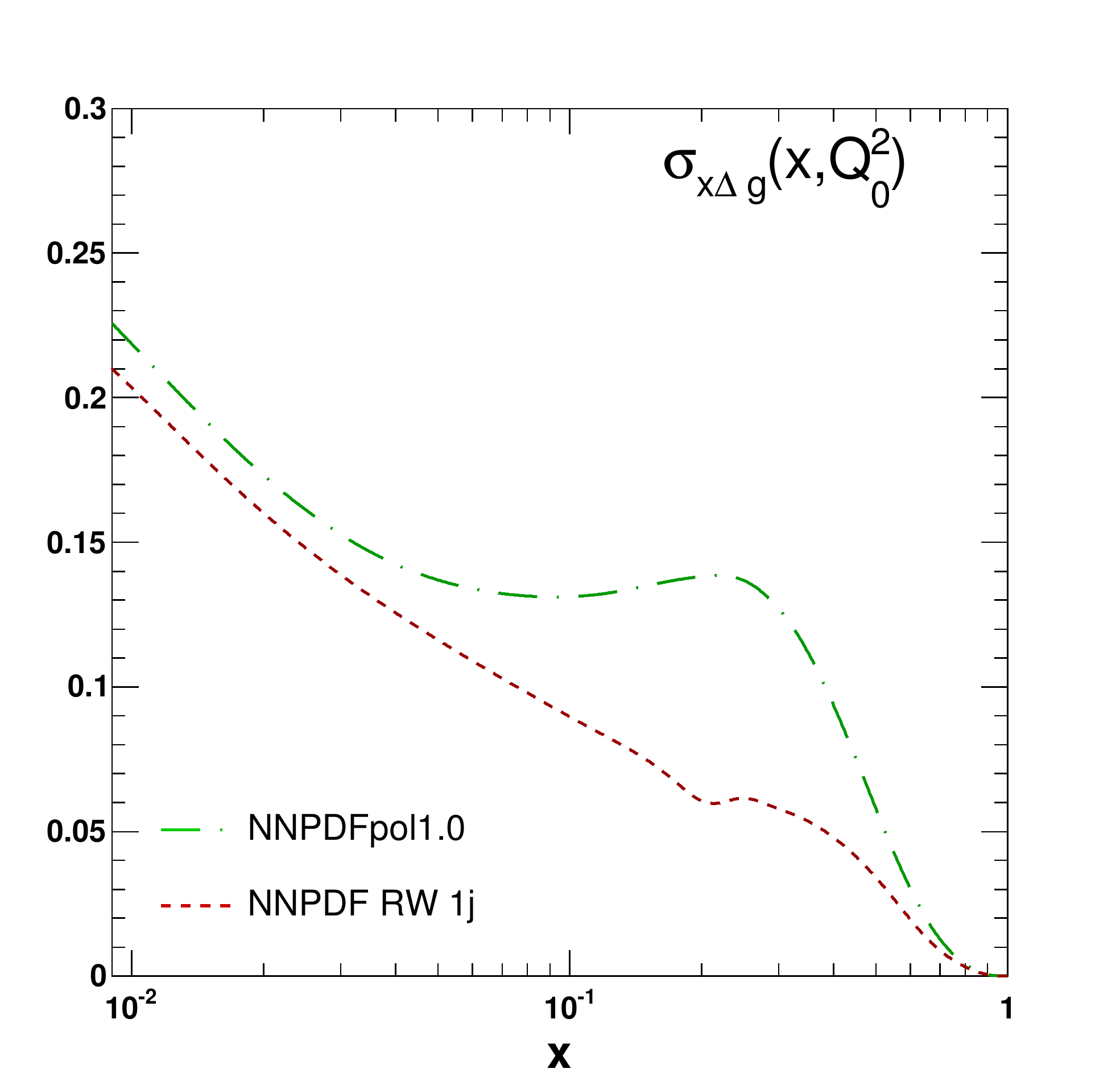}
\epsfig{width=0.32\textwidth,figure=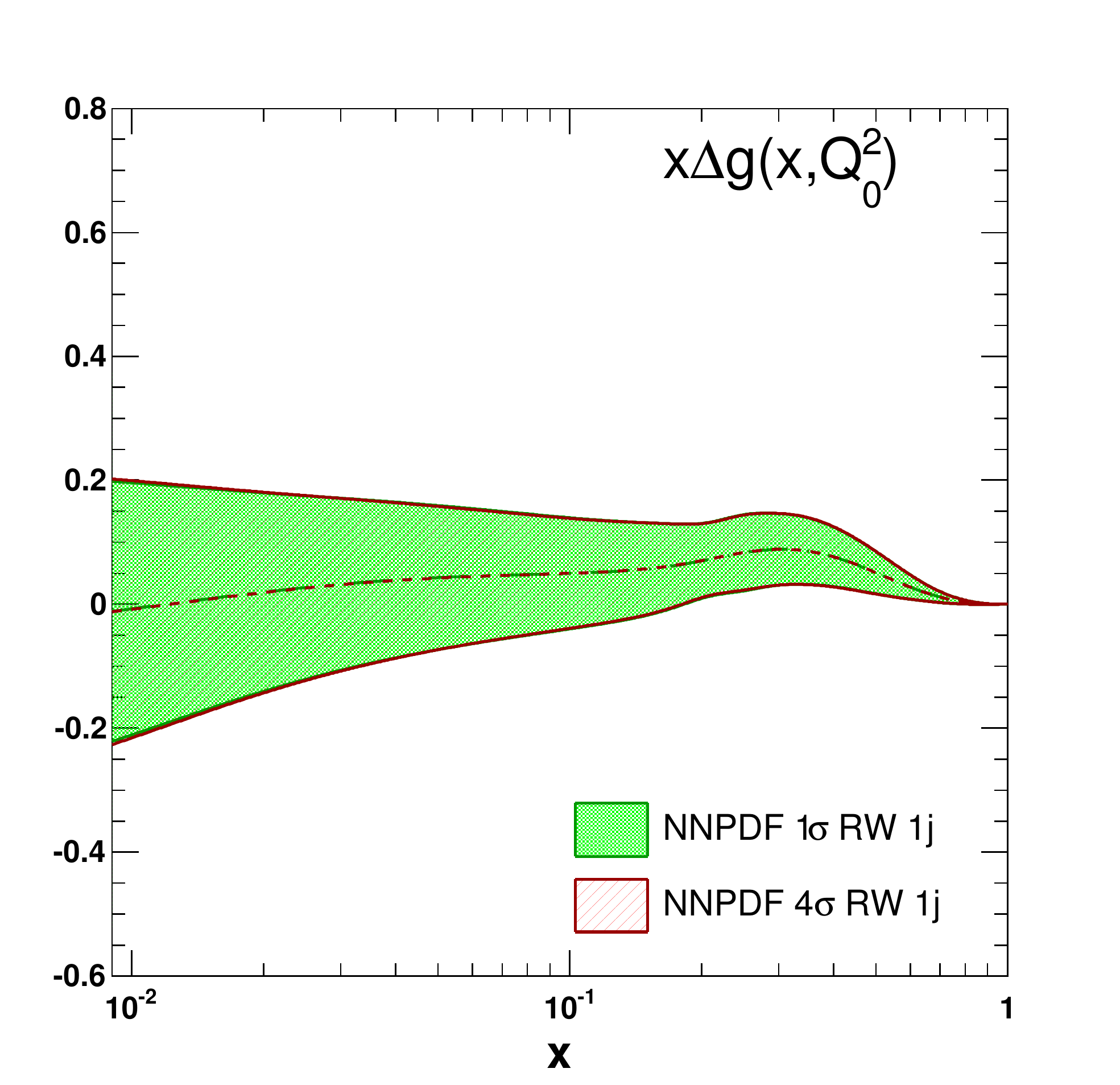}
\mycaption{(First panel) Comparison between the polarized gluon PDF from 
\texttt{NNPDFpol1.0} parton set and the result obtained 
by reweighting with RHIC jet data. (Second panel) the same comparison,
but for absolute PDF uncertainty. (Third panel) Comparison between the
reweighted gluon PDF obtained from the $1\sigma$ and the $4\sigma$ prior
PDF ensembles.}
\label{fig:gluonrw-jets}
\end{figure}

Finally, in Fig.~\ref{fig:gluonrw-jets} we compare the polarized gluon PDF from 
the \texttt{NNPDFpol1.0} parton set~\cite{Ball:2013lla} with its counterpart
from the $1\sigma$ prior reweighted with RHIC jet data. 
We also show the corresponding one-sigma absolute error and a comparison
between the reweighted results obtained with extremum prior PDF sets,
namely $1\sigma$ and $4\sigma$. 
We observe that, in the kinematic range probed at RHIC, the polarized gluon
PDF becomes positive and its uncertainty is reduced.
This effect is consistent with what was found in 
Ref.~\cite{Aschenauer:2013woa}, where these data were included in the
\texttt{DSSV++} parton set. 
Again, we stress that, in the case of jet data, results are independent
of the choice of any prior PDF ensemble discussed in Sec.~\ref{sec:prior11}:
the stability of our results is clearly visible from both 
Tabs.~\ref{tab:global1}-\ref{tab:global2} and from the third panel of
Fig.~\ref{fig:gluonrw-jets}. 
Hence, results obtained from the reweighting of any prior
are equivalent, and can be used indifferently to describe the polarized gluon
PDF. We have explicitly checked that other PDFs, not shown 
in Fig.~\ref{fig:gluonrw-jets} are unaffected by jet data, as expected.

\subsubsection{\texorpdfstring{$W$}{W} boson production at STAR}
\begin{figure}[t]
\begin{center}
\epsfig{width=0.40\textwidth,figure=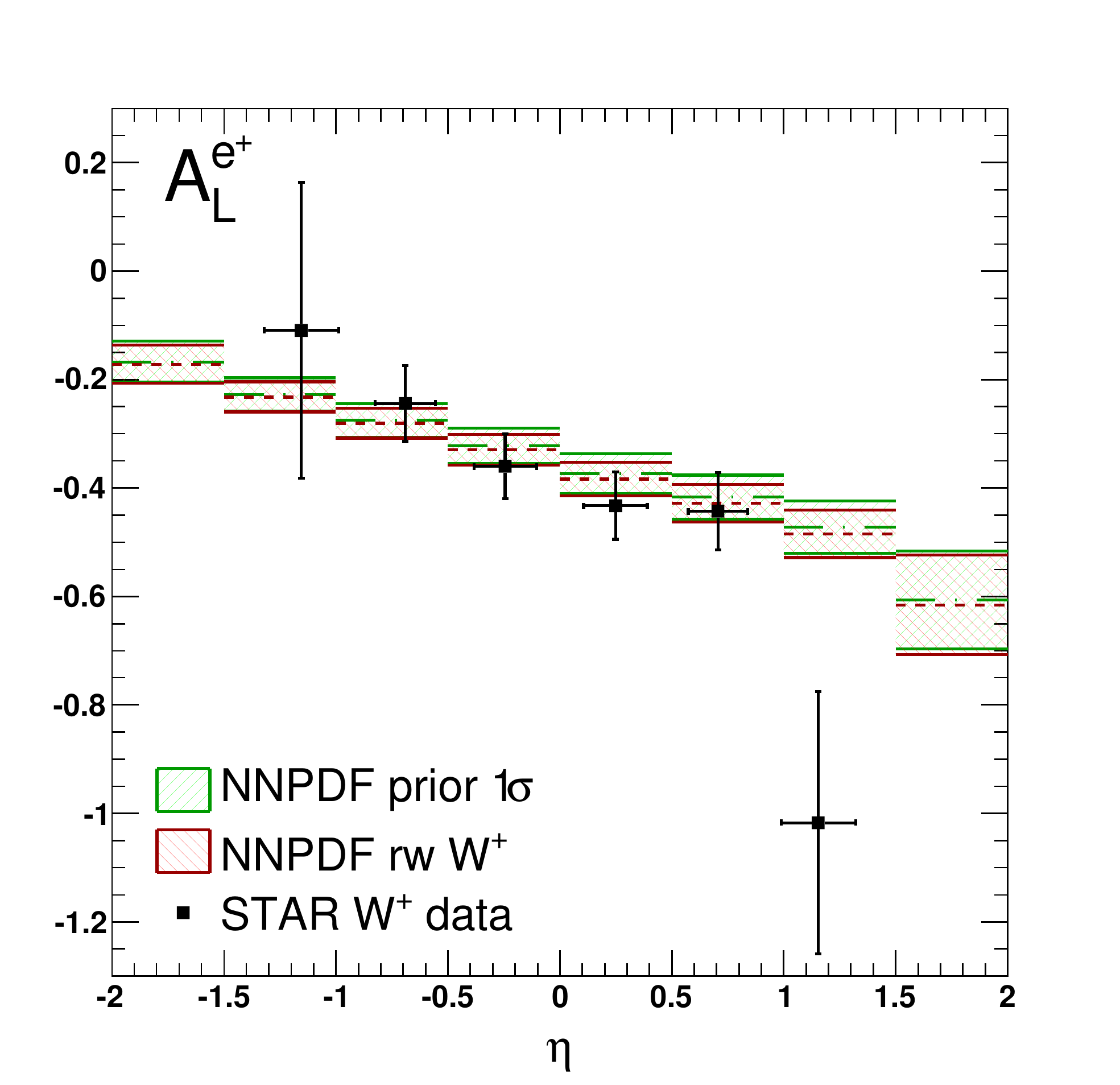}
\epsfig{width=0.40\textwidth,figure=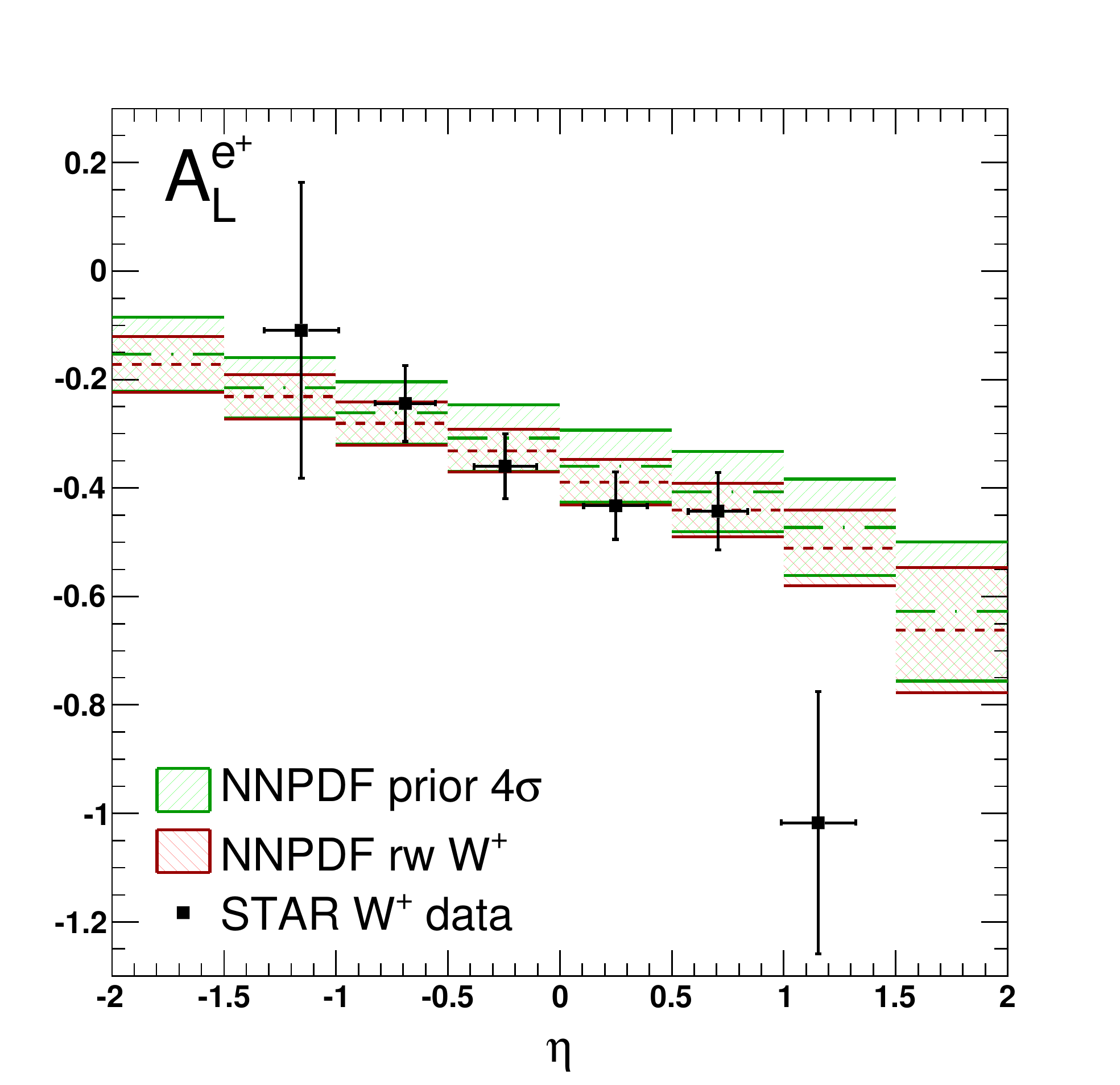}\\
\epsfig{width=0.40\textwidth,figure=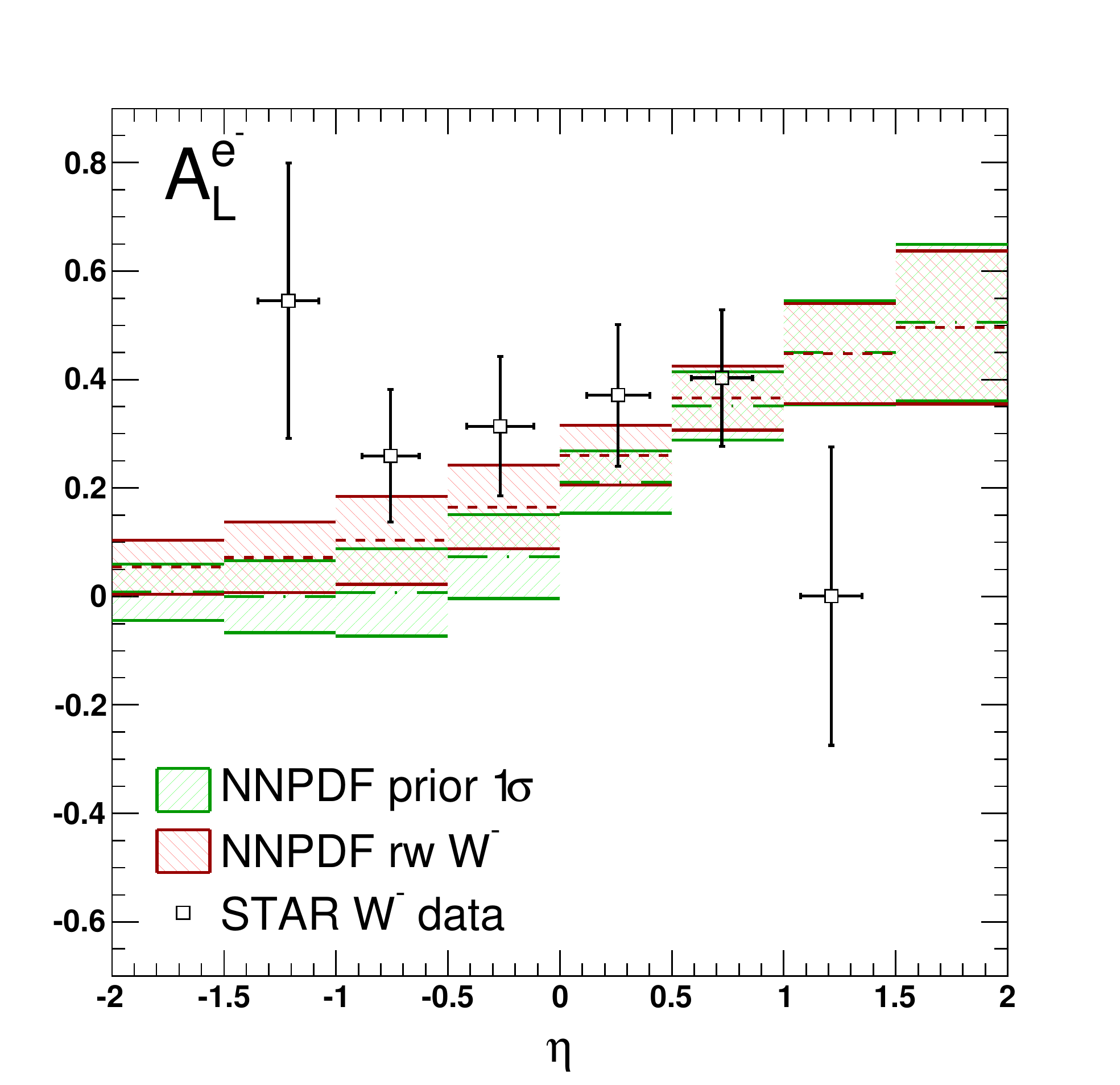}
\epsfig{width=0.40\textwidth,figure=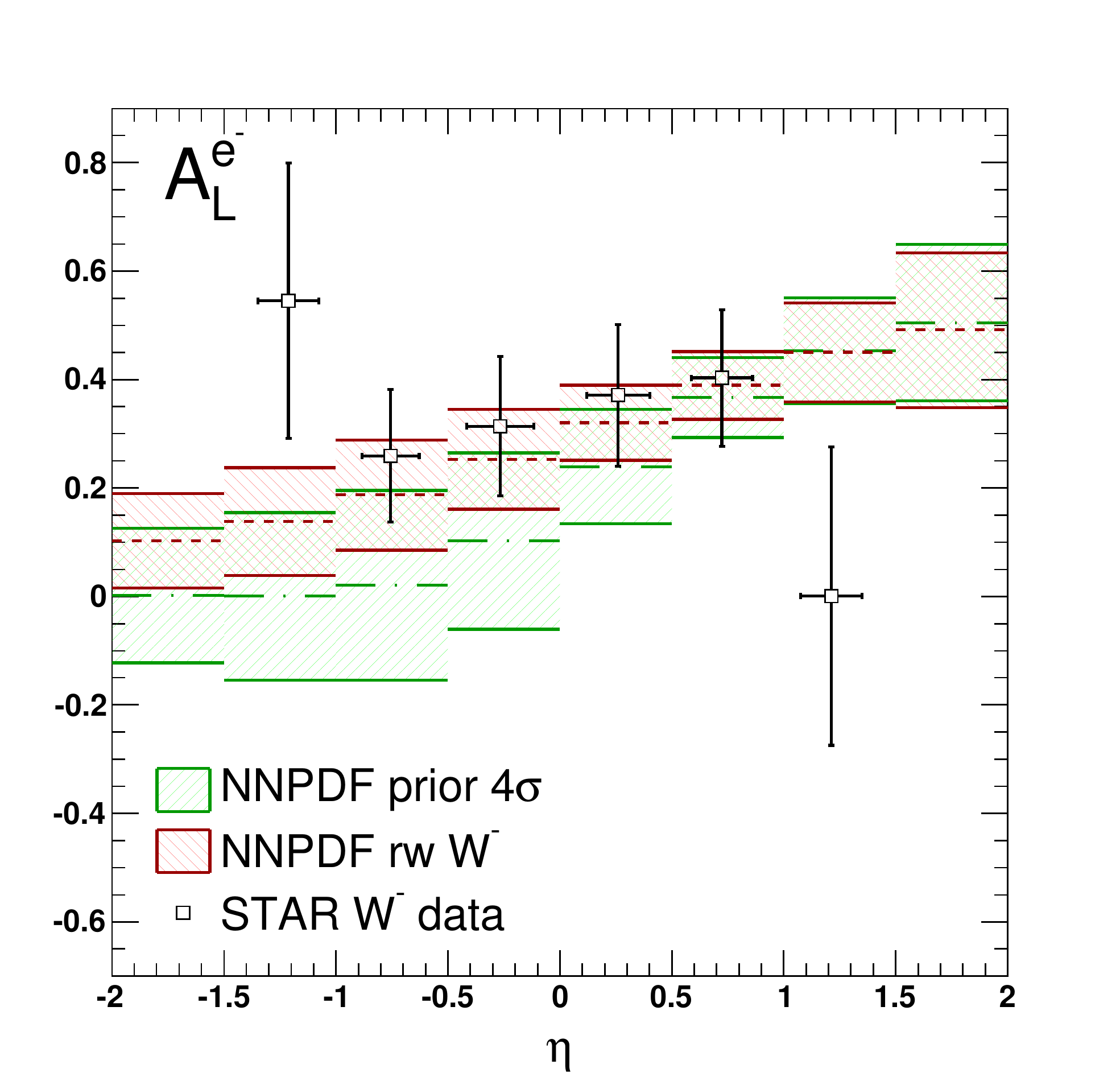}\\
\mycaption{Predictions for the longitudinal positron (upper plots) and electron 
(lower plots) single spin asymmetry $A_L^{e^+}$ and $A_L^{e^-}$ 
before and after reweighting with STAR data~\cite{Stevens:2013raa}.
Results from $1\sigma$ (left) and $4\sigma$ (right) prior PDF ensembles 
are shown.
Curves are obtained at NLO with the CHE code~\cite{deFlorian:2010aa}.
Experimental points are also shown, uncertainties are statistical only.}
\label{fig:STAR_afterrw}
\end{center}
\end{figure}

Predictions for the longitudinal single-spin asymmetry, Eq.~(\ref{eq:Wasy}),
are computed at NLO using the program of Ref.~\cite{deFlorian:2010aa},
which was consistently modified to handle NNPDF parton sets.
In Fig.~\ref{fig:STAR_afterrw}, we show our predictions for 
the longitudinal positron (electron) single spin asymmetry
$A_L^{e^+}$ ($A_L^{e^-}$) from production and decay of $W^{+(-)}$
bosons in bins of the rapidity $\eta$, 
compared to STAR data~\cite{Stevens:2013raa}. 
Results are diplayed for the $1\sigma$ and $4\sigma$
prior PDF ensembles, and we have checked that intermediate results
are obtained from the other priors constructed in Sec.~\ref{sec:prior11}.
Unlike open-charm and jet production observables 
discussed above, the longitudinal single spin asymmetry, 
Eq.~(\ref{eq:Wasy}), is sensitive to separate quark and antiquark
PDFs. For this reason, the corresponding theoretical predictions 
made from the $4\sigma$ prior are affected by larger uncertainties than 
those obtained from the $1\sigma$ prior.

In Tab.~\ref{tab:global1}, we summarize the agreement between experimental data and 
parton sets before reweighting, as quantified by the value of $\chi^2/N_{\mathrm{dat}}$:
as we can see, it is not very good, and all prior PDF sets fail 
to describe these data sets with sufficient accuracy.
The discrepancy between data and corresponding theoretical predictions
is particularly noticeable for the $W^-$ set.
This may reflect some tension between semi-inclusive data, 
included in the \texttt{DSSV08} global fit, and $W$ production data.
Indeed, the experimental information included in \texttt{DSSV08} 
was inherited by NNPDF prior ensembles, due to the way they were constructed.
After reweighting with STAR data, we should demonstrate that this
dependence has been removed and check that reweighted PDFs properly describe
$W^\pm$ data sets.

\begin{figure}[p]
\flushleft{$1\sigma$ prior}\\
\centering
\epsfig{width=0.315\textwidth,figure=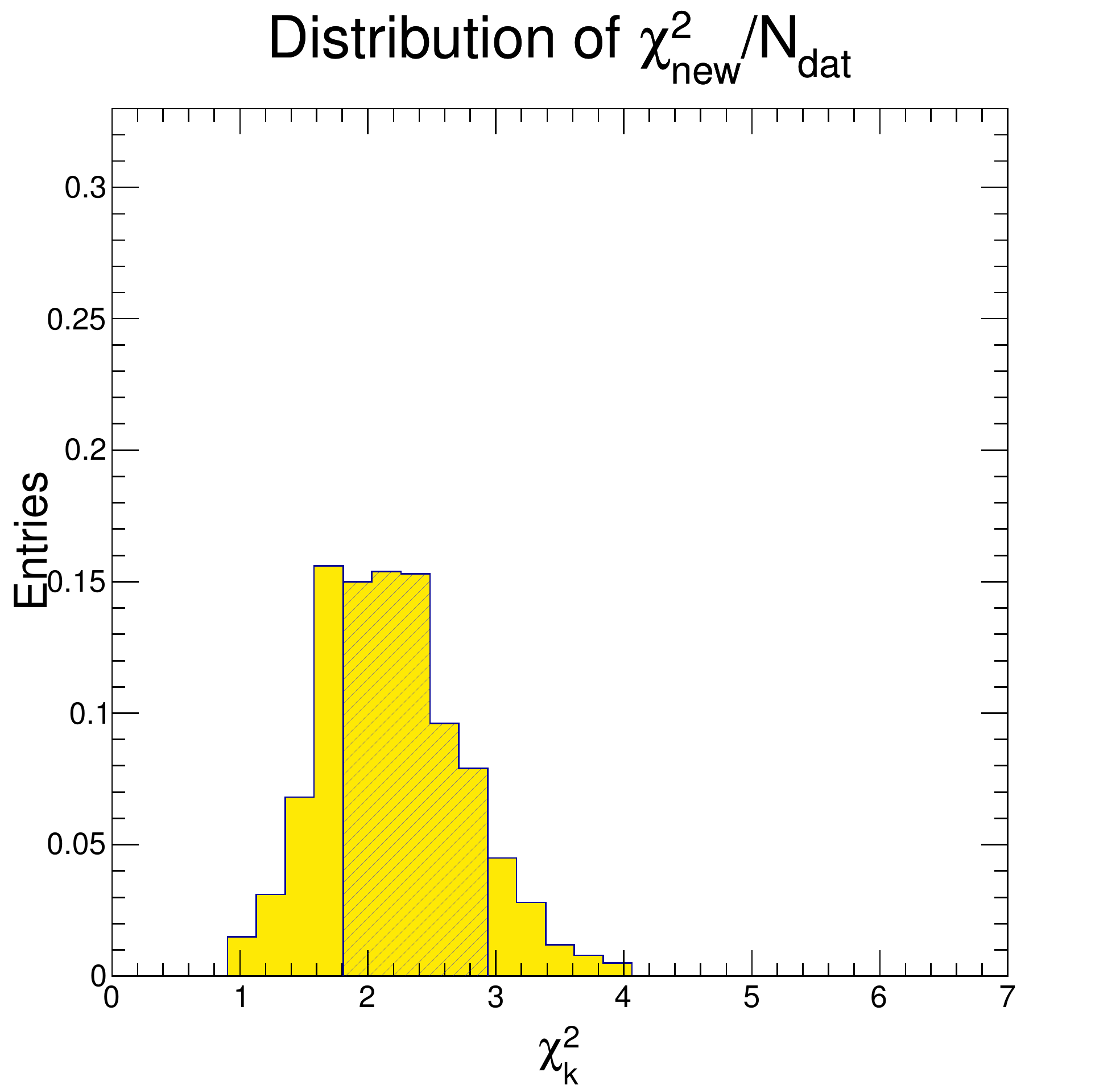}
\epsfig{width=0.315\textwidth,figure=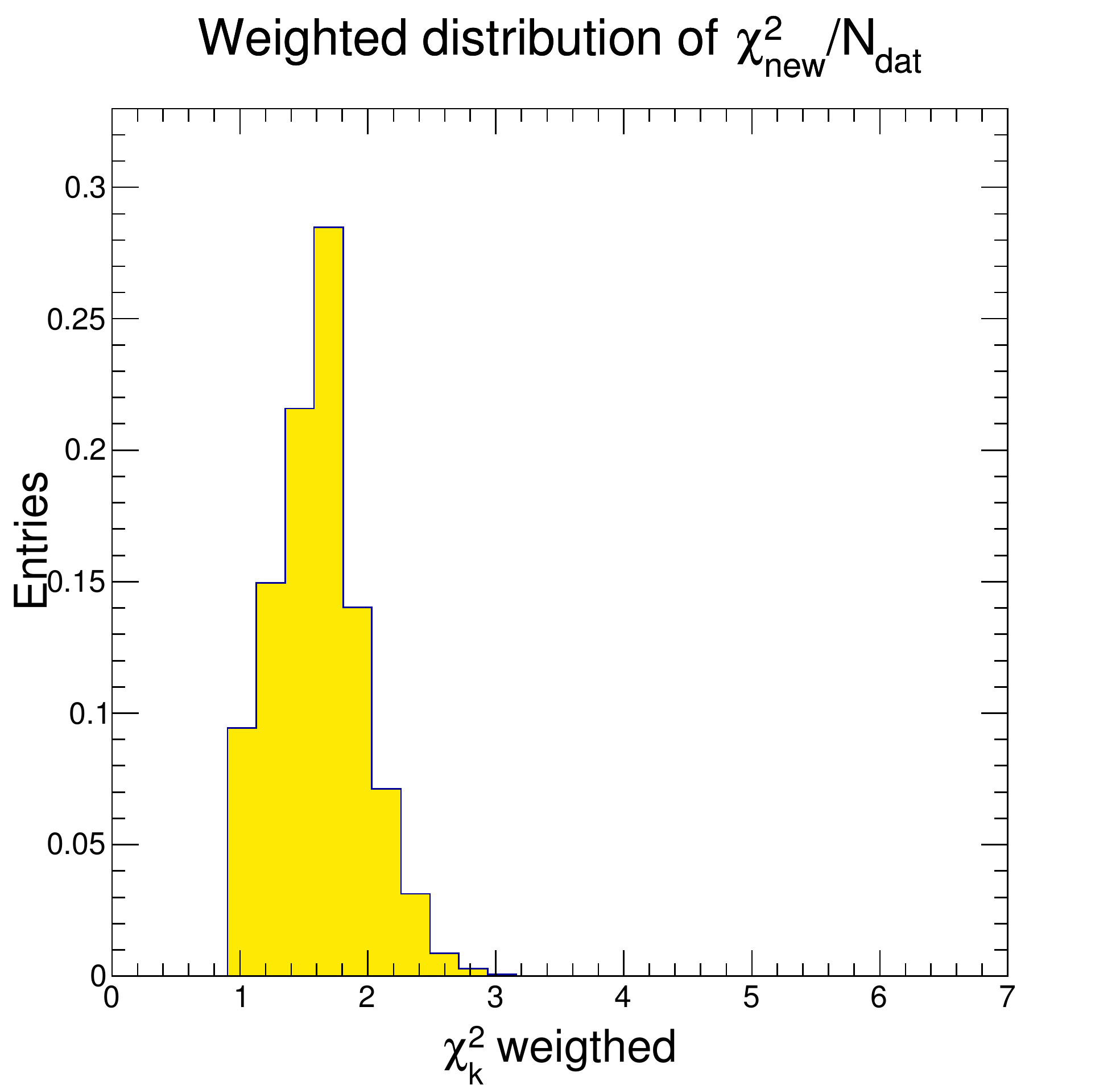}
\epsfig{width=0.315\textwidth,figure=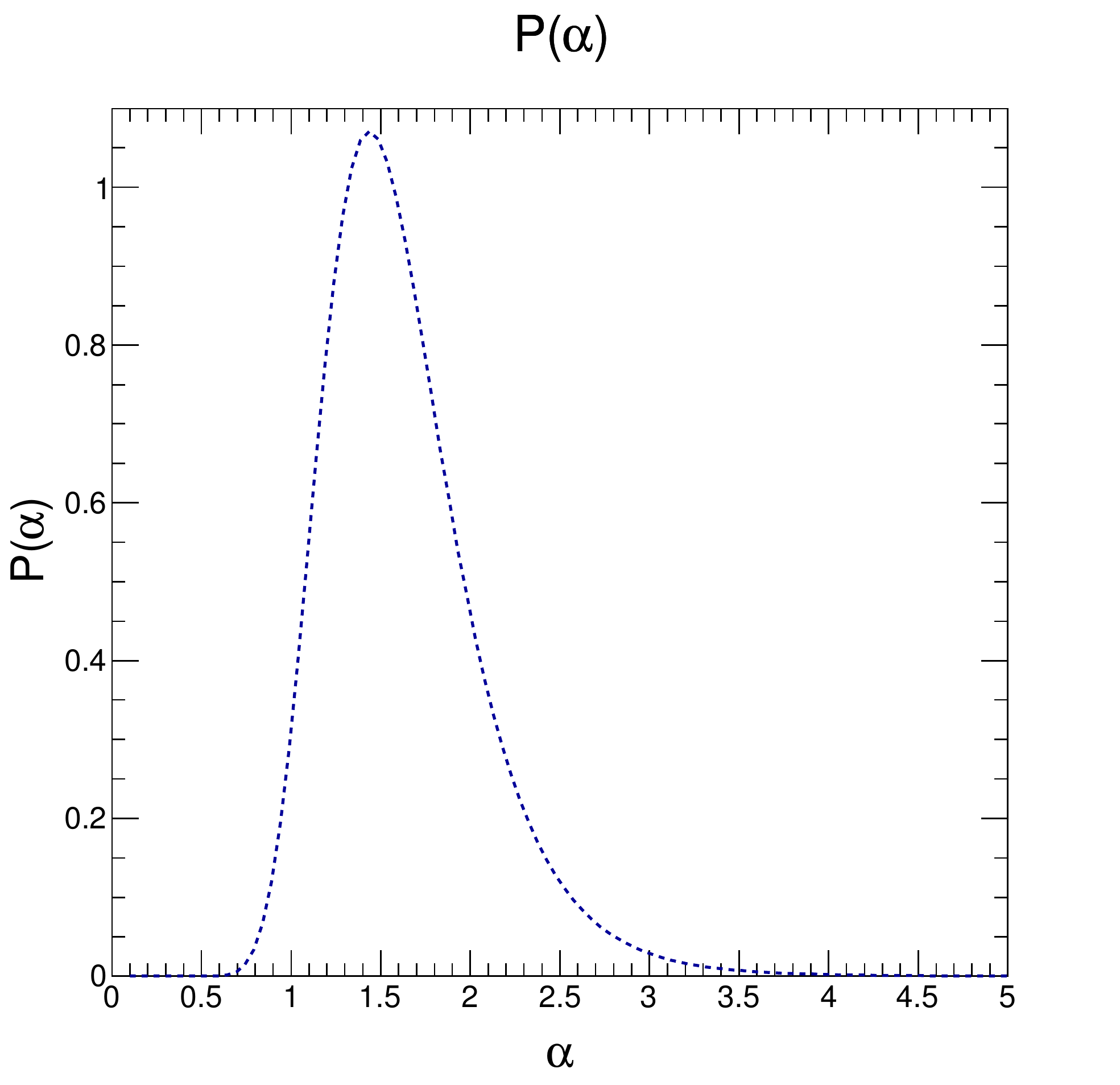}\\
\flushleft{$2\sigma$ prior}\\
\centering
\epsfig{width=0.315\textwidth,figure=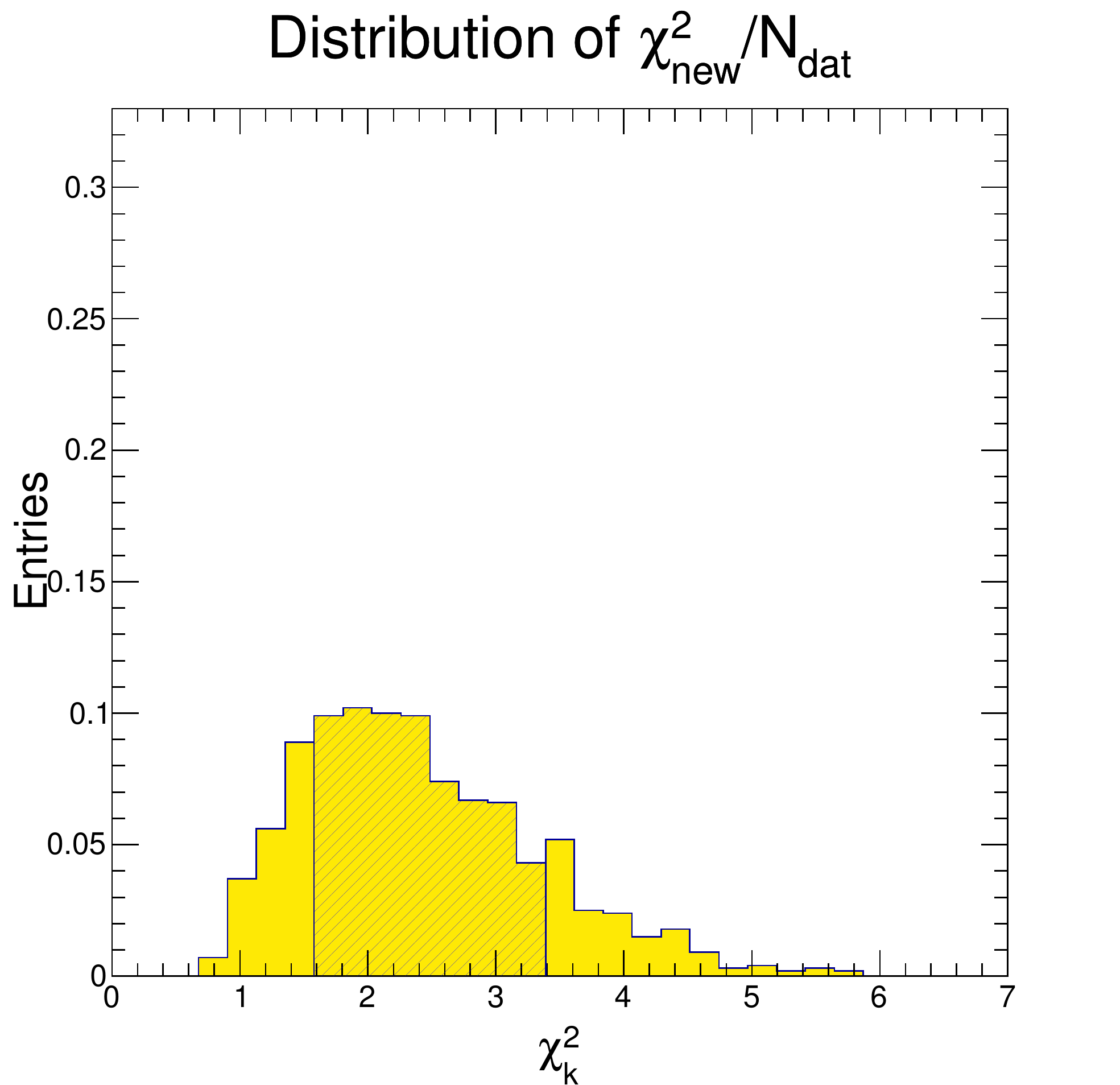}
\epsfig{width=0.315\textwidth,figure=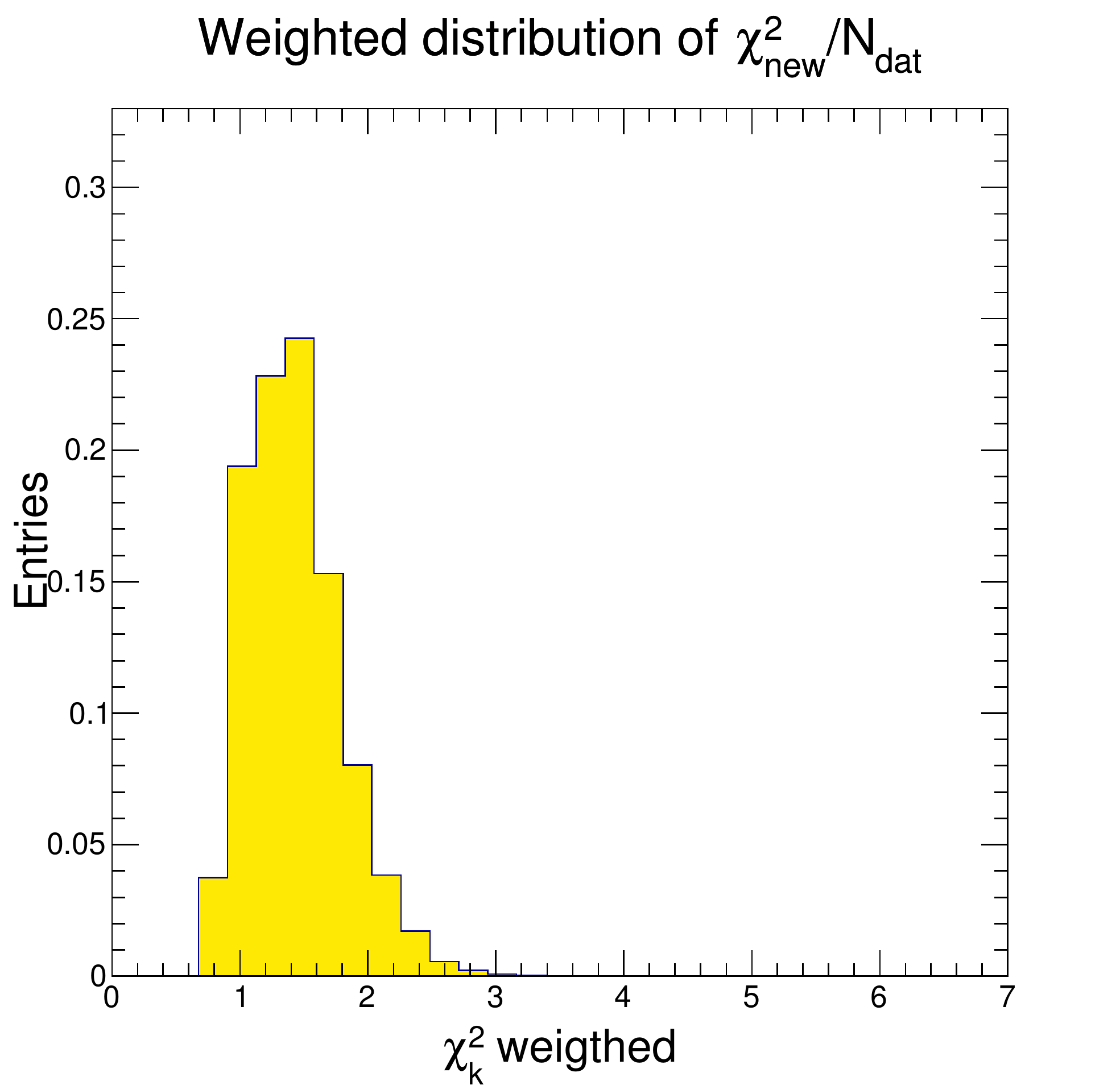}
\epsfig{width=0.315\textwidth,figure=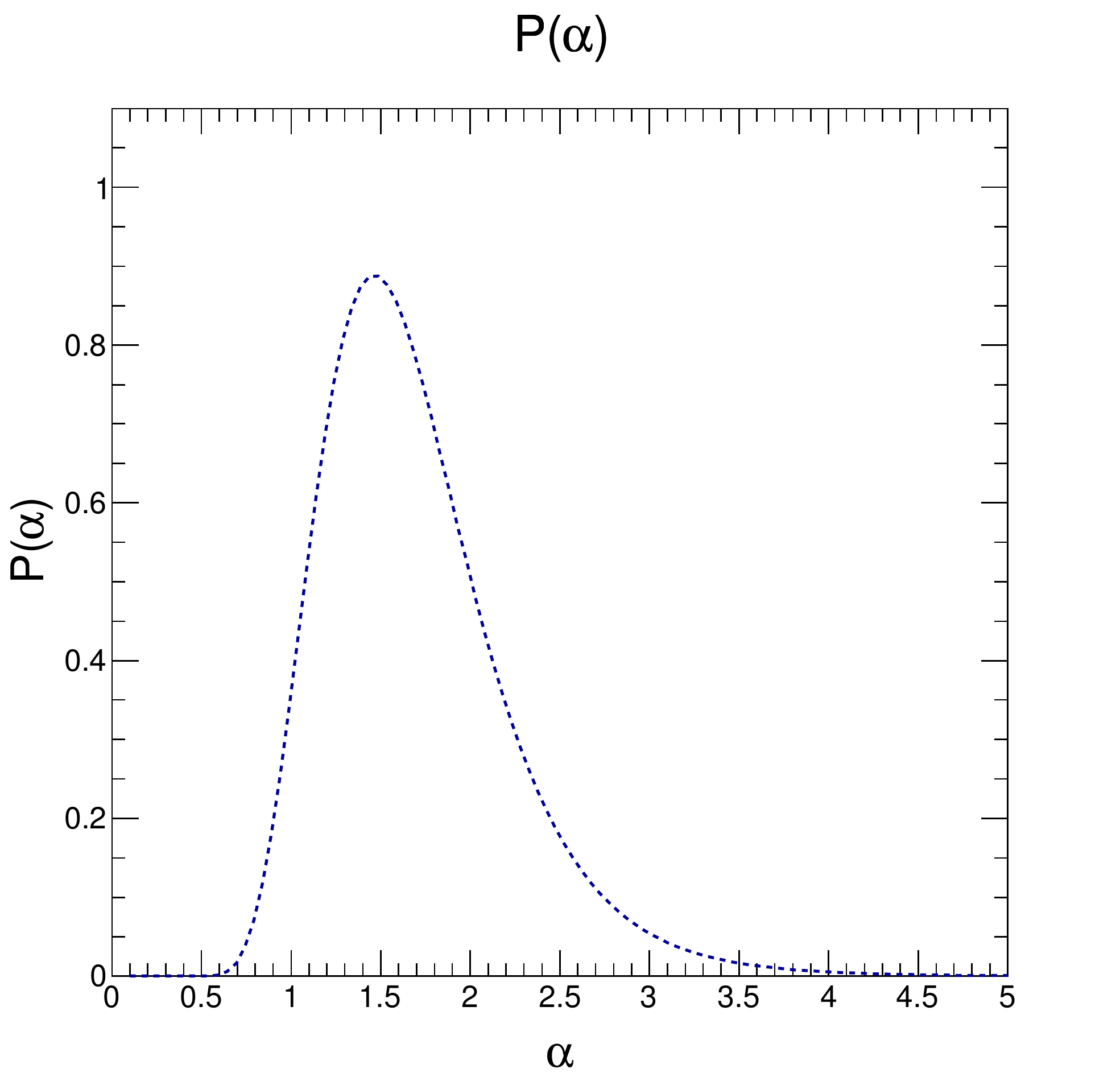}\\
\flushleft{$3\sigma$ prior}\\
\centering
\epsfig{width=0.315\textwidth,figure=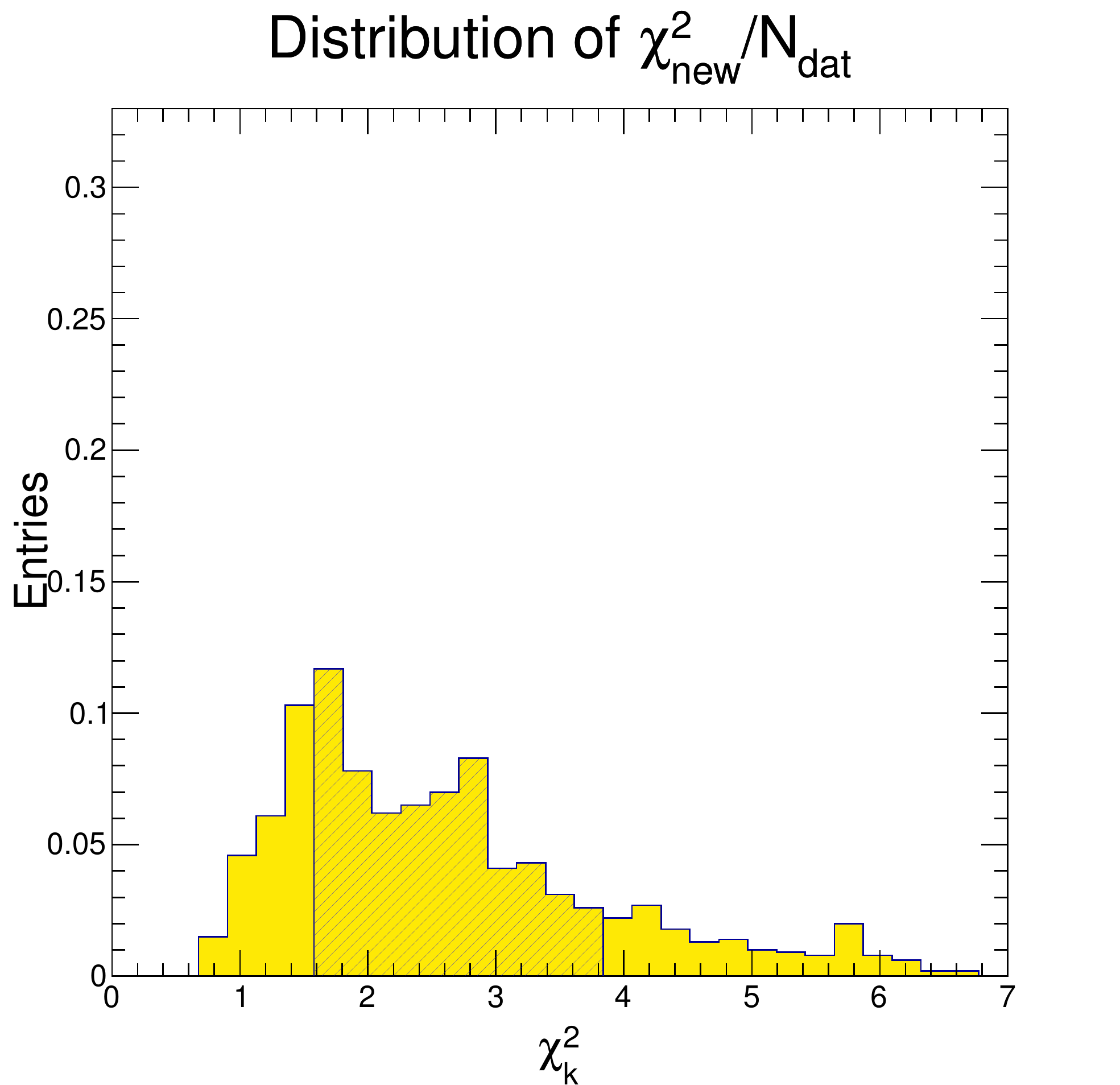}
\epsfig{width=0.315\textwidth,figure=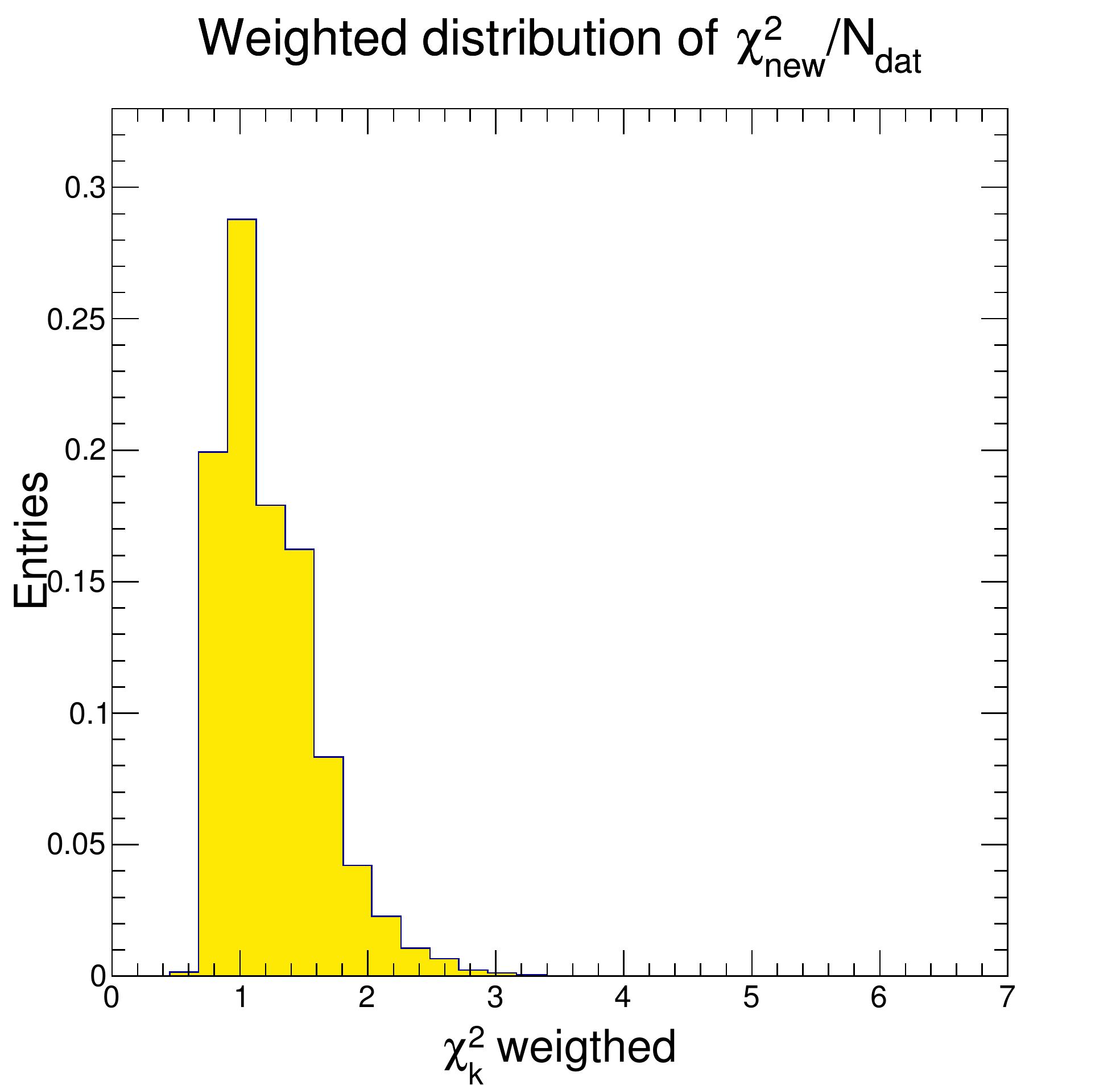}
\epsfig{width=0.315\textwidth,figure=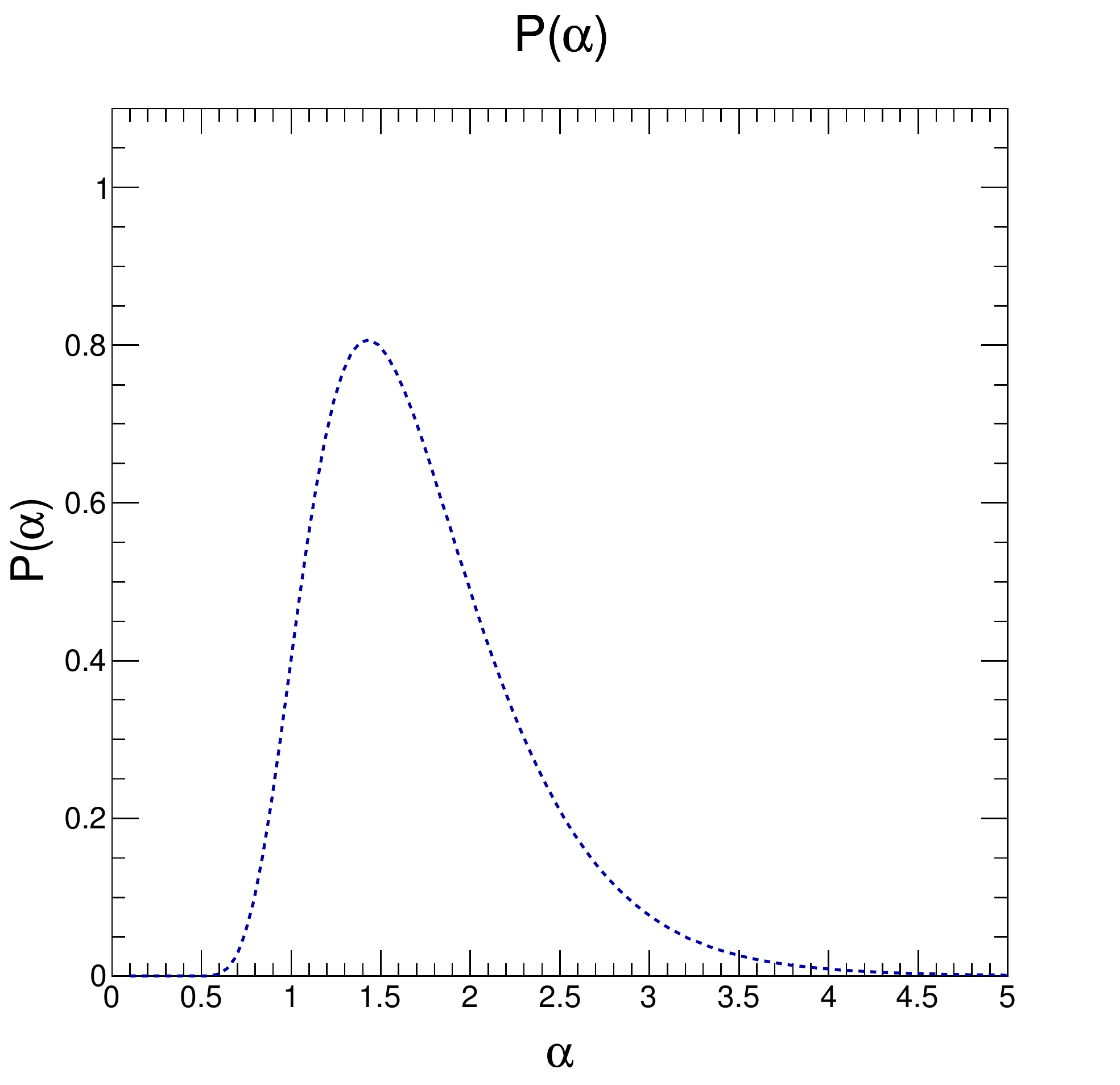}\\
\flushleft{$4\sigma$ prior}\\
\centering
\epsfig{width=0.315\textwidth,figure=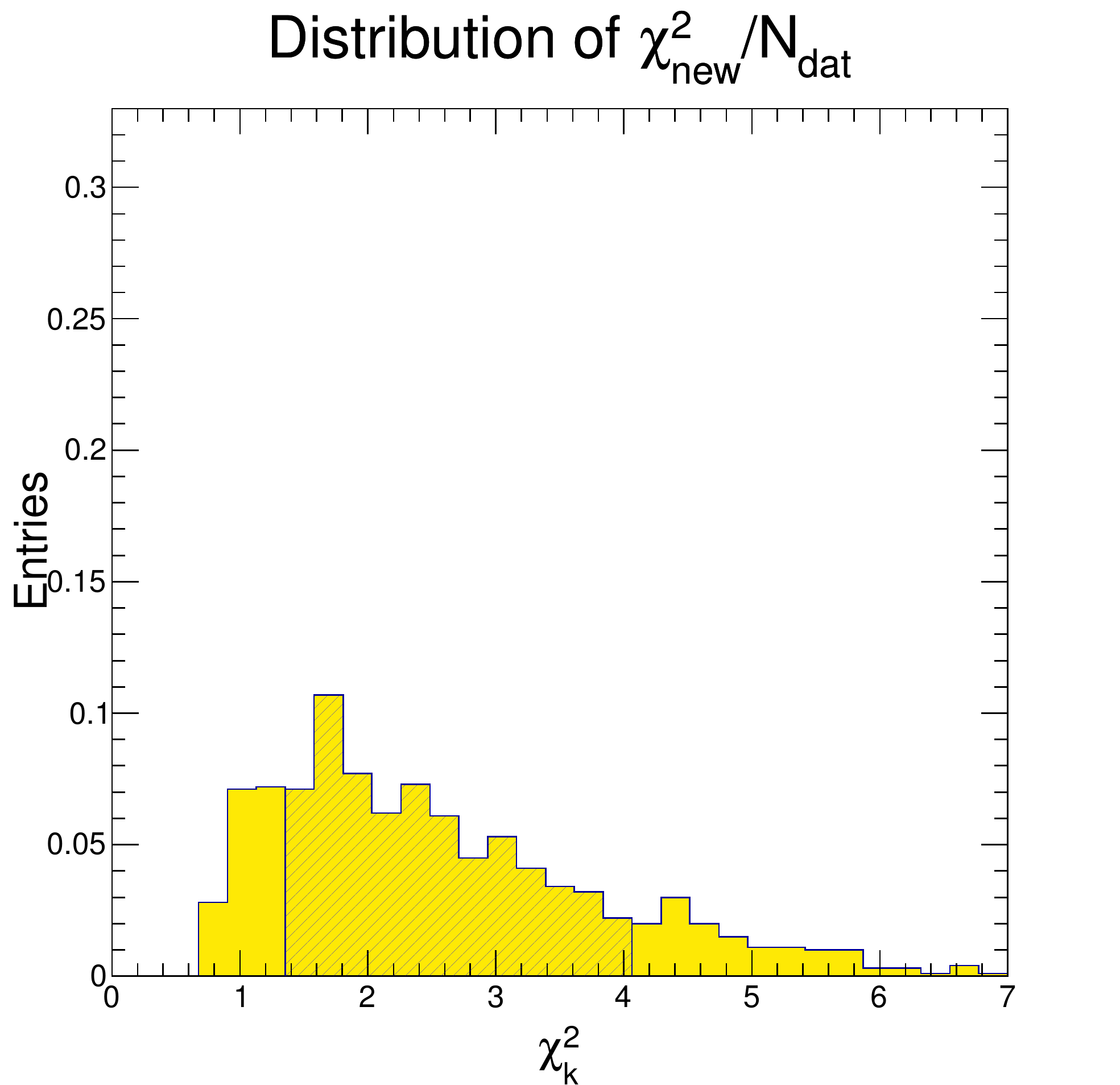}
\epsfig{width=0.315\textwidth,figure=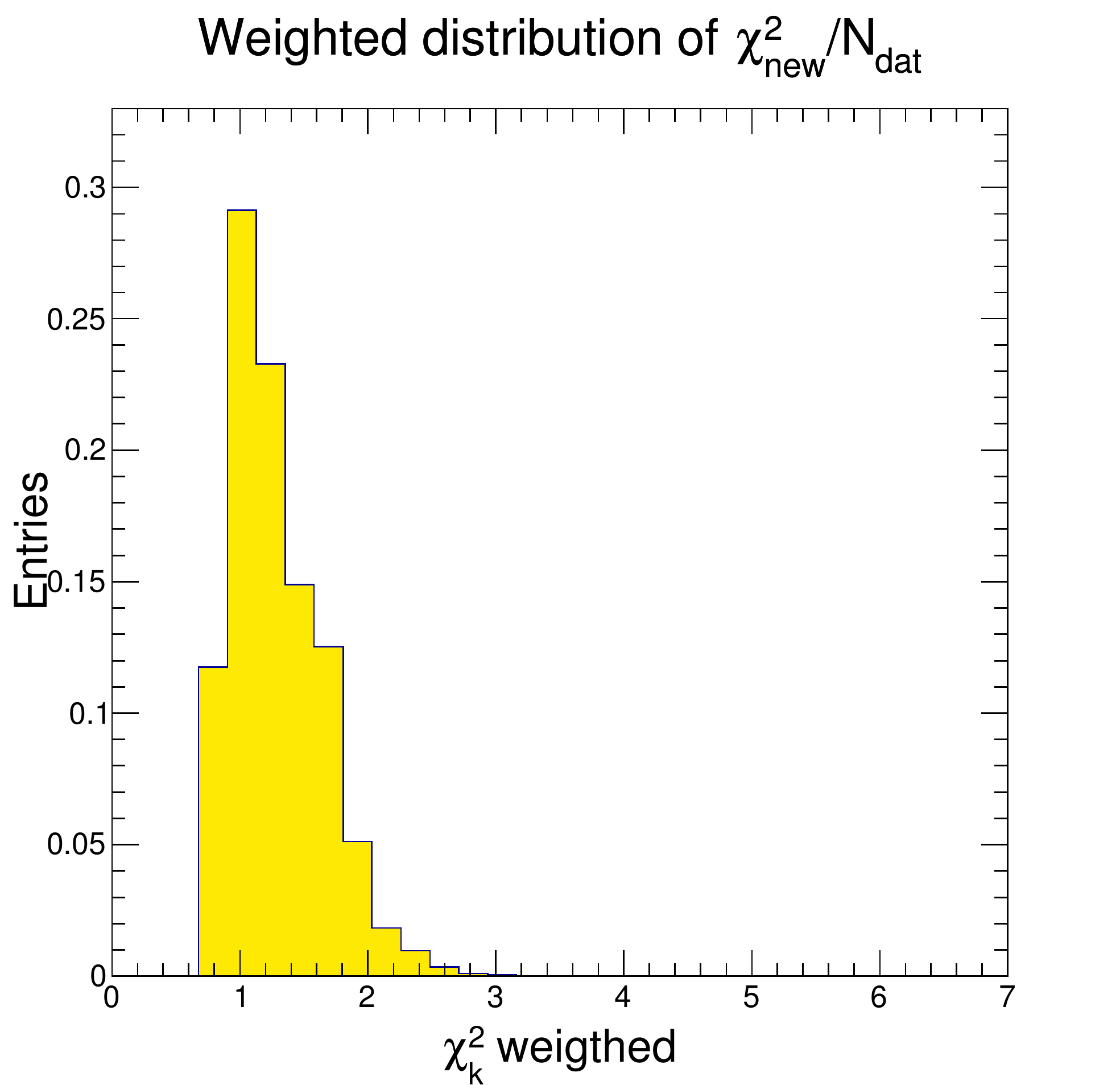}
\epsfig{width=0.315\textwidth,figure=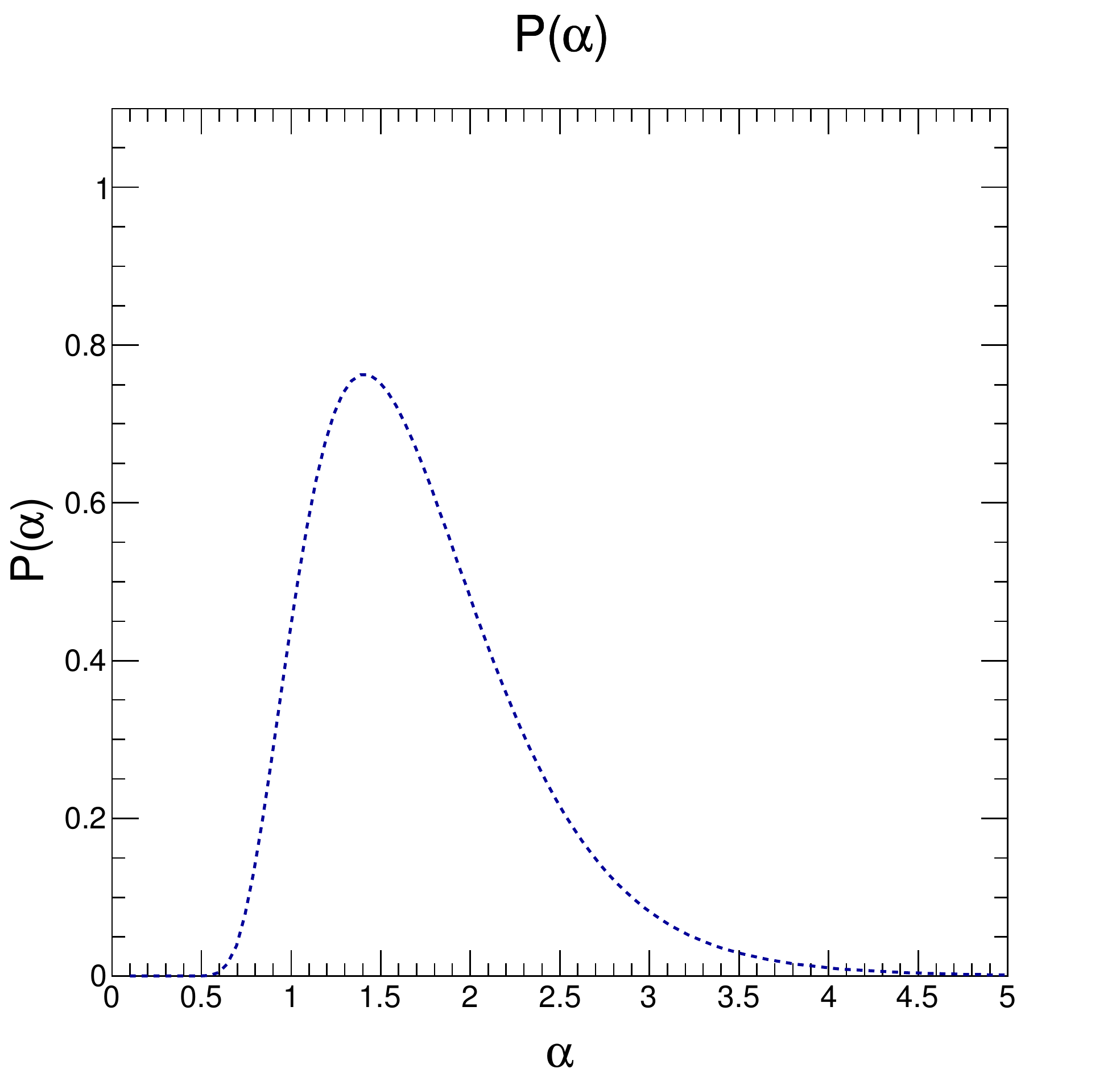}\\
\mycaption{Same as Fig.~\ref{fig:COMPASS-rwest1}, but for combined 
$W^+$ and $W^-$ STAR data sets~\cite{Stevens:2013raa} 
and for each prior.
}
\label{fig:STARest_tot}
\end{figure}

As for the other observables, we proceed to reweight the different
prior ensembles with STAR $W$ production data.
Results are collected in Tabs.~\ref{tab:global1}-\ref{tab:global2}
for both separate and combined STAR data sets,
and compared to the analogous quantity before reweighting. 
In Fig.~\ref{fig:STARest_tot}
we display, for each prior PDF ensemble, and for combined $W^+$ and
$W^-$ STAR data sets, the unweighted distribution of $\chi^2/N_{\mathrm{dat}}$,
the corresponding weighted distribution of $\chi^2_{\mathrm{rw}}/N_{\mathrm{dat}}$, 
and the $\mathcal{P}(\alpha)$ distribution.
Finally, in Fig.~\ref{fig:STAR_afterrw} we compare the asymmetry
before and after reweighting.

We see that, after reweighting, our predictions agree
with STAR data, for both the $W^+$ and the $W^-$ final states, 
but the goodness of this agreement actually depends on the prior.
In general, we observe that the $\chi^2/N_{\mathrm{dat}}$
value decreases after reweighting (see Tab.~\ref{tab:global1}),
and that the $\chi^2$ distribution tends to properly peak at one,
being more narrow than before reweighting (compare, for each prior
PDF ensemble, plots in the first and second column of 
Fig.~\ref{fig:STARest_tot}). 
The value of the effective number of replicas after reweighting, $N_{\mathrm{eff}}$,
is always larger than $N_{\mathrm{rep}}=100$, 
thus the size of the initial prior sample was large enough.
The modal value of the $\mathcal{P}(\alpha)$ distribution,
$\langle \alpha \rangle$ is close to one for
the $W^+$ data but almost 1.4 for the $W^{-}$ data, suggesting that in the 
latter case some experimental uncertainties might be somewhat underestimated.
This might also explain why the STAR-$W^-$ data set has a much smaller value
of $N_{\mathrm{eff}}$ than STAR-$W^+$, while we expected the two sets to have
a similar impact.
Finally, the reweighted observable nicely agrees with experimental data and
its uncertainty is reduced with respect to the prior, 
as clearly shown in Fig.~\ref{fig:STAR_afterrw}.

\begin{figure}[!t]
\centering
\epsfig{width=0.40\textwidth,figure=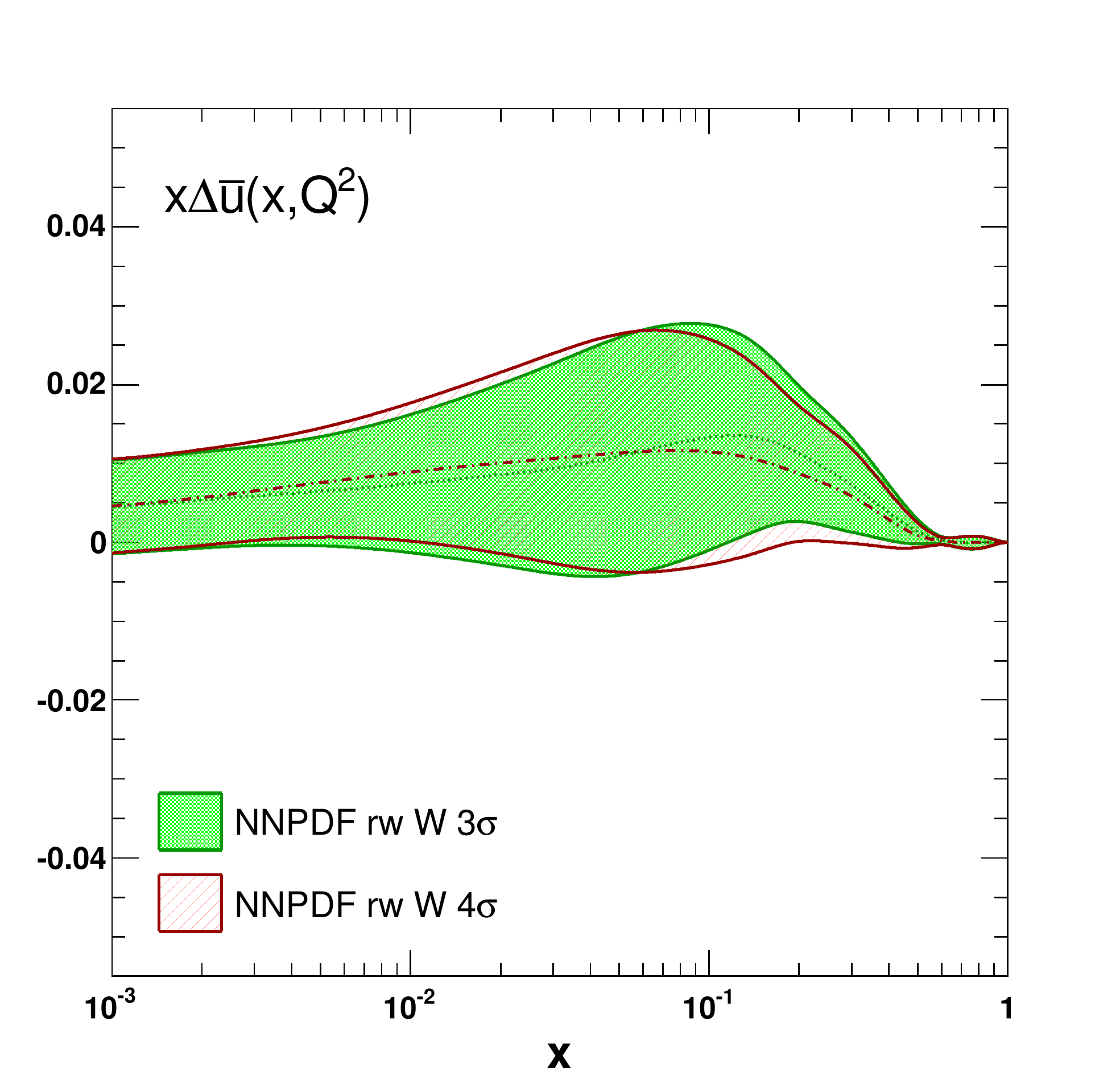}
\epsfig{width=0.40\textwidth,figure=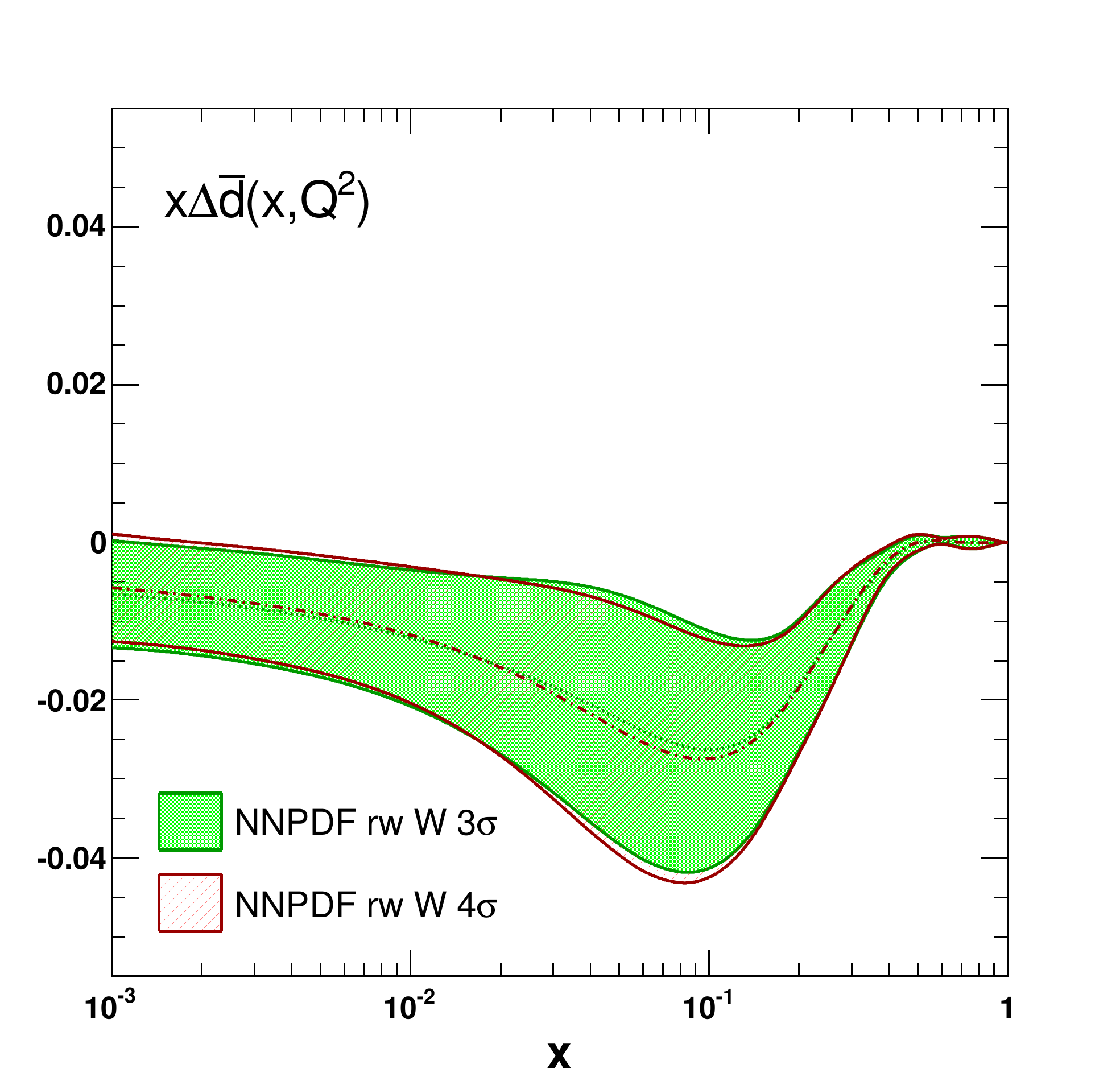}
\mycaption{Comparison between the $\Delta\bar{u}$ ($\Delta\bar{d}$)
PDF obtained from the reweighting of the $3\sigma$ and $4\sigma$ prior
ensembles with STAR $W$ data at $Q^2=10$ GeV$^2$.}
\label{fig:34sigma-bench}
\end{figure}

As emphasized in Sec.~\ref{sec:prior11}, in order to get reliable results
we must require the reweighted PDFs to be independent 
of the prior PDF ensemble. In other words, we should discard results
which are not stable upon the choice of different prior PDF ensembles
and, if needed, we should construct new priors, assuming a different ansatz 
on the quark-antiquark PDF separation, until this independence is effectively achieved.
In this case, we can verify that both the $3\sigma$ and $4\sigma$ prior PDF ensembles
lead to fully equivalent results, as can be seen at the level of 
the $\chi^2$ value per data point after reweighting
(see Tab.~\ref{tab:global1}), single-spin asymmetries 
and PDFs: the explicit comparison of these two distributions
from the $3\sigma$ and $4\sigma$ samples is shown
in Fig.~\ref{fig:34sigma-bench} at $Q_0^2=1$ GeV$^2$.
Then, we can conclude that results from both $3\sigma$ and $4\sigma$
prior PDF ensembles are statistically indistinguishable and that they can 
both be used to provide a robust determination of the 
light antiquark PDFs, $\Delta\bar{u}$
and $\Delta\bar{d}$. 

Results displayed in Fig.~\ref{fig:34sigma-bench} refer to the
simultaneous reweighting with both $W^+$ and $W^-$ 
data sets; we have explicitly checked that reweighting with $W^+$
($W^-$) data set separately probe $\Delta\bar{d}$ ($\Delta\bar{u}$) PDF.
Parton distributions not shown 
in Fig.~\ref{fig:34sigma-bench}, including strangeness, are not
affected by reweighting with $W$ data, as we have explicitly checked.
The situation is rather different from the unpolarized case, where, 
instead, $W$ production data also provide some information on 
strangeness. However, we notice that in the unpolarized case 
Drell-Yan data from both fixed-target experiments 
and colliders are available: in particular, the former are provided 
by E605 and E688 experiments at Fermilab~\cite{Moreno:1990sf,
Webb:2003ps,Webb:2003bj,Towell:2001nh},
while the latter by CDF and D0 at 
the Tevatron~\cite{Aaltonen:2009ta,Aaltonen:2010zza,Abazov:2007jy} 
and by ATLAS, CMS and LHCb at the 
LHC~\cite{Aad:2011dm,Chatrchyan:2012xt,Aaij:2012vn}.
These data sets span about three orders of magnitude in the 
energy scale $Q^2$, from $Q^2\sim 20 - 250$ GeV$^2$ for 
fixed-target experiments, up to the $W$ and $Z$ masses 
for collider experiments.
Hence, the effects of the evolution, enhanced by the rather
wide $Q^2$ lever-arm of experimental data, 
combine with the contributions from the
Cabibbo-favored partonic subprocesses, initiated by a 
$s\bar{c}$ or $\bar{s}c$ pair: this way, they
provide some constraints on the strangeness.
Of course, since in the polarized case only collider data 
are available, we cannot observe this effect.

\subsubsection{Combining COMPASS, STAR and PHENIX data}
The goal of the present analysis is to deliver a polarized parton set
including the experimental information coming from the complete 
piece of information provided by data discussed in 
Sec.~\ref{sec:expinput11}.
To this purpose, we perform a global reweighting of our prior polarized 
PDF ensembles generated in Sec.~\ref{sec:prior11} with
all the relevant data from the
COMPASS, STAR and PHENIX experiments simultaneously.
In Tab.~\ref{tab:global1}, we show the values of the $\chi^2$ per
data point before ($\chi^2/N_{\mathrm{dat}}$) and after 
($\chi_{\mathrm{rw}}^2/N_{\mathrm{dat}}$) reweighting
for each prior PDF ensemble. 
In Tab.~\ref{tab:global2}, we also quote the number of effective replicas
left after reweighting, $N_{\mathrm{eff}}$, and the modal
value of the $\mathcal{P}(\alpha)$ distribution.
In both Tabs.~\ref{tab:global1}-\ref{tab:global2}
the corresponding values for separate experiments and data sets 
are also provided.
\begin{table}[!t]
 \centering
 \footnotesize
 \begin{tabular}{llccccccccc}
   \toprule
   \multirow{2}*{Experiment} 
&  \multirow{2}*{Set}
&  \multirow{2}*{$N_{\mathrm{dat}}$} 
&  \multicolumn{4}{c}{$\chi^2/N_{\mathrm{dat}}$}
&  \multicolumn{4}{c}{$\chi_{\mathrm{rw}}^2/N_{\mathrm{dat}}$}\\
& & & $1\sigma$ & $2\sigma$ & $3\sigma$ & $4\sigma$ 
    & $1\sigma$ & $2\sigma$ & $3\sigma$ & $4\sigma$\\ 
\midrule
COMPASS &                 & $45$ & $1.23$ & $1.23$ & \textbf{1.23} & $1.23$ 
                                 & $1.23$ & $1.23$ & \textbf{1.23} & $1.23$\\
        & COMPASS~$K1\pi$ & $15$ & $1.27$ & $1.27$ & \textbf{1.27} & $1.27$
                                 & $1.27$ & $1.27$ & \textbf{1.27} & $1.27$\\
        & COMPASS~$K2\pi$ & $15$ & $0.51$ & $0.51$ & \textbf{0.51} & $0.51$
                                 & $0.51$ & $0.51$ & \textbf{0.51} & $0.51$\\
        & COMPASS~$K3\pi$ & $15$ & $1.90$ & $1.90$ & \textbf{1.90} & $1.90$
                                 & $1.89$ & $1.89$ & \textbf{1.89} & $1.89$\\
\midrule
STAR    &                 & $30$ & $1.19$ & $1.19$ & \textbf{1.20} & $1.20$
                                 & $0.79$ & $0.79$ & \textbf{0.79} & $0.79$\\
        & STAR 1j-05      & $10$ & $1.04$ & $1.04$ & \textbf{1.04} & $1.04$
                                 & $1.01$ & $1.02$ & \textbf{1.02} & $1.01$\\
        & STAR 1j-06      &  $9$ & $0.75$ & $0.75$ & \textbf{0.75} & $0.76$
                                 & $0.59$ & $0.59$ & \textbf{0.59} & $0.59$\\
        & STAR 1j-09      & $11$ & $1.69$ & $1.69$ & \textbf{1.70} & $1.71$
                                 & $1.02$ & $1.02$ & \textbf{1.03} & $1.04$\\
PHENIX & & & & & & & & & & \\
        &  PHENIX 1j      &  $6$ & $0.24$ & $0.24$ & \textbf{0.24} & $0.24$
                                 & $0.24$ & $0.24$ & \textbf{0.24} & $0.24$ \\
\midrule
 STAR   &                 & $12$ & $1.93$ & $1.97$ & \textbf{1.97} & $1.91$ 
                                 & $1.43$ & $1.12$ & \textbf{1.03} & $1.02$\\
        & STAR-$W^+$      &  $6$ & $1.15$ & $1.14$ & \textbf{1.22} & $1.25$
                                 & $1.04$ & $1.01$ & \textbf{1.02} & $1.01$\\
        & STAR-$W^-$      &  $6$ & $2.71$ & $2.80$ & \textbf{2.59} & $2.43$
                                 & $1.78$ & $1.25$ & \textbf{1.04} & $1.02$\\
\midrule
\multicolumn{2}{l}{GLOBAL REWEIGHTING} 
                          & $63$ & $1.22$ & $1.25$ & \textbf{1.25} & $1.24$ 
                                 & $1.07$ & $1.04$ & \textbf{1.02} & $1.02$ \\
   \bottomrule
 \end{tabular}
\mycaption{The value of the $\chi^2$ per data point
$\chi^2/N_{\mathrm{dat}}$ ($\chi^2_{\mathrm{rw}}/N_{\mathrm{dat}}$) 
before (after) global reweighting
with all data sets and for each prior PDF ensemble discussed in the text.}
\label{tab:global1}
\end{table}
\begin{table}[!t]
 \centering
 \footnotesize
 \begin{tabular}{llccccccccc}
   \toprule
   \multirow{2}*{Experiment} 
&  \multirow{2}*{Set}
&  \multirow{2}*{$N_{\mathrm{dat}}$} 
&  \multicolumn{4}{c}{$N_{\mathrm{eff}}$}
&  \multicolumn{4}{c}{$\langle\alpha\rangle$}\\
& & & $1\sigma$ & $2\sigma$ & $3\sigma$ & $4\sigma$ 
    & $1\sigma$ & $2\sigma$ & $3\sigma$ & $4\sigma$\\
   \midrule
COMPASS &                 & $45$ & $980$ & $980$ & \textbf{980} & $980 $ 
                                 & $1.07$ & $1.07$ & \textbf{1.07} & $1.07$\\
        & COMPASS~$K1\pi$ & $15$ & $990$ & $990$ & \textbf{990} & $990$
                                 & $1.15$ & $1.15$  & \textbf{1.15} & $1.15$\\
        & COMPASS~$K2\pi$ & $15$ & $990$ & $990$ & \textbf{990} & $990$
                                 & $0.72$ &  $0.72$ &  \textbf{0.72} &  $0.72$\\
        & COMPASS~$K3\pi$ & $15$ & $970$ & $970$ & \textbf{970} & $970$
                                 & $1.36$ & $1.36$ & \textbf{1.36} & $1.36$\\
\midrule
STAR    &                 & $19$ & $299$ & $301$ & \textbf{297} & $300$
                                 & $1.01$ & $1.10$ & \textbf{1.02} & $1.01$\\
        & STAR 1j-05      & $10$ & $931$ & $931$ & \textbf{930} & $931$
                                 & $1.03$ & $1.02$ & \textbf{1.05} & $1.06$ \\
        & STAR 1j-06      &  $9$ & $621$ & $623$ & \textbf{622} & $621$
                                 & $0.99$ & $1.20$ & \textbf{0.92} & $0.98$ \\
        & STAR 1j-09      & $11$ & $233$ & $235$ & \textbf{235} & $234$ 
                                 & $1.12$ & $1.14$ & \textbf{1.10} & $1.11$\\
PHENIX & & & & & & & & & & \\
        &  PHENIX 1j      &  $6$ & $740$ & $740$ & \textbf{740} & $741$
                                 & $0.55$ & $0.50$ & \textbf{0.50} & $0.55$ \\
\midrule
STAR    &                 & $12$ & $395$ & $335$ & \textbf{346} & $350$
                                 & $1.45$ & $1.43$ & \textbf{1.40} & $1.39$ \\
        & STAR-$W^+$      &  $6$ & $896$ & $850$ & \textbf{739} & $715$
                                 & $1.11$ & $1.13$ & \textbf{1.15} & $1.14$\\
        & STAR-$W^-$      &  $6$ & $337$ & $298$ & \textbf{376} & $412$
                                 & $1.60$ & $1.47$ & \textbf{1.38} & $1.35$\\
\midrule
\multicolumn{2}{l}{GLOBAL REWEIGHTING}  
                          & $63$ & $224$ & $197$ & \textbf{177} & $176$  
                                 & $1.22$ & $1.21$ & \textbf{1.23} & $1.22$\\
   \bottomrule
 \end{tabular}
\mycaption{The number of replicas left after reweighting, 
$N_{\mathrm{eff}}$, and the modal value
of the $\mathcal{P}(\alpha)$ distribution, $\langle\alpha\rangle$. 
Results refer to global reweighting
with all data sets and for each prior PDF ensemble discussed in the text.}
\label{tab:global2}
\end{table}

\begin{figure}[t]
\begin{center}
\epsfig{width=0.90\textwidth,figure=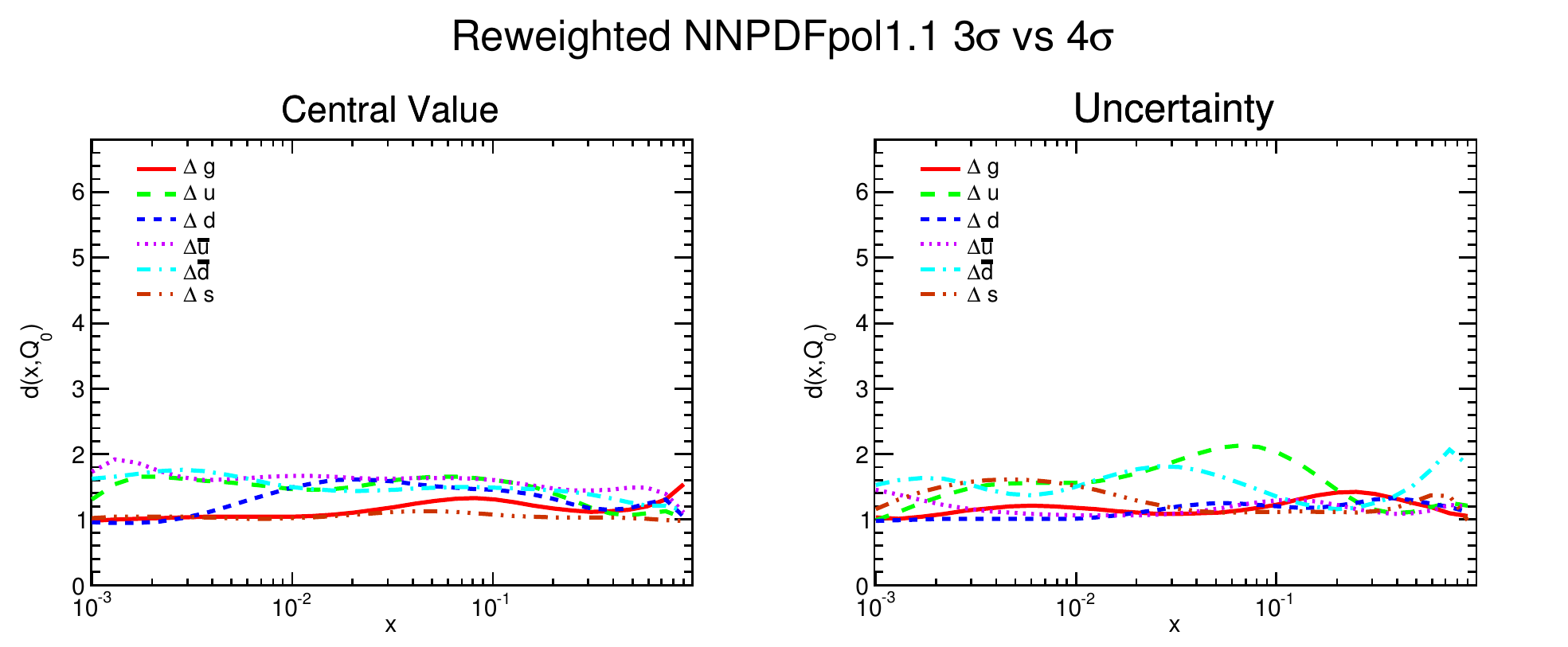}
\end{center}
\mycaption{Distances between parton sets obtained via global reweighting of 
$3\sigma$ and $4\sigma$ prior PDF ensembles at $Q^2=10$ GeV$^2$.}
\label{fig:distances-34}
\end{figure}

As discussed in Sec.~\ref{sec:prior11}, we must retain only the results
which are stable upon the choice of different prior PDF ensembles.
Looking at the values of the $\chi^2$ per data point after reweighting,
we argue that this stability is achieved for results obtained 
starting from the $3\sigma$ and $4\sigma$ priors, 
see Tab.~\ref{tab:global1}. Conversely, the results obtained from
the reweighting of the $1\sigma$ and $2\sigma$ prior ensembles must be
discarded, since they actually depend on the assumptions on quark-antiquark
separation we made for constructing the corresponding priors.
In order to quantitatively check that the results of the global reweighting
of the $3\sigma$ and $4\sigma$ samples are indeed statistically equivalent, 
we compute the distances $d(x,Q^2)$ between the respective PDFs. 
We recall that the distance is a statistical estimator which has the 
value $d\sim 1$ when the two samples of $N_{\mathrm{rep}}$ replicas 
are extracted from the same underlying distribution, while it is 
$d=\sqrt{N_{\mathrm{rep}}}$ when the two samples are extracted from 
two distributions which differ on average by one standard deviation 
(see Sec.~\ref{sec:stab} and Appendix~\ref{sec:appB} for further details).
The distances are plotted at $Q^2=10$ GeV$^2$ in Fig.~\ref{fig:distances-34}.
As $d\sim 2$, we conclude that the two ensembles obtained from the reweighting 
of the $3\sigma$ or $4\sigma$ priors describe the same underlying PDF 
probability distribution.
We choose the reweighted set obtained from the $3\sigma$ prior for reference.

The overall agreement between new data and the corresponding
theoretical predictions obtained with this reweighted PDF set is very good, 
as quantified by the value of the $\chi^2$ per data point, 
$\chi^2_{\mathrm{rw}}/N_{\mathrm{dat}}=1.02$. The effective 
number of replicas is $N_{\mathrm{eff}} \sim 180$, 
so the size of the initial prior ($N_{\mathrm{rep}}=1000$) 
was large enough even when the information
from all data sets is simultaneously combined.
The modal value of the $\mathcal{P}(\alpha)$ distribution 
is $\langle\alpha\rangle \sim 1.2 $, thus quantifying the good agreement
between inclusive DIS data in \texttt{NNPDFpol1.0} 
and the new data included in \texttt{NNPDFpol1.1}.

\subsection{Unweighting: the \texorpdfstring{\texttt{NNPDFpol1.1}}{NNPDFpol1.1} parton set}
\label{sec:unweighting11}

After global reweighting of the $3\sigma$ prior PDF ensemble, 
the unweighting procedure described in Sec.~\ref{sec:NNPDFapproach}
is used to produce a polarized PDF set made of
$N_{\mathrm{rep}}=100$ replicas, \texttt{NNPDFpol1.1},
statistically equivalent to the reweighted PDF ensemble, but in which all
PDF replicas are equally probable and hence do not need to be used 
with corresponding weights. 
In comparison to \texttt{NNPDFpol1.0}, 
the new polarized parton set provides a meaningful
determination of sea flavor PDFs $\Delta\bar{u}$ and $\Delta\bar{d}$,
though based on a small set of $W$ boson production data 
(but with the advantage to be free of any bias, including poorly known
fragmentation functions) and a determination of the gluon PDF
$\Delta g$ which is improved by open-charm and, particularly, jet data. 
In order to study the compatibility of the new data with the DIS sets,
included in \texttt{NNPDFpol1.0}, in Tab.~\ref{tab:chi2oldnew}
we show the $\chi^2$ of each of the 
experiments included in the \texttt{NNPDFpol1.0} analysis evaluated 
with both the old \texttt{NNPDFpol1.0} and the new \texttt{NNPDFpol1.1}
parton sets. We observe that DIS data are described by the two 
parton sets with comparable accuracy, as we already noticed from the 
modal value of the $\mathcal{P}(\alpha)$ distribution, 
$\langle\alpha\rangle\sim 1.2$ (see Tab.~\ref{tab:global2}).
\begin{table}[t]
\centering
\footnotesize
\begin{tabular}{lcc}
\toprule
Experiment & \texttt{NNPDFpol1.0} & \texttt{NNPDFpol1.1}\\
\midrule
EMC       & 0.44 & 0.43\\
SMC       & 0.93 & 0.90\\
SMClowx   & 0.97 & 0.97\\
E143      & 0.64 & 0.67\\
E154      & 0.40 & 0.45\\
E155      & 0.89 & 0.85\\
COMPASS-D & 0.65 & 0.70\\
COMPASS-P & 1.31 & 1.38\\
HERMES97  & 0.34 & 0.34\\
HERMES    & 0.79 & 0.82\\
\bottomrule
\end{tabular}
\mycaption{The $\chi^2$ per data point of all the experiments included in the 
\texttt{NNPDFpol1.0} analysis evaluated with \texttt{NNPDFpol1.0} and
\texttt{NNPDFpol1.1} parton sets}
\label{tab:chi2oldnew}
\end{table}

\begin{figure}[!t]
\begin{center}
\epsfig{width=0.40\textwidth,figure=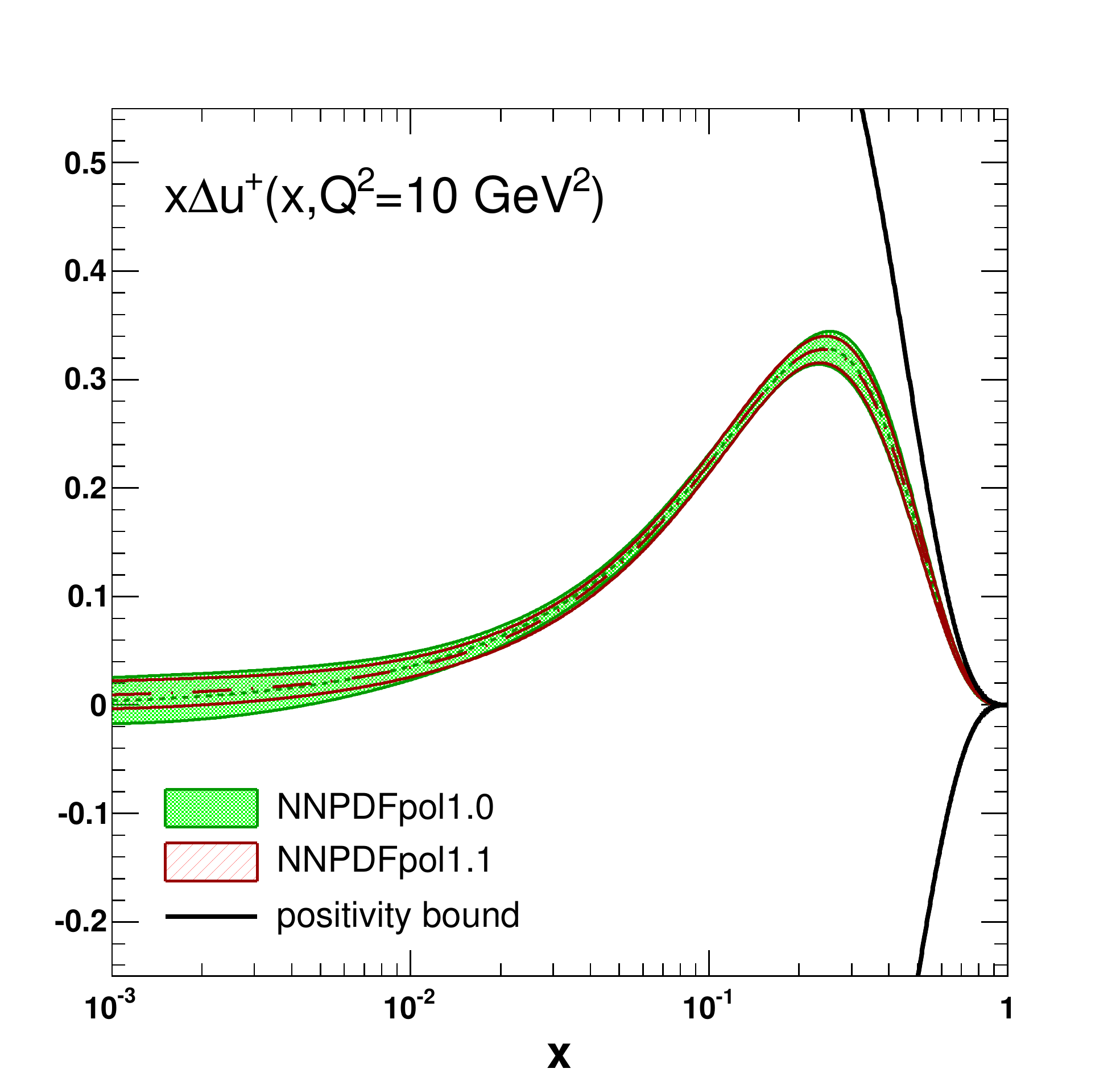}
\epsfig{width=0.40\textwidth,figure=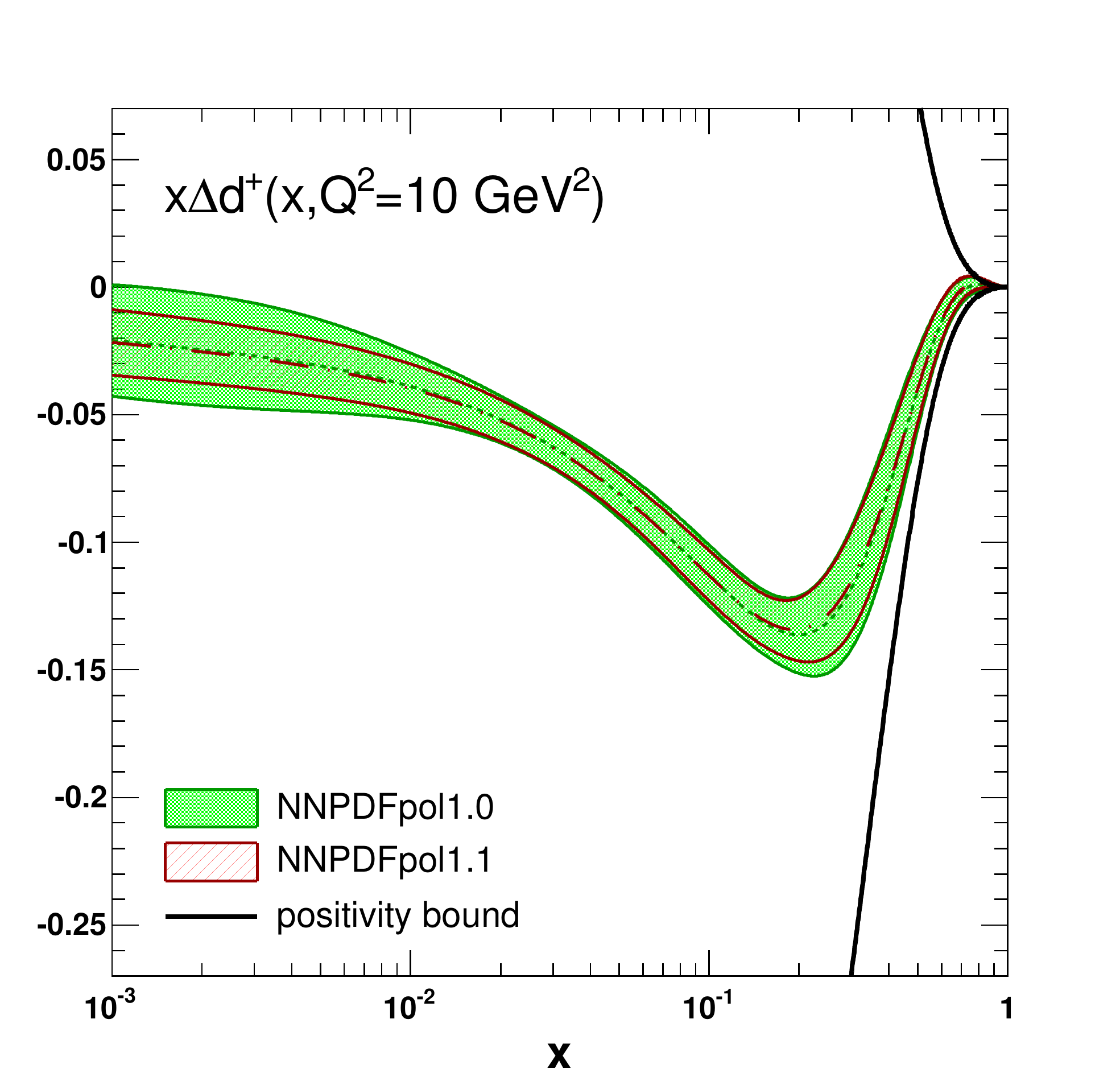}\\
\epsfig{width=0.40\textwidth,figure=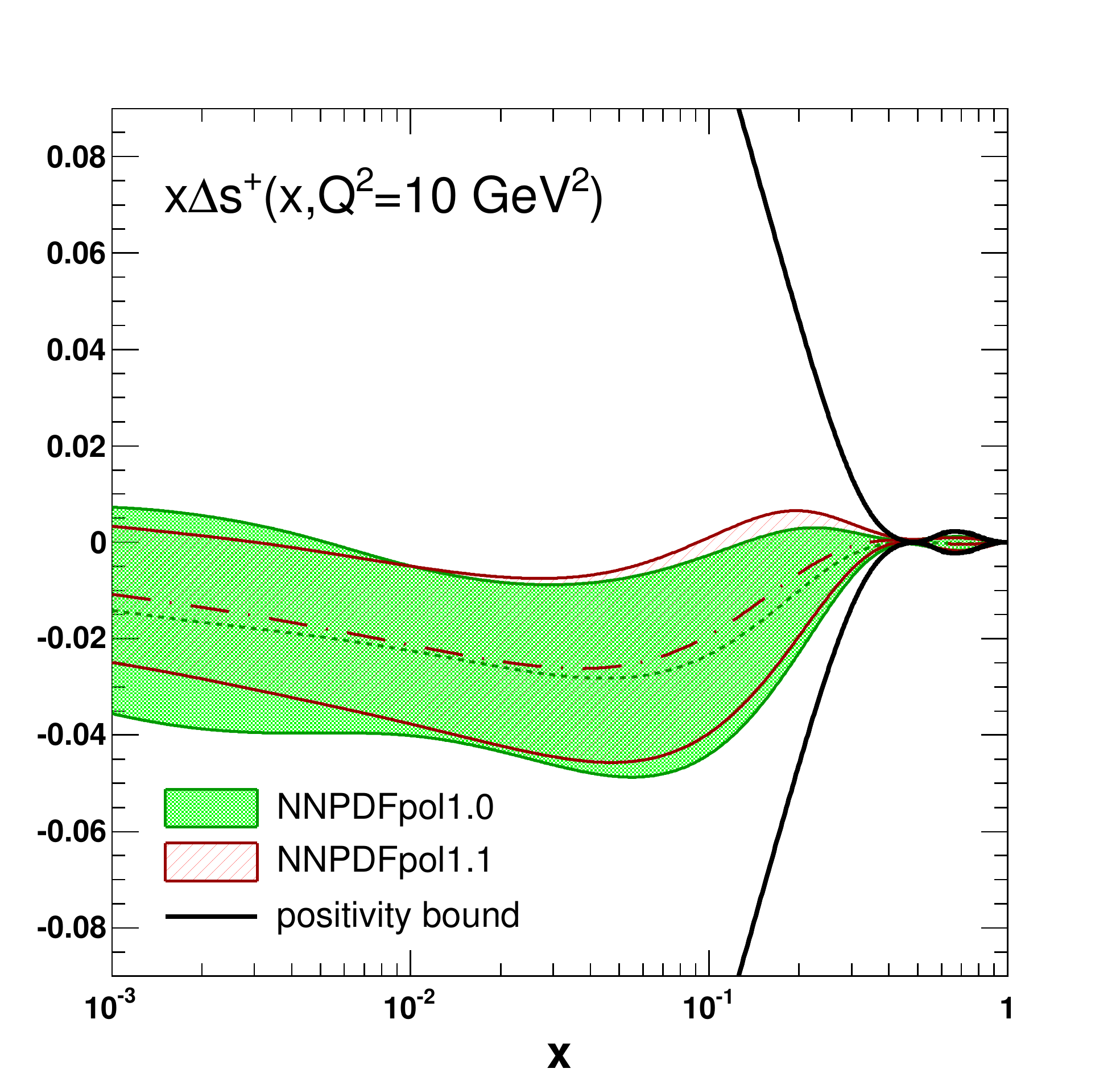}
\epsfig{width=0.40\textwidth,figure=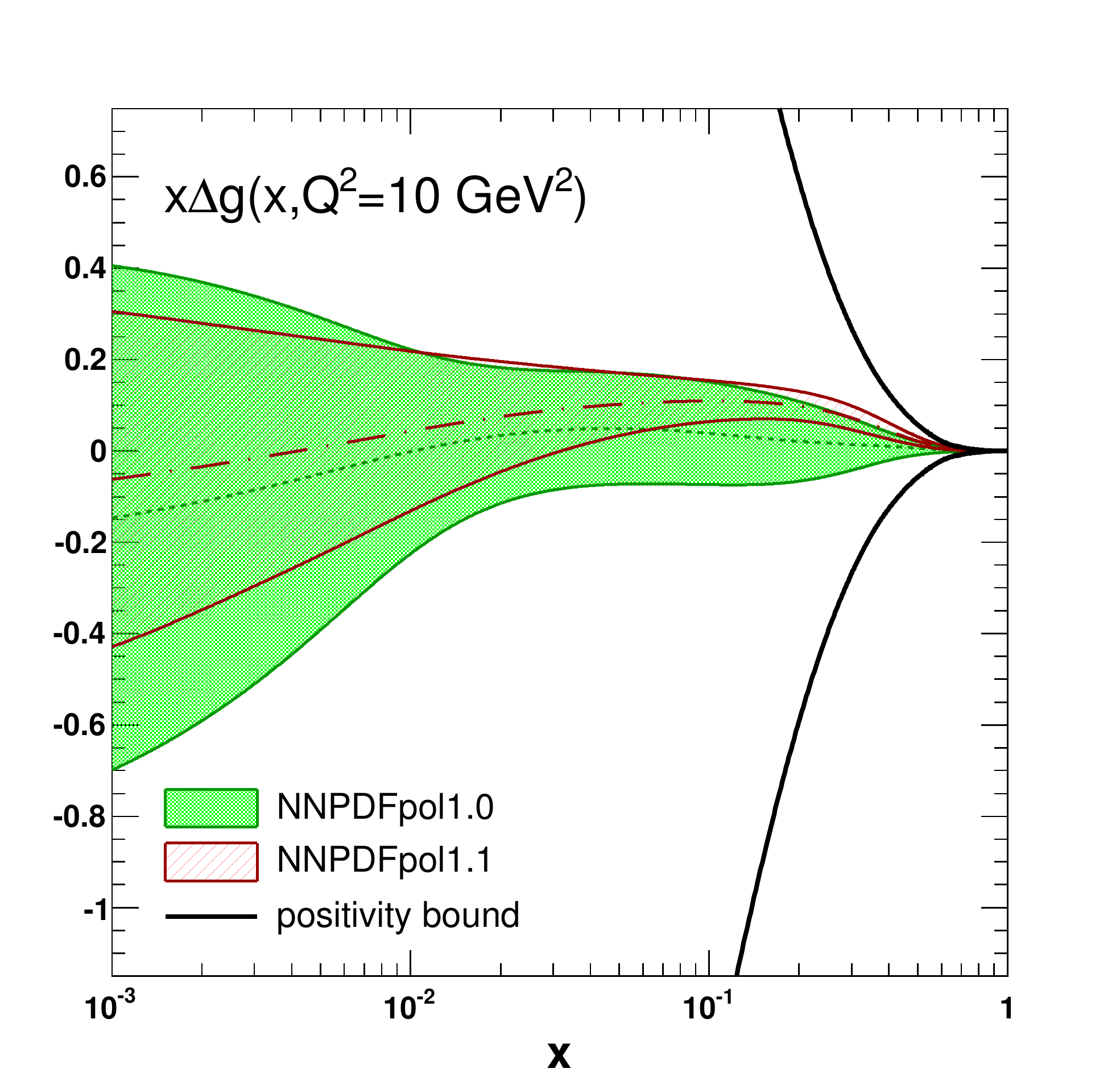}\\
\end{center}
\mycaption{Comparison between \texttt{NNPDFpol1.0} and \texttt{NNPDFpol1.1}
parton sets at $Q^2=10$ GeV$^2$.}
\label{fig:pdfs11}
\end{figure}

\begin{figure}[!t]
\begin{center}
\epsfig{width=0.90\textwidth,figure=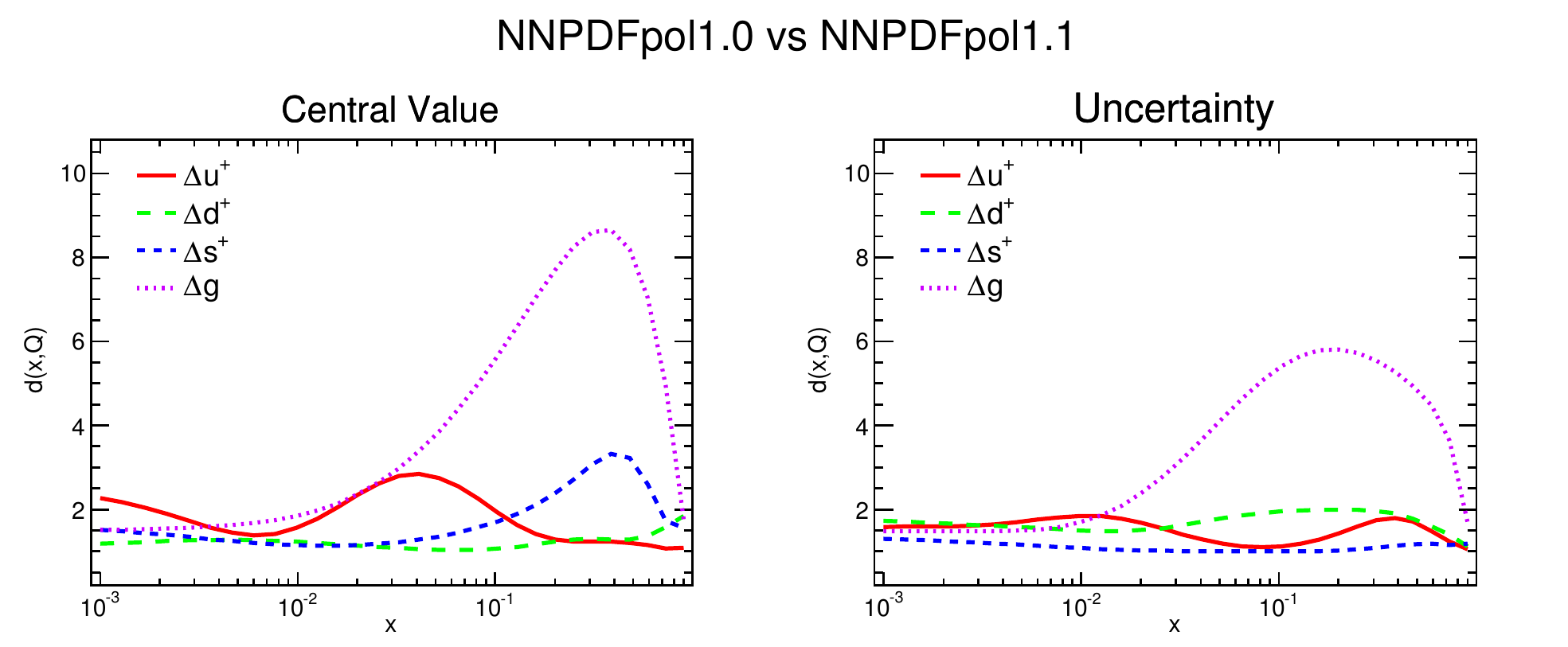}
\end{center}
\mycaption{Distances between \texttt{NNPDFpol1.0} and \texttt{NNPDFpol1.1} parton
determinations at $Q^2=10$ GeV$^2$.}
\label{fig:distances-oldnew}
\end{figure}

In Fig.~\ref{fig:pdfs11}, we compare the total PDF combinations
$\Delta q^+\equiv\Delta q+\Delta\bar{q}$, 
$q=u,d,s$,
and the gluon PDF $\Delta g$ from \texttt{NNPDFpol1.1} and \texttt{NNPDFpol1.0} 
at $Q^2=10$ GeV$^2$.
Since the latter is a fit to inclusive DIS data, only these PDFs can be 
compared meaningfully between the two parton sets.
In order to quantitatively asses the difference between them,
we plot the corresponding distances $d(x,Q)$ at $Q^2=10$ GeV$^2$ 
in Fig.~\ref{fig:distances-oldnew}.

We observe that the total quark PDF combinations from the two
determinations are statistically equivalent, since the distance for both 
their central value and uncertainty is not larger than two 
(Fig.~\ref{fig:distances-oldnew}) and differences between them 
are hardly noticeable (Fig.~\ref{fig:pdfs11}). 
On the other hand, the
gluon PDF shows significant differences between the two NNPDF parton
determinations, in particular in the $x$ region probed by STAR jet data,
$0.05\lesssim x\lesssim 0.2$. In this region, the gluon from the
\texttt{NNPDFpol1.1} parton set is definitely positive and has
a much reduced error band with respect to its \texttt{NNPDFpol1.0}
counterpart. The polarized gluon PDF in the two determinations 
actually samples different underlying probability distributions,
which may differ up to one sigma, as the distance grows up to
$d\sim 10$. At lower values of $x$, where no new data are included,
\texttt{NNPDFpol1.0} and \texttt{NNPDFpol1.1} are again 
statistically equivalent, as expected. 

\begin{figure}[!t]
\begin{center}
\epsfig{width=0.40\textwidth,figure=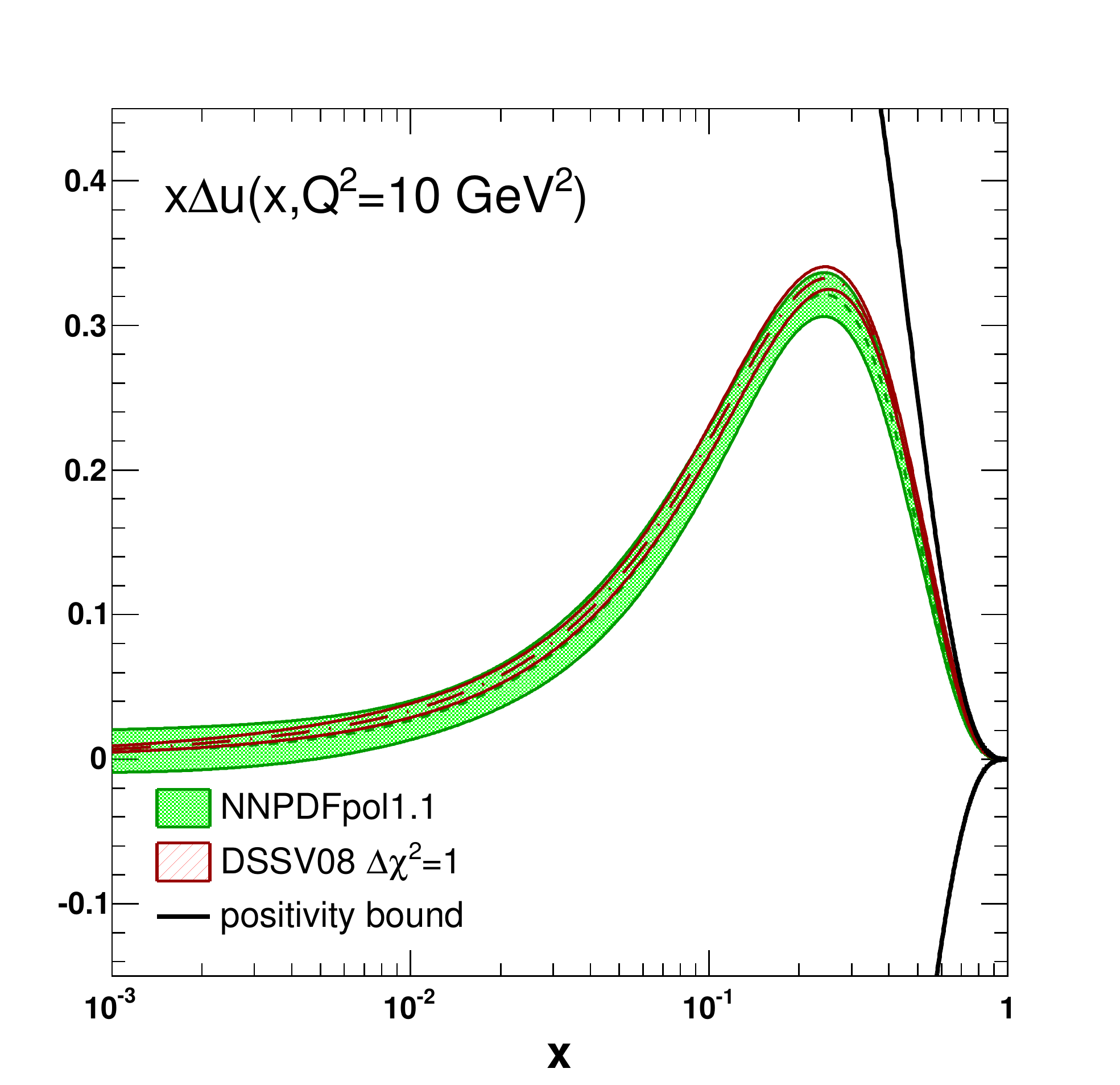}
\epsfig{width=0.40\textwidth,figure=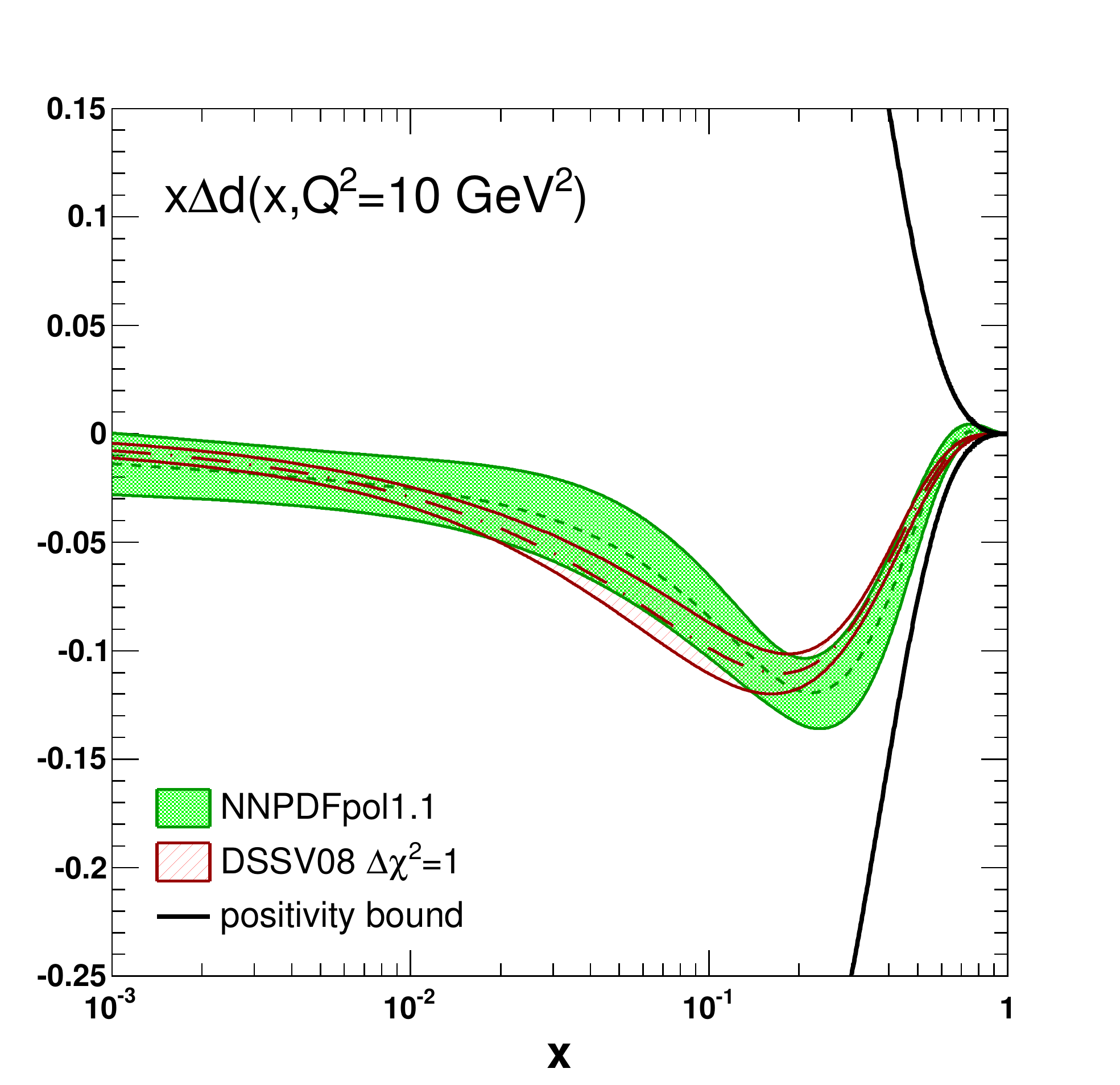}\\
\epsfig{width=0.40\textwidth,figure=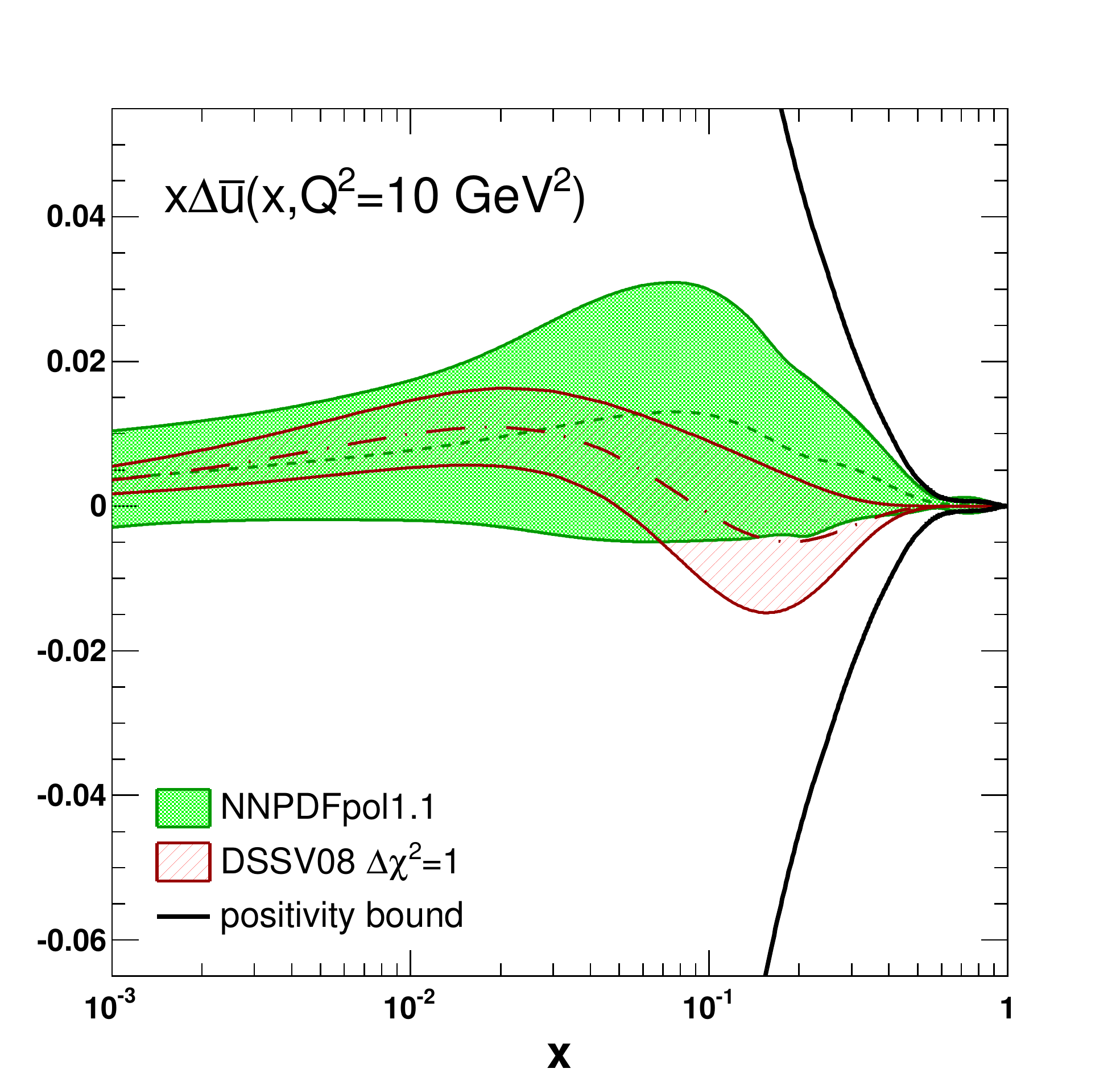}
\epsfig{width=0.40\textwidth,figure=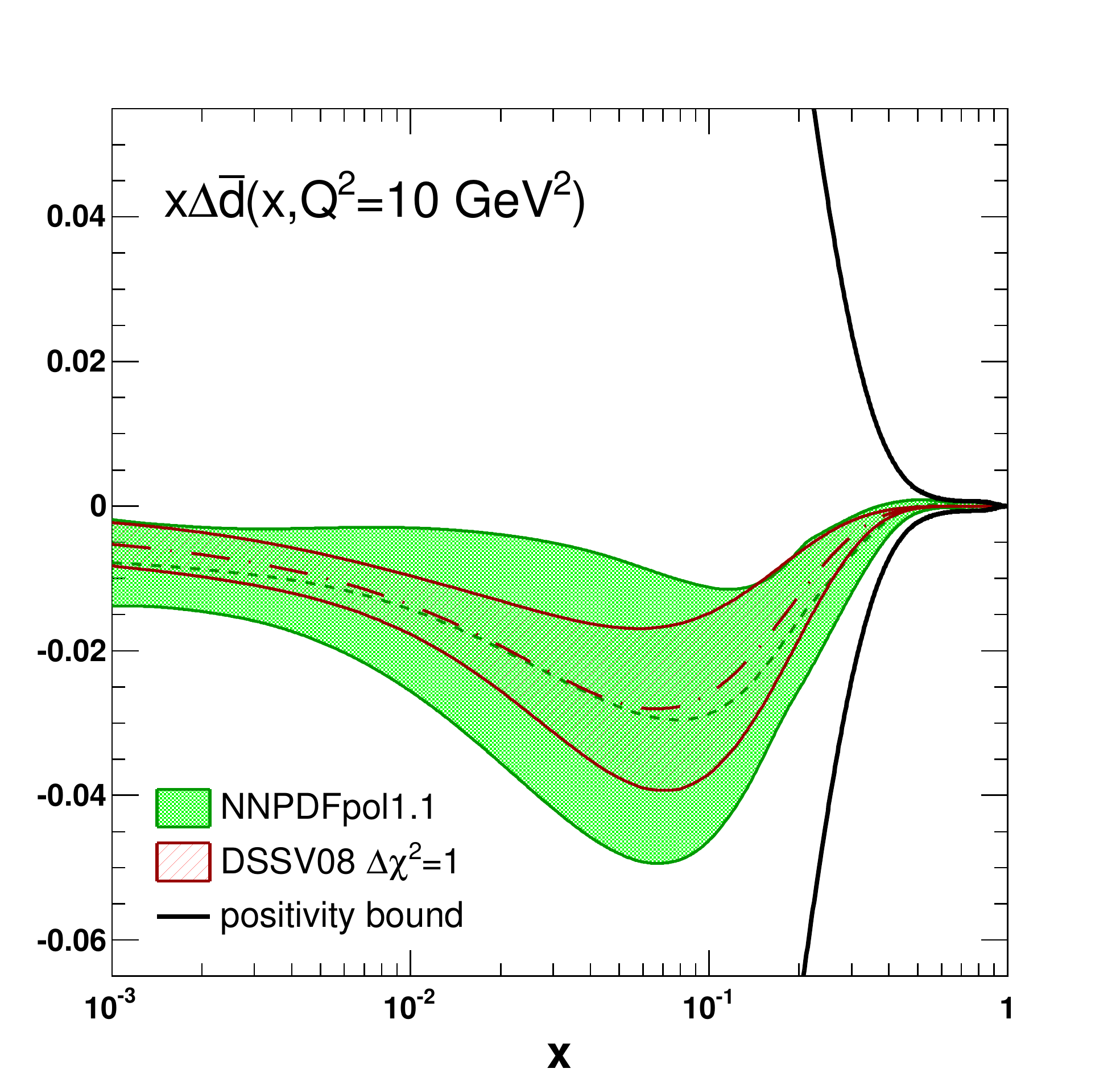}\\
\epsfig{width=0.40\textwidth,figure=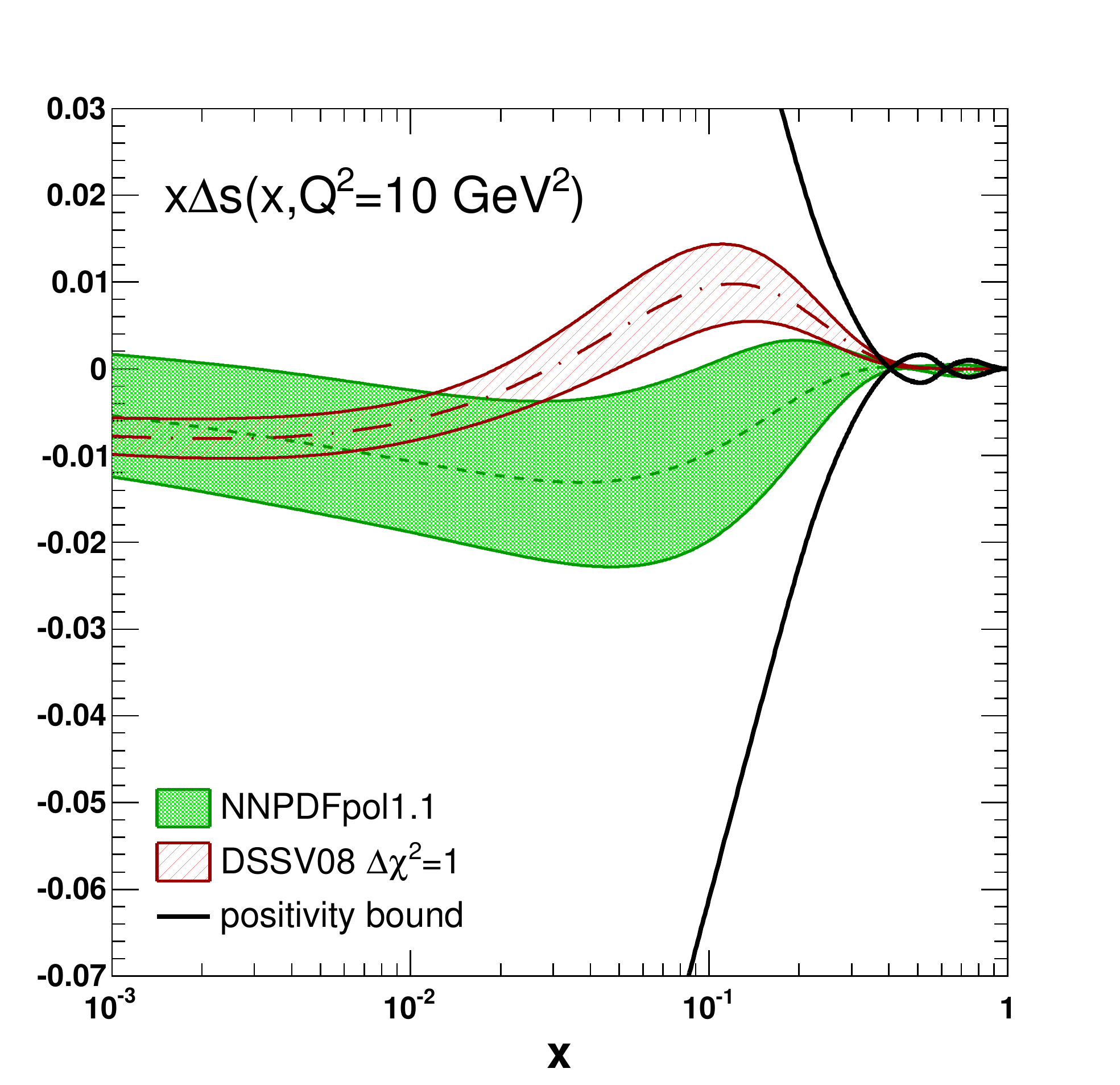}
\epsfig{width=0.40\textwidth,figure=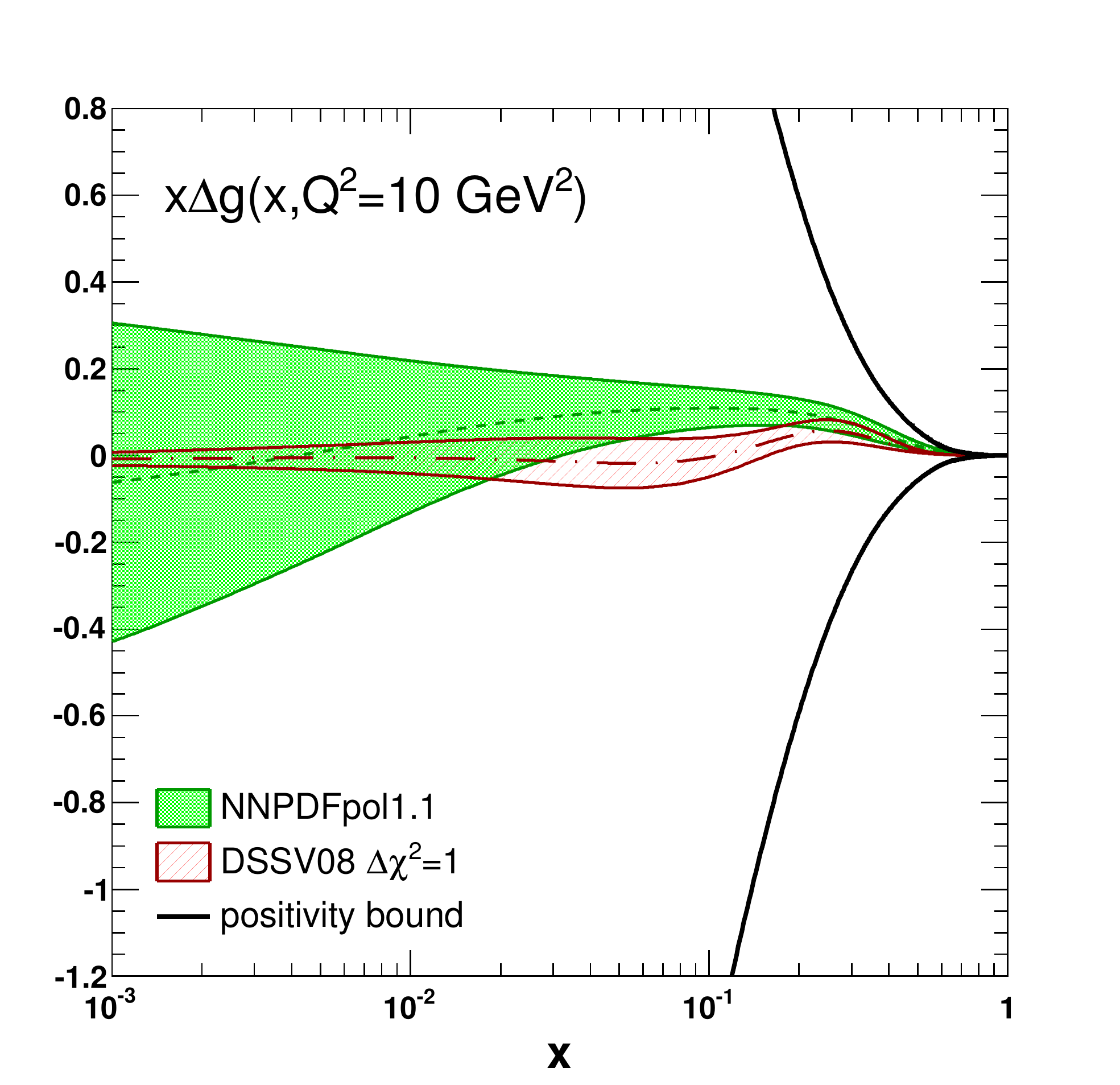}
\mycaption{The \texttt{NNPDFpol1.1} parton set
 compared to \texttt{DSSV08}~\cite{deFlorian:2009vb} at $Q^2=10$ GeV$^2$.}
\label{fig:xpdfs}
\end{center}
\end{figure}

In Fig.~\ref{fig:xpdfs}, we compare PDFs from \texttt{NNPDFpol1.1}
with those from the global \texttt{DSSV08} fit~\cite{deFlorian:2009vb}: 
we display $x\Delta u$, $x\Delta d$, 
$x\Delta\bar{u}$, $x\Delta\bar{d}$, $x\Delta s$  and $x\Delta g$ 
at $Q^2=10$ GeV$^2$. 
Uncertainties are nominal one-sigma error bands for NNPDF sets, while they are 
Hessian uncertainties ($\Delta\chi^2=1$) for \texttt{DSSV08}. 
This choice was made 
in Ref.~\cite{deFlorian:2009vb} with some caution, since it may lead to
underestimated PDF ucertainties in some $(x,Q^2)$ regions, particularly where 
constraints from experimental data are weak, see also Sec.~\ref{sec:generalstrategy}. 

The main conclusions of the comparison in Fig.~\ref{fig:xpdfs} 
are the following.
\begin{itemize}
 \item Consistent results are found in the two parton determinations for 
$\Delta u$ and $\Delta d$ PDFs, though the NNPDF uncertainties are slightly
larger, especially at small-$x$ values, where experimental data are lacking.
 \item The NNPDF polarized gluon PDF is in perfect agreement with its DSSV 
counterpart in the large-$x$ region, $x\gtrsim 0.2$, where they show 
similar uncertainties. However, for $x<0.2$, $\Delta g$ has a node in the
\texttt{DSSV08} determination, while it is clearly positive 
from \texttt{NNPDFpol1.1}.
This result is driven in particular by the most recent and precise 
jet production data from STAR (labelled as STAR 1j-09 above), 
which were not available at the time of the original \texttt{DSSV08} 
analysis~\cite{deFlorian:2009vb} shown in Fig.~\ref{fig:xpdfs}.
Actually, only the data sets labelled as STAR 1j-05 and STAR 1j-06
were included there. An update of the \texttt{DSSV08} 
fit including also preliminar STAR 1j-09,
called \texttt{DSSV++}~\cite{Aschenauer:2013woa}, 
pointed to a positive $\Delta g$ consistent with the result 
of our analysis. 
 \item Related to the polarized sea quarks, a slight discrepancy is clearly noticeable 
for the $\Delta\bar{u}$ distribution above $x\sim 3\cdot10^{-2}$
between the two parton sets. 
We recall that $W^\pm$ data were not included
in the \texttt{DSSV08} global fit~\cite{deFlorian:2009vb} 
shown in Fig.~\ref{fig:xpdfs},
hence the differences in the $\Delta\bar{u}$ distribution may suggest 
some tension between $W^\pm$ and SIDIS data. 
This discrepancy may be explained by our poor knowledge 
of fragmentation functions.  
A shift away the $\Delta\bar{u}$ central curve towards positive values,
as observed in our analysis, was
also found in a preliminary global fit including STAR data in the DSSV
framework~\cite{stratmann:DIS2013}. 
 \item Since $W$ boson production data in the kinematic regime probed by STAR
are not sensitive to strangeness, the discrepancy 
between the NNPDF and DSSV determinations of $\Delta s$, already found in 
\texttt{NNPDFpol1.0}, is still present. As discussed in 
Sec.~\ref{sec:results}, in the NNPDF analysis 
the polarized strange PDF is obtained from inclusive DIS data through its 
$Q^2$ evolution and assumptions about flavor symmetry of the proton sea,
enforced by experimentally measured baryon octet decay constants,
see Sec.~\ref{sec:sumrule}.
On the other hand, the \texttt{DSSV08} determination of polarized PDFs also 
includes semi-inclusive data with identified kaons in final states, 
which are directly sensitive to strangeness, but are likely to 
introduce an uncertainty difficult to quantify due to the
poor knowledge of the kaon fragmentation function.
Finally, we made the choice $\Delta s=\Delta\bar{s}$, 
like all other polarized 
analyses~\cite{deFlorian:2009vb,Leader:2010rb,Hirai:2008aj,Blumlein:2010rn,Jimenez-Delgado:2013boa}: 
actually, this is not justified by any physical reason, but experimental data
do not allow for a determination of $\Delta s$ and $\Delta\bar{s}$ 
separately. 
\end{itemize}

\section{Phenomenology of the nucleon spin structure}
\label{sec:pheno11}

In this Section, we use the \texttt{NNPDFpol1.1} parton set to 
reevalute the first moments of the polarized PDFs, separately
for each quark flavor-antiflavor and for the gluon, in 
light of the new data sets included in this analysis. 
Then, we produce some predictions for single-hadron production
spin asymmetries at RHIC; we compare them with available measured 
results in order to qualitatively gauge their potential in
pinning down the gluon uncertainty. 

\subsection{The spin content of the proton revisited}
The first moments of the polarized PDFs can be directly related 
to the fraction of the proton spin carried by partons, as 
explained in Sec.~\ref{sec:sf-factorization}. 
This reason has especially motivated our efforts in an accurate and
unbiased determination of polarized PDF uncertainties.
In Sec.~\ref{sec:spinmom}, we presented a detailed 
analysis of the first moments of the total quark PDF combinations,
$\Delta u^+$, $\Delta d^+$, $\Delta s^+$ and of the gluon PDF, 
$\Delta g$, from a fit to polarized inclusive DIS data only.
Their potential improvement at a future Electron-Ion Collider
was then studied in Sec.~\ref{sec:spinEIC}. Now, we use 
the \texttt{NNPDFpol1.1} parton set to reassess the determination 
of the first moments of the polarized parton distributions 
in order to quantify the impact of new data on the proton's spin
content.
We recall thet the (truncated) first moments of the polarized
PDFs $\Delta f(x,Q^2)$ in the region $[x_{\mathrm{min}},x_{\mathrm{max}}]$
are defined as
\begin{equation}
 \langle \Delta f(Q^2)\rangle^{[x_{\mathrm{min}},x_{\mathrm{max}}]}
 =
 \int_{x_{\mathrm{min}}}^{x_{\mathrm{max}}}dx \Delta f(x,Q^2)
 \,\mbox{.}
 \label{eq:moments11}
\end{equation}
We consider both \textit{full}
moments, \textit{i.e.} $\langle\Delta f(Q^2)\rangle^{[0,1]}$, 
and \textit{truncated} moments in the $x$ region covered by experimental data, 
roughly $\langle\Delta f(Q^2)\rangle^{[10^{-3},1]}$.

Let us begin by looking at polarized quarks and antiquarks.
We compute Eq.~(\ref{eq:moments11}) for the total quark-antiquark 
combinations, \textit{i.e.} $\Delta f=\Delta u^+,\,\Delta d^+$, for sea quarks,
\textit{i.e.} $\Delta f=\Delta\bar{u},\,\Delta\bar{d}$, for the polarized 
strangeness, \textit{i.e.} $\Delta f=\Delta s$, and for the singlet PDF 
combination, \textit{i.e.} $\Delta f=\Delta\Sigma=\sum_{q=u,d,s}\Delta q^+$.
The corresponding central values and one-sigma
PDF uncertainties obtained from the
$N_{\mathrm{rep}}=100$ replicas of the \texttt{NNPDFpol1.1} parton set 
at $Q^2=10$ GeV$^2$ are collected in Tab.~\ref{tab:qqmomenta}.
We compare our results to both \texttt{NNPDFpol1.0} and
\texttt{DSSV08}.
In the latter case, we quote the 
conservative uncertainty estimate obtained in Ref.~\cite{deFlorian:2009vb}
using the Lagrange multiplier method with $\Delta\chi^2/\chi^2=2\%$. In
parenthesis, we also show the uncertainty due to the extrapolation
outside the region covered by experimental data,  
estimated as the difference between the full first moment and its
truncated counterpart in the region $[10^{-3},1]$, quoted in 
Ref.~\cite{deFlorian:2009vb}. 
\begin{table}[t]
 \centering
 \footnotesize
 \begin{tabular}{cccccc}
 \toprule
 & \multicolumn{2}{c}{$\langle\Delta f(Q^2)\rangle^{[0,1]}$}
 & \multicolumn{3}{c}{$\langle\Delta f(Q^2)\rangle^{[10^{-3},1]}$}\\
 $\Delta f$ 
 & \texttt{NNDPFpol1.0}
 & \texttt{NNPDFpol1.1}
 & \texttt{NNDPFpol1.0}
 & \texttt{NNPDFpol1.1}
 & \texttt{DSSV08}~\cite{deFlorian:2009vb}\\
 \midrule
 $\Delta u + \Delta\bar{u}$
 & $+0.77\pm 0.10$
 & $+0.79\pm 0.06$
 & $+0.76\pm 0.06$
 & $+0.76\pm 0.03$
 & $+0.793^{+0.028}_{-0.034}(\pm 0.020)$\\
 $\Delta d+\Delta\bar{d}$
 & $-0.46\pm 0.10$
 & $-0.47\pm 0.06$
 & $-0.41\pm 0.06$
 & $-0.41\pm 0.04$
 & $-0.416^{+0.035}_{-0.025}(\pm 0.042)$\\
 $\Delta\bar{u}$
 & ---
 & $+0.06\pm 0.05$ 
 & ---
 & $+0.05\pm 0.05$
 & $+0.028^{+0.059}_{-0.059}(\pm 0.008)$\\
 $\Delta\bar{d}$
 & ---
 & $-0.12\pm 0.07$
 & ---
 & $-0.10\pm 0.05$
 & $-0.089^{+0.090}_{-0.080}(\pm 0.026)$\\
 $\Delta\bar{s}$
 & $-0.07\pm 0.06$
 & $-0.06\pm 0.05$
 & $-0.06\pm 0.04$
 & $-0.05\pm 0.04$
 & $-0.006^{+0.028}_{-0.031}(\pm 0.051)$\\
 $\Delta\Sigma$
 & $+0.16\pm 0.30$
 & $+0.20\pm 0.18$
 & $+0.23\pm 0.15$
 & $+0.25\pm 0.10$
 & $+0.366^{+0.042}_{-0.062}(\pm 0.124)$\\
 \bottomrule
 \end{tabular}
 \mycaption{Full and truncated first moments of the polarized quark 
distributions at $Q^2=10$ GeV$^2$ for the \texttt{NNPDFpol1.1} set
compared to \texttt{NNPDFpol1.0}
and \texttt{DSSV08}~\cite{deFlorian:2009vb}. The uncertainty 
quoted in parenthesis for \texttt{DSSV08} is due to the extrapolation in 
the unintegrated region as discussed in the text.}
 \label{tab:qqmomenta}
\end{table}
\begin{table}[t]
 \centering
 \footnotesize
 \begin{tabular}{lccc}
 \toprule
 & $\langle \Delta g(Q^2)\rangle^{[0,1]}$
 & $\langle \Delta g(Q^2)\rangle^{[10^{-3},1]}$
 & $\langle \Delta g(Q^2)\rangle^{[0.05,0.2]}$ \\
 \midrule
 \texttt{NNPDFpol1.0}
 & $-0.95\pm 3.87$
 & $-0.06\pm 1.12$
 & $+0.05\pm 0.15$\\
 \texttt{NNPDFpol1.1}
 & $-0.13\pm 2.60$
 & $+0.31\pm 0.77$
 & $+0.15\pm 0.06$\\ 
 \texttt{DSSV08}~\cite{deFlorian:2009vb}
 & ---
 & $0.013^{+0.702}_{-0.314}(\pm 0.097)$
 & $0.005^{+0.129}_{-0.164}$\\
 \texttt{DSSV++}~\cite{Aschenauer:2013woa}
 & ---
 & ---
 & $0.10^{+0.06}_{-0.07}$\\ 
 \bottomrule
 \end{tabular}
 \mycaption{Full and truncated first moments of the polarized gluon
distributions at $Q^2=10$ GeV$^2$ for the \texttt{NNPDFpol1.1} set
compared to \texttt{NNPDFpol1.0}
and various fits of the DSSV family. The uncertainty 
quoted in parenthesis for \texttt{DSSV08} is due to the extrapolation in 
the unintegrated region as discussed in the text.}
 \label{tab:gmomenta}
\end{table}

Results from Tab.~\ref{tab:qqmomenta} clearly show that first moments obtained
with \texttt{NNPDFpol1.1} and \texttt{NNPDFpol1.0} are perfectly consistent
with each other, as we already knew from the corresponding agreement at the
level of polarized PDFs, see Fig.~\ref{fig:pdfs11}.
Besides, the sensitivity to quark-antiquark separation introduced
by $W$ data allows for a reduction of the uncertainty up to $50\%$ in the
\texttt{NNPDFpol1.1} determination with respect to \texttt{NNPDFpol1.0}.
The comparison between NNPDF full and truncated moments shows that the 
relative contribution to the total PDF uncertainty from the
small-$x$ extrapolation region is roughly the same
in both \texttt{NNPDFpol1.0} and
\texttt{NNPDFpol1.1}, and it is about 
two times larger than the uncertainty in the measured  
$x$ region. Therefore, we can conclude that
the contribution to the total uncertainty from the
extrapolation region has reduced by almost a half 
in \texttt{NNPDFpol1.1} with respect to \texttt{NNPDFpol1.0}.
This is likely because the new data,
supplemented by the smoothness provided by the neural network parametrization,
decrease the number of acceptable small-$x$ behaviors of the
polarized quark PDFs. Nevertheless, the uncertainty from the 
small-$x$ extrapolation region is still dominant and it could be
finally pinned down only by accurate measurements in this 
region. These may be performed at a future Electron-Ion Collider and
they were demonstrated to largely keep under control
the extrapolation uncertainties in Sec.~\ref{sec:spinEIC}.

Coming now to the comparison between \texttt{NNPDFpol1.1} and 
\texttt{DSSV08}~\cite{deFlorian:2009vb}, we notice that truncated 
first moments are in perfect agreement, both central values and uncertainties.
Slight differences are found for $\Delta\bar{u}$ and $\Delta\bar{s}$, due to
the different shape of the corresponding PDFs (see Fig.~\ref{fig:xpdfs}).
On the other hand, when considering full first moments the 
\texttt{NNPDFpol1.1} uncertainties are somewhat larger than those
found in the \texttt{DSSV08} analysis whenever the extrapolation uncertainty
is included. This is the major effect of our more flexible PDF parametrization,
as already discussed in Sec.~\ref{sec:spinmom}. 

Let us now move on to discuss the first moment of the polarized gluon
$\Delta g$. 
Results for full and truncated moments at $Q^2=10$ GeV$^2$
are presented in Tab.~\ref{tab:gmomenta}.
There, we compare the predictions from \texttt{NNPDFpol1.1}
with those from \texttt{NNPDFpol1.0}, and two fits from the DSSV family:
we consider both the original \texttt{DSSV08} parton 
set~\cite{deFlorian:2009vb} and its update, 
\texttt{DSSV++}~\cite{Aschenauer:2013woa}, 
which includes the same jet production data in \texttt{NNPDFpol1.1}.
As for quarks, we compute both the full 
and the truncated moments in the
measured region $[10^{-3},1]$.
In order to quantify the impact of the RHIC
inclusive jet data, we also provide
results for the truncated first moment restricted to
the region $x \in [0.05,0.2]$, which corresponds to the
range covered by these data, see
Fig.~\ref{fig:NNPDFpol11-kin}.

The results quoted in Tab.~\ref{tab:gmomenta} show the 
substantial improvement in the PDF uncertainties
of the gluon first moment in \texttt{NNPDFpol1.1}
as compared to \texttt{NNPDFpol1.0},
due to the constraints on $\Delta g$ provided
by RHIC jet data, see Fig.~\ref{fig:pdfs11}.
This is further illustrated by the truncated first moment
in the region covered by these data,
$\langle \Delta g(Q^2)\rangle^{[0.05,0.2]}$, 
where the PDF uncertainty is reduced by a factor close to three, and
where its central value is clearly positive, almost three sigma away from zero.
It is clear that the RHIC jet data strongly suggest
a positive polarized gluon first moment in the region
$x\in \left[ 0.05,0.2\right]$, unfortunately the absence of
other direct constraints outside this region still lead to
a quite large value of the gluon full first moment.
In addition, our results for  $\langle \Delta g(Q^2)\rangle^{[0.05,0.2]}$,
in terms of both central value and uncertainty, turn out to be very close
to those obtained in the \texttt{DSSV++} analysis, which is based on 
the same set of inclusive jet data~\cite{Aschenauer:2013woa}.

As in the previous \texttt{NNPDFpol1.0} analysis, 
the uncertainty due to the extrapolation outside the region
covered by experimental data
is substantial and dominates the total
uncertainty of the full first moment $\langle \Delta g(Q^2)\rangle^{[0,1]}$. 
The only way to further reduce this uncertainty is
to provide measurements which probe $\Delta g$ at smaller
values of $x$ than those that are available now.
In this respect, additional jet data 
from RHIC taken at higher center-of-mass energy, up to $\sqrt{s}=500$
GeV$^2$, may be helpful.
However, to really pin down the small-$x$ behavior of
the polarized PDFs and thus be able to finally reach
an accurate determination of $\langle \Delta g(Q^2)\rangle^{[0,1]}$,
one will have to resort to the Electron-Ion Collider,
as quantified in detail in Chap.~\ref{sec:chap4}.

Our main conclusion on the partons' contribution
to the proton spin is then twofold.
On the one hand, we have found that their
first moments are rather well determined in the 
kinematic region covered by experimental data and in good 
agreement with the values obtained in 
\texttt{DSSV++}~\cite{Aschenauer:2013woa},
the only analysis in which collider data are included.
In particular, the singlet full first moment is less than a 
half of the proton spin
within its uncertainty (see tab.~\ref{tab:qqmomenta});
the gluon first moment is definitely positive, though 
rather small, in the region
constrained by recent STAR jet data, roughly
$0.05\lesssim x \lesssim 0.2$.
On the other hand, we emphasize that the uncertainty on both the 
singlet and the gluon full first moments coming from the
extrapolation to the unmeasured, small-$x$, region dominates
their total uncertainty. For this reason, large values of 
the gluon first moment are not completely ruled out: 
within our accurate determination of uncertainties,  
the almost vanishing value for the singlet axial charge observed
in the experiment may still be explained as a cancellation between 
a rather large quark contribution and the anomalous gluon
contribution, as discussed in Sec.~\ref{sec:QCDevol}.
Of course, more experimental data, such as those
available at a future Electron-Ion Collider, 
are needed to discriminate which is the behavior
of the polarized gluon in the unmeasured small-$x$ region,
in particular whether it is actually small, as 
it is commonly believed.

\subsection{Predictions for single-hadron production asymmetries at RHIC}
\label{sec:pipredictions}

The parton set presented in Sec.~\ref{sec:unweighting11} does not include 
information from semi-inclusive hadron production spin asymmetries at colliders.
As discussed in Sec.~\ref{sec:PPprobes}, the analysis of these data 
requires the usage of fragmentation functions, whose poor knowledge entails
an additional source of theoretical uncertainty on the extracted PDFs, 
which is difficult to quantify, (see Sec.~\ref{sec:SIDISprobes}).
Nevertheless, in view of the experimental program which is ongoing at RHIC, 
it is interesting to compare some predictions for these spin asymmetries
with the experimental data, which are now available with significant 
statistics. The size of the uncertainties of theoretical predictions
will then fix, at least qualitatively, the 
experimental precision required for data to further pin down 
the uncertainties on PDFs, once included in a global fit.

\begin{figure}[!t]
\begin{center}
\epsfig{width=0.40\textwidth,figure=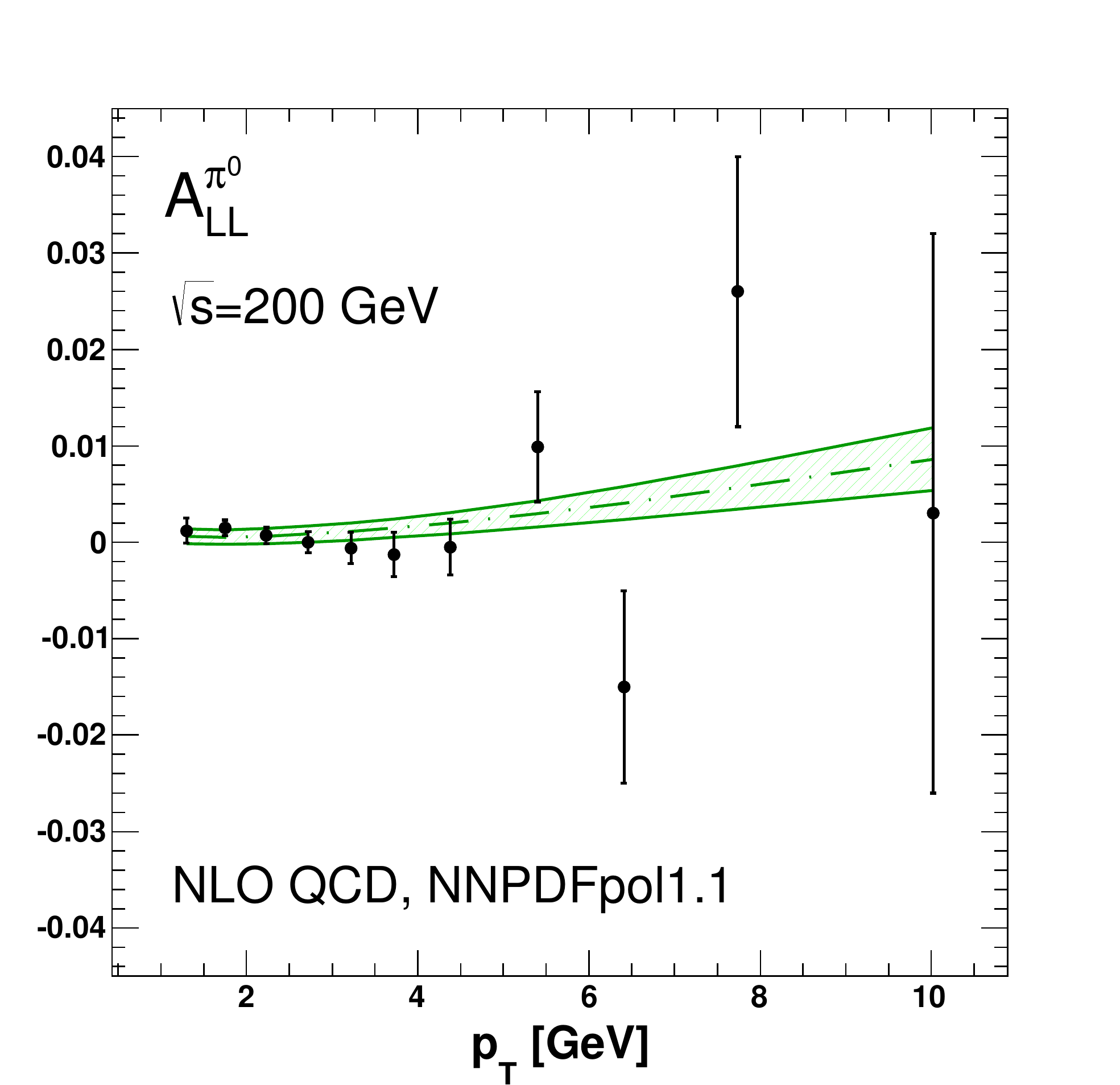}
\epsfig{width=0.40\textwidth,figure=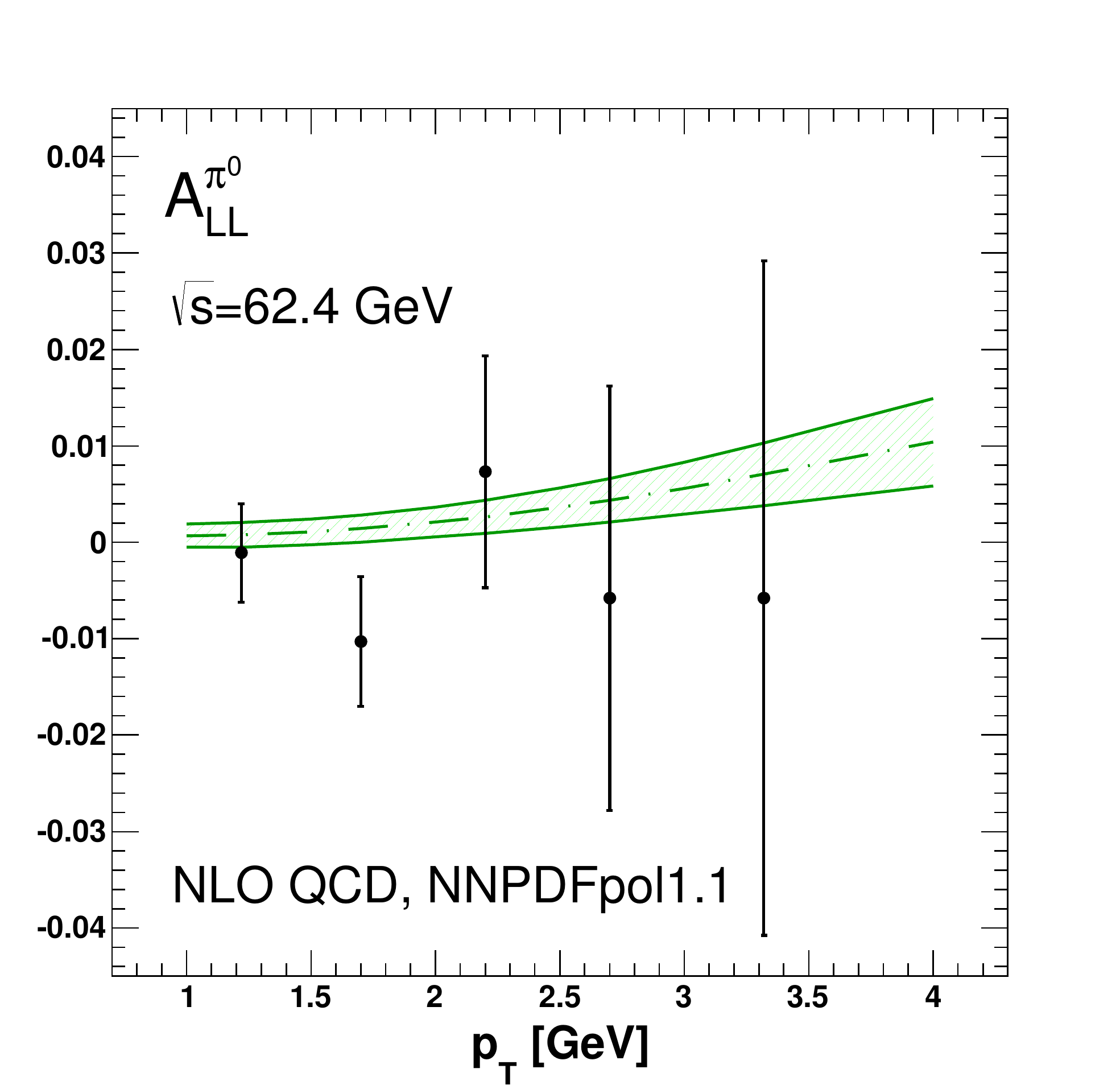}\\
\epsfig{width=0.40\textwidth,figure=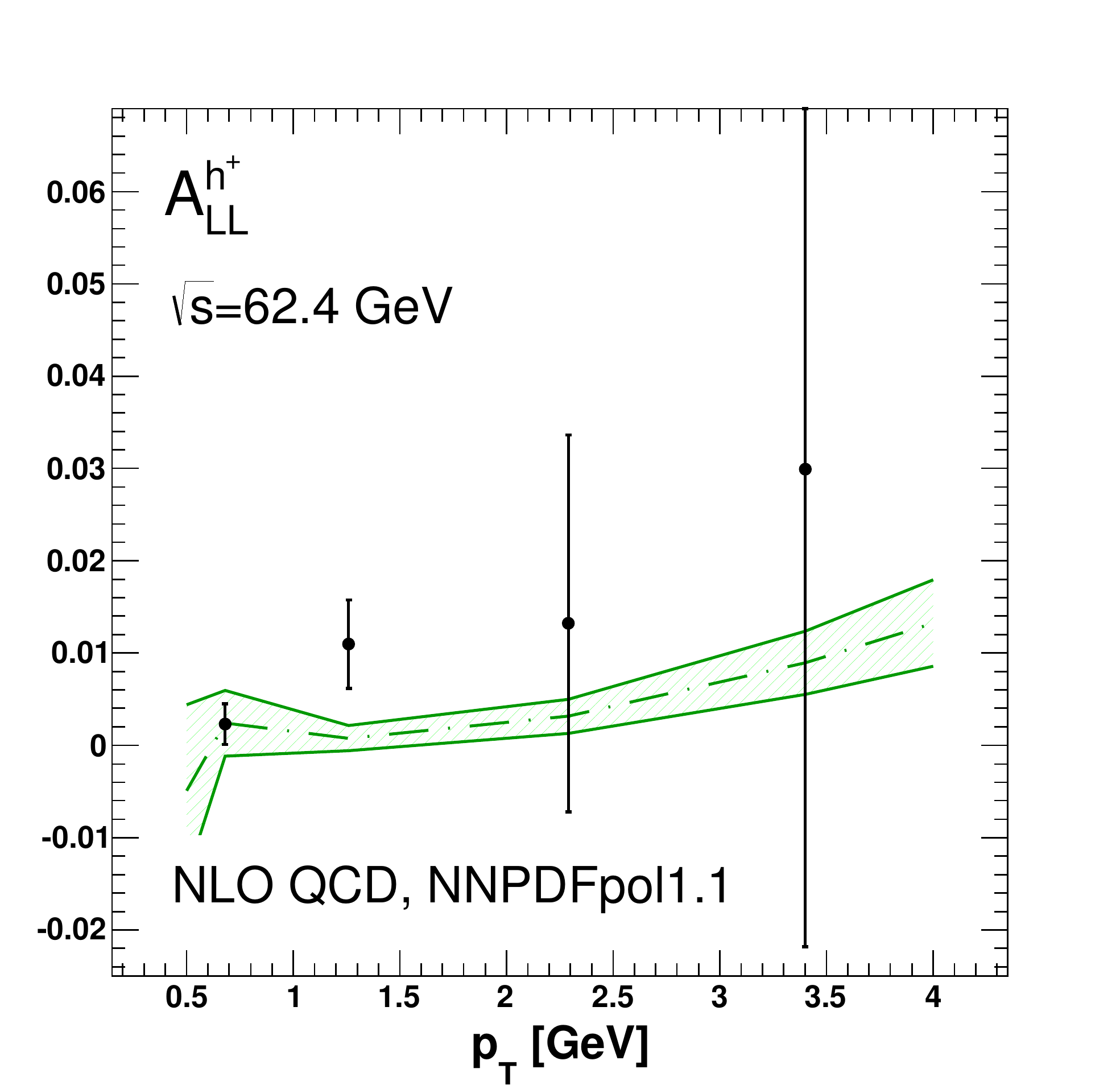}
\epsfig{width=0.40\textwidth,figure=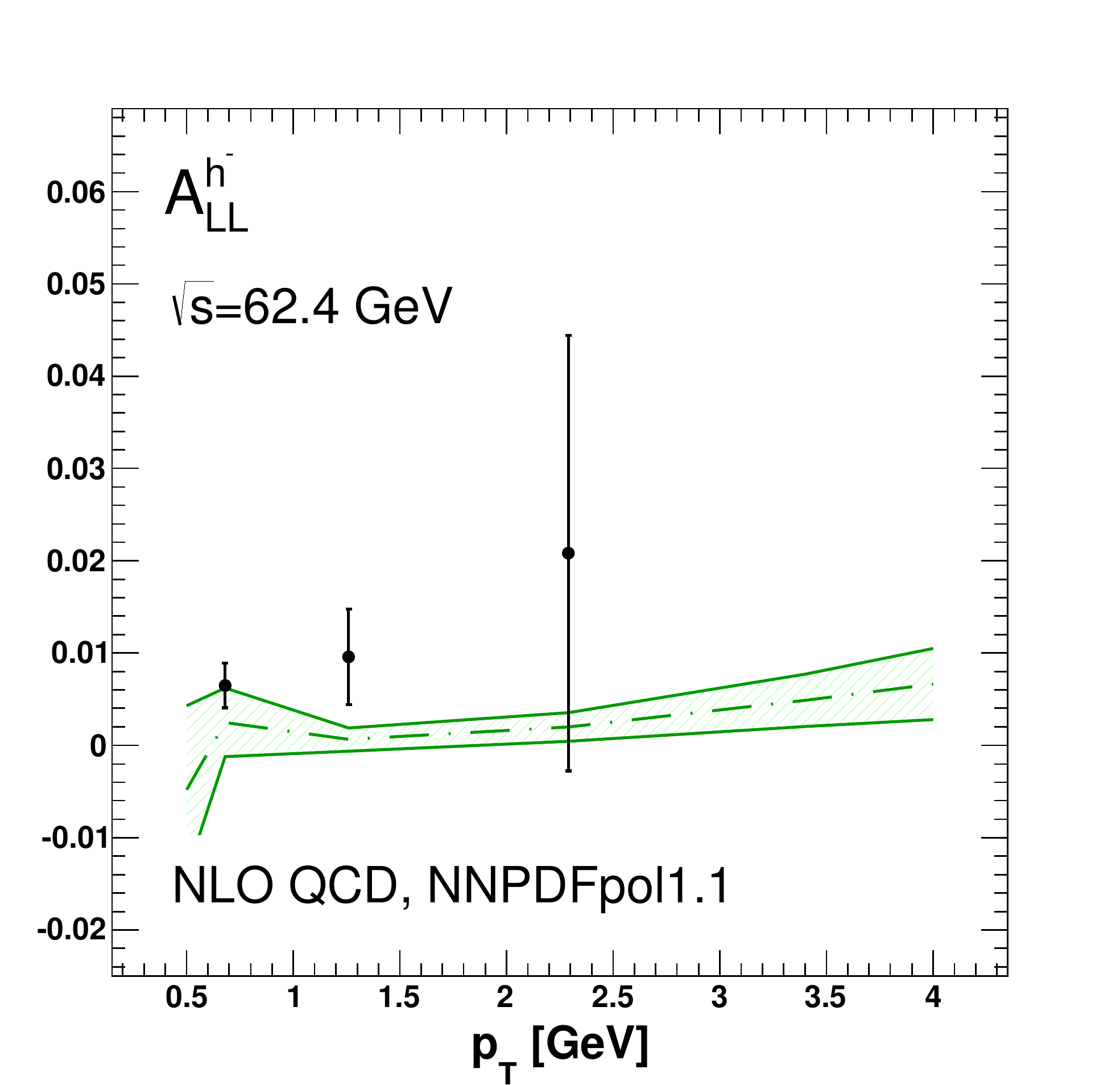}
\mycaption{Predictions for the neutral-pion (upper plots) 
and charged hadron (lower plots)
spin asymmetries computed at NLO accuracy with the \texttt{NNPDFpol1.1}
and \texttt{NNPDF2.3} parton sets, compared to measured data from 
PHENIX~\cite{Adare:2008aa,Adare:2008qb,Adare:2012nq}.}
\label{fig:phenixasy1}
\end{center}
\end{figure}

\begin{figure}[t]
\begin{center}
\epsfig{width=0.40\textwidth,figure=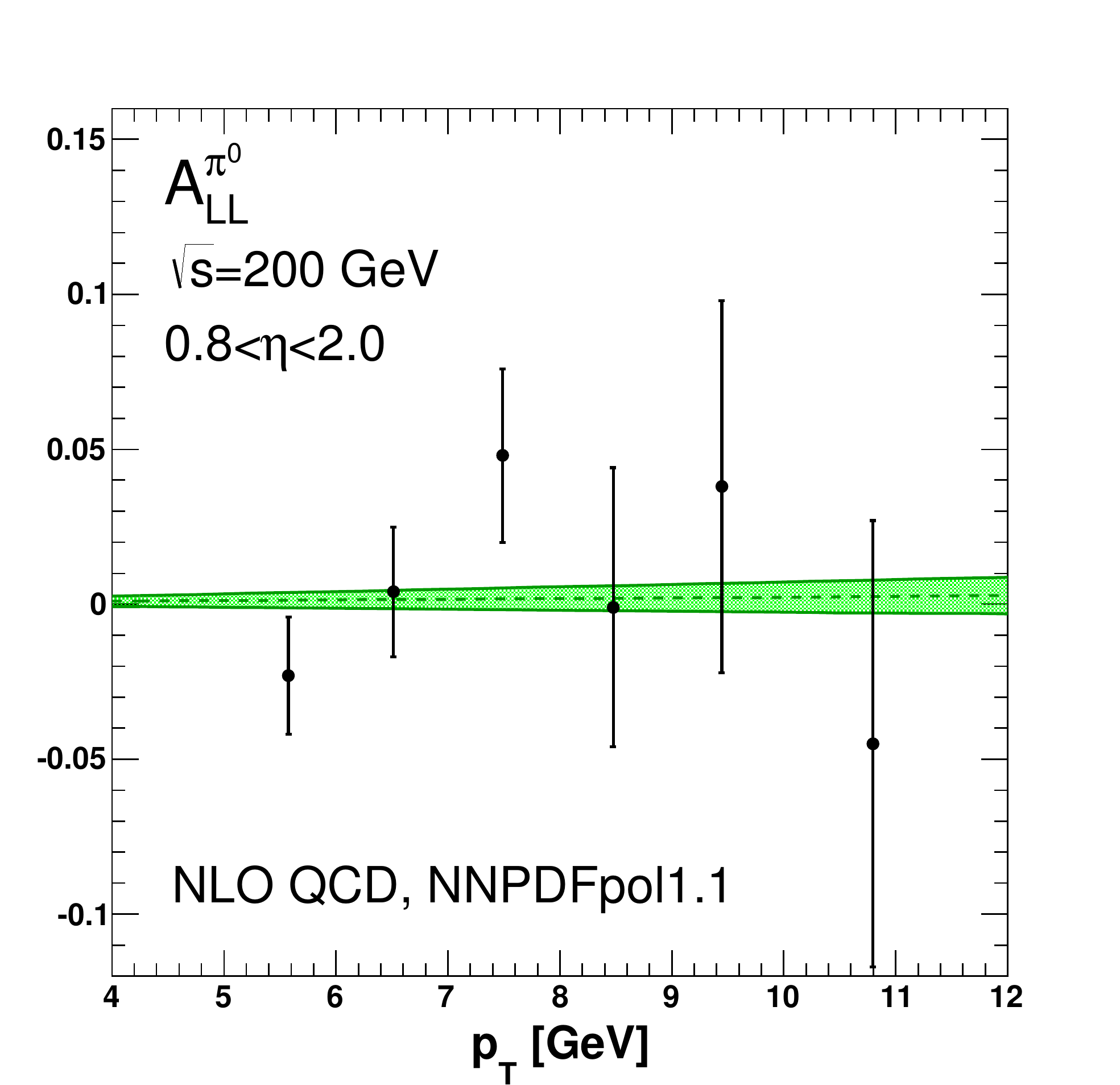}
\epsfig{width=0.40\textwidth,figure=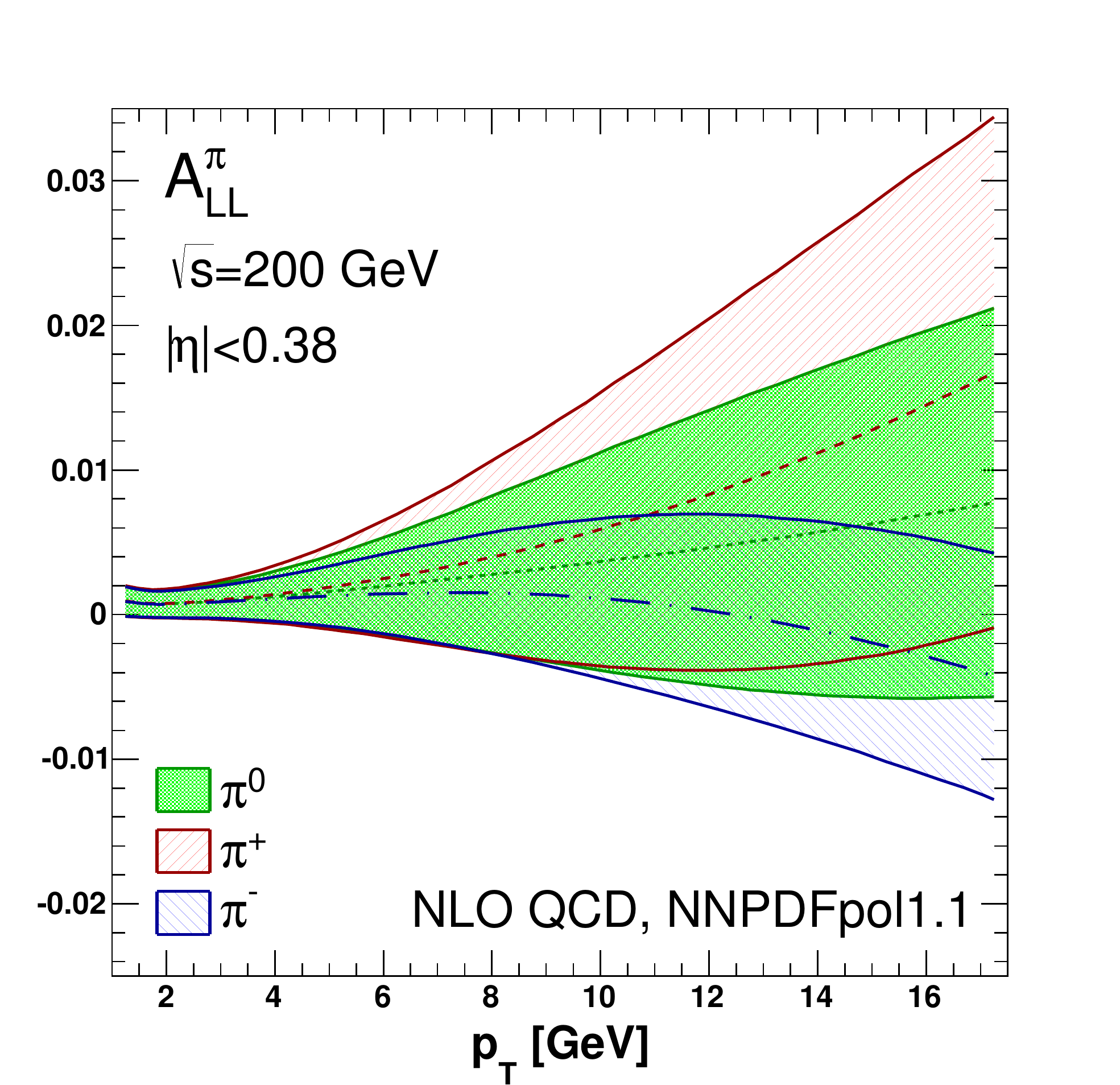}\\
\mycaption{(Left panel) Prediction for the neutral-pion spin asymmetry 
compared to data measured by STAR~\cite{Adamczyk:2013yvv}.
(Right panel) Prediction for the neutral- and charged-pion spin
asymmetries in the kinematic range accessed by the upcoming 
PHENIX measurements. All theoretical predictions are obtained from
the \texttt{NNPDFpol1.1} and \texttt{NNPDF2.3} 
parton sets at NLO accuracy.}
\label{fig:phenixasy2}
\end{center}
\end{figure}

In Fig.~\ref{fig:phenixasy1}, we show the double-spin asymmetry for 
single-hadron production in polarized proton-proton collisions,
Eq.~(\ref{eq:RHICasy}), compared to experimental 
measurements from PHENIX. In particular,
we provide predictions for 
neutral-pion production at center-of-mass energy
$\sqrt{s}=200$ GeV~\cite{Adare:2008aa} and 
$\sqrt{s}=62.4$ GeV~\cite{Adare:2008qb}, and 
mid-rapidity ($|\eta|<0.35$) charged hadron production at $\sqrt{s}=62.4$ 
GeV~\cite{Adare:2012nq}. 
Earlier measurments with 
neutral-pions~\cite{Adler:2004ps,Adler:2006bd,Adare:2007dg}, with 
significantly larger uncertainties, are not considered here.
In Fig.~\ref{fig:phenixasy2}, we compare the double-spin asymmetry
for neutral-pion production at 
forward rapidity ($0.8<\eta<2.0$) 
with recent STAR data
at $\sqrt{s}=200$ GeV~\cite{Adamczyk:2013yvv}.
In Fig.~\ref{fig:phenixasy2}, we also show the predictions
for neutral- and charged-pion production spin asymmetries
for which data from the PHENIX experiment will be soon available.

The asymmetries are computed as illustrated in 
Sec.~\ref{sec:reweighting-separate}: the (polarized) numerator 
in Eq.~(\ref{eq:RHICasy}) is computed for each replica in the
\texttt{NNPDFpol1.1} parton set ($N_{\mathrm{rep}}=100$), while the
(unpolarized) denominator is computed only once taking the average 
PDFs from the \texttt{NNPDF2.3}~\cite{Ball:2012cx} parton set at NLO.
In both the numerator and the denominator, we use the best fit
fragmentation functions from the \texttt{DSS07} set~\cite{deFlorian:2007aj}.
The central value and the uncertainty of the prediction are then
obtained as the mean and the standard deviations computed from 
the $N_{\mathrm{rep}}=100$ results for each replica in the polarized PDF set.
Hence, the estimated uncertainty on our prediction only takes into 
account the uncertainty of the polarized PDFs. 
While we expect the uncertainty on unpolarized PDFs has a negligible
impact, conversely the additional theoretical uncertainty due to 
the choice of a set of fragmentation functions may have a rather large impact.
We finally note that our predictions are
made using the code presented in Ref.~\cite{Jager:2002xm},
which we have modified to handle NNPDF parton sets.

Results in Figs.~\ref{fig:phenixasy1}-\ref{fig:phenixasy2}
show that the asymmetry remains very small in the measured
$p_T$ range. Experimental data are in good agreement with predictions
and seems to reinforce the conclusion that the polarized gluon 
is small in the
measured kinematic range. However, we notice that, except for PHENIX
neutral-pion production at small $p_T$, experimental uncertainties are
rather large in comparison to those of the corresponding 
theoretical predictions: 
this is particularly evident for STAR data in Fig.~\ref{fig:phenixasy2}, 
which, however, cover a rapidity range larger than that measured by PHENIX.
The mutual size of experimental and theoretical uncertainties 
is similar to that observed with 
COMPASS open-charm data in Sec.~\ref{sec:reweighting-separate}. 
However, in the present case, data show a more definite trend towards 
a growing asymmetry as $p_T$ increases, partly reproduced by the  
behavior of the corresponding theoretical prediction. For this reason, 
we expect that data on semi-inclusive particle production presented here 
will have a moderate impact on pinning down the size
of the polarized gluon uncertainty, once included in a global PDF determination.

\chapter{Conclusions and outlook}
\label{sec:conclusions}

In this Thesis, we have presented the first unbiased determination of 
spin-dependent, or polarized, Parton Distribution Functions 
(PDFs) of the proton.
These distributions are defined as the momentum densities of 
partons polarized along or opposite the
direction of the parent nucleon and are usually denoted
as $\Delta f(x,Q^2)$, where $f$ may refer either 
to individual quark or antiquark
flavors, or to a combination of them, or to the gluon. 
Parton distributions depend on both the 
Bjorken scaling variable $x$, the fraction of the proton 
momentum carried by the parton, and on the energy scale $Q^2$
with which the proton is probed. While the first is a 
non-perturbative dependence to be determined from experimental 
data, the latter is fully predictable in perturbative 
Quantum Chromodynamics (QCD), the theory of strong interaction.

In the framework of perturbative QCD, polarized parton distributions 
are essential ingredients
for any phenomenological study of hard scattering processes involving 
polarized hadrons in initial states.
The description of these processes in terms of expressions
in which the perturbative and the non-perturbative parts are factorized
is a powerful success of QCD itself. In such a picture, parton distributions 
are the fundamental objects encoding the information on the 
inner structure of the nucleon; in particular, polarized parton distributions
are related to its spin structure, since their integrals over the 
Bjorken scaling variable are interpreted as fractions of the proton spin.

The interest in the determination of polarized PDFs of the nucleon
is largely related to the experimental discovery, in the late 80s, 
that the singlet axial charge of the proton is anomalously 
small~\cite{Ashman:1987hv,Ashman:1989ig},
soon followed by the theoretical
realization~\cite{Altarelli:1988nr,Altarelli:1990jp} that the
perturbative behavior of polarized PDFs deviates from parton model
expectations, according to which gluons decouple in the asymptotic limit.
An accurate determination of polarized PDFs is then needed to precisely
assess which fraction of the nucleon spin is carried by quark and gluon
spins. The residual part of the nucleon spin which possibly would not be 
accounted for by quarks and gluons may be explained by resorting 
to their intrinsic Fermi motion and orbital angular 
momenta~\cite{Jaffe:1989jz,Ji:1996ek,Bakker:2004ib}
(for a recent discussion on the spin 
decomposition see also Ref.~\cite{Leader:2013jra}).
In addition to the investigation of the nucleon's spin structure,
polarized PDFs have been recently shown to be useful
in the probe of different beyond-standard-model scenarios~\cite{Bozzi:2004qq}
and in the determination of the Higgs boson spin in the diphoton decay channel,
by means of the linear polarization of gluons in an unpolarized 
proton~\cite{Boer:2013fca}.

Polarized parton distributions are presently known with much less accuracy 
than their unpolarized counterparts. 
As pointed out several times in this Thesis, this is mostly due to the 
experimental data they rely on, 
which are both less abundant and less accurate than 
those available in the unpolarized case.
Several polarized PDF sets have been determined in the last few 
years~\cite{Hirai:2008aj,deFlorian:2008mr,deFlorian:2009vb,Blumlein:2010rn,
Leader:2010rb,Ball:2013lla,Jimenez-Delgado:2013boa,Khorramian:2010qa,Arbabifar:2013tma}, 
but they are all based on the 
\textit{standard} Hessian methodology for PDF fitting and uncertainty 
estimation. This approach is known~\cite{Forte:2010dt} 
to potentially lead to an underestimation 
of PDF uncertainties, due to the limitations in the 
linear propagation of errors and to PDF parametrization
in terms of fixed functional forms,
both assumed in the \textit{standard} methodology.
These issues are especially delicate when the experimental information 
is scarce, like in the case of polarized data.

In light of these considerations, an unbiased determination
of polarized PDFs is crucial in order to provide an adequate 
estimate of the uncertainty with which quarks and gluons 
can actually contribute to the nucleon spin.
In particular, such a determination allows for scrutinizing 
the common belief that the anomalous gluon contribution is too small 
to compensate a reasonably large singlet spin contribution
into the almost vanishing axial charge observed in experiments.
Providing the first unbiased determination of polarized parton distributions 
has precisely been the goal of the present Thesis.

\section{Summary of the main results}
\label{sec:summaryofresults}

In this Thesis, the determination of polarized parton sets has been carried
out within the NNPDF methodology. This uses a robust set of statistical 
tools, devised for a statistically sound determination
of PDFs and their uncertainties, which include Monte Carlo methods for error propagation, 
neural networks for PDF parametrization and genetic algorithms 
for their minimization.
This methodology has already been successfully applied in the 
unpolarized case~\cite{DelDebbio:2004qj,DelDebbio:2007ee,
Ball:2008by,Ball:2009qv,Ball:2009mk,
Ball:2010de,Ball:2010gb,Ball:2011mu,Ball:2011uy,
Ball:2011eq, Ball:2011gg,Ball:2012cx,Ball:2013hta}, 
where the NNPDF
sets are routinely used by the LHC collaborations in their data 
analysis and data-theory comparison.
It has been extended here to polarized PDFs for the first time.
In more detail, the main achievements presented in this Thesis are summarized
below.
\begin{itemize}
\item Based on world-available data from polarized inclusive 
Deep-Inelastic Scattering, we determined a first polarized parton set
at next-to-leading order accuracy, \texttt{NNPDFpol1.0}. 
We reviewed in detail the theory
and phenomenology of polarized DIS, in particular focusing on the
features of the data included in our analysis. We discussed
how the NNPDF methodology has been adapted to the polarized case
and which strategies have been devised to face some issues,
like the proper implementation of target mass corrections and positivity
constraints in the fitting algorithm.

Our analysis showed that some PDF uncertainties are likely to be underestimated
in other existing determinations, based on the standard methodology,
due to their less flexible parametrization.
This is particularly the case of the gluon, 
which is left almost unconstrained by inclusive DIS data: hence, 
its contribution to the nucleon spin is still largely uncertain,
unless one makes strong assumptions on the PDF functional form in 
the small-$x$ ($x\lesssim 10^{-3}$) extrapolation region, where experimental
data are presently lacking. 
For the same reason, we also showed that a determination of 
the strong coupling $\alpha_s$ from the Bjorken sum rule is not competitive, 
again because the nonsinglet structure function
in the unmeasured small-$x$ region is largely uncertain.

These conclusions were supported 
by a careful analysis of the stability of our results upon the 
variation of a number of theoretical and methodological assumptions,
in particular related to the effects of target mass corrections, 
sum rules, and positivity constraints. 
First, we found that inclusive DIS data, 
with our kinematic cuts, do not show sensitivity to
finite nucleon mass effects, neither in terms of fit quality, nor in terms 
of the effect on PDFs. Second, we concluded that our fit 
results are quite stable upon variations of the treatment of sum rules
dictated by hyperon decays. Finally, we emphasized that positivity 
significantly affects PDFs in the region where no data
are available, in particular their large-$x$ behavior.

\item The proposed Electron-Ion 
Collider (EIC)~\cite{Deshpande:2005wd,Boer:2011fh,Accardi:2012hwp} 
is expected to enlarge the kinematic coverage of data,
which is presently rather limited, by at least two
orders of magnitude in both $x$ and $Q^2$. 
This will reduce the uncertainty due to PDF 
extrapolation to small-$x$ values and will allow for a better 
determination of the polarized gluon PDF through scaling violations, 
thanks to a larger $Q^2$ lever arm.
Using simulated pseudodata for two realistic scenarios at
an EIC, with increasing energy of both the lepton and 
hadron beams, we have studied its potential impact on the determination of 
polarized PDFs. 
We found that inclusive DIS data at an EIC
would entail a considerable reduction in the gluon PDF
uncertainty and also provide evidence of a possible large gluon 
contribution to the nucleon spin, though the latter goal would still be reached
with a sizable residual uncertainty.   

The measurement of the charm contribution to the
proton structure function, $g_1^{p,c}$, which is directly sensitive 
to the gluon, might provide more
information on the corresponding distribution. We showed that 
$g_1^{p,c}$, though being small, could be as much larger
as $10-20\%$ of the total structure function $g_1^p$ 
in the kinematic region probed by an EIC: hence, in order to 
further pin down the gluon uncertainty from
intrinsic charm effects, one should be able to measure the corresponding
contribution to the $g_1$ structure function within this accuracy.

\item The Relativistic Heavy Ion Collider (RHIC) is the first facility
in the world to collide polarized proton beams. Measurements on 
inclusive jet and $W$ boson production asymmetries have been recently
presented: we studied their potential in constraining the 
polarized gluon and in separating light quark and antiquark PDFs, respectively.
This new piece of experimental information was included in our polarized
parton set by means of Bayesian reweighting of suitable Monte Carlo
PDF ensembles~\cite{Ball:2010gb,Ball:2011gg}. 
This method, which consist of updating the underlying 
PDF probability distribution of a prior ensemble according to the conditional 
probability for the old PDFs with respect to new data, 
allows for the inclusion of new
data in a PDF set without the need of a global refitting; hence, it
could be used to quickly update a PDF set with any new 
piece of experimental information.
This way, we were able to provide
the first global polarized PDF set obtained within the NNPDF framework,
\texttt{NNPDFpol1.1}. In comparison to \texttt{NNPDFpol1.0}, 
the new polarized parton set provides a meaningful
determination of sea flavor PDFs $\Delta\bar{u}$ and $\Delta\bar{d}$,
based on $W$ boson production data
(otherwise not determined by inclusive DIS data or determined in SIDIS, 
but with the bias introduced by poorly known fragmentation functions),
and a determination of the gluon PDF
$\Delta g$ which is improved by open-charm and, particularly, jet data. 

We should also notice that, from a conceptual point of view, the methodology
we followed to determine this parton set is in itself particularly valuable.
Indeed, we have explicitly shown how a PDF set can be succesfully obtained by
including all data through reweighting of a first unbiased guess,
as originally proposed in Refs.~\cite{Giele:1998gw,Giele:2001mr}.

The main conclusion on the partons' contributions to the nucleon spin 
based on the \texttt{NNPDFpol1.1} parton set is twofold. 
On the one hand, we have found that PDF first moments are
rather well determined in the 
kinematic region covered by experimental data and are in good 
agreement with the values obtained in the only available 
analysis including the same collider data~\cite{Aschenauer:2013woa}.
In particular, in the region constrained by data, 
the singlet full first moment is less than a 
half of the proton spin within its uncertainty, while
the gluon first moment is definitely positive, though 
rather small. The determination of the gluon 
is more accurate in \texttt{NNPDFpol1.1} than in 
\texttt{NNPDFpol1.0}, mostly thanks to jet data, 
located in the region $0.05\lesssim x \lesssim 0.2$.
On the other hand, we emphasize that the uncertainty on both the 
singlet and the gluon full first moments coming from the
extrapolation to the unmeasured, small-$x$, region dominates
their total uncertainty. For this reason, large values of 
the gluon first moment are not completely ruled out: 
within our accurate determination of uncertainties,  
the almost vanishing value for the singlet axial charge observed
in the experiment may still be completely explained as a cancellation between 
a rather large quark and the anomalous gluon contributions.

\item  We developed a \textit{Mathematica} package which allows for fast
and interactive usage of any available NNPDF parton set, both unpolarized 
and polarized, see Appendix~\ref{sec:appC}. 
This interface includes all the features already available 
through LHAPDF~\cite{Whalley:2005nh,web:LHAPDF}, 
but they can be profitably combined together with 
those provided by \textit{Mathematica}. 
The software we developed was tailored to
the users who are not familiar with Fortran or C++ languages used by 
the LHAPDF interface 
and who can benefit from the more direct usage of PDFs within a 
\textit{Mathematica} notebook. 

\end{itemize} 

The \texttt{NNPDFpol1.0} and \texttt{NNPDFpol1.1} polarized PDF sets, 
with $N_{\mathrm{rep}}=100$ replicas, are publicly available from the NNPDF website
\begin{center}
{\textbf{\url{http://nnpdf.hepforge.org/}}~}.
\end{center}
The \textit{Mathematica} interface, as well as \texttt{FORTRAN} and \texttt{C++}
stand-alone codes for handling these parton distributions, are also available 
from the same source.

\section{Future directions}
\label{sec:future}

The \texttt{NNPDFpol1.1} parton set
is based on all the relevant and up-to-date experimental information 
from deep-inelastic scattering and proton-proton collisions
which do not depend on the fragmentation of the struck quark into 
final observed hadrons. Further data are expected from PHENIX and STAR 
in the upcoming years, which will further improve the accuracy
of polarized PDF determinations. As further refinements of polarized PDFs 
will be achieved, they will become more and more appealing for the 
experimental collaborations to be used in
their analysis and for data-theory comparison. 
In this sense, efforts will be devoted to make the NNPDF
polarized parton sets the \textit{gold standard}, 
as their unpolarized counterparts are quickly becoming.

In order to obtain additional information on the spin structure of the proton,
it will be certainly beneficial to include
a wide range of semi-inclusive measurements, namely semi-inclusive 
DIS in fixed-target 
experiments~\cite{Adeva:1997qz,Airapetian:2004zf,Alekseev:2007vi,
Alekseev:2009ac,Alekseev:2010ub}, and semi-inclusive particle 
production in polarized collisions at
RHIC~\cite{Adare:2008aa,Adare:2008qb,Adare:2012nq,Adamczyk:2013yvv}.
However, a consistent inclusion of these data in a \textit{global} fit
requires first of all the corresponding determination 
of fragmentation functions using the NNPDF methodology.
Indeed, available fragmentation function sets~\cite{Kretzer:2000yf,Kniehl:2000fe,
Bourhis:2000gs,Hirai:2007cx,Albino:2005me,deFlorian:2007aj,deFlorian:2007hc,Albino:2008fy} 
suffer from several limitations due to their too rigid 
parametrization. It was recently shown that none of these sets can
describe the most updated inclusive charged-particle
spectra data at the LHC satisfactorily~\cite{d'Enterria:2013vba}.
Therefore, a determination of fragmentation functions using the
NNPDF methodology is highly desirable by itself,
and may be important in various areas of phenomenology~\cite{Albino:2008gy}; 
in particular, it will pave
the way to use a large data set of semi-inclusive polarized
data in future NNPDF analyses. 

Finally, we notice that the methods illustrated
here apply to the determination of any non-perturbative object from 
experimental data. Hence, even though 
a phenomenological study of either TMDs or GPDs
was beyond the scope of this Thesis, the NNPDF methodology
may be used as well to provide the determination of such 
distributions in the future, when relevant 
experimental data will reach more and more abundance and accuracy.

In summary, in this Thesis not only we have extended the NNPDF framework
to the determination of spin-dependent parton sets, but we have also
reached the state-of-the-art in our unbiased understanding
of the proton's spin content, as allowed by available experimental data.
Further constraints will be provided by a variety of semi-inclusive 
measurements, which in turn will require the development of a set of 
parton fragmentation functions using the NNPDF methodology.
In the long term, the final word on the spin content of the proton will
require brand new facilities such as an Electron-Ion Collider, 
as we have also extensively discussed in this Thesis.
Indeed, it could finally bring polarized PDF determinations to a 
similar level of accuracy as the one reached for their unpolarized counterparts. 
We hope that the NNPDF collaboration will play a leading role in this 
exciting game.

\appendix
\chapter{Statistical estimators}
\label{sec:appB}

In this Appendix, we collect the definitions of the statistical estimators
used in the NNPDF analyses presented in 
Chaps.~\ref{sec:chap2}-\ref{sec:chap4}-\ref{sec:chap5}.
Despite they were already described 
in Refs.~\cite{Forte:2002fg,DelDebbio:2004qj,Ball:2010de},
we find it useful to give them for completeness and ease of 
reference here. In the following, we denote with $\mathcal{O}$ a generic 
quantity depending on replicas in a Monte Carlo ensemble of PDFs; it may be
a PDF, a linear combination of PDFs, or a physical observable. We also
denote as $\langle\mathcal{O}\rangle_{\mathrm{rep}}$ the mean computed over the
$N_{\mathrm{rep}}$ replicas in the ensemble, and as 
$\langle\mathcal{O}\rangle_{\mathrm{dat}}$ the mean computed over the 
$N_{\mathrm{dat}}$ experimental data for a fixed replica in the ensemble.
\begin{itemize}[leftmargin=*]
\item Central value
\begin{equation}
\left\langle \mathcal{O}\right\rangle_{\mathrm{rep}}
=
\frac{1}{N_{\mathrm{rep}}}\sum_{k=1}^{N_{\mathrm{rep}}}\mathcal{O}^{(k)}
\,\mbox{.}
\end{equation}
\item Variance
\begin{equation}
\sigma
=
\sqrt
{
\left\langle \mathcal{O}^2\right\rangle_{\mathrm{rep}}
-
\left\langle \mathcal{O} \right\rangle^2_{\mathrm{rep}}
}
\,\mbox{.}
\label{eq:variance}
\end{equation}
\item Elements of the correlaton matrix
\begin{equation}
\rho_{ij}
=
\frac{\left\langle \mathcal{O}_i\mathcal{O}_j\right\rangle_{\mathrm{rep}}
-\left\langle\mathcal{O}_i\right\rangle_{\mathrm{rep}}
\left\langle\mathcal{O}_j\right\rangle_{\mathrm{rep}}}
{\sigma_i\sigma_j}
\,\mbox{.}
\end{equation}
\item Elements of the covariance matrix
\begin{equation}
\mathrm{cov}_{ij}
=
\rho_{ij}\sigma_i\sigma_j
\,\mbox{.}
\end{equation}
\item Percentage error over the $N_{\mathrm{dat}}$ data points
\begin{equation}
\left\langle\mathrm{PE}\left[\langle\mathcal{O}\rangle_{\mathrm{rep}}\right] 
\right\rangle
=
\frac{1}{N_{\mathrm{dat}}}\sum_{i=1}^{N_{\mathrm{dat}}}
\left[
\frac{\langle\mathcal{O}_i\rangle_{\mathrm{rep}}-\mathcal{O}_i}{\mathcal{O}_i}
\right]
\,\mbox{.}
\end{equation}
\item Scatter correlation between two quantities 
\begin{equation}
r(\mathcal{O}_1,\mathcal{O}_2)
=
\frac{\left\langle\mathcal{O}_1\mathcal{O}_2\right\rangle_{\mathrm{dat}}
-
\left\langle\mathcal{O}_1\right\rangle_{\mathrm{dat}}
\left\langle\mathcal{O}_2\right\rangle_{\mathrm{dat}}}
{\sigma_1\sigma_2}
\,\mbox{,}
\end{equation}
where $\mathcal{O}_{1,2}$ may be obtained as averages over Monte Carlo replicas.
\item Square distance between central value estimates from two PDF ensembles
\begin{equation}
d^2\left(\langle\mathcal{O}^{(1)}\rangle, \langle\mathcal{O}^{(2)}\rangle\right)
=
\frac
{
[\langle\mathcal{O}^{(1)}\rangle - \langle\mathcal{O}^{(2)}\rangle]^2
}
{
\sigma^2\left[\langle\mathcal{O}^{(1)}\rangle\right]
+
\sigma^2\left[\langle\mathcal{O}^{(2)}\rangle\right]
}
\,\mbox{,}
\end{equation}
where the variance of the mean is given by
\begin{equation}
\sigma^2[\langle\mathcal{O}^{(i)}\rangle]
=
\frac{1}{N_{\mathrm{rep}}^{(i)}}\sigma^2[\mathcal{O}^{(i)}]
\label{eq:dist2}
\end{equation}
in terms of the variance $\sigma[\mathcal{O}^{(i)}]$ 
of the quantities $\mathcal{O}^{(i)}$, estimated as the variance
of the replica sample, Eq.~(\ref{eq:variance}). In our notation $i=1,2$.
\item Square distance between square uncertainty estimates from two 
PDF ensembles
\begin{equation}
d^2\left(\sigma^2[\mathcal{O}^{(1)}], \sigma^2[\mathcal{O}^{(2)}]\right)
=
\frac
{\left(\bar{\sigma}_{(1)}^2-\bar{\sigma}_{(2)}^2\right)^2}
{\sigma^2[\bar{\sigma}_{(1)}^2]+\sigma^2[\bar{\sigma}_{(2)}^2]}
\,\mbox{,}
\label{eq:dist1}
\end{equation}
where we have defined $\bar{\sigma}_{(i)}^2\equiv\sigma^2[\mathcal{O}^{(i)}]$,
$i=1,2$.
In practice, for small-size replica samples the distances defined in 
Eqs.~(\ref{eq:dist1})-(\ref{eq:dist2}) display sizable statistical fluctuations.
In order to stabilize the result, all distances computed in this Thesis
are determined as follows: we randomly pick $N^{(i)}_{\mathrm{rep}}/2$ out of the
$N^{(i)}_{\mathrm{rep}}$ replicas for each of the two subsets. The
computation of the square distance Eq.~(\ref{eq:dist1}) or 
Eq.~(\ref{eq:dist2}) is then repeated for 
$N_\mathrm{part}=100$ (randomly generated) choices of $N^{(i)}_{\mathrm{rep}}/2$ 
replicas, and the result is averaged: 
this is sufficient to bring the statistical fluctuations of
the distance at the level of a few percent.
\end{itemize}

\chapter{A \textit{Mathematica} interface to NNPDF parton sets}
\label{sec:appC}

In this Appendix, we present a package for handling 
both unpolarized and polarized NNPDF parton sets
within a \textit{Mathematica} notebook file~\cite{Hartland:2012ia}. 
This allows for performig PDF manipulations easily and quickly,
thanks to the powerful features of the \textit{Mathematica} software.
The package was tailored to the users who are not familiar with 
\texttt{FORTRAN} or \texttt{C++} programming codes, on which the standard
available PDF interface, LHAPDF~\cite{Whalley:2005nh,web:LHAPDF}, is based.
However, since our \textit{Mathematica} package includes all the features
available in the LHAPDF interface, any user can benefit from the interactive
usage of PDFs within \textit{Mathematica}.

The NNPDF \textit{Mathematica} package can be downloaded from the NNPDF
web page
\begin{center}
{\url{http://nnpdf.hepforge.org/}}
\end{center}
together with sample notebooks containing a step by step 
explanation of the NNPDF usage within
{\textit{Mathematica}}, as well as a variety of examples.
The procedure to download and run our \textit{Mathematica} package is rather 
simple:
\begin{enumerate}
\item \textbf{Download}\newline
\texttt{wget http://nnpdf.hepforge.org/math$\_$package.tgz}
\item \textbf{Unpack}\newline
\texttt{tar -xvzf math$\_$package.tgz }
\item \textbf{Run the tutorial}\newline
\texttt{mathematica Demo-unpol.nb}\newline
\texttt{mathematica Demo-pol.nb}     
\end{enumerate}
The input for our \textit{Mathematica} package is any \texttt{.LHgrid} file
delivered by the NNPDF Collaboration as the final result of a fit. These
files are publicly available from the NNPDF hepforge 
website or from the LHAPDF library and should be downloaded 
separately from the \textit{Mathematica} package.

The functions implemented in the package are summarized in 
Tab.~\ref{tab:itemfunc}. In the following, we briefly
demonstrate the NNDPF \textit{Mathematica} package by examining 
the \texttt{NNPDF2.3} parton determination at NLO~\cite{Ball:2012cx}.
\begin{list}{}{\leftmargin=0pt}
 \item {\textbf{Compute PDF central value and variance.}} 
We have defined proper functions to keep the computation 
of PDF central value and variance very easy. 
These built-in functions only need $x$, $Q^2$ and PDF flavour as input.
The user can also specify the confidence level to which central value 
and variance should be computed.
 \item {\textbf{Make PDF plots.}} 
\textit{Mathematica} enables a wide range of plotting options. 
As a few examples, we
show the 3D plot and the contour plot of the singlet PDF combination from 
the \texttt{NNPDF2.3} parton set at NLO (see Fig.~\ref{fig:mathematica}).
\begin{figure}[t]
\epsfig{width=0.50\textwidth,figure=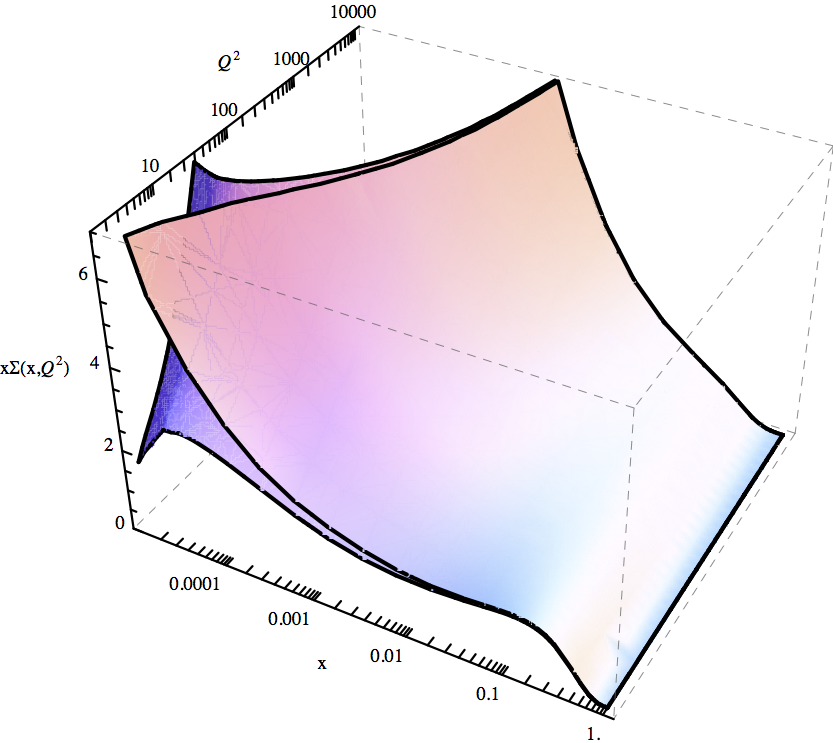}
\ \ \ \ \  
\epsfig{width=0.45\textwidth,figure=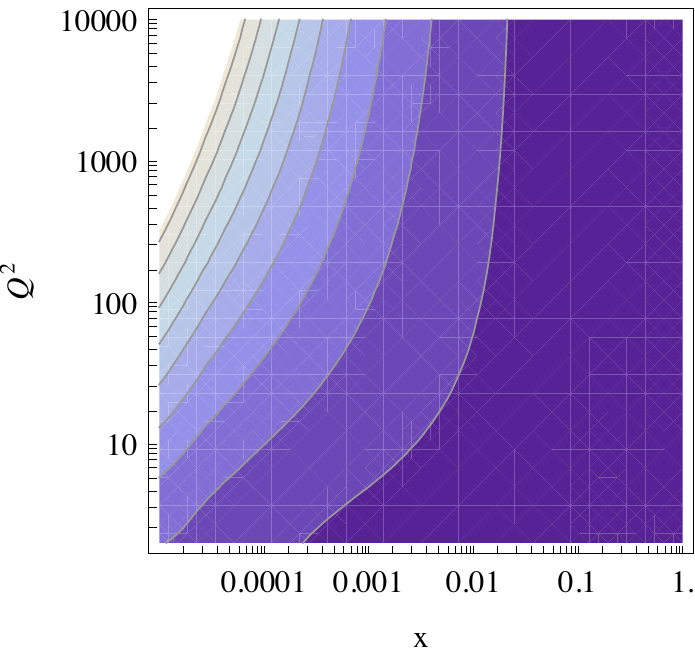}
\mycaption{(Left) Simultaneous $(x,Q^2)$ dependence of the
singlet PDF and its one-sigma error band from the NLO 
\texttt{NNPDF2.3} parton set. 
(Right) Contour plot of the of the singlet PDF from the \texttt{NNPDF2.3}
parton set at NLO in the $(x,Q^2)$ plane.}
\label{fig:mathematica}
\end{figure}
\item {\textbf{Perform computations involving PDFs.}} 
PDF manipulation can be carried out straighforwardly
since we have defined functions which handle either 
single replicas or the whole Monte Carlo ensemble. 
The user can then easily perform any computation which involves PDFs.
For example, we show in Fig.~\ref{fig:snapshot} a snapshot 
of a typical {\textit{Mathematica}} notebook in which we use our interface to
check the momentum and valence sum rules from the \texttt{NNPDF2.3} 
parton set at NLO.
\end{list}
\begin{sidewaystable}[p]
\centering
\footnotesize
\begin{tabularx}{\textwidth}{lX}
\toprule
Function & Description\\
\midrule
\texttt{InitializePDFGrid[path, namegrid]} & 
It reads the \texttt{.LHgrid} file into memory specified by \texttt{namegrid}
(either unpolarized or polarized) from the location specified by \texttt{path}. 
It also performs the PDF interpolation in 
the ($x$,$Q^2$) space by means of a built-in \textit{Mathematica} interpolation 
algorithm.\\
\midrule
\texttt{xPDFcv[x,Q2,f]} &
It returns \texttt{x} times the central value of the PDF with flavor  
\texttt{f} at a given momentum fraction \texttt{x} and scale \texttt{Q2} 
(in GeV$^2$). Note that \texttt{f} must be an integer, 
and \texttt{x} and \texttt{Q2} must be numeric quantities. 
For the unpolarized case, and polarized \texttt{NNPDFpol1.1} onwards, 
the LHAPDF convention is used for the flavor 
\texttt{f}, that is, \texttt{f}=-6, -5, -4, -3, -2, -1, 0, 1, 2, 3, 4, 5, 6 
corresponds to $\bar{t}$, $\bar{b}$, $\bar{c}$, $\bar{s}$, $\bar{u}$, $\bar{d}$,
$g$, $d$, $u$, $s$, $c$, $b$, $t$.
For the polarized case, \texttt{NNPDFpol1.0}, 
the following convention is used for the flavor \texttt{f}: 
\texttt{f}=0, 1, 2, 3, 4 corresponds to $\Delta g$, $\Delta u +\Delta\bar{u}$,
$\Delta d +\Delta\bar{d}$, $\Delta s +\Delta\bar{s}$.\\
\midrule
\texttt{xPDFEnsemble[x,Q2,f]} &
It returns \texttt{x} times the vector of PDF replicas of flavor \texttt{f} at 
a given momentum fraction \texttt{x} and scale \texttt{Q2} (in GeV$^2$).\\
\midrule
\texttt{xPDFRep[x,Q2,f,irep]} &
It returns \texttt{x} times the \texttt{irep} PDF replica of flavor 
\texttt{f} at a given momentum fraction \texttt{x} and scale 
\texttt{Q2} (in GeV$^2$). For \texttt{irep}=0, the mean value over the PDF
ensemble is returned.\\
\midrule
\texttt{xPDF[x,Q2,f]} &
It returns \texttt{x} times the value of the PDF of flavor \texttt{f} at a 
given momentum fraction \texttt{x} and scale \texttt{Q2} (in GeV$^2$) 
with its standard deviation.\\
\midrule
\texttt{xPDFCL[ensemble,x,Q2,f,CL]} &
It returns \texttt{x} times the value of the PDF of flavor \texttt{f} at a 
given momentum fraction \texttt{x} and scale \texttt{Q2} (in GeV$^2$) 
with its standard deviation. These values are computed over the PDF ensemble
specified by \texttt{ensemble} at a confidence level specified by \texttt{CL}.
The variable \texttt{ensemble} must be a function of \texttt{x}, \texttt{Q2}
and \texttt{f}: it can be either the function \texttt{xPDFEnsemble} or any
PDF ensemble defined by the user. The variable \texttt{CL} must be a real
number between 0 and 1.\\
\midrule
\texttt{NumberPDF[]} &
It returns the number of PDF members in the set.\\
\midrule
\texttt{Infoalphas[]} &
It returns information on $\alpha_s$ evolution used in the QCD analysis.\\
\midrule
\texttt{alphas[Q2,ipo,imodev]} &
It returns the QCD strong coupling constant $\alpha_s$. 
The inputs are:
\begin{itemize}[leftmargin=*]
\item \texttt{Q2}: the energy scale $Q^2$, in GeV$^2$, at which $\alpha_s$ 
is computed;
\item \texttt{ipo}: the perturbative order at which $\alpha_s$ is computed;
0: LO; 1: NLO; 2: NNLO;
\item \texttt{imodev}: the evolution mode with which $\alpha_s$ is computed;
\newline
0: $\alpha_s$ is computed as a function of $\alpha_s$ at the $Z$ 
mass as given in the \texttt{.LHgrid} file; 
\newline
1: exact solution of the QCD $\beta$-function equation using 
Runge-Kutta algorithm.
\end{itemize}\\
\midrule
\texttt{alphasMZ[]} &
It returns the list of $\alpha_s$ values at the $Z$ mass used 
in the QCD analysis for each replica.\\
\midrule
\texttt{mCharm[]}, \texttt{mBottom[]}, \texttt{mTop[]}, \texttt{mZ[]} &
They return the charm, bottom, top quark mass (in GeV) and the 
$Z$ boson mass (in GeV) used in the QCD analysis\\
\midrule
\texttt{Lam4[]}, \texttt{Lam5[]} &
They return the $\Lambda_{\mathrm{QCD},4}$, $\Lambda_{\mathrm{QCD},5}$ 
used in the QCD analysis\\
\midrule
\texttt{xMin[]}, \texttt{xMax[]}, \texttt{Q2Min[]}, \texttt{Q2Max[]} &
They return the minimum or maximum value for $x$ or $Q^2$ (in GeV$^2$)
used in the input \texttt{.LHgrid} grid.\\
\bottomrule
\end{tabularx}
\mycaption{Description of the functions available in the \textit{Mathematica}
interface to NNPDF partons sets.}
\label{tab:itemfunc}
\end{sidewaystable}
\begin{sidewaysfigure}[p]
\centering
\includegraphics[scale=0.37]{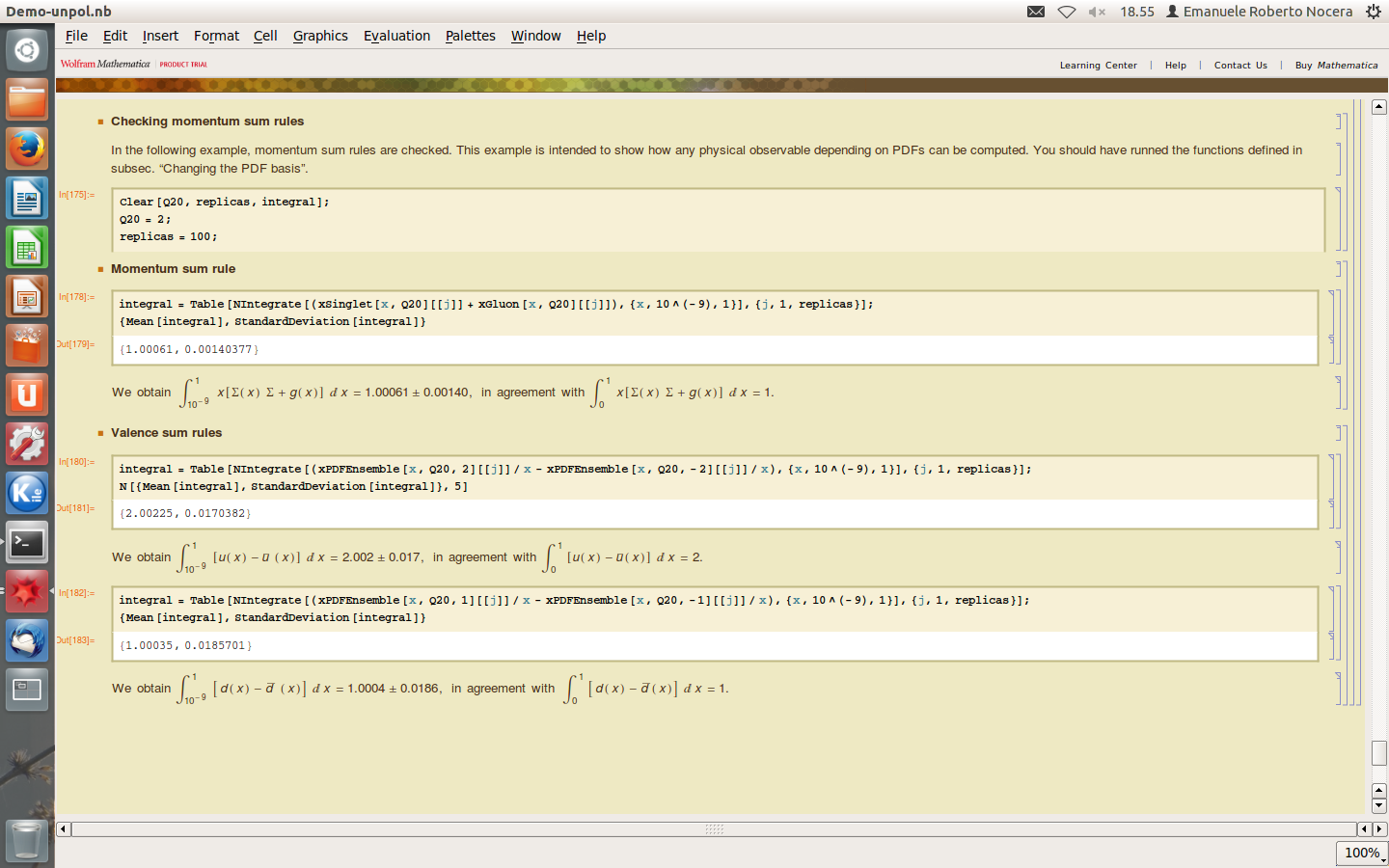}
\mycaption{A snapshot from a \textit{Mathematica} notebook in which the NNPDF
interface is used: here momentum and valence sum rules are checked from
the \texttt{NNPDF2.3} parton set at NLO.}
\label{fig:snapshot}
\end{sidewaysfigure}

\chapter{The FONLL scheme for \texorpdfstring{$g_1(x,Q^2)$}{g1}
up to \texorpdfstring{$\mathcal{O}(\alpha_s)$}{O(alpha)}}
\label{sec:appA}

In this Appendix, we collect the relevant explicit formulae for the practical
computation of the polarized DIS structure function $g_1(x,Q^2)$ within the
FONLL approach~\cite{Forte:2010ta} up to $\mathcal{O}(\alpha_s)$. 
In particular, we restrict to the heavy charm quark contribution, $g_{1_c}$,
to the polarized proton structure function $g_1$, 
which may be of interest for studies at an
Electron-Ion Collider, see Sec.~\ref{sec:g1charm}.
The formulae below extend to the polarized case those collected in 
Appendix~A of Ref.~\cite{Forte:2010ta}.

For $g_{1_c}$, up to $\mathcal{O}(\alpha_s)$, the relevant equations,
to be compared to Eqs.~(88)-(92) in Ref.~\cite{Forte:2010ta}, are
expressed in terms of $n_l=3$ light flavours:
\begin{eqnarray}
g_{1_c}^{\mathrm{FONLL}}(x,Q^2)
& = &
g_{1_c}^{(d)}(x,Q^2) + g_{1_{c}}^{n_l} (x,Q^2)
\,\mbox{,}
\label{eq:eq1}\\
g_{1_c}^{(d)}(x,Q^2)
& = &
g_{1_c}^{(n_l+1)}(x,Q^2) - g_{1_{c}}^{n_l,0}(x,Q^2)
\,\mbox{,}
\label{eq:eq2}\\
g_{1_c}^{(n_l+1)}(x,Q^2)
& = &
e_c^2\int_x^1\frac{dz}{z}
\left\{
C_q(z,\alpha_s(Q^2))
\Delta c^+\left(\frac{x}{z},Q^2\right)
\right.\\
& & 
\left.
+ \frac{\alpha_s(Q^2)}{4\pi} C_g^{(1)}(z)\Delta g \left(\frac{x}{z},Q^2\right)
\right\}
\,\mbox{,}
\label{eq:eq3}\\
g_{1_c}^{(n_l)}(x,Q^2)
& = &
e_c^2\frac{\alpha_s(Q^2)}{4\pi}
\int_{ax}^1\frac{dz}{z} H_g^{(1)}(Q^2,m_c^2,z)
\Delta g\left(\frac{x}{z}, Q^2 \right)
\,\mbox{,}
\label{eq:eq4}\\
g_{1_c}^{(n_l,0)}(x,Q^2)
& = &
e_c^2\frac{\alpha_s(Q^2)}{4\pi}
\int_x^1\frac{dz}{z} H_g^{(1),0}\left(\ln\left(\frac{Q^2}{m_c^2} \right),z \right)
\Delta g\left(\frac{x}{z}, Q^2 \right)
\,\mbox{,}
\label{eq:eq5}
\end{eqnarray}
where the strong coupling $\alpha_s(Q^2)$, the polarized gluon distribution
$\Delta g(x,Q^2)$ and the charm combination 
$\Delta c^+(x,Q^2)= \Delta c(x,Q^2) + \Delta\bar{c}(x, Q^2)$
are expressed in the same decoupling $n_f=n_l=3$ scheme.
The charm fractional charge squared is
$e_c^2=4/9$ and $a=1+4m_c^2/Q^2$.
The massive coefficient functions and their massless limits (labelled with 
the superscript $0$) in Eqs.~(\ref{eq:eq3})-(\ref{eq:eq5}) are taken
from Ref.~\cite{Steffens:1997da,Zijlstra:1993sh,Gluck:1990in}\footnote{We 
notice two misprintings in 
Ref.~\cite{Steffens:1997da}: Eq.~(8) should read
$sq=\sqrt{1-4\frac{z}{1-z}\frac{m_c^2}{Q^2}}$ and the first term in the 
second line of
Eq.~(12) should read $-2\frac{1+z^2}{1-z}\ln z$.} 
and read:
\begin{itemize}[leftmargin=*]
\item for the gluon
\begin{eqnarray}
C_{g}^{(1)}(z)
& = &
T_f
\left[
4(2z-1)\ln\left(\frac{1-z}{z} \right) + 4(3-4z)
\right]
\label{eq:eq6}\\
H_g^{(1)}(Q^2,m_c^2,z)
& = &
T_f
\left[
4(2z-1)\ln\left(\frac{1+\beta}{1-\beta} \right) + 4(3-4z)\beta
\right]
\label{eq:eq7}\\
H_g^{(1),0}\left(\ln\left(\frac{Q^2}{m_c^2} \right),z \right)
& = &
T_f
\left[
4(2z-1)\ln\left(\frac{1-z}{z}\frac{Q^2}{m_c^2} \right) + 4(3-4z)
\right]
\label{eq:eq8}
\end{eqnarray}
with $T_f=1/2$ and
\begin{equation}
\beta = \sqrt{1-4\frac{z}{1-z}\frac{m_c^2}{Q^2}}
\label{eq:eq9}
\end{equation}
\item for the quark
\begin{equation}
C_q(z,\alpha_s(Q^2)) = \delta(1-z) + \frac{\alpha_s(Q^2)}{4\pi} C_q^{(1)}(z)
\label{eq:eq10}
\end{equation}
with
\begin{eqnarray}
C_q^{(1)}(z)
& = &
C_F
\left\{
4\left[\frac{\ln(1-z)}{1-z}\right]_+
-3\left[\frac{1}{1-z}\right]_+-2(1+z)\ln(1-z)
\right.
\nonumber
\\
& & \left.
-2\frac{1+z^2}{1-z}\ln z +4 +2z +\delta(1-z)[-4\zeta(2)-9]
\right\}
\label{eq:eq11}
\end{eqnarray}
and $C_F=4/3$.
\end{itemize}

The formulae listed in this Appendix, should be implemented in the
\texttt{FastKernel} framework~\cite{Ball:2010de}, as already succesfully 
performed in the unpolarized case~\cite{Ball:2011mu}, in order to
gauge the impact of heavy charm flavor on polarized PDFs. As
discussed in Sec.~\ref{sec:g1charm} this may be 
interesting in future studies at an Electron-Ion Collider.

\bibliographystyle{h-elsevier3}
{\small\bibliography{Tesi}}

\backmatter


\backmatter






\end{document}